\documentclass{article}
\usepackage{springer}
\usepackage{cite}
\usepackage{psfig}
\usepackage{makeidx}

\let\secref\ref
\renewcommand{\ref}[1]{(\secref{#1})}

\newcommand{\leqapprox}{ \raisebox{-0.5ex}{$\; \stackrel{<}{\sim} \;$}}
\newcommand{\geqapprox}{ \raisebox{-0.5ex}{$\; \stackrel{>}{\sim} \;$}}

\makeindex



\def\mpicplace#1#2{\vbox{\hbox to #1{\vrule width
\fboxrule height #2\hfil}}}

\def\contentsname{Contents}

\begin{document}


\title{Bosonization of Interacting Fermions in Arbitrary Dimensions}

\author{Peter Kopietz}
\title{Bosonization of Interacting Fermions in Arbitrary Dimensions}
\subtitle{This review is a summary of my work (partially in collaboration
with Kurt Sch\"{o}nhammer) on higher-dimensional bosonization during
the years 1994-1996. It has been published as a book entitled
`` Bosonization of interacting fermions in arbitrary dimensions" by Springer Verlag
(Lecture Notes in Physics m48,
Springer, Berlin, 1997). 
}

\thispagestyle{empty}
\mbox{}
\vspace{5cm}

{\begin{center}
{\huge{\bf{
Bosonization of Interacting Fermions in Arbitrary Dimensions}
}} 

\end{center}
}

\begin{center}
{\large{\bf{Peter Kopietz}}}
\end{center}

\begin{center}
{\large{\bf{
 Institut f\"{u}r Theoretische Physik, Universit\"{a}t
  Frankfurt,  Max-von-Laue Strasse 1, 60438 Frankfurt, Germany}}}
\end{center}

\begin{center}
{\large{\bf{May 15, 2006}}}
\end{center}

This review is a summary of my work (partially  in collaboration
with Kurt Sch\"{o}nhammer) on higher-dimensional bosonization during
the years 1994-1996. It has been published as a book entitled
`` Bosonization of interacting fermions in arbitrary dimensions" by Springer Verlag
(Lecture Notes in Physics m48,
Springer, Berlin, 1997). I have not revised this review, so that
there is no reference to the literature after 1996.
However, the basic ideas underlying the functional bosonization approach
outlined in this review are still valid today.

\pagebreak

\pagenumbering{Roman}

\chapter*{Preface}

In this book we describe a new non-perturbative approach
to the fermionic many-body problem, which can be considered as 
a generalization to arbitrary dimensions
of the well-known bosonization
technique for one-dimensional fermions. 
Our approach is based on the
direct calculation of correlation functions 
of interacting Fermi systems with dominant
forward scattering
via functional integration and Hubbard-Stratonovich transformations --
we do not attempt to generalize 
one-dimensional operator identities 
between the fermionic and bosonic field operators to
higher dimensions.
The fundamental ideas of higher-dimensional bosonization have first
been formulated by A. Luther (1979) and by F. D. M. Haldane (1992).
In this book we shall go beyond these ideas 
and develop a powerful and systematic method for
bosonizing interacting fermions on a {\it{curved}} Fermi surface.
We then apply our method to a number
of problems of physical interest which are very difficult -- and in some
cases impossible --
to solve by means of conventional diagrammatic perturbation theory.

The restriction to dominant forward scattering 
means that in real space the effective interaction between the fermions
must be sufficiently long-range. Physical examples are
the Coulomb interaction  at high densities, or the effective current-current
interaction mediated by transverse gauge fields.
On the other hand, short-range interactions like the local
Hubbard-interaction cannot be directly treated within
our approach. It seems, however, that our method 
can be generalized to include (at least perturbatively)
scattering processes with large momentum transfer.
Although we shall restrict ourselves to {\it{normal}} Fermi systems,
with our functional approach 
it should be straightforward to take also
spontaneous symmetry breaking into account.
We would like to encourage interested readers to contribute
to the further development of our method.
At the end of each chapter we have therefore mentioned open 
research problems, which might be solvable
with the help of extensions of the methods developed in this book.

I would like to thank at this point everyone who -- directly
or indirectly -- has helped me to complete this book.
First of all, I am grateful to 
Kurt Sch\"{o}nhammer for  numerous collaborations and discussions,
for getting me interested in bosonization shortly after I had moved to
G\"{o}ttingen, and for giving me the freedom I needed to pursue my own ideas.
The formal development of the functional bosonization approach was 
partially carried out in collaboration with Kurt, and without him
this approach would have never been formulated in this simplicity and clarity.
More recently I have been collaborating with my friend
Guillermo Castilla, on whom I could always count whenever I needed encouragement, advise,
or help.  We communicate mainly via E-mail,  but
my information exchange with Guillermo has been
almost as intense as during our common time as graduate students at UCLA.

I am also grateful to
Sudip Chakravarty and Konstantin Efetov
for being my teachers.
Under Sudip's guidance I have learnt to do independent research.
He has taught me to distinguish interesting physics from 
empty mathematics, and his very intuitive way
of thinking about physical problems has strongly influenced 
my personal style of choosing and solving my own research projects.
I have enjoyed very much being a postdoc in 
Konstantin Efetov's international and very active group at the {\it{Max-Planck-Institut 
f\"{u}r Festk\"{o}rperforschung}} at Stuttgart.
During this time I could broaden my horizon and become familiar with
the physics of disordered Fermi systems.
I have greatly profited from Konstantin's profound knowledge in this field.

I would like to thank Peter W\"{o}lfle for
comments on the manuscript, and for
pointing out some references related to gauge fields.
In one way or the other, I have also profited 
from discussions and collaborations with
Lorenz Bartosch, 
Jim ``Claude'' Cochran,
Fabian ``Fabman'' Essler,
Jens Fricke, 
Lev Gehlhoff, 
Ralf Hannappel, 
Joachim Hermisson, 
Jens Kappey, 
Stefan Kettemann, 
Volker Meden,
Walter Metzner,
Jacob Morris, 
Ben Sauer,
Peter Scharf,
Axel V\"{o}lker,
and Roland Zeyher.

Although I sometimes tend to ignore it, I
know very well that there are more important things in life than physics.
This book is dedicated to my girlfriend Cornelia Buhrke
for helping me to keep in touch with the real world during the
nearly two years of writing, and for much more...

\vspace{5mm}

\noindent
G\"{o}ttingen, December 1996 \hfill {\it{Peter Kopietz}}

\tableofcontents

\newpage
\pagenumbering{arabic}

\part{Development of the formalism}
%
%
%

\chapter{Introduction}
\label{chap:intro}

{\it{ \ldots in which we try to explain why we have written this book.}}

\section{Perturbation theory and quasi-particles}
\label{sec:thelimitations}

{\it{Perturbation theory for the single-particle Green's function 
of an interacting Fermi system usually works  
as long as the quasi-particle picture is valid.}}

\vspace{7mm}

\noindent
The long-wavelength and low-energy behavior of the single-particle 
Green's function\index{Green's function}
$G ( {\vec{k}} , \omega )$
of an interacting  many-body system is directly related to
the nature of its ground state and low lying
excited states \cite{Matsubara55,Martin59,Abrikosov63,Kadanoff62,Nozieres64,Fetter71}.
Because the qualitative features of the low-energy 
spectrum of a many-body Hamiltonian are usually
determined by certain universal parameters such as 
dimensionality, symmetries,  and conservation laws \cite{Pines89}, 
the infrared behavior of the single-particle Green's function can be used to  
classify interacting many-body systems.
Moreover, if $G ( {\vec{k}} , \omega )$
is known for all wave-vectors ${\vec{k}}$ and frequencies $\omega$,
one can in principle calculate all thermodynamic properties of the system \cite{Fetter71}.
Unfortunately, in almost all physically interesting cases 
it is impossible to calculate the Green's function exactly, so that
one has to resort to approximate methods.
The most naive approach would be the direct expansion of $G ( {\vec{k}} , \omega )$ in
powers of the interaction. 
It is well known, however,  that even for small interactions
such an expansion is not valid for all wave-vectors and frequencies, 
because $G ( {\vec{k}} , \omega )$ usually has poles or other singularities,
in the vicinity of which a power series expansion of $G ( {\vec{k}} , \omega )$ is not possible.
In many cases this
problem can be avoided 
if one introduces the irreducible
self-energy\index{self-energy} $\Sigma ( {\vec{k}} , \omega )$ 
via the Dyson equation\index{Dyson equation},
 \begin{equation}
 [ G ( {\vec{k}} , \omega ) ]^{-1} = [ G_{0} ( {\vec{k}} , \omega ) ]^{-1} - \Sigma ( {\vec{k}} , \omega )
 \; \; \; ,
 \label{eq:Dysoneq}
 \end{equation}
and calculates $\Sigma ( {\vec{k}} , \omega )$ 
instead of $G ( {\vec{k}} , \omega )$ in powers of the interaction.
Here $G_{0} ( {\vec{k}} , \omega )$ is the Green's function of a suitably defined non-interacting system, which
can be calculated exactly.
It is important to stress that the Dyson equation does not simply express one unknown quantity 
$G ( {\vec{k}} , \omega )$ in terms of another unknown $\Sigma ( {\vec{k}} , \omega )$, but tells us
that the 
{\it{inverse}} Green's function 
should be expanded in powers of the interaction.

In so-called {\it{Landau Fermi liquids}}\index{Landau Fermi liquid}\index{Fermi liquid!Landau}  the above
perturbative approach can indeed be used to calculate
the Green's function. Of course, for strong interactions
infinite orders in perturbation theory have to be summed, but the
integrals generated in the perturbative expansion
are free of divergencies and lead to a finite expression for the self-energy.
The theory of Fermi liquids was advanced by Landau \cite{Landau56} in 1956 as
a phenomenological theory to describe the static and dynamic properties of
a large class of interacting fermions \cite{Baym91}. 
The most important physical realization of a Fermi liquid are
electrons in clean three-dimensional metals, but also
liquid ${^3}$He is a Fermi liquid \cite{Vollhardt90}.
Simultaneously with Landau's pioneering ideas the powerful 
machinery of quantum field theory was 
developed and applied to condensed matter 
systems \cite{Matsubara55,Martin59,Abrikosov63,Kadanoff62,Nozieres64}, 
and a few years later his phenomenological theory
was put on a solid theoretical basis \cite{Baym91}.
The retarded single-particle 
Green's function\index{Green's function!retarded}\footnote{
We denote the Fourier transform of the {\it{time-ordered}} Green's function 
at wave-vector ${\vec{k}}$ and frequency $\omega$ by
$G ( {\vec{k}} , \omega )$. The corresponding retarded Green's function will
be denoted by
$G ( {\vec{k}} , \omega + \I 0^{+})$, and the advanced one by
$G ( {\vec{k}} , \omega - \I 0^{+})$.}
of a Fermi liquid
is for wave-vectors ${\vec{k}}$ in the vicinity of the Fermi surface 
and small frequencies $\omega$
to a good approximation given by\index{Green's function!Landau Fermi liquid}
 \begin{equation}
 G ( {\vec{k}} , \omega + \I 0^{+} ) \approx \frac{ Z_{\vec{k}} }{ \omega -
 \tilde{\xi}_{\vec{k}}  + \I \gamma_{\vec{k}} }
 \label{eq:GFLdef}
 \; \; \; ,
 \end{equation}
where the number $Z_{\vec{k}}$ is the so-called 
{\it{quasi-particle residue}}\index{quasi-particle residue},
and the energy $\tilde{\xi}_{\vec{k}}$ 
is the single-particle excitation energy.
Because by definition Landau Fermi liquids are metals,
the excitation energy $\tilde{\xi}_{\vec{k}}$ must be gapless. 
This means that there exists a surface in ${\vec{k}}$-space 
where $\tilde{\xi}_{\vec{k}} = 0$.
In a Fermi liquid this equation can be used to {\it{define}} 
the Fermi surface\index{Fermi surface}.
The positive energy $\gamma_{\vec{k}} $ 
in Eq.\ref{eq:GFLdef} can be identified with
the quasi-particle damping\index{damping}, and is assumed to vanish faster than
$\tilde{\xi}_{\vec{k}}$ when the wave-vector ${\bf{k}}$
approaches the Fermi surface.
Note that in the complex $\omega$-plane 
$G ( {\vec{k}} , \omega + \I 0^{+} )$ has a simple
pole at $\omega = \tilde{\xi}_{ {\vec{k}} } - \I \gamma_{\vec{k}}$ 
with residue $Z_{\vec{k}}$.
Obviously,  the Green's function of
{\it{non-interacting}} fermions can be obtained as a special case of
Eq.\ref{eq:GFLdef}, namely by setting
$Z_{\vec{k}} = 1$, $\gamma_{\vec{k}} = 0^{+}$, and 
identifying
$\tilde{\xi}_{\vec{k}}$ with the non-interacting energy dispersion  
measured relative to the chemical potential.
Then
the pole at $\omega = \tilde{\xi}_{\vec{k}} - \I 0^{+}$
with unit residue
is a consequence of the undamped propagation 
of a particle with energy dispersion
$\tilde{\xi}_{\vec{k}}$ through the system.
The corresponding pole in the Green's function of an interacting Fermi liquid 
is associated with a so-called {\it{quasi-particle}}\index{quasi-particle}.
The important point is that
in the vicinity of the quasi-particle pole
the Green's function of a Fermi liquid 
has {\it{qualitatively}} the same structure
as the Green's function of free fermions.
In renormalization group language, the interacting Fermi liquid and the free Fermi gas
correspond to the same fixed point in the infinite-dimensional
parameter space spanned by all possible scattering processes \cite{Ma76,Shankar94}.
As explained in detail in Chap.~\secref{chap:abasic},
in a Landau Fermi liquid the quantities 
$Z_{\vec{k}}$, $\tilde{\xi}_{\vec{k}}$ and $\gamma_{\vec{k}}$ can be 
calculated from the derivatives of the self-energy $\Sigma ( {\vec{k}} , \omega )$.\\
\indent
In some cases, however, the application of the standard machinery of many-body
theory leads to divergent integrals 
in the perturbative expansion of $\Sigma ( {\vec{k}} , \omega )$.
The breakdown of perturbation theory is a manifestation of the fact that
the interacting Green's function is not any more
related in a simple way to the non-interacting one. 
In this case the system cannot be a Fermi liquid.
A well known example are electrons
in one spatial dimension with regular interactions, 
which under quite general conditions 
show {\it{Luttinger liquid}} behavior \cite{Solyom79,Emery79,Haldane81}\index{Luttinger liquid}.
In contrast to a Fermi liquid,
the Green's function of a Luttinger liquid 
does not have simple poles in the complex frequency plane, but 
exhibits only
branch cut singularities involving non-universal power laws\footnote{
In Chap.~\secref{sec:Green1}
we shall discuss the behavior of the Green's function of 
Luttinger liquids in some detail.}.
As a consequence, in a Luttinger liquid
$[G ( \vec{k} , \omega )]^{-1}$
cannot be calculated by simple perturbation theory around 
$[G_0 ( \vec{k} , \omega )]^{-1}$.
Hence, non-perturbative methods are necessary
to calculate the Green's function of interacting
fermions in $d=1$ dimension.
Besides the Bethe ansatz \cite{Essler94} and renormalization group methods \cite{Solyom79},
the bosonization approach has been applied 
to one-dimensional Fermi systems with great success \cite{Solyom79,Emery79,Haldane81}. 
Over the past $30$ years  
numerous interesting results have been obtained with this  non-perturbative method. 
The so-called Tomonaga-Luttinger model\index{Tomonaga-Luttinger model} is a paradigm
for an exactly solvable non-trivial many-body system which exhibits
all the characteristic Luttinger liquid 
properties, such as the absence of a quasi-particle peak in the
single-particle Green's function, anomalous scaling, and spin-charge
separation \cite{Tomonaga50,Luttinger63,Mattis65}.
Even now interesting new results on the Tomonaga-Luttinger model 
are reported in the literature \cite{Meden92,Voit93}.
For an up-to-date overview  and extensive references 
on bosonization in $d=1$ we would like to refer the reader to the recent 
reprint volume by M. Stone \cite{Stone94}.
The central topic of this book is the generalization of the
bosonization approach to arbitrary dimensions.

\section{A brief history of bosonization in $d > 1$}
\label{subsec:Abrief}

{\it{We apologize in advance if we should have forgotten someone. Maybe some Russians
have bosonized higher-dimensional Fermi systems long time ago, 
and we just don't know about their work \ldots }}

\vspace{7mm}

\noindent
The discovery of the
high-temperature superconductors and Anderson and co-workers 
suggestion \cite{Anderson90a,Anderson90b} that
the normal-state properties of these materials
are a manifestation of non-Fermi liquid behavior in dimensions $d>1$ has revived the
interest to develop non-perturbative methods for analyzing interacting fermions in 
$d > 1$.
Note, however, 
that for regular interactions in $d > 1$ 
perturbation theory is consistent in the sense that
within the framework of
perturbation theory itself
there is no signal for its breakdown \cite{Engelbrecht90,Fabrizio91}.
Nevertheless, consistency of perturbation theory
does not imply that the perturbative result must be correct.
It is therefore highly desirable to analyze interacting Fermi systems  
by means of a non-perturbative approach which does not assume
{\it{a priori}} that the system is a Fermi liquid.
The recently developed higher-dimensional generalization
of bosonization seems to be the most promising analytical
method which satisfies this criterion in $d > 1$.

In one dimension
bosonization is based on the observation that,
after proper rescaling,
the operators describing  density fluctuations 
obey canonical bosonic commutation relations \cite{Solyom79,Emery79,Haldane81}.
But also in $d=3$ density fluctuations in an interacting 
Fermi system  behave in many respects like bosonic degrees of 
freedom \cite{Bohm53,Pines61}.
The first serious attempt to formalize this observation and exploit it 
to develop a generalization of the one-dimensional bosonization approach 
to arbitrary dimensions was due to Luther \cite{Luther79}.
However, Luther's pioneering work has not received
much attention until Haldane \cite{Haldane92} added the grain of salt
that was necessary to turn higher-dimensional bosonization 
into a practically useful non-perturbative 
approach to the fermionic many-body problem.
Haldane's crucial insight was that
the degrees of freedom in the vicinity of the Fermi surface should be
subdivided into boxes of {\it{finite cross section}}, such that 
{\it{the motion of particle-hole pairs can be described without
taking momentum-transfer between different boxes into account.}}
In Luther's formulation only the motion normal to the Fermi surface
can be described in such a simple way.
The first applications of Haldane's bosonization ideas to problems
of physical interest were given by
Houghton, Marston and Kwon \cite{Houghton93}, and
independently by Castro Neto and Fradkin \cite{Castro94}.
These approaches follow closely the usual bosonization procedure
in one-dimensional systems, and 
are based on higher-dimensional generalizations
of the Kac-Moody algebra that is
{\it{approximately}}
satisfied by charge and spin current operators.
Just like in $d=1$, it is possible to map with this method
the fermionic many-body Hamiltonian onto an effective
non-interacting bosonic Hamiltonian.
The potential of these operator bosonization approaches
is certainly not yet exhausted \cite{Houghton94,Castro95}. 
However, unlike recent claims in the literature \cite{Castro95}, bosonization
in $d > 1$ is {\it{not exact}}.
For example, scattering processes that transfer momentum between different 
boxes on the Fermi surface and non-linear terms in the energy dispersion
definitely give rise to corrections to the
free-boson approximation for the Hamiltonian.
The problem of calculating these corrections within the conventional operator approach
seems to be very difficult and so far has not been solved.
\newline
\indent
In the present book we shall develop an alternative 
generalization of the bosonization approach to arbitrary dimensions,
which is based on functional integration and 
Hubbard-Stratonovich transformations.  In this way we avoid
the algebraic considerations of commutation relations
which form the basis of the operator bosonization approaches \cite{Houghton93,Castro94}.
The functional integral formulation of higher-dimensional bosonization
has been developed by the author 
in collaboration with Kurt Sch\"{o}nhammer \cite{Kopietz94}
during spring 1994.
Since then we have considerably refined this 
method \cite{Kopietz95,Kopietz95d,Kopietz96bac}
and applied it to various problems of physical interest.
A coherent and detailed presentation of these results will be given in 
this book.
A similar functional bosonization method,
which emphasizes more the mathematical aspects of bosonization,
has been developed independently by Fr\"{o}hlich and 
collaborators \cite{Frohlich94,Chen95}. 
In the context of the one-dimensional Tomonaga-Luttinger model
the functional bosonization technique has first been discussed by
Fogedby \cite{Fogedby76}, and later by Lee and Chen \cite{Lee88}.
\newline
\indent
Compared with the more conventional operator 
bosonization \cite{Houghton93,Castro94,Houghton94,Castro95},
the functional bosonization approach has several advantages.
The most important advantage is that within our functional  
integral approach it is possible 
{\it{to handle the non-linear terms in the energy dispersion}}
(and hence in $d>1$ the {\it{curvature}} of the Fermi surface).
Note that the linearization of the energy dispersion
close to the Fermi surface is one of the crucial 
(and a priori uncontrolled) approximations
of conventional bosonization; even in $d=1$
it is very difficult to calculate
systematically the
corrections due to the non-linear terms 
in the expansion of the dispersion relation
close to the Fermi surface \cite{Haldane81,Hermissondipl}.
A practically useful method for doing this will be developed
in this book.
In Chap.~\secref{chap:a4bos}
we shall explicitly calculate the leading correction 
to the free bosonized Hamiltonian and the density-density correlation function.
Moreover, in Chap.~\secref{sec:eik}
we shall show how the 
bosonization result for the
single-particle Green's function for fermions with linearized energy dispersion
is modified by the quadratic term in the expansion 
of the energy dispersion close to the Fermi surface.
In this way the approximations 
inherent in higher-dimensional bosonization become
very transparent.

\begin{sloppypar}
Another advantage of the functional integral formulation
of higher-dimensional bosonization is that
it can be applied in a straightforward way to physical problems
where non-locality and retardation are essential. 
It is well-known \cite{Feynman65} that 
these important many-body effects
can be described in the most simple
and general way via functional integrals and effective actions\index{effective action}.
In fact, the complicated effective dynamics 
of a quantum mechanical system that is coupled to another subsystem can sometimes
only be described by means of a non-local effective action, and not by a 
Hamiltonian \cite{Feynman63}.
For example, the effective retarded interaction between electrons
that is mediated via phonons or photons cannot be represented in terms of a conventional
Hamiltonian.  It is therefore advantageous to use functional integrals
and the concept of an effective action
as a basis to generalize the bosonization approach to dimensions larger than one.
%
%

Alternative formulations of higher-dimensional bosonization have
also been proposed by Schmelzer and Bishop \cite{Schmelzer93}, by
Khveshchenko and collaborators \cite{Khveshchenko94,Khveshchenko94b}, 
and by Li \cite{Li95}.
In particular, Khveshchenko \cite{Khveshchenko94b} has also developed a
formal method to include the curvature of the Fermi surface into higher-dimensional
bosonization. However, so far his method has not been proven to be useful
in practice.
We shall not further discuss the above works in this book, because we believe that
our functional bosonization technique leads to a more transparent and
practically more useful approach to the bosonization
problem in arbitrary dimensions.
Finally, it should be mentioned that
recently
Castellani, Di Castro and Metzner \cite{Castellani94,Castellani94b,Metzner95hab}
have proposed
another non-perturbative  approach to the
fermionic many-body problem in $d > 1$.
Their method is based on Ward identities and \index{Ward identity}
sums exactly the same infinite number of Feynman diagrams in the 
perturbation series as higher-dimensional bosonization
with linearized energy dispersion.  
We shall derive the precise relation between the Ward identity approach
and bosonization in Chap.~\secref{sec:ward}.
\end{sloppypar}

\section{The scope of this book}

We have subdivided this book into two parts.
Part I comprises the first five chapters and 
is devoted to the formal development of the
functional bosonization approach.
We begin by reminding the reader in
Chap.~\secref{chap:abasic} 
of some basic facts about interacting fermions. We also describe in some detail
various ways of subdividing the momentum space
in the vicinity of the Fermi surface into sectors.
These geometric constructions are the
key to the generalization of the bosonization approach
to arbitrary dimensions.
In Chap.~\secref{chap:ahub} we introduce two 
Hubbard-Stratonovich transformations which 
directly lead to the bosonization result for the
single-particle Green's function and the boson representation
of the Hamiltonian.
The explicit calculation of the bosonic Hamiltonian 
is presented in Chap.~\secref{chap:a4bos}, 
where we also show that the problem of bosonizing the Hamiltonian is
essentially equivalent with the problem of calculating the
density-density correlation function. 
We also show that the non-Gaussian terms in the
bosonic Hamiltonian are closely related to the local field corrections
to the random-phase approximation.
Chapter~\secref{chap:agreen} is devoted to the calculation of the
single-particle Green's function.
This is the most 
important chapter of this book, because here
we describe in detail our non-perturbative method for
including the non-linear terms 
in the expansion of the energy dispersion
for wave-vectors close to the Fermi surface
into the bosonization procedure.
Note that in $d > 1$ 
the local {\it{curvature}} of the Fermi surface
can only be described if the quadratic term
in the energy dispersion is retained.
Our method is based on 
a generalization of the Schwinger ansatz for the Green's function in a
given external field, an imaginary-time eikonal expansion, and
diagrammatic techniques borrowed from the theory of disordered 
systems.

In Part II we shall use our formalism 
to calculate and classify the long-wavelength and low-energy behavior of 
a number of normal fermionic quantum liquids.
In most cases we shall concentrate on
parameter regimes where conventional perturbation theory is not applicable.
In particular, we discuss fermions with singular density-density 
interactions (Chap.~\secref{chap:a7sing}), 
quasi-one-dimensional metals (Chap.~\secref{chap:apatch}), 
electron-phonon interactions (Chap.~\secref{chap:aph}),
electrons in a dynamic random medium (Chap.~\secref{chap:adis}), 
and fermions that are coupled to transverse gauge fields 
(Chap.~\secref{chap:arad}.). 
Finally, in the Appendix we summarize some useful
results on screening and collective modes in arbitrary
dimensions.

Because the method described in this book is rather new, much
remains to be done to establish higher-dimensional bosonization
as a generally accepted, practically useful non-perturbative
tool for studying strongly correlated Fermi systems.
We would like to encourage all readers to 
actively  participate in the process of further
developing this method.
For this purpose we have given at the end of each chapter
a brief summary of the main
results, together with a list of open problems and possible
directions for further research.

\section{Notations and assumptions}
\label{subsec:conv}

Let us briefly summarize the conventions that will
be used throughout this work. 
We shall measure temperature $T$ and frequencies $\omega$ in units of energy, which amounts to formally
setting the Boltzmann constant $k_{{\rm{B}}}$  and Planck's constant $\hbar$ equal to unity.
Note that in these units it is not necessary to distinguish between
wave-vectors and momenta.
The charge of the electron will be denoted by  $-e$, and the 
fine structure constant\index{fine structure constant} is
$\alpha = \frac{e^{2} }{ c} \approx \frac{1}{137}$. The velocity of light $c$ will
not be set equal to unity, because in our discussion of
transverse gauge fields in Chap.~\secref{chap:arad} it is useful to explicitly see the ratio
$v_{\rm F} / c $, where $v_{\rm F}$ is the Fermi velocity.
The inverse temperature will be denoted by $\beta = 1/T$, and the volume of the
system by $V$. Although at intermediate steps the volume of space-time
$V \beta$ will be held finite, we are eventually interested in the limits of
infinite volume $(V \rightarrow \infty$) and zero temperature ($\beta \rightarrow \infty$). 
As pointed out by Kohn, Luttinger, and Ward \cite{Kohn60},
in case of ambiguities the limit $ V \rightarrow \infty$ should be taken 
before the limit $\beta \rightarrow \infty$. However, we shall 
ignore the subtleties associated with the infinite volume limit that have recently
been discussed by Metzner and Castellani \cite{Metzner94}. 
Although we are interested in the zero-temperature limit,
we shall use the Matsubara formalism and work 
at intermediate steps at finite temperatures. 
In this way we also eliminate
possible unphysical ``anomalous'' terms \cite{Kohn60}
which sometimes appear in a zero-temperature formalism,
but are avoided if the Matsubara sums are performed at finite
temperature and the $T \rightarrow 0$ limit is carefully taken afterwards.

We shall denote bosonic Matsubara frequencies\index{Matsubara frequency} by
$\omega_{m} = 2 \pi m T$, $m =0 , \pm 1 , \pm 2 , \ldots$, and put an extra tilde over
fermionic ones,
$\tilde{\omega}_{n} = 2 \pi [ n + \frac{1}{2} ] T$,
$ n = 0 , \pm 1, \pm 2 , \ldots$. 
To simplify the notation, we introduce composite labels for
wave-vectors and Matsubara frequencies:
$k \equiv [ {\vec{k}} , \I \tilde{\omega}_{n} ]$,
$q \equiv [ {\vec{q}} , \I {\omega}_{m} ]$,
and
$\tilde{q} \equiv [ {\vec{q}} , \I \tilde{\omega}_{n} ]$.
Note that the label $q$ is associated with bosonic frequencies,
whereas $k$ and $\tilde{q}$ involve fermionic frequencies.

%
%

%
%
%

\chapter{Fermions and the Fermi surface}
\setcounter{equation}{0}
\label{chap:abasic}

{\it{
We summarize some basic facts
about interacting fermions and introduce notations that will be 
used throughout this book. 
We also describe Haldane's 
way of partitioning the Fermi surface into patches and
generalize it such that the curvature of the
Fermi surface can be taken into account.
}}

\section{The generic many-body Hamiltonian}
\label{sec:Thegeneric}

{\it{We first introduce the many-body Hamiltonian
for interacting fermions and point out 
some subtleties associated with ultraviolet cutoffs.}}

\vspace{7mm}

\begin{figure}
\sidecaption
\psfig{figure=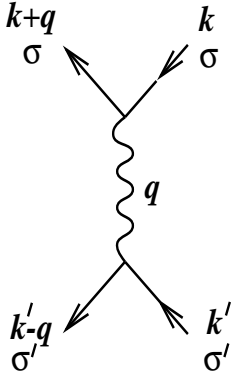,width=3cm,height=5cm}
\caption[Feynman diagram representing a generic two-body interaction]{
Feynman diagram representing the interaction
$f_{\vec{q}}^{ {\vec{k}} \sigma {\vec{k}}^{\prime} \sigma^{\prime} }$
in Eq.\ref{eq:H1def}.}
\label{fig:Feynmanscatt}
\end{figure}

\noindent
The starting point of conventional many-body theory
is a second-quantized Hamiltonian\index{Hamiltonian!second-quantized} of the form
 \begin{eqnarray}
 \hat{H}_{\rm mat} & = &
\hat{H}_{0} + \hat{H}_{\rm int}
\; \; \; ,
\label{eq:Hdef}
\\
\hat{H}_{0} & = & \sum_{\vec{k} } \sum_{\sigma} \epsilon_{\vec{k}} \hat{\psi}^{\dagger}_{\vec{k} \sigma }
\hat{\psi}_{\vec{k} \sigma }
\; \; \; ,
\label{eq:H0def}
\\
\hat{H}_{\rm int} & = & \frac{1}{2 {{V}}} 
\sum_{  \vec{q} \vec{k}  \vec{k}^{\prime}  }
\sum_{\sigma \sigma^{\prime} }
 f_{\vec{q}}^{\vec{k} \sigma  \vec{k}^{\prime} \sigma^{\prime} }
 \hat{\psi}^{\dagger}_{\vec{k+q} \sigma } 
 \hat{\psi}^{\dagger}_{\vec{k^{\prime}-q} \sigma^{\prime} } 
 \hat{\psi}_{\vec{k}^{\prime} \sigma^{\prime}}
 \hat{\psi}_{\vec{k} \sigma }
 \label{eq:H1def}
\; \; \; ,
\end{eqnarray}
where $\hat{\psi}_{\vec{k} \sigma}$ and
$\hat{\psi}_{\vec{k} \sigma}^{\dagger}$
are canonical annihilation  and creation
operators for fermions with wave-vector ${\vec{k}}$
and spin $\sigma$, which satisfy the anti-commutation relations
 \begin{equation}
[ \hat{\psi}_{ {\vec{k}} \sigma } ,
\hat{\psi}^{\dagger}_{ {\vec{k}}^{\prime} \sigma^{\prime} } ]_{+}
= 
 \hat{\psi}_{ {\vec{k}} \sigma } 
\hat{\psi}^{\dagger}_{ {\vec{k}}^{\prime} \sigma^{\prime} } 
+
\hat{\psi}^{\dagger}_{ {\vec{k}}^{\prime} \sigma^{\prime} } 
 \hat{\psi}_{ {\vec{k}} \sigma } 
= \delta_{ {\vec{k} }  {\vec{k}}^{\prime} } \delta_{\sigma  \sigma^{\prime} }
\label{eq:canncomm}
\; \; \; .
\end{equation}
The quantities $f_{\vec{q}}^{ \vec{k} \sigma \vec{k}^{\prime} \sigma^{\prime}}$ 
are the so-called {\it{Landau interaction parameters}}\index{Landau interaction parameter},
describing the scattering of two particles from initial states with
quantum numbers $({\vec{k}} , \sigma )$ and $({\vec{k}}^{\prime} , \sigma^{\prime} )$
into final states with quantum numbers
$({\vec{k}} + {\vec{q}} , \sigma )$ and $({\vec{k}}^{\prime} - {\vec{q}}, \sigma^{\prime} )$.
This process can be represented graphically by the Feynman diagram shown in 
Fig.~\secref{fig:Feynmanscatt}.
%
%
Quantum many-body theory is usually formulated in the
grand canonical ensemble,  where the relevant combination  is
$\hat{H}_{\rm mat} - \mu \hat{N}$.
Here $\hat{N} = \sum_{{\vec{k}} } \sum_{\sigma}
\hat{\psi}^{\dagger}_{\vec{k} \sigma }
\hat{\psi}_{\vec{k} \sigma }$ is
the particle number operator, and 
$\mu$ is the chemical potential\index{chemical potential}.
Thus, the energy dispersion $\epsilon_{\vec{k}}$ appears exclusively in the
combination
 \begin{equation}
\xi_{\vec{k}} \equiv
\epsilon_{\vec{k}} - \mu 
\; \; \; .
 \label{eq:xikdef}
 \end{equation}
The value of $\mu$ at zero temperature is also called 
the Fermi energy\index{Fermi energy} $E_{\rm F}$.
Although in most physical applications we are interested in 
three dimensions,
it is very useful and instructive to formulate
the theory in arbitrary dimension $d$.
Then the equation
 \begin{equation}
\xi_{\vec{k}} 
  = 0
 \label{eq:FSdef}
 \end{equation}
defines a $d-1$-dimensional surface in momentum space,
the {\it{non-interacting}} Fermi surface.
The precise definition of the {\it{interacting}} Fermi surface\index{Fermi surface!non-interacting} 
will be given in Sect.~\secref{sec:Thesingle}. 
Note that in $d=1$ the non-interacting Fermi surface consists 
of two distinct points $\pm k_{\rm F}$, where $k_{ \rm F}$ is the
Fermi wave-vector. In higher dimensions the Fermi surface is a $d-1$-dimensional manifold,
the topology of which depends on the form of $\xi_{\vec{k}}$.
There is actually a subtle point hidden in the above definition:
although the energy $\epsilon_{\vec{k}}$ is a parameter of the non-interacting
Hamiltonian $\hat{H}_{0}$, the chemical potential $\mu$ is by definition
{\it{the exact chemical potential of the interacting many-body system}}.
Of course, the precise value of $\mu$ remains unknown 
unless we can solve the many-body problem, but
fortunately it is not necessary to know $\mu$ in order
to calculate physical correlation functions. 
By defining $\mu$ to be the chemical potential of the interacting
many-body system, one implicitly adds a suitable counter-term \index{counter-term}
to the bare chemical potential which eliminates, order by order
in perturbation theory, all 
terms which would otherwise contribute to $\Sigma ( {\vec{k}} , 0)$ 
for wave-vectors ${\vec{k}}$ on the Fermi surface. \index{chemical potential!renormalization}
In particular, all Feynman diagrams of the Hartree type are cancelled by the counter-term.
Such a procedure is familiar from perturbative quantum 
field theory \cite{ZinnJustin89}. 
The consistency for such a construction is by no means obvious, and has
recently been questioned by Anderson \cite{Anderson93}.
For a thorough discussion and partial solution of this problem see  
\cite{Salmhofer96}.

It should be emphasized that
Eqs.\ref{eq:Hdef}--\ref{eq:H1def}
can be interpreted in three distinct ways, 
which can be classified according to the effective ultraviolet cutoff
for the wave-vector sums.

(a) {\it{Homogeneous electron gas.}}\index{homogeneous electron gas}
First of all, we may define
$\hat{H}_{\rm mat}$ to be the Hamiltonian of
the homogeneous electron gas in $d$ dimensions. 
For example the 
Coulomb-interaction in $d=3$ dimensions corresponds to
$\epsilon_{\vec{k}} =  {\vec{k}}^2/({2m})$ and
 $f_{\vec{q}}^{\vec{k} \sigma  \vec{k}^{\prime} \sigma^{\prime} } = 
 { 4 \pi e^2}/{ {\vec{q}}^2 }$,
where $m$ is the mass of the electrons.
In this case there is no intrinsic short-distance cutoff for the wave-vector sums. 
%

(b) {\it{Relevant band of a lattice model.}}
Because in realistic materials the electrons feel the periodic potential
due to the ions, the allowed energies in the absence of interactions are subdivided
into energy bands, and the interaction has interband matrix elements.
But if there exists only a single band in the vicinity of the Fermi surface, then
it is allowed to ignore all other bands as long as one is interested in
energy scales small compared with the interband gap. 
In this case the Hamiltonian defined in
Eqs.\ref{eq:Hdef}--\ref{eq:H1def} 
should be considered as
the {\it{effective Hamiltonian}}\index{Hamiltonian!effective}
for the band
in the vicinity of the Fermi energy.
In this model the wave-vector sums have a cutoff of the order of
$ { 2 \pi}/{a}$, where $a$ is the distance between the ions. 
The energy dispersion $\epsilon_{\vec{k}}$ in Eq.\ref{eq:H0def} incorporates 
then by definition the effects of the underlying lattice, 
which in general leads also to a renormalization of the effective 
mass of the electrons.
\newline
\indent
(c) {\it{Effective Hamiltonian for degrees of freedom close to the
Fermi surface.}}
Finally, we may define $\hat{H}_{\rm mat}$ to be the 
effective Hamiltonian for the low-energy degrees of freedom
in the vicinity of the Fermi surface, assuming that all degrees of freedom
outside a thin shell with radial thickness $\lambda \ll k_{\rm F}$ 
have been integrated out via functional integration and renormalization
group methods \cite{Shankar94}.
Of course, the operation of integrating out the high-energy degrees of freedom 
will also generate three-body and higher order interactions, which are
ignored in Eqs.\ref{eq:Hdef}--\ref{eq:H1def}.
The quantities $\epsilon_{\vec{k}}$
and $f_{\vec{q}}^{\vec{k} \sigma  \vec{k}^{\prime} \sigma^{\prime} }$
should then be considered as effective parameters, which take the
finite renormalizations due to the high-energy degrees of freedom into account.
In this picture the ${\vec{k}}$- and ${\vec{k}}^{\prime}$-sums in 
Eqs.\ref{eq:H0def} and \ref{eq:H1def} are
confined to a thin shell of thickness $\lambda$ around the Fermi surface, while 
the ${\vec{q}}$-sum in Eq.\ref{eq:H1def} is restricted to the regime
$| {\vec{q}} | \leq \lambda$.


All three interpretations of the many-body Hamiltonian \ref{eq:Hdef}--\ref{eq:H1def}
are useful. First of all, the model (a) has the advantage that it contains no
free parameters, so that it can be the starting point of a 
first principles microscopic calculation. 
The model (b) is more realistic, 
although the effects of the underlying lattice are only included on a phenomenological level.
Finally, the model (c) has the advantage that it contains explicitly only the low-energy degrees
of freedom close to the Fermi surface, so that, to a first approximation,
we may locally linearize the energy
dispersion at the Fermi surface. 
Evidently the model (c) cannot be used
for the calculation of the precise numerical value of
physically measurable quantities that depend 
on fluctuations on all length scales.
Furthermore, the integration over the 
degrees of freedom far away 
from the Fermi surface 
usually cannot be explicitly carried out.

\section{The single-particle Green's function}
\label{sec:Thesingle}

{\it{We define the single-particle 
Green's function\index{Green's function} and the Fermi surface\index{Fermi surface} 
of an interacting Fermi system. 
We then discuss in some detail the low-energy behavior of the Green's function
in a Landau Fermi liquid.}}

\vspace{7mm}

\noindent
Because in the rest of this book the spin degree of freedom will
not play any role, we shall
from now on simply ignore the spin index.
Formally, the spin is easily taken
into account by defining ${\vec{k}}$ and ${\vec{k}}^{\prime}$ to be collective
labels for wave-vector and spin.
For practical calculations we prefer
to work with the Matsubara formalism, because
in this way we avoid the problem of 
regularizing formally divergent integrals by means of pole prescriptions,
which arises in the real time zero-temperature formulation of quantum many-body theory.
Furthermore, the Matsubara Green's function  
can be represented as an imaginary time functional integral 
\cite{Popov83,Popov87,Negele88,Kapusta89},
so that the entire many-body problem can be reformulated in the language of
path integrals.
In this work we shall make extensive use of this modern 
approach to the many-body problem. 

\subsection{Definition of the Green's function}

The single particle Matsubara Green's function\index{Green's function!Matsubara}
$G ( k )$ of an interacting Fermi  system 
is defined by
 \begin{eqnarray}
 \hspace{-3mm}
 G ( k ) \equiv G ( {\vec{k}} , \I \tilde{\omega}_{n} )  & = &
 -
 \frac{1}{\beta  } \int_{0}^{\beta}  \D \tau 
 \int_{0}^{\beta} \D \tau^{\prime}
 \E^{ - \I \tilde{\omega}_{n}  ( \tau - \tau^{\prime}) }
 < {\cal{T}} \left[ \hat{\psi}_{\vec{k}} ( \tau )
 \hat{\psi}^{\dagger}_{ {\vec{k} } }  ( \tau^{\prime} ) \right] >
 \; ,
 \nonumber
 \\
 & &
 \label{eq:GMatsubaradef}
 \end{eqnarray}
where for fermions the time-ordering\index{time-ordering!fermionic} operator ${\cal{T}}$ 
in imaginary time is defined by
 \begin{eqnarray}
  {\cal{T}} \left[ \hat{\psi}_{\vec{k}} ( \tau )
 \hat{\psi}^{\dagger}_{ {\vec{k} } }  ( \tau^{\prime} ) \right] 
 & = & \Theta ( \tau - \tau^{\prime} - 0^{+})
  \hat{\psi}_{\vec{k}} ( \tau )
 \hat{\psi}^{\dagger}_{ {\vec{k} } }  ( \tau^{\prime} ) 
 \nonumber
 \\
 & - & \Theta ( \tau^{\prime} - \tau + 0^{+})
 \hat{\psi}^{\dagger}_{ {\vec{k} } }  ( \tau^{\prime} ) 
  \hat{\psi}_{\vec{k}} ( \tau )
  \; \; \; ,
  \label{eq:timeordF}
  \end{eqnarray}
and the average in Eq.\ref{eq:GMatsubaradef} denotes grand canonical 
thermal average with respect to all degrees of freedom
in the system.
For any operator $\hat{O}$ the time evolution
in imaginary time is defined by
 \begin{equation}
 \hat{O} ( \tau ) = \E^{ \tau ( \hat{H}_{\rm mat} - \mu \hat{N} ) }
\hat{O} \E^{ - \tau ( \hat{H}_{\rm mat} - \mu \hat{N} ) }
\label{eq:Ohatev}
\; \; \; ,
\end{equation}
where $\hat{H}_{\rm mat}$ is given in Eqs.\ref{eq:Hdef}--\ref{eq:H1def}.
The Matsubara Green's function of a system of non-interacting 
fermions with Hamiltonian $\hat{H}_{0}$ (see Eq.\ref{eq:H0def}) is given by
 \begin{equation}
 G_{0} ( k ) = \frac{1}{ \I \tilde{\omega}_{n}  - \xi_{\vec{k}} }
 \label{eq:G0Matdef}
 \; \; \; ,
 \end{equation}
where the subscript $_{0}$ indicates the absence of interactions.
Once the imaginary-frequency Green's function is known,
we can obtain the corresponding retarded zero-temperature Green's function
by  analytic continuation\index{analytic continuation} 
in the complex frequency plane
just above the real axis,
$ \I \tilde{\omega}_{n} \rightarrow \omega + \I 0^{+}$.
For the non-interacting retarded Green's function\index{Green's function!retarded} we obtain
 \begin{equation}
 G_{0} ( {\vec{k}} , \omega + \I 0^{+} ) = 
 \frac{1}{ \omega  - \xi_{\vec{k}} + \I 0^{+} }
 \label{eq:G0def}
 \; \; \; .
 \end{equation}
This function has a pole
at $\omega = \xi_{\vec{k}} - \I 0^{+}$ with residue $Z_{\vec{k}} = 1$. 
The infinitesimal imaginary part shifts the pole below the real axis, so that
the retarded Green's function is analytic in the upper half of the
complex frequency plane \cite{Nozieres64,Fetter71}. The corresponding advanced 
Green's function\index{Green's function!advanced} 
$G_{0} ( {\vec{k}} , \omega - \I 0^{+} )$ is analytic
in the lower half of the frequency plane, while the
time-ordered Green's function\index{Green's function!time-ordered},
 \begin{equation}
 G_{0} ( {\vec{k}} , \omega ) = 
 \frac{1}{ \omega  - \xi_{\vec{k}} +  \I 0^{+} {\rm sgn} ( \omega )  }
 \; \; \; ,
 \end{equation}
agrees for
$\omega > 0$ with the retarded Green's function, and for
$\omega < 0$ with the advanced one.
The analytic structure of the
time-ordered Green's function $G ( {\vec{k}} , \omega )$
of the interacting many-body system
is similar \cite{Nozieres64,Fetter71}: It has cuts above the real negative axis and below
the real positive axis, a branch point at $\omega = 0$, and poles in the neighboring
Riemann sheets.
The simple pole structure of 
the non-interacting Matsubara Green's function
$G_{0} ( k )$ makes the analytic continuation trivial.
In general it can be quite difficult to perform
the analytic continuation of the interacting Matsubara Green's function
to obtain the corresponding real frequency function.
Nevertheless, we prefer to work with the Matsubara formalism, because
Euclidean time-ordering leads to the very simple
result \ref{eq:G0Matdef} for the non-interacting Green's function.
Note that the denominator in Eq.\ref{eq:G0Matdef} can never vanish,
so that we avoid in this way the singular integrands 
with poles on the real frequency axis
that appear in a zero-temperature formalism.

\subsection{Definition of the interacting Fermi surface}
\label{subsec:defFSint}

We define the Fermi surface of an interacting Fermi system\index{Fermi surface!interacting} 
as the set of points in momentum space where,
in the limit of zero temperature,
the momentum distribution\index{momentum distribution}
$n_{\vec{k}}$ has some kind of non-analyticity. 
The momentum distribution
can be expressed in terms 
of the exact Matsubara Green's function as
 \begin{equation}
 n_{\vec{k}} = \frac{1}{\beta} \sum_{n} G ( {\vec{k}} , \I \tilde{\omega}_{n} )
 \; \; \; .
 \label{eq:nkdef}
 \end{equation}
In the absence of interactions we have
 $n_{\vec{k}} = f ( \xi_{\vec{k}} )$,
where
 \begin{equation}
 f (  E ) = \frac{1}{\beta} \sum_{n} \frac{ 1 }{ \I \tilde{\omega}_{n} - E }
 = \frac{1}{ \E^{\beta E} + 1 }
 \label{eq:fermifuncdef}
 \end{equation}
is the Fermi function\footnote{The Matsubara sum in Eq.\ref{eq:fermifuncdef}
is formally divergent, and should be regularized by 
inserting a convergence factor $\E^{\I \tilde{\omega}_n 0^{+} }$, see
\cite{Fetter71}.}\index{Fermi function}.
In the  zero-temperature limit $f ( E ) \rightarrow \Theta ( - E )$, so that
the momentum distribution reduces to a step function,
 $n_{\vec{k}} = \Theta ( - \xi_{\vec{k}} ) $.
Because $\Theta ( x )$ is not analytic at $x = 0$, 
we recover in the absence of interactions 
the definition of the non-interacting Fermi surface
given in Eq.\ref{eq:FSdef}.
We would like to emphasize that it is by no means clear that
the momentum distribution
of an interacting Fermi system 
has always non-analyticities.
In fact, in Chap.~\secref{subsec:thelim} we shall
give an example for a quantum liquid where $n_{\vec{k}}$ is analytic.
In this case the interacting system simply does not have a sharp Fermi surface.
To avoid misunderstandings, we shall from now on
reserve the word {\it{Fermi surface}}
for the non-interacting Fermi surface, as defined
in Eq.\ref{eq:FSdef}.

\subsection{Landau Fermi liquids\index{Landau Fermi liquid}\index{Fermi liquid!Landau}}
\label{subsec:LandauFL}

\begin{figure}[h]
\sidecaption
\psfig{figure=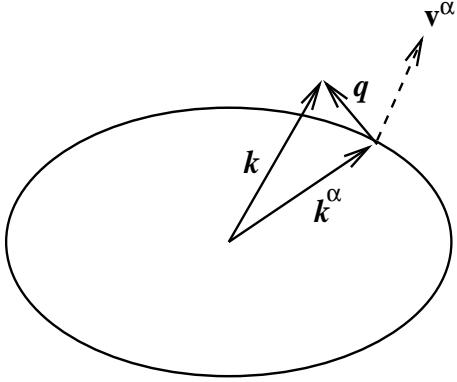,width=6cm,height=5cm}
\caption[Local coordinate system centered on the Fermi surface.]
{Local coordinate system centered at point $ {\vec{k}}^{\alpha}$ 
on an elliptic Fermi surface. Note that 
in general the local Fermi velocity\index{Fermi velocity} ${\vec{v}}^{\alpha}$ is not
parallel to ${\vec{k}}^{\alpha}$.
}
\label{fig:geomlocal}
\end{figure}

As already mentioned in Chap.~\secref{chap:intro},
for wave-vectors ${\vec{k}}$ sufficiently close to the Fermi surface
and sufficiently small energies, 
the Green's function of a {\it{Landau Fermi liquid}}
has qualitatively the same pole structure as the non-interacting Green's function.
The pole represents an elementary excitation of the
system which approximately behaves like a free particle.
This is the {\it{quasi-particle}}\index{quasi-particle}.
To formulate the quasi-particle concept in
precise mathematical language,
consider a point ${\vec{k}}^{\alpha}$ on the Fermi surface 
(i.e. $\xi_{\vec{k}^{\alpha}} = 0$), and let us measure
wave-vectors locally with respect to this point. 
The geometry is shown in Fig.~\secref{fig:geomlocal}.
%
%
The energy dispersion is then given by 
 \begin{equation}
  \xi^{\alpha}_{\vec{q}} \equiv \xi_{ \vec{k}^{\alpha} + \vec{q} }
  \label{eq:xialphadefdef}
  \; \; \; .
  \end{equation} 
We now expand for small ${\vec{q}}$,
 \begin{equation}
 \xi^{\alpha}_{ {\vec{q}}} = {\vec{v}}^{\alpha} \cdot {\vec{q}} +
 \frac{1}{2} \sum_{ij = 1}^{d } q_{i} c_{ij}^{\alpha} q_{j} + O (  | \vec{q} |^3 )
 \label{eq:xialphaqexp}
 \; \; \; , 
 \end{equation}
where
 \begin{equation}
 {\vec{v}}^{\alpha}  =    
 \left. \nabla_{\vec{k}}
 \epsilon_{\vec{k}} \right|_{\vec{k} = {\vec{k}}^{\alpha} }
 \; \; \; , \; \; \; 
  c_{ij}^{\alpha} = \left. \frac{\partial^2 \epsilon_{\vec{k}}}{\partial k_{i} \partial k_{j}} 
   \right|_{\vec{k} = \vec{k}^{\alpha} }
  \label{eq:v0cijdef}
  \; \; \; .
  \end{equation}
Similarly, we expand the retarded self-energy\index{self-energy} 
$\Sigma ( {\vec{k}} , \omega + \I 0^{+} )$
defined in Eq.\ref{eq:Dysoneq},
 \begin{eqnarray}
 \Sigma ( {\vec{k}}^{\alpha} + {\vec{q}} , \omega + \I 0^{+} )
 & =  & {\vec{q}} \cdot  
 \left. \nabla_{\vec{k}} \Sigma ( {\vec{k}} , \I 0^{+})
 \right|_{ {\vec{k}} = {\vec{k}}^{\alpha} }
   + 
 \omega  
 \left. \frac{\partial \Sigma ( {\vec{k}}^{\alpha} , \omega + \I 0^{+} ) }{\partial \omega }
 \right|_{\omega = 0} 
 \nonumber
 \\
 & + &
 \delta{\Sigma}^{\alpha} ( {\vec{q}} , \omega + \I 0^{+} )
 \; \; \; ,
 \label{eq:sigmaexp}
 \end{eqnarray}
where in a Fermi liquid
$\delta{\Sigma}^{\alpha} ( {\vec{q}} , \omega + \I 0^{+} )$ is by assumption for small
${\vec{q}}$ and $ \omega $ quadratic in these quantities.
Note that in this expansion we have set
 \begin{equation}
 \Sigma ( {\vec{k}}^{\alpha} , \I 0^{+} ) = 0
 \label{eq:sigmachempot}
 \; \; \; ,
 \end{equation}
assuming that the chemical potential\index{chemical potential}
$\mu $
is chosen such that Eq.\ref{eq:sigmachempot} is satisfied for all
points ${\vec{k}}^{\alpha}$ on the Fermi surface.
As already mentioned in Sect.~\secref{sec:Thegeneric}, this is a 
non-trivial assumption \cite{Anderson93,Salmhofer96}.

\begin{center}
{\bf{The quasi-particle residue\index{quasi-particle residue}}}
\end{center}
\noindent
Substituting Eq.\ref{eq:sigmaexp} into the Dyson equation \ref{eq:Dysoneq}, we see that 
the Green's function of the interacting system can be written as
 \begin{equation}
 G ( {\vec{k}}^{\alpha} + {\vec{q}} , \omega + \I 0^{+}) = \frac{Z^{\alpha}}
 { \omega -  \xi^{\alpha}_{\vec{q}} - \delta{{\vec{v}}}^{\alpha} \cdot {\vec{q}} 
 -   Z^{\alpha} \delta \Sigma^{\alpha} ( {\vec{q }} , \omega + \I 0^{+} ) }
 \label{eq:Gdysonexp}
 \; \; \; ,
 \end{equation}
where the so called {\it{quasi-particle residue}} $Z^{\alpha}$ is given by
 \begin{equation}
 Z^{\alpha}  =  \frac{1}
 {1 -
 \left. \frac{\partial \Sigma ( {\vec{k}}^{\alpha} , \omega + \I 0^{+} ) }{\partial \omega }
 \right|_{\omega = 0}  }
 \; \; \; ,
 \label{eq:Zdef}
 \end{equation}
and the renormalization of the  Fermi velocity at point ${\vec{k}}^{\alpha}$ is
 \begin{eqnarray}
 \delta {\vec{v}}^{\alpha}  & = & (Z^{\alpha} - 1 ) {\vec{v}}^{\alpha}
 + Z^{\alpha}
 \left.
 \nabla_{\vec{k}} 
 \Sigma ( {\vec{k}} , \I 0^{+} ) \right|_{ {\vec{k}} = {\vec{k}}^{\alpha} }
 \label{eq:valpharen2}
 \; \; \; .
 \end{eqnarray}
Thus, the effective Fermi velocity at ${\vec{k}}^{\alpha}$ is
 \begin{eqnarray}
 \tilde{\vec{v}}^{\alpha}  =  {\vec{v}}^{\alpha} + \delta {\vec{v}}^{\alpha} & = &
 Z^{\alpha} \left[ 
  {\vec{v}}^{\alpha}
 + 
 \left.
 \nabla_{\vec{k}} 
 \Sigma ( {\vec{k}} , \I 0^{+} ) \right|_{ {\vec{k}} = {\vec{k}}^{\alpha} }
 \right]
 \nonumber
 \\
 & = & 
 \frac{
  {\vec{v}}^{\alpha}
 + 
 \left.
 \nabla_{\vec{k}} 
 \Sigma ( {\vec{k}} , \I 0^{+} ) \right|_{ {\vec{k}} = {\vec{k}}^{\alpha} }}{
 1 -
 \left. \frac{\partial \Sigma ( {\vec{k}}^{\alpha} , \omega + \I 0^{+} ) }{\partial \omega }
 \right|_{\omega = 0}  }
 \; \; \; .
 \label{eq:valpharen}
 \end{eqnarray}
The finite temperature generalization of Eq.\ref{eq:Zdef} 
is \cite{Serene91} 
 \begin{equation}
 Z^{\alpha} ( T )  =  \frac{1}
 {1 -
  \frac{{\rm Im} \Sigma ( {\vec{k}}^{\alpha} ,  \I  \tilde{\omega}_{0} ) }{  \tilde{ \omega}_{0} }
  }
 \; \; \; ,
 \label{eq:ZdefT}
 \end{equation}
where $\tilde{\omega}_{0} = \pi T$ is the zeroth fermionic Matsubara frequency. 
The quasi-particle residue determines at $T = 0$ the discontinuity of the
momentum distribution\index{momentum distribution!discontinuity} $n_{\vec{k}}$ 
when $\vec{k}$ crosses the Fermi surface.
To calculate the change in 
the momentum distribution at point ${\vec{k}}^{\alpha}$ on the
Fermi surface, consider
 \begin{equation}
 \delta n^{\alpha}_{ {\vec{q}} } =
 n_{ \vec{k}^{\alpha} - {\vec{q}} } -
 n_{ \vec{k}^{\alpha} + {\vec{q}} } 
 \; \; \; .
 \label{eq:nkdiscondef}
 \end{equation}
For small enough ${\vec{q}}$ 
we may approximate $\xi^{\alpha}_{\vec{q}} \approx {\vec{v}}^{\alpha} \cdot {\vec{q}}$ and
ignore the correction term 
$\delta \Sigma^{\alpha}$
in Eq.\ref{eq:Gdysonexp}. At finite temperatures we obtain then
 \begin{equation}
 \delta n^{\alpha}_{ {\vec{q}} } =
 Z^{\alpha} ( T) \left[ 
 f ( - \tilde{{\vec{v}}}^{\alpha} \cdot {\vec{q}}  )
   -
 f (  \tilde{{\vec{v}}}^{\alpha} \cdot {\vec{q}}  ) \right]
 \; \; \; .
 \label{eq:momentumZ}
 \end{equation}
In the  zero-temperature limit $f ( E ) \rightarrow \Theta ( - E )$  and 
$Z^{\alpha} ( T ) \rightarrow Z^{\alpha}$, so that
 $\delta n^{\alpha}_{ {\vec{q}} } =
 Z^{\alpha}  {\rm sgn} (  \tilde{\vec{v}}^{\alpha} \cdot {\vec{q}} )
 $.
%
Note that $\delta n^{\alpha}_{\vec{q}}$ depends only on
the  projection of ${\vec{q}}$ that is normal to the Fermi surface, 
because this corresponds to a crossing
of the Fermi surface and can thus give rise to a discontinuity.

\begin{center}
{\bf{
The effective mass\index{effective mass!renormalization} renormalization}}
\end{center}
\noindent
If $ \left. \nabla_{\vec{k}} \Sigma ( {\vec{k}} , \I 0^{+} ) 
 \right|_{\vec{k} = {\vec{k}}^{\alpha} }$ is parallel to
${\vec{v}}^{\alpha}$ (for example, for
spherical Fermi surfaces and rotationally invariant interactions
this is the case), 
we see from Eq.\ref{eq:valpharen} that
the renormalized 
Fermi velocity $\tilde{ \vec{v}}^{\alpha}$ associated with
point ${\vec{k}}^{\alpha}$ on the Fermi surface can be written as
 \begin{equation}
 \tilde{{\vec{v}}}^{\alpha}
 =
  Z_{\rm m}^{\alpha} {\vec{v}}^{\alpha}
 \label{eq:Zmdef}
 \; \; \; ,
 \end{equation}
where the {\it{effective mass renormalization factor}}
 $Z_{\rm m}^{\alpha}$ is given by
 \begin{eqnarray}
 Z^{\alpha}_{\rm m} & = & Z^{\alpha}
 \left[ 1 +   
 \frac{ \hat{\vec{v}}^{\alpha} \cdot
  \nabla_{\vec{k}} 
 \left.
 \Sigma ( {\vec{k}} , \I 0^{+} ) \right|_{ {\vec{k}} = {\vec{k}}^{\alpha} }}{ | {\vec{v}}^{\alpha} | }
 \right]
 \nonumber
 \\
 & = & \frac{
  1 + 
 \frac{ \hat{\vec{v}}^{\alpha} \cdot
  \nabla_{\vec{k}} 
 \left.
 \Sigma ( {\vec{k}} , \I 0^{+} ) \right|_{ {\vec{k}} = {\vec{k}}^{\alpha} }}{ | {\vec{v}}^{\alpha} | }
  }
 {1 -
 \left. \frac{\partial \Sigma ( {\vec{k}}^{\alpha} , \omega + \I 0^{+} ) }{\partial \omega }
 \right|_{\omega = 0}  }
 \label{eq:Zmres}
 \; \; \; ,
 \end{eqnarray}
with ${\hat{\vec{v}}}^{\alpha} = {\vec{v}}^{\alpha} / | {\vec{v}}^{\alpha} |$.
At finite temperatures, Eq.\ref{eq:Zmres} should again 
be generalized as follows,
 \begin{equation}
 Z^{\alpha}_{\rm m} (T ) = 
  \frac{
  1 + 
 \frac{ \hat{\vec{v}}^{\alpha} \cdot
  \nabla_{\vec{k}} 
 \left. {\rm{Re }}
 \Sigma ( {\vec{k}} , \I \tilde{\omega}_{0}  ) \right|_{ {\vec{k}} = {\vec{k}}^{\alpha} }}{ | {\vec{v}}^{\alpha} | }
  }
 {1 -
  \frac{ {\rm Im } \Sigma ( {\vec{k}}^{\alpha} , \I \tilde{\omega}_{0}  ) }{\tilde{\omega}_{0} }
 }
 \label{eq:ZmresT}
 \; \; \; .
 \end{equation}
The effective mass $\tilde{m}^{\alpha}$ is defined in terms of the
bare mass $m$ via $ \tilde{m}^{\alpha} \tilde{\vec{v}}^{\alpha} = {m} {\vec{v}}^{\alpha}$, so that
 \begin{equation}
 Z^{\alpha}_{\rm m} = \frac{m}{ \tilde{m}^{\alpha}  }
 = \frac{ | \tilde{{\vec{v}}}^{\alpha} | }{ | {\vec{v}}^{\alpha} | }
 \; \; \; .
 \label{eq:Zmbare}
 \end{equation}
In other words, 
a {\it{small}} value of $Z^{\alpha}_{\rm m}$ corresponds to a 
{\it{large}} effective mass.
One of the fundamental properties of a Fermi liquid is that
the renormalization factors $Z^{\alpha}$ and $Z^{\alpha}_{\rm m}$ are 
finite\footnote{
This working definition is sufficient for most physically 
interesting systems, although in some rather special cases it is not
accurate enough. For example, if we retain
only so-called $g_4$-processes in the one-dimensional
Tomonaga-Luttinger model \cite{Tomonaga50,Luttinger63} with spin and set $g_2 =0$ 
(the spinless model is discussed in Chap.~\secref{sec:Green1}), 
then $Z^{\alpha}$ and $Z^{\alpha}_{\rm m}$ are finite,
but the Green's function exhibits spin-charge
separation\index{spin-charge separation}, which does not occur in Fermi liquids.
I would like to thank Walter Metzner for pointing this out to me.}.

\begin{center}
{\bf{The quasi-particle 
damping\index{damping!quasi-particle}\index{quasi-particle damping}}}
\end{center}
\noindent
%
Eq.\ref{eq:Gdysonexp} is formally exact, provided 
Eq.\ref{eq:sigmaexp} is taken as the {\it{definition}} of
$\delta \Sigma^{\alpha} ( {\vec{q}} , \omega + \I 0^{+} )$, expressing one unknown 
quantity in terms of another one.
Of course, this parameterization is only useful if 
the correction $\delta \Sigma^{\alpha}$ becomes negligible, at least for wave-vectors 
in the vicinity of the Fermi surface.
In Landau Fermi liquids\index{Landau Fermi liquid!self-energy} 
$\delta \Sigma^{\alpha} ( {\vec{q}} , \omega + \I 0^{+} )$
is by assumption analytic, so that
for small ${\vec{q}} $ and for frequencies $\omega$
close to $\tilde{{\vec{v} }}^{\alpha} \cdot {\vec{q}}$ we may approximate
 \begin{equation}
 Z^{\alpha} 
\delta \Sigma^{\alpha} ( {\vec{q}} , \omega + \I 0^{+} ) \approx
 Z^{\alpha} 
\delta \Sigma^{\alpha} ( {\vec{q}} , \tilde{{\vec{v}}}^{\alpha} \cdot {\vec{q}} + \I 0^{+} ) 
\approx  \frac{1}{2} \sum_{ij=1}^{d} q_{i} \delta c^{\alpha}_{ij} q_{j}
\label{eq:poleclose}
\; \; \; ,
\end{equation}
where $ \delta c^{\alpha}_{ij}$ is a complex matrix that is determined by the various second partial
derivatives of the self-energy.
Defining the renormalized  second-derivative matrix
 \begin{equation}
 \tilde{c}_{ij}^{\alpha} = c_{ ij}^{\alpha} + \delta c_{ij}^{\alpha}
 \; \; \; ,
 \end{equation}
the real part of the renormalized energy dispersion for wave-vectors close to
${\vec{k}}^{\alpha}$ is
 \begin{equation}
 \tilde{\xi}^{\alpha}_{\vec{q}}  =  \tilde{{\vec{v}}}^{\alpha} \cdot {\vec{q}} +
 \frac{1}{2} 
 \sum_{ij=1}^{d} q_{i} [ {\rm Re} \tilde{c}^{\alpha}_{ij}] q_{j}
 + O (  | \vec{q} |^3 )
 \label{eq:tildexialphadef}
 \; \; \; .
 \end{equation}
Although $c_{ij}^{\alpha}$ is real, the matrix $\delta c_{ij}^{\alpha}$ is 
in general complex, so that
the renormalized energy dispersion acquires an imaginary part due to the 
interactions. Defining
 \begin{equation}
 \gamma^{\alpha}_{\vec{q}}  = 
 - \frac{ 1}{2} 
 \sum_{ij=1}^{d} q_{i} [ {\rm Im} \delta c^{\alpha}_{ij}] q_{j}
 \label{eq:gammaalphadef}
 \; \; \; ,
 \end{equation}
the interacting retarded Green's function\index{Green's function!Landau Fermi liquid} 
of the many-body system  
is for sufficiently small ${\vec{q}}$  and $\omega$
 given by
 \begin{equation}
 G ( {\vec{k}}^{\alpha} + {\vec{q}} , \omega + \I 0^{+})
 \approx \frac{ Z^{\alpha}}{ \omega - \tilde{{\xi}}^{\alpha}_{\vec{q}} + \I \gamma^{\alpha}_{\vec{q}}
 }
 \label{eq:GQP}
 \; \; \; ,
 \end{equation}
which is equivalent\footnote{
Recall that
wave-vectors are now measured with respect to the local coordinate
system centered at ${\vec{k}}^{\alpha}$ on the Fermi surface,
so that in Eqs.\ref{eq:GFLdef} and \ref{eq:GQP}
we should identify $ Z_{ {\vec{k}}^{\alpha}} = Z^{\alpha}$,
$ \tilde{\xi}_{ {\vec{k}}^{\alpha} + {\vec{q}}} =
\tilde{\xi}^{\alpha}_{\vec{q}} $, and
$\gamma_{ {\vec{k}}^{\alpha} + {\vec{q}} } =
\gamma^{\alpha}_{\vec{q}}$.}
with Eq.\ref{eq:GFLdef}. 
This expression has a pole in the complex frequency plane at
$\omega = \tilde{\xi}^{\alpha}_{\vec{q}} - \I \gamma^{\alpha}_{\vec{q}}$
with residue given by $Z^{\alpha}$.
By contour integration \cite{Fetter71} it is easy to see that the
pole contribution to the real time Fourier transform of the Green's function is
 \begin{eqnarray}
 G ( {\vec{k}}^{\alpha} + {\vec{q}} , t ) & = & 
 \int_{- \infty}^{\infty}  \frac{ \D \omega}{2 \pi} \E^{ - \I \omega t }
 G ( {\vec{k}}^{\alpha} + {\vec{q}} , \omega + \I 0^{+})
 \nonumber
 \\
 & = & - \I \Theta ( t ) Z^{\alpha} 
 \E^{ - \I \tilde{\xi}^{\alpha}_{\vec{q}} t } \E^{- \gamma^{\alpha}_{ {\vec{q}} } t }
 \label{eq:Gqreal}
 \; \; \; .
 \end{eqnarray}
If the damping
$\gamma^{\alpha}_{\vec{q}}$ is small compared with the
real part
 $\tilde{\xi}^{\alpha}_{\vec{q}} $
of the energy,
then the behavior of the interacting Green's function is, up to times of 
order $1/ \gamma^{\alpha}_{\vec{q}}$, qualitatively 
similar to the behavior  of the non-interacting Green's function.
The pole is therefore said to 
represent a {\it{quasi-particle}}\index{quasi-particle}.
Actually, at times shorter than $1/ \tilde{\xi}^{\alpha}_{\vec{q}}$ 
it is not allowed to keep only the pole contribution in
Eq.\ref{eq:Gqreal}, so that quasi-particle behavior can only be
observed in the intermediate time domain \cite{Fetter71}
 \begin{equation}
 1/ \tilde{\xi}^{\alpha}_{ \vec{q} }  \ll t \leqapprox 1 / \gamma^{\alpha}_{\vec{q}}
 \label{eq:QPsee}
 \; \; \; .
 \end{equation}

\section{The density-density correlation function}
\label{sec:Thedensity}

{\it{
We define the density-density correlation function $\Pi ( q )$, 
the dynamic structure factor $S ( {\vec{q}} , \omega )$, and the dielectric function
$\epsilon ( q )$ of an interacting
Fermi system. We also explain what is meant 
by ``random-phase approximation''.}}

\vspace{7mm}

\noindent
Besides the single-particle Green's function, we are interested in the
density-density correlation function\index{density-density correlation function}
$ \Pi ( q ) \equiv \Pi ( {\vec{q}} , \I \omega_{m} )$,
which is for $ {\vec{q}} \neq 0$ defined by\footnote{
At ${\vec{q}} = 0$ we should subtract from the time-ordered product in
Eq.\ref{eq:cordens} the 
term $< \hat{\rho}_{\vec{q}} ( \tau ) > 
< \hat{\rho}_{\vec{-q}} ( \tau^{\prime} ) > $, which in a
translationally invariant system vanishes for any ${\vec{q}} \neq 0$.
Because in the present work we are only interested in
the ${\vec{q}} \neq 0$ part of
the density-density correlation function, we shall omit this term.}
 \begin{equation}
 \Pi ( q )  
  =  \frac{1}{\beta V } \int_{0}^{\beta}  \D \tau 
 \int_{0}^{\beta} \D \tau^{\prime}
 \E^{ - \I \omega_{m}  ( \tau - \tau^{\prime}) }
 \left< {\cal{T}} \left[ \hat{\rho}_{\vec{q}} ( \tau )
 \hat{\rho}_{ -{\vec{q} } }  ( \tau^{\prime} ) \right] \right>
 \nonumber
 \label{eq:cordens}
 \; \; \; ,
 \end{equation}
where the operator
 \begin{equation}
\hat{\rho}_{\vec{q}} = \sum_{\vec{k}} \hat{\psi}^{\dagger}_{\vec{k}} 
\hat{\psi}_{\vec{k}+\vec{q}}
\label{eq:denstotop}
\end{equation}
represents the Fourier components of the total 
density, and ${\cal{T}}$ denotes {\it{bosonic}} time-ordering\index{time-ordering!bosonic}, 
i.e. 
 \begin{eqnarray}
  {\cal{T}} \left[ \hat{\rho}_{\vec{q}} ( \tau )
 \hat{\rho}_{ -{\vec{q} } }  ( \tau^{\prime} ) \right] 
 & = & \Theta ( \tau - \tau^{\prime} - 0^{+})
   \hat{\rho}_{\vec{q}} ( \tau )
 \hat{\rho}_{ -{\vec{q} } }  ( \tau^{\prime} )  
 \nonumber
 \\
 & + &  \Theta ( \tau^{\prime} - \tau + 0^{+})
 \hat{\rho}_{ -{\vec{q} } }  ( \tau^{\prime} )  
   \hat{\rho}_{\vec{q}} ( \tau )
  \; \; \; .
  \label{eq:timeordB}
  \end{eqnarray}
Note that, in contrast to Eq.\ref{eq:timeordF}, there is no
minus sign associated with a permutation, so that
$\Pi ( q )$ depends on
bosonic  Matsubara frequencies $\omega_{m} = 2 \pi m T$.
We shall also refer to $\Pi ( q )$ as the {\it{polarization function}}\index{polarization}, or
simply the {\it{polarization}}.
In the absence of interactions $\Pi ( q )$ reduces to the imaginary frequency 
Lindhard function\index{Lindhard function},
 \begin{equation}
 \Pi_{0}
 ( q )
 = - \frac{1}{\beta V } \sum_{k} G_{0} ( k ) G_{0} ( k + q )
    =   - \frac{1}{ {V}} \sum_{\vec{k}}
\frac{ f (  \xi_{\vec{k+q}}  )
 - f (  \xi_{\vec{k}} ) }
{ \xi_{\vec{k+q}} - \xi_{\vec{k}} - \I \omega_{m} }
 \label{eq:Pi0total}
 \; \; \; .
 \end{equation} 
The corresponding real frequency function can be obtained
via analytic continuation\index{analytic continuation}.
The discontinuity of $\Pi ( {\vec{q}} , z )$ across
the real axis defines the {\it{dynamic structure 
factor}}\index{dynamic structure factor} 
\index{polarization!relation to dynamic structure factor}
$S ( {\vec{q}} , \omega )$ \cite{Pines89}
 \begin{equation}
 {\rm Im} \Pi ( {\vec{q}} ,  \omega + \I 0^{+} ) =
 \pi \left[  S ( {\vec{q}} , \omega )
 -  S ( {\vec{q}} , - \omega ) \right]
 \label{eq:Srealaxis}
 \; \; \; .
 \end{equation}
In terms of the exact eigenstates $|n \rangle$ and eigen-energies $E_{n}$
of the operator $\hat{H}_{\rm mat} - \mu \hat{N}$
defined in Eqs.\ref{eq:Hdef}--\ref{eq:H1def},
$S ( {\vec{q}} , \omega )$ 
has the spectral representation
 \begin{equation}
 S ( {\vec{q}} , \omega ) = 
 \frac{1}{V} \sum_{nm} \frac{\E^{- \beta E_{m} }}{ \cal{Z}} 
 | \langle n| \hat{\rho}_{\vec{q}}^{\dagger} | m \rangle |^2 \delta ( \omega -
 (E_{n} - E_{m} ) )
 \label{eq:Sstrucspec}
 \; \; \; ,
 \end{equation}
where ${\cal{Z}}$ is the exact grand canonical partition function.
From this expression  it is obvious that $S ( {\vec{q}} , \omega )$ is real and positive,
and satisfies the detailed balance condition\index{detailed balance}
 \begin{equation}
 S ( {\vec{q}} , - \omega ) = \E^{- \beta \omega } S ( {\vec{q}} , \omega )
 \; \; \; .
 \end{equation}
Using
$ \frac{1}{ 1 - \E^{- \beta \omega} } = 1 + \frac{1}{\E^{\beta \omega} - 1}$,
it is easy to see that the imaginary part of $\Pi ( {\vec{q}} , \omega + \I 0^{+})$ and the
dynamic structure factor are related via 
 \begin{equation}
 S ( {\vec{q}} , \omega ) = 
 \left[ 1 + \frac{1}{\E^{\beta \omega} - 1 } \right]
 \frac{1}{\pi} {\rm Im } \Pi ( {\vec{q}} , \omega + \I 0^{+} )
 \; \; \; .
 \label{eq:SPi}
 \end{equation}
This relation is called the {\it{fluctuation-dissipation 
theorem}}\index{fluctuation-dissipation theorem}.
For arbitrary complex frequencies $z$ we have \cite{Pines89}, 
 \begin{eqnarray}
 \Pi ( {\vec{q}} , z ) & =  &
 \frac{1}{\pi} \int_{- \infty}^{\infty} \D \omega 
 \frac{ {\rm{Im}} \Pi ( {\vec{q}} , \omega + \I 0^{+} )}{ \omega - z }
 \nonumber
 \\
 & = &
 \int_{0}^{\infty} \D \omega  [ 1 - \E^{ - \beta \omega } ] S ( {\vec{q}} , \omega )
 \left[ \frac{1}{  \omega - z } + \frac{1}{  \omega + z } \right]
 \nonumber
 \\
 & = & \int_{0}^{\infty} \D \omega 
 [ 1 - \E^{ - \beta \omega } ] 
 S ( {\vec{q}} , \omega )
  \frac{ 2 \omega }{  \omega^{2} - z^2 } 
 \label{eq:dynstruc}
 \; \; \; .
 \end{eqnarray}
  
A widely used approximation for the density-density correlation function
is the so-called {\it{random-phase approximation}} \cite{Mattuck67}, which
we shall abbreviate by RPA\index{random-phase approximation}.
If the quasi-particle interaction in Eq.\ref{eq:Hdef}
depends only on the
momentum-transfer ${\vec{q}}$ (and not on the
momenta $\vec{k}$ and $\vec{k}^{\prime}$ of the incoming particles),
the density-density correlation function
within RPA is approximated by
 \begin{equation}
 \Pi_{\rm RPA} ( q) = \frac{ \Pi_{0} ( q )}{ 1 + f_{\vec{q}} \Pi_{0} ( q ) }
 \label{eq:PiRPA}
 \; \; \; ,
 \end{equation}
or equivalently
 \begin{equation}
 [ \Pi_{\rm RPA} ( q ) ]^{-1} = [ \Pi_{0} (q ) ]^{-1} + f_{\vec{q}}
 \label{eq:PiRPA2}
 \; \; \; .
 \end{equation}
Corrections to the RPA are usually parameterized in terms
of a local field correction\index{local field correction} $g ( q )$, which is defined
by writing the exact $\Pi ( q )$ as \cite{Mahan81,Gorabchenko89} 
 \begin{equation}
 [ \Pi ( q ) ]^{-1} = [ \Pi_{0} (q ) ]^{-1} + f_{\vec{q}} - g ( q )
 \label{eq:Piexact2}
 \; \; \; .
 \end{equation}
Defining the {\it{proper polarization}}\index{polarization!proper polarization} 
$\Pi_{\ast} ( q )$ via
 \begin{equation}
 [ \Pi_{\ast} ( q ) ]^{-1} = [ \Pi_{0} (q ) ]^{-1}   - g ( q )
 \label{eq:Piproper}
 \; \; \; ,
 \end{equation} 
we have
 \begin{equation}
 \Pi ( q ) = \frac{ \Pi_{\ast} (q ) }{ 1 + f_{\vec{q}} \Pi_{\ast} ( q ) }
 = \frac{ \Pi_{\ast} ( q ) }{ \epsilon ( q ) }
 \label{eq:PiDyson}
 \; \; \; ,
 \end{equation}
where the dimensionless function
 \begin{equation}
 \epsilon ( q ) = 1 + f_{\vec{q}} \Pi_{\ast} ( q )
 \label{eq:dielectricdef}
 \end{equation}
is called the {\it{dielectric function}}\index{dielectric function}.
Using Eqs.\ref{eq:PiDyson} and \ref{eq:dynstruc}, we may also write
 \begin{equation}
 \frac{1}{\epsilon (q )} = 1 - f_{{\vec{q}}} \Pi ( q ) =
 1 - f_{{\vec{q}}}
 \int_{0}^{\infty} \D \omega [ 1 - \E^{- \beta \omega } ] S ( {\vec{q}} , \omega )
  \frac{ 2 \omega }{  \omega^{2} + \omega_{m}^2 } 
  \; \; \; .
  \end{equation}
Note that Eq.\ref{eq:Piproper} has
the structure $G^{-1} = G_{0}^{-1} - \Sigma$, i.e. it resembles the
Dyson equation for the single-particle Green's function of a bosonic problem, with the
proper polarization and the local field factor playing the role of the exact Green's function
and the irreducible self-energy.
Although this analogy has been noticed many times in the 
literature \cite{Mahan81,Gorabchenko89,Singwi68,Rajagopal72,Dharma76,Holas79}, 
it has not been thoroughly exploited as a guide to develop 
systematic methods to calculate corrections to the RPA. 
In Chap.~\secref{chap:a4bos} we shall show that
higher-dimensional bosonization gives a natural explanation for this analogy
and yields a new procedure for 
calculating the dielectric function beyond the RPA.

\section{Patching the Fermi surface\index{Fermi surface}}
\label{subsec:patch}

{\it{ 
We now discuss
Haldane's version of subdividing the 
degrees of freedom close to the Fermi surface into
boxes. This geometric construction
opens the way for generalizing 
the bosonization approach to arbitrary dimensions.}}

\subsection{Definition of the patches\index{patch} and boxes\index{squat box}}

{\it{This leads to the bosonization of the potential energy in arbitrary dimensions.}}

\vspace{7mm}

\begin{figure}
\sidecaption
\psfig{figure=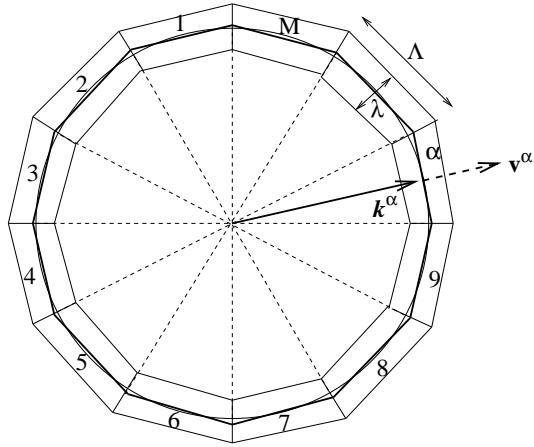,width=7cm}
\caption[Subdivision of a two-dimensional Fermi surface into patches\index{patch}]
{\begin{sloppypar}
Subdivision of a two-dimensional spherical Fermi surface into $M=12$ patches
${P}^{\alpha}_{\Lambda}$, $ \alpha = 1 , \ldots , 12$,
and associated boxes  $K^{\alpha}_{\Lambda , \lambda}$.
The vector ${\vec{k}}^{\alpha}$ has length $k_{\rm F}$ and points to the center of 
the patch ${P}^{\alpha}_{\Lambda}$. The dashed arrow represents the local Fermi velocity
${\vec{v}}^{\alpha}$ 
associated with patch ${P}^{\alpha}_{\Lambda}$.
\end{sloppypar}
} 
\label{fig:patch2dexample}
\end{figure}

\noindent
Following Haldane \cite{Haldane92}, we partition 
the Fermi surface into a finite number $M$ of
disjoint patches with volume $\Lambda^{d-1}$. \index{patch} 
The precise shape of the patches and the size of $\Lambda$ should be chosen
such that, to a first approximation, within a given patch the curvature of the Fermi surface can
be locally neglected. 
We introduce a label $ \alpha $ to enumerate the patches in some convenient ordering and 
denote the patch with label $\alpha$ by ${P}^{\alpha}_{\Lambda}$.
For example, a possible subdivision of a two-dimensional Fermi surface
into $M=12$ patches is shown in Fig.~\secref{fig:patch2dexample}.
\begin{figure}
\sidecaption
\psfig{figure=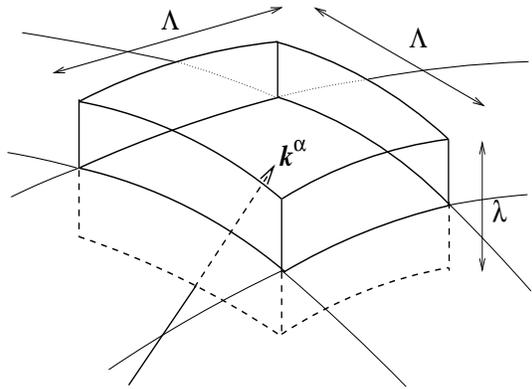,width=7cm,height=5.1cm}
\caption[A squat box\index{box!squat box} on a three-dimensional Fermi surface]
{Graph of a squat box
$K^{\alpha}_{\Lambda , \lambda}$ with patch cutoff $\Lambda$ and 
radial cutoff $\lambda$ in three dimensions.
${\vec{k}}^{\alpha}$ points to the center
of patch ${P}^{\alpha}_{\Lambda}$ on the Fermi surface.
}
\label{fig:patch}
\end{figure}
\noindent
By definition ${P}^{\alpha}_{\Lambda}$ is 
a subset of the Fermi surface, i.e.
a $d-1$-dimensional manifold.
To cover the entire momentum space in the vicinity
of the Fermi surface,\index{patch} \index{box} \index{sector}
each patch ${P}^{\alpha}_{\Lambda}$ is extended
into a $d$-dimensional box (or sector) $K^{\alpha}_{\Lambda , \lambda}$ 
such that the union $\bigcup_{\alpha=1}^{M} K^{\alpha}_{\Lambda , \lambda }$ 
comprises all degrees of freedom in the 
system. 

The definition of the boxes requires the introduction of an additional
radial cutoff $\lambda$. 
If we assume that the degrees of freedom with wave-vectors outside
a thin shell of radial thickness $\lambda \ll k_{\rm F}$ 
around the Fermi surface have been integrated out, then 
in two dimensions the sectors $K^{\alpha}_{\Lambda , \lambda}$
can be chosen in form of the trapezoids shown 
in Fig.~\secref{fig:patch2dexample}, while in $d=3$
a convenient choice of
the  $K^{\alpha}_{\Lambda , \lambda}$ are the squat boxes shown in
Fig.~\secref{fig:patch}.
The difference between Haldane's \cite{Haldane92} 
and Luther's \cite{Luther79} way of subdividing the degrees  
of freedom close to the Fermi surface is that Luther 
takes $\Lambda = O ( V^{- 1/d} ) $, so that his sectors
are actually thin tubes, with a cross section that covers only
a few discrete ${\vec{k}}$-points.
This has the obvious disadvantage that
the motion {\it{parallel to the Fermi surface}} cannot
be described without taking scattering between different
tubes into account.
Haldane's crucial idea was to choose boxes with finite cross section.
In this case scattering processes
that transfer momentum between different sectors\footnote{These
so-called around-the-corner processes  
are difficult to handle
within higher-dimensional bosonization, 
see Sect.~\secref{subsec:proper}\index{around-the-corner processes}
and Chap.~\secref{subsec:invdiag}.}
can be ignored
{\it{as long as the width $\Lambda$ of the boxes
is large compared with the typical momentum-transfer $| {\vec{q}} |$ of the interaction.}}

To bosonize the potential energy,\index{bosonization!potential energy}
we decompose the Fourier components $\hat{\rho}_{ {\vec{q}} }$
of the density operator into the contributions from the various boxes, 
 \begin{equation}
 \hat{ \rho}_{ {\vec{q}} }
 = \sum_{\alpha = 1}^{M}
 \hat{\rho}^{\alpha}_{\vec{q}} 
 \label{eq:rhototsum}
 \; \; \; ,
 \end{equation}
where $\hat{\rho}^{\alpha}_{\vec{q}} $
is the contribution from wave-vectors $\vec{k}$ in
sector $K^{\alpha}_{\Lambda , \lambda }$ 
to the total density, 
 \begin{equation}
 \hat{\rho}^{\alpha}_{\vec{q}} = \sum_{\vec{k}} \Theta^{\alpha} ( {\vec{k}} )
 \hat{\psi}^{\dagger}_{\vec{k}} \hat{\psi}_{\vec{k} + \vec{q} }
 \label{eq:rhoopdef}
 \; \; \; .
 \end{equation}
The cutoff function\index{cutoff!function} $\Theta^{\alpha} ( {\vec{k}} )$ is defined by
 \begin{equation}
 \Theta^{\alpha} ( {\vec{k}} ) =
 \left\{
 \begin{array}{cl}
 1 & \mbox{ if ${\vec{k}} \in K^{\alpha}_{\Lambda , \lambda}$ } \\
 0 & \mbox{ else }
 \end{array}
 \right.
 \; \; \; ,
 \label{eq:Thetadef}
 \end{equation}
and satisfies
\index{patch} \index{box} \index{sector}
 \begin{equation}
 \sum_{\alpha = 1}^{M} \Theta^{\alpha} ( {\vec{k}} ) = 1
 \label{eq:Thetasum}
 \; \; \; ,
 \end{equation}
because by construction
the union of all $K^{\alpha}_{\Lambda , \lambda}$  
agrees with
the total relevant ${\vec{k}}$-space. 
We shall refer to 
$\hat{\rho}^{\alpha}_{\vec{q}} $
as {\it{sector density}}. \index{sector!density}
Note that Eq.\ref{eq:Thetasum} insures that the sum of
all sector densities 
yields again the full density $\hat{\rho}_{\vec{q}}$, see Eq.\ref{eq:rhototsum}.  
In terms of the sector density operators the
interaction part \ref{eq:H1def}
of the many-body Hamiltonian can be written as
 \begin{equation}
 \hat{H}_{\rm int} = \frac{ 1}{2V} \sum_{\vec{q}}  \sum_{ \alpha \alpha^{\prime} }
 {{f}}_{\vec{q}}^{\alpha \alpha^{\prime}}  
 : \hat{\rho}_{ - \vec{q}}^{\alpha} \hat{\rho}_{\vec{q}}^{\alpha^{\prime}} :
 \label{eq:Hintcourse}
 \; \; \; ,
 \end{equation}
where $: \ldots :$ denotes normal ordering,
and it is assumed that the variations of $f^{ {\vec{k}} {\vec{k}}^{\prime} }_{\vec{q}}$ 
are negligible if
${\vec{k}}$ and ${\vec{k}}^{\prime}$ are restricted to given boxes,
so that it is allowed to introduce coarse-grained interaction functions
 \begin{equation}
 {{f}}_{\vec{q}}^{ \alpha \alpha^{\prime}} = 
 \frac{ \sum_{ {\vec{k}} {\vec{k}}^{\prime}  }
  \Theta^{\alpha} ( {\vec{k}} )
  \Theta^{\alpha^{\prime} } ( {\vec{k}}^{\prime} )
  f^{ \vec{k k^{\prime}} }_{\vec{q}}
  }{ 
  \sum_{ {\vec{k}}  {\vec{k}}^{\prime} }  
  \Theta^{\alpha} ( {\vec{k}} )
  \Theta^{\alpha^{\prime} } ( {\vec{k}}^{\prime} )
  }
   \label{eq:Landaufuncdef}
   \; \; \; .
 \end{equation} 
The motivation for introducing 
the operators $\hat{\rho}^{\alpha}_{\vec{q}}$ is that
these operators obey approximately 
(up to an overall  scale factor)
bosonic commutation relations among each other \cite{Houghton93,Castro94}.
{\it{Thus, Eq.\ref{eq:Hintcourse} is already the bosonized potential energy.}}

It should be mentioned that the usefulness of the geometric construction
described above is not restricted to higher-dimensional bosonization.
A very similar construction has recently been used by Feldman {\it{et al.}} \cite{Feldman93} 
to devise a ${1}/{M}$-expansion for 
interacting Fermi systems. Furthermore,
sectorizations of this type play an important role
in modern renormalization group approaches 
to the fermionic many-body problem \cite{Chen95}.

\subsection{Linearization\index{linearization of energy dispersion} 
of the energy dispersion}
\label{subsec:locallin}

{\it{
In order to bosonize the full Hamiltonian, we should also
obtain a boson representation for the kinetic energy.
This is only possible in a simple way \index{bosonization!kinetic energy}
if the energy dispersion is linearized at the Fermi surface.}}

\vspace{7mm}
\index{patch} \index{box} \index{sector}

\noindent
The crucial advantage of the subdivision of the Fermi surface into patches is
that it opens the way for a {\it{linearization of the non-interacting energy dispersion}}. 
In first-quantized notation this means
that the kinetic
energy operator $\hat{H}_{0}$ is replaced by 
an operator involving only first order spatial derivatives.
Then it is not difficult to show
that the operators $\hat{\rho}^{\alpha}_{\vec{q}}$
defined in Eq.\ref{eq:rhoopdef} have in the high-density limit
simple commutation relations with the kinetic energy operator $\hat{H}_{0}$. 
Together with the bosonized potential energy in Eq.\ref{eq:Hintcourse},
this directly 
leads to the free boson representation of the Hamiltonian \cite{Houghton93,Castro94}.
In Chap.~\secref{chap:a4bos} we shall discuss
the bosonization of the Hamiltonian and the
underlying approximations within the framework of our
functional integral approach.

\begin{figure}[t]
\sidecaption
\psfig{figure=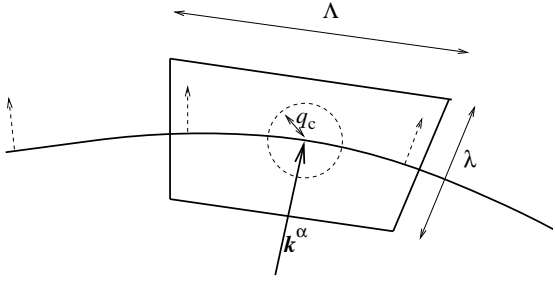,width=8.4cm}
\caption[Proper choice of the cutoffs.]
{Proper choice of the cutoffs. The patch cutoff
$\Lambda$ should be chosen small enough so that within a given box
the variations in the direction of the local normal vectors 
to the Fermi surface (dashed arrows) can be neglected.
On the other hand,
both cutoffs $\Lambda$ and $\lambda$ should be chosen large
compared with the range $q_{\rm c}$ of the interaction in momentum space
(dashed  circle).
Only in this case physical correlation functions 
at distances $ | {\vec{r}} | \geqapprox q_{c}^{-1}$
do not explicitly depend on the unphysical cutoffs $\Lambda$ and $\lambda$.
} 
\label{fig:qc}
\end{figure}

Due to the non-trivial topology of the Fermi surface,
it is impossible to linearize the energy dispersion
globally in a fixed coordinate system. 
However, if the size of the patches is chosen sufficiently small,
we may {\it{locally}} linearize the energy dispersion within each sector separately.
To do this,
let us denote by $ {\vec{k}}^{\alpha} $, $\alpha = 1, \ldots , M$, 
the set of vectors on the 
Fermi surface ($\xi_{ {\vec{k}}^{\alpha} }= 0 $)
that point to the  centers of the patches\index{patch} ${P}^{\alpha}_{\Lambda}$ 
(see Figs.~\secref{fig:patch2dexample} and \secref{fig:patch}).
Let us then identify 
the vectors ${\vec{k}}^{\alpha} $ 
with the origins 
of local coordinate systems on the Fermi surface\footnote{Such a collection
of coordinate systems is also called an {\it{atlas}}\index{atlas} \cite{Nash83}.},
and measure any given wave-vector $\vec{k}$ with respect to that
reference point $\vec{k}^{\alpha}$ for which $ | \vec{k} - \vec{k}^{\alpha} |$ 
assumes a minimum.
The corresponding geometry has already been discussed in 
Sect.~\secref{subsec:LandauFL}, see Fig.~\secref{fig:geomlocal}.
Formally, we use Eq.\ref{eq:Thetasum} and write
 \begin{equation}
 \epsilon_{\vec{k}} - \mu \equiv \xi_{\vec{k}} 
 = \sum_{\alpha} 
 \Theta^{\alpha} ( {\vec{k}} ) 
  \xi_{\vec{k}}
 =
 \sum_{\alpha} 
 \Theta^{\alpha} ( {\vec{k}} ) 
 \xi^{\alpha}_{ {\vec{k}} - {\vec{k}}^{\alpha}  }
 \label{eq:partition}
 \; \; \; ,
 \end{equation}
\index{patch} \index{box} \index{sector}
where the functions $\xi^{\alpha}_{\vec{q}}$ are simply defined 
by $\xi^{\alpha}_{\vec{q}} = \xi_{ {\vec{k}}^{\alpha} + {\vec{q}} }$,
see Eqs.\ref{eq:xikdef} and \ref{eq:xialphadefdef}.
Suppose now that the cutoff $\Lambda$ 
is chosen sufficiently small so that within a given patch the 
curvature of the Fermi surface can be neglected.
As shown in Fig.~\secref{fig:qc}, this means that the variations in the direction of the local
normal vector to the Fermi surface must be small within a given patch.
In general $\Lambda$ should be chosen small compared
with the typical momentum scale on which the Fermi surface
changes its shape.
For spherical Fermi surfaces this means that
 \begin{equation}
 \Lambda \ll k_{\rm F}
 \label{eq:Lambdaup}
 \; \; \; .
 \end{equation}
On the other hand, for intrinsically flat Fermi surfaces  
the size of $\Lambda$ can be chosen comparable to $k_{\rm F}$.
We shall discuss Fermi surfaces of this type in some detail
in Chap.~\secref{chap:apatch}.
In the opposite limit, 
when the Fermi surface has certain critical areas where 
its shape changes on some other characteristic scale $k_{0} \ll k_{\rm F}$,  we should choose
$\Lambda \ll k_{0}$.
Note that in the case of
Van Hove singularities\index{Van Hove singularities} $k_{0} \rightarrow 0$, 
so that we have to exclude this possibility 
if we insist on the linearization of the energy dispersion.
For sufficiently small $ | {\vec{q}} | = | {\vec{k}}  - {\vec{k}}^{\alpha} |$ we may then
ignore the quadratic and higher order corrections in
Eq.\ref{eq:xialphaqexp}, and approximate
 \begin{equation}
 \xi^{\alpha}_{\vec{q}} \approx   {\vec{v}}^{\alpha} \cdot  {\vec{q}}  
 \; \; \; .
 \label{eq:xilin}
 \end{equation}
Note that for energy dispersions 
that are intrinsically almost linear\footnote{
For example,
for some peculiar form of the band structure 
the coefficients $c^{\alpha}_{ij}$ in Eq.\ref{eq:v0cijdef} might be small.}
the quadratic corrections to Eq.\ref{eq:xilin}
are small even for $| {\vec{q}} | = O( k_{\rm F})$.  In most cases, however,
Eq.\ref{eq:xilin} will only be a good approximation
for the calculation of quantities that are determined by the
degrees of freedom in the vicinity of the Fermi surface.

\subsection{Around-the-corner processes and the proper choice \hspace{20mm}
of the cutoffs\index{cutoff!choice of sector cutoffs}
\index{around-the-corner processes}}
\label{subsec:proper}

{\it{The sector cutoffs $\Lambda$ and $\lambda$ 
should not be chosen too small, but also not too large. The proper choice depends on the
shape of the Fermi surface and on the nature of the interaction.}}

\vspace{7mm}
\index{patch} \index{box} \index{sector}

\noindent
Although the variations in the direction of the local
normal vector can always be
reduced by choosing a sufficiently small patch cutoff $\Lambda$, 
this cutoff cannot be made arbitrarily small. The reason is that
for practical calculations the sectorization
turns out to be only useful if scattering processes
that transfer momentum between different boxes 
(so called {\it{around-the-corner processes}}\index{around-the-corner processes}) 
can be neglected.
This will only be the case if 
the Fourier transform of the interaction 
is dominated by momentum-transfers $ | {\vec{q}} | \leqapprox q_{\rm c} $, where
$q_{\rm c}$ is some physical interaction cutoff satisfying\index{cutoff!interaction}
 \begin{equation}
 q_{\rm c}  \ll {\rm min} \{ \Lambda , \lambda \}
 \label{eq:qccond}
 \; \; \; .
 \end{equation}
In other words, the interaction must be dominated by forward scattering\index{forward scattering}.
As illustrated in Fig.~\secref{fig:qc}, the volume 
in momentum space 
swept out by the interaction
is then small compared with the 
volume $\Lambda^{d-1} \lambda$ of the boxes, so that
boundary effects can be neglected.
For example, 
in case of the long-range part of the Coulomb potential
the cutoff $q_{\rm c}$  can be identified with the usual
Thomas-Fermi screening wave-vector\index{Thomas-Fermi wave-vector} $\kappa$. In this case the condition
$\kappa \ll k_{\rm F}$ is satisfied at high densities
(see Chap.~\secref{subsec:Cbnice} and
Appendix~\secref{subsubsec:Cb}).
Of course, the Coulomb potential has also a non-vanishing short-range part, 
which cannot be treated explicitly within our bosonization approach.
Fortunately, there exist physically interesting quantities (for example the quasi-particle residue or 
the leading behavior of the momentum distribution in the vicinity of the Fermi surface,
see Chap.~\secref{sec:singular}) which are completely determined
by long-wavelength fluctuations 
with wave-vectors $ | {\vec{q}}  | \leqapprox \kappa$. In this case our bosonization approach
leads to cutoff-independent results that involve only physical quantities,
because the condition \ref{eq:qccond} insures that
the numerical value of momentum integrals
is independent of the unphysical cutoffs $\Lambda$ and $\lambda$.

Finally, let us consider the radial cutoff $\lambda$.
If we would like to linearize the energy dispersion, then we should choose
$\lambda$ small enough such that
it does not matter whether the energy dispersion is linearized
precisely at the Fermi surface, or at the
top (or bottom) of the boxes $K^{\alpha}_{\Lambda , \lambda}$.
For a spherical Fermi surface this 
condition is satisfied if
 \begin{equation}
 \lambda \ll k_{\rm F}
 \label{eq:lambdaup}
 \; \; \; .
 \end{equation}
However, by introducing such a radial cutoff
we are assuming that the high-energy degrees of freedom
have already been integrated out.
As discussed in Sect.~\secref{sec:Thegeneric},
the parameters which define our  model (such as
the local Fermi velocities ${\vec{v}}^{\alpha}$ or the
physical cutoff $q_{\rm c}$) must then incorporate the 
finite renormalizations due to the high-energy degrees of freedom.
Therefore these parameters
{\it{depend implicitly on the
cutoff $\lambda$}}.
Although the precise form of this cutoff dependence remains unknown
unless we can explicitly perform the integration over the
high-energy degrees of freedom, 
these physical parameters can in principle be determined
from experiments,  
for example by measuring the density of states at the Fermi energy or the
screening length.
Such a procedure is familiar
from renormalizable quantum field theories, where all cutoff
dependence can be lumped onto a finite number of experimentally
measurable parameters \cite{ZinnJustin89,Sterman93}.
But also in field theory approaches to condensed matter systems this strategy has been
adopted with great success \cite{Chakravarty89}.

\section{Curved patches and reduction of the patch number}
\label{sec:sectors}

{\it{If we do not require that the energy dispersion 
should be linearized, we are free to subdivide the
Fermi surface into a small number of curved patches.
In some special cases
we may completely abandon the patching construction, and formally
identify the entire momentum space with a single sector.
Then the around-the-corner processes 
simply do not exist.}}

\begin{figure}
\psfig{figure=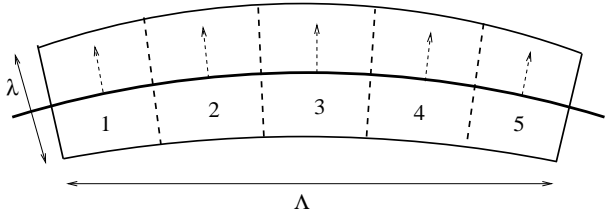,width=8cm}
\caption[
Sector with non-negligible curvature parameter.]
{Sector ${K^{\alpha}_{\Lambda , \lambda }}$ on the Fermi surface
(thick solid line)
with non-negligible curvature.\index{curvature of Fermi surface}
The dashed arrows are the local normal vectors 
at the Fermi surface. 
The dashed lines separate the  smaller boxes which are more
appropriate if the energy dispersion is linearized.
Note that around-the-corner processes corresponding to momentum transfer between
the smaller boxes $1 \leftrightarrow 2$, $2 \leftrightarrow 3$, 
$3 \leftrightarrow 4$ and $4 \leftrightarrow 5$ are
automatically taken into account in $K^{\alpha}_{\Lambda , \lambda}$.}
\label{fig:patchlarger}
\end{figure}

\begin{figure}
\sidecaption
\psfig{figure=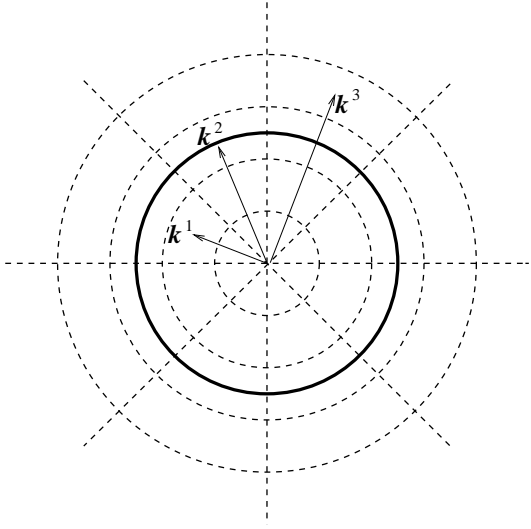,width=7cm}
\caption[
Subdivision of the entire momentum space of a 
two-dimensional system with a spherical Fermi surface 
into sectors.]
{Subdivision of the entire momentum space of a 
two-dimensional system with a spherical Fermi surface (thick solid circle)
into sectors.
The solid arrows point to the origins ${\vec{k}}^{\alpha}$ of 
local coordinate systems associated with the sectors.
Note that only for sectors at the Fermi surface
we may choose ${\vec{k}}^{\alpha}$ such that
$\epsilon_{ {\vec{k}}^{\alpha} } = \mu $.
For example
$\epsilon_{ {\vec{k}}^{2} } = \mu $,
but 
$\epsilon_{ {\vec{k}}^{1} } $ and 
$\epsilon_{ {\vec{k}}^{3} }$ 
are different from $\mu$.
}
\label{fig:patchall}
\end{figure}

\vspace{7mm}
\noindent
Because in this book we shall 
develop a systematic method for including the non-linear terms 
of the energy dispersion into higher-dimensional bosonization,
we shall ultimately drop
the requirement that the variation of the
local normal vector within a given patch must be negligible.
We then have the freedom of choosing much larger patches 
${{P}^{\alpha}_{\Lambda}}$ and sectors
${K^{\alpha}_{\Lambda , \lambda}}$  
than  for linearized energy dispersion. 
For example, in  Fig.~\secref{fig:patchlarger} we show
a sector $K^{\alpha}_{\Lambda , \lambda}$ that is 
constructed from five smaller boxes.
Clearly, by choosing larger boxes with finite curvature we automatically
take into account all around-the-corner processes\index{around-the-corner processes} between
the smaller sub-boxes used in the linearized theory!
Note that the {\it{curvature}} of the Fermi surface
is described by the non-linear terms in the
expansion of the non-interacting energy dispersion
close to the Fermi surface, see Eq.\ref{eq:xialphaqexp}.
For our purpose, it will be sufficient to 
assume that the expansion of 
$\epsilon_{ {\vec{k}}^{\alpha} + {\vec{q}} }$ 
for small ${\vec{q}}$ truncates
at the quadratic order. By a proper orientation of the axes of the
local coordinate system centered at ${\vec{k}}^{\alpha}$, we can always 
diagonalize the second-derivative tensor $c^{\alpha}_{ij}$
in Eq.\ref{eq:xialphaqexp}, so that
the energy dispersion relative to the chemical potential for
wave-vectors close to ${\vec{k}}^{\alpha}$ becomes
 \begin{equation}
 \epsilon_{ {\vec{k}}^{\alpha} + {\vec{q}} } - \mu 
 = \epsilon_{ {\vec{k}}^{\alpha} } - \mu +  
 \xi^{\alpha}_{\vec{q}}
 \label{eq:energyquad}
 \; \; \; ,
 \end{equation}
where 
 \begin{equation}
 \xi^{\alpha}_{\vec{q}} =
 {\vec{v}}^{\alpha} \cdot {\vec{q}}
 + \sum_{i = 1 }^{d} \frac{q_i^2}{ 2 m^{\alpha}_i}
 \label{eq:chiquad}
 \end{equation}
is the excitation energy relative to $\epsilon_{ {\vec{k}}^{\alpha} }$, and 
the inverse effective masses $1/m_i^{\alpha}$ \index{effective mass} are the eigenvalues of the
second derivative tensor $c^{\alpha}_{ij}$ defined in Eq.\ref{eq:v0cijdef}.
So far we have always chosen  ${\vec{k}}^{\alpha}$ such that
 $\epsilon_{ {\vec{k}}^{\alpha} } - \mu =0$,
in which case the first two terms on the right-hand side 
of Eq.\ref{eq:energyquad} cancel and
$ \epsilon_{ {\vec{k}}^{\alpha} + {\vec{q}} } - \mu = \xi^{\alpha}_{\vec{q}}$.
More generally, we may
subdivide the entire momentum space into
sectors centered at points ${\vec{k}}^{\alpha}$ 
which are not necessarily located on the Fermi surface. 
Of course, in this case
$\epsilon_{ {\vec{k}}^{\alpha} } - \mu $ does in general not vanish, so that
we should distinguish between the quantities
$\epsilon_{ {\vec{k}}^{\alpha} + {\vec{q}} } - \mu $ and
the {\it{excitation energy}}
$ \xi^{\alpha}_{\vec{q}} =
\epsilon_{ {\vec{k}}^{\alpha} + {\vec{q}} }
-
\epsilon_{ {\vec{k}}^{\alpha} }$ given in
Eq.\ref{eq:chiquad}. However, as long as we keep track of this 
difference,\index{cutoff!choice of sector cutoffs}
we may partition all degrees of freedom into sectors
as shown in Fig.~\secref{fig:patchall}.
Note that in general it will also be convenient 
to allow for sector-dependent cutoffs $\Lambda^{\alpha}$ and
$\lambda^{\alpha}$ in order to match the special geometry of the
Fermi surface.
\begin{figure}[htb]
\sidecaption
\psfig{figure=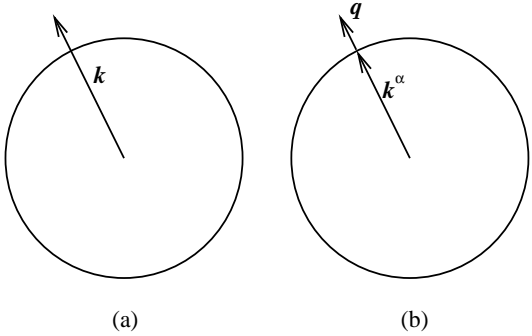,width=7cm,height=4.4cm}
\caption[Spherical Fermi surface and wave-vector
$\vec{k}$ measured with respect to the origin of the Fermi sphere.]
{(a) 
Spherical Fermi surface and wave-vector
$\vec{k}$ close to the Fermi surface.
(b) If we are interested in
$G ( \vec{k} , \I \tilde{\omega}_n )$, we choose
the coordinate origin $\vec{k}^{\alpha}$
such that ${\vec{k}} = \vec{k}^{\alpha} + \vec{q}$, with
${\vec{q}}$ parallel to ${\vec{k}}^{\alpha}$.
Note that in this case $| \vec{k} - \vec{k}^{\alpha} |$ assumes
the smallest possible value.
}
\label{fig:coordgood}
\end{figure}
As discussed in Sect.~\secref{subsec:patch}, the
sectors cutoffs should be chosen large compared with
the range $q_{\rm c}$ of the interaction in momentum space, so that
the final result for the Green's function at distances
large compared with $q_{\rm c}^{-1}$ is independent of
the unphysical sector cutoffs.
In fact, it is advantageous to choose the sectors 
as large as possible in order to avoid corrections 
due to the around-the-corner processes. \index{around-the-corner processes}
Hence, as soon as we include the non-linear terms
in the energy dispersion,
the only condition which puts an upper
limit to the sector size is the
requirement that within a given sector 
$K^{\alpha}_{\Lambda , \lambda}$ the effective masses
$m_i^{\alpha}$ and
the coarse-grained Landau parameters $f_{\vec{q}}^{\alpha \alpha^{\prime}}$
(see Eq.\ref{eq:Landaufuncdef}) should be well-defined.
If these conditions are satisfied, we may use our
formalism to calculate the single-particle
Green's function $G ( {\vec{k}}^{\alpha} + {\vec{q}} , \I \tilde{\omega}_n )$ for
all wave-vectors ${\vec{q}}$ that are small compared with the
sector cutoffs. 
Obviously, in the extreme case of Fermi surfaces
that have constant curvature 
(at least within the range $q_{\rm c}$ of the interaction),
and for Landau parameters 
that are independent of the
momenta of the incoming particles (i.e. 
$f_{\vec{q}}^{\alpha \alpha^{\prime}} = f_{\vec{q}}$,
such as the long-range tail of the
Coulomb interaction),
we may identify the
entire momentum space with a single sector.
In other words, there is no need any more for subdividing the
degrees of freedom into several sectors. 
In this case the problem of around-the-corner processes is solved trivially.
However, given the fact that our main interest is the calculation of the
single particle Green's function in the vicinity of the Fermi surface,
it is still advantageous to work with a coordinate system 
centered on the Fermi surface, as shown in Fig.~\secref{fig:coordgood}.
Once we know the function
$G^{\alpha} ( {\vec{q}} , \I \tilde{\omega}_n ) \equiv G ( {\vec{k}}^{\alpha}
+ \vec{q} , \I \tilde{\omega}_n )$ for wave-vectors of the form
${\vec{q}} = q^{\alpha}_{\|} \hat{\vec{k}}^{\alpha}$
(where $\hat{\vec{k}}^{\alpha}$ is a unit vector in the direction
of ${\vec{k}}^{\alpha}$), we 
may use the symmetry of the Fermi surface to reconstruct
$G ( {\bf{k}} , \I \tilde{\omega}_n )$.
For spherically symmetric systems
we simply have to substitute
$q^{\alpha}_{\|} \rightarrow | {\vec{k}} | - k_{\rm{F}}$
in the result for
$G^{\alpha} ( q^{\alpha}_{\|} \hat{\vec{k}}^{\alpha} , \I \tilde{\omega}_n ) $.

\section{Summary and outlook}

In the first three sections of this chapter we have 
summarized some basic facts about the fermionic many-body problem, mainly
to introduce the notation and to set the stage for the calculations that
follow.  In Sect.\secref{subsec:patch} we have given a detailed description of
the geometric patching construction in momentum space, which forms the basis of
higher-dimensional bosonization with linearized energy dispersion.
This construction has first been suggested by Haldane \cite{Haldane92}, and has
been discussed in some detail in the work of
Houghton, Kwon, and Marston \cite{Houghton93,Houghton94}, and later
in \cite{Kopietz95}.
Note that Haldane's way of subdividing
the degrees of freedom close to the Fermi surface into sectors
is a generalization of an earlier suggestion due to Luther \cite{Luther79},
who used thin tubes.

We have also pointed out that scattering processes that transfer
momentum between different sectors (the around-the-corner processes)
are difficult to handle within higher-dimensional bosonization.
It is therefore desirable to choose
the size of the sectors as large as possible.
In Sect.\secref{sec:sectors} we have further generalized
the patching construction by defining larger patches with finite
curvature, anticipating that in this book we shall present a systematic
method for including curvature effects into bosonization.
In other words, Haldane's boxes, which can be considered as  
the union of a large number of Luther's narrow tubes, 
have merged into a small number of sectors, within which
the curvature of the Fermi surface cannot be neglected.
In the case of a spherical Fermi surface  
and rotationally invariant interactions
we shall formally identify the entire momentum space with a 
single sector, thus completely abandoning the patching construction.

Finally, we would like to draw the attention of the reader to 
the problem of {\bf{Van Hove singularities}}\index{Van Hove singularities},
which will not be further 
discussed in this book, although non-perturbative methods for 
analyzing this problem
will be developed in the following chapters\footnote{
The fact that so far I have not 
studied this problem by myself with the help of
the technique described in this book
does not necessarily mean that this problem
is very difficult or requires conceptually new ideas.
I simply have not found the time to work on this problem.
This is also true for the other research problems mentioned in the
concluding sections of the following chapters.
}.
As discussed at the end of Sect.~\secref{subsec:locallin}, at a Van Hove singularity
the local Fermi velocity $\vec{v}^{\alpha}$ vanishes,
so that the leading term
in the expansion of the energy dispersion close to the Fermi surface
is quadratic.
Obviously, the effect of Van Hove singularities 
on the low-energy behavior
of the Green's function 
cannot be studied 
within an approximation that relies on the linearization of the
energy dispersion close to the Fermi surface.
However,
our more refined functional bosonization approach developed in Chaps.~\secref{sec:beyond} and
\secref{sec:eik} retains the quadratic term in the energy dispersion,
so that our method might
shed some new light on the problem of
Van Hove singularities 
in strongly correlated Fermi systems.

%
%

%
%
%

\chapter{Hubbard-Stratonovich transformations}
\label{chap:ahub}

{\it{Our functional bosonization approach is 
based on two Hubbard-Stratonovich transformations, which are  
described in detail in this chapter.}}

\vspace{7mm}

\noindent
We start with the imaginary time functional integral formulation of quantum statistical 
mechanics. This modern approach to the many-body
problem has recently been described in excellent 
textbooks \cite{Popov83,Popov87,Negele88,Kapusta89},
so that we can be rather brief here and simply summarize
the relevant representations of
fermionic correlation functions as Grassmannian functional integrals. 
We then eliminate the Grassmann fields in favour of collective bosonic fields
by means of suitable Hubbard-Stratonovich transformations \cite{Stratonovich57}.
These can be viewed as a clever change of variables to collective coordinates in functional
integrals, which exhibit the physically most relevant 
degrees of freedom.  The associated Jacobians
define the effective actions for the 
Hubbard-Stratonovich fields.
It turns out that the non-perturbative bosonization result
for the single-particle Green's function can be obtained
with the help of a conventional Hubbard-Stratonovich transformation that
involves a space- and time-dependent  auxiliary field $\phi^{\alpha}$.
This transformation will be discussed in Sect.~\secref{sec:HS1}.  
On the other hand, for the calculation of the
boson representation of the Hamiltonian or the
density-density correlation function
we need a generalized two-field Hubbard-Stratonovich transformation, 
which involves besides the $\phi^{\alpha}$-field  another
bosonic field $\tilde{\rho}^{\alpha}$. 
Section~\secref{sec:HS2} is devoted to a detailed description of this transformation.

\section{Grassmannian functional integrals}

{\it{ 
Fermionic correlation functions
can be represented as Grassmannian functional integrals.
These representations are particularly convenient 
for our purpose, because they can be directly 
manipulated via Hubbard-Stratonovich transformations.}}

\vspace{7mm}

\noindent
The grand canonical partition function ${\cal{Z}}$ of our
many-body Hamiltonian defined in Eqs.\ref{eq:Hdef}, \ref{eq:H0def} 
and \ref{eq:Hintcourse} can be written as
an imaginary time (i.e. Euclidean) functional integral 
over a Grassmann field\index{Grassmann field} $\psi$ \cite{Popov83,Popov87,Negele88,Kapusta89},
 \begin{equation}
 \frac{ {\cal{Z}} }{ {\cal{Z}}_{0} }
 = 
 \frac{ 
 \int {\cal{D}} \left\{ \psi \right\} 
 \E^{-S_{\rm mat} \{ \psi \} } 
 }
 { \int {\cal{D}} \left\{ \psi \right\} 
 \E^{-S_{0} \{ \psi \} } }
 \; \; \; ,
 \end{equation}
where ${\cal{Z}}_{0}$ is the grand canonical partition function
in the absence of interactions, and
the Euclidean action $S_{\rm mat} \{ \psi \}$ is
given by
 \begin{eqnarray}
 S_{\rm mat} \{ \psi \} & = &
S_{0} \{ \psi \}  + S_{\rm int} \{ \psi \}
\; \; \; ,
\label{eq:Smatpsidef}
\\
 S_{0} \left\{ \psi \right\}
  & =  &
  \beta \sum_{ k }  \left[ - \I \tilde{\omega}_{n} + 
 \xi_{\vec{k}}  \right] \psi^{\dagger}_{k} \psi_{ k }
 \; \; \; ,
 \label{eq:S0psidef}
 \\
 S_{\rm int} \{ \psi \} & = & \frac{ \beta }{2V} \sum_{{q}}  \sum_{ \alpha \alpha^{\prime} }
 {{f}}_{{q}}^{\alpha \alpha^{\prime}}  
 {\rho}_{ - {q}}^{\alpha} {\rho}_{q}^{\alpha^{\prime}} 
 \label{eq:Sintpsidef}
 \; \; \; .
 \end{eqnarray}
Here
 \begin{equation}
 \rho_{q}^{\alpha}  =  \sum_{k} 
 \Theta^{\alpha} ( {\vec{k}} )
 \psi^{\dagger}_{k} \psi_{k+q}
 \label{eq:rhoalpha}
 \end{equation}
is the Grassmann representation of the
sector density operator $\hat{\rho}^{\alpha}_{\vec{q}}$ defined
in Eq.\ref{eq:rhoopdef}.
Note that the $k$- and $q$-sums in these expressions are over
wave-vectors {\it{and}} Matsubara frequencies. 
Although the Landau parameters\index{Landau interaction parameter} $f^{\alpha \alpha^{\prime}}_{\vec{q}}$ 
that appear in
the Hamiltonian $\hat{H}_{\rm int}$ in Eq.\ref{eq:Hintcourse}
depend only on the wave-vector ${\vec{q}}$, we have
replaced them in Eq.\ref{eq:Sintpsidef}
by more general frequency-dependent parameters
$f^{\alpha \alpha^{\prime}}_{q} \equiv
f^{\alpha \alpha^{\prime}}_{{\vec{q}} , \I \omega_{m}}$.
In our functional integral approach
the frequency-dependence does not introduce
any additional complications. 
In  physical applications the frequency-dependence 
is due to the fact that the underlying microscopic mechanism responsible for
the effective interaction between the electrons
is the exchange of some particle with a finite
velocity, such as phonons\footnote{
The coupled electron-phonon system will be discussed
in detail in Chap.~\secref{chap:aph}.}.
Moreover, even in the case of electromagnetism the
effective interaction becomes frequency-dependent 
if the corrections of higher order in ${v_{\rm F}}/{c}$ are retained.
The static Coulomb potential is just the 
${v_{\rm F}}/{c} = 0$ limit.
The leading correction 
is a retarded current-current interaction mediated  
by the transverse radiation field, which will be
discussed in Chap.~\secref{chap:arad}.

\vspace{7mm}

The time-ordered Matsubara Green's function\index{Green's function!representation as
functional integral} 
defined in Eq.\ref{eq:GMatsubaradef}
can be represented as the functional integral average
of $\psi_{k} \psi^{\dagger}_{k}$, 
\begin{equation}
 G ( k ) 
 =
 - \beta \frac{ 
 \int {\cal{D}} \left\{ \psi \right\} 
 \E^{-S_{\rm mat} \{ \psi \} } 
 \psi_{k}  \psi^{\dagger}_{k} 
 }
 { \int {\cal{D}} \left\{ \psi \right\} 
 \E^{-S_{\rm mat} \{ \psi \} } }
 \label{eq:GmatFourier}
 \; \; \; .
 \end{equation}
In absence of interactions 
this reduces to
 \begin{equation}
 G_{0} ( k ) 
  =  - \beta \frac{ 
 \int {\cal{D}} \left\{ \psi \right\} 
 \E^{-S_{0} \{ \psi \} } 
 \psi_{k} \psi^{\dagger}_{k}
 } 
 { \int {\cal{D}} \left\{ \psi \right\} 
 \E^{-S_{0} \{ \psi \} } } 
  =  \frac{1 }{ \I \tilde{\omega}_{n} - \xi_{\vec{k}}  }
 \label{eq:G0def2}
 \; \; \; ,
 \end{equation}
in agreement with Eq.\ref{eq:G0Matdef}.
From the Matsubara Green's function we can obtain the real space imaginary time
Green's function via Fourier transformation\index{Fourier transformation}, 
\begin{equation}
 G ( {\vec{r}} , \tau  ) = \frac{1}{\beta V} 
 \sum_{k} \E^{  \I ( {\vec{k}} \cdot {\vec{r}} - \tilde{\omega}_{n} \tau )}
 G ( k ) 
 \; \; \; .
 \label{eq:FTGreal}
 \end{equation}
Defining
 \begin{equation}
 \psi ( {\vec{r}} , \tau )   =  \frac{1}{ \sqrt{ V } }
 \sum_{ k } \E^{ \I ( {\vec{k}} \cdot {\vec{r}} - \tilde{\omega}_{n} \tau )} 
 \psi_{ k }
 \; \; \; ,
 \end{equation}
we can also write
 \begin{equation}
 G ( {\vec{r}} - {\vec{r}}^{\prime} , \tau - \tau^{\prime} )
 = 
 - \frac{ 
 \int {\cal{D}} \left\{ \psi \right\} 
 \E^{-S_{\rm mat} \{ \psi \} } 
 \psi ( {\vec{r}} , \tau ) \psi^{\dagger} ( {\vec{r}}^{\prime} , \tau^{\prime} )
 }
 { \int {\cal{D}} \left\{ \psi \right\} 
 \E^{-S_{\rm mat} \{ \psi \} } }
 \label{eq:Gdef}
 \; \; \; .
 \end{equation}

Two-particle Green's functions can also 
be represented as functional integral averages.
The density-density correlation function\index{density-density correlation
function!functional integral representation} defined in Eq.\ref{eq:cordens}
can be written as
 \begin{equation}
 \Pi ( q ) =
 \frac{\beta}{ V} \frac{ 
 \int {\cal{D}} \left\{ \psi \right\} 
 \E^{-S_{\rm mat} \{\psi \}} 
 \rho_{q }  \rho_{-q} 
 }
 { \int {\cal{D}} \left\{ \psi \right\} 
 \E^{-S_{\rm mat} \{ \psi \}} }
 \label{eq:cordensfuncgen}
 \; \; \; ,
 \end{equation}
where the composite Grassmann field corresponding to the Fourier components of the
total density is
(see Eq.\ref{eq:denstotop})
 \begin{equation}
 \rho_{q} = 
 \sum_{\alpha} \rho_{q}^{\alpha} 
 =
 \sum_{k} \psi^{\dagger}_{k} \psi_{k+q}
 \; \; \; .
 \label{eq:rhocomptot}
 \end{equation}
Using Eq.\ref{eq:Thetasum} 
we may also write
 \begin{equation}
 \Pi ( q) = \sum_{\alpha \alpha^{\prime}} \Pi^{\alpha \alpha^{\prime}} ( q )
 \label{eq:Pitotdecompose}
 \; \; \; ,
 \end{equation}
where for $q \neq 0$
 \begin{eqnarray}
 \Pi^{\alpha  \alpha^{\prime} } ( q )
 & = & \frac{1}{\beta V } \int_{0}^{\beta}  \D \tau 
 \int_{0}^{\beta} \D \tau^{\prime}
 \E^{ - \I \omega_{m}  ( \tau - \tau^{\prime}) }
 \langle {\cal{T}} \left[ \hat{\rho}^{\alpha}_{\vec{q}} ( \tau )
 \hat{\rho}^{ \alpha^{\prime} }_{ -{\vec{q} } } ( \tau^{\prime} ) 
 \right] \rangle 
 \nonumber
 \\
 & = &
 \frac{\beta}{ V} \frac{ 
 \int {\cal{D}} \left\{ \psi \right\} 
 \E^{-S_{\rm mat} \{\psi \}} 
 \rho^{\alpha}_{q }  \rho^{\alpha^{\prime}}_{-q} 
  } 
 { \int {\cal{D}} \left\{ \psi \right\} 
 \E^{-S_{\rm mat} \{ \psi \} } }
 \label{eq:cordensloc}
 \; \; \; .
 \end{eqnarray}
We shall refer to $\Pi ( q )$ as the {\it{global}} or {\it{total}}
density-density correlation function, and to $\Pi^{\alpha \alpha^{\prime} } ( q )$ as the
{\it{local}} or {\it{sector}} density-density correlation 
function\index{density-density correlation function!sector}.
In the non-interacting limit Eq.\ref{eq:cordensloc} reduces to 
 \begin{equation}
 \Pi_{0}^{\alpha \alpha^{\prime} }
 ( q )
    =   - \frac{1}{ {V}} \sum_{\vec{k}}
    \Theta^{\alpha} ( {\vec{k}} )
    \Theta^{\alpha^{\prime}} ( {\vec{k}} + {\vec{q}} )
\frac{ f (  \xi_{\vec{k+q}}  )
 - f (  \xi_{\vec{k}} ) }
{ \xi_{\vec{k+q}} - \xi_{\vec{k}} - \I \omega_{m} }
 \label{eq:Pi0loc}
 \; \; \; .
 \end{equation} 
By relabeling ${\vec{k}} + {\vec{q}} \rightarrow {\vec{k}}$
it is easy to see that
 \begin{equation}
 \Pi_{0}^{\alpha \alpha^{\prime} } ( q )
 =
 \Pi_{0}^{\alpha^{\prime} \alpha } ( - q )
 \; \; \; .
 \label{eq:Pi0sym}
 \end{equation}
Substituting Eq.\ref{eq:Pi0loc} into Eq.\ref{eq:Pitotdecompose}
and using $\sum_{\alpha} \Theta^{\alpha} ( {\vec{k}} ) = 1$,  
we recover the non-interacting Lindhard function given in Eq.\ref{eq:Pi0total}.
We would like to emphasize that in the above functional integral representations
of the correlation
functions the precise normalization for the integration measure
${\cal{D}} \{ \psi \}$ is irrelevant, because the measure appears always
in the numerator as well as in the denominator.

\section{The first Hubbard-Stratonovich transformation}
\label{sec:HS1}

{\it{
We decouple the two-body interaction between the fermions 
with the help of a Hubbard-Stratonovich\index{Hubbard-Stratonovich transformation!first
(background field $\phi^{\alpha}$)} field $\phi^{\alpha}$.
After integrating over the Fermi fields,
the single-particle Green's function
can then be written as a quenched average 
with probability distribution given by
the effective action of the $\phi^{\alpha}$-field.}}

\subsection{Decoupling of the interaction}
\label{subsec:HS1sub2}

The generalized Landau parameters $f_{q}^{\alpha \alpha^{\prime}}$ in Eq.\ref{eq:Sintpsidef} have units of
energy $\times$ volume. Because we would like
to work with dimensionless Hubbard-Stratonovich fields,
it is useful to introduce  the dimensionless\index{Landau interaction parameter!dimensionless
($\tilde{f}_{q}^{\alpha \alpha^{\prime}}$)}
Landau parameters 
 \begin{equation}
 \tilde{f}_{q}^{\alpha \alpha^{\prime}} = \frac{\beta}{V} f_{q}^{\alpha \alpha^{\prime}}
 \; \; \; .
 \label{eq:Landaudimless}
 \end{equation}
The interaction part of our Grassmannian  action 
can then be written as
 \begin{equation}
 S_{\rm int} \{ \psi \}  =  \frac{ 1 }{2} \sum_{{q}}  \sum_{ \alpha \alpha^{\prime} }
 {\tilde{f}}_{{q}}^{\alpha \alpha^{\prime}}  
 {\rho}_{ - {q}}^{\alpha} {\rho}_{q}^{\alpha^{\prime}} 
 \label{eq:Sintpsi2}
 \; \; \; .
 \end{equation}
Using the invariance
of the sum in Eq.\ref{eq:Sintpsi2} under simultaneous relabelling
$ \alpha \leftrightarrow \alpha^{\prime}$ and $q \rightarrow -q$,
it is easy to see that, without loss of generality, we may assume that
 \begin{equation}
 {\tilde{f}}_{q}^{ \alpha \alpha^{\prime}} =
 {\tilde{f}}_{  -q}^{ \alpha^{\prime} \alpha} 
 \label{eq:ftildeherm}
 \; \; \; ,
 \end{equation}
which is analogous to Eq.\ref{eq:Pi0sym}.
We now decouple this action by means of the following Hubbard-Stratonovich transformation
involving a dimensionless bosonic auxiliary field $\phi^{\alpha}_{q}$,
 \begin{eqnarray}
 \exp \left[ - S_{\rm int} \{ \psi \} \right] & \equiv &
 \exp \left[ 
  - \frac{1}{2} \sum_{q}  \sum_{\alpha \alpha^{\prime} }
  [ \underline{\tilde{f}}_{ {{q}} }  ]^{\alpha \alpha^{\prime} } 
  \rho^{\alpha}_{-q} \rho^{\alpha^{\prime}}_{q}
  \right ] 
  \nonumber
  \\
 & &\hspace{-30mm}  =
 \frac{  
 \int {\cal{D}} \left\{ \phi^{\alpha} \right\}
 \exp \left[ -
  \frac{1}{2} \sum_{q}  
  \sum_{\alpha \alpha^{\prime} }
  [ \underline{\tilde{f}}_{ {{q}} }^{-1} ] ^{\alpha \alpha^{\prime} } 
  \phi^{\alpha}_{-q} \phi^{\alpha^{\prime}}_{q}
 - \I  \sum_{q} \sum_{\alpha} \phi^{\alpha}_{-q} \rho_{q}^{\alpha}   \right]   }
 {
 \int {\cal{D}} \left\{ \phi^{\alpha} \right\}
 \exp \left[ -
  \frac{1}{2} \sum_{q}  
  \sum_{\alpha \alpha^{\prime} }
  [ \underline{\tilde{f}}_{ {{q}} }^{-1} ] ^{\alpha \alpha^{\prime}  }
  \phi^{\alpha}_{-q} \phi^{\alpha^{\prime}}_{q} \right] }
  \label{eq:HStrans1}
  \; .
  \end{eqnarray}
Here $\underline{ \tilde{f}}_{q}$ is a matrix in the patch indices, with
matrix elements given by
 \begin{equation}
 [ \underline{ \tilde{f}}_{q}]^{\alpha \alpha^{\prime} }
 = \tilde{f}_{q}^{\alpha \alpha^{\prime}} = \frac{\beta}{V} f_{q}^{\alpha \alpha^{\prime}}
 \label{eq:ftildematdef}
 \; \; \; .
 \end{equation}
Throughout this work we shall use the convention that
all underlined quantities are
matrices in the patch indices.
Eq.\ref{eq:HStrans1} is easily proved by shifting\index{shift transformation} the
$\phi^{\alpha}$-field in the numerator of the right-hand side according to
 \begin{equation}
 \phi^{\alpha}_{q} \rightarrow  \phi^{\alpha}_{q} 
 - \I \sum_{\alpha^{\prime}} [ \underline{\tilde{f}}_{q} ]^{\alpha \alpha^{\prime} }
 \rho_{q}^{\alpha^{\prime}}
 \; \; \; ,
 \label{eq:phialphashift}
 \end{equation}
and using Eq.\ref{eq:ftildeherm}.
For later convenience, let us fix the 
measure for the $\phi^{\alpha}$-integration such that
 \begin{equation}
 \int {\cal{D}} \left\{ \phi^{\alpha} \right\}
 \exp \left[ 
 -
  \frac{1}{2} \sum_{q}  
  \sum_{\alpha \alpha^{\prime} }
  [ \underline{\tilde{f}}_{ {{q}} }^{-1} ] ^{\alpha \alpha^{\prime} } 
  \phi^{\alpha}_{-q} \phi^{\alpha^{\prime}}_{q}
  \right] 
  = \prod_{q} 
   \det ( \underline{ \tilde{f}}_{q} ) 
   \label{eq:measure1}
  \; ,
  \end{equation}
where $\det$ denotes the determinant with respect to the patch indices.
Note that our complex auxiliary field
satisfies $\phi^{\alpha}_{-q} = ( \phi^{\alpha}_{q} )^{\ast}$, 
because it couples to the Fourier components of the density, which have also this symmetry.
Of course, mathematically the $\phi^{\alpha}$-integrals in Eq.\ref{eq:HStrans1} and
\ref{eq:measure1} are only well defined if the matrix
$\underline{ \tilde{f}}_{q}$ is positive definite.
However, Eqs.\ref{eq:phialphashift} and \ref{eq:ftildeherm} are 
sufficient to proof Eq.\ref{eq:HStrans1} as an algebraic 
identity, so that we shall use this transformation for intermediate algebraic manipulations 
even if the matrix $\underline{ \tilde{f}}_{q}$ is not positive definite.
Possible infinities due to vanishing (or even negative) eigenvalues
of the matrix $\underline{\tilde{f}}_{q}$ cancel between the denominator and numerator of
Eq.\ref{eq:HStrans1}.
For example, if all matrix elements of a $M \times M$-matrix have the same (non-zero) value,
then $M-1$ of its eigenvalues are equal to zero,
so that for constant matrices $\underline{\tilde{f}}_{q}$ 
we implicitly assume that the Gaussian integrations in Eq.\ref{eq:HStrans1}
have been regularized in some convenient way.
Note also that the appearance of
$\underline{\tilde{f}}_{{q}}^{-1}$ 
is only an intermediate
step in our calculation. The final expressions for physical correlation functions 
can be written entirely in terms of
$\underline{\tilde{f}}_{q}$, and remain finite even if
this matrix is not positive definite.
Such a rather loose use of mathematics is quite common 
in statistical field theory,
although for mathematicians it is certainly not acceptable.
Formally, the appearance of
$\underline{\tilde{f}}_{{q}}^{-1}$ 
at intermediate steps can be avoided with the help
of the two-field Hubbard-Stratonovich transformation
discussed in Sect.~\secref{sec:HS2},
see Eq.\ref{eq:HStrans3} below\footnote{
Other formal ways to avoid this problem are briefly discussed
in the books by Amit \cite[\mbox{p. $24$}]{Amit84}, and by
Itzykson and Drouffe \cite[\mbox{p. $ 153$}]{Itzykson89}. On the other hand,
Zinn-Justin mentions this problem \cite[\mbox{p. $518$}]{ZinnJustin89}, 
but does not hesitate to perform 
a transformation of the form \ref{eq:HStrans1} for a general matrix
$\underline{\tilde{f}}_{q}$.
Moreover, in the book by Negele and Orland \cite[\mbox{p. $198$}]{Negele88} as well as
in Parisi's book \cite[\mbox{p. $209$}]{Parisi88} 
this transformation is used without further comment.
I would like to thank Kurt
Sch\"{o}nhammer for giving me a copy of his notes with 
a summary and discussion of the relevant references.}.

\subsection{Transformation of the single-particle Green's function}

Applying  the Hubbard-Stratonovich transformation \ref{eq:HStrans1}
to the functional integral representation \ref{eq:GmatFourier}
of the single-particle Green's function, we obtain
 \begin{equation}
 G ( k )
 = - \beta \frac{ 
 \int {\cal{D}} \left\{ \psi \right\} 
 {\cal{D}} \left\{ \phi^{\alpha} \right\}
 \E^{- {S} \{ \psi , \phi^{\alpha} \} } 
 \psi_{k} \psi^{\dagger}_{k}
 }
 { \int {\cal{D}} \left\{ \psi \right\} 
 {\cal{D}} \left\{ \phi^{\alpha} \right\}
 \E^{- {S} \{ \psi , \phi^{\alpha} \} }}
 \label{eq:Gpatchdef}
 \; \; \; ,
 \end{equation}
where the decoupled action is given by
 \begin{equation}
 {S}  \{ \psi , \phi^{\alpha} \}
 = {S}_{0} \left\{ \psi \right\} + {S}_{1} \left\{ \psi , \phi^{\alpha} \right\}
 + {S}_{2} \left\{ \phi^{\alpha} \right\}
 \label{eq:Sdecoupdef}
 \; \; \; ,
 \end{equation}
with
 \begin{eqnarray}
 {S}_{1} \{ \psi , \phi^{\alpha} \}
 &  = &   \sum_{ q } \sum_{\alpha} 
 \I \rho^{\alpha}_{q}  \phi_{ -q}^{\alpha}
 \; \; \; ,
 \label{eq:S1decoupdef}
 \\
 {S}_{2} \{ {{\phi}}^{\alpha}  \} & = &
 \frac{1}{2} \sum_{q} \sum_{\alpha \alpha^{\prime} }
  [ \underline{\tilde{f}}_{ {{q}} }^{-1} ]^{ \alpha \alpha^{\prime} }
 \phi_{-q}^{\alpha} \phi_{q}^{\alpha^{\prime}}
 \label{eq:S2decoupdef}
 \; \; \; .
 \end{eqnarray}
Thus, the fermionic two-body interaction has
disappeared. Instead, we have the problem of a coupled field theory
in which a dynamic bosonic field $\phi^{\alpha}$ is coupled
linearly to the fermionic density.
The $\phi^{\alpha}$-field mediates the interaction between the
fermionic matter in the sense that
integration over the $\phi^{\alpha}$-field
(i.e. undoing the Hubbard-Stratonovich transformation)
generates an effective fermionic two-body interaction. 
In fact, because all interactions in nature can be viewed as the result
of the exchange of some sort of particles, it is more general and fundamental
to define the problem of interacting fermions in terms of an action 
that does not contain the fermionic two-body interaction explicitly, but
involves the linear coupling of the fermionic density to another bosonic field. 
This point of view has been emphasized by
Feynman and Hibbs \cite{Feynman65}.
We shall come back to the
physical  meaning of the Hubbard-Stratonovich field $\phi^{\alpha}$ 
in Chap.~\secref{subsec:Definitionofmodel}, where we shall show that
for the Maxwell action the $\phi^{\alpha}$-field
can be identified physically with the
scalar potential of electromagnetism.

In a functional integral we 
have the freedom of performing the integrations
in any convenient order.
Let us now perform the fermionic 
integration over the $\psi$-field in Eq.\ref{eq:Gpatchdef} before 
integrating over the $\phi^{\alpha}$-field. 
To do this, we write
 \begin{equation}
 {S}_{0} \left\{ \psi \right\}    +
 {S}_{1} \left\{ \psi , \phi^{\alpha} \right\}   
 = - \beta \sum_{k k^{\prime} } \psi^{\dagger}_{k} 
 [  {\hat{G}}^{-1}]_{  k  k^{\prime} }  
 \psi_{k^{\prime}}
 \; \; \; ,
 \label{eq:SmatG}
 \end{equation}
where $\hat{G}^{-1}$ is an infinite matrix in momentum and frequency space, with
matrix elements given by the formal Dyson equation
 \begin{equation}
 [ \hat{G}^{-1} ]_{k k^{\prime}} =
 [ \hat{G}^{-1}_{0} ]_{k k^{\prime}} - [ \hat{V} ]_{k k^{\prime}}
 \label{eq:GhatDyson}
 \; \; \; .
 \end{equation}
Here $\hat{G}_{0}$ is the non-interacting Matsubara Green's function matrix,
 \begin{equation}
 [ \hat{G}_{0} ]_{ k k^{\prime} }
   =  \delta_{k k^{\prime} } G_{0} ( k) 
   \; \; \; , \; \; \;
 G_{0} ( k ) = \frac{1}{ \I \tilde{\omega}_{n} - \xi_{\vec{k}} }
 \label{eq:hatG0}
 \; \; \; ,
 \end{equation}
and the self-energy matrix $\hat{V}$ is defined by
 \begin{equation}
 [ \hat{V} ]_{ k k^{\prime} }   =
 \sum_{\alpha} 
   \Theta^{\alpha} ( {\vec{k}} ) V^{\alpha}_{k- k^{\prime}}
   \; \; \; , \; \; \;
  V^{\alpha}_{q} =
      \frac{\I}{\beta} 
  {\phi}^{\alpha}_{q} 
 \label{eq:hatVphi}
 \; \; \; . 
 \end{equation}
Recall that $k$ denotes wave-vector and frequency, so that
$\delta_{k k^{\prime} } = \delta_{ {\vec{k}}  {\vec{k}}^{\prime}} \delta_{n n^{\prime}}$.
Choosing the normalization of the integration measure 
${\cal{D}} \{ \psi \}$ suitably, the ``trace-log'' formula \cite{Popov83}\index{trace-log formula}
yields
 \begin{eqnarray}
 \int
 {\cal{D}} \left\{ \psi \right\} 
 \exp \left[ { 
 - {S}_{0} \left\{ \psi \right\} - {S}_{1} \left\{ \psi , \phi^{\alpha} \right\} 
 }  \right]
 = \det \hat{G}^{-1} 
 \nonumber
 \\
 &  & \hspace{-50mm} = \E^{ {\rm Tr} \ln \hat{G}^{-1} }
 = \E^{ {\rm Tr} \ln \hat{G}_{0}^{-1} }
 \E^{ {\rm Tr} \ln [ 1 - \hat{G}_{0} \hat{V} ] } 
 \label{eq:Fermiint}
 \; \; \; ,
 \\
 - \beta \int
   {\cal{D}} \left\{ \psi \right\} 
  \psi_{k} \psi^{\dagger}_{k}
 \exp \left[ {
 - {S}_{0} \left\{ \psi \right\} - {S}_{1} \left\{ \psi , \phi^{\alpha} \right\} }
 \right]
 \nonumber
 \\
 &   &
 \hspace{-50mm} =
 [ \hat{G} ]_{k k }
 \E^{{\rm Tr} \ln \hat{G}_{0}^{-1}} \E^{ {\rm Tr} \ln [ 1 - \hat{G}_{0} \hat{V} ]  }
 \label{eq:Fermiint3}
 \; \; \; .
 \end{eqnarray}
Hence, after integration over the fermions
the exact interacting Green's function \ref{eq:Gpatchdef}
can be written as a quenched average\index{Green's function!representation
as quenched average} of the diagonal element $[ \hat{G} ]_{kk}$, 
 \begin{eqnarray}
 G( k )  = 
 \int {\cal{D}} \{ \phi^{\alpha} \} 
 {\cal{P}} \{ \phi^{\alpha} \} 
 [ \hat{G} ]_{kk}
 \equiv 
 \left< [ \hat{G}]_{  k  k }   \right>_{ S_{\rm eff} }
 \label{eq:avphi}
 \; \; \; .
 \end{eqnarray}
Note that 
$ [ \hat{G}]_{  k  k }$ is in general a very complicated
functional of the field $\phi^{\alpha}$.
The normalized probability distribution\index{probability distribution!background field}
 ${\cal{P}} \{ \phi^{\alpha} \} $ is 
 \begin{equation}
 {\cal{P}} \{ \phi^{\alpha} \} 
 = 
 \frac{  
 \E^{ - {S}_{\rm eff} \{ \phi^{\alpha} \} }  }
 {
 \int {\cal{D}} \left\{ \phi^{\alpha} \right\} 
 \E^{ - {S}_{\rm eff} \{ \phi^{\alpha} \} }  }
 \; \; \; ,
 \label{eq:probabphidef}
 \end{equation}
where the effective action\index{effective action!for $\phi^{\alpha}$-field} 
for the $\phi^{\alpha}$-field 
contains, in addition to the action
${S}_{2} \{ \phi^{\alpha} \} $
defined in Eq.\ref{eq:S2decoupdef}, 
a contribution due to the coupling to the electronic degrees of freedom,
 \begin{equation}
 {S}_{\rm eff} \{ \phi^{\alpha} \} 
 = {S}_{2} \{ \phi^{\alpha} \} 
 +  {S}_{\rm kin} \{ \phi^{\alpha} \} 
 \label{eq:Seffphidef}
 \; \; \; ,
 \end{equation}
with
 \begin{equation}
{S}_{\rm kin} \{ \phi^{\alpha} \}  =
 -  {\rm Tr} \ln [ 1 - \hat{G}_{0} \hat{V} ]  
 \label{eq:Skinphidef}
 \; \; \; .
 \end{equation}
Note that in Eq.\ref{eq:avphi} one first calculates the Green's function
for a frozen configuration of the
$\phi^{\alpha}$-field, and then averages the resulting expression over all
configurations this field, with the probability distribution given
in Eq.\ref{eq:probabphidef}.
Such a procedure closely resembles the
{\it{background field method}}\index{background field method}, which is well-known in 
the field theory literature \cite{DeWitt67}.
Following this terminology, we shall also
refer to our auxiliary field $\phi^{\alpha}_q$
as the {\it{background field}}.

The above transformations are exact. Of course, 
in practice  it is impossible to calculate the
interacting Green's function  from Eq.\ref{eq:avphi},
because (a) the matrix $\hat{G}^{-1}$ cannot be inverted exactly,
(b) the kinetic energy contribution
$S_{\rm kin} \{ \phi^{\alpha} \}$ to the effective action
of the $\phi^{\alpha}$-field can only be calculated perturbatively,
and (c) the probability distribution ${\cal{P}} \{ \phi^{\alpha} \}$ 
in Eq.\ref{eq:probabphidef} is not Gaussian, so that the averaging
procedure cannot be carried out exactly.
{\it{The amazing fact is now that there exists a physically interesting limit
where the difficulties (a), (b) and (c) can all be overcome.}} 
The above method leads then to a new and non-perturbative
approach to the fermionic many-body problem.
The detailed description of this method and its application to
physical problems is the central topic of this book. 
The highly non-perturbative character of this approach is evident
from the fact that in $d=1$ the 
well-known bosonization result
for the Green's function of the
Tomonaga-Luttinger model \cite{Mattis65} can be obtained 
with this method \cite{Lee88}. This will be explicitly shown 
in Chap.~\secref{sec:Green1}.

\section{The second Hubbard-Stratonovich transformation}
\label{sec:HS2}

{\it{
In order to introduce collective bosonic density fields, we  
perform another change of variables in the functional integral
by means of a second Hubbard-Stratonovich transformation\index{Hubbard-Stratonovich
transformation!second (density field $\tilde{\rho}^{\alpha}$)}. 
In this way we arrive at the  general definition of the bosonized kinetic energy.}}

\vspace{7mm}

\noindent
From Eq.\ref{eq:S1decoupdef} we see that
after the first Hubbard-Stratonovich transformation
the composite Grassmann field $\rho^{\alpha}$
couples linearly to the $\phi^{\alpha}$-field. 
Evidently the $\phi^{\alpha}$-field is related to the $\rho^{\alpha}$-field
in a very similar fashion as
the chemical potential is related to the particle number.
In other words, the $\phi^{\alpha}$-field is 
the {\it{conjugate field}} to the sector density $\rho^{\alpha}$.
We now use a second Hubbard-Stratonovich transformation
to eliminate the composite Grassmann field $\rho^{\alpha}$ 
in favour of a collective bosonic field
$\tilde{\rho}^{\alpha}$, which 
can then be identified physically with the bosonized density fluctuation.
This additional transformation is 
useful for the calculation of quantities that can be written
in terms of collective density fluctuations, such as the
density-density correlation function or the bosonized
Hamiltonian. On the other hand, for the calculation of the {\it{single-particle Green's function}} 
the first Hubbard-Stratonovich transformation introduced in the previous section
is sufficient.

\subsection[Transformation of the density-density \mbox{\hspace{20mm}} correlation function]
{Transformation of the density-density correlation function}
 
Applying the first Hubbard-Stratonovich transformation 
\ref{eq:HStrans1} to the Grassmannian functional
integral representation \ref{eq:cordensloc}
of the sector density-density
correlation function\index{density-density correlation function!sector}, 
we obtain
 \begin{equation}
 \Pi^{\alpha  \alpha^{\prime} } ( q )
 = \frac{\beta}{V}
 \frac{
 \int {\cal{D}} \left\{ \psi \right\} 
 {\cal{D}} \left\{ \phi^{\alpha} \right\}
 \rho^{\alpha}_{q} \rho^{\alpha^{\prime}}_{-q}
 \exp \left[ -  {S} \left\{ \psi , \phi^{\alpha} \right\} \right]
 } 
 { \int {\cal{D}} \left\{ \psi \right\} 
 {\cal{D}} \left\{ \phi^{\alpha} \right\}
 \exp \left[ - {S} \left\{ \psi , \phi^{\alpha} \right\} \right] }
 \label{eq:PiHS1}
 \; \; \; .
 \end{equation}
We now decouple the quadratic action $S_{2} \{ \phi^{\alpha} \}$
in this expression by means of an integration over another bosonic
field $\tilde{\rho}^{\alpha}_{q}$,
 \begin{eqnarray}
 \exp \left[ - S_{2} \{ \phi^{\alpha} \} \right] & \equiv &
 \exp \left[ 
  - \frac{1}{2} \sum_{q}  \sum_{\alpha \alpha^{\prime} }
  [ \underline{\tilde{f}}_{ {{q}} }^{-1}  ]^{\alpha \alpha^{\prime} } 
  \phi^{\alpha}_{-q} \phi^{\alpha^{\prime}}_{q}
  \right ] 
  \nonumber
  \\
 & & \hspace{-27mm}  =
 \frac{  
 \int {\cal{D}} \left\{ \tilde{\rho}^{\alpha} \right\}
 \exp \left[ -
  \frac{1}{2} \sum_{q}  
  \sum_{\alpha \alpha^{\prime} }
  [ \underline{\tilde{f}}_{ {{q}} } ] ^{\alpha \alpha^{\prime} } 
  \tilde{\rho}^{\alpha}_{-q} \tilde{\rho}^{\alpha^{\prime}}_{q}
 + \I  \sum_{q} \sum_{\alpha} \phi^{\alpha}_{-q} \tilde{\rho}_{q}^{\alpha}   \right]   }
 {
 \int {\cal{D}} \left\{ \tilde{\rho}^{\alpha} \right\}
 \exp \left[ -
  \frac{1}{2} \sum_{q}  
  \sum_{\alpha \alpha^{\prime} }
  [ \underline{\tilde{f}}_{ {{q}} } ] ^{\alpha \alpha^{\prime}  }
  \tilde{\rho}^{\alpha}_{-q} \tilde{\rho}^{\alpha^{\prime}}_{q} \right] }
  \label{eq:HStrans2}
  \;  .
  \end{eqnarray}
It is convenient to define the integration measure
for the $\tilde{\rho}^{\alpha}$-integral such that
 \begin{equation}
 \int {\cal{D}} \left\{ \tilde{\rho}^{\alpha} \right\}
 \exp \left[ 
  - \frac{1}{2} \sum_{q}  \sum_{\alpha \alpha^{\prime} }
  [ \underline{\tilde{f}}_{ {{q}} }  ]^{\alpha \alpha^{\prime} } 
  \tilde{\rho}^{\alpha}_{-q} \tilde{\rho}^{\alpha^{\prime}}_{q}
  \right]
   =  \prod_{q} \left[ { \det ( \underline{ \tilde{f} }_{q} )} \right]^{-1}
  \label{eq:measure2}
  \; \; \; ,
  \end{equation}
so that with Eq.\ref{eq:measure1}  we have
 \begin{eqnarray}
 \int {\cal{D}} \left\{ \tilde{\rho}^{\alpha} \right\}
 \int {\cal{D}} \left\{ \phi^{\alpha} \right\}
 \exp \left[ 
  - \frac{1}{2} \sum_{q}  \sum_{\alpha \alpha^{\prime} }
  [ \underline{\tilde{f}}_{ {{q}} }  ]^{\alpha \alpha^{\prime} } 
  \tilde{\rho}^{\alpha}_{-q} \tilde{\rho}^{\alpha^{\prime}}_{q}
  \right.
  & &
  \nonumber
  \\
  & &
  \hspace{-40mm}
  \left.
 -
  \frac{1}{2} \sum_{q}  
  \sum_{\alpha \alpha^{\prime} }
  [ \underline{\tilde{f}}_{ {{q}} }^{-1} ] ^{\alpha \alpha^{\prime} } 
  \phi^{\alpha}_{-q} \phi^{\alpha^{\prime}}_{q}
  \right] 
 = 1
 \; \; \; .
 \label{eq:measure3}
 \end{eqnarray}
Then our two-field decoupling\index{Hubbard-Stratonovich transformation!two-field} 
of the original fermionic two-body interaction reads
 \begin{eqnarray}
 \exp \left[ - S_{\rm int} \{ \psi \} \right] \equiv
 \exp \left[ 
  - \frac{1}{2} \sum_{q}  \sum_{\alpha \alpha^{\prime} }
  [ \underline{\tilde{f}}_{ {{q}} }  ]^{\alpha \alpha^{\prime} } 
  \rho^{\alpha}_{-q} \rho^{\alpha^{\prime}}_{q}
  \right ] 
  =
 \int {\cal{D}} \left\{ \tilde{\rho}^{\alpha} \right\}
 \int {\cal{D}} \left\{ \phi^{\alpha} \right\}
  \nonumber
  \\
  & & \hspace{-107mm}
  \times 
 \exp \left[ 
  - \frac{1}{2} \sum_{q}  \sum_{\alpha \alpha^{\prime} }
  [ \underline{\tilde{f}}_{ {{q}} }  ]^{\alpha \alpha^{\prime} } 
  \tilde{\rho}^{\alpha}_{-q} \tilde{\rho}^{\alpha^{\prime}}_{q}
 + \I  \sum_{q} \sum_{\alpha} \left[ \tilde{{\rho}}_{q}^{\alpha} -
 {\rho}_{q}^{\alpha} \right] \phi_{-q}^{\alpha}  \right]  
  \; .
  \label{eq:HStrans3}
  \end{eqnarray}
Note that $\rho^{\alpha}_{q} = \sum_{k} \Theta^{\alpha} ( {\vec{k}} )
\psi_{k}^{\dagger} \psi_{k+q }$ on the left-hand side
of this equation is a composite Grassmann field, while
$\tilde{\rho}^{\alpha}_{q}$ on the right-hand side is a complex collective bosonic field.
Eq.\ref{eq:HStrans3} can be viewed as a functional generalization of the
elementary identity
 \begin{equation}
 \E^{-x^2} = \int_{- \infty}^{\infty} \D y \int_{- \infty}^{\infty} \frac{ \D \phi}{2 \pi}
 \E^{ - y^2 + \I ( y-x) \phi}
 \label{eq:HS1d}
 \; \; \; .
 \end{equation}
Let us also point out that the two-field Hubbard-Stratonovich transformation
\ref{eq:HStrans3} does not involve the inverse
of the matrix $\underline{\tilde{f}}_{q}$, so that 
it is perfectly well defined for matrices with constant elements.

Applying the Hubbard-Stratonovich transformation \ref{eq:HStrans2}
to the denominator in Eq.\ref{eq:PiHS1} and integrating over the
fermionic $\psi$-field, we obtain with the help of
the ``trace-log'' formula \ref{eq:Fermiint}, 
 \begin{eqnarray}
  \int {\cal{D}} \left\{ \psi \right\} 
 {\cal{D}} \left\{ \phi^{\alpha} \right\}
 \exp \left[ 
 - {S} \left\{ \psi , \phi^{\alpha} \right\} \right]
 =
 \frac{ \E^{ {\rm Tr} \ln \hat{G}_{0}^{-1} }}
 { \int {\cal{D}} \left\{ \tilde{\rho}^{\alpha} \right\}  
 \E^{ - \tilde{S}_{2} \{ \tilde{\rho}^{\alpha} \}  } }
 &  &
 \nonumber
 \\
 & &  \hspace{-88mm}  \times
 \int 
 {\cal{D}} \left\{ \tilde{\rho}^{\alpha} \right\} 
 {\cal{D}} \left\{ \phi^{\alpha} \right\}
 \exp \left[ - \tilde{{S}}_{2} \left\{ \tilde{\rho}^{\alpha} \right\} 
  + \I   \sum_{q \alpha} \phi^{\alpha}_{-q} \tilde{\rho}^{\alpha}_{q} 
  - S_{\rm kin} \{ \phi^{\alpha} \} \right]
  \; ,
  \label{eq:denomint}
  \end{eqnarray}
where the interaction contribution to the effective action of the
collective $\tilde{\rho}^{\alpha}$-field is
 \begin{equation}
 \tilde{S}_{2} \left\{ \tilde{\rho}^{\alpha} \right\}   = 
 \frac{1}{2} \sum_{q} \sum_{\alpha \alpha^{\prime}} 
 [ \underline{\tilde{f}}_{q} ]^{ \alpha \alpha^{\prime} } 
 \tilde{\rho}^{\alpha}_{-q} \tilde{\rho}^{\alpha^{\prime}}_{q}
 \; \; \; ,
 \label{eq:Sintrhodef}
 \end{equation}
and the action $ S_{\rm kin} \{ \phi^{\alpha} \}$ is defined in Eq.\ref{eq:Skinphidef}. 
The relation analogous to Eq.\ref{eq:denomint} for the numerator in Eq.\ref{eq:PiHS1}
is
 \begin{eqnarray}
 { \int {\cal{D}} \left\{ \psi \right\} 
 {\cal{D}} \left\{ \phi^{\alpha} \right\}
 \rho^{\alpha}_{q} \rho^{\alpha^{\prime}}_{-q}
 \exp \left[ - {S} \left\{ \psi , \phi^{\alpha} \right\} \right] }  
 & = &
 \frac{ \E^{ {\rm Tr} \ln \hat{G}_{0}^{-1} }}
 { \int {\cal{D}} \left\{ \tilde{\rho}^{\alpha} \right\}  
 \E^{ - \tilde{S}_{2} \{ \tilde{\rho}^{\alpha} \}  } }
 \nonumber
 \\
 &  &  \hspace{-72mm} \times
 \int 
 {\cal{D}} \left\{ \tilde{\rho}^{\alpha} \right\} 
 {\cal{D}} \left\{ \phi^{\alpha} \right\}
 \tilde{\rho}^{\alpha}_{q} \tilde{\rho}^{\alpha^{\prime}}_{-q}
 \exp \left[ - \tilde{S}_{2} \{ \tilde{\rho}^{\alpha} \} 
  + \I   \sum_{q \alpha} \phi^{\alpha}_{-q} \tilde{\rho}^{\alpha}_{q} 
  - S_{\rm kin} \{ \phi^{\alpha} \} \right]
  \label{eq:enumint}
  \; .
  \end{eqnarray}
To proof Eq.\ref{eq:enumint}, we introduce the generating functional
 \begin{equation}
 {\cal{F}} \{  \tilde{\phi}^{\alpha} \}
 = 
 \int
 {\cal{D}} \left\{ \psi \right\}
 {\cal{D}} \left\{ \phi^{\alpha} \right\}
 \exp \left[ 
  - {S} \left\{ \psi , \phi^{\alpha} \right\}
  + \I  \sum_{q \alpha} 
 \tilde{\phi}^{\alpha}_{-q}  
 \rho^{\alpha}_{q}
 \right] 
 \; \; \; ,
 \label{eq:generatingdef}
 \end{equation}
which depends on external bosonic fields
$\tilde{\phi}^{\alpha}$ and generates via differentiation (up to
a constant factor) the left-hand side of Eq.\ref{eq:enumint},
 \begin{eqnarray}
 \left.
 \frac{ \partial^2 {\cal{F}} \{ \tilde{\phi}^{\alpha} \} }{
 \partial \tilde{\phi}^{\alpha}_{-q} 
 \partial \tilde{\phi}^{\alpha^{\prime}}_{q} }
 \right|_{ \tilde{\phi}^{\alpha} = 0 }
 & = &
  \I^{2}
 \int 
 {\cal{D}} \left\{ \psi  \right\} 
 {\cal{D}} \left\{ \phi^{\alpha} \right\}
 {\rho}^{\alpha}_{q} {\rho}^{\alpha^{\prime}}_{-q}
 \exp \left[ - {S} \left\{ \psi , \phi^{\alpha} \right\} \right]
  \label{eq:generate}
  \; \; \; .
  \end{eqnarray}
Applying the Hubbard-Stratonovich transformation \ref{eq:HStrans2} 
to our generating functional, we obtain
 \begin{eqnarray}
{\cal{ F}} \{  \tilde{\phi}^{\alpha} \}
 & = & 
 \frac{ 1}
 { \int {\cal{D}} \left\{ \tilde{\rho}^{\alpha} \right\}  
 \E^{ - \tilde{S}_{2} \{ \tilde{\rho}^{\alpha} \}  } }
 \int 
 {\cal{D}} \left\{ \tilde{\rho}^{\alpha} \right\} 
 {\cal{D}} \left\{ \phi^{\alpha} \right\}
 \E^{  
 - \tilde{S}_{2} \left\{ \tilde{\rho}^{\alpha} \right\} }
 \nonumber
 \\
 &  & \hspace{-15mm} \times
 \int
 {\cal{D}} \left\{ \psi \right\}
 \exp \left[ - {S}_{0} \{ \psi \}  
 + \I  \sum_{q \alpha} [ \phi^{\alpha}_{-q} \tilde{\rho}^{\alpha}_{q}   
 -  ( \phi^{\alpha}_{-q} - \tilde{\phi}^{\alpha}_{-q} )  \rho^{\alpha}_{q} ]
 \right] 
 \label{eq:generateman}
 \;  .
 \end{eqnarray}
Shifting the integration over the $\phi^{\alpha}$-field according to
$\phi^{\alpha}_{q} \rightarrow \phi^{\alpha}_{q} + \tilde{\phi}^{\alpha}_{q}$,
we replace in the last term of the exponential in Eq.\ref{eq:generateman}
 \begin{equation}
 \phi^{\alpha}_{-q} \tilde{\rho}^{\alpha}_{q}   
 -  ( \phi^{\alpha}_{-q} - \tilde{\phi}^{\alpha}_{-q} )  \rho^{\alpha}_{q} 
 \rightarrow
 \tilde{\phi}^{\alpha}_{-q} \tilde{\rho}^{\alpha}_{q}   
 +  \phi^{\alpha}_{-q} (  \tilde{\rho}^{\alpha}_{ q} -  \rho^{\alpha}_{q} ) 
 \label{eq:shiftreplace}
 \; \; \; ,
 \end{equation}
so that after the shift the derivatives with respect to the external $\tilde{\phi}^{\alpha}$-field
generate factors of the collective bosonic density field $\tilde{\rho}^{\alpha}$.
Performing now the fermionic integration and taking two derivatives with
respect to the external field, we conclude from Eq.\ref{eq:generateman} that
 \begin{eqnarray}
 \left.
 \frac{ \partial^2 {\cal{F}} \{ \tilde{\phi}^{\alpha} \} }{
 \partial \tilde{\phi}^{\alpha}_{-q} 
 \partial \tilde{\phi}^{\alpha^{\prime}}_{q} }
 \right|_{ \tilde{\phi}^{\alpha} = 0 }
 & = &
  \I^2 
 \frac{ \E^{ {\rm Tr} \ln \hat{G}_{0}^{-1} }}
 { \int {\cal{D}} \left\{ \tilde{\rho}^{\alpha} \right\}  
 \E^{ - \tilde{S}_{2} \{ \tilde{\rho}^{\alpha} \}  } }
 \int 
 {\cal{D}} \left\{ \tilde{\rho}^{\alpha} \right\} 
 {\cal{D}} \left\{ \phi^{\alpha} \right\}
 \nonumber
 \\
 &   & \hspace{-20mm} \times
 \tilde{\rho}^{\alpha}_{q} \tilde{\rho}^{\alpha^{\prime}}_{-q}
 \exp \left[ - \tilde{S}_{2} \{ \tilde{\rho}^{\alpha} \} 
  + \I   \sum_{q \alpha} \phi^{\alpha}_{-q} \tilde{\rho}^{\alpha}_{q} 
  - S_{\rm kin} \{ \phi^{\alpha} \} \right]
  \label{eq:generate2}
  \; .
  \end{eqnarray}
Comparing the right-hand sides of Eqs.\ref{eq:generate} and \ref{eq:generate2}, 
the validity of Eq.\ref{eq:enumint} is evident.
In summary, with the help of Eqs.\ref{eq:denomint}, \ref{eq:enumint} and
\ref{eq:PiHS1}
the sector density-density correlation function \ref{eq:cordensloc}
can be represented as
 \begin{eqnarray}
 \Pi^{\alpha  \alpha^{\prime} } ( q ) =
 \nonumber
 \\
 & & \hspace{-20mm} 
  \frac{\beta}{V}
 \frac{
 \int 
 {\cal{D}} \left\{ \tilde{\rho}^{\alpha} \right\} 
 \E^{- \tilde{S}_{2} \left\{ \tilde{\rho}^{\alpha} \right\} }
 \tilde{\rho}^{\alpha}_{q} \tilde{\rho}^{\alpha^{\prime}}_{-q}
 \int {\cal{D}} \left\{ \phi^{\alpha} \right\}
 \exp \left[ 
  \I   \sum_{q \alpha} \phi^{\alpha}_{-q} \tilde{\rho}^{\alpha}_{q} 
  - S_{\rm kin} \left\{ \phi^{\alpha} \right\}  \right]
} 
 { \int
 {\cal{D}} \left\{ \tilde{\rho}^{\alpha} \right\} 
 \E^{- \tilde{S}_{2} \left\{ \tilde{\rho}^{\alpha} \right\} }
 \int {\cal{D}} \left\{ \phi^{\alpha} \right\}
 \exp \left[ 
  \I   \sum_{q \alpha} \phi^{\alpha}_{-q} \tilde{\rho}^{\alpha}_{q} 
 - S_{\rm kin} \left\{ \phi^{\alpha} \right\}  \right]
 }
 \; .
 \nonumber
 \\
 \label{eq:Piphirho}
 \end{eqnarray}

\subsection{Definition of the bosonized kinetic energy\index{bosonization!kinetic energy}}

In complete analogy with Eqs.\ref{eq:avphi}--\ref{eq:Seffphidef},
let us rewrite Eq.\ref{eq:Piphirho} as
 \begin{equation}
 \Pi^{\alpha  \alpha^{\prime} } ( q )
 = \frac{\beta}{V}
 \int {\cal{D}} \{ \tilde{\rho}^{\alpha} \} 
 \tilde{{\cal{P}}} \{ \tilde{\rho}^{\alpha} \} 
 \tilde{\rho}^{\alpha}_{q} \tilde{\rho}^{\alpha^{\prime}}_{-q}
 \equiv 
  \frac{\beta}{V}
 \left< 
 \tilde{\rho}^{\alpha}_{q} \tilde{\rho}^{\alpha^{\prime}}_{-q}
 \right>_{ \tilde{S}_{\rm eff} }
 \label{eq:avrho}
 \; \; \; ,
 \end{equation}
where the normalized probability distribution\index{probability distribution!density field}
 $\tilde{{\cal{P}}} \{ \tilde{\rho}^{\alpha} \}$  
for the collective density field $\tilde{\rho}^{\alpha}$ is
 \begin{equation}
 \tilde{\cal{P}} \{ \tilde{\rho}^{\alpha} \} 
 = 
 \frac{  
 \E^{ - \tilde{S}_{\rm eff} \{ \tilde{\rho}^{\alpha} \} }  }
 {
 \int {\cal{D}} \left\{ \tilde{\rho}^{\alpha} \right\} 
 \E^{ - \tilde{S}_{\rm eff} \{ \tilde{\rho}^{\alpha} \} }  }
 \; \; \; .
 \label{eq:probabrhodef}
 \end{equation}
The effective action\index{effective action!for $\tilde{\rho}^{\alpha}$-field} 
of the $\tilde{\rho}^{\alpha}$-field has again two contributions,
 \begin{equation}
 \tilde{S}_{\rm eff} \left\{ \tilde{\rho}^{\alpha} \right\}
 = 
 \tilde{S}_{2} \left\{ \tilde{\rho}^{\alpha} \right\}
 + \tilde{S}_{\rm kin} \left\{ \tilde{\rho}^{\alpha} \right\}
 \label{eq:Seffrhodef}
 \; \; \; ,
 \end{equation}
with 
 $\tilde{S}_{2} \left\{ \tilde{\rho}^{\alpha} \right\}$
given in Eq.\ref{eq:Sintrhodef}, and
 \begin{equation}
 \tilde{S}_{\rm kin} \left\{ \tilde{\rho}^{\alpha} \right\} 
 = - \ln \left(
 \int {\cal{D}} \left\{ \phi^{\alpha} \right\}
  \exp \left[ \I   \sum_{q \alpha} \phi^{\alpha}_{-q} \tilde{\rho}^{\alpha}_{q} 
  - S_{\rm kin} \left\{ \phi^{\alpha} \right\}  \right] \right)
 \label{eq:Skinrhodef}
 \; \; \; .
 \end{equation}
Note that 
 $\tilde{S}_{\rm kin} \left\{ \tilde{\rho}^{\alpha} \right\} $ is
 related to
$S_{\rm kin} \{  \phi^{\alpha} \}$ via a 
{\it{functional Fourier transformation}}\index{Fourier transformation!functional},
while the quadratic action $\tilde{S}_{2} \{ \tilde{\rho}^{\alpha} \}$
is simply obtained from 
$S_{\rm int} \{ \psi \}$ in
Eq.\ref{eq:Sintpsidef} by replacing the
composite Grassmann field $\rho^{\alpha}$ by the
collective bosonic field $\tilde{\rho}^{\alpha}$.
In this way the effect of the electron-electron interaction
is taken into account exactly,
while the contribution 
$\tilde{S}_{\rm kin} \left\{ \tilde{\rho}^{\alpha} \right\}$
due to the kinetic energy
can in general only be calculated approximately.
In the next chapter
we shall show that 
in the limit of long wavelengths and low energies
the effective action
$\tilde{S}_{\rm eff} \{ \tilde{\rho}^{\alpha} \}$
in Eq.\ref{eq:Seffrhodef}
is equivalent with the bosonized Hamiltonian of the interacting Fermi system.
Obviously
$\tilde{S}_{\rm eff} \{ \tilde{\rho}^{\alpha} \}$
is in general {\it{not}} quadratic, so that the 
equivalent bosonized Hamiltonian contains terms
describing interactions between the bosons.
However, under certain conditions, which will be 
described in detail in Chap.~\secref{sec:closedloop},
$\tilde{S}_{\rm eff} \{ \tilde{\rho}^{\alpha} \}$ can be approximated
by a quadratic form. In this case
bosonization enormously simplifies the many-body problem.
In a sense, the
collective density fields $\tilde{\rho}^{\alpha}$ are the ``correct coordinates''
to parameterize the low-energy excitations of the system.

\section{Summary and outlook}

In this chapter we have used well-known representations of
fermionic correlation functions as Grassmannian functional
integrals and Hubbard-Stratonovich transformations 
to eliminate the fermionic degrees of freedom in favour of bosonic ones.
The only new feature of these transformations is that
our Hubbard-Stratonovich fields carry
not only a momentum-frequency label $q$,
but also a label $\alpha$ 
that refers to the sectors $K^{\alpha}_{\Lambda , \lambda}$ defined
in Chaps.~\secref{subsec:patch} and \secref{sec:sectors}. 
Although our manipulations are formally exact, 
at this point the reader is perhaps rather skeptical
whether they will turn out to be useful to 
obtain truly non-perturbative information about the
interacting many-body system\footnote{As already mentioned, in $d=1$ we have the 
ambitious goal to reproduce the exact solution of the Tomonaga-Luttinger model.}.
After all, the use of Hubbard-Stratonovich transformation is
a well-known technique in the theory of strongly 
correlated systems \cite{Evenson70,Hertz76,Gomes77,Kotliar86,Schulz90}, 
and in practice it is very difficult
to go beyond the saddle point approximation. 
An important exception is a beautiful
paper by Hertz \cite{Hertz76},
which has inspired the development of our functional bosonization approach.
Hertz
used a Hubbard-Stratonovich transformation to derive 
quantum Landau-Ginzburg-Wilson functionals for
interacting Fermi systems,
which form then the basis for a renormalization group analysis.
Our fields $\phi^{\alpha}_q$ are closely related to the
Hubbard-Stratonovich
fields introduced by Hertz; the only difference is that
our fields carry an extra patch index $\alpha$.
As will be shown in Chap.~\secref{chap:agreen}, in this book we shall
be able to {\it{treat the full quantum
dynamics of the Hubbard-Stratonovich field non-perturbatively}} --
we shall neither rely on saddle point approximations, 
nor on the naive perturbative calculation of fluctuation corrections
around saddle points!

%
%

%
%
%

\chapter[Bosonization of the Hamiltonian 
and the density-density \mbox{\hspace{5mm}} 
correlation function] 
{Bosonization of the Hamiltonian \mbox{\hspace{20mm}} 
and the density-density correlation function} 
\chaptermark{Bosonization of the Hamiltonian and $\ldots$}

\label{chap:a4bos}

\setcounter{equation}{0}

{\it{We use our functional integral formalism 
to bosonize the Hamiltonian of an interacting Fermi system with
two-body density-density interactions.
At the level of the Gaussian approximation 
the problem of deriving the bosonic
representation of the Hamiltonian is closely related to the
problem of calculating the density-density correlation function within the RPA.
We develop a general formalism for  
obtaining corrections to the Gaussian approximation, and
show that these are nothing but the local-field corrections
to the RPA. Some of the results presented in this chapter
has been published in \cite{Kopietz95}.
}}

\vspace{7mm}

\noindent
In order to obtain the bosonized effective action
$\tilde{S}_{\rm eff} \{ \tilde{\rho}^{\alpha} \}$
defined in Eq.\ref{eq:Seffrhodef},
it is necessary to calculate
first the effective action $S_{\rm eff} \{ {\phi}^{\alpha} \}$ 
of the $\phi^{\alpha}$-field given in Eq.\ref{eq:Seffphidef}.
Note that the electron-electron interaction
is taken into account exactly via $S_{2} \{ \phi^{\alpha} \}$, so that the
difficulty lies in the calculation of the
kinetic energy contribution  $S_{\rm kin} \{ \phi^{\alpha} \}$.
Similarly, the interaction part $\tilde{S}_{2} \{ \tilde{\rho}^{\alpha} \}$
of the effective action $\tilde{S}_{\rm eff} \{ \tilde{\rho}^{\alpha} \}$ 
for the collective density field
can be obtained trivially by replacing  
$\rho^{\alpha}_{q} \rightarrow 
\tilde{\rho}^{\alpha}_{q} $ 
in the Grassmannian action $S_{\rm int} \{ \psi \}$ 
defined in Eq.\ref{eq:Sintpsidef}.
On the other hand, to obtain the bosonized kinetic energy
$\tilde{S}_{\rm kin} \{ \tilde{\rho}^{\alpha} \}$
it is necessary to perform
the functional Fourier transformation
of $\exp [ - S_{\rm kin} \{  \phi^{\alpha} \}]$ 
in Eq.\ref{eq:Skinrhodef}.

Of course, in general the above kinetic energy contributions
can only be calculated perturbatively by expanding
 \begin{equation}
 S_{\rm kin} \{ \phi^{\alpha} \}  \equiv
 - {\rm Tr} \ln [ 1 - \hat{G}_{0} \hat{V} ]
 = \sum_{n=1}^{\infty} 
 \frac{1}{n} {\rm Tr} \left[ \hat{G}_{0} \hat{V} \right]^n
 \equiv 
 \sum_{n=1}^{\infty} 
 S_{{\rm kin} , n } \{ \phi^{\alpha} \}
 \; ,
 \label{eq:tracelogexp}
 \end{equation}
and truncating the expansion at some finite order.
The functional Fourier transformation in
Eq.\ref{eq:Skinrhodef} should then also be performed perturbatively to this order.
Within the {\it{Gaussian approximation}}\index{Gaussian approximation}  all terms
with $ n \geq 3$ in Eq.\ref{eq:tracelogexp} are neglected, so that
one sets
 \begin{equation}
 S_{\rm kin} \{ \phi^{\alpha} \} \approx   
   {\rm Tr} \left[ \hat{G}_{0} \hat{V} \right]
  + \frac{1}{2}
  {\rm Tr} \left[ \hat{G}_{0} \hat{V} \right]^2
  \label{eq:Trloggauss}
  \; \; \; .
  \end{equation}
Because within this approximation
$S_{\rm kin} \{ \phi^{\alpha} \}$ is a quadratic functional
of the $\phi^{\alpha}$-field\footnote{
As shown in Eq.\ref{eq:Skin1} below,
the term 
   ${\rm Tr} [ \hat{G}_{0} \hat{V} ]$ 
in Eq.\ref{eq:Trloggauss}
gives rise to a contribution that is proportional to the
$q=0$ component of the $\phi^{\alpha}$-field, 
which renormalizes the
$q=0$ component of the collective density field
$\tilde{\rho}^{\alpha}_{q}$ 
(see Eqs.\ref{eq:FTren} and \ref{eq:rhoredef}).
In this work we shall restrict ourselves to the calculation
of zero temperature correlation functions at finite $q $, in which case
possible subtleties associated with these $q=0$ components of the
Hubbard-Stratonovich fields can be ignored.
For the calculation of the free energy a more careful treatment
of these terms is certainly necessary.},
the functional Fourier transformation \ref{eq:Skinrhodef} 
reduces to a trivial Gaussian integration.   Evidently
the effective action $\tilde{S}_{\rm eff} \{ \tilde{\rho}^{\alpha} \}$ of the 
collective density field is then also quadratic.
Note that in the work by 
Houghton {\it{et al.}} \cite{Houghton93,Houghton94} and
Castro Neto and Fradkin \cite{Castro94,Castro95} it is {\it{implicitly assumed
that the Gaussian approximation is justified}}. 
However, in none of these works  the corrections to the Gaussian
approximation have been considered, so that the small parameter which actually
controls the accuracy of the Gaussian approximation has not
been determined.

On the other hand, in the exactly solvable
one-dimensional Tomonaga-Luttinger model\index{Tomonaga-Luttinger model} \cite{Tomonaga50,Luttinger63} 
the bosonized Hamiltonian is known to be quadratic, so that
the expansion in Eq.\ref{eq:tracelogexp} truncates at the second order. In this case
we have {\it{exactly}}
 \begin{equation}
 - {\rm Tr} \ln [ 1 - \hat{G}_{0} \hat{V} ]
 =
   {\rm Tr} \left[ \hat{G}_{0} \hat{V} \right]
  + \frac{1}{2}
  {\rm Tr} \left[ \hat{G}_{0} \hat{V} \right]^2
  \label{eq:Trlogexact1d}
  \; \; \; .
  \end{equation}
All higher order terms
vanish identically due to a large scale cancellation between
self-energy and vertex corrections, which has 
been discovered by Dzyaloshinskii and Larkin \cite{Dzyaloshinskii74}.
A few years later T. Bohr gave a much more readable proof of this
cancellation \cite{Bohr81}, and formulated it 
as a theorem, which he called the {\it{closed loop theorem}}.
In $d=1$ there are certainly alternative (but equivalent) approaches
to the bosonization problem, which do not explicitly make use
of the closed loop theorem \cite{Fogedby76,Lee88}. 
However, we find it advantageous to start from the closed loop theorem,
because then it is very easy to see
that the cancellations responsible for
the validity of Eq.\ref{eq:Trlogexact1d}
in $d = 1$ {\it{exist also in higher dimensions}}, and control 
in the limit of high densities and small momentum-transfers 
the accuracy of the Gaussian approximation in arbitrary $d$.
Following the terminology coined by T. Bohr \cite{Bohr81}, we shall 
describe the mechanism responsible for this cancellation
in terms of a theorem, which we call the
{\it{generalized closed loop theorem}}.

\section{The generalized closed loop theorem\index{closed loop theorem}}
\label{sec:closedloop}

{\it{This is the fundamental reason why bosonization works.}}

\vspace{7mm}

\begin{figure}
\sidecaption
\psfig{figure=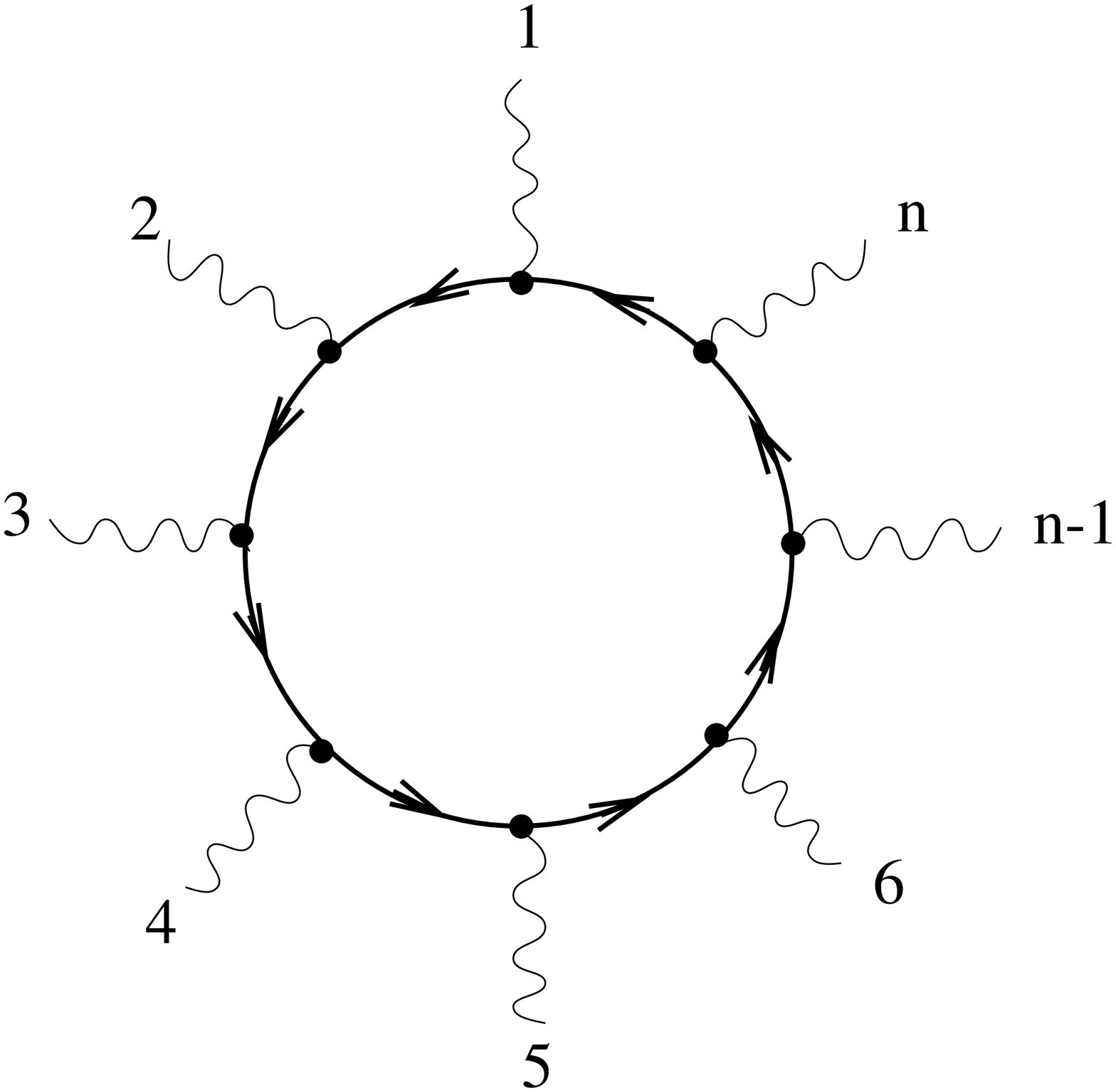,width=5.5cm}
\caption[Feynman diagram representing ${ {\rm Tr} [ \hat{G}_{0} \hat{V} ]^{n}}$.] 
{Feynman diagram representing 
${\rm Tr}[ \hat{G}_{0} \hat{V} ]^{n}$ with $n=8$, see Eqs.\ref{eq:tracelogexp} and
\ref{eq:Seffphin}.
The lines with arrows denote non-interacting fermionic Green's functions, the dots
represent the bare vertex, and
the wavy lines denote external $\phi^{\alpha}$-fields.}
\label{fig:closedloop}
\end{figure}

\noindent
Graphically, the traces ${\rm Tr} [ \hat{G}_{0} \hat{V} ]^n$ 
in Eq.\ref{eq:tracelogexp} can be represented as closed 
fermion loops with $n$ external $\phi^{\alpha}$-fields, as shown in 
Fig.~\secref{fig:closedloop}.
Performing the 
trace of $n^{\rm th}$ order term 
in Eq.\ref{eq:tracelogexp}, we obtain
 \begin{equation}
 S_{{\rm kin},n} \left\{ { { {\phi}^{\alpha}}} \right\}  =  
 \frac{1}{n} \sum_{q_{1}  \ldots q_{n} }
 \sum_{\alpha_{1} \ldots \alpha_{n} } 
 U_{n} ( 
 q_1 \alpha_{1} \ldots q_{n} \alpha_{n}  ) {\phi}^{\alpha_{1}}_{q_{1}} \cdots
 {\phi}^{\alpha_{n}}_{q_{n}}
 \; \; \; ,
 \label{eq:Seffphin}
 \end{equation}
where the dimensionless vertices $U_{n}$ are given by \cite{Hertz74}
 \begin{eqnarray}
 U_{n} ( q_1  \alpha_{1}  \ldots q_{n} \alpha_{n}  ) 
 & = &
\delta_{ {q}_{1} + \ldots
+ {{q}}_{n} , 0 } 
 \left( \frac{\I}{\beta} \right)^n
 \frac{1}{n!} \sum_{P(1 \ldots n)} \sum_{k} 
 \Theta^{\alpha_{P_1}} ( {\vec{k}} )
 \nonumber
 \\
 &  & \hspace{-15mm} \times
 \Theta^{\alpha_{P_2}} ( {\vec{k}} + {\vec{q}}_{P_2} )
 \cdots
 \Theta^{\alpha_{P_n}} ( {\vec{k}} + {\vec{q}}_{P_2} + \ldots + {\vec{q}}_{P_{n}} )
 \nonumber
 \\
 & & \hspace{-15mm} \times 
 G_{0} ( k )
 G_{0} ( k + q_{P_2} ) \cdots
 G_{0} ( k + q_{P_2} + \ldots + q_{P_{n}} )
 \; .
 \label{eq:Uvertex}
 \end{eqnarray}
Here $\delta_{ {q}_{1} + \ldots + q_{n} , 0 }$ denotes a 
Kronecker-$\delta$ in wave-vector and frequency space.
We have used the invariance of $S_{{\rm kin}, n} \left\{ {\phi}^{\alpha} \right\}$
under relabeling of the fields to symmetrize\index{symmetrization} the vertices $U_{ n }$
with respect to the interchange of any two
labels. The sum $\sum_{P(1 \ldots n)}$ is over the $n!$ permutations of $n$ integers, and
$P_{i}$ denotes the image of $i$ under the permutation.
Note that the vertices $U_{n}$ are 
uniquely determined by the energy dispersion 
$\epsilon_{\vec{k}} - \mu$.
The amazing fact is now that there exists a physically interesting limit where
all higher order vertices $U_{n}$ with $n \geq 3$ vanish. 
This limit is characterized by the requirement that the following two approximations $(A1)$ and $(A2)$
become accurate:

\begin{center}
{\bf{
(A1): Diagonal-patch approximation\index{diagonal-patch approximation}}}
\end{center}

\noindent
Let us assume that there exists 
a cutoff $q_{\rm c} \ll  k_{\rm F}$ such
that the contribution from fields $\phi^{\alpha}_{q}$
with $|{\vec{q}} | \geqapprox q_{\rm c}$ to physical observables becomes
negligibly small\footnote{ 
As already mentioned in Chap.~\secref{subsec:proper}, in the case of the
long-range part of the Coulomb interaction $q_{\rm c}$ can be 
identified with the Thomas-Fermi 
wave-vector $\kappa$, which is small compared with $k_{\rm F}$ at high densities.}.
Because the fields $\phi^{\alpha}_{q}$ mediate the interaction 
between the fermions, this 
condition is equivalent with  the requirement that the 
nature of the {\it{bare interaction}}
$\underline{\tilde{f}}_{q}$ should be such that the 
resulting {\it{effective screened interaction}}
(which takes into account the modification of the bare interaction
between two particles due to the presence of all other particles)
is negligibly small for $|{\vec{q}} | \geqapprox q_{\rm c}$.
If this condition is satisfied, we
may approximate in Eq.\ref{eq:Uvertex}
 \begin{eqnarray}
 \lefteqn{\Theta^{\alpha_{P_1}} ( {\vec{k}} )
 \Theta^{\alpha_{P_2}} ( {\vec{k}} + {\vec{q}}_{P_2} )
 \cdots
 \Theta^{\alpha_{P_n}} ( {\vec{k}} + {\vec{q}}_{P_2} + \ldots + {\vec{q}}_{P_{n}} ) }
 \nonumber
 \\
 & &
 \approx 
 \delta^{ \alpha_{P_{1}} \alpha_{P_{2}} }
 \delta^{ \alpha_{P_{1}} \alpha_{P_{3}} }
 \cdots
 \delta^{ \alpha_{P_{1}} \alpha_{P_{n}} }
 \Theta^{\alpha_{P_1}} ( {\vec{k}} )
 \label{eq:highdens}
 \; \; \; ,
 \end{eqnarray}
because the ${\vec{k}}$-sum in Eq.\ref{eq:Uvertex} is
dominated by wave-vectors of the order of $k_{\rm F}$.
This approximation is correct to leading order in
$q_{\rm c} / k_{\rm F}$, and becomes exact in the limit
$q_{\rm c} / k_{\rm F} \rightarrow 0$.
Note that this limit is approached either at high densities,
where $k_{\rm F} \rightarrow \infty$ at constant $q_{\rm c}$, or in the limit
that the range $q_{\rm c}$ of the effective interaction in momentum space
approaches zero while $k_{\rm F}$ is held constant.
It follows that, up to higher order corrections in $q_{\rm c} / k_{\rm F}$, the vertex
 $U_{n} ( q_1  \alpha_{1}  \ldots q_{n} \alpha_{n}  ) $ is diagonal in all
 patch labels, 
 \begin{equation}
 U_{n} ( q_1  \alpha_{1}  \ldots q_{n} \alpha_{n}  ) 
 = \delta^{\alpha_{1} \alpha_{2} } \cdots \delta^{\alpha_{1} \alpha_{n}}
 U_{n}^{\alpha_{1}} ( q_{1} \ldots q_{n} )
 \label{eq:Undiag}
 \; \; \; ,
 \end{equation}
with
 \begin{eqnarray}
 U_{n}^{\alpha} ( q_{1} \ldots q_{n} )
 &  =  &
\delta_{ {q}_{1} + \ldots
+ {{q}}_{n} , 0 } 
 \left( \frac{\I}{\beta} \right)^n
 \frac{1}{n!} \sum_{P(1 \ldots n)} \sum_{k} 
 \Theta^{\alpha} ( {\vec{k}} )
 \nonumber
 \\
 & \times &
 G_{0} (k) 
 G_{0} ( k + q_{P_2} ) \cdots
 G_{0} ( k + q_{P_2} + \ldots + q_{P_{n}} )
 \; \; \; .
 \label{eq:Uvertexdiag}
 \end{eqnarray}
Below we shall refer to the approximation \ref{eq:highdens} as the
{\it{diagonal-patch approximation}}.
It is important to note that at finite $q_{\rm c} / k_{\rm F}$ this approximation
can only become exact  in $d=1$, because in this case the Fermi surface
consists of two widely separated points. 
Except for special cases
(see Chap.~\secref{chap:apatch}), in higher dimensions
the covering of the Fermi surface involves always some adjacent patches, 
which can be connected by arbitrarily small
momentum-transfers ${\vec{q}}$. 
These around-the-corner processes\index{around-the-corner processes} 
are ignored within the diagonal-patch approximation $(A1)$. 
As discussed in detail in Chap.~\secref{subsec:proper}, this is only
justified if the sector cutoffs $\Lambda$ and $\lambda$ are chosen 
large compared with $q_{\rm c}$.

\begin{center}
{\bf{(A2): Local linearization\index{linearization of energy dispersion} 
of the energy dispersion}}
\end{center}

\noindent
Suppose we put the origins ${\vec{k}}^{\alpha}$ of our local
coordinate systems on the Fermi surface
(so that $\epsilon_{\vec{k}^{\alpha} } = \mu$), and locally linearize
the energy dispersion, 
$\xi^{\alpha}_{\vec{q}} \equiv 
\epsilon_{ {\vec{k}}^{\alpha} + {\vec{q}} } -  \mu
\approx {\vec{v}}^{\alpha} \cdot {\vec{q}}$ (see
Eq.\ref{eq:xialphaqexp}).
Inserting 
unity in the form \ref{eq:partition}
into the non-interacting matter action
$S_{0} \{ \psi \}$ defined in Eq.\ref{eq:S0psidef},
we see that the linearization amounts to replacing
 \begin{equation}
 S_{0} \{ \psi \} \approx \beta \sum_{k} \sum_{\alpha}
 \Theta^{\alpha} ( {\vec{k}} ) [ - \I \tilde{\omega}_{n} +
 {\vec{v}}^{\alpha} \cdot ( {\vec{k}} - {\vec{k}}^{\alpha} ) ]
 \psi^{\dagger}_{k} \psi_{k}
 \label{eq:S0psiapprox}
 \; \; \; .
 \end{equation}
Thus, the Fermi surface is
approximated by a collection of {\it{flat}}  $d-1$-dimensional
hyper-surfaces, i.e. planes in $d=3$ and straight lines in $d=2$.
The corresponding non-interacting 
Green's function 
is then approximated by
 \begin{equation}
 G_{0} ( {\vec{k}}^{\alpha} + {\vec{q}} , \I \tilde{\omega}_{n} )
 \equiv
 G^{\alpha}_{0} ( \tilde{q})  
 \approx 
  \frac{1}{ \I \tilde{\omega}_{n} -  {\vec{v}}^{\alpha} \cdot {\vec{q}} }
 \label{eq:G0alphashiftdef}
 \; \; \; .
 \end{equation}
Shifting the summation wave-vector in Eq.\ref{eq:Uvertexdiag} according to
${\vec{k}} = {\vec{k}}^{\alpha} + {\vec{q}}$,
we obtain
 \begin{eqnarray}
 U_{n}^{\alpha} ( q_{1} \ldots q_{n} )
  & = &
\delta_{ {q}_{1} + \ldots
+ {{q}}_{n} , 0 } 
 \left( \frac{\I}{\beta} \right)^n
 \frac{1}{n!} \sum_{P(1 \ldots n)} \sum_{ \tilde{q}} 
 \Theta^{\alpha} ( {\vec{k}}^{\alpha} + {\vec{q}} )
 \nonumber
 \\
 & \times &
 G_{0}^{\alpha} ( \tilde{q}) 
 G_{0}^{\alpha} ( \tilde{q} + q_{P_2} ) \cdots
 G_{0}^{\alpha} ( \tilde{q} + q_{P_2} + \ldots + q_{P_{n}} )
 \; \; \; .
 \label{eq:Uvertexdiag2}
 \end{eqnarray}
Recall that we have introduced  
the convention that $\tilde{q} = [ {\vec{q}} , \I \tilde{\omega}_{n} ]$ 
labels
{\it{fermionic}} Matsubara frequencies,
while $q = [ {\vec{q}} , \I \omega_{m} ]$ 
labels {\it{bosonic}} ones.
Because the sum of a bosonic and a fermionic
Matsubara frequency is a fermionic one, the external
labels $q_{1}, \ldots , q_{n}$ in Eq.\ref{eq:Uvertexdiag2} 
depend on bosonic frequencies.

\vspace{7mm}

Having made the approximations $(A1)$ and $(A2)$, we are 
now ready to show that {\it{in arbitrary dimensions}}
the vertices
 $U_{n}^{\alpha} ( q_{1} \ldots q_{n} ) $
with $n \geq 3$ vanish in the limit
$q_{\rm c} \rightarrow 0$,
so that in this limit {\it{the Gaussian approximation becomes exact!}}
As already mentioned,
in the context of the Tomonaga-Luttinger model the vanishing of the
$U_{n}$ for $n \geq 3$ has been called {\it{closed loop theorem}}, 
and is discussed
and proved in unpublished lecture notes by T. Bohr \cite{Bohr81}.
Under the assumptions $(A1)$ and $(A2)$ 
the proof goes through in any dimension without changes.
Note that the validity of $(A1)$ and $(A2)$ is implicitly built into the Tomonaga-Luttinger model 
by definition. 
The vanishing of $U_{n}$ for $n \geq 3$ is equivalent with the statement that 
the RPA for the density-density correlation function is exact in this model.
This is due to a complete cancellation between self-energy and vertex
corrections \cite{Dzyaloshinskii74}.
In \cite{Bohr81} the proof is formulated in the space-time domain, but
for our purpose 
it is more convenient to work in momentum space, 
because here the Fermi surface and the patching construction are defined.
The following two
properties of our linearized  non-interacting Green's function
in Eq.\ref{eq:G0alphashiftdef} are essential,
 \begin{equation}
 G_{0}^{\alpha} ( -  \tilde{q} ) =  - G_{0}^{\alpha } ( \tilde{q} )
 \; \; \; ,
 \label{eq:godd}
 \end{equation}
 \begin{equation}
 G_{0}^{\alpha} ( \tilde{q} ) G_{0}^{\alpha} ( \tilde{q} + q^{\prime} )
 = G^{\alpha}_{0} ( q^{\prime} )
 \left[ G_{0}^{\alpha} ( \tilde{q} ) - G_{0}^{\alpha} ( \tilde{q} + q^{\prime} )
 \right]
 \; \; \; .
 \label{eq:Gpartialfrac}
 \end{equation}
Note that Eq.\ref{eq:godd} follows trivially from the definition 
\ref{eq:G0alphashiftdef},  
while Eq.\ref{eq:Gpartialfrac} is nothing but the partial fraction decomposition
of the product of two rational functions.
To show that the odd vertices $U_{3}, U_{5}, \ldots$ vanish, we only need Eq.\ref{eq:godd} and the
fact that the sector $K^{\alpha}_{\Lambda , \lambda}$ in Eq.\ref{eq:Uvertexdiag2} has inversion symmetry
with respect to ${\vec{k}}^{\alpha}$, so that the domain for the ${\vec{q}}$-sum is invariant 
under ${\vec{q}} \rightarrow - {\vec{q}}$.
Then it is easy to see that the contribution from a given permutation
$(P_{1} P_{2} \ldots  P_{n-1} P_{n})$ is exactly cancelled
by the contribution from the permutation 
$(P_{n} P_{n-1}  \ldots P_{2} P_{1} )$ in which
the loop is traversed in the 
opposite direction.
As already pointed out by T. Bohr \cite{Bohr81}, the vanishing of the odd vertices
is a direct consequence of Furry's theorem \cite{Itzykson80}.
To show that the even vertices $U_{n}$, $n=4,6, \ldots$ vanish,
we use Eq.\ref{eq:Gpartialfrac} $n$-times for the pairs
 \begin{equation}
 \begin{array}{l}
 G_{0}^{\alpha} ( \tilde{q} ) G_{0}^{\alpha} ( \tilde{q} + q_{P_{2}} ) \; \; \; , \\
 G_{0}^{\alpha} ( \tilde{q} +q_{P_{2}} ) G_{0}^{\alpha} ( \tilde{q}+ q_{P_{2}} +q_{P_{3}} ) 
 \; \; \; ,  \\
 \ldots \; \; \; , \\
 G_{0}^{\alpha} ( \tilde{q} +q_{P_{2}} + \ldots + q_{P_{n-1}}) 
 G_{0}^{\alpha} ( \tilde{q} +q_{P_{2}} + \ldots + q_{P_{n}}) 
 \; \; \; , \\
 G_{0}^{\alpha} ( \tilde{q} +q_{P_{2}} + \ldots + q_{P_{n}}) G_{0}^{\alpha} ( \tilde{q})
 \; \; \; ,
 \end{array}
 \label{eq:G0pair}
 \end{equation}
and take into account that we may replace
 $q_{P_{2}} + \ldots + q_{P_{n}} = - q_{P_{1}}$ because of overall
energy-momentum conservation. 
Using the fact that
in Eq.\ref{eq:Uvertexdiag2} we sum over all permutations, it is
easy to show that under the summation sign the second line
in Eq.\ref{eq:Uvertexdiag2} can be replaced by
 \begin{eqnarray}
 \lefteqn{ G_{0}^{\alpha} ( \tilde{q}) 
 G_{0}^{\alpha} ( \tilde{q} + q_{P_2} ) \cdots
 G_{0}^{\alpha} ( \tilde{q} + q_{P_2} + \ldots + q_{P_{n}} ) }
 \nonumber
 \\
 &  & \hspace{-7mm} \rightarrow
 \frac{1}{n} 
 \left[ G_{0}^{\alpha} ( q_{P_1} ) - G_{0}^{\alpha} ( q_{P_2} ) \right] 
 \left\{
 G_{0}^{\alpha} ( \tilde{q} + q_{P_2} ) \cdots
 G_{0}^{\alpha} ( \tilde{q} + q_{P_2} + \ldots + q_{P_{n}} )
 \right\}
 \label{eq:Uevencancel}
 \; .
 \end{eqnarray}
Substituting Eq.\ref{eq:Uevencancel} in Eq.\ref{eq:Uvertexdiag2},
noting that after the shift
$\tilde{q} \rightarrow \tilde{q} -q_{P_{2}} + q_{P_{1}}$ 
of the summation label the factor in the curly braces
in Eq.\ref{eq:Uevencancel}
can be replaced by the symmetrized (with respect to
$q_{P_1} \leftrightarrow q_{P_2}$) expression
 \begin{equation}
 \frac{1}{2} \left\{
 G_{0}^{\alpha} ( \tilde{q} + q_{P_2} ) \cdots
 G_{0}^{\alpha} ( \tilde{q} + q_{P_2} + \ldots + q_{P_{n}} )
  + \left[ q_{P_1} \leftrightarrow q_{P_2} \right] 
 \right\}
 \label{eq:shiftsymmetrize}
 \; \; \; ,
 \end{equation}
and finally using again
the fact that we may rename 
$q_{P_{1}} \leftrightarrow q_{P_{2}}$ 
because we sum
over all permutations, 
it is easy to see that the
resulting expression vanishes due to the antisymmetry
of the first factor on the right-hand side of Eq.\ref{eq:Uevencancel}.
This argument is not valid for $n=2$, because
in this case $G^{\alpha}_{0} ( q_{P_{1}} )  - G^{\alpha}_{0} ( q_{P_{2}} )
= 2 
G^{\alpha}_{0} ( q_{P_{1}} ) $
due to energy-momentum conservation. 
We shall discuss the vertex $U_{2}$ in detail in Sect.~\secref{subsec:gaussphi}.
Note that the shift
${\tilde{q}} \rightarrow {\tilde{q}} - {{q}}_{P_2} + {{q}}_{P_1}$ affects
also the patch cutoff, $\Theta^{\alpha} ( {\vec{k}}^{\alpha} + {\vec{q}} )
\rightarrow
\Theta^{\alpha} ( {\vec{k}}^{\alpha} + {\vec{q}} 
- {\vec{q}}_{P_2} + {\vec{q}}_{P_1} )$, 
but this leads to corrections of higher order in $q_{\rm c}$.
Because we have already ignored
higher order terms in $ q_{\rm c} $ by making the diagonal-patch approximation $(A1)$,
it is consistent to ignore this 
shift.
We would like to encourage the reader to explicitly verify
the above manipulations for the simplest
non-trivial case $n = 4$.

In fermionic language, the vanishing of the higher order vertices is due to
a {\it{complete cancellation between self-energy and 
vertex corrections}}.\index{vertex corrections!cancellation}
This cancellation is automatically incorporated in our 
bosonic formulation via the symmetrization of the vertices $U_{n}$.
We would like to emphasize again that 
this remarkable cancellation happens 
not only in $d=1$ \cite{Dzyaloshinskii74,Bohr81}
but in arbitrary dimensions\footnote{
It should be mentioned that
recently W. Metzner has independently given an alternative proof
of the generalized closed loop theorem in $d > 1$ \cite{Metzner95}. His approach is based on
operator identities for the
sector density operators $\hat{\rho}^{\alpha}_{\vec{q}}$ defined in Eq.\ref{eq:rhoopdef},
and the resulting consequences for time-ordered expectation values of 
products of these operators.}.
The existence of these cancellations in the perturbative
calculation of the dielectric function of the homogeneous
electron gas\index{homogeneous electron gas} in $d=3$ has already been noticed by
Geldart and Taylor more than 20 years ago \cite{Geldart70},
although the origin for this cancellation has not been
identified. The generalized closed loop theorem 
discussed here gives a clear mathematical explanation
for this cancellation {\it{to all orders in perturbation theory}}.
It is important to stress that the
cancellation does not depend on the nature
of the external fields that enter the closed loop; 
in particular, it occurs  also in models
where the fermionic current density is coupled
to transverse gauge fields (see Chap.~\secref{chap:arad}).
The one-loop corrections to the RPA for the gauge invariant two-particle Green's functions of
electrons interacting with transverse gauge fields
have recently been calculated by Kim {\it{et al.}} \cite{Kim94}.
They found that at long wavelengths and low frequencies the
leading self-energy and vertex corrections cancel.
In the light of the generalized closed loop theorem 
this cancellation is not surprising. However,
the generalized closed loop theorem is a much
stronger statement, because it implies
a cancellation between the leading self-energy and vertex corrections
to all orders in perturbation theory.

\section{The Gaussian approximation\index{Gaussian approximation}}
\label{sec:Gauss}

{\it{
We now calculate the density-density correlation
function and the bosonized Hamiltonian within the Gaussian approximation.
We also show that at long wavelengths the resulting bosonized Hamiltonian  
agrees with the corresponding Hamiltonian 
derived via the conventional 
operator approach \cite{Houghton93,Castro95}.}}

\subsection{The effective action for the $\phi^{\alpha}$-field}
\label{subsec:gaussphi}

Within the Gaussian 
approximation\index{effective action!background field, Gaussian approximation} 
the expansion for
the kinetic energy contribution $S_{\rm kin} \{ \phi^{\alpha} \}$ 
to the effective action for the $\phi^{\alpha}$-field in Eq.\ref{eq:tracelogexp}
is truncated at the second order (see Eq.\ref{eq:Trloggauss}),
so that the effective action \ref{eq:Seffphidef} is approximated by
 \begin{eqnarray}
 S_{\rm eff} \{ \phi^{\alpha} \} &  \approx &
 S_{2} \{ \phi^{\alpha} \} +
 S_{{\rm kin},1} \{ \phi^{\alpha} \} + S_{{\rm kin},2} \{ \phi^{\alpha} \}
 \nonumber
 \\
 & =  & 
 \frac{1}{2} \sum_{q} \sum_{\alpha \alpha^{\prime} }
  [ \underline{\tilde{f}}_{ {{q}} }^{-1} ]^{ \alpha \alpha^{\prime} }
 \phi_{-q}^{\alpha} \phi_{q}^{\alpha^{\prime}}
 \nonumber
 \\
 & + &
 \sum_{q} \sum_{\alpha} U_{1} ( q \alpha )
 \phi_{q}^{\alpha}
  + \frac{1}{2} \sum_{q_{1} q_{2}} \sum_{\alpha \alpha^{\prime}} U_{2}
 ( q_{1} \alpha , q_{2} \alpha^{\prime} ) \phi^{\alpha}_{q_{1}} \phi^{\alpha^{\prime}}_{q_{2}}
 \; \; \; .
 \label{eq:Seffphigauss}
 \end{eqnarray}
The generalized closed loop theorem implies that
the Gaussian approximation is justified
in a parameter regime where the approximations $(A1)$ and $(A2)$ 
discussed in Sect.~\secref{sec:closedloop}
are accurate. 

We now calculate the vertices $U_{1}$ and $U_{2}$.
From Eq.\ref{eq:Uvertex} we obtain 
 \begin{equation}
 U_{1} ( q \alpha )  =  \delta_{q,0} \frac{\I}{\beta} \sum_{k} 
 \Theta^{\alpha}  ( {\vec{k}} ) 
 \frac{1}{ \I \tilde{\omega}_{n} - \xi_{\vec{k}}  }
  =  
  \I \delta_{q,0}  N^{\alpha}_{0}
 \label{eq:U1}
 \; \; \; ,
 \end{equation}
where
 \begin{equation}
 N^{\alpha}_{0} =
 \sum_{\vec{k}} 
 \Theta^{\alpha } ( {\vec{k}} ) 
 f (  \xi_{\vec{k}}  )
 \label{eq:N0def}
 \end{equation}
is the number of occupied states in 
sector $K^{\alpha}_{\Lambda , \lambda}$ in the non-interacting limit.
Thus, 
 \begin{equation}
 S_{{\rm kin},1} \left\{ \phi^{\alpha} \right\} = 
 \I  \sum_{\alpha}
  \phi^{\alpha}_{0} N^{\alpha}_{0}
 \; \; \; .
 \label{eq:Skin1}
 \end{equation}
The second-order vertex is given by
 \begin{eqnarray}
 U_{2} ( q_{1} \alpha , q_{2} \alpha^{\prime} )
 & =  & - \delta_{q_{1} + q_{2} , 0 }
 \frac{1}{2 \beta^2} \sum_{k} 
 \left[ 
 \Theta^{\alpha} ( {\vec{k}} ) \Theta^{\alpha^{\prime}} ( {\vec{k}}+ {\vec{q}}_{2} ) 
 G_{0} ( k ) G_{0} ( k + q_{2} )
 \right.
 \nonumber
 \\
 & & +
 \left.
 \Theta^{\alpha^{\prime}} ( {\vec{k}} ) \Theta^{\alpha} ( {\vec{k}}+ {\vec{q}}_{1} ) 
 G_{0} ( k ) G_{0} (  k + q_{1} )
 \right]
 \label{eq:U2def}
 \; \; \; .
 \end{eqnarray}
Performing the frequency sum 
we obtain
 \begin{equation}
 U_{2}( -q \alpha ,q \alpha^{\prime} ) = 
\frac{V}{\beta}  
\Pi_{0}^{\alpha \alpha^{\prime} } (q) 
\equiv \tilde{\Pi}_{0}^{\alpha \alpha^{\prime}} ( q )
\label{eq:U2res}
\; \; \; ,
\end{equation}
where $\Pi_{0}^{\alpha \alpha^{\prime} } ( q )$ is the
non-interacting sector polarization\index{polarization!sector}, see Eq.\ref{eq:Pi0loc}.
We conclude that $S_{{\rm kin},2} \{ \phi^{\alpha} \}$ is given by
 \begin{equation}
 S_{{\rm kin},2} \{ \phi^{\alpha} \} =
 \frac{1}{2 } \sum_{q} \sum_{\alpha \alpha^{\prime}} 
 \tilde{\Pi}_{0}^{\alpha \alpha^{\prime}} (q)
 \phi^{\alpha}_{-q} \phi^{\alpha^{\prime}}_{q}
 \label{eq:Skin2phires}
 \; \; \; .
 \end{equation}
For $|{\vec{q}}| \ll k_{\rm F}$ the diagonal-patch approximation
$(A1)$ is justified, 
so that we may replace 
$ \Theta^{\alpha} ( {\vec{k}} ) \Theta^{ \alpha^{\prime} }
( {\vec{k+q}} )  \approx 
\delta^{\alpha \alpha^{\prime}}
\Theta^{\alpha} ( {\vec{k}} )  $. 
To leading order in $|{\vec{q}} | / k_{\rm F}$ we have therefore in any dimension
 \begin{equation}
 \Pi^{\alpha \alpha^{\prime} }_{0} ( q ) \approx
 \delta^{\alpha \alpha^{\prime} } \Pi^{\alpha}_{0} ( q )
 \; \; \; , \; \; \; 
 \Pi^{\alpha}_{0} ( q) =
 \nu^{\alpha} 
 \frac{  {\vec{v}}^{\alpha} \cdot {\vec{q}} }
 {  {\vec{v}}^{\alpha} \cdot {\vec{q}} - \I \omega_{m} }
 \; \; \; ,
 \label{eq:Pilong}
 \end{equation}
where
 \begin{equation}
 \nu^{\alpha} = \frac{1}{V} \frac{ \partial N^{\alpha}_{0} }{\partial \mu }
 =
  \frac{1}{V} \sum_{\vec{k}} \Theta^{\alpha} ( {\vec{k}} ) 
  \left[ - \frac{ \partial f ( \xi_{\vec{k}} ) }{\partial \xi_{\vec{k}} } \right]
 \label{eq:nualphadef}
 \end{equation}
is the {\it{local}} (or patch) density of states\index{density of states!local (or patch)} 
associated with sector $K^{\alpha}_{\Lambda , \lambda }$, 
and ${\vec{v}}^{\alpha}$ is the local Fermi velocity\index{Fermi velocity}
(see Eq.\ref{eq:v0cijdef}).
Note that the approximation \ref{eq:Pilong} is 
valid  for small $|{\vec{q}}| / k_{\rm F}$ but for {\it{arbitrary}} frequencies.
The patch density of states $\nu^{\alpha}$ is proportional to
$\Lambda^{d-1}$, i.e. in dimensions $d > 1$ it is a cutoff-dependent quantity.
To see this more clearly, we take the limit
$\beta \rightarrow \infty$, $V \rightarrow \infty$ in Eq.\ref{eq:nualphadef} and convert the
volume integral over the $\delta$-function into a surface integral in the usual way,
 \begin{equation}
 \nu^{\alpha} = \int_{  K^{\alpha}_{\Lambda , \lambda} }
 \frac{ \D {\vec{k}} }{ ( 2 \pi )^{d} } \delta ( \xi_{ \vec{k}} )
 = \int_{  P^{\alpha}_{\Lambda} }
 \frac{ \D S_{\vec{k}} }{ ( 2 \pi )^d} \frac{1}{ | \nabla_{\vec{k}} \xi_{\vec{k}} | }
 \label{eq:patchdenscutoff}
 \; \; \; ,
 \end{equation}
where the $d-1$-dimensional surface integral is over the patch
$P^{\alpha}_{\Lambda}$, i.e. the intersection
of the sector $K^{\alpha}_{\Lambda , \lambda}$ with the Fermi surface.
Using now the fact that 
 $| \nabla_{\vec{k}} \xi_{\vec{k}} |  \approx | {\vec{v}}^{\alpha} |$
for  ${\vec{k}} \in P^{\alpha}_{\Lambda}$, and that for linearized energy
dispersion the area of $P^{\alpha}_{\Lambda}$ is by construction
given by $\Lambda^{d-1}$, we have in $d$ dimensions
 \begin{equation}
 \nu^{\alpha} \approx \frac{ \Lambda^{d-1}}{ ( 2 \pi )^d | {\vec{v}}^{\alpha} | }
 \; \; \; .
 \label{eq:patchdenscutoff2}
 \end{equation}
On the other hand, we shall show in this work that physical quantities
depend only on the global density of states\index{density of states!total}
(or some weighted average of the $\nu^{\alpha}$),
 \begin{equation}
 \nu = \sum_{\alpha} \nu^{\alpha} = 
 \frac{1}{V} \sum_{\vec{k}}
 \left[ - \frac{ \partial f ( \xi_{\vec{k}} ) }{ \partial \xi_{\vec{k}} }
 \right]
 \; \; \; ,
 \label{eq:nudef}
 \end{equation}
which is manifestly cutoff-independent.

In summary, within the Gaussian approximation the effective action 
of the $\phi^{\alpha}$-field is given by
 \begin{equation}
 S_{\rm eff} \{ \phi^{\alpha} \}    \approx 
 \I  \sum_{\alpha}
  \phi^{\alpha}_{0} N^{\alpha}_{0}
  + S_{{\rm eff},2} \{ \phi^{\alpha} \}
  \label{eq:Seffphigaussres}
  \; \; \; ,
  \end{equation}
with
 \begin{eqnarray}
   S_{{\rm eff},2} \{ \phi^{\alpha} \}
   & \equiv & S_{2} \{ \phi^{\alpha} \}
   + S_{{\rm kin},2} \{ \phi^{\alpha} \}
   \nonumber
   \\
   & = & 
 \frac{1}{2} \sum_{q} \sum_{\alpha \alpha^{\prime} }
 [    \underline{\tilde{f}}_{ {{q}} }^{-1} 
 + \underline{\tilde{\Pi}}_{0} (q) ]^{ \alpha \alpha^{\prime} }
 \phi_{-q}^{\alpha} \phi_{q}^{\alpha^{\prime}}
 \label{eq:Seff2phigaussres}
 \; \; \; ,
 \end{eqnarray}
where the elements of the matrix $\underline{ \tilde{\Pi}}_{0} (q)$ 
are defined by
$ [ \underline{ \tilde{\Pi}}_{0} (q) ]^{\alpha \alpha^{\prime}}
= \tilde{\Pi}_{0}^{\alpha \alpha^{\prime}} ( q )$, 
with $\tilde{\Pi}_{0}^{\alpha \alpha^{\prime}} ( q )$ 
given in Eq.\ref{eq:U2res}.

\subsection{The Gaussian propagator 
of the $\phi^{\alpha}$-field}
\label{subsec:theRPAint}

{\it{\ldots which is also known under the name RPA interaction.}}

\vspace{7mm}

\noindent
Within the Gaussian approximation the propagator of the 
$\phi^{\alpha}$-field\index{Gaussian propagator!of $\phi^{\alpha}$-field} 
is simply given by
 \begin{equation} 
 \left< \phi^{\alpha}_{q} \phi^{\alpha^{ \prime}}_{-q} \right>_{S_{{\rm eff},2}}
 = 
 \left[  [    \underline{\tilde{f}}_{ {{q}} }^{-1} 
 + \underline{\tilde{\Pi}}_{0} (q) ]^{-1} \right]^{ \alpha \alpha^{\prime} } 
 \label{eq:gaupropphi}
 \; \; \; ,
 \end{equation}
where the averaging $ < \ldots >_{S_{{\rm eff} ,2 }}$ is defined as
in Eqs.\ref{eq:avphi} and \ref{eq:probabphidef}, with
$S_{\rm eff} \{ \phi^{\alpha } \}$ approximated by
$S_{{\rm eff},2} \{ \phi^{\alpha } \}$.
As already mentioned in the footnote after Eq.\ref{eq:Trloggauss}, the first
term in Eq.\ref{eq:Seffphigaussres} involving the
$q=0$ component of the $\phi^{\alpha}$-field does not contribute
to correlation functions at finite $q$.
From Eqs.\ref{eq:ftildematdef} and \ref{eq:U2res} we have
 $\underline{\tilde{f}}_{q}  =  \frac{\beta}{V} \underline{f}_{q}$ and
 $\underline{\tilde{\Pi}}_{0} ( q )  =  \frac{V}{\beta} \underline{\Pi}_{0} ( q )$,
so that Eq.\ref{eq:gaupropphi} implies
 \begin{equation}
 \left< \phi^{\alpha}_{q} \phi^{\alpha^{ \prime}}_{-q} \right>_{S_{{\rm eff},2}}
 =  \frac{\beta}{V} [ \underline{f}^{{\rm RPA}}_{q} ]^{\alpha \alpha^{\prime}}
  \label{eq:phiphiprop}
  \; \; \; ,
  \end{equation}
where the RPA interaction matrix\index{random-phase approximation!effective interaction} 
$ \underline{f}^{{\rm RPA}}_{ q} $ is defined via
 \begin{equation}
  \underline{f}^{{\rm RPA}}_{q}
  = \left[ 
  \underline{{f}}_{ {{q}} }^{-1} 
 + \underline{{\Pi}}_{0} (q) \right]^{-1} 
 =
  \underline{{f}}_{ {{q}} } 
 \left[ 1 + 
  \underline{{\Pi}}_{0} (q) 
 \underline{{f}}_{q}  \right]^{-1} 
 \label{eq:frpapatchdef}
 \; \; \; .
 \end{equation}
Thus, the Gaussian propagator of the
$\phi^{\alpha}$-field is (up to a factor of ${\beta}/{V}$)
given by the RPA interaction matrix $\underline{f}^{{\rm RPA}}_{q}$.
In the special case that all matrix elements of the
bare interaction are identical, $[ \underline{f}_{q} ]^{\alpha \alpha^{\prime}}
= f_{q}$, the matrix elements 
$[ \underline{f}^{{\rm RPA}}_{{q}} ]^{\alpha \alpha^{\prime}}$ are also
independent of the patch indices, and can be identified with the usual
RPA interaction.
To see this, we expand Eq.\ref{eq:frpapatchdef} as a Neumann series
 \begin{equation}
 {\underline{f}}^{{\rm RPA}}_{q} = 
  \underline{{f}}_{q}
 - \underline{{f}}_{q} \underline{{\Pi}}_{0} (q )
  \underline{{f}}_{q}
  +
 \underline{{f}}_{q} \underline{{\Pi}}_{0} (q )
 \underline{{f}}_{q} \underline{{\Pi}}_{0} ( q)
  \underline{{f}}_{q}
  - \ldots
  \; \; \; ,
  \label{eq:Neumann2}
  \end{equation}
and then take matrix elements term by term. Using the fact that
all matrix elements of $\underline{f}_{q}$ are
identically given by $f_{q}$, we may sum the series again and obtain
the usual RPA interaction,
 \begin{equation}
 [ {\underline{f}}^{{\rm RPA}}_{q} ]^{\alpha \alpha^{\prime}} = 
 f^{{\rm RPA}}_{ q } \equiv
 \frac{ f_{q} }{1 + f_{q} \Pi_{0} (q) }
 \label{eq:frpatot}
 \; \; \; ,
 \end{equation}
where
 \begin{equation}
 \Pi_{0} (q)    
 = \sum_{\alpha \alpha^{\prime}} \Pi_{0}^{\alpha \alpha^{\prime}} (q)
 \label{eq:pi0tot}
 \end{equation}
is the {\it{total}} non-interacting polarization
(see Eq.\ref{eq:Pitotdecompose}).

\subsection{The effective 
action for the $\tilde{\rho}^{\alpha}$-field}
\label{subsec:gaussrho}

According to Eq.\ref{eq:Skinrhodef} the
kinetic energy contribution to the effective action for the collective density field is within
the Gaussian approximation 
given by\index{effective action!density field, Gaussian approximation}
 \begin{eqnarray}
 \tilde{S}_{\rm kin} \{ \tilde{\rho}^{\alpha} \}
 & \approx &
 \nonumber
 \\
 & & \hspace{-20mm}
   - \ln \left[
 \int {\cal{D}} \{ \phi^{\alpha} \}
 \exp \left( 
  \I   \sum_{q \alpha} \phi^{\alpha}_{-q} \tilde{\rho}^{\alpha}_{q} 
 - S_{{\rm kin},1} \left\{ \phi^{\alpha} \right\}  
 - S_{{\rm kin},2} \left\{ \phi^{\alpha} \right\}  \right)
 \right]
 \label{eq:Skinrhogaussdef}
 \; .
 \end{eqnarray}
Using Eq.\ref{eq:Skin1}, the first two terms in the
exponent can be combined as follows,
 \begin{equation}
 \I \sum_{q} \sum_{\alpha} \phi^{\alpha}_{-q} \tilde{\rho}^{\alpha}_{q}
 - S_{{\rm kin},1} \left\{ \phi^{\alpha} \right\}
 = 
 \I \sum_{q} \sum_{\alpha} \phi^{\alpha}_{-q} \left[ \tilde{\rho}^{\alpha}_{q}
 - \delta_{q,0} N^{\alpha}_{0} \right]
 \label{eq:FTren}
 \; \; \; ,
 \end{equation}
so that it is obvious that
the first order term
 $S_{{\rm kin},1} \{ \phi^{\alpha} \}$ simply shifts the 
collective density field $\tilde{\rho}^{\alpha}$ according to
 \begin{equation}
 \tilde{\rho}^{\alpha}_{q} \rightarrow \tilde{\rho}^{\alpha}_{q} - \delta_{q,0} N^{\alpha}_{0}
 \label{eq:rhoredef}
 \; \; \; ,
 \end{equation}
i.e. the uniform component is shifted.
Hence, $\tilde{S}_{\rm kin} \left\{ \tilde{\rho}^{\alpha} \right\}$ 
in Eqs.\ref{eq:Skinrhodef} and \ref{eq:Skinrhogaussdef}
is actually a functional of the shifted field. For simplicity we shall from now on
redefine the collective density field according to Eq.\ref{eq:rhoredef}. 
Note that the $q = 0$ term in the interaction part 
$\tilde{S}_{2} \left\{ \tilde{\rho}^{\alpha} \right\}$
given in Eq.\ref{eq:Sintrhodef}
is usually excluded due to charge neutrality, so that 
the effective action $\tilde{S}_{\rm eff} \{ \tilde{\rho}^{\alpha} \}$ 
depends exclusively on the shifted field.
The integration in Eq.\ref{eq:Skinrhogaussdef}
yields the usual Debye-Waller factor, so that
within the Gaussian approximation
 \begin{equation}
 \tilde{S}_{\rm kin} \{ \tilde{\rho}^{\alpha} \}  \approx 
 \tilde{S}_{{\rm kin},0}^{(0)} + 
 \tilde{S}_{{\rm kin},2}^{(0)} \{ \tilde{\rho}^{\alpha} \}
 \label{eq:Skinrhogauss}
 \; \; \; ,
 \end{equation}
where 
 \begin{equation}
\tilde{S}_{{\rm kin},0}^{(0)}
 =   - \ln \left[
 \int {\cal{D}} \left\{ \phi^{\alpha} \right\}
 \E^{
 - S_{{\rm kin},2} \left\{ \phi^{\alpha} \right\}  }
 \right]
 \label{eq:S00def}
 \end{equation}
is a constant independent of the $\tilde{\rho}^{\alpha}$-field, and
 \begin{equation}
 \tilde{S}_{{\rm kin},2}^{(0)} \{ \tilde{\rho}^{\alpha} \}
 =
  \frac{1}{2} \sum_{q} \sum_{\alpha \alpha^{\prime}}
  \Gamma^{\alpha \alpha^{\prime}} (q) \tilde{\rho}^{\alpha}_{-q} 
  \tilde{\rho}^{\alpha^{\prime}}_{q} 
  \label{eq:Skinrhogaussres}
  \; \; \; .
  \end{equation}
Here $\Gamma^{\alpha \alpha^{\prime} }(q)$ is the propagator of the $\phi^{\alpha}$-field
with respect to the 
quadratic action $S_{{\rm kin},2} \{ \phi^{\alpha} \}$ 
defined in Eq.\ref{eq:Skin2phires}, i.e.
 \begin{eqnarray}
 \Gamma^{\alpha \alpha^{\prime}} (q)
  & = &
 \frac{ \int {\cal{D}} \{ \phi^{\alpha} \} \E^{ - S_{{\rm kin},2} \{ \phi^{\alpha} \}}
  \phi^{\alpha}_{q} \phi^{\alpha^{\prime} }_{-q} }
 { \int {\cal{D}} \{ \phi^{\alpha} \} \E^{ - S_{{\rm kin},2} \{ \phi^{\alpha} \}}}
 \equiv  \left<
  \phi^{\alpha}_{q} \phi^{\alpha^{\prime} }_{-q} \right>_{S_{{\rm kin},2}}
 \nonumber
 \\
 & =  &
 [ \underline{ \tilde{\Pi}}_{0}^{-1} ( q ) ]^{\alpha \alpha^{\prime}}
 \label{eq:Gammapropdef}
 \; \; \; .
 \end{eqnarray}
Note that
$\Gamma_{q}^{\alpha \alpha^{\prime}} (q)$ 
is (up to a factor of ${\beta}/{V}$) given by
the matrix inverse of the non-interacting sector polarization
$\Pi_0^{\alpha \alpha^{\prime}} ( q )$.
In summary, within the Gaussian approximation the effective action 
of the $\tilde{\rho}^{\alpha}$-field is given by
 \begin{equation}
 \tilde{S}_{\rm eff} \{ \tilde{\rho}^{\alpha} \}    \approx 
  \tilde{S}_{{\rm kin},0}^{(0)}
  + \tilde{S}^{(0)}_{{\rm eff},2} \{ \tilde{\rho}^{\alpha} \}
  \label{eq:Seffrhogaussres}
  \; \; \; ,
  \end{equation}
with
 \begin{eqnarray}
  \tilde{S}^{(0)}_{{\rm eff},2} \{ \tilde{\rho}^{\alpha} \}
   & \equiv & \tilde{S}_{2} \{ \tilde{\rho}^{\alpha} \}
   + \tilde{S}_{{\rm kin},2}^{(0)} \{ \tilde{\rho}^{\alpha} \}
   \nonumber
   \\
   & =  &
 \frac{1}{2} \sum_{q} \sum_{\alpha \alpha^{\prime} }
 [    \underline{\tilde{f}}_{ {{q}} } 
 + \underline{\Gamma} (q) ]^{ \alpha \alpha^{\prime} }
 \tilde{\rho}_{-q}^{\alpha} \tilde{\rho}_{q}^{\alpha^{\prime}}
 \label{eq:Seff2rhogaussres}
 \; ,
 \end{eqnarray}
where  
$ \underline{ \Gamma} (q) = \underline{\tilde{\Pi}}^{-1}_{0} ( q )$.
In contrast to $S_{{\rm eff},2} \{ \phi^{\alpha} \}$, the
corresponding Gaussian action of the
collective density field 
$\tilde{S}^{(0)}_{{\rm eff},2} \{ \tilde{\rho}^{\alpha} \}$
carries an extra superscript $^{(0)}$, which
indicates that higher
order corrections will renormalize the parameters of
$\tilde{S}^{(0)}_{{\rm eff},2} \{ \tilde{\rho}^{\alpha} \}$.
In the case of  $S_{{\rm eff}, 2} \{ \phi^{\alpha} \}$
corrections of this type do not exist. 
In Sect.~\secref{sec:beyond} we shall explicitly calculate
the leading correction to the Gaussian approximation.

\subsection{The Gaussian 
propagator of the $\tilde{\rho}^{\alpha}$-field} 
\label{subsec:gaussdens}

{\it{\ldots which is nothing but the RPA 
polarization\index{random-phase approximation!polarization}.}}

\vspace{7mm}

\noindent
Having determined the effective action for the collective density field,
we may calculate the density-density correlation function
from Eq.\ref{eq:avrho} by performing the bosonic integration
over the $\tilde{\rho}^{\alpha}$-field. Because within the Gaussian
approximation $\tilde{S}_{\rm eff} \{ \tilde{\rho}^{\alpha} \}$ is
quadratic, the integration can be carried out trivially, and we obtain
 \begin{equation}
 \left< 
 \tilde{\rho}^{\alpha}_{ q} \tilde{\rho}^{\alpha^{\prime}}_{-q} 
 \right>_{ \tilde{S}_{{\rm eff},2}^{(0)} } 
 =
 \left[ \left[  \underline{\tilde{f}}_{q} 
 + \underline{\Gamma} (q) \right]^{-1} \right]^{\alpha \alpha^{\prime}}
 \label{eq:rhopropgauss}
 \; \; \; .
 \end{equation}
Using again
 $\underline{\tilde{f}}_{q}  =  \frac{\beta}{V} \underline{f}_{q}$ and
 $\underline{\Gamma} (q)  =  \frac{\beta}{V} \underline{\Pi}_{0}^{-1} ( q )$,
we conclude that within the Gaussian approximation
the sector density-density correlation function is approximated by
 \begin{equation}
 \Pi^{\alpha \alpha^{\prime}}(q) \approx 
  [ \underline{\Pi}_{\rm RPA} ( q ) ]^{\alpha \alpha^{\prime}}
 \label{eq:densrpa}
 \; \; \; ,
 \end{equation}
where the matrix 
$ \underline{\Pi}_{\rm RPA} ( q ) $ is given by
 \begin{equation}
  \underline{\Pi}_{\rm RPA} ( q ) 
  =
 \left[
 \underline{\Pi}_{0}^{-1} ( q )  +
 \underline{f}_{q} \right]^{-1}
 =
  \underline{\Pi}_{0} ( q ) 
  \left[ 1 + 
  \underline{f}_{q} 
  \underline{\Pi}_{0} ( q ) 
  \right]^{-1}
 \label{eq:densrpadef}
 \; \; \; .
 \end{equation}
Eq.\ref{eq:densrpadef} is nothing but the RPA
for the sector density-density correlation function.
Thus, the Gaussian 
propagator\index{Gaussian propagator!of $\tilde{\rho}^{\alpha}$-field} 
of the $\tilde{\rho}^{\alpha}$-field is simply given 
by the RPA polarization matrix $\underline{\Pi}_{\rm RPA} ( q )$.

To obtain the standard RPA result for the total density-density
correlation function, we should 
sum Eq.\ref{eq:densrpa} over both patch labels,
 \begin{equation}
 \Pi_{\rm RPA} ( q) = 
 \sum_{\alpha \alpha^{\prime}} 
 \left[
 \left[ \underline{\Pi}_{0}^{-1} ( q ) + \underline{f}_{q}  \right]^{-1} 
 \right]^{\alpha \alpha^{\prime}}
 \; \; \; ,
 \label{eq:Ptotdecompose1}
 \end{equation}
see Eq.\ref{eq:Pitotdecompose}. 
For simplicity let us assume that 
$[ \underline{f}_{q} ]^{\alpha \alpha^{\prime} } = {f}_{q}$ is
independent of the patch indices.
Expanding
 \begin{equation}
  [ \underline{\Pi}_{0}^{-1} (q) + \underline{f}_{q} ]^{-1}
 = \underline{\Pi}_{0} (q)  - 
 \underline{ \Pi}_{0} (q) \underline{f}_{{q}} \underline{\Pi}_{0} ( q) +
 \ldots
  \label{eq:Neumann}
  \; \; \; ,
  \end{equation}
and taking matrix elements, we see that Eqs.\ref{eq:Ptotdecompose1} and \ref{eq:Neumann} 
reduce to the usual RPA result \ref{eq:PiRPA},
 \begin{equation}
 \Pi_{\rm RPA} ( q) = \frac{ \Pi_{0} (q) }{1 + f_{q} \Pi_{0} (q) }
 \label{eq:PiRPAtot}
 \; \; \; ,
 \end{equation}
where the total non-interacting polarization $\Pi_{0} (q )$ is given in Eq.\ref{eq:pi0tot}. 
We would like to emphasize that up to this point we have not linearized the
energy dispersion, so that Eq.\ref{eq:PiRPAtot} is the exact RPA result
for all wave-vectors, 
including the short-wavelength regime.

\subsection{The bosonized Hamiltonian\index{Hamiltonian!bosonized, Gaussian approximation}}

To make contact with the operator approach to bosonization \cite{Houghton93,Castro95},
let us now derive a bosonic Hamiltonian 
that at long wavelengths is
equivalent with our Gaussian action 
$\tilde{S}_{{\rm eff},2}^{(0)} \{ \tilde{\rho}^{\alpha} \}$
in Eq.\ref{eq:Seff2rhogaussres}.
The key observation is that,
{\it{in the limit of high densities and long wavelengths}}
(i.e. in the limit where the {\it{diagonal-patch approximation}} $(A1)$ is correct),
the sector polarization
is diagonal in the sector indices, and is to leading order 
given in Eq.\ref{eq:Pilong}.
It follows that the matrix elements of $\underline{\Gamma} (q)$ (which according to
Eqs.\ref{eq:U2res} and \ref{eq:Gammapropdef} is proportional to the 
inverse non-interacting polarization) are in the above limit given by
 \begin{equation}
 \Gamma^{\alpha \alpha^{\prime} } (q) \approx \delta^{ \alpha \alpha^{\prime}}
 \frac{ \beta}{V \nu^{\alpha}}
 \frac{  {\vec{v}}^{\alpha} \cdot {\vec{q}} - \I \omega_{m} }
 {  {\vec{v}}^{\alpha} \cdot {\vec{q}} }
 \label{eq:Gammalong}
 \; \; \; .
 \end{equation}
Hence the Gaussian action \ref{eq:Seff2rhogaussres} can be written as
\begin{equation}
 \tilde{S}^{(0)}_{{\rm eff},2} \{ \tilde{\rho}^{\alpha} \}   = 
 \frac{\beta}{2V} \sum_{q} \sum_{\alpha \alpha^{\prime}} 
 \left[ {f}_{q}^{ \alpha \alpha^{\prime} } 
  +
  \delta^{\alpha \alpha^{\prime}}
 \frac{  {\vec{v}}^{\alpha} \cdot {\vec{q}} - \I \omega_{m} }
 { \nu^{\alpha}  {\vec{v}}^{\alpha} \cdot {\vec{q}} }
  \right]
 \tilde{\rho}^{\alpha}_{-q} \tilde{\rho}^{\alpha^{\prime}}_{q}
 \label{eq:Seffgauss2}
 \; \; \; .
 \end{equation}
The term proportional to $\I \omega_{m}$ 
defines the dynamics of the $\tilde{\rho}^{\alpha}$-field.
We now recall that in the 
functional integral for canonically quantized bosons
{\it{the coefficient of the term proportional to $-\I \omega_{m}$ should be
precisely $\beta$.}}  
Any other value of this coefficient 
would describe operators with non-canonical
commutation relations \cite{Kopietz89a}.
In a different context such a rescaling has 
also been performed in \cite{Kopietz89b}.
Thus, to write our effective action in terms of a canonical
boson field $b_{q}^{\alpha}$, we should rescale the $\tilde{\rho}^{\alpha}$-field 
accordingly.
This is achieved by substituting
in Eq.\ref{eq:Seffgauss2} 
 \begin{equation}
 \tilde{\rho}^{\alpha}_{q} = ({ {{V}} \nu^{\alpha} |   
 {\vec{v}}^{\alpha} \cdot {\vec{q}} | })^{1/2}
 \left[
 \Theta ( {\vec{v}}^{\alpha} \cdot {\vec{q}} ) b^{ \alpha}_{q} +
 \Theta ( - {\vec{v}}^{\alpha} \cdot {\vec{q}} ) b^{\dagger \alpha}_{-q} 
 \right]
 \; \; \; .
 \label{eq:rhobscale}
 \end{equation}
The $\Theta$-functions are necessary to make the coefficient
of $- \I \omega_{m}$ equal to $\beta$ for all patches, 
because the sign of $ \I \omega_{m}$ in 
Eq.\ref{eq:Seffgauss2} depends on the sign of ${ {\vec{v}}^{\alpha} \cdot {\vec{q}}}$.
Our final result for the
bosonized action 
$ \tilde{S}_{\rm b} 
\{ b^{\alpha} \} \equiv \tilde{S}^{(0)}_{{\rm eff},2} 
\{ \tilde{\rho}^{\alpha} (b^{\alpha}) \}$
is
 \begin{eqnarray}
 \tilde{S}_{\rm b} \{ b^{\alpha} \}
 & = &
  \beta \sum_{q} \sum_{\alpha  } \Theta ( {\vec{v}}^{\alpha } \cdot {\vec{q}} )
  ( - \I \omega_{m} )
 b^{ \alpha \dagger }_{q} b^{\alpha}_{q}  
 \nonumber
 \\
 & + & 
  \beta  \left[ H_{ {\rm b }, {\rm kin}} \{ b^{\alpha} \}
 +  H_{{\rm b}, {\rm int} }
 \{ b^{\alpha} \} \right]
 \; \; \; ,
 \label{eq:Sb2}
 \end{eqnarray}
 \begin{eqnarray}
 H_{\rm b, kin} \{ b^{\alpha} \} & = & 
   \sum_{q} \sum_{\alpha  } \Theta ( {\vec{v}}^{\alpha } \cdot {\vec{q}} )
  {\vec{v}}^{\alpha} \cdot {\vec{q}} 
 b^{\alpha \dagger }_{q} b^{\alpha}_{q} 
 \; \; \; ,
 \label{eq:Hkin}
 \\
 H_{\rm b, int} \{ b^{\alpha} \} & = & 
 \frac{1}{2} \sum_{q} \sum_{\alpha \alpha^{\prime} } 
 \Theta ( {\vec{v}}^{\alpha } \cdot {\vec{q}} )
  \sqrt{ |  {\vec{v}}^{\alpha} \cdot {\vec{q}} |
 |  {\vec{v}}^{\alpha^{\prime}} \cdot {\vec{q}} |}
 \nonumber
 \\
 &  & \hspace{-11mm} \times \left[
 \Theta ( {\vec{v}}^{\alpha^{\prime} } \cdot {\vec{q}} )
 \left(
 {F}_{\vec{q}}^{\alpha \alpha^{\prime} }
 b^{\alpha \dagger }_{q} b^{\alpha^{\prime}}_{q}
 +
 {F}_{\vec{q}}^{\alpha^{\prime} \alpha }
 b^{ \alpha^{\prime} \dagger }_{q} b^{\alpha}_{q}
 \right)
 \right.
 \nonumber
 \\
 &  &  \hspace{-9mm} 
 \left.
 +  \Theta ( - {\vec{v}}^{\alpha^{\prime} } \cdot {\vec{q}} )
 \left(
 {F}_{\vec{q}}^{\alpha \alpha^{\prime} }
 b^{ \alpha \dagger }_{q} b^{ \alpha^{\prime} \dagger }_{-q}
 +
 {F}_{\vec{q}}^{\alpha^{\prime} \alpha }
 b^{\alpha^{\prime}}_{-q} b^{\alpha}_{q}
 \right)
 \right]
 \label{eq:Hint}
 \;  \; ,
 \end{eqnarray}
where ${F}_{\vec{q}}^{\alpha \alpha^{\prime} } = 
 \sqrt{ \nu^{\alpha} \nu^{\alpha^{\prime}} } f^{\alpha \alpha^{\prime}}_{\vec{q}}$ are
dimensionless couplings, and
we have assumed that the bare interaction depends only on ${\vec{q}}$.
For frequency-dependent bare interactions
it is not possible to write down a conventional Hamiltonian
that is equivalent to the effective action in Eq.\ref{eq:Seffgauss2}.
The functional integral for the $b^{\alpha}$-field 
is now formally identical with a
standard bosonic functional integral.
The corresponding second-quantized bosonic Hamiltonian 
is therefore $\hat{H}_{\rm b} = 
\hat{H}_{\rm b, kin} + \hat{H}_{\rm b, int}$, 
where $\hat{H}_{\rm b, kin}$ and $\hat{H}_{\rm b, int}$ are simply 
obtained by replacing 
the bosonic fields $b^{\alpha}_{q}$ 
in Eqs.\ref{eq:Hkin} and \ref{eq:Hint} by operators
$\hat{b}^{\alpha}_{\vec{q}}$ satisfying
$[ \hat{b}^{\alpha}_{\vec{q}} , \hat{b}^{\alpha^{\prime} \dagger }_{ {\vec{q}}^{\prime} }
 ] = \delta^{ \alpha \alpha^{\prime} } \delta_{ {\vec{q}}  {\vec{q}}^{\prime} }$.
The resulting $\hat{H}_{\rm b}$ agrees with 
the bosonized Hamiltonian derived
in \cite{Houghton93,Castro95} 
by means of an operator approach.

Note, however, 
that the above identification with a canonical bosonic Hamiltonian is only
possible in the limit of long wavelengths and high densities, 
so that our parameterization \ref{eq:Seff2rhogaussres}
of the effective Gaussian action is more general.
Moreover, for practical calculations the
substitution \ref{eq:rhobscale} is not very useful, because 
it maps the very simple form \ref{eq:Seff2rhogaussres}
of $\tilde{S}_{{\rm eff},2}^{(0)} \{ \tilde{\rho}^{\alpha} \}$ onto the complicated effective
action $\tilde{S}_{\rm b} \{ b^{\alpha} \}$ in 
Eqs.\ref{eq:Sb2}--\ref{eq:Hint} without containing new information.

\section{Beyond the Gaussian approximation}
\label{sec:beyond}

{\it{
We develop a systematic method for calculating 
the corrections to the Gaussian approximation, 
and then explicitly evaluate the one-loop correction.
In this way we determine the hidden small parameter which 
determines the range of validity of the Gaussian approximation.
We also show that bosonization leads to a new method for calculating 
the density-density correlation function beyond the RPA.}}

\subsection{General expansion of the bosonized kinetic energy}
\label{subsec:linkedcluster}

{\it{
The bosonized kinetic energy 
$\tilde{S}_{\rm kin} \{ \tilde{\rho}^{\alpha} \}$
is calculated via a linked cluster expansion\index{bosonization!linked cluster expansion of
kinetic energy} 
of the functional Fourier transformation 
in Eq.\ref{eq:Skinrhodef}.}}

\vspace{7mm}

\noindent
Defining $S_{\rm kin}^{\prime} \{ \phi^{\alpha} \}$ to be the sum
of all non-Gaussian terms in the expansion 
\ref{eq:tracelogexp} 
of $S_{\rm kin} \{ {\phi}^{\alpha} \}$,
 \begin{equation}
 S_{\rm kin}^{\prime} \{ \phi^{\alpha} \} = \sum_{n=3}^{\infty} S_{{\rm kin},n}
 \{ \phi^{\alpha} \}
 =
 \sum_{n=3}^{\infty} 
 \frac{1}{n} {\rm Tr} \left[ \hat{G}_{0} \hat{V} \right]^n
 \label{eq:Skin3}
 \; \; \; ,
 \end{equation}
we may write
 \begin{equation}
 \E^{ - \tilde{S}_{\rm kin} \{ \tilde{\rho}^{\alpha} \}}
  = 
  \exp \left[ - \tilde{S}_{{\rm kin},0}^{(0)} - 
  \tilde{S}_{{\rm kin},2}^{(0)} \{ \tilde{\rho}^{\alpha} \} \right]
 \left< \E^{ -
 S_{\rm kin}^{\prime} \{ \phi^{\alpha} \} }  
 \right>^{\tilde{\rho} }_{{S}_{{\rm kin},2}}
 \label{eq:Skinrho2}
 \; ,
 \end{equation}
where  according to
Eqs.\ref{eq:S00def} and \ref{eq:Skinrhogaussres}, 
 \begin{eqnarray}
  \exp \left[ - \tilde{S}_{{\rm kin},0}^{(0)} - 
  \tilde{S}_{{\rm kin},2}^{(0)} \{ \tilde{\rho}^{\alpha} \} \right]
 & = &
 \nonumber
 \\
 & & \hspace{-20mm}
 \int {\cal{D}} \{ \phi^{\alpha} \}
  \exp \left[ 
  \I   \sum_{q \alpha} \phi^{\alpha}_{-q} \tilde{\rho}^{\alpha}_{q} 
  - S_{ {\rm kin},2} \{ \phi^{\alpha} \} \right] 
  \; \; \; ,
  \end{eqnarray}
and for any functional ${\cal{F}} \{ \phi^{\alpha} \}$
the averaging in Eq.\ref{eq:Skinrho2} is defined as follows,
 \begin{equation}
 \hspace{-5mm}
 \left< {\cal{F}} \{ \phi^{\alpha} \} \right>_{S_{{\rm kin},2} }^{ \tilde{\rho}}
 =
 \frac{
 \int {\cal{D}} \left\{ \phi^{\alpha} \right\}
  {\cal{F}} \left\{ \phi^{\alpha} \right\} 
 \exp \left[ 
  \I   \sum_{q \alpha} \phi^{\alpha}_{-q} \tilde{\rho}^{\alpha}_{q} 
  - S_{{\rm kin},2} \{ \phi^{\alpha} \}  \right]
  }
 {
 \int {\cal{D}} \{ \phi^{\alpha} \}
 \exp \left[ 
  \I   \sum_{q \alpha} \phi^{\alpha}_{-q} \tilde{\rho}^{\alpha}_{q} 
  - S_{{\rm kin},2} \{ \phi^{\alpha} \}  \right]
  }
  \; .
 \label{eq:funcrhodef}
 \end{equation}
Performing in this expression the shift transformation\index{shift transformation} 
 \begin{equation}
 \phi^{\alpha}_{q} \rightarrow {\phi}^{\alpha}_{q} + 
 \I \sum_{\alpha^{\prime}}
 \Gamma^{\alpha \alpha^{\prime} } (q) \tilde{\rho}_{q}^{\alpha^{\prime}}
 \label{eq:shift1}
 \; \; \; ,
 \end{equation}
it is easy to see that 
 \begin{eqnarray}
 \left< {\cal{F}} \{ \phi^{\alpha} \} \right>^{\tilde{\rho} }_{S_{{\rm kin},2}}
 & = &
 \frac{
 \int {\cal{D}} \{ \phi^{\alpha} \}
 {\cal{F }}
 \{ \phi^{\alpha} + \I 
 \sum_{\alpha^{\prime}}
 \Gamma^{\alpha \alpha^{\prime} } \tilde{\rho}^{\alpha^{\prime}}
 \} 
 \exp \left[ - S_{{\rm kin},2} \{ \phi^{\alpha} \}  \right]
  } 
 {
 \int {\cal{D}} \{ \phi^{\alpha} \}
 \exp \left[
  - S_{{\rm kin},2} \{ \phi^{\alpha} \}  \right]
  }
  \nonumber
  \\
  & \equiv &
 \left< 
 {\cal{F }}
 \{ \phi^{\alpha} + \I 
 \sum_{\alpha^{\prime}}
 \Gamma^{\alpha \alpha^{\prime} } \tilde{\rho}^{\alpha^{\prime}}
 \} 
 \right>_{S_{{\rm kin},2} }
 \label{eq:funcrho2}
 \; \; \; .
 \end{eqnarray}
In our case we have to calculate 
 \begin{equation}
 \left< \E^{ -
 S_{\rm kin}^{\prime} \{ \phi^{\alpha} \}  } 
 \right>^{\tilde{\rho} }_{S_{{\rm kin},2}}
 =
 \left< \E^{ -
 S_{\rm kin}^{\prime} \{ \phi^{\alpha} 
 + \I \sum_{\alpha^{\prime}}
 \Gamma^{\alpha \alpha^{\prime} } \tilde{\rho}^{\alpha^{\prime}} \} 
  } \right>_{ S_{{\rm kin},2} }
 \label{eq:Skinexprho}
 \; \; \; .
 \end{equation}
Consider first the term of order $(\phi^{\alpha} )^n$ in the expansion \ref{eq:Skin3}
of $S^{\prime}_{\rm kin} \{ \phi^{\alpha} \}$. 
Clearly the substitution
$ \phi^{\alpha} \rightarrow 
 \phi^{\alpha} 
 + \I \sum_{\alpha^{\prime}}
 \Gamma^{\alpha \alpha^{\prime} 
 } \tilde{\rho}^{\alpha^{\prime}}$ generates (among many other terms) a term of order
$(\tilde{\rho}^{\alpha})^n$, 
which does not depend on the
$\phi^{\alpha}$-field and can be pulled out of the average in Eq.\ref{eq:Skinexprho}.
Let us denote this contribution by 
$\tilde{S}_{{\rm kin},n}^{(0)} \{ \tilde{\rho}^{\alpha} \}$. 
From Eq.\ref{eq:Seffphin} it is easy to see that 
$\tilde{S}_{{\rm kin},n}^{ (0)} \{ \tilde{\rho}^{\alpha} \}$ 
is obtained by replacing
$ \phi^{\alpha}_{q} \rightarrow 
  \I \sum_{\alpha^{\prime}}
 \Gamma^{\alpha \alpha^{\prime} 
 } (q) \tilde{\rho}^{\alpha^{\prime}}_{q}$ 
in $S_{{\rm kin},n} \{ \phi^{\alpha} \}$, so that
it is given by
 \begin{eqnarray}
 \tilde{S}_{{\rm kin},n}^{(0)} 
 \left\{ { { \tilde{\rho}^{\alpha}}} \right\} & = & 
S_{{\rm kin},n} \{ 
  \I \sum_{\alpha^{\prime}}
 \Gamma^{\alpha \alpha^{\prime} 
 } (q) \tilde{\rho}^{\alpha^{\prime}}_{q} 
\} 
\nonumber
\\
&  & \hspace{-19mm}  =
 \frac{1}{n} \sum_{q_{1}  \ldots q_{n} }
 \sum_{\alpha_{1} \ldots \alpha_{n} } 
 \Gamma_{n}^{(0)} ( 
 q_1 \alpha_{1} \ldots q_{n} \alpha_{n}  ) \tilde{\rho}^{\alpha_{1}}_{q_{1}} \cdots
 \tilde{\rho}^{\alpha_{n}}_{q_{n}}
 \; \; \; ,
 \label{eq:Seffrhon0}
 \end{eqnarray}
where for $n \geq 3$ the vertices $\Gamma_{n}^{(0)}$ are 
 \begin{eqnarray}
 \Gamma_{n}^{(0)} ( 
 q_1 \alpha_{1} \ldots q_{n} \alpha_{n}  ) 
 & = & \I^{n} \sum_{\alpha_{1}^{\prime} \ldots \alpha_{n}^{\prime} }
 U_{n} ( 
 q_1 \alpha_{1}^{\prime} \ldots q_{n} \alpha_{n}^{\prime}  ) 
 \nonumber
 \\
 &  & \hspace{10mm} \times
 \Gamma^{\alpha_{1}^{\prime} \alpha_{1}} ( q_1 ) \ldots 
 \Gamma^{\alpha_{n}^{\prime} \alpha_{n}} (q_n )
 \label{eq:Gammandef0}
 \; \; \; .
 \end{eqnarray}
Recall that $\Gamma^{\alpha \alpha^{\prime}} ( q )$ is
according to Eq.\ref{eq:Gammapropdef} proportional
to the matrix inverse of the non-interacting sector polarization
$\Pi_0^{\alpha \alpha^{\prime}} ( q )$.
Obviously the Gaussian action 
$\tilde{S}^{(0)}_{{\rm kin},2} \{ \tilde{\rho}^{\alpha} \}$
in Eq.\ref{eq:Skinrhogaussres} is also of the form \ref{eq:Seffrhon0}, with
 \begin{equation}
 \Gamma_{2}^{(0)} ( q_{1} \alpha_{1} q_{2} \alpha_{2} )
 = \delta_{q_{1} + q_{2} , 0} \Gamma^{\alpha_{1} \alpha_{2} } ({q_{2}})
 \; \; \; .
 \label{eq:Gamma20Gauss}
 \end{equation}
The vertex $U_{1}$ has been absorbed into the redefinition
of $\tilde{\rho}^{\alpha}_{q}$ (see Eq.\ref{eq:rhoredef}), so that
$\tilde{S}_{{\rm kin},1}^{(0)} \{ \tilde{\rho}^{\alpha}_{q} \} = 0$.
Defining 
 \begin{eqnarray}
 \tilde{S}_{\rm kin}^{(0)}
 \left\{ { { \tilde{\rho}^{\alpha}}} \right\}  &   = &
 \tilde{S}_{{\rm kin},0}^{(0)} +
 \sum_{n=2}^{\infty} \tilde{S}_{{\rm kin},n}^{(0)}
 \left\{ { { \tilde{\rho}^{\alpha}}} \right\}    
 \; \; \; ,
 \label{eq:Skin0rho}
 \\
 S^{\prime \prime}_{\rm kin} \left\{ \phi^{\alpha} , \tilde{\rho}^{\alpha} \right\}
 & = &
 S_{\rm kin}^{\prime} \{ \phi^{\alpha} 
 + \I \sum_{\alpha^{\prime}}
 \Gamma^{\alpha \alpha^{\prime} 
 } \tilde{\rho}^{\alpha^{\prime}} \} 
 - 
S_{\rm kin}^{\prime} \{ 
  \I \sum_{\alpha^{\prime}}
 \Gamma^{\alpha \alpha^{\prime} } \tilde{\rho}^{\alpha^{\prime}} 
\}
 \label{eq:Sprimeprime}
 \; \; \; ,
 \end{eqnarray}
the general perturbative expansion for $\tilde{S}_{\rm kin} 
\{ \tilde{\rho}^{\alpha} \}$ is 
 \begin{equation}
 \hspace{-3mm}
\tilde{S}_{\rm kin} \left\{ \tilde{\rho}^{\alpha} \right\}
= \tilde{S}^{(0)}_{\rm kin}
\left\{ \tilde{\rho}^{\alpha} \right\}
- \ln \left[ 1 +
 \sum_{n=1}^{\infty} \frac{(-1)^{n}}{n ! } 
 \left< \left[ 
 S_{\rm kin}^{\prime \prime} \{ \phi^{\alpha} , \tilde{\rho}^{\alpha} \} 
 \right]^n \right>_{S_{{\rm kin},2}}
 \right]
 \label{eq:generalpert}
 \; .
 \end{equation}
According to the linked cluster theorem \cite{Mahan81}\index{linked cluster theorem} the 
logarithm eliminates all disconnected diagrams, so that Eq.\ref{eq:generalpert}
can also be written as
 \begin{equation}
\tilde{S}_{\rm kin} \left\{ \tilde{\rho}^{\alpha} \right\}
= \tilde{S}^{(0)}_{\rm kin}
\left\{ \tilde{\rho}^{\alpha} \right\}
- 
 \sum_{n=1}^{\infty} \frac{(-1)^{n}}{n  } 
 \left< 
 \left[ 
 S_{\rm kin}^{\prime \prime} \{ \phi^{\alpha} , \tilde{\rho}^{\alpha} \} 
 \right]^n \right>_{S_{{\rm kin},2}}^{\rm con}
 \label{eq:generalpert1}
 \; \; \; ,
 \end{equation}
where the superscript $^{\rm con}$ means that all different connected diagrams should
be retained \cite{Mahan81}.
From this expression it is easy to see that 
$\tilde{S}_{\rm kin} \{ \tilde{\rho}^{\alpha} \}$ is in general of the following form
 \begin{equation}
 \tilde{S}_{\rm kin} \left\{ \tilde{\rho}^{\alpha} \right\}
 =  \tilde{S}_{{\rm kin},0} +  \sum_{n=1}^{\infty} \tilde{S}_{{\rm kin} , n}
 \left\{ \tilde{\rho}^{\alpha} \right\}
 \label{eq:Skinn}
 \; \; \; ,
 \end{equation}
where $\tilde{S}_{{\rm kin},0}$ is a constant
independent of the fields that cancels in the calculation of correlation functions, and
for $n \geq 1$
 \begin{equation}
 \tilde{S}_{{\rm kin},n} \left\{ { { \tilde{\rho}^{\alpha}}} \right\}  =  
 \frac{1}{n} \sum_{q_{1}  \ldots q_{n} }
 \sum_{\alpha_{1} \ldots \alpha_{n} } 
 \Gamma_{n} ( 
 q_1 \alpha_{1} \ldots q_{n} \alpha_{n}  ) \tilde{\rho}^{\alpha_{1}}_{q_{1}} \cdots
 \tilde{\rho}^{\alpha_{n}}_{q_{n}}
 \; \; \; ,
 \label{eq:Seffrhon}
 \end{equation}
where the vertices $\Gamma_{n}$ have an expansion of the form
 \begin{equation}
 \Gamma_{n} ( 
 q_1 \alpha_{1} \ldots q_{n} \alpha_{n}  ) 
 = \sum_{m = 0}^{\infty}
 \Gamma_{n}^{(m)} ( 
 q_1 \alpha_{1} \ldots q_{n} \alpha_{n}  ) 
 \; \; \; .
 \label{eq:Gammank}
 \end{equation}
Here $\Gamma_{n}^{(m)}$ 
is the interaction vertex between $n$ collective density fields $\tilde{\rho}^{\alpha}$,
that is generated from all diagrams in the linked cluster expansion \ref{eq:generalpert1}
containing $m$ internal loops of the $\phi^{\alpha}$-field.
Note that the vertices $\Gamma_{n}^{(0)}$ in Eq.\ref{eq:Gammandef0}
are the tree-approximation\index{tree-approximation} for the exact vertices $\Gamma_{n}$,
because they do not involve any
internal $\phi^{\alpha}$-loops. 
Each internal $\phi^{\alpha}$-loop attached to a vertex $U_{n}$ reduces the number
of external $\phi^{\alpha}$-fields by $2$, so that for $m \geq 1$ the vertices
$\Gamma_{n}^{(m)}$ can only by determined by vertices $U_{n^{\prime}}$ with 
$n^{\prime} > n$.
Within the Gaussian approximation all $U_{n}$ with $n \geq 3$ are set equal to zero,
while the contribution from $U_{1}$ can be absorbed into the redefinition
of $\tilde{\rho}^{\alpha}_{0}$, see Eq.\ref{eq:rhoredef}.
Hence the Gaussian approximation amounts to setting
 \begin{equation}
 \Gamma_{2} ( -q \alpha , q \alpha^{\prime} ) \approx
 \Gamma^{(0)}_{2} ( -q \alpha , q \alpha^{\prime} ) = 
 \Gamma^{\alpha \alpha^{\prime} } (q)
 \label{eq:gaussgamma1}
 \; \; \; ,
 \end{equation}
 \begin{equation}
 \Gamma_{n}^{(m)} = 0 \; \; \; , \; \; \; \mbox{for $n > 2$ or $m > 0$} 
 \label{eq:gaussgamma2}
 \; \; \; ,
 \end{equation}
where $\Gamma^{\alpha \alpha^{\prime} }(q)$ is defined in Eq.\ref{eq:Gammapropdef}.
Although $\Gamma_{1} = 0$ within the Gaussian approximation, the higher
order terms will in general lead to a finite value of
$\Gamma_{1}$, which describes the fluctuations of the total number
of occupied states in the sectors $K^{\alpha}_{\Lambda , \lambda}$.
As already pointed out in the footnote after Eq.\ref{eq:Trloggauss},
at zero temperature these terms do not contribute to
correlation functions at finite $q$, but they are certainly important
for the calculation of the free energy.

\subsection{The leading correction 
to the effective action
\index{Hamiltonian!bosonized, non-Gaussian corrections}}
\label{subsec:Explicit}

{\it{We now show that our formalism can indeed be used
in practice for a systematic calculation of the
corrections to the non-interacting 
boson approximation.}}

\vspace{7mm}

\noindent
The leading correction to the Gaussian approximation is obtained from the
one-loop approximation for our effective
bosonic theory, which amounts to a two-loop calculation at the fermionic level.
Note that we have mapped the problem of calculating
a two-particle Green's function of the original fermionic model onto the
problem of calculating a one-particle Green's function of an effective
bosonic model. The latter is conceptually  simpler, because the
symmetrized vertices $U_{n}$ and $\Gamma_{n}^{(m)}$ automatically contain
the relevant self-energy and vertex corrections of the underlying
fermionic problem. \index{vertex corrections}
This will become evident below.

At one-loop order, it is sufficient to truncate the
expansion of the interaction part $S^{\prime}_{\rm kin} \{ \phi^{\alpha} \}$ of 
the effective action \ref{eq:Skin3} 
of the $\phi^{\alpha}$-field at the fourth 
order\index{effective action!background field, non-Gaussian corrections}, 
 \begin{eqnarray}
 S^{\prime}_{\rm kin} \{ \phi^{\alpha} \} & \approx &
 S_{{\rm kin},3} \{ \phi^{\alpha} \} +
 S_{{\rm kin},4} \{ \phi^{\alpha} \} 
 \nonumber
 \\
 &  & \hspace{-15mm} =
 \frac{1}{3} \sum_{q_{1}  q_{2} q_{3} }
 \sum_{\alpha_{1} \alpha_{2} \alpha_{3} } 
 U_{3} ( 
 q_1 \alpha_{1} q_{2} \alpha_{2} q_{3} \alpha_{3}  ) {\phi}^{\alpha_{1}}_{q_{1}} 
 {\phi}^{\alpha_{2}}_{q_{2}}
 {\phi}^{\alpha_{3}}_{q_{3}}
 \nonumber
 \\
 &  & \hspace{-15mm} +
 \frac{1}{4} \sum_{q_{1}  q_{2} q_{3} q_{4}}
 \sum_{\alpha_{1} \alpha_{2} \alpha_{3} \alpha_{4}} 
 U_{4} ( 
 q_1 \alpha_{1} q_{2} \alpha_{2} q_{3} \alpha_{3}  q_{4} \alpha_{4} ) {\phi}^{\alpha_{1}}_{q_{1}} 
 {\phi}^{\alpha_{2}}_{q_{2}}
 {\phi}^{\alpha_{3}}_{q_{3}}
 {\phi}^{\alpha_{4}}_{q_{4}}
 \label{eq:Skinprimeapprox}
 \; \; \; ,
 \end{eqnarray}
where the vertices $U_{3}$ and $U_{4}$ are defined in Eq.\ref{eq:Uvertex}.
According to the general formalism outlined above, 
the bosonized kinetic energy 
$\tilde{S}_{\rm kin} \{ \tilde{\rho}^{\alpha} \}$ 
is obtained by calculating
the functional Fourier transform of $S_{\rm kin} \{ \phi^{\alpha } \}$.
Within the one-loop approximation it is sufficient to retain
only the term $n=1$ in the linked cluster expansion \ref{eq:generalpert1}, so that 
 \begin{equation}
\tilde{S}_{\rm kin} \left\{ \tilde{\rho}^{\alpha} \right\}
\approx \tilde{S}^{(0)}_{\rm kin}
\left\{ \tilde{\rho}^{\alpha} \right\}
+
 \left< 
 S_{\rm kin}^{\prime \prime} \{ \phi^{\alpha} , \tilde{\rho}^{\alpha} \} 
 \right>_{S_{{\rm kin},2}}^{\rm con}
 \; \; \; ,
 \label{eq:Skinrho1loop}
 \end{equation}
where
 \begin{eqnarray}
 \tilde{S}^{(0)}_{\rm kin}
\left\{ \tilde{\rho}^{\alpha} \right\}
& \approx  &
 \tilde{S}_{{\rm kin},0}^{(0)} +
  \frac{1}{2} \sum_{q} \sum_{\alpha \alpha^{\prime}}
  \Gamma^{\alpha \alpha^{\prime}} (q) \tilde{\rho}^{\alpha}_{-q} \tilde{\rho}^{\alpha^{\prime}}_{q} 
 \nonumber
 \\
&  & \hspace{-15mm} +
 \frac{1}{3} \sum_{q_{1}  q_{2} q_{3} }
 \sum_{\alpha_{1} \alpha_{2} \alpha_{3} } 
 \Gamma_{3}^{(0)} ( 
 q_1 \alpha_{1} q_{2} \alpha_{2} q_{3} \alpha_{3}  ) {\tilde{\rho}}^{\alpha_{1}}_{q_{1}} 
 {\tilde{\rho}}^{\alpha_{2}}_{q_{2}}
 {\tilde{\rho}}^{\alpha_{3}}_{q_{3}}
 \nonumber
 \\
 &  & \hspace{-15mm} +
 \frac{1}{4} \sum_{q_{1}  q_{2} q_{3} q_{4}}
 \sum_{\alpha_{1} \alpha_{2} \alpha_{3} \alpha_{4}} 
 \Gamma_{4}^{(0)} ( 
 q_1 \alpha_{1} q_{2} \alpha_{2} q_{3} \alpha_{3}  q_{4} \alpha_{4} ) {\tilde{\rho}}^{\alpha_{1}}_{q_{1}} 
 {\tilde{\rho}}^{\alpha_{2}}_{q_{2}}
 {\tilde{\rho}}^{\alpha_{3}}_{q_{3}}
 {\tilde{\rho}}^{\alpha_{4}}_{q_{4}}
 \label{eq:Skinapproxrho}
 \; \; \; ,
 \end{eqnarray}
with
 \begin{eqnarray}
 \Gamma_{3}^{(0)} ( 
 q_1 \alpha_{1} q_{2} \alpha_{2} q_{3} \alpha_{3}  ) 
  & = &
    - \I \sum_{\alpha_{1}^{\prime} \alpha_{2}^{\prime} \alpha_{3}^{\prime} }
 U_{3} ( 
 q_1 \alpha_{1}^{\prime} q_{2} \alpha_{2}^{\prime} q_{3} \alpha_{3}^{\prime}  ) 
 \nonumber
 \\
 & & \times
 \Gamma^{\alpha_{1}^{\prime} \alpha_{1}}  (  q_1 )
 \Gamma^{\alpha_{2}^{\prime} \alpha_{2}}  (q_2 )
 \Gamma^{\alpha_{3}^{\prime} \alpha_{3}} (q_3 )
 \; \; \; ,
 \label{eq:Gamma03}
 \\
 \Gamma_{4}^{(0)} ( 
 q_1 \alpha_{1} q_{2} \alpha_{2} q_{3} \alpha_{3}  q_{4} \alpha_{4} ) 
 & = &
 \sum_{\alpha_{1}^{\prime} \alpha_{2}^{\prime} \alpha_{3}^{\prime} \alpha_{4}^{\prime}}
 U_{4} ( 
 q_1 \alpha_{1}^{\prime} q_{2} \alpha_{2}^{\prime} q_{3} \alpha_{3}^{\prime}  q_{4} \alpha_{4}^{\prime}) 
 \nonumber
 \\
 & & \hspace{-15mm} \times
 \Gamma^{\alpha_{1}^{\prime} \alpha_{1}}  (q_1 )
 \Gamma^{\alpha_{2}^{\prime} \alpha_{2}}  (q_2 )
 \Gamma^{\alpha_{3}^{\prime} \alpha_{3}} (q_3 ) 
 \Gamma^{\alpha_{4}^{\prime} \alpha_{4}} (q_4 ) 
 \; \; \; .
 \label{eq:Gamma04}
 \end{eqnarray}
The correction term due to one internal $\phi^{\alpha}$-loop is 
 \begin{equation}
 \left< 
 S_{\rm kin}^{\prime \prime} \{ \phi^{\alpha} , \tilde{\rho}^{\alpha} \} 
 \right>_{S_{{\rm kin},2}}^{\rm con} = \tilde{S}_{{\rm kin},0}^{(1)} 
 + \tilde{S}_{{\rm kin},1}^{(1)} \left\{ \tilde{\rho}^{\alpha} \right\}
 + \tilde{S}_{{\rm kin},2}^{(1)}
  \left\{ \tilde{\rho}^{\alpha} \right\}
 \label{eq:Skinloop}
 \; \; \; ,
 \end{equation}
where 
 \begin{eqnarray}
\tilde{S}_{{\rm kin},0}^{(1)} & = &
\frac{3}{2} \sum_{q q^{\prime}}
 \sum_{\alpha_{1} \alpha_{2} \alpha_{3} \alpha_{4}}
 U_{4} ( - q \alpha_{1} , q \alpha_{2} , -q^{\prime} \alpha_{3} , q^{\prime} \alpha_{4} )
 \nonumber
 \\
 & & \hspace{20mm} \times
 \Gamma^{\alpha_{2} \alpha_{1}} (q)
 \Gamma^{\alpha_{4} \alpha_{3}} ( q^{\prime} )
 \; \; \; ,
 \label{eq:Skin01}
 \\
 \tilde{S}_{{\rm kin},1}^{(1)} 
  \left\{ \tilde{\rho}^{\alpha} \right\}
 &  = & \sum_{\alpha} \Gamma_{1}^{(1)} (0 \alpha ) \tilde{\rho}^{\alpha}_{0}
 \; \; \; ,
 \label{eq:Skin11}
 \\
 \tilde{S}_{{\rm kin},2}^{(1)} 
  \left\{ \tilde{\rho}^{\alpha} \right\}
 &  = & 
  \frac{1}{2} \sum_{q} \sum_{\alpha \alpha^{\prime}}
  \Gamma^{(1)}_{2} ( - q {\alpha} , q \alpha^{\prime} ) \tilde{\rho}^{\alpha}_{-q} 
  \tilde{\rho}^{\alpha^{\prime}}_{q} 
 \label{eq:Skin12}
 \; \; \; ,
 \end{eqnarray}
with
 \begin{eqnarray}
 \Gamma_{1}^{(1)} (0 \alpha ) & = & \I \sum_{q} 
 \sum_{\alpha_{1} \alpha_{2} \alpha_{3} }
 U_{3} ( -q \alpha_{1} , q \alpha_{2} , 0 \alpha_{3} ) 
 \nonumber
 \\
 & & \hspace{15mm} \times
 \Gamma^{ \alpha_{2} \alpha_{1} } (q)
 \Gamma^{\alpha \alpha_{3} } (0)
 \; \; \; ,
 \label{eq:Gamma11def}
 \\
  \Gamma^{(1)}_{2} ( - q {\alpha} , q \alpha^{\prime} )  
  & = &
  - 3 \sum_{q^{\prime}}
 \sum_{\alpha_{1} \alpha_{2} \alpha_{3} \alpha_{4}}
 U_{4} ( - q \alpha_{1} , q \alpha_{2} , -q^{\prime} \alpha_{3} , q^{\prime} \alpha_{4} )
 \nonumber
 \\
 & & \hspace{15mm} \times
 \Gamma^{\alpha \alpha_{1}}  (q )
 \Gamma^{\alpha_{2} \alpha^{\prime}} (q)
 \Gamma^{\alpha_{4} \alpha_{3}}  (q^{\prime} )
 \; \; \; .
 \label{eq:Gamma21def}
 \end{eqnarray}
Recall that the superscript $^{(1)}$ indicates that these terms contain one internal bosonic loop.
Thus, within the one-loop approximation the constant in Eq.\ref{eq:Skinn}
is $\tilde{S}_{{\rm kin},0} =  \tilde{S}_{{\rm kin},0}^{(0)} + 
\tilde{S}_{{\rm kin},0}^{(1)}$
(see Eqs.\ref{eq:S00def} and \ref{eq:Skin01}),
and the vertices $\Gamma_{n}$ in Eq.\ref{eq:Gammank} are approximated by
 \begin{eqnarray}
 \Gamma_{1} ( q \alpha ) & = & \Gamma_{1}^{(1)} ( q \alpha )
 \; \; \; ,
 \\
 \Gamma_{2} ( -q \alpha , q \alpha^{\prime} ) & = & 
 \Gamma^{\alpha \alpha^{\prime} } (q)
 + \Gamma_{2}^{(1)} ( -q \alpha , q \alpha^{\prime} )
 \; \; \; ,
 \\
 \Gamma_{3} ( q_{1} \alpha_{1} q_{2} \alpha_{2} q_{3} \alpha_{3} )
 & = &
 \Gamma_{3}^{(0)} ( q_{1} \alpha_{1} q_{2} \alpha_{2} q_{3} \alpha_{3} )
 \; \; \; ,
 \\
 \Gamma_{4} ( q_{1} \alpha_{1} q_{2} \alpha_{2} q_{3} \alpha_{3} q_{4} \alpha_{4} )
 & = &
 \Gamma_{4}^{(0)} ( q_{1} \alpha_{1} q_{2} \alpha_{2} q_{3} \alpha_{3} q_{4} \alpha_{4} )
 \; \; \; ,
 \end{eqnarray}
and all $\Gamma_{n}$ with $n \geq 5$ are set equal to zero.
The term with $\Gamma_{1}$ can again be ignored for a calculation of correlation functions at finite $q$, because
it involves only the $q=0$ component of the density fields. Furthermore, 
for our one-loop calculation we may also ignore
the vertex $\Gamma_{3}$, because the Gaussian expectation value of a product of three
$\tilde{\rho}^{\alpha}$-fields vanishes.
Combining the relevant contributions from the kinetic energy with the
interaction contribution, we finally arrive at 
the effective action\index{effective action!density field, non-Gaussian corrections}
 \begin{eqnarray}
 \tilde{S}_{\rm eff} \left\{ \tilde{\rho}^{\alpha} \right\}   & \approx &
 \frac{1}{2} \sum_{q} \sum_{\alpha \alpha^{\prime}} 
 \left[  [ \underline{\tilde{f}}_{q} ]^{ \alpha \alpha^{\prime} } 
   + \Gamma^{\alpha \alpha^{\prime}}   (q )
  \right]
 \tilde{\rho}^{\alpha}_{-q} 
 \tilde{\rho}^{\alpha^{\prime}}_{q}
 \nonumber
 \\
 &  & \hspace{-10mm} +
  \frac{1}{2} \sum_{q} \sum_{\alpha \alpha^{\prime}}
  \Gamma^{(1)}_{2} ( - q {\alpha} , q \alpha^{\prime} ) \tilde{\rho}^{\alpha}_{-q} 
 \tilde{\rho}^{\alpha^{\prime}}_{q}
  \nonumber
  \\
  &  & \hspace{-10mm} +
 \frac{1}{4} \sum_{q_{1}  q_{2} q_{3} q_{4}}
 \sum_{\alpha_{1} \alpha_{2} \alpha_{3} \alpha_{4}} 
 \Gamma_{4}^{(0)} ( 
 q_1 \alpha_{1} q_{2} \alpha_{2} q_{3} \alpha_{3}  q_{4} \alpha_{4} ) {\tilde{\rho}}^{\alpha_{1}}_{q_{1}} 
 {\tilde{\rho}}^{\alpha_{2}}_{q_{2}}
 {\tilde{\rho}}^{\alpha_{3}}_{q_{3}}
 {\tilde{\rho}}^{\alpha_{4}}_{q_{4}}
 \label{eq:Seff1loop}
 \; \; \; ,
 \end{eqnarray}
which should be compared with the Gaussian action 
in Eq.\ref{eq:Seff2rhogaussres}.
We emphasize that this effective action is only good for the purpose of calculating the 
one-loop corrections to the Gaussian approximation. 
At two-loop order one should also retain the terms with $\Gamma_{3}$ and $\Gamma_{6}$.
The last two terms in Eq.\ref{eq:Seff1loop} contain the one-loop corrections to the
non-interacting boson approximation for the bosonized collective density fluctuations.
In the limit of long wavelengths we may again write down an equivalent effective Hamiltonian 
of canonically quantized bosons by using the substitution \ref{eq:rhobscale}.
However, we shall not even bother writing down this complicated expression, because
this mapping is only valid at long wavelengths and high densities, and does not
lead to any simplification.
For all practical purposes the parameterization
in terms of the $\tilde{\rho}^{\alpha}$-field is superior.
We shall now use this parameterization to calculate the
leading correction to the free bosonic propagator, 
and in this way determine the hidden small parameter
which controls the range of validity of the Gaussian approximation.

\subsection{The leading correction to the bosonic propagator}
\label{subsec:firstorder}

{\it{The calculation in this section takes non-linearities in the energy dispersion
as well as momentum-transfer between different patches (i.e. around-the-corner processes)
into account.}}

\vspace{7mm}

\noindent
Let us define a dimensionless proper 
self-energy matrix\index{self-energy!bosonic} $\underline{\Sigma}_{\ast}(q)$ via
 \begin{equation}
 \left< \tilde{\rho}^{\alpha}_{q} \tilde{\rho}^{\alpha^{\prime} }_{-q} 
 \right>_{\tilde{S}_{\rm eff}}
 = 
 \left[ \left[   \underline{\tilde{f}}_{q} + \underline{\Gamma} (q ) - 
 \underline{\Sigma}_{\ast} (q ) \right]^{-1} 
 \right]^{\alpha \alpha^{\prime}}
 \label{eq:rhoprop}
 \; \; \; ,
 \end{equation}
where the probability distribution for the average is determined by
the exact effective action $\tilde{S}_{\rm eff} \{ \tilde{\rho}^{\alpha} \}$, see
Eqs.\ref{eq:avrho}--\ref{eq:Skinrhodef}.
From Eq.\ref{eq:rhopropgauss} it is clear 
that the self-energy $\underline{\Sigma}_{\ast} (q )$ 
contains by definition all corrections to the RPA.
Introducing the exact proper polarization\index{polarization!proper polarization} 
matrix $\underline{\Pi}_{\ast} (q)$ via
 \begin{equation}
  \underline{\Pi}_{\ast }^{-1} ( q )   
  =   \underline{\Pi}_{0}^{-1} (q )  - 
 \underline{g} ( q )
 \; \; \; , \; \; \; 
 \underline{g} ( q ) = \frac{V}{\beta} \underline{\Sigma}_{\ast} (q )
  \; \; \; ,
 \label{eq:Piast}
  \end{equation}
the exact total density-density correlation function can be written as
 \begin{equation}
 \Pi ( q ) = \sum_{\alpha \alpha^{\prime}}
 \left[ \left[   \underline{\Pi}_{\ast }^{-1} ( q)  +  \underline{f}_{q} 
 \right]^{-1} \right]^{\alpha \alpha^{\prime} }
 \; \; \; .
 \label{eq:Piexact}
 \end{equation}
If all matrix elements of $\underline{f}_{q}$ are identical and equal
to $f_{q}$, we may repeat the manipulations in 
Eqs.\ref{eq:Ptotdecompose1}--\ref{eq:PiRPAtot}, so that
Eq.\ref{eq:Piexact} reduces to  Eq.\ref{eq:PiDyson}, with
 \begin{equation}
 \Pi_{\ast} ( q ) = \sum_{\alpha \alpha^{\prime}}
 [ \underline{\Pi}_{\ast} ( q ) ]^{\alpha \alpha^{\prime}}
 \; \; \; .
 \label{eq:Piglobprop}
 \end{equation}
Comparing Eq.\ref{eq:Piast} with Eq.\ref{eq:Piproper}, we see that
the quantities
$[ \underline{\Sigma}_{\ast} ({q}) ]^{\alpha \alpha^{\prime}}$
can be identified physically with 
generalized local field corrections\index{local field correction} 
$ [\underline{g}(q) ]^{\alpha \alpha^{\prime}}$, 
which differentiate between the contributions from the
various sectors.

\begin{figure}
\psfig{figure=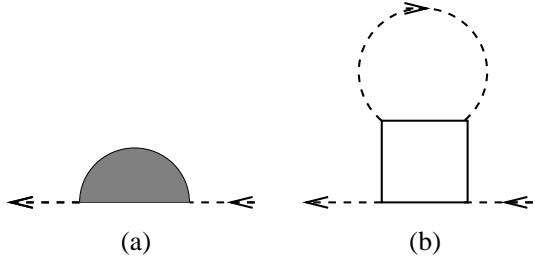,width=7cm}
\caption[Leading self-energy corrections to the Gaussian propagator
of the collective density field.]
{Leading self-energy corrections to the Gaussian  propagator of the collective
$\tilde{\rho}^{\alpha}$-field.
Diagram (a) represents the first term
in Eq.\ref{eq:self1}, while diagram (b) represents the second term.
Dashed arrows denote collective density fields $\tilde{\rho}^{\alpha}$,
and the dashed loop is the Gaussian propagator of the $\tilde{\rho}^{\alpha}$-field,
see Eq.\ref{eq:rhopropgauss}.
The vertex $\Gamma_{2}^{(1)}$ 
is represented by the shaded semi-circle. 
The shading indicates that this vertex
involves an internal bosonic loop summation,
see Eq.\ref{eq:Gamma21def}.
The tree-level vertex $\Gamma_{4}^{(0)}$ given in Eq.\ref{eq:Gamma04} 
is represented by an empty square.
}
\label{fig:self1}
\end{figure}

We now calculate the irreducible bosonic self-energy
to first order in an expansion in the number of bosonic loops.
To this order we simply have to add the two diagrams
shown in Fig.~\secref{fig:self1}.
Note that according to Eq.\ref{eq:Gamma21def} the shaded 
semi-circle vertex in the diagram (a) implicitly involves one 
internal loop summation. 
Hence, within the one-loop approximation,
the diagram (a) should be added to the diagram (b), which
explicitly contains a bosonic loop.
Because we have symmetrized the vertices, the diagram (b) has a combinatorial factor of
three, so that at one-loop order we obtain $\underline{\Sigma}_{\ast} (q ) \approx
\underline{\Sigma}_{\ast}^{(1)} (q )$, with
 \begin{eqnarray}
 [ \underline{\Sigma}_{\ast}^{(1)} (q ) ]^{\alpha \alpha^{\prime} }
 & = & - \Gamma_{2}^{(1)} ( -q \alpha , q \alpha^{\prime} )
 -   3 \sum_{q^{\prime}} \sum_{\alpha_{3} \alpha_{4} }
 \Gamma_{4}^{(0)} ( - q \alpha , q
 \alpha^{\prime} , -q^{\prime} \alpha_{3} ,q^{\prime} \alpha_{4} ) 
 \nonumber
 \\
 & \times &
 \left[ \left[   \underline{\tilde{f}}_{q^{\prime}} + 
 \underline{\Gamma} ({q^{\prime}}) \right]^{-1} \right]^{\alpha_{4} \alpha_{3}}
 \label{eq:self1}
 \; \; \; .
 \end{eqnarray}
Using the definitions of $\Gamma_{2}^{(1)}$ and $\Gamma_{4}^{(0)}$
(see Eqs.\ref{eq:Gamma21def} and \ref{eq:Gamma04}), 
it is easy to show that Eq.\ref{eq:self1} can also be written as
 \begin{eqnarray}
 [ \underline{\Sigma}_{\ast}^{(1)} ( q ) ]^{\alpha \alpha^{\prime} }
 & = &  3 
 \sum_{q^{\prime}} \sum_{\alpha_{1} \alpha_{2} \alpha_{3} \alpha_{4} }
 U_{4} ( - q \alpha_{1} , q
 \alpha_{2} , -q^{\prime} \alpha_{3} , q^{\prime} \alpha_{4} ) 
 \Gamma^{\alpha \alpha_{1}} (q)
 \Gamma^{\alpha_{2} \alpha^{\prime}} (q )
 \nonumber
 \\
 & \times &
 \left[ 
 \underline{\Gamma} ({q^{\prime}}) -
 \underline{\Gamma} ({q^{\prime}})
 \left[   \underline{\tilde{f}}_{q^{\prime}} + \underline{\Gamma} ({q^{\prime}}) \right]^{-1} 
 \underline{\Gamma} ({q^{\prime}})
 \right]^{ \alpha_{4} \alpha_{3} }
 \label{eq:self2}
 \; \; \; .
 \end{eqnarray}
A simple manipulation shows that the matrix in the last line 
of Eq.\ref{eq:self2}
can be identified with
 $\frac{ \beta}{V}  \underline{f}^{\rm RPA}_{q}$,
where $\underline{f}^{\rm RPA}_{q}$ is the RPA interaction
matrix defined in Eq.\ref{eq:frpapatchdef}.
We conclude that
 \begin{eqnarray}
 [ \underline{\Sigma}_{\ast}^{(1)} (q ) ]^{\alpha \alpha^{\prime} }
  & = &  3 \frac{ \beta}{V}
 \sum_{q^{\prime}} \sum_{\alpha_{1} \alpha_{2} \alpha_{3} \alpha_{4} }
 \Gamma^{\alpha \alpha_{1}} (q)
 \Gamma^{\alpha_{2} \alpha^{\prime}} (q)
 \nonumber
 \\
 & \times &
 U_{4} ( - q \alpha_{1} , q
 \alpha_{2} , -q^{\prime} \alpha_{3} , q^{\prime} \alpha_{4} ) 
 \left[ \underline{f}^{\rm RPA}_{q^{\prime} } 
 \right]^{ \alpha_{4} \alpha_{3} }
 \label{eq:self3}
 \; \; \; .
 \end{eqnarray}
Note that $\underline{\Sigma}_{\ast}^{(1)}(q)$ is proportional to the
RPA screened interaction and vanishes in the non-interacting limit,
as it should. Eq.\ref{eq:self3} is the general result for the
leading correction to the Gaussian approximation 
due to non-linearities in the energy dispersion and around-the-corner
processes for arbitrary sectorizations
and bare interaction matrices $\underline{f}_{q}$.

\subsection{The hidden small parameter\index{hidden small parameter}}
\label{subsec:hidden}

{\it{We now neglect the around-the-corner processes, 
but keep the non-linearities in the energy dispersion.}}

\vspace{7mm}

\begin{figure}
\sidecaption
\psfig{figure=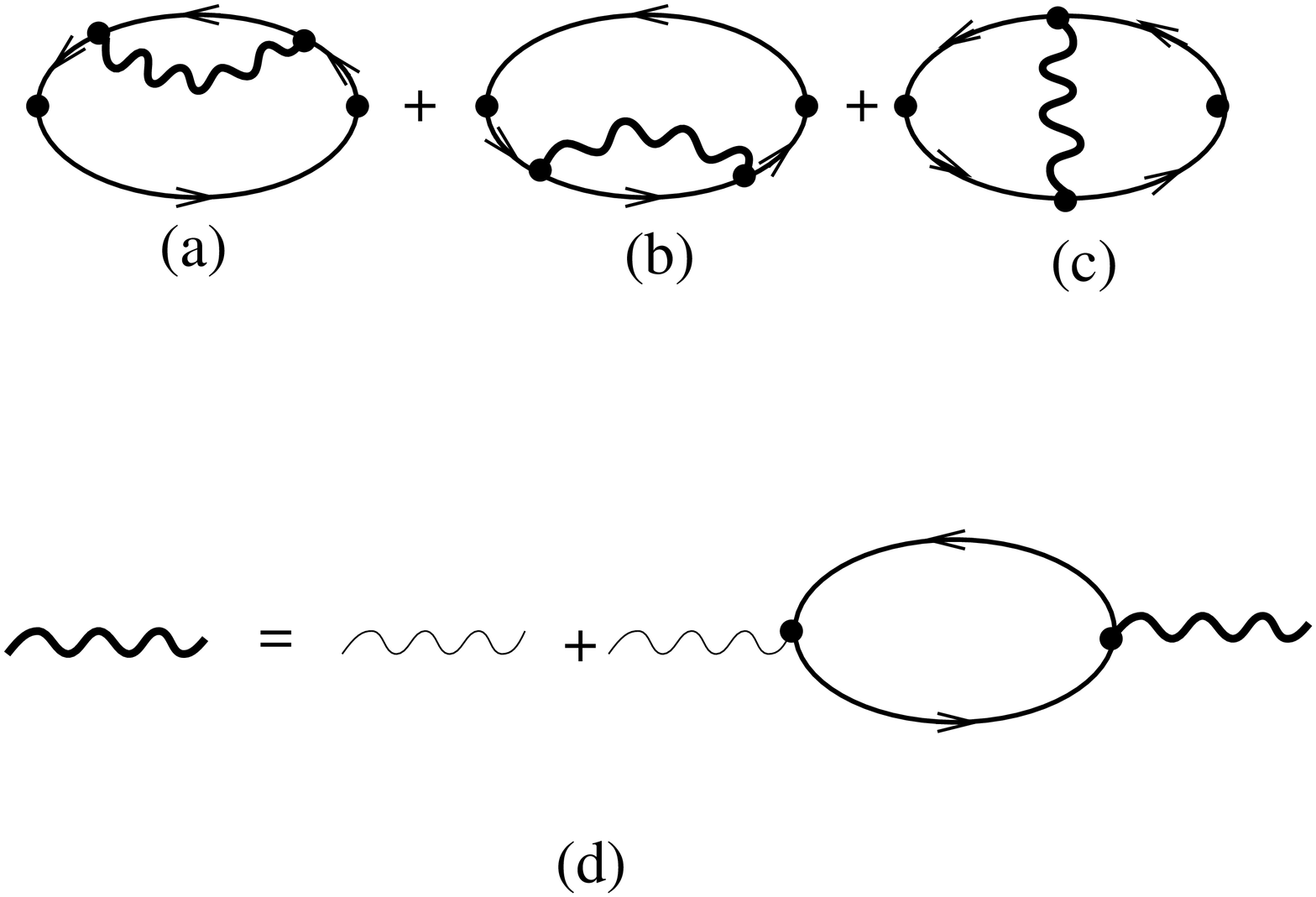,width=7.0cm}
\caption[Leading local field corrections to the non-interacting polarization.]
{\begin{sloppypar}
Leading local field corrections to the non-interacting polarization.
(a) and (b) are the leading self-energy corrections (see Eq.\ref{eq:sigmaF1}), while
(c) is the leading vertex correction (see Eq.\ref{eq:gammaF1}).
The thick wavy line denotes the RPA interaction, as defined in (d).
The thin wavy line represents the bare interaction.
\end{sloppypar}}
\label{fig:bubblecor}
\end{figure}

\noindent
To make further progress, we shall 
ignore from now on scattering processes that transfer momentum between
different sectors, i.e. the around-the-corner processes.
As discussed in Chap.~\secref{sec:sectors}, for 
non-linear energy dispersion we are free to choose rather large 
patches with finite curvature, so that the 
neglect of the around-the-corner processes\index{around-the-corner processes}
is not a serious restriction.
Moreover,
this approximation is always justified if there exists a cutoff 
$q_{\rm c} \ll \Lambda , \lambda \ll k_{\rm F}$ 
such that for wave-vectors $|{\vec{q}} | \geqapprox q_{\rm c}$ 
the effective interaction $\underline{f}_{q}^{\rm RPA}$ becomes negligibly small.
Choosing also the magnitude of the external wave-vector 
${\vec{q}}$ in Eq.\ref{eq:self3} small compared with the cutoffs $\Lambda $ and $\lambda$,
the diagonal-patch approximation $(A1)$ is justified, so that
$\underline{\Gamma} ( q )$ and
 $U_{4} ( - q \alpha_{1} , q
 \alpha_{2} , -q^{\prime} \alpha_{3} , q^{\prime} \alpha_{4} ) $ 
are diagonal in the patch indices, see Eqs.\ref{eq:Gammalong} 
and \ref{eq:Undiag}.
Then Eq.\ref{eq:self3} reduces to
 \begin{equation}
 [ \underline{\Sigma}_{\ast}^{(1)} ( q ) ]^{\alpha \alpha^{\prime} }
 = \delta^{\alpha \alpha^{\prime}} \frac{ \beta}{V \nu^{\alpha}}
 \left( \frac{  {\vec{v}}^{\alpha} \cdot {\vec{q}} - \I \omega_{m} }
 {  {\vec{v}}^{\alpha} \cdot {\vec{q}} } \right)^2 A^{\alpha}_{q}
 \; \; \; ,
 \label{eq:Sigma1diag}
 \end{equation}
where the dimensionless function $A^{\alpha}_{q}$ is given by
 \begin{equation}
 A^{\alpha}_{q} =  \frac{ 3}{ \nu^{\alpha} } \left( \frac{\beta}{V} \right)^2
 \sum_{q^{\prime}} U^{\alpha}_{4} ( - q  , q , - q^{\prime} , q^{\prime}  )
 \left[ \underline{f}^{\rm RPA}_{q^{\prime}} \right]^{\alpha \alpha}
 \label{eq:Aalphadef}
 \; \; \; ,
 \end{equation}
with $U^{\alpha}_{4}$ defined in Eq.\ref{eq:Uvertexdiag}.
We thus obtain in the limit of high densities and long wavelengths 
to first order in the screened interaction
 \begin{eqnarray}
  \Gamma^{\alpha \alpha^{\prime} } (q )
 - [ \underline{ \Sigma }_{\ast}^{(1)} (q) ]^{\alpha \alpha^{\prime} }
 & = &
 \frac{\beta}{V}
 [ \underline{\Pi}_{\ast}^{-1} ( q )  ]^{\alpha \alpha^{\prime}}
 \nonumber
 \\
 &  & \hspace{-38mm} =
 \delta^{\alpha \alpha^{\prime}} 
 \frac{\beta}{V \nu^{\alpha}} 
 \frac{
  ( 1 - A^{\alpha}_{q} )
   {\vec{v}}^{\alpha} \cdot {\vec{q}} 
  - ( 1 - 2 A^{\alpha}_{q} ) \I \omega_{m} - A^{\alpha}_{q} 
  \frac{ ( \I \omega_{m} )^2}{  {\vec{v}}^{\alpha} \cdot {\vec{q}} }
  }
  { 
     {\vec{v}}^{\alpha} \cdot {\vec{q}} }
  \label{eq:Pi1def}
  \; \; \; .
  \end{eqnarray}
Comparing this expression with Eq.\ref{eq:Gammalong}, it is evident that
the non-interacting boson approximation is quantitatively correct provided the condition
 $| A^{\alpha}_{q} | \ll 1$
is satisfied for all $\alpha$,
because then the corrections to the propagator of the  collective
density field $\tilde{\rho}^{\alpha}$ are small.
Using the general definition of the vertices $U_{n}$ in Eq.\ref{eq:Uvertex}, it is easy to show that
 \begin{eqnarray}
 A^{\alpha}_{q} 
 & = &
 - \frac{1}{\nu^{\alpha}  \beta V }
 \sum_{k}
 \Theta^{\alpha} ( {\vec{k}} )
 \left\{     G_{0} ( k) \Sigma^{(1)} ( k )  G_{0} ( k ) [ G_{0} ( k + q )  + G_{0} ( k- q ) ]
 \right.
 \nonumber
 \\
 & & +
 \left. \frac{1}{2}
  G_{0} ( k )   [ \Lambda^{(1) } ( k ; q )  G_{0} ( k + q ) 
  + \Lambda^{(1)} ( k ; -q ) G_{0}  ( k -q ) ]
 \right\} 
 \label{eq:Aalphaqres}
 \; \; \; ,
 \end{eqnarray}
with
 \begin{eqnarray}
 \Sigma^{(1)} ( k ) & = &
 - \frac{1}{\beta V } \sum_{q^{\prime}} 
 f^{\rm RPA}_{{q}^{\prime}} G_{0} ( k+q^{\prime} )
 \; \; \; ,
 \label{eq:sigmaF1}
 \\
 \Lambda^{(1)} ( k ; q) & = &
 - \frac{1}{  \beta V } \sum_{q^{\prime}} 
 f^{\rm RPA}_{{q}^{\prime}} 
 G_{0} ( k+q^{\prime})  G_{0} ( k + q^{\prime} + q )
 \label{eq:gammaF1}
 \; \; \; .
 \end{eqnarray}
Note  that $A^{\alpha}_{-q} = A^{\alpha}_{q}$ due to the symmetrization of the
vertex $U_{4}$ with respect to the interchange of any two labels.
It is now obvious that the vertices of our effective bosonic 
action automatically contain the relevant self-energy and vertex corrections 
of the underlying fermionic problem \cite{Geldart70}. \index{vertex corrections}
The first term in Eq.\ref{eq:Aalphaqres} corresponds to the
self-energy corrections to the non-interacting polarization
bubble shown in Fig.~\secref{fig:bubblecor} (a) and (b), while the last term
is due to the vertex correction\index{vertex function} shown in Fig.~\secref{fig:bubblecor}~(c).

In order to determine the range of validity of the non-interacting boson approximation,
we have to
calculate the dependence of  $A^{\alpha}_{q}$ on the various parameters in the problem.
In the limit of long wavelengths and low energies,
it is to leading order in $ {\vec{v}}^{\alpha} \cdot {\vec{q}}$ and 
$\omega_{m}$  consistent to replace in  Eq.\ref{eq:Pi1def}
$A^{\alpha}_{q} \rightarrow A^{\alpha}_{0}$.
Actually, the $q \rightarrow 0$ limit of $A^{\alpha}_{q}$ should be taken in such a way that
the ratio $  \I \omega_{m}  / (    {\vec{v}}^{\alpha} \cdot {\vec{q}}  )$ 
is held constant, because 
in this case we obtain the low-energy behavior of 
$A^{\alpha}_{q}$ close to the poles of the Gaussian propagator.
However, since we are only interested in the 
order of magnitude of $A^{\alpha}_{q}$ for small $\omega_{m}$ and
${\vec{q}}$, it is sufficient to consider the ``${\vec{q}}$-limit''
$A^{\alpha}_{0} = \lim_{\vec{q} \rightarrow 0} [ \lim_{\omega_{m} \rightarrow 0 } A^{\alpha}_{q} ]$.  
For simplicity, let us assume that $[ \underline{f}^{\rm RPA}_{q} ]^{\alpha \alpha^{\prime} }
= f^{\rm RPA}_{\vec{q}}$ is independent of the sector labels and depends only on the
wave-vector. This amounts to the static approximation for the dielectric function, which
seems reasonable to obtain the correct order of magnitude of $A^{\alpha}_{0}$.
The ``${\vec{q}}$-limit'' is obtained by
setting $q=0$ under the summation sign and performing the 
Matsubara sums before doing the wave-vector integrations.
For $\beta \rightarrow \infty$ we obtain
 \begin{eqnarray}
 A^{\alpha}_{0} & = & \frac{1}{\nu^{\alpha} V^2}
 \sum_{ {\vec{k}} {\vec{q}} }
 \Theta^{\alpha} ( {\vec{k}} )
 f^{\rm RPA}_{\vec{q}}
 \left\{ f (  \xi_{\vec{k+q}} )  
 \frac{ \partial^{2}}{\partial \mu^2} f ( \xi_{\vec{k}}  )
 \right.
 \nonumber
 \\
 & & \hspace{30mm}
 \left. 
 + \frac{\partial}{\partial \mu } f ( \xi_{ \vec{k} + \vec{q} }  )
 \frac{\partial}{\partial \mu } f ( \xi_{ \vec{k} }  )
 \right\}
 \label{eq:A0res1}
 \; \; \; .
 \end{eqnarray}
Because the
${\vec{k}}$-sum extends over the entire sector $K^{\alpha}_{\Lambda , \lambda}$ and
by assumption the ${\vec{q}}$-sum is cut off by the interaction
at $q_{\rm c} \ll  \Lambda , \lambda $,
we may set $\xi_{\vec{k} + \vec{q}} \approx \xi_{\vec{k}}$ in the Fermi functions
of Eq.\ref{eq:A0res1}. Then the summations factorize,
and we obtain 
 \begin{equation}
 A^{\alpha}_{0} = 
 \left[ \frac{1}{V} \sum_{\vec{q} }
 f^{\rm RPA}_{\vec{q}} \right]
  \frac{1}{ \nu^{\alpha} } \int_{- \infty}^{\infty} \D \xi
 \nu^{\alpha} ( \xi )
  \left[ f ( \xi ) f^{\prime \prime} ( \xi ) 
  + f^{\prime} ( \xi ) f^{\prime} ( \xi )  \right]
  \label{eq:A0res2}
  \; \; \; ,
  \end{equation}
where $\nu^{\alpha} ( \xi )$ is the
energy-dependent patch density of states, \index{density of states!local (or patch)}
 \begin{equation}
 \nu^{\alpha} ( \xi ) = \frac{1}{V} \sum_{\vec{k}} 
 \Theta^{\alpha} ( {\vec{k}} ) \delta (  \xi - \xi_{\vec{k}}  )
 \label{eq:nualphae}
 \; \; \; .
 \end{equation}
Note that from the definition \ref{eq:nualphadef}
of $\nu^{\alpha}$ it follows that
 \begin{equation}
 \nu^{\alpha} = \int_{- \infty}^{\infty} \D \xi 
 \nu^{\alpha} ( \xi ) \left[ -  f^{\prime} ( \xi )
 \right]
 \label{eq:nunu}
 \; \; \; .
 \end{equation}
Integrating by parts and taking the limit
$\beta \rightarrow \infty$, the integral in Eq.\ref{eq:A0res2} 
can be written as 
 \begin{equation}
  \int_{- \infty}^{\infty} \D \xi
 \nu^{\alpha} ( \xi ) \frac{\partial}{\partial \xi}
  \left[ f ( \xi ) f^{\prime} ( \xi ) \right]
 = \frac{1}{ 2 } \frac{ \partial \nu^{\alpha}}{ \partial \mu }
 \label{eq:intebyparts1}
 \; \; \; .
 \end{equation}
Because by assumption $f^{\rm RPA}_{\vec{q}}$ 
becomes negligibly small for $|{\vec{q}}  | \geqapprox q_{\rm c}$,
the first factor in Eq.\ref{eq:A0res2}
is for $V \rightarrow \infty$ given by
 \begin{equation}
  \frac{1}{V} \sum_{\vec{q} }
 f^{\rm RPA}_{\vec{q}} = q_{\rm c}^{d} \langle {f}^{\rm RPA} \rangle
 \label{eq:firstfac}
 \; \; \; ,
 \end{equation}
where $\langle {f}^{\rm RPA} \rangle $ is some suitably defined measure for the
average strength of the screened interaction.
Ignoring a numerical factor of the order of unity,
the final result for $A^{\alpha}_{0}$ can be written as
 \begin{equation}
A^{\alpha}_{0} = \frac{ q_{\rm c}^d \langle {f}^{\rm RPA}\rangle }{\mu } C^{\alpha}
 \label{eq:Afinal1}
 \; \; \; ,
 \end{equation}
where the dimensionless parameter 
 \begin{equation}
 C^{\alpha} = \frac{ \mu}{\nu^{\alpha}} \frac{ \partial \nu^{\alpha} }{\partial \mu }
 =  
 \frac{  \mu  \partial^2 N_{0}^{\alpha} / {\partial \mu^2}}{
  \partial N_{0}^{\alpha} / \partial \mu}
 \label{eq:g1def}
 \end{equation}
is for $d > 1$ a measure for the {\it{local curvature of the Fermi 
surface\index{Fermi surface!curvature} in 
patch ${P}^{\alpha}_{\Lambda}$}}.
Although the patch density of states $\nu^{\alpha}$ 
is proportional to $\Lambda^{d-1}$ (see Eq.\ref{eq:patchdenscutoff2}),
the cutoff-dependence cancels in Eq.\ref{eq:g1def}, because it
appears in the numerator as well as in the denominator.
Therefore $C^{\alpha}$ is a cutoff-independent quantity.
In fact, writing $\nu^{\alpha}$ as a surface integral over
the curved patch $P^{\alpha}_{\Lambda}$
(see Eq.\ref{eq:patchdenscutoff}), simple geometric
considerations lead to the result
 \begin{equation}
 C^{\alpha} = \frac{ \langle k_{\rm F} \rangle }{ m^{\alpha} | {\vec{v}}^{\alpha} | }
 \label{eq:Ccurveresult}
 \; \; \; ,
 \end{equation}
where $\langle k_{\rm F} \rangle$ is some suitably defined average
radius of the Fermi surface, and $m^{\alpha}$ is the effective mass close to
${\bf{k}}^{\alpha}$, see Eq.\ref{eq:chiquad}.
Note that 
$\langle k_{\rm F} \rangle$ characterizes the {\it{global}} geometry of the
Fermi surface, while $m^{\alpha}$ and ${\vec{v}}^{\alpha}$ depend
on the {\it{local}} shape of the Fermi surface in patch $P^{\alpha}_{\Lambda}$.
Evidently
$C^{\alpha}$ vanishes 
if we linearize the
energy dispersion in patch ${P}^{\alpha}_{\Lambda}$, 
because the linearization  amounts to taking the limit
$|m^{\alpha} | \rightarrow  \infty$ while keeping
$\langle k_{\rm F} \rangle$ finite.
Then there is no correction to the Gaussian approximation.
Of course, we already know from the closed loop theorem that 
the Gaussian approximation is exact if
the energy dispersion is linearized and the
around-the-corner\index{around-the-corner processes} 
processes are neglected.

As usual, we introduce the 
dimensionless interaction $\langle {F}^{\rm RPA} \rangle = \nu 
\langle {f}^{\rm RPA} \rangle $, 
which measures the strength of the potential energy relative to the
kinetic energy.
Using the fact that the global density of states is in $d$
dimensions proportional to $ k_{\rm F}^{d-2}$ (see Eq.\ref{eq:nurelation}), 
we conclude that the Gaussian approximation is 
quantitatively accurate as long as 
for all patches ${P}^{\alpha}_{\Lambda}$
 \begin{equation}
 | A^{\alpha}_{0} | \equiv  \left( \frac{ q_{\rm c}}{k_{\rm F} } \right)^{d} | 
 \langle {F}^{\rm RPA} \rangle |  | C^{\alpha}| \ll 1
 \label{eq:gausscorrect}
 \; \; \; .
 \end{equation}
The appearance of three parameters that control the accuracy
of the Gaussian approximation has a very simple intuitive interpretation.
First of all, if everywhere on the Fermi surface the
curvature is intrinsically small 
(i.e. $| C^{\alpha} | \ll 1$ for all $\alpha$)
then the corrections to the linearization of the energy dispersion are
negligible, and hence the Gaussian approximation becomes accurate. 
Note that in the one-dimensional Tomonaga-Luttinger model
$C^{\alpha} = 0$, 
because the energy dispersion is linear by definition. 
However, in $d > 1$ and for realistic energy dispersions of the form
$\epsilon_{\vec{k}} = { {\vec{k}}^{2}}/({2 m})$ the
dimensionless curvature parameter
$C^{\alpha}$ is of the order of unity. 
But even then the
Gaussian approximation is accurate, provided the nature of the interaction 
is such that it involves only small momentum-transfers. 
This is also intuitively obvious, because in this case
the scattering processes probe only a thin shell around the
Fermi surface and do not feel the deviations from linearity.
Finally, it is clear that also the strength of the effective interaction should
determine the range of validity of Gaussian approximation, 
because in the limit that
the strength of the interaction approaches zero
all corrections to the Gaussian approximation vanish. 
We would like to emphasize, however, that we have not explicitly
calculated the corrections to the Gaussian approximation due
to around-the-corner processes, although our general result
for the bosonic self-energy in Eq.\ref{eq:self3}
includes also these corrections.
Nevertheless, the around-the-corner processes can to a large extent be
eliminated by subdividing the Fermi surface into a small number of
curved patches, as discussed in Chap.~\secref{sec:sectors}.

\subsection{Calculating corrections to the RPA\index{random-phase approximation!corrections} 
via bosonization}
\label{subsec:diagram}

{\it{Here comes the first practical application of our formalism.}}

\vspace{7mm}

\noindent
For simplicity let us assume that the diagonal-patch approximation $(A1)$ 
is justified, so that Eq.\ref{eq:Piast} reduces to an
equation for the diagonal elements,
 \begin{eqnarray}
 \Pi_{\ast}^{ \alpha} ( q ) & = & \frac{ \Pi_{0}^{\alpha} ( q ) }{ 1 -
 g^{\alpha} ( q ) \Pi_{0}^{\alpha} ( q) }
 \nonumber
 \\
 & \approx &
 \Pi_{0}^{\alpha} ( q ) + 
 \Pi_{0}^{\alpha} ( q )    
 g^{ \alpha} (q) \Pi_{0}^{\alpha} ( q) 
 + \ldots 
 \; \; \; .
 \label{eq:Dysonexp}
 \end{eqnarray}
Here $ \Pi_{\ast}^{ \alpha} ( q ) = [ \underline{\Pi}_{\ast} ( q ) ]^{\alpha \alpha }$,
$ g^{\alpha} (q ) = [ \underline{g} (q) ]^{\alpha \alpha }$, and
$\Pi_{0}^{\alpha} ( q )$ is at long wavelengths given
in Eq.\ref{eq:Pilong}. 
Note that
our approach is based on the perturbative
calculation of the {\it{inverse}} proper polarization, while in the
naive perturbative approach 
the corrections to the proper polarization are obtained by 
direct expansion of $\Pi_{\ast} ( q )$ in powers of the interaction \cite{Holas79}.
Such a procedure does not correspond to the perturbative calculation
of the {\it{irreducible self-energy}} in our effective bosonic problem, but
is equivalent to a direct expansion of the {\it{Green's function}}.
As discussed in Chap.~\secref{sec:thelimitations}, 
close to the poles of the Green's function this expansion cannot be expected to be reliable.
To first order, only the leading correction in the
expansion of the Dyson equation\index{Dyson equation!for proper polarization}
(i.e. the second line in Eq.\ref{eq:Dysonexp}) is kept in this method,
so that the total proper polarization is approximated by
 \begin{eqnarray}
 \Pi_{\ast} ( q ) & \approx & \sum_{\alpha} 
 \left[ 
 \frac{  
  \nu^{\alpha}  
  {\vec{v}}^{\alpha} \cdot {\vec{q}} }
 {  {\vec{v}}^{\alpha} \cdot {\vec{q}} - \I \omega_{m} }
 + \nu^{\alpha} A^{\alpha}_{q} \right]
 \nonumber
 \\
 & = & \Pi_{0} ( q ) - \frac{1}{\beta V } 
 \sum_{k}
 \left\{     G_{0} ( k) \Sigma^{(1)} ( k )  G_{0} ( k ) [ G_{0} ( k + q )  + G_{0} ( k- q ) ]
 \right.
 \nonumber
 \\
 & &  \hspace{-10mm} +
 \left. \frac{1}{2}
  G_{0} ( k )   [ \Lambda^{(1) } ( k ; q )  G_{0} ( k + q ) 
  + \Lambda^{(1)} ( k ; -q ) G_{0}  ( k -q ) ]
 \right\} 
 \; \; \; ,
 \label{eq:Pipropres1}
 \end{eqnarray}
see Eqs.\ref{eq:Aalphaqres}--\ref{eq:gammaF1}.
For the Coulomb interaction in $d=3$ the correction term in Eq.\ref{eq:Pipropres1} has been 
discussed by Geldart and Taylor \cite{Geldart70}, as well 
as by Holas {\it{et al.}} \cite{Holas79}. Note, however, that these 
authors have evaluated the
fermionic self-energy $\Sigma^{(1)} ( k )$ and vertex correction
$\Lambda^{(1)} ( k ;q )$ with the bare Coulomb interaction.
Holas {\it{et al.}} \cite{Holas79} have also pointed 
out that the expansion \ref{eq:Pipropres1} 
leads to unphysical singularities in the dielectric function close to the
plasmon poles. The origin for these singularities is easy to understand
within our bosonization approach.
The crucial point is that the problem of calculating the corrections to the
RPA can be completely mapped onto an effective bosonic problem: 
our functional bosonization method allows us
to explicitly construct 
the {\it{interacting}} bosonic Hamiltonian.
Once we accept the validity of this mapping, 
standard many-body theory tells us that
the corrections to the propagator 
of this effective bosonic theory should be calculated 
by expanding its irreducible self-energy
$\underline{\Sigma}_{\ast} ({q})$ in the number of internal bosonic loops,
and then resumming the perturbation series
by means of the Dyson equation.
A similar resummation has been suggested in \cite{Rajagopal72,Dharma76,Holas79}, but 
it is not so easy to justify this procedure at the fermionic level.
Our bosonization approach provides the natural justification for this
resummation.
The unphysical singularities \cite{Holas79} that are encountered
in the naive perturbative approach are easy to understand from the
point of view of bosonization:
they are most likely due to the fact that one attempts to
calculate a bosonic single-particle Green's function by direct expansion.
This expansion is bound to fail close to the poles of the Green's function, i.e.
close to the plasmon poles!

Based on the insights gained from our bosonization approach,
we would like to suggest that
corrections to the RPA should be calculated by 
expanding the generalized local field corrections 
$\underline{g} ( q )$ in powers of the RPA interaction.
We suspect that in this way
unphysical singularities in the dielectric function can be avoided. 
From the first line in Eq.\ref{eq:Dysonexp} we obtain
in our method for the total proper polarization
at long wave-lengths
within the diagonal-patch approximation
 \begin{equation}
 \Pi_{\ast} ( q ) = \sum_{\alpha} 
 \frac{  
 \frac{ \nu^{\alpha} }{ 1 - A^{\alpha}_{q} } 
  {\vec{v}}^{\alpha} \cdot {\vec{q}} }
 {  {\vec{v}}^{\alpha} \cdot {\vec{q}} - \I \omega_{m} ( 1 - {B}^{\alpha}_{q} )
 - \frac{ ( \I \omega_{m} )^2 }{  {\vec{v}}^{\alpha} \cdot {\vec{q}} } {{B}}^{\alpha}_{q} }
 \; \; \; ,
 \label{eq:Pipropres}
 \end{equation}
with
 ${B}^{\alpha}_{q} =  A^{\alpha}_{q} / ( 1 - A^{\alpha}_{q} )$.
Recall that the around-the-corner processes have been neglected
in the derivation of Eq.\ref{eq:Pipropres},
so that it is expected to be accurate for sufficiently small
${\vec{q}}$ and for interactions that are dominated by small
momentum-transfers.

\section{Summary and outlook}

In this chapter we have developed a general formalism which allows us to
bosonize the Hamiltonian of fermions interacting with
two-body density-density forces in arbitrary dimensions. 
We have also shown that the bosonization of the Hamiltonian
is closely related to the problem of calculating the density-density correlation
function.  In general,
the bosonized system is described by an effective action of
collective density fields which contains also multiple-particle interactions between
the bosons. However, the generalized closed loop theorem
discussed in Sect.~\secref{sec:closedloop} guarantees that
in certain parameter regimes the 
vertices describing the interactions are small.
To leading order, the collective density fields
can then be treated as non-interacting bosons. 
The relevant small parameter justifying this
approximation has been explicitly calculated, and is given in 
Eqs.\ref{eq:Ccurveresult} and \ref{eq:gausscorrect}.

From the practical point of view, higher-dimensional bosonization
might lead to a new systematic method for
{\bf{calculating corrections to the RPA}}.
This is an old problem, which in the context of the
homogeneous electron gas has been discussed
thoroughly by Geldart and Taylor \cite{Geldart70} long time ago. 
These authors already observed partial
cancellations between the leading corrections to the RPA.
We now know that these cancellations occur to all orders in
perturbation theory, and are a direct consequence of the
generalized closed loop theorem.
The calculation of the local field corrections to the RPA is
still an active area of research \cite{Fleszar95,Moroni95},
which could  get some fresh momentum from the
non-perturbative insights gained via higher-dimensional bosonization.
Note that the corrections to the RPA 
describe the damping of the collective density
oscillations.  This and other effects can in principle be obtained
from Eq.\ref{eq:Pipropres} and the resulting
dielectric function $\epsilon ( q ) = 1 + f_{q} \Pi_{\ast} ( q )$.
This calculation
requires a careful analysis of the analytic properties of the 
function $A^{\alpha}_{q}$ defined in Eq.\ref{eq:Aalphaqres}, 
and still remains to be done.
The possibility that higher-dimensional bosonization might
lead to a new systematic method for calculating corrections to the RPA
has also been suggested by Khveshchenko \cite{Khveshchenko94b}.

%
%

%
%

\chapter{The single-particle Green's function}
\setcounter{equation}{0}
\label{chap:agreen}

{\it{
In this central chapter of this book
we calculate the single-particle Green's function
by means of the background field method\index{background field method} 
outlined in Chap.~\secref{sec:HS1}.
We carefully examine
the approximations and limitations inherent in higher-dimensional bosonization,
and develop a new systematic method for
including the non-linear terms in the expansion
of the energy dispersion close to the Fermi surface into the bosonization
procedure.  Short accounts of
the results and methods developed in this chapter have
been published in \cite{Kopietz94,Kopietz95d,Kopietz96bac}.
}}

\vspace{7mm}

\noindent
According to Eq.\ref{eq:avphi}
the Matsubara Green's function $G ( k ) \equiv G ( {\vec{k}} , \I \tilde{\omega}_n )$ can be
{\it{exactly}} written as\index{background field method}
 \begin{eqnarray}
 G( k )  = 
 \int {\cal{D}} \{ \phi^{\alpha} \} 
 {\cal{P}} \{ \phi^{\alpha} \} 
 [ \hat{G} ]_{kk}
 \equiv 
 \left< [ \hat{G}]_{  k  k }   \right>_{ S_{\rm eff} }
 \label{eq:avphi3}
 \; \; \; ,
 \end{eqnarray}
where the probability distribution\index{probability distribution!background field} 
${\cal{P}} \{ \phi^{\alpha} \} $ is defined in Eq.\ref{eq:probabphidef},
and the matrix elements of the {\it{inverse}} of the infinite matrix
$\hat{G}$ are given by
 \begin{equation}
 [ \hat{G}^{-1} ]_{k k^{\prime}} 
 =
 \sum_{\alpha} 
 \Theta^{\alpha} ( {\vec{k}} )
 \left[  \delta_{k k^{\prime}} ( \I \tilde{\omega}_{n} - 
 \epsilon_{\vec{k}} + \mu )
 - V^{\alpha}_{k -k^{\prime} } \right] 
 \; \; \; ,
 \label{eq:GhatDysonshift2}
 \end{equation}
with $V^{\alpha}_{q} = \frac{\I}{\beta} \phi^{\alpha}_{q}$, 
see Eq.\ref{eq:hatVphi}.
The cutoff function
$\Theta^{\alpha} ( {\vec{k}} )$ refers either to the
boxes intersecting
the Fermi surface discussed in Chap.~\secref{subsec:patch},
or to the more general sectors introduced in
Chap.~\secref{sec:sectors}, which by construction cover the entire
momentum space.
Note also that Eq.\ref{eq:GhatDysonshift2} includes
the special case (discussed at the end of
Chap.~\secref{sec:sectors}) that the entire momentum space 
is identified with a single sector.
Then the $\alpha$-sum contains only a single term, and by definition
we may replace the cutoff-function $\Theta^{\alpha} ( {\vec{k}} )$ 
by unity.

\section[The Gaussian approximation with linearized energy dispersion]
{The Gaussian approximation \mbox{\hspace{40mm}} with linearized energy dispersion}
\label{sec:Derivation}

{\it{
We show how for linearized energy dispersion the calculation of the Green's function 
from Eq.\ref{eq:avphi3} is carried out in practice.
We first discuss the inversion problem of the
infinite matrix
$\hat{G}^{-1}$. The averaging of the diagonal elements
$[ \hat{G} ]_{kk}$ with respect to the Gaussian probability
distribution ${\cal{P}}_{2} \{ \phi^{\alpha} \}$
yields then a simple Debye-Waller factor.}}

\vspace{7mm}
\noindent
In a parameter regime where the approximations $(A1)$ and $(A2)$
discussed in Chap.~\secref{sec:closedloop} are justified,
the generalized closed loop theorem guarantees
that the Gaussian approximation is very accurate.
As shown in Chap.~\secref{subsec:gaussphi}, 
the effective action $S_{\rm eff} \{ \phi^{\alpha} \}$ of the
$\phi^{\alpha}$-field is then to a good approximation given 
by\index{effective action!Gaussian approximation for $\phi^{\alpha}$-field}
(see Eqs.\ref{eq:Seffphigaussres} and \ref{eq:Seff2phigaussres}), 
 \begin{equation}
 S_{\rm eff} \{ \phi^{\alpha} \}     \approx 
 \I  \sum_{\alpha}
  \phi^{\alpha}_{0} N^{\alpha}_{0}
  + S_{{\rm eff},2} \{ \phi^{\alpha} \}
 \label{eq:Seffphigreen}
 \; \; \; ,
 \end{equation}
where the quadratic part is
   \begin{equation}
   S_{{\rm eff},2} \{ \phi^{\alpha} \}
   = 
 \frac{V}{2 \beta } \sum_{q} \sum_{\alpha \alpha^{\prime} }
 [    [\underline{{f}}_{ {{q}} }^{-1} ]^{\alpha \alpha^{\prime}}
 + \delta^{\alpha \alpha^{\prime}} {\Pi}_{0}^{\alpha} (q) ]
 \phi_{-q}^{\alpha} \phi_{q}^{\alpha^{\prime}}
 \label{eq:Seffphigreen2}
 \; \; \; ,
 \end{equation}
with $\Pi^{\alpha}_{0} ( q )$ given in Eq.\ref{eq:Pilong}.
The probability distribution ${\cal{P}} \{ \phi^{\alpha} \}$
associated with the $\phi^{\alpha}$-field in Eq.\ref{eq:probabphidef} is then 
Gaussian\index{probability distribution!background field},
 \begin{equation}
 {\cal{P}} \{ \phi^{\alpha} \} 
 \approx 
 {\cal{P}}_{2} \{ \phi^{\alpha} \} 
 \equiv
 \frac{  
 \E^{ - {S}_{{\rm{eff}},2} \{ \phi^{\alpha} \} }  }
 {
 \int {\cal{D}} \left\{ \phi^{\alpha} \right\} 
 \E^{ - {S}_{{\rm eff},2} \{ \phi^{\alpha} \} }  }
 \; \; \; .
 \label{eq:probabphigauss}
 \end{equation}
The first term in Eq.\ref{eq:Seffphigreen} involving $\phi^{\alpha}_{0}$
can be ignored for the calculation of correlation functions at $q \neq 0$.
Although within the Gaussian approximation the density-density correlation
function is given by the usual RPA result,
the {\it{single-particle Green's function}} in Eq.\ref{eq:avphi3}
can exhibit a large variety of behaviors, which range from conventional Fermi liquids over
Luttinger liquids to even more exotic quantum liquids.
Which of these possibilities is realized depends crucially
on the dimensionality of the system, on the nature of the interaction, 
and on the symmetry of the Fermi surface.

Of course, in general it is impossible to
invert $\hat{G}^{-1}$ exactly, so that one usually has to
use some sort of perturbation theory to calculate the matrix elements
$[\hat{G}]_{kk}$. 
However,  in the parameter regime where the conditions $(A1)$ and $(A2)$
are accurate, it is possible
to {\it{calculate the matrix elements $[\hat{G}]_{kk}$ exactly as functionals
of the $\phi^{\alpha}$-field.}}
Note that the conditions $(A1)$ and $(A2)$ imply also the
validity of the closed loop theorem, which in turn insures that the probability
distribution ${\cal{P}} \{ \phi^{\alpha} \}$ is Gaussian.
In other words, the conditions under which 
${\cal{P}} \{ \phi^{\alpha} \}$  
can be approximated by a Gaussian are 
also sufficient to guarantee that 
$\hat{G}^{-1}$ can be inverted exactly.


\subsection{The Green's function for fixed background field}
\label{subsec:invdiag}

{\it{To invert $\hat{G}^{-1}$, we proceed in two steps. We first
show that the condition $(A1)$ discussed in Chap.~\secref{sec:closedloop}
implies that
$\hat{G}^{-1}$ is approximately
block diagonal, with diagonal blocks $( \hat{G}^{\alpha} )^{-1}$ labelled
by the sector (or patch) indices.
Therefore the problem of inverting $\hat{G}^{-1}$ can be reduced
to the problem of inverting each diagonal block separately.
We then show that, after linearization of the energy dispersion,
each block $( \hat{G}^{\alpha})^{-1}$ can be inverted exactly.
}}

\begin{center}
{\bf{Block diagonalization\index{Green's function!block diagonalization}}}
\end{center}

\noindent
The quadratic form defining the matrix elements
$[ \hat{G}^{-1} ]_{k k^{\prime}}$  
in Eq.\ref{eq:GhatDyson} can be written as
 \begin{equation}
 {S}_{0} \{ \psi \}    +
 {S}_{1} \{ \psi , \phi^{\alpha} \}   
 = 
 - \beta \sum_{k q}  \psi^{\dagger}_{k+q}
 [ \hat{G}^{-1} ]_{k+q , k} \psi_{k}
 \; \; \; ,
 \label{eq:SmatG2}
 \end{equation}
with
 \begin{equation}
 [ \hat{G}^{-1} ]_{k +q , k} 
 =
 \sum_{\alpha} 
 \Theta^{\alpha} ( {\vec{k}} )
 \left[  \delta_{q,0} ( \I \tilde{\omega}_{n}
 - \xi^{\alpha}_{ {\vec{k}} - {\vec{k}}^{\alpha} }  
 - \epsilon_{ \vec{k}^{\alpha}} + \mu
 )
 - V^{\alpha}_{q } \right] 
 \; \; \; ,
 \label{eq:GhatDysonshift}
 \end{equation}
where $\xi^{\alpha}_{\vec{q}}= \epsilon_{ {\vec{k}}^{\alpha} + {\vec{q}}} - 
\epsilon_{\vec{k}^{\alpha}}  $ is 
the excitation energy relative to 
the energy at $\vec{k}^{\alpha}$ (see Eq.\ref{eq:energyquad}),
and $V^{\alpha}_{q} = \frac{\I}{\beta} \phi^{\alpha}_{q}$ (see 
Eq.\ref{eq:hatVphi}).
The cutoff function $\Theta^{\alpha} ( {\vec{k}} )$
groups the matrix elements of the infinite matrix 
$\hat{G}^{-1}$
into rows labelled by the patch index $\alpha$.
{
\begin{figure}
\sidecaption
\psfig{figure=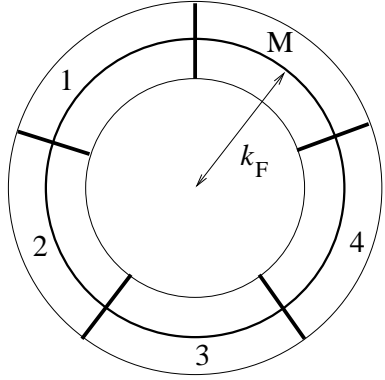,width=5cm}
\caption[Subdivision of momentum space in $d=2$ into boxes.]
{
Subdivision of momentum space close to a spherical Fermi surface in $d=2$ into $M=5$ 
sectors $K^{\alpha}_{\Lambda , \lambda}$, $\alpha = 1 , \ldots , M$.}
\label{fig:patch3}
\end{figure}
To see this more clearly,
consider for simplicity a spherical Fermi surface in $d=2$. We partition  the
degrees of freedom in the vicinity of the Fermi surface into $M $ sectors $K^{\alpha}_{\Lambda , \lambda}$,
and label neighboring sectors in increasing order, as shown in Fig.~\secref{fig:patch3}.
The group of matrix elements corresponding to a given label $\alpha$ 
in Eq.\ref{eq:GhatDysonshift}
can be found in the horizontal stripes in the 
schematic representation of the matrix $\hat{G}^{-1}$ shown
in Fig.~\secref{fig:band}(a).
\begin{figure}
\begin{minipage}{7.2cm}
\psfig{figure=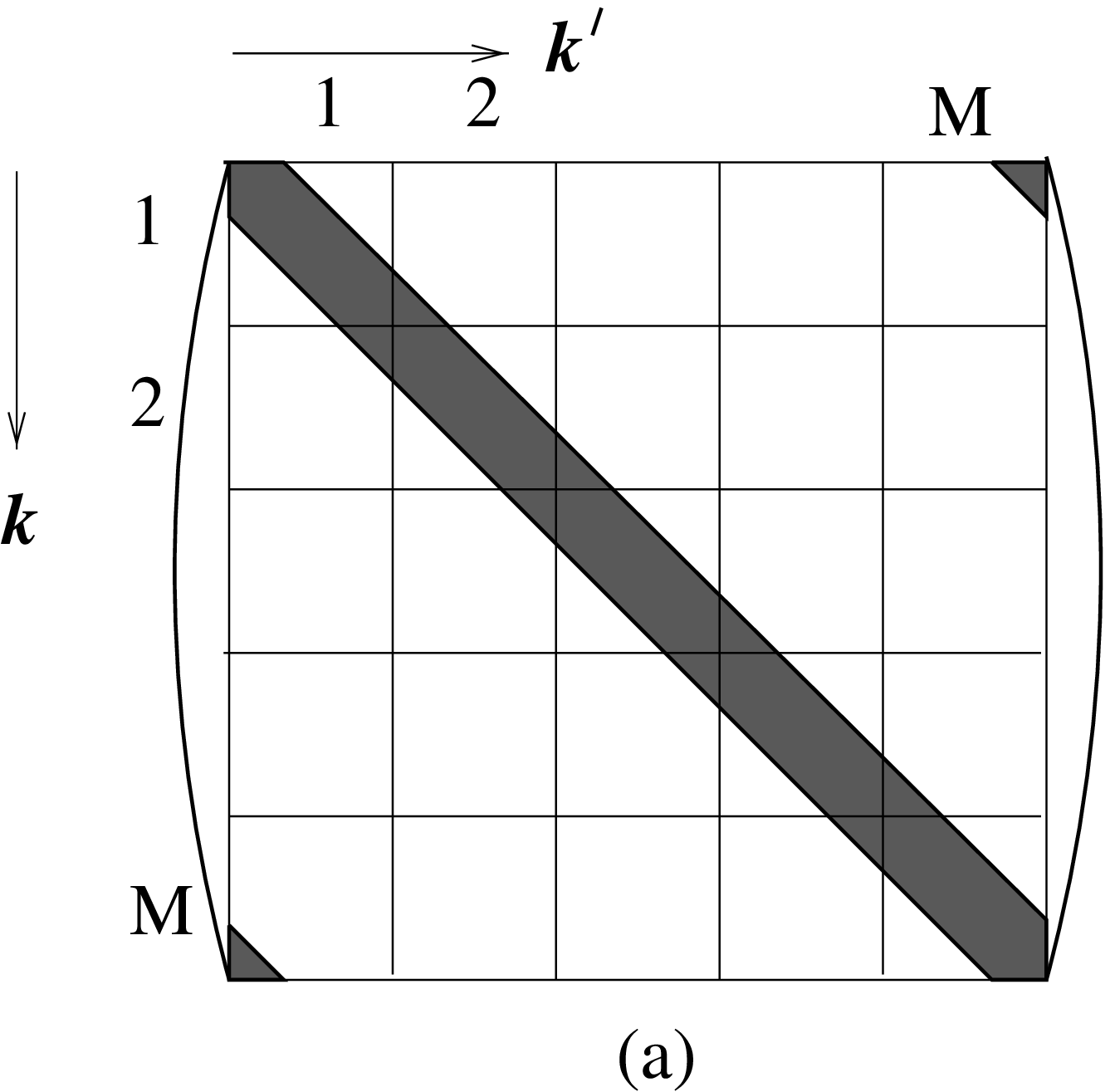,width=5.2cm}
\end{minipage}
\hspace{1cm}
\begin{minipage}{7.2cm}
\psfig{figure=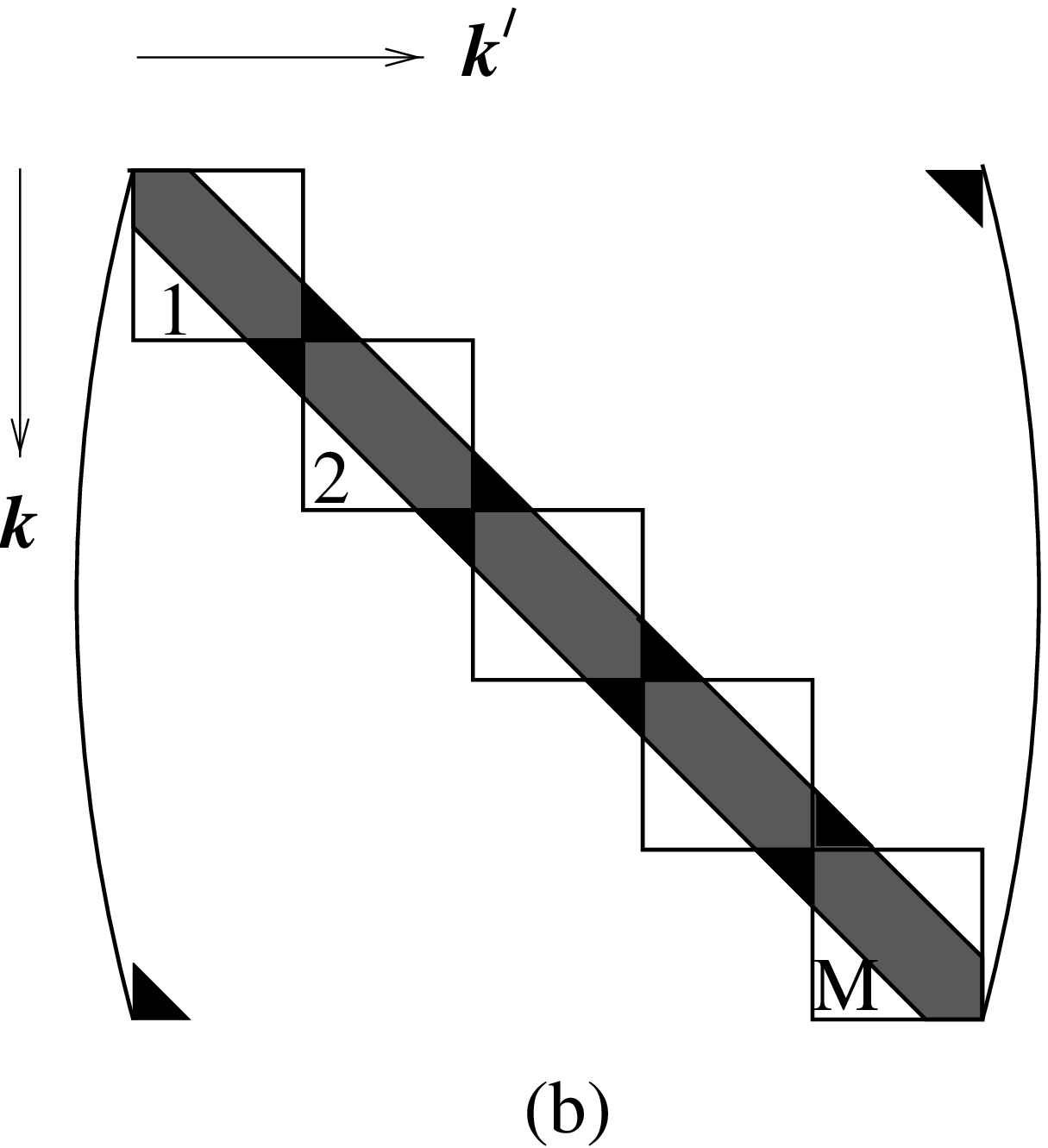,width=4.7cm}
\end{minipage}
\caption[Diagonal blocks and around-the-corner processes of the
inverse matrix Green's function.]
{
(a) Schematic representation of the matrix $\hat{G}^{-1}$ defined in 
Eq.\ref{eq:GhatDysonshift} for $d=2$.
Only the wave-vector index is shown, i.e. each matrix-element is an infinite
matrix in frequency space. 
Regions with non-zero matrix elements are shaded.
The triangles in the upper right and lower left corner
represent scattering processes between sectors $1$ and $M$.
Because these sectors are adjacent, they can be connected by small
momentum-transfers. 
(b) Diagonal blocks and 
around-the-corner processes (represented by black triangles).}
\label{fig:band}
\end{figure}  
}
The width of the diagonal band with non-zero matrix elements is
determined by the range $q_{\rm c}$ of the interaction in momentum 
space, because 
the vanishing of the interaction
$ f_{q}^{\alpha \alpha^{\prime}} $ for
$|{\vec{q}} | \geqapprox q_{\rm c}$ implies 
that the field $V^{\alpha}_{q} $ mediating
this interaction must also vanish.
But $q_{\rm c} \ll k_{\rm F}$
by assumption $(A1)$ in Chap.~\secref{sec:closedloop}, 
so that we have the freedom of choosing
the sector cutoffs $\Lambda$ and $\lambda$ such that
$ q_{\rm c}  \ll  \Lambda , \lambda   \ll k_{\rm F}$
(see Eqs.\ref{eq:Lambdaup}, \ref{eq:qccond}, and \ref{eq:lambdaup}).
As shown in Fig.~\secref{fig:band}(b), in this way
$\hat{G}^{-1}$ is subdivided
into block matrices associated with
the sectors such that $\hat{G}^{-1}$ is
{\it{approximately block diagonal}}.
The block diagonalization is not exact, because
non-zero matrix elements are also located in the black triangles of Fig.~\secref{fig:band}(b).
These matrix elements represent scattering processes that
transfer momentum between different sectors 
(the around-the-corner  
processes\index{around-the-corner processes} mentioned in Chap.~\secref{subsec:proper}).
The crucial approximation 
is now to neglect these  processes.
This is precisely the diagonal-patch
approximation\index{diagonal-patch approximation} $(A1)$ 
discussed in Chap.~\secref{sec:closedloop},
which is also necessary to insure that the
probability distribution ${\cal{P}} \{ \phi^{\alpha} \}$ can be approximated by a Gaussian.
The justification for this step is that the relative number of
matrix elements representing momentum-transfer between different sectors
is small as long as $q_{\rm c} \ll \Lambda , \lambda$.
In $d > 1 $ dimensions the 
relative number of around-the-corner matrix elements 
associated with a given sector $K^{\alpha}_{\Lambda, \lambda}$ is
of the order of $q_{\rm c}^d / (  \Lambda^{d-1} \lambda )$.
Note that this approximation makes
only sense {\it{if the sector cutoffs are kept 
finite and large compared with the range of the
interaction in momentum space.}}

Although the relative number of matrix elements
describing around-the-corner processes 
is small, we have to make one important caveat:
Possible non-perturbative effects 
that depend on the {\it{global topology}} of the Fermi surface
cannot be described within this approximation.
For example, in $d=2$ each sector has two neighbors, but the
first and the last sector are
adjacent, so that there exist also around-the-corner
processes connecting the sectors $1$ and $M$, which give rise
to the off-diagonal
triangles in the lower left and upper right corners
of the matrix shown in Fig.~\secref{fig:band}.
More generally, in higher dimensions 
the number of non-zero blocks  corresponding to around-the-corner 
processes in
each row or column of ${\hat{G}}^{-1}$  
is equal to the coordination number $z_{d-1}$ of the patches ${P}^{\alpha}_{\Lambda}$
on the Fermi surface. For example, for hyper-cubic patches on a Fermi surface in
$d$ dimensions the coordination number is $z_{d-1} = 2 (d-1)$.
Hence, the total number of blocks corresponding to
around-the-corner processes is $N_{\rm c} = M z_{d-1}$, where
$M$ is the total number of patches that cover the Fermi surface\footnote{
In Fig.~\secref{fig:band} we have $M=5$, $z_{1} = 2$ and $N_{\rm c} = 10$}.
Note that $N_{\rm c} = 0$ for $d=1$, 
because around-the-corner processes
are absent due to the widely  separated Fermi points.
On the other hand, in any dimension the number of diagonal blocks is 
equal to the number $M$ of the patches,
so that in $d > 2$ only a small number
of the around-the-corner triangles can be found in the vicinity of the
diagonal band. The case $d=2$ shown in Fig.~\secref{fig:band} is special, 
because there exist only two off-diagonal around-the-corner blocks, independently
of the number $M$ of the patches.
In higher dimensions, however, the off-diagonal
around-the corner blocks are distributed in 
a complicated manner over the matrix $\hat{G}^{-1}$.
The precise position of the blocks depends on
the way in which the patches are labelled on the Fermi surface.
The effect of these sparsely distributed around-the-corner blocks is
difficult to estimate, and we are assuming that
they do not lead to qualitatively new effects. 
This is an important {\it{assumption}}
which is implicitly made in all of the following calculations.
In systems  where the {\it{topological structure}} of the Fermi surface
is crucial, this assumption may not be justified.
We would like to emphasize that this assumption is implicitly also
made in the operator bosonization approach \cite{Houghton93,Castro94}, 
as well as in the Ward identity approach by Castellani, Di Castro and
Metzner \cite{Castellani94,Castellani94b,Metzner95hab}.

Once we have disposed of the around-the-corner matrix elements,
the matrix $\hat{G}^{-1}$ is a direct sum of diagonal
blocks $( \hat{G}^{\alpha})^{-1}$, $\alpha = 1 , \ldots , M$. Hence,
 \begin{equation}
 [ \hat{G}^{-1} ]_{k k^{\prime}} = \sum_{\alpha} 
 \Theta^{\alpha} ( {\vec{k}} ) \Theta^{\alpha} ( {\vec{k}}^{\prime} )
 [ ( \hat{G}^{ \alpha })^{-1} ]_{k  k^{\prime} }
 \; \; \; ,
 \label{eq:Galphaexp}
 \end{equation}
where the matrix $ ( \hat{G}^{ \alpha })^{-1}$ is the diagonal block 
of $\hat{G}^{-1}$ associated with sector $K^{\alpha}_{\Lambda , \lambda}$,
with matrix elements given by
 \begin{equation}
 [ ( \hat{G}^{ \alpha } )^{ -1} ]_{k k^{\prime} }= 
 \delta_{k k^{\prime}}
 [ \I \tilde{\omega}_{n} - \xi^{\alpha}_{ {\vec{k}} - {\vec{k}}^{\alpha} }  
 - \epsilon_{\vec{k}^{\alpha}} + \mu ]
 - V^{\alpha}_{k - k^{\prime} }  
 \label{eq:Galphadef}
 \; \; \; .
 \end{equation}
Thus, {\it{the problem of inverting $\hat{G}^{-1}$ is reduced to the
problem of inverting each diagonal block separately}}.
The diagonal elements of $\hat{G}$ 
are then simply 
 \begin{equation}
 [ \hat{G} ]_{k k} = \sum_{\alpha} 
 \Theta^{\alpha} ( {\vec{k}} ) 
 [ \hat{G}^{ \alpha } ]_{k k }
 \; \; \; .
 \label{eq:Galphaexpinv}
 \end{equation}
Note that $\hat{G}^{\alpha}$ is still an infinite matrix in
frequency space, so that the quantum dynamics is 
fully taken into account.

\begin{center}
{\bf{Inversion of the diagonal blocks}}
\end{center}

\noindent
Up to this point we have {\it{not}} linearized the energy 
dispersion\index{linearization of energy dispersion}, so that
the above block diagonalization is valid for arbitrary
dispersion $\xi^{\alpha}_{\vec{q}}$.
The crucial advantage of the subdivision of
$\hat{G}^{-1}$ into blocks is that, to a first approximation,
within a given block we may linearize the energy 
dispersion, $\xi^{\alpha}_{\vec{q}} \approx 
{\vec{v}}^{\alpha} \cdot {\vec{q}}$ 
(see Eqs.\ref{eq:xialphaqexp} and \ref{eq:v0cijdef}).
It is also convenient to choose the centers ${\vec{k}}^{\alpha}$ of the
sectors such that $\epsilon_{\vec{k}^{\alpha}} = \mu$, so that the
last two terms in the square brace of Eq.\ref{eq:Galphadef} cancel.
In Chap.~\secref{subsec:locallin} we have argued\footnote{As will be discussed
in Sect.~\secref{sec:eik} and in more detail
in Chap.~\secref{chap:arad}, in $d > 1$ the linearization 
of the energy dispersion is not always a good approximation, because
in $d > 1$
the condition ${\vec{v}}^{\alpha} \cdot {\vec{q}} = 0$
defines hyper-planes in momentum space on which the leading term
in the expansion of $\xi^{\alpha}_{\vec{q}}$ is quadratic
in $\vec{q}$. Linearization is only allowed 
if the contribution from these hyper-planes to the
physical quantity of interest is negligible.
Whether this is really the case depends also on the
nature of the interaction. For example,
in physically relevant models of
transverse gauge fields that couple to the electronic
current density (to be discussed
in Chap.~\secref{chap:arad}) the linearization is
{\it{not}} allowed.}
that the linearization is justified
if the sectors are sufficiently small, so that within a given sector
the variation of the local normal vector to the Fermi surface
is small.  On the other hand, 
as discussed in detail in
Chap.~\secref{subsec:proper},
the cutoffs must be kept large
compared with $q_{\rm c}$ in order to guarantee that 
the patching construction leads to an approximate
block diagonalization of $\hat{G}^{-1}$.

Once the linearization has been made, it is possible to invert
the diagonal blocks $( \hat{G}^{\alpha} )^{-1}$ exactly.
Note that $\hat{G}^{\alpha}$ is still an infinite matrix in
frequency space. 
Shifting the wave-vector labels
according to ${\vec{k}} = {\vec{k}}^{\alpha} + {\vec{q}}$ and
${\vec{k}}^{\prime} = {\vec{k}}^{\alpha} + {\vec{q}}^{\prime}$, the 
diagonal block $\hat{G}^{\alpha}$ is determined by
the equation
 \begin{equation}
 \sum_{\tilde{q}^{\prime}} \left[ \delta_{\tilde{q}  \tilde{q}^{\prime}  } 
 [ G^{\alpha}_{0} ( \tilde{q} ) ]^{-1}
 - V^{\alpha}_{ \tilde{q} - \tilde{q}^{\prime} } \right] 
 [ \hat{G}^{\alpha} ]_{ \tilde{q}^{\prime} \tilde{q}^{\prime \prime} }
 = \delta_{ \tilde{q}  \tilde{q}^{\prime \prime} }
 \label{eq:Galphadif}
 \; \; \; ,
 \end{equation}
where $[G_{0}^{ \alpha} ( \tilde{q} )]^{-1} = \I \tilde{\omega}_{n}
- {\vec{v}}^{\alpha} \cdot {\vec{q}}$, see Eq.\ref{eq:G0alphashiftdef}.
The important point is now that Eq.\ref{eq:Galphadif}
is first order and linear, and can therefore be solved exactly by means of a trivial generalization
of a method due to Schwinger \cite{Schwinger62}.  
Defining
 \begin{eqnarray}
 {\cal{G}}^{\alpha} ( {\vec{r}} , {\vec{r}}^{\prime} , \tau , \tau^{\prime} )
 & = & \frac{1}{ \beta V} \sum_{ \tilde{q} \tilde{q}^{\prime}}
 \E^{ \I ( {\vec{q}} \cdot {\vec{r}} - \tilde{\omega}_{n} \tau )} 
 \E^{ - \I ( {\vec{q}}^{\prime} \cdot {\vec{r}}^{\prime} - \tilde{\omega}_{n^{\prime}} \tau^{\prime} )} 
 [ \hat{G}^{\alpha} ]_{\tilde{q} \tilde{q}^{\prime} }
 \; \; \; ,
 \label{eq:calGdef}
 \\
 V^{\alpha} ( {\vec{r}} , \tau ) & = & \sum_{q}
 \E^{ \I ( {\vec{q}} \cdot {\vec{r}} - {\omega}_{m} \tau )}  V^{\alpha}_{q}
 \label{eq:Vrt}
 \; \; \; ,
 \end{eqnarray}
it is easy to see that Eq.\ref{eq:Galphadif} is equivalent with 
 \begin{equation}
  \left[- {\partial}_{ \tau} + \I   
  {\vec{v}}^{\alpha} \cdot \nabla_{\vec{r}} - V^{\alpha}  ( {\vec{r}} , \tau )
  \right]
 {\cal{G}}^{\alpha} ( {\vec{r}} , {\vec{r}}^{\prime} , \tau , \tau^{\prime} )
 = \delta ( {\vec{r}} - {\vec{r}}^{\prime} ) \delta^{\ast} ( \tau - \tau^{\prime} )
 \;  ,
 \label{eq:Galphadifrt}
 \end{equation}
where
 \begin{equation}
 \delta^{\ast} ( \tau - \tau^{\prime} )
 = \frac{1}{\beta} \sum_{n} \E^{ - \I \tilde{\omega}_{n} ( \tau - \tau^{\prime} ) }
 \label{eq:deltaastdirac}
 \end{equation}
is the antiperiodic $\delta$-function.
Note that the Fourier transformation in Eq.\ref{eq:calGdef} involves
fermionic Matsubara frequencies,  because
 ${\cal{G}}^{\alpha} ( {\vec{r}} , {\vec{r}}^{\prime} , \tau , \tau^{\prime} )$
has to satisfy the Kubo-Martin-Schwinger (KMS)\index{KMS boundary conditions}
boundary condition \cite{Kubo57,Martin59}
 \begin{equation}
 {\cal{G}}^{\alpha} ( {\vec{r}} , {\vec{r}}^{\prime} , \tau + \beta , \tau^{\prime} )
 =
 {\cal{G}}^{\alpha} ( {\vec{r}} , {\vec{r}}^{\prime} , \tau , \tau^{\prime} + \beta )
 =
 - {\cal{G}}^{\alpha} ( {\vec{r}} , {\vec{r}}^{\prime} , \tau , \tau^{\prime} )
 \label{eq:KMS}
 \; \; \; .
 \end{equation}
In contrast, $V^{\alpha} ( {\vec{r}} , \tau )$ 
is by definition a periodic function of $\tau$,
so that  the sum in Eq.\ref{eq:Vrt} involves bosonic Matsubara 
frequencies\footnote{ 
The $q=0$ component of the interaction requires a special treatment, and
should be excluded from the $q$-sum in Eq.\ref{eq:Vrt}. 
Formally this condition can be taken into account by
setting $\phi^{\alpha}_{q=0} = 0$, so that the $q=0$ term
in the sum \ref{eq:Vrt} (as well as in all subsequent
$q$-sums in this chapter) is automatically excluded.
Note that this is equivalent with 
$\int \D {\vec{r}} \int_{0}^{\beta} \D \tau V^{\alpha} ( {\vec{r}} , \tau ) = 0$.
Any finite value of this integral
can be absorbed into a redefinition of the chemical potential, which
has disappeared in Eq.\ref{eq:Galphadifrt}, because by assumption we have linearized
the energy dispersion at the true chemical 
potential. \index{chemical potential!renormalization}
See also the footnote after Eq.\ref{eq:Trloggauss}
in Chap.~\secref{chap:a4bos}.
}.
Following Schwinger \cite{Schwinger62}\index{Schwinger ansatz}, let us make the ansatz
 \begin{equation}
 {\cal{G}}^{\alpha} ( {\vec{r}} , {\vec{r}}^{\prime} , \tau , \tau^{\prime} )
  = 
 {{G}}^{\alpha}_{0} ( {\vec{r}} - {\vec{r}}^{\prime} , \tau - \tau^{\prime} )
 \E^{ \Phi^{\alpha} ( {\vec{r}} , \tau ) - \Phi^{\alpha} ( {\vec{r}}^{\prime} , \tau^{\prime} ) }
 \label{eq:Ansatz}
 \; \; \; ,
 \end{equation}
where 
 ${{G}}^{\alpha}_{0} ( {\vec{r}} - {\vec{r}}^{\prime} , \tau - \tau^{\prime} )$ satisfies
 \begin{equation}
  \left[- {\partial}_{ \tau} +   \I
  {\vec{v}}^{\alpha} \cdot \nabla_{\vec{r}} 
  \right]
 {{G}}^{\alpha}_{0} ( {\vec{r}} - {\vec{r}}^{\prime} , \tau - \tau^{\prime} )
 = \delta ( {\vec{r}} - {\vec{r}}^{\prime} ) \delta^{\ast} ( \tau - \tau^{\prime} )
 \; \; \; .
 \label{eq:Galpha0difrt}
 \end{equation}
To take the KMS boundary condition \ref{eq:KMS} into account, we require that
 ${{G}}^{\alpha}_{0} ( {\vec{r}} - {\vec{r}}^{\prime} , \tau - \tau^{\prime} )$ 
should be antiperiodic in $\tau$ and $\tau^{\prime}$,
while $\Phi^{\alpha} ( {\vec{r}} , \tau )$ should be a 
{\it{periodic}} function of $\tau$,
 \begin{equation}
 \Phi^{\alpha} ( {\vec{r}} , \tau + \beta ) = \Phi^{\alpha} ( {\vec{r}} , \tau )
 \label{eq:Phiboundary}
 \; \; \; .
 \end{equation}
Substituting Eq.\ref{eq:Ansatz} into Eq.\ref{eq:Galphadifrt}, 
it is easy to show that
 \begin{eqnarray}
  \lefteqn{  \left[- {\partial}_{ \tau} +    \I
  {\vec{v}}^{\alpha} \cdot \nabla_{\vec{r}} - V^{\alpha}  ( {\vec{r}} , \tau )
  \right]
 {\cal{G}}^{\alpha} ( {\vec{r}} , {\vec{r}}^{\prime} , \tau , \tau^{\prime} )
 = 
  \delta ( {\vec{r}} - {\vec{r}}^{\prime} ) \delta^{\ast} ( \tau - \tau^{\prime} )
 } 
 \nonumber
 \\
 & &
 + 
 {\cal{G}}^{\alpha} ( {\vec{r}} , {\vec{r}}^{\prime} , \tau , \tau^{\prime} )
 \left\{ \left[ - {\partial}_{ \tau} +   \I
  {\vec{v}}^{\alpha} \cdot \nabla_{\vec{r}} 
  \right] \Phi^{\alpha} ( {\vec{r}} , \tau ) - V^{\alpha} ( {\vec{r}} , \tau ) \right\}
  \;  .
  \label{eq:Ansatzins}
  \end{eqnarray}
Comparing Eq.\ref{eq:Ansatzins} with Eq.\ref{eq:Galphadifrt}, we see that our ansatz is
consistent provided $\Phi^{\alpha} ( {\vec{r}} , \tau )$ satisfies
 \begin{equation}
  \left[ - {\partial}_{ \tau} +   \I
  {\vec{v}}^{\alpha} \cdot \nabla_{\vec{r}} 
  \right] \Phi^{\alpha} ( {\vec{r}} , \tau ) = V^{\alpha} ( {\vec{r}} , \tau ) 
  \label{eq:phidif}
  \; \; \; .
  \end{equation}
Eqs.\ref{eq:Galpha0difrt} and \ref{eq:phidif} 
are first order
linear partial differential equations. 
The solution with the correct boundary condition
is easily obtained via Fourier transformation,
 \begin{eqnarray}
 {{G}}^{\alpha}_{0} ( {\vec{r}}  , \tau  )
 & = &
 \frac{1}{\beta V} \sum_{ \tilde{q}} 
 \frac{ \E^{ \I ( {\vec{q}} \cdot {\vec{r}} - \tilde{\omega}_{n} \tau )}}{ \I \tilde{\omega}_{n}
 -  {\vec{v}}^{\alpha} \cdot {\vec{q}} } 
 \; \; \; ,
 \label{eq:G0res}
 \\
 \Phi^{\alpha} ( {\vec{r}} , \tau ) & = &
  \sum_{q} 
 \frac{ \E^{ \I ( {\vec{q}} \cdot {\vec{r}} - \omega_{m} \tau )}}{ \I \omega_{m}
 -  {\vec{v}}^{\alpha} \cdot {\vec{q}} } V^{\alpha}_{q}
 \; \; \; .
 \label{eq:Phires}
 \end{eqnarray}
Let us emphasize again that the Matsubara sum in
Eq.\ref{eq:Phires} involves bosonic frequencies 
because we have to satisfy the boundary condition \ref{eq:Phiboundary}.
Having determined ${{G}}^{\alpha}_{0} ( {\vec{r}} , \tau )$ and
$\Phi^{\alpha} ( {\vec{r}} , \tau )$, the  
diagonal blocks $( \hat{G}^{\alpha})^{-1}$ have been inverted, so that
$\hat{G}^{\alpha}$ is known as functional of the
$\phi^{\alpha}$-field. The
diagonal elements are explicitly given by
 \begin{eqnarray}
 [ \hat{G}^{\alpha} ]_{kk}
 & = & \frac{1}{\beta V} \int \D {\vec{r}} \int \D {\vec{r}}^{\prime} 
 \int_{0}^{\beta} \D \tau 
 \int_{0}^{\beta} \D \tau^{\prime} 
 \E^{ - \I [ ( {\vec{k}} - {\vec{k}}^{\alpha} ) \cdot ( {\vec{r}} - {\vec{r}}^{\prime} )
- \tilde{\omega}_{n} ( \tau - \tau^{\prime} ) ] }
 \nonumber
 \\
 & & \hspace{-14mm} \times
 G^{\alpha}_{0} ( {\vec{r}} - {\vec{r}}^{\prime} , \tau - \tau^{\prime} )
 \exp \left[ { \frac{\I}{\beta} \sum_{q}
 \frac{ \E^{ \I ( {\vec{q}} \cdot {\vec{r}} - \omega_{m} \tau)} -
  \E^{  \I ( {\vec{q}} \cdot {\vec{r}}^{\prime} - \omega_{m} \tau^{\prime} )} }
  { \I \omega_{m} - {\vec{v}}^{\alpha} \cdot {\vec{q}} } \phi^{\alpha}_{q} } \right]
 \; .
 \label{eq:Gkkinv}
 \end{eqnarray}

\subsection{Gaussian averaging: calculation of the Debye-Waller }
\label{subsec:greenav}

{\it{This is the easy part of the calculation, 
because we have to average an exponential of $\phi^{\alpha}$ with
respect to a Gaussian probability distribution. This yields, of course, a
Debye-Waller factor\index{Debye-Waller factor!density-density interactions}!
}}

\vspace{7mm}

\noindent
Combining Eqs.\ref{eq:avphi3}, \ref{eq:Galphaexpinv} and \ref{eq:Gkkinv}, 
and using the fact that averaging restores translational invariance in space
and time, we conclude that
the interacting Matsubara Green's function is given by
 \begin{eqnarray}
 G (k)  & = & \sum_{\alpha} \Theta^{\alpha} ( {\vec{k}} )
 \int \D {\vec{r}} \int_{0}^{\beta} \D \tau 
 \E^{ - \I [ ( {\vec{k}} - {\vec{k}}^{\alpha} ) \cdot  {\vec{r}}
 - \tilde{\omega}_{n}  \tau  ] }
 \nonumber
 \\
 & & \times
 {{G}}^{\alpha}_{0} ( {\vec{r}}  , \tau  )
 \left< \E^{ \Phi^{\alpha} ( {\vec{r}} , \tau ) - \Phi^{\alpha} ( 0 , 0 ) } 
 \right>_{S_{{\rm eff},2}}
 \; \; \; .
 \label{eq:Gkres1}
 \end{eqnarray}
Using Eqs.\ref{eq:hatVphi} and \ref{eq:Phires}, we may write
 \begin{equation}
 \Phi^{\alpha} ( {\vec{r}} , \tau ) - \Phi^{\alpha} ( 0 , 0 )  = 
  \sum_{q} 
  {\cal{J}}^{\alpha}_{-q} ( {\vec{r}} , \tau )
  \phi^{\alpha}_{q} 
  \label{eq:Phidifer}
  \; \; \; ,
  \end{equation}
with
 \begin{equation}
  {\cal{J}}^{\alpha}_{q} ( {\vec{r}} , \tau )
  =  \frac{\I}{\beta} 
  \left[ \frac{ 1 - \E^{ - \I ( {\vec{q}} \cdot {\vec{r}} - \omega_{m} \tau )} }
  { \I \omega_{m} - {\vec{v}}^{\alpha} \cdot {\vec{q}} } \right]
  \label{eq:Jdef}
  \; \; \; .
  \end{equation}
The problem of calculating the interacting Green's function is now reduced to a
trivial Gaussian integration, which simply yields the usual
{\it{Debye-Waller factor}},
 \begin{eqnarray}
 \left< \E^{ \Phi^{\alpha} ( {\vec{r}} , \tau ) - \Phi^{\alpha} ( 0 , 0 ) } 
 \right>_{S_{{\rm eff},2}}
 & =  & \left< \E^{
  \sum_{q} 
  {\cal{J}}^{\alpha}_{-q} ( {\vec{r}} , \tau )
  \phi^{\alpha}_{q} } \right>_{ S_{{\rm eff},2}}
  \nonumber
  \\
  &  &  \hspace{-40mm} = \exp \left[  
 \frac{1}{2} \sum_{q} 
 \left< \phi^{\alpha}_{q} \phi^{\alpha}_{-q} \right>_{S_{{\rm eff},2}}
 {\cal{J}}^{\alpha}_{-q}  ( {\vec{r}} , \tau )
 {\cal{J}}^{\alpha}_{q}  ( {\vec{r}} , \tau )
  \right] 
  =
  \E^{ Q^{\alpha} ( {\vec{r}} , \tau ) }
 \label{eq:DebyeWaller}
 \; \; \; ,
 \end{eqnarray}
with
 \begin{equation}
 Q^{\alpha} ( {\vec{r}} , \tau ) 
 =
 \frac{\beta}{2V} \sum_{q} 
 [ \underline{f}^{\rm RPA}_{q} ]^{\alpha \alpha}
 {\cal{J}}^{\alpha}_{-q}  ( {\vec{r}} , \tau )
 {\cal{J}}^{\alpha}_{q}  ( {\vec{r}} , \tau )
 \label{eq:Q1gaussnew}
 \; \; \; .
 \end{equation}
We have used the fact that the Gaussian propagator of the
$\phi^{\alpha}$-field is according to Eq.\ref{eq:phiphiprop}
proportional to the RPA interaction.
For consistency, 
in Eq.\ref{eq:Q1gaussnew} 
the  polarization contribution to 
$[ \underline{f}^{\rm RPA}_{q} ]^{\alpha \alpha}$ 
should be approximated by
its leading long-wavelength limit given in Eq.\ref{eq:Pilong}, because
in deriving Eq.\ref{eq:DebyeWaller} we have neglected
momentum transfer between different sectors (i.e.
the around-the-corner processes represented by the
black triangles in Fig.~\secref{fig:band}(b)). Using
 \begin{equation}
 {\cal{J}}^{\alpha}_{-q} ( {\vec{r}} , \tau )
 {\cal{J}}^{\alpha}_{q} (  {\vec{r}} , \tau )
 =   \frac{2}{\beta^2}  \frac{ 1 - \cos ( {\vec{q}} \cdot {\vec{r}} - \omega_{m} \tau )}{
 ( \I \omega_{m} - {\vec{v}}^{\alpha} \cdot {\vec{q}} )^2 }
 \; \; \; ,
 \label{eq:JJ}
 \end{equation}
we conclude that
 \begin{equation}
 Q^{\alpha}
 ( {\vec{r}} , \tau )  = 
 R^{\alpha} 
 - S^{\alpha} ( {\vec{r}} , \tau )
 \;  \; \; ,
 \label{eq:Qlondef}
 \end{equation}
with
 \begin{eqnarray}
 R^{\alpha} 
  &  =  & 
 \frac{1}{\beta {{V}}} \sum_{ q }  
 \frac{   f^{{\rm RPA}, \alpha}_{q}
 }
 {
 ( \I \omega_{m} - {\vec{v}}^{\alpha} \cdot {\vec{q}} )^{2 }}
   =  S^{\alpha} ( 0 , 0 ) 
  \label{eq:Rlondef}
 \; \; \; ,
 \\
 S^{\alpha} 
 ( {\vec{r}}  , \tau   ) 
  &  =  &
 \frac{1}{\beta {{V}}} \sum_{ q }  
 \frac{   f^{{\rm RPA}, \alpha}_{q}
  \cos ( {\vec{q}} \cdot  {\vec{r}} 
  - {\omega}_{m}  \tau  ) 
 }
 {
 ( \I \omega_{m} - {\vec{v}}^{\alpha} \cdot {\vec{q}} )^{2 }}
  \label{eq:Slondef}
 \; \; \; .
 \end{eqnarray}
Here $f^{{\rm RPA} , \alpha}_{q}$
is the diagonal element\index{random-phase approximation!effective interaction}
of the RPA interaction matrix defined in Eq.\ref{eq:frpapatchdef},
 \begin{equation}
 f^{ {\rm RPA} , \alpha}_{q} \equiv
 [ \underline{f}^{\rm RPA}_{q} ]^{\alpha \alpha}
 =
  \left[ \underline{{f}}_{ {{q}} } 
 \left[ 1 + 
  \underline{{\Pi}}_{0} (q) 
 \underline{{f}}_{q}  \right]^{-1} \right]^{\alpha \alpha}
 \label{eq:frpadiagonaldef}
 \; \; \; .
 \end{equation}
An important special case is a patch-independent bare
interaction, i.e.
$ [ \underline{f}_{q} ]^{\alpha \alpha^{\prime}} = f_{q}$
for all $\alpha$ and $\alpha^{\prime}$. 
From Eq.\ref{eq:frpatot} we know that 
in this case $f_{q}^{{\rm RPA} , \alpha}$
can be identified with the usual RPA interaction, 
 \begin{equation}
 f_{q}^{{\rm RPA} , \alpha}  = f_{q}^{\rm RPA} \equiv \frac{f_{q}}{ 1 +  f_{q} \Pi_{0} ( q )  }
 \; \; \; , \; \; \; 
 \mbox{if 
 $ \; \; \; [ \underline{f}_{q} ]^{\alpha \alpha^{\prime}} = f_{q}$}
 \; \; \; .
 \label{eq:frpaalphaspecial}
 \end{equation}
In summary, the averaged diagonal blocks are given 
by\index{Green's function!bosonization result for density-density interactions}
 \begin{equation}
 \left< [ \hat{G}^{\alpha} ]_{k k} \right>_{S_{{\rm eff},2}}
 =
 \int \D {\vec{r}} \int_{0}^{\beta} \D \tau 
 \E^{ - \I [ ( {\vec{k}} - {\vec{k}}^{\alpha} ) \cdot  {\vec{r}}
 - \tilde{\omega}_{n}  \tau  ] }
 {{G}}^{\alpha} ( {\vec{r}}  , \tau  )
 \label{eq:avdiagonal}
 \; \; \; ,
 \end{equation}
with
 \begin{equation}
 {{G}}^{\alpha} ( {\vec{r}}  , \tau  )
  =  
 {{G}}^{\alpha}_{0} ( {\vec{r}}  , \tau  )
 \E^{ Q^{\alpha} ( {\vec{r}} , \tau ) }
 \label{eq:Galphartdef}
 \; \; \; .
 \end{equation}
From Eqs.\ref{eq:avphi3} and \ref{eq:Galphaexpinv} we finally obtain 
for the Matsubara Green's function of the
interacting many-body system 
 \begin{equation}
 G (k)   =  
 \sum_{\alpha} \Theta^{\alpha} ( {\vec{k}} )
 G^{\alpha} ( {\vec{k}} - {\vec{k}}^{{\alpha}} , \I \tilde{\omega}_{n} )
 \; \; \; ,
 \label{eq:Gkres2}
 \end{equation}
where
 \begin{equation}
 G^{\alpha} ( \tilde{q} )   
 \equiv  G^{\alpha} ( {\vec{q}} , \I \tilde{\omega}_{n} ) =
 \int \D {\vec{r}} \int_{0}^{\beta} \D \tau 
 \E^{ - \I (  {\vec{q}}  \cdot  {\vec{r}}
 - \tilde{\omega}_{n}  \tau  ) }
 G^{\alpha} ( {\vec{r}} , \tau )
 \label{eq:Galphaqtildedef}
 \; \; \; .
 \end{equation}
Shifting in Eq.\ref{eq:Gkres2}
${\vec{k}} = {\vec{k}}^{\alpha^{\prime}} + {\vec{q}}$ and choosing $| {\vec{q}} |$
small compared with the cutoffs $\Lambda$ and $\lambda$ that 
determine the size of the sector $K^{\alpha}_{\Lambda , \lambda}$, it is easy to see that
only the term $\alpha^{\prime} = \alpha$ in the sum \ref{eq:Gkres2}
contributes, so that
(after renaming again $\alpha^{\prime} \rightarrow \alpha$)
 \begin{equation}
 G ( {\vec{k}}^{\alpha} + {\vec{q}} , \I \tilde{\omega}_{n} )
 = G^{\alpha} ( {\vec{q}} , \I \tilde{\omega}_{n} )
 \; \; \; , \; \; \; | {\vec{q}} | \ll \Lambda , \lambda
 \label{eq:Gkalphashiftres}
 \; \; \; .
 \end{equation}

\subsection{The Green's function in real space}
\label{subsec:Greal}

{\it{
The real space Green's function\index{Green's function!real space} $G ( {\vec{r}} , \tau )$ 
should not be confused with the
sector Green's function $G^{\alpha} ( {\vec{r}} , \tau )$
in Eq.\ref{eq:Galphartdef}. 
Here we derive the precise relation between these functions.}}

\vspace{7mm}

\noindent
Given the exact Matsubara Green's function $G ( k )$, we can 
use Eq.\ref{eq:FTGreal}
to reconstruct the exact real space imaginary time Green's function
$G ( {\vec{r}} , \tau )$ by inverse Fourier transformation.
Substituting Eqs.\ref{eq:Gkres2} and \ref{eq:Galphaqtildedef} into
Eq.\ref{eq:FTGreal}, we obtain
 \begin{equation}
 G ( {\vec{r}} , \tau ) = \sum_{\alpha}   
 \int \D {\vec{r}}^{\prime} 
 \E^{  \I  {\vec{k}}^{\alpha}  \cdot {\vec{r}}^{\prime} }
 \frac{1}{V} \sum_{\vec{k}} 
 \Theta^{\alpha} ( {\vec{k}} )
 \E^{ \I {\vec{k}} \cdot ( {\vec{r}} - {\vec{r}}^{\prime} ) }
 G^{\alpha} ( {\vec{r}}^{\prime} , \tau )
 \label{eq:Grealpatchreal}
 \; \; \; .
 \end{equation}
At distances $ | \vec{r} |$ that are
large compared with the inverse sector cutoffs 
$\Lambda^{-1}$ and $\lambda^{-1}$ we may approximate
\begin{equation}
 \frac{1}{V} \sum_{\vec{k}} 
 \Theta^{\alpha} ( {\vec{k}} )
 \E^{ \I {\vec{k}} \cdot ( {\vec{r}} - {\vec{r}}^{\prime} ) }
 \approx 
 \frac{1}{V} \sum_{\vec{k}} 
 \E^{ \I {\vec{k}} \cdot ( {\vec{r}} - {\vec{r}}^{\prime} ) }
= 
 \delta  ( {\vec{r}} - {\vec{r}}^{\prime} ) 
 \; \; \; ,
 \end{equation}
so that Eq.\ref{eq:Grealpatchreal} reduces to
 \begin{equation}
 G ( {\vec{r}} , \tau ) =  \sum_{\alpha}   
 \E^{  \I  {\vec{k}}^{\alpha}  \cdot {\vec{r}} }
 G^{\alpha} ( {\vec{r}} , \tau )
 \; \; \; ,
 \label{eq:Grealtotal}
 \end{equation}
which is the real space imaginary time version of
Eq.\ref{eq:Gkres2}.

To see the role of the cutoffs more clearly\index{cutoff!choice of sector cutoffs},
it is instructive to calculate
the non-interacting sector Green's function $G^{\alpha}_{0} ( {\vec{r}} , \tau )$
defined in Eq.\ref{eq:G0res}.
Performing the fermionic Matsubara sum
we obtain 
 \begin{equation}
 G^{\alpha}_{0} ( {\vec{r}}  , \tau )
  =  
 \frac{1}{ V } \sum_{\vec{q}} 
 \E^{ \I {\vec{q}} \cdot  {\vec{r}} }
 G^{\alpha}_{0} ( {\vec{q}}  , \tau )
 \; \; \; ,
 \label{eq:Grtqt}
 \end{equation}
with
 \begin{equation}
 \hspace{-3mm}
 G^{\alpha}_{0} ( {\vec{q}}  , \tau )
 = \E^{- {\vec{v}}^{\alpha} \cdot {\vec{q}}  \tau  }
 \left[ f ( {\vec{v}}^{\alpha} \cdot {\vec{q}} ) \Theta ( - \tau + 0^{+} )
 - f ( - {\vec{v}}^{\alpha} \cdot {\vec{q}} ) \Theta ( \tau - 0^{+} ) \right]
 \label{eq:Gpatchrealq0}
 \; .
 \end{equation}
Because 
$G^{\alpha}_{0} ( {\vec{q}}  , \tau )$ depends on ${\vec{q}}$ only via
the component $ {\vec{v}}^{\alpha} \cdot {\vec{q}}$,
it is convenient to choose the orientation of 
the local coordinate system attached to sector $K^{\alpha}_{\Lambda , \lambda}$ 
such that one of its axes matches the direction 
$ \hat{\vec{v}}^{\alpha} = {{\vec{v}}^{\alpha} }/{ | {\vec{v}}^{\alpha} |}$
of the local Fermi velocity.  
In this coordinate system we have the decomposition
 \begin{eqnarray}
 {\vec{q}} & = &  q^{ \alpha}_{\|} \hat{\vec{v}}^{\alpha} 
 + 
 \vec{q}^{\alpha}_{\bot} 
 \; \;  , \; \;  q^{\alpha}_{\|} = {\vec{q}} \cdot \hat{\vec{v}}^{\alpha} 
 \; \;  ,  \; \; \vec{q}^{\alpha}_{\bot}  \cdot \hat{\vec{v}}^{\alpha} = 0
 \; \; \; ,
 \label{eq:qlocaldecomp}
 \\
 {\vec{r}} & = &  r^{ \alpha}_{\|} \hat{\vec{v}}^{\alpha} 
 + 
 \vec{r}^{\alpha}_{\bot} 
 \; \;  , \; \;  r^{\alpha}_{\|} = {\vec{r}} \cdot \hat{\vec{v}}^{\alpha} 
 \; \;  ,  \; \; \vec{r}^{\alpha}_{\bot}  \cdot \hat{\vec{v}}^{\alpha}  = 0
 \; \; \; .
 \label{eq:rlocaldecomp}
 \end{eqnarray}
For $ \beta \rightarrow \infty$ and $V \rightarrow \infty$
we obtain then after a  simple calculation 
 \begin{equation}
 G^{\alpha}_{0} ( {\vec{r}}  , \tau  )
 = 
 \delta^{(d-1)} ( {\vec{r}}^{\alpha}_{\bot}  )
 \left( \frac{ - \I}{2 \pi} \right)
 \frac{1}
 { 
 r^{\alpha}_{\|}  
 + \I | {\vec{v}}^{\alpha} |  \tau }
 \; \; \; ,
 \label{eq:Gpatchreal1}
 \end{equation}
with
 \begin{equation}
 \delta^{(d-1)} ( {\vec{r}}^{\alpha}_{\bot}  )
 = \int \frac{ \D {\vec{r}}^{\alpha}_{\bot} }{ (2 \pi)^{d-1} }
 \E^{ \I {\vec{q}}^{\alpha}_{\bot} \cdot \vec{r}^{\alpha}_{\bot} }
 \; \; \; ,
 \label{eq:deltabotdef}
 \end{equation}
where the integral is over the $d-1$ components of ${\vec{r}}$ that are
perpendicular to $\vec{v}^{\alpha}$.
In deriving Eq.\ref{eq:Gpatchreal1} we have not been very careful 
about cutoffs. In order not to over-count the
degrees of freedom,
the ${\vec{q}}$-summations should be restricted  
to the sectors $K^{\alpha}_{\Lambda , \lambda}$. Hence, there is a hidden cutoff 
function $\Theta^{\alpha} ( {\vec{k}}^{\alpha} + {\vec{q}} )$ in all
${\vec{q}}$-sums, which we have not explicitly written out.
However, we may ignore
this cutoff function
as long as we are interested in length scales
 \begin{equation}
 | {\vec{r}}^{\alpha}_{\bot} | \gg \Lambda^{-1}
 \; \; \; , \; \; \;  
  | r^{\alpha}_{\|} | \gg \lambda^{-1}
 \label{eq:qcutoff}
 \; \; \; ,
 \end{equation}
because the oscillating exponential in Eq.\ref{eq:Grtqt}
cuts off the ${\vec{q}}$-summations before the boundaries of the sectors are reached.
It should be kept in mind, however, that
Eq.\ref{eq:Gpatchreal1} is only correct 
if the conditions \ref{eq:qcutoff} are satisfied.
More precisely, in a finite system of volume $V = L^d$ the $\delta$-function
in Eq.\ref{eq:Gpatchreal1} should be replaced by
the cutoff-dependent function
 \begin{equation}
 \delta^{(d-1)}_{\Lambda} ( {\vec{r}}^{\alpha}_{\bot}  )
 = \frac{1}{L^{d-1}}
 \sum_{  {\vec{q}}_{\bot}^{\alpha} }
 \Theta^{\alpha} ( {\vec{k}}^{\alpha} + {\vec{q}}^{\alpha}_{\bot} )
 \E^{ \I {\vec{q}}^{\alpha}_{\bot} \cdot \vec{r}^{\alpha}_{\bot} }
 \label{eq:deltaLambda}
 \; \; \; .
 \end{equation}
We conclude that for length scales $| \vec{r}^{\alpha}_{\bot} | \gg \Lambda^{-1}$ 
we may replace ${\vec{r}} \rightarrow r^{\alpha}_{\|} \hat{\vec{v}}^{\alpha}$ in the
argument of the Debye-Waller factor, so that Eq.\ref{eq:Galphartdef}
becomes
 \begin{equation}
 {{G}}^{\alpha} ( {\vec{r}}  , \tau  )
 =
 {{G}}^{\alpha}_{0} ( {\vec{r}}  , \tau  )
 \E^{ Q^{\alpha} ( r^{\alpha}_{\|} \hat{\vec{v}}^{\alpha} , \tau ) }
 \label{eq:Galphartreplace}
 \; \; \; .
 \end{equation}
From Eq.\ref{eq:Grealtotal} we obtain then for the
real space Green's function of the interacting system,
 \begin{equation}
 G ( {\vec{r}} , \tau ) =  
 \frac{ - \I}{2 \pi}
 \sum_{\alpha}   
 \delta^{(d-1)} ( {\vec{r}}^{\alpha}_{\bot}  )
 \frac{ 
 \exp \left[  \I  {\vec{k}}^{\alpha}  \cdot {\vec{r}} + 
 Q^{\alpha} ( r^{\alpha}_{\|} \hat{\vec{v}}^{\alpha} , \tau ) \right] }
 { r^{\alpha}_{\|}  
 + \I | {\vec{v}}^{\alpha} |  \tau }
 \; \; \; .
 \label{eq:Grealtotalfull}
 \end{equation}
Note that this expression 
has units of ${V}^{-1}$, as expected from a real space
single-particle Green's function in $d$ dimensions.

\subsection{The underlying asymptotic Ward identity}
\label{sec:ward}

{\it{
Our bosonization formula \ref{eq:Galphartdef} for the Green's function 
is the result of an infinite resummation of the perturbation series, but
it is not clear which type of diagrams have been summed.
In this section we shall clarify this point.
We first derive from Eq.\ref{eq:Galphartdef}
an integral equation, which is
exactly equivalent with the
integral equation derived by
Castellani, Di Castro and Metzner \cite{Castellani94}.
We then combine the integral equation 
with the Dyson equation to obtain a Ward identity. In this way 
we see the relation between bosonization 
and diagrammatic perturbation theory.
}}

\begin{center}
{\bf{The integral equation}}
\end{center}

\noindent
Let us apply the differential operator
$ - \partial_{\tau} + \I {\vec{v}}^{\alpha} \cdot \nabla_{\vec{r}}$
to the bosonization result \ref{eq:Galphartdef}
for the sector Green's function $G^{\alpha} ( {\vec{r}} , \tau )$. 
Using the fact that 
according to Eqs.\ref{eq:Galpha0difrt} and \ref{eq:deltaastdirac} 
the application of this operator to $G^{\alpha}_{0} ( {\vec{r}} , \tau )$
generates as $\delta$-function, it is easy to show that
 \begin{eqnarray}
 \left[ 
 - \partial_{\tau} + \I {\vec{v}}^{\alpha} \cdot \nabla_{\vec{r}}
 + X^{\alpha} ( {\vec{r}} - {\vec{r}}^{\prime} , \tau - \tau^{\prime} )
 \right] G^{\alpha} ( 
  {\vec{r}} - {\vec{r}}^{\prime} , \tau - \tau^{\prime} )
  & = &
  \nonumber
  \\
  & & \hspace{-30mm}
  \delta 
  ( {\vec{r}} - {\vec{r}}^{\prime}) \delta^{\ast} (  \tau - \tau^{\prime} )
  \label{eq:Gdif1}
  \; \; \; ,
  \end{eqnarray}
with
 \begin{equation}
  X^{\alpha} 
  ( {\vec{r}} - {\vec{r}}^{\prime} , \tau - \tau^{\prime} )
  = - 
  [ - \partial_{\tau} + \I {\vec{v}}^{\alpha} \cdot \nabla_{\vec{r}}]
  Q^{\alpha} 
  ( {\vec{r}} - {\vec{r}}^{\prime} , \tau - \tau^{\prime} )
  \label{eq:Xalphadef}
  \; \; \; .
  \end{equation}
From the explicit expression for $Q^{\alpha} ( {\vec{r}} , \tau )$
given in Eqs.\ref{eq:Qlondef}--\ref{eq:Slondef} we find
that the function $X^{\alpha}  ( {\vec{r}} , \tau )$ has the Fourier expansion
 \begin{equation}
 X^{\alpha} ( {\vec{r}} , \tau )  =  \frac{1}{\beta V} \sum_{q} 
 \E^{ \I ( {\vec{q}} \cdot {\vec{r}} - \omega_{m} \tau ) }
 X^{\alpha}_{q}
 \; \; \; ,
 \label{eq:XalphaFourier}
 \end{equation}
with Fourier coefficients given by
 \begin{equation}
 X^{\alpha}_{q}
  =  \frac{ f_{q}^{{\rm RPA}, \alpha}}{ \I \omega_{m} - {\vec{v}}^{\alpha} \cdot {\vec{q}} }
 \label{eq:Xalphaqdef}
 \; \; \; .
 \end{equation}
In Fourier space Eq.\ref{eq:Gdif1} becomes\index{integral equation}
 \begin{equation}
 [ \I \tilde{\omega}_{n} - {\vec{v}}^{\alpha} \cdot {\vec{q}} ]
 G^{\alpha} ( \tilde{q} ) + 
 \frac{1}{\beta V} \sum_{ \tilde{q}^{\prime}} X^{\alpha}_{ \tilde{q} - \tilde{q}^{\prime} }
 G^{\alpha} ( \tilde{q}^{\prime} ) = 1
 \label{eq:Gdif2FT}
 \; \; \; ,
 \end{equation}
or equivalently
 \begin{eqnarray}
 [ \I \tilde{\omega}_{n} - {\vec{v}}^{\alpha} \cdot {\vec{q}} ]
 G^{\alpha} ( {\vec{q}} , \I \tilde{\omega}_{n} )
 & = & 1 
 \nonumber
 \\
 & & \hspace{-15mm} 
 - \frac{1}{\beta V } \sum_{{\vec{q}}^{\prime} , n^{\prime}}
 \frac{ f^{{\rm RPA} , \alpha}_{ {\vec{q}} - {\vec{q}}^{\prime} , 
 \I {\omega}_{n - n^{\prime}} } }
 { 
 \I {\omega}_{n - n^{\prime} }
 - {\vec{v}}^{\alpha} \cdot ( {\vec{q}} - {\vec{q}}^{\prime} ) }
 G^{\alpha} ( {\vec{q}}^{\prime} , \I \tilde{\omega}_{n^{\prime}} )
 \label{eq:Gdif3FT}
 \; \; \; .
 \end{eqnarray}
Because the difference between two fermionic Matsubara frequencies is a 
bosonic one, the kernel 
$X^{\alpha}_{ \tilde{q} - \tilde{q}^{\prime}}$ in  Eq.\ref{eq:Gdif2FT} 
depends on bosonic frequencies.
In the limit $\beta \rightarrow \infty$
Eq.\ref{eq:Gdif3FT} is equivalent with the integral equation given in
Eq.(13) of the work \cite{Castellani94} 
by Castellani, Di Castro and Metzner.
Our bosonization approach maps the solution
of Eq.\ref{eq:Gdif3FT} onto the
standard problem of solving a linear partial differential
equation (Eq.\ref{eq:Ansatzins}) and
calculating a Debye-Waller factor in a Gaussian integral.
The solution for the Green's function is given in 
Eqs.\ref{eq:Galphartdef}--\ref{eq:Galphaqtildedef}, with the Debye-Waller factor given in 
Eqs.\ref{eq:Qlondef}--\ref{eq:Slondef}.
On the other hand, Castellani, Di Castro and Metzner
do not directly solve Eq.\ref{eq:Gdif3FT}
but {\it{first}} perform an angular
averaging operation on this integral equation and {\it{then}} solve
the resulting averaged equation.
Although in general the operations of averaging and solving
integral equations do not commute (i.e. the solution of the
averaged integral equation is not necessarily identical with
the average of the solution of the integral equation), in the 
particular case of interest the final result seems to be
equivalent, at least up to re-definitions of cutoffs.

\begin{center}
{\bf{The Ward identity}}
\end{center}

\begin{figure}
\sidecaption
\psfig{figure=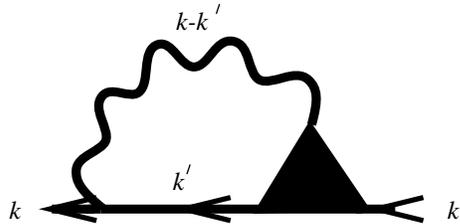,width=6cm}
\caption[Skeleton diagram for the irreducible self-energy.]
{
Skeleton diagram for the irreducible self-energy.
The thick wavy line denotes the exact screened effective
interaction $f^{\ast}_{q}$, the shaded triangle is the exact
three-legged vertex $\Lambda ( k; k - k^{\prime} )$, 
and the solid line with arrow is the exact Green's function.}
\label{fig:skelself}
\end{figure}

\noindent
In modern many-body theory it is sometimes
convenient \cite{Nozieres64,Fetter71}
to define so-called skeleton diagrams in order to exhibit the structure
of the perturbation series more clearly. The skeleton diagram\index{self-energy!skeleton diagram} for
the exact self-energy is shown in Fig.~\secref{fig:skelself}.
In the Matsubara formalism this diagram represents the following
expression,
 \begin{equation}
 \Sigma ( k ) =  - \frac{1}{\beta V} \sum_{k^{\prime}}
 f^{\ast}_{ k - k^{\prime}} \Lambda ({k ; k - k^{\prime}} ) G  ( k^{\prime} )
 \; \; \; .
 \label{eq:selfexact}
 \end{equation}
The exact effective interaction  $f^{\ast}_{q} $ is related to the bare interaction
$f_{\vec{q}}$ via the dielectric function,
$f^{\ast}_{q} = {f_{\vec{q}} }/{ \epsilon ( q )}$, 
which in turn can be expressed in terms of the
exact proper polarization via $\epsilon ( q ) = 1 + f_{\vec{q}} \Pi_{\ast} ( q )$, see
Eqs.\ref{eq:Piproper}--\ref{eq:dielectricdef}.
By definition, the vertex function\index{vertex function} $\Lambda ( k ; q )$ is the
sum of all diagrams with three external ends corresponding to
two solid lines and a single interaction line.
To lowest non-trivial order we have
$\Lambda  ( k ; q )  \approx 1 + \Lambda^{(1)} ( k ; q )$, with 
$\Lambda^{(1)} ( k ; q )$ given in Eq.\ref{eq:gammaF1}.
Because $G ( k^{\prime} )$ on the right-hand side of Eq.\ref{eq:selfexact} depends
again on $\Sigma ( k^{\prime} )$ via the Dyson equation, Eq.\ref{eq:selfexact} is
a complicated integral equation, which can only be solved approximately.
Moreover, the formal kernel 
$f^{\ast}_{k - k^{\prime}} \Lambda ({k}; k - k^{\prime} )$
of this integral equation is again a functional of $G ( k )$, so that
it cannot be calculated exactly unless the entire perturbation series
has been summed.
In practice one therefore replaces the effective interaction
$f^{\ast}_{k - k^{\prime}}$ and the
vertex $\Lambda ({k; k - k^{\prime}} )$
by some ``reasonable'' approximation.

For better comparison with the self-energy calculated within our
bosonization approach, let us shift  
again ${\vec{k}} = {\vec{k}}^{\alpha} + {\vec{q}}$ and
${\vec{k}}^{\prime} = {\vec{k}}^{\alpha} + {\vec{q}}^{\prime}$, so that
wave-vectors are measured with respect to the local coordinate
system with origin in sector $K^{\alpha}_{\Lambda , \lambda}$.
Defining 
 \begin{eqnarray}
G ( {\vec{k}}^{\alpha} + {\vec{q}} , \I \tilde{\omega}_{n} )
  & =  &
 G^{\alpha} ( \tilde{q} )
 \; \; \; ,
 \label{eq:Glocalalpha}
 \\
 \Sigma ( {\vec{k}}^{\alpha} + {\vec{q}} , \I \tilde{\omega}_{n} )
  & =  &
 \Sigma^{\alpha} ( \tilde{q} )
 \; \; \; ,
 \label{eq:selflocalalpha}
 \\
 \Lambda ( {\vec{k}}^{\alpha} + {\vec{q}} , \I \tilde{\omega}_{n} ;
 {\vec{q}} - {\vec{q}}^{\prime} , \I \omega_{n - n^{\prime}} )
 & = & \Lambda^{\alpha} ( \tilde{q} ; \tilde{q } - \tilde{q}^{\prime} )
 \; \; \; ,
 \label{eq:vertlocalalpha}
 \end{eqnarray}
the skeleton equation \ref{eq:selfexact} reads
 \begin{equation}
 \Sigma^{\alpha} ( \tilde{q} ) =  - \frac{1}{\beta V} \sum_{\tilde{q}^{\prime}}
 f^{\ast}_{ \tilde{q} - \tilde{q}^{\prime}} 
 \Lambda^{\alpha} ({\tilde{q} ; \tilde{q} - \tilde{q}^{\prime}} ) G^{\alpha}  ( \tilde{q}^{\prime} )
 \; \; \; ,
 \label{eq:selfexactlocal}
 \end{equation}
while the Dyson equation\index{Dyson equation} can be written as
 \begin{equation}
 [ G^{\alpha} ( \tilde{q} ) ]^{-1}
 =  [ G^{\alpha}_{0} ( \tilde{q} ) ]^{-1}
 - \Sigma^{\alpha} ( \tilde{q} )
 \label{eq:Dysonalpha}
 \; \; \; .
 \end{equation}

Let us now determine the skeleton parameters that correspond to
our bosonization result for the Green's function.
Starting point is the integral equation
\ref{eq:Gdif2FT}. 
Keeping in mind that after linearization we may  write
$ \I \tilde{\omega}_{n} - {\vec{v}}^{\alpha} \cdot {\vec{q}}  =
[ G^{\alpha}_{0} ( \tilde{q} ) ]^{-1}$ and  
dividing both sides of Eq.\ref{eq:Gdif2FT} by
$G^{\alpha} ( \tilde{q} )$, we obtain
 \begin{equation}
 [ G^{\alpha} ( \tilde{q} ) ]^{-1} =
 [ G^{\alpha}_{0} ( \tilde{q} ) ]^{-1} +
  \frac{1}{\beta V} \sum_{ \tilde{q}^{\prime}} 
 \frac{ X^{\alpha}_{ \tilde{q} - \tilde{q}^{\prime} }}{G^{\alpha} ( \tilde{q} ) }
 G^{\alpha} ( \tilde{q}^{\prime} ) 
 \label{eq:Gdif4FT}
 \; \; \; .
 \end{equation}
Comparing this with Eq.\ref{eq:Dysonalpha}, we conclude that
in our bosonization approach
the self-energy satisfies
 \begin{equation}
  \Sigma^{\alpha} ( \tilde{q} ) = -
  \frac{1}{\beta V} \sum_{ \tilde{q}^{\prime}} 
 \frac{ X^{\alpha}_{ \tilde{q} - \tilde{q}^{\prime} }}{G^{\alpha} ( \tilde{q} ) }
 G^{\alpha} ( \tilde{q}^{\prime} ) 
 \label{eq:selfbos1}
 \; \; \; .
 \end{equation}
From Eqs.\ref{eq:selfexactlocal} and \ref{eq:selfbos1}
we finally obtain
 \begin{equation}
 f^{\ast}_{ \tilde{q} - \tilde{q}^{\prime}} 
 \Lambda^{\alpha} ({\tilde{q} ; \tilde{q} - \tilde{q}^{\prime}} ) =
  \frac{ X^{\alpha}_{ \tilde{q} - \tilde{q}^{\prime} }}{G^{\alpha} ( \tilde{q} ) }
 =
  \frac{ f^{{\rm RPA} , \alpha}_{ \tilde{q} - \tilde{q}^{\prime}  } }
 { [ \I \omega_{n - n^{\prime}} - {\vec{v}}^{\alpha} \cdot ( {\vec{q}} - {\vec{q}}^{\prime} ) ]
 G^{\alpha} ( \tilde{q} ) }
 \; \; \; .
 \label{eq:lambdaidentify}
 \end{equation}
Note that in the skeleton equation \ref{eq:selfexact} it is assumed that the
bare interaction depends only on the momentum-transfer. 
From Eq.\ref{eq:frpatot} we know that in this case
$f^{{\rm RPA} , \alpha }_{q} = f^{\rm RPA}_{q}$, the usual RPA interaction.
Then we see from Eq.\ref{eq:lambdaidentify} that 
the approximations inherent in our bosonization 
approach amount to replacing the
exact effective interaction $f^{\ast}_{q}$ by the RPA interaction 
$f^{\rm RPA}_{q}$, and setting the vertex function 
equal to
 \begin{equation}
 \Lambda^{\alpha} ({\tilde{q} ; \tilde{q} - \tilde{q}^{\prime}} )  = 
 \frac{ 1}
 { [ \I \omega_{n - n^{\prime}} - {\vec{v}}^{\alpha} \cdot ( {\vec{q}} - {\vec{q}}^{\prime} ) ]
 G^{\alpha} ( \tilde{q} ) }
 \label{eq:Ward1}
 \; \; \; ,
 \end{equation}
which is equivalent with
 \begin{equation}
 \left[ \frac{1}{ G_{0}^{\alpha} ( \tilde{q} ) }
- \frac{1}{ G_{0}^{\alpha} ( \tilde{q}^{\prime} ) } \right]
 \Lambda^{\alpha} ({\tilde{q} ; \tilde{q} - \tilde{q}^{\prime}} )  = 
  \frac{1}{G^{\alpha} ( \tilde{q})}
 \label{eq:Ward2}
 \; \; \; ,
 \end{equation}
or, after shifting 
$ \tilde{q} - \tilde{q}^{\prime} \rightarrow
 q^{\prime} \equiv [  {\vec{q}}^{\prime} ,  \I \omega_{m^{\prime}} ]$,
 \begin{equation}
 [ \I \omega_{m^{\prime}} - {\vec{v}}^{\alpha} \cdot {\vec{q}}^{\prime}]
 \Lambda^{\alpha} ( \tilde{q} ;  q^{\prime} ) = 
 [ G^{\alpha} ( \tilde{q}) ]^{-1}
 \; \; \; .
 \label{eq:Ward3}
 \end{equation}
In terms of the symmetrized vertex function
 \begin{equation}
 \tilde{\Lambda}^{\alpha} ( \tilde{q} ; \tilde{q}^{\prime} )
 =
 \Lambda^{\alpha} (\tilde{q} ; \tilde{q} - \tilde{q}^{\prime} )  + 
 \Lambda^{\alpha} ( \tilde{q}^{\prime} ; \tilde{q}^{\prime} - \tilde{q} )   
 \label{eq:Lambdaverttsym}
 \; \; \; ,
 \end{equation}
Eq.\ref{eq:Ward2} can also be rewritten in the more symmetric form
 \begin{equation}
 \left[ \frac{1}{ G_{0}^{\alpha} ( \tilde{q} ) }
- \frac{1}{ G_{0}^{\alpha} ( \tilde{q}^{\prime} ) } \right]
 \tilde{\Lambda}^{\alpha} ({\tilde{q} ; \tilde{q}^{\prime}} )  = 
  \left[
  \frac{1}{G^{\alpha} ( \tilde{q})}
  - \frac{1}{G^{\alpha} ( \tilde{q}^{\prime} )}
  \right]
 \label{eq:Ward4}
 \; \; \; .
 \end{equation}
The important point is that the right-hand side of Eq.\ref{eq:Ward4} 
depends again on the exact Green's function. Such a relation between
a vertex function and a Green's function is called a 
{\it{Ward identity}}\index{Ward identity}.
In the limit $\beta \rightarrow \infty$ 
Eq.\ref{eq:Ward4} 
is equivalent with the Ward identity derived in \cite{Castellani94}.
Of course, in $d>1$ or for non-linear energy dispersion 
this Ward identity is not exact.
In Sect.~\secref{sec:eik}
we shall develop a powerful method 
for calculating in a controlled and quantitative way the corrections
neglected in Eq.\ref{eq:Ward4}.

In summary, although within our 
bosonization approach the dielectric function is approximated by the
RPA expression,  bosonization does not simply
reproduce the usual RPA self-energy, because it sums in addition infinitely many
other diagrams by means of a non-trivial Ward identity for the vertex function.
The analytic expressions for these  diagrams
can be generated order by order in the RPA interaction by 
iterating the integral equation \ref{eq:Gdif3FT}.

\vspace{7mm}

Finally, let us compare the skeleton equation 
\ref{eq:selfexact} with the 
dynamically screened exchange diagram, the so-called
GW approximation\index{GW approximation} for the self-energy \cite{Hedin65}.
In this approximation the effective interaction
$f^{\ast}_{q}$ is approximated by the RPA interaction,
just like in our bosonization approach.
The crucial difference with bosonization is
that vertex corrections are completely ignored
within the GW approximation, so that one sets
$\Lambda ( k , k - k^{\prime} ) \rightarrow 1$.
Then Eq.\ref{eq:selfexactlocal} reduces to
the simpler integral equation
 \begin{equation}
 \Sigma^{\alpha} ( \tilde{q} )
=  - \frac{1}{\beta V} \sum_{ \tilde{q}^{\prime}} 
f^{\rm RPA }_{ \tilde{q} - \tilde{q}^{\prime} }
G^{\alpha} ( \tilde{q}^{\prime} )
\; \; \; , \; \; \; \mbox{GW approximation}
\label{eq:GW}
\; \; \; .
\end{equation}
If we replace 
the interacting Green's function 
on the right-hand side of Eq.\ref{eq:GW} 
by the non-interacting one,
we recover the lowest order self-energy 
correction given in Eq.\ref{eq:sigmaF1}.
The self-consistent solution of Eq.\ref{eq:GW} contains
infinite orders in perturbation theory.
It seems, however, that the only reason for ignoring
vertex corrections is that
one is unable to calculate them in a controlled way.
As recently pointed out by Farid \cite{Farid95}, 
the errors due to the omission of vertex corrections seem to
be partially cancelled by the replacement
$G \rightarrow G_{0}$ on the right-hand side of 
Eq.\ref{eq:GW}. In other words, non-self-consistent GW 
can be better than self-consistent GW!
Evidently such an approximation cannot be systematic.
On the other hand, for interactions that are dominated by forward scattering
the bosonization approach uses the small parameter
$q_{\rm c} / k_{\rm F}$ to sum the dominant terms
of the entire perturbation series.

\subsection{The Fermi liquid renormalization 
factors $Z^{\alpha}$ and $Z^{\alpha}_{\rm m}$}
\label{sec:Identification}

{\it{ 
We show how in a Fermi liquid
the quasi-particle residue
$Z^{\alpha}$ and the effective mass renormalization
$Z^{\alpha}_{\rm m}$ can be obtained from
the Debye-Waller factor 
$Q^{\alpha} ( {\vec{r}} , \tau )$.}}

\begin{center}
{\bf{The quasi-particle residue $Z^{\alpha}$}}
\end{center}

\noindent
As shown in Chap.~\secref{subsec:LandauFL} 
(see Eq.\ref{eq:nkdiscondef}), the
quasi-particle residue\index{quasi-particle residue} $Z^{\alpha}$ of 
a Fermi liquid can be identified from the
discontinuity $\delta n^{\alpha}_{\vec{q}}$ 
of the momentum distribution\index{momentum distribution!discontinuity}
at the Fermi surface.
Hence, in order to relate our Debye-Waller factor
$Q^{\alpha} ( {\vec{r}} , \tau )$ given in Eq.\ref{eq:Qlondef} to the
quasi-particle residue, we simply  
have to calculate $\delta n^{\alpha}_{ {\vec{q}} }$ 
from Eq.\ref{eq:Galphartdef}.
Substituting Eqs.\ref{eq:Galphartdef},
\ref{eq:Galphaqtildedef}
and \ref{eq:Gkalphashiftres} into Eq.\ref{eq:nkdef} we obtain
 \begin{equation}
 n_{ {\vec{k}}^{\alpha}+ \vec{q} } = 
 \int \D {\vec{r}} 
 \E^{ - \I  {\vec{q}}  \cdot  {\vec{r}} }
 {{G}}^{\alpha }_{0} ( {\vec{r}}  , 0  )
 \E^{ Q^{\alpha} ( {\vec{r}} , 0 ) }
 \; \; \; ,
 \label{eq:occup2}
 \end{equation}
so that the change
$\delta n^{\alpha}_{\vec{q}}$ of the momentum distribution defined in Eq.\ref{eq:nkdiscondef}
is given by
\begin{equation}
 \delta n^{\alpha}_{ {\vec{q}} } 
 = 2  \I
 \int \D {\vec{r}} 
 \sin ( {\vec{q}}  \cdot  {\vec{r}} )
 {{G}}^{\alpha }_{0} ( {\vec{r}}  , 0  )
 \E^{ Q^{\alpha} ( {\vec{r}} , 0 ) }
 \; \; \; .
 \label{eq:occup3}
 \end{equation}
From Eq.\ref{eq:Gpatchreal1} we obtain for the
non-interacting sector Green's function
at equal times
 \begin{equation}
 G^{\alpha}_{0} ( {\vec{r}}  , 0  )
 = 
 \delta^{(d-1)} ( {\vec{r}}^{\alpha}_{\bot}  )
 \frac{- \I}
 { 2 \pi 
 r^{\alpha}_{\|}  
  }
 \; \; \; ,
 \label{eq:G0patchreal2equal}
 \end{equation}
so that 
\begin{equation}
 \delta n^{\alpha}_{ {\vec{q}} }  = 
  \frac{2}{\pi}     
  \int_{ 0 }^{\infty}
 \D {x} \frac{ \sin ( q^{\alpha}_{\|} x  ) }{ x }
 \E^{ Q^{\alpha} ( x \hat{\vec{v}}^{\alpha} , 0 ) }
 \; \; \; ,
 \label{eq:occup4}
 \end{equation}
where we have renamed $r^{\alpha}_{\|} = x$.
As discussed in Chap.~\secref{subsec:proper},
bosonization should lead to cutoff-independent results
if the interaction is dominated by wave-vectors
$ | {\vec{q}} | \leqapprox q_{\rm c} \ll \Lambda , \lambda$.
Hence, in real space
the bosonization result for the Green's function is accurate
at length scales $x \gg q_{\rm c}^{-1}$.
We therefore separate from 
Eq.\ref{eq:occup4} the non-universal short-distance regime,
 \begin{equation}
 \hspace{-5mm}
 \delta n^{\alpha}_{ {\vec{q}} }  = 
  \frac{2}{\pi }     
  \left[
  \int_{ 0 }^{ q_{\rm c }^{-1} }
 \D {x} 
 \frac{ \sin (  q^{\alpha}_{\|} x  ) }{ x }
 \E^{ Q^{\alpha} ( x \hat{\vec{v}}^{\alpha} , 0 ) }
  +
  \int_{    q_{\rm c}^{-1} }^{\infty}
 \D x \frac{ \sin  ( q^{\alpha}_{\|} x )  }{ x }
 \E^{ Q^{\alpha} (     x \hat{\vec{v}}^{\alpha}  , 0 ) }
 \right]
 \; .
 \label{eq:occup4b}
 \end{equation}
For $|q^{\alpha}_{\|}| \ll q_{\rm c}$
it is allowed to expand the sine-function in the first term. 
Evidently, this yields
an analytic contribution 
to $\delta n^{\alpha}_{\vec{q}}$, which 
to leading order is proportional to
$  q^{\alpha}_{\|}  / q_{\rm c}$.
In a Fermi liquid this term is negligible compared with
the contribution from the second term in Eq.\ref{eq:occup4b},
so that we obtain to leading order 
 \begin{equation}
 \delta n^{\alpha}_{ {\vec{q}} }  \sim
 {\rm sgn} ( q^{\alpha}_{\|} )
  \frac{2}{  \pi}  
  \int_{     | q^{\alpha}_{\|} | / q_{\rm c} }^{\infty}
 \D x^{\prime} \frac{ \sin  ( x^{\prime} )  }{ x^{\prime} }
 \E^{ Q^{\alpha} (      x^{\prime}  \hat{\vec{v}}^{\alpha} / | q^{\alpha}_{\|} |  , 0 ) }
 \; \; \; ,
 \label{eq:occup4c}
 \end{equation}
where we have rescaled $x =  x^{\prime} / | q^{\alpha}_{\|} |$.
For $q^{\alpha}_{\|} \rightarrow 0$
we may set $ |q^{\alpha}_{\|}| / q_{\rm c} = 0$ in the lower limit 
and replace 
 $Q^{\alpha} (     x^{\prime} \hat{\vec{v}}^{\alpha} / | q^{\alpha}_{\|} | , 0 ) $ by
its asymptotic expansion for large $x^{\prime}  / | q^{\alpha}_{\|}|$. 
Assuming that the limit
 \begin{equation}
 \lim_{r^{\alpha}_{\|} \rightarrow \infty} 
 Q^{\alpha} ( r^{\alpha}_{\|} \hat{\vec{v}}^{\alpha} , 0 )  \equiv Q^{\alpha}_{\infty}
 \label{eq:Qalphainfty}
 \end{equation}
exists, we obtain  for $q^{\alpha}_{\|} \rightarrow 0$
 \begin{equation}
 \delta n^{\alpha}_{ {\vec{q}} }  \sim
 {\rm sgn} ( q^{\alpha}_{\|} )
  \frac{2}{  \pi}   
  \int_{     0 }^{\infty}
 \D x^{\prime} \frac{ \sin  ( x^{\prime} )  }{ x^{\prime} }
  \E^{Q^{\alpha}_{\infty} }
  =
 {\rm sgn} ( q^{\alpha}_{\|} )
  \E^{Q^{\alpha}_{\infty} }
  \; \; \; .
  \label{eq:eQQ}
  \end{equation}
But from Eqs.\ref{eq:Qlondef}--\ref{eq:Slondef} we 
have
 $Q^{\alpha} ( r^{\alpha}_{\|} \hat{\vec{v}}^{\alpha} , 0 ) = R^{\alpha} 
 -
 S^{\alpha} ( r^{\alpha}_{\|} \hat{\vec{v}}^{\alpha} , 0 ) $, where the
constant term is simply given by
$ R^{\alpha} = S^{\alpha} ( 0 , 0 ) $. A sufficient condition
for the existence of the limit $Q^{\alpha}_{\infty}$ 
is the existence of $R^{\alpha}$.
Recall that according to Eq.\ref{eq:Rlondef}
$R^{\alpha}$ is for
$\beta, V  \rightarrow \infty$ given by
 \begin{equation}
 R^{\alpha} 
   = 
   \int \frac{ \D {\vec{q}}}{ ( 2 \pi)^d}
   \int_{- \infty}^{\infty} \frac{\D \omega}{2 \pi}
 \frac{   f^{{\rm RPA}, \alpha}_{ {\vec{q}} , \I \omega }}
 {
 ( \I \omega - {\vec{v}}^{\alpha} \cdot {\vec{q}} )^{2 }}
  \label{eq:Rlonthermo}
 \; \; \; .
 \end{equation}
If this integral exists, 
then the Fourier integral theorem \cite{Henrici91} implies that 
the integral
 \begin{equation}
 S^{\alpha} ( r^{\alpha}_{\|} \hat{\vec{v}}^{\alpha} , 0 )
   = 
   \int \frac{ \D {\vec{q}}}{ ( 2 \pi)^d}
   \int_{- \infty}^{\infty} \frac{\D \omega}{2 \pi}
 \frac{   f^{{\rm RPA}, \alpha}_{ {\vec{q}} , \I \omega } 
 \cos ( r^{\alpha}_{\|} \hat{\vec{v}}^{\alpha} \cdot {\vec{q}}  )}
 {
 ( \I \omega - {\vec{v}}^{\alpha} \cdot {\vec{q}} )^{2 }}
  \label{eq:Slonthermo}
 \end{equation}
exists as well, {\it{and vanishes\footnote{As 
shown in \cite{Henrici91}, a sufficient condition
for the vanishing of the Fourier transform\index{Fourier integral theorem}
$G ( \omega ) = \int_{- \infty}^{\infty} \D \omega \E^{- \I \omega \tau } F ( \tau )$
of a function $F( \tau )$ for $\omega \rightarrow \pm \infty$
is that $F ( \tau )$ is (at least improperly) integrable on every finite interval, and that
$\int_{- \infty}^{\infty} | F ( \tau ) | \D \tau < \infty$.
In our case these conditions have to be satisfied by the function
$F ( q^{\alpha}_{\|} ) = \int \D \vec{q}^{\alpha}_{\bot}
\int_{- \infty}^{\infty} \D \omega f^{{\rm RPA} , \alpha}_{ {\vec{q}} , \I \omega}
( \I \omega - | {\vec{v}}^{\alpha} | q^{\alpha}_{\|} )^{-2}$,
where $q^{\alpha}_{\|}$ and $\vec{q}^{\alpha}_{\bot}$ are defined 
as in Eq. \ref{eq:qlocaldecomp}.
Due to the rather singular structure of the integrand, it is by no means
obvious that for arbitrary interactions the Fourier integral theorem is applicable.
However, in all physical applications
discussed in the second part of this book 
we find that $S^{\alpha} ( r^{\alpha}_{\|} \hat{\vec{v}}^{\alpha} , 0 )$
indeed vanishes as $r^{\alpha}_{\|} \rightarrow \pm \infty$.}
in the limit $r^{\alpha}_{\|} \rightarrow \infty$}}.
Hence, the finiteness of the integral
in Eq.\ref{eq:Rlonthermo} implies 
that $Q^{\alpha}_{\infty} = R^{\alpha}$.
In this case we obtain 
 \begin{equation}
 \delta n^{\alpha}_{ {\vec{q}} }  \sim  Z^{\alpha} \mbox{sgn} ( q^{\alpha}_{\|} ) 
 \; \; \; , \; \; \; q^{\alpha}_{\|} \rightarrow 0
 \label{eq:FLdisc}
 \; \; \; ,
 \end{equation}
with the quasi-particle residue given by
 \begin{equation}
 Z^{\alpha} = \E^{R^{\alpha}}
 \label{eq:ZRrelation}
 \; \; \; .
 \end{equation}
Because we know that $Z^{\alpha}$ must be a real number
between zero and unity, $R^{\alpha}$ should be real and negative.
In Chap.~\secref{sec:exactman} we shall show 
with the help of the dynamic structure factor that 
this is indeed the case.

\newpage

\begin{center}
{\bf{The effective mass 
renormalization\index{effective mass!renormalization} $Z^{\alpha}_{\rm m}$}}
\end{center}


\noindent
Because the spatial dependence 
of the Debye-Waller factor
$ Q^{\alpha} ( r^{\alpha}_{\|} \hat{\vec{v}}^{\alpha} , \tau )$  
enters only via the projection $r^{\alpha}_{\|} = \hat{\vec{v}}^{\alpha} \cdot {\vec{r}}$,
the direction of the renormalized Fermi velocity $\tilde{\vec{v}}^{\alpha}$ is
{\it{for linearized energy dispersion}}
always parallel to the 
direction of the bare Fermi velocity $ {\vec{v}}^{\alpha}$.
From Eq.\ref{eq:Zmdef} we see that in this case the
effective mass renormalization factor associated with 
patch $P^{\alpha}_{\Lambda}$ is given by
 $Z^{\alpha}_{\rm m} = | \tilde{\vec{v}}^{\alpha} | / | {\vec{v}}^{\alpha} |$.
The renormalized Fermi velocity can be directly obtained from the real space
imaginary time sector Green's function  $G^{\alpha} ( {\vec{r}} , \tau )$
by taking {\it{first}} the limit $\tau \rightarrow \infty$ and
{\it{then}} the limit $r^{\alpha}_{\|} \rightarrow \infty$.
From Eq.\ref{eq:Gpatchreal1} we obtain 
for the non-interacting sector Green's function in this limit
 \begin{equation}
 G^{\alpha}_{0} ( {\vec{r}} , \tau )
 \sim
 -  \delta^{(d-1)} ( {\vec{r}}^{\alpha}_{\bot}  ) \frac{1}{ 2 \pi | {\vec{v}}^{\alpha} | \tau }
 \; \; \; , \; \; \; \frac{\tau}{ r^{\alpha}_{\|} } \rightarrow  \infty
 \; \; \; , \; \; \; r^{\alpha}_{\|}  \rightarrow \infty 
 \; \; \; .
 \label{eq:Gpatchrealtauinf}
 \end{equation}
Assuming the existence of the limit
 \begin{equation}
 S^{\alpha}_{\infty} =
 \lim_{r^{\alpha}_{\|} \rightarrow \infty}
 \left[ \lim_{\tau \rightarrow \infty}
 S^{\alpha} ( r^{\alpha}_{\|} \hat{\vec{v}}^{\alpha} , \tau ) \right]
 \; \; \; ,
 \label{eq:Salphaeffmass}
 \end{equation}
the relation analogous to Eq.\ref{eq:Gpatchrealtauinf}
for the interacting sector Green's function 
given in Eq.\ref{eq:Galphartreplace}  is
 \begin{equation}
 G^{\alpha} ( {\vec{r}} , \tau )
 \sim
 -  \delta^{(d-1)} ( {\vec{r}}^{\alpha}_{\bot}  ) \frac{ Z^{\alpha} \E^{ - S^{\alpha}_{\infty}}}
 { 2 \pi | {\vec{v}}^{\alpha} | \tau }
 \; \; \; , \; \; \; \frac{\tau}{ r^{\alpha}_{\|} } \rightarrow  \infty
 \; \; \; , \; \; \; r^{\alpha}_{\|}  \rightarrow \infty 
 \; \; \; .
 \label{eq:Ginttauinf}
 \end{equation}
But in an interacting  Fermi liquid we should have
 \begin{equation}
 G^{\alpha} ( {\vec{r}} , \tau )
 \sim
 - \delta^{(d-1)} ( {\vec{r}}^{\alpha}_{\bot}  ) \frac{Z^{\alpha}}
 { 2 \pi | \tilde{\vec{v}}^{\alpha} | \tau }
 \; \; \; , \; \; \; \frac{\tau}{ r^{\alpha}_{\|} } \rightarrow  \infty
 \; \; \; , \; \; \; r^{\alpha}_{\|}  \rightarrow \infty 
 \; \; \; ,
 \label{eq:GinttauinfFL}
 \end{equation}
where $\tilde{\vec{v}}^{\alpha}$ is the renormalized Fermi velocity.
Comparing Eqs.\ref{eq:Ginttauinf} and \ref{eq:GinttauinfFL}
with \ref{eq:Zmbare}, we conclude that
the effective mass renormalization factor is given by
 \begin{equation}
 Z^{\alpha}_{\rm m} = \E^{S^{\alpha}_{\infty} }
 \label{eq:ZSrelation}
 \; \; \; .
 \end{equation}

The analytic evaluation of the limit
$S^{\alpha}_{\infty}$ in Eq.\ref{eq:Salphaeffmass} 
is rather difficult.
We have not been able to obtain for general interactions a simple analytic
expression for $S^{\alpha}_{\infty}$,
which explicitly contains 
only the parameters $\underline{f}_{q}$ and $\xi_{\vec{k}}$ that
appear in the definition of the original action\footnote{
In the case of $Z^{\alpha}$ such an expression is given in
Eqs.\ref{eq:ZRrelation} and \ref{eq:Rlonthermo}.}.
Note that the naive application of the  Fourier integral theorem \cite{Henrici91}
to $S^{\alpha} ( 0 , \tau )$ implies that $S^{\alpha}_{\infty} $ should vanish, so that
bosonization with linearized energy dispersion does not 
incorporate effective mass renormalizations.
To examine this point more carefully, let us
substitute the Dyson equation \ref{eq:Dysonalpha} 
into Eq.\ref{eq:selfbos1}, and then solve for 
$\Sigma^{\alpha} ( \tilde{q} )$ as functional of
$G^{\alpha} ( \tilde{q} )$. After some trivial algebra we obtain
 \begin{equation}
  \Sigma^{\alpha} ( \tilde{q} ) = [ \I \tilde{\omega}_{n} - {\vec{v}}^{\alpha} \cdot {\vec{q}} ]
  \frac{  T^{\alpha}_{ \tilde{q}} }{1 + T^{\alpha}_{\tilde{q}} }
 \label{eq:sigmaeffmass}
 \; \; \; ,
 \end{equation}
with
 \begin{equation}
 T^{\alpha}_{\tilde{q}} =
  \frac{1}{\beta V} \sum_{ \tilde{q}^{\prime}} 
  X^{\alpha}_{  \tilde{q}^{\prime}  - \tilde{q}   }
 G^{\alpha} ( \tilde{q}^{\prime} ) 
 \label{eq:Yrenormdef}
 \; \; \; ,
 \end{equation}
where we have used $X^{\alpha}_{- q} = - X^{\alpha}_{q}$, 
see Eqs.\ref{eq:XalphaFourier} and \ref{eq:Xalphaqdef}. 
At ${q} = 0$ Eq.\ref{eq:Yrenormdef} reduces to
 \begin{equation}
 T^{\alpha}_{0} =
  \frac{1}{\beta V} \sum_{ {\vec{q}} , m  }
  \frac{ f^{{\rm RPA}, \alpha}_{{\vec{q}} , \I \omega_{m}}}{ ( \I \omega_{m} - {\vec{v}}^{\alpha}
  \cdot {\vec{q}} )}
 G^{\alpha} ( {\vec{q}} , \I \tilde{\omega}_{m}  ) 
 \label{eq:Yrenormdef2}
 \; \; \; ,
 \end{equation}
which should be compared with
 \begin{equation}
 R^{\alpha} 
   = 
   \frac{1}{\beta V } \sum_{ {\vec{q}} , m}
 \frac{   f^{{\rm RPA}, \alpha}_{ {\vec{q}} , \I \omega_{m} }}
 {
 ( \I \omega_{m} - {\vec{v}}^{\alpha} \cdot {\vec{q}} )}
 \frac{1}{( \I \omega_{m} - {\vec{v}}^{\alpha} \cdot {\vec{q}} )}
  \label{eq:Rlonthermo1b}
 \; \; \; .
 \end{equation}
Obviously 
the only difference between $T^{\alpha}_{0}$ and $R^{\alpha}$
is that the full Green's function $G^{\alpha} ( {\vec{q}} , \I \tilde{\omega}_{m} )$
on the right-hand side of Eq.\ref{eq:Yrenormdef2}
is replaced
by a factor of $( \I \omega_{m} - {\vec{v}}^{\alpha} \cdot {\vec{q}} )^{-1}$ in 
Eq.\ref{eq:Rlonthermo1b}.
Keeping in mind that
in a Fermi liquid the integral in Eq.\ref{eq:Rlonthermo1b} remains finite
in the limit $\beta , V \rightarrow \infty$
(recall Eq.\ref{eq:Rlonthermo}),
it is tempting to speculate that 
the finiteness of $R^{\alpha}$ implies
that also the expression for $T^{\alpha}_{0}$ in Eq.\ref{eq:Yrenormdef2}
must be finite. Defining the retarded function
$ T^{\alpha} ( {\vec{q}} , \omega ) = 
T^{\alpha}_{ {\vec{q}} , \I \tilde{\omega}_{n} } 
|_{ \I \tilde{\omega}_{n} \rightarrow \omega + \I 0^{+}}$,
let us now {\it{assume}} that
$T^{\alpha} ( 0 , 0)$ is  finite, and that 
for small $ {\vec{q}}$ and $\omega$
the corrections vanish with some positive power, 
 \begin{equation}
 T^{\alpha} ( {\vec{q}} , \omega ) \sim T^{\alpha} (0,0) + O ( | \vec{q} |^{{\mu}_{1}} , 
 | \omega | ^{{\mu}_{2}} )
 \; \; \; , \; \; \;  \mu_{1} , \mu_{2} > 0
 \; \; \; .
 \label{eq:Pasym}
 \end{equation}
We would like to emphasize that at this point
Eq.\ref{eq:Pasym} should be considered as an assumption, 
which is motivated by the similarity between Eqs.\ref{eq:Yrenormdef2} and
\ref{eq:Rlonthermo1b}, and by the fact that in a Fermi liquid $R^{\alpha}$ is finite.
From Eq.\ref{eq:sigmaeffmass} we see that the retarded self-energy 
can then be written as
 \begin{equation}
 \Sigma^{\alpha} ( {\vec{q}} , \omega + \I 0^{+} )
 =
 [ \omega  - {\vec{v}}^{\alpha} \cdot {\vec{q}} ]
 \frac{  T^{\alpha} ( {\vec{q}} , \omega  ) }{1 + T^{\alpha} ( {\vec{q}} , \omega )}
 \; \; \; ,
 \label{eq:SigmaretP}
 \end{equation}
and satisfies
 \begin{equation}
 \left. \frac{ \partial \Sigma^{\alpha} ( 0 , \omega + \I 0^{+} ) }
 {\partial \omega} \right|_{\omega = 0} = \frac{  T^{\alpha} ( 0 , 0) }{ 1 + T^{\alpha} (0,0) }
 \; \; \; ,
 \label{eq:sigmaYderiv}
 \end{equation}
so that according to Eq.\ref{eq:Zdef} the quasi-particle residue 
exists and is given by  
 \begin{equation}
 Z^{\alpha} = {\E}^{R^{\alpha}} = 1 + T^{\alpha} ( 0 , 0)
 \label{eq:ZY}
 \; \; \; .
 \end{equation}
From Eq.\ref{eq:ZY} we see that
the replacement of the last factor 
$( \I \omega_{m} - {\vec{v}}^{\alpha} \cdot {\vec{q}} )^{-1}$
in Eq.\ref{eq:Rlonthermo1b} by $G^{\alpha} ( {\vec{q}} , \I \tilde{\omega}_{m} )$
in Eq.\ref{eq:Yrenormdef2} amounts to an exponentiation.
Substituting now Eqs.\ref{eq:ZY} and \ref{eq:Yrenormdef} 
into Eq.\ref{eq:valpharen} we obtain for the renormalization
of the Fermi velocity
 \begin{eqnarray}
 \delta {\vec{v}}^{\alpha}  & = &  (Z^{\alpha} - 1 ) {\vec{v}}^{\alpha}
 + Z^{\alpha}
 \left.
 \nabla_{\vec{q}} 
 \Sigma^{\alpha} ( {\vec{q}} , i 0^{+} ) \right|_{ {\vec{q}} = 0 }
 \nonumber
 \\
 & = &
 T^{\alpha} (0,0) {\vec{v}}^{\alpha} - ( 1 + T^{\alpha} (0,0) ) {\vec{v}}^{\alpha}
 \frac{ T^{\alpha} ( 0,0)}{ 1 + T^{\alpha} (0,0) } = 0
 \label{eq:vnorenorm}
 \; \; \; .
 \end{eqnarray}
Hence, under the assumption \ref{eq:Pasym} 
the Fermi velocity is not renormalized, so that  $Z^{\alpha}_{\rm m}=1$.
Although we have not proven Eq.\ref{eq:Pasym},
the similarity between Eqs.\ref{eq:Yrenormdef2} and
\ref{eq:Rlonthermo1b} strongly suggests that it is indeed correct.
This is also in accordance with the Fourier integral theorem, which implies
that $S^{\alpha}_{\infty}$ should vanish if $S^{\alpha} ( 0 , 0 ) = R^{\alpha}$ exists.
We thus conclude that higher-dimensional bosonization 
{\it{with linearized energy dispersion}} does not contain
effective mass renormalizations.
We shall come back to this point
in Chap.~\secref{subsec:Thetermslin},
where we shall show that this is closely related to the fact
that for linearized energy dispersion the Fermi surface is 
approximated by a {\it{finite}} number $M$ of completely flat patches.

\section{Beyond the Gaussian approximation}
\label{sec:eik}

{\it{
We now describe a general method for including
the non-linear terms in the energy dispersion into our
background field approach.
This enables us to include the effects
of the curvature of the Fermi surface into our 
non-perturbative expression for the single-particle
Green's function. A brief description of our method
has been published in the Letter \cite{Kopietz96bac}.
Here we present for the first time the details.}}

\vspace{7mm}

\noindent
One of the main approximations in 
Sect.~\secref{sec:Derivation}
was the replacement of the Fermi surface by a collection of
flat hyper-planes, which amounts to 
setting $1 / m^{\alpha} = 0$ in the
expansion \ref{eq:chiquad} of the energy dispersion
close to the Fermi surface.
Although we have intuitively justified
this approximation for sufficiently long-range interactions,
we have not given a quantitative estimate of the corrections due to
non-linear terms in the energy dispersion.
Recall that for the density-density correlation function
such a quantitative estimate has been given in
Chap.~\secref{sec:beyond}; in this case the  corrections due to the
non-linear terms could be explicitly calculated, and in Eq.\ref{eq:gausscorrect}
we have identified the relevant small parameter.

\index{Fermi surface!curvature}
In the context of conventional one-dimensional bosonization
Haldane \cite{Haldane81} has speculated that it should
be possible to develop some kind of perturbation theory around the non-perturbative
bosonization solution for linearized energy dispersion, using the inverse
effective mass $1 / m^{\alpha}$ as a small parameter.
However, even in $d=1$ a practically useful formulation of such a perturbation
theory has not been developed. This seems to be due to the fact
that the naive expansion of the conventional 
bosonization formula for the Green's function 
in powers of $1 / m^{\alpha}$ 
becomes rather awkward in the absence of interactions \cite{Hermissondipl}, 
because in this case we can trivially write down
the exact solution $G_0^{\alpha} =  [ \I \tilde{\omega}_n
- \xi^{\alpha}_{\vec{q}}]^{-1}$. 
This expression contains infinite
orders in $1 / m^{\alpha}$,
so that it can only be recovered 
by means of the $1 / m^{\alpha}$-expansion suggested
in \cite{Haldane81} if all terms in the series are summed.
This is of course an impossible task.\index{impossible task}
In this chapter we shall develop a new method for including
the non-linear terms in the energy dispersion
into the bosonization procedure,
which {\it{in the non-interacting limit
reproduces the exact free Green's function.}} 
Thus, our method is {\it{not}}
based on a direct expansion in powers of $1 / m^{\alpha}$.
It should be mentioned that
in the special case of one dimension an alternative algebraic bosonization approach,
which includes arbitrary non-linear terms in the dispersion relation,
has recently been developed by Zemba and collaborators \cite{Caracciolo95}.
In higher dimensions
the non-linear terms in the energy dispersion
have also been discussed by
Khveshchenko \cite{Khveshchenko94b} within his
``geometric'' bosonization  approach.
However, his formalism
is based on a rather complicated  mathematical construction, and
so far has not been of practical use for the 
explicit calculation of curvature effects
on the bosonization result for the Green's function
with linearized energy dispersion.

In dimensions $d > 1$ 
it is certainly more important  to retain the
non-linear terms in the energy dispersion than in $d=1$,
because only in higher dimensions the Fermi surface has a curvature.
To see this more clearly, let us assume for the moment that
locally the Fermi surface can be approximated by a
quadratic form, and that in an appropriately oriented coordinate
system the energy dispersion $\xi^{\alpha}_{\vec{q}}$
defined in Eq.\ref{eq:xialphaqexp} can be written as
 \begin{equation}
 \xi^{\alpha}_{\vec{q}} = {\vec{v}}^{\alpha} \cdot {\vec{q}} 
 + \frac{( \hat{\vec{v}}^{\alpha} \cdot {\vec{q}} )^2 }{ 2 m^{\alpha}_{\|}}
 + \frac{ ( {\vec{q}}^{\alpha}_{\bot} )^2}{2 m^{\alpha}_{\bot} }
 \label{eq:xisphere}
 \; \; \; ,
 \end{equation}
where 
${\vec{q}}^{\alpha}_{\bot} = {\vec{q}} -  
( {\vec{q}} \cdot \hat{\vec{v}}^{\alpha})
\hat{\vec{v}}^{\alpha} $, and $m^{\alpha}_{\|}$ and
$m^{\alpha}_{\bot}$ are the effective masses for the motion parallel
and perpendicular to the local normal $\hat{\vec{v}}^{\alpha}$.
The important point is now that only the last term
in Eq.\ref{eq:xisphere} describes the curvature of the patches. 
In other words,
for $1 / m^{\alpha}_{\bot} = 0$ but finite $1 / m^{\alpha}_{\|}$ 
we still have completely flat patches.
Obviously in $d > 1$ there exist hyper-planes in momentum space (defined by
$\hat{\vec{v}}^{\alpha} \cdot \vec{q} = 0$) where
the last term in Eq.\ref{eq:xisphere} is the
dominant contribution in the expansion of the energy dispersion.
As already mentioned in the second footnote in Sect.~\secref{subsec:invdiag}, 
a priori it is not clear whether the contribution from these hyper-planes
to some physical quantity of interest is negligible or not.
From the previous section we expect that
the curvature of the Fermi surface will certainly play an important
role to obtain the correct effective mass renormalization
in a Fermi liquid
(recall the discussion in Sect.~\secref{sec:Identification}). 
As we shall see in Chap.~\secref{subsec:Thetermslin},
this problem is closely  related to the 
existence of a {\it{double pole}}\index{double pole} in the integrand of the
linearized bosonization result for the Debye-Waller factor,
see Eqs.\ref{eq:Rlondef} and \ref{eq:Slondef}.

Let us point out two more rather peculiar features of the
higher-dimensional bosonization result for the
Green's function with linearized energy dispersion.
First of all,
for any finite number
$M$ of patches the real space Green's 
function is of the form
$G ( {\vec{r}} , \tau ) = \sum_{\alpha = 1}^{M} \E^{\I {\vec{k}}^{\alpha} \cdot {\vec{r}} }
G^{\alpha} ( {\vec{r}}, \tau )$  
where $G^{\alpha} ( {\vec{r}}, \tau )$ is proportional 
to a $d-1$-dimensional 
$\delta$-function\footnote{
As discussed in Sect.~\secref{subsec:Greal},
at short distances the $\delta$-function should actually be
replaced by the cutoff-dependent function
$\delta^{(d-1)}_{\Lambda} ( \vec{r}_{\bot}^{\alpha} )$
defined in Eq.\ref{eq:deltaLambda}.}
$\delta^{(d-1)} ( \vec{r}_{\bot}^{\alpha} )$
of the components of ${\vec{r}}$ that are perpendicular to the local
Fermi velocity ${\vec{v}}^{\alpha}$
(see Eqs.\ref{eq:Grealtotal} and
\ref{eq:Grealtotalfull}). 
As a consequence, we may replace ${\vec{r}} \rightarrow  
( {\vec{r}} \cdot \hat{\vec{v}}^{\alpha} ) \hat{\vec{v}}^{\alpha}$ in the 
expression for the Debye-Waller factor $Q^{\alpha} ( {\vec{r}} , \tau )$
(see Eq.\ref{eq:Galphartreplace}).
If we naively take the limit of infinite patch number $M \rightarrow \infty$, then
the patch summation is turned into a $d-1$-dimensional integral over the
Fermi surface, so that in this limit the singular function 
$\delta^{(d-1)}_{\Lambda} ( \vec{r}_{\bot}^{\alpha} )$ appears under a 
$d-1$-dimensional integral, and the final result for the
real space Green's function does not exhibit any singularities.
However, because $M \rightarrow \infty$ implies 
a vanishing patch cutoff, $\Lambda \rightarrow 0$, and because
the approximations made in deriving the  above result can
formally only be justified if $\Lambda$ is held finite
and large compared with the range $q_{\rm c}$ of the interaction in momentum space
(see Fig.~\secref{fig:qc}), one may wonder whether 
the above limiting procedure is justified.
Of course, in momentum space this problem remains hidden,
because the function
$\delta^{(d-1)}_{\Lambda} ( \vec{r}_{\bot}^{\alpha} )$ 
is eliminated trivially via the Fourier transformation. 
Consequently the interacting Green's function for wave-vectors close to
${\vec{k}}^{\alpha}$ is simply
$ G ( {\vec{k}}^{\alpha} + {\vec{q}} , \I \tilde{\omega}_{n} )
 = G^{\alpha} ( {\vec{q}} , \I \tilde{\omega}_{n} )$
(see Eq.\ref{eq:Gkalphashiftres}),
where $G^{\alpha} ( {\vec{q}} , \I \tilde{\omega}_n )$ is the
Fourier transform of
$ G^{\alpha} ( {\vec{r}}, \tau )$.
Nevertheless, it is legitimate to
ask how the Green's function looks in real space, and
the prediction of 
higher-dimensional bosonization with linearized energy dispersion
is not quite satisfactory.
Another shortcoming of the linearized theory will
be discussed in detail in Chap.~\secref{sec:nesting}: the 
replacement of a curved Fermi surface by a finite number of
flat patches can give rise to unphysical nesting singularities.

It is intuitively obvious that the problems mentioned above 
are related to the fact that we have ignored the curvature of the 
Fermi surface within a given patch. 
To cure these drawbacks of higher-dimensional bosonization,
we shall now generalize our background field approach
to the case of finite masses $m_i^{\alpha}$.

\subsection{The Green's function for fixed background field}
\label{sec:backfixed}

{\it{
We develop an imaginary time eikonal expansion
for the single-particle Green's function 
${\cal{G}}^{\alpha} (  {\vec{r}} , {\vec{r}}^{\prime} , \tau , \tau^{\prime} )$
at fixed background field, which takes the non-linear terms
in the energy dispersion non-perturbatively into account.
In this way we obtain
the generalization of the
Schwinger ansatz given in Eq.\ref{eq:Ansatz}  
for non-linear energy dispersions.}}

\begin{center}
{\bf{Generalization of the Schwinger ansatz}}
\end{center}

\noindent
We would like to invert the infinite
matrix $\hat{G}^{-1}$ in Eq.\ref{eq:GhatDysonshift2} for general
energy dispersion $\epsilon_{\vec{k}}$.
As explained in Sect.~\secref{subsec:invdiag}, 
it is convenient to 
measure wave-vectors with respect to 
a coordinate system centered at $\vec{k}^{\alpha}$
and define
 \begin{equation}
 [\hat{G}]_{ {\vec{k}}^{\alpha} + \vec{q} , \I \tilde{\omega}_n ;
{\vec{k}}^{\alpha} + \vec{q}^{\prime} , \I \tilde{\omega}_{n^{\prime}} }
 = [ \hat{G}^{\alpha} ]_{ \vec{q} , \I \tilde{\omega}_n;
  \vec{q}^{\prime} , \I \tilde{\omega}_{ n^{\prime} } }
  \equiv [\hat{G}^{\alpha}]_{\tilde{q}  \tilde{q}^{\prime} }
  \; \; \; .
  \end{equation}
Then the infinite matrix 
$[ \hat{G}^{\alpha}]_{\tilde{q}  \tilde{q}^{\prime}}$ is determined by an equation
of the form \ref{eq:Galphadif}, with
$[ G_0^{\alpha} ( \tilde{q} )]^{-1}$ now given by
$[ G_0^{\alpha} ( \tilde{q} )]^{-1} =
\I \tilde{\omega}_n - \epsilon_{ \vec{k}^{\alpha} + \vec{q}} + \mu $. 
Note that the above transformations are 
valid for an arbitrary sectorization of momentum space
(see Sect.~\secref{sec:sectors}), including the special
case that we identify the entire momentum space with a single
sector (then we just shift the coordinate origin 
in momentum space to $\vec{k}^{\alpha}$, as shown in Fig.~\secref{fig:coordgood}.).
Defining the Fourier transforms
$ {\cal{G}}^{\alpha} ( {\vec{r}} , {\vec{r}}^{\prime} , \tau , \tau^{\prime} )$ 
and $V^{\alpha} ( {\vec{r}} , \tau )$
of $ [\hat{G}^{\alpha}]_{\tilde{q} \tilde{q}^{\prime}}$ 
and  $V^{\alpha}_q$
as in Eqs.\ref{eq:calGdef} and \ref{eq:Vrt}, it is easy to see that 
$ {\cal{G}}^{\alpha} ( {\vec{r}} , {\vec{r}}^{\prime} , \tau , \tau^{\prime} )$ 
satisfies the partial differential equation
 \begin{equation}
 \hspace{-3mm}
  \left[- {\partial}_{ \tau} -    
  {\epsilon}_{ {\vec{k}}^{\alpha} +  {\vec{P}}_{\vec{r}} }  + \mu - 
  V^{\alpha}  ( {\vec{r}} , \tau )
  \right]
 {\cal{G}}^{\alpha} ( {\vec{r}} , {\vec{r}}^{\prime} , \tau , \tau^{\prime} )
     =
 \delta ( {\vec{r}} - {\vec{r}}^{\prime} ) \delta^{\ast} ( \tau - \tau^{\prime} )
 \; ,
 \label{eq:Galphadifrt2}
 \end{equation}
where ${\vec{P}}_{\vec{r}} = - \I \nabla_{\vec{r}}$ is the momentum operator.
Eq.\ref{eq:Galphadifrt2} is the
generalization of Eq.\ref{eq:Galphadifrt} 
to arbitrary energy dispersions $\epsilon_{\vec{k}}$.
This partial differential equation together with the
Kubo-Martin-Schwinger boundary condition \ref{eq:KMS} 
uniquely determines the function\index{KMS boundary conditions}
${\cal{G}}^{\alpha} ( {\vec{r}} , {\vec{r}}^{\prime} , \tau , \tau^{\prime} )$.
We now truncate the expansion of
$\epsilon_{\vec{k}^{\alpha} + \vec{q} } - \mu$ for small $\vec{q}$ at the
second order, see Eqs.\ref{eq:energyquad} and \ref{eq:chiquad}. 
Then Eq.\ref{eq:Galphadifrt} is of second order in the spatial
derivatives. 
Note that for free  fermions
with energy dispersion $\epsilon_{\vec{k}} =  \vec{k}^2/ ({2 m})$
the truncation at the second order is exact,
but for more complicated Fermi surfaces
we are assuming that the sectors have
been chosen sufficiently small such that the {\it{local curvature}}
can be approximated by a constant.
Linearization of the energy dispersion amounts to ignoring
the quadratic terms in the expansion of $\epsilon_{\vec{k}^{\alpha} + \vec{q}}$
for small $\vec{q}$, 
in which case the Schwinger ansatz \ref{eq:Ansatz} 
solves Eq.\ref{eq:Galphadifrt2}.
It is not difficult to see that for non-linear energy dispersion 
this ansatz does {\it{not}} lead to a consistent solution of Eq.\ref{eq:Galphadifrt}.
In order to develop a systematic method for treating the non-linear terms
in the energy dispersion in a non-perturbative way, we need a generalization
of the Schwinger ansatz \ref{eq:Ansatz} which in the limit
$1 / m^{\alpha}_i \rightarrow 0$ reduces to the solution of the linearized 
differential equation.  The crucial observation is that
the quantity ${\cal{G}}^{\alpha} ( {\vec{r}} , {\vec{r}} , \tau , \tau )$
(which is obtained by setting
${\vec{r}} = {\vec{r}}^{\prime}$ and $\tau = \tau^{\prime}$ in
the solution of Eq.\ref{eq:Galphadifrt2})
represents physically a contribution to the {\it{density of the system}}.
Moreover, on physical grounds it is also clear that
the external potential $V^{\alpha} ( \vec{r} , \tau )$ 
should lead to a deviation of the density from its
equilibrium value. Evidently the Schwinger ansatz \ref{eq:Ansatz} 
predicts that the external potential does not lead to any modulation
of the density, which is of course an unphysical artefact of the linearization.
For non-linear energy dispersion, our generalized Schwinger ansatz should
allow for density fluctuations. The simplest possible way 
to incorporate the physics of density fluctuations
without changing the important exponential 
factor in the Schwinger ansatz\index{Schwinger ansatz!generalization} 
is to set\footnote{I would have never tried this
ansatz without a hint from Lorenz Bartosch.}
 \begin{equation}
 {\cal{G}}^{\alpha} ( {\vec{r}} , {\vec{r}}^{\prime} , \tau , \tau^{\prime} )
  = 
 {\cal{G}}^{\alpha}_1 ( {\vec{r}} , {\vec{r}}^{\prime} , \tau , \tau^{\prime} )
 \E^{ \Phi^{\alpha} ( {\vec{r}} , \tau ) - \Phi^{\alpha} ( {\vec{r}}^{\prime} , \tau^{\prime} ) }
 \label{eq:Ansatz2}
 \; \; \; .
 \end{equation}
The KMS boundary conditions\index{KMS boundary conditions} are satisfied by requiring that
$\Phi^{\alpha} ( {\vec{r}} , \tau )$ should be  periodic in $\tau$, while
${\cal{{G}}}^{\alpha}_1 ( {\vec{r}} , {\vec{r}}^{\prime} , \tau , \tau^{\prime} )$ 
should be antiperiodic in $\tau$ and $\tau^{\prime}$.
Setting ${\vec{r}} = {\vec{r}}^{\prime}$ and $\tau = \tau^{\prime}$,
we conclude that
${\cal{G}}^{\alpha}_1 ( {\vec{r}} , {\vec{r}} , \tau , \tau)$
is the contribution from states  with momenta 
in sector $K^{\alpha}_{\Lambda , \lambda}$ 
to the density of the system.
From the arguments given above it is therefore clear that
for non-linear energy dispersion
 ${\cal{G}}^{\alpha}_1 ( {\vec{r}} , {\vec{r}}^{\prime} , \tau , \tau^{\prime} )$ 
must be a non-trivial function of the external potential.
Of course, 
in Eq.\ref{eq:Ansatz2} we could always choose
$\Phi^{\alpha} = 0$ and ${\cal{G}}^{\alpha} = {\cal{G}}_1^{\alpha}$, so that
nothing would be gained. The crucial point is, however, that there exists
another non-trivial choice of 
$\Phi^{\alpha} $ and ${\cal{G}}_1^{\alpha}$ which
leads to the natural generalization of the Schwinger ansatz \ref{eq:Ansatz}
to systems with energy dispersions of the type \ref{eq:energyquad} and
\ref{eq:chiquad}.
To see this, we substitute
Eq.\ref{eq:Ansatz2} into Eq.\ref{eq:Galphadifrt2} 
and obtain after a simple calculation 
 \begin{eqnarray}
 \lefteqn{
  \left[ - {\partial}_{ \tau} -    
  {\epsilon}_{ {\vec{k}}^{\alpha} + {\vec{P}}_{\vec{r}} } + \mu
  - {\vec{u}}^{\alpha} ({  {\vec{r}} , \tau } )\cdot {\vec{P}}_{\vec{r}} 
  \right]
 {\cal{G}}^{\alpha}_1 ( {\vec{r}} , {\vec{r}}^{\prime} , \tau , \tau^{\prime} )
 =
  \delta ( {\vec{r}} - {\vec{r}}^{\prime} ) \delta^{\ast} ( \tau - \tau^{\prime} )
  \; \;  }
 \nonumber
 \\
 &  &
 +
 {\cal{G}}^{\alpha}_1 ( {\vec{r}} , {\vec{r}}^{\prime} , \tau , \tau^{\prime} )
 \left\{
 \left[ - {\partial}_{ \tau} -    
 {\xi}^{\alpha}_{ {\vec{P}}_{\vec{r}}}
 \right]
 \Phi^{\alpha} ( {\vec{r}} , \tau ) - 
 {V}^{\alpha}  ( {\vec{r}} , \tau )
 \right.
 \nonumber
 \\
 & &
 \left.
 \hspace{25mm}
   -   
   [{\vec{P}}_{\vec{r}} \Phi^{\alpha} ( {\vec{r}} , \tau )] 
    (2 {\tens{M}}^{\alpha}  )^{-1}
   [{\vec{P}}_{\vec{r}} \Phi^{\alpha} ( {\vec{r}} , \tau )  ]
  \right\} 
  \;  .
  \label{eq:Ansatzsu}
  \end{eqnarray}
where the $d \times d$-matrix ${\tens{M}}^{\alpha}$
contains the effective masses,\index{effective mass} 
 \begin{equation}
 [{\tens{M}}^{\alpha} ]_{ij} = \delta_{ij} m^{\alpha}_i
 \label{eq:Meffdef}
 \; \; \; ,
 \end{equation}
and the components $u_i^{\alpha} ( {\vec{r}} , \tau )$
of the velocity 
$ {\vec{u}}^{\alpha}({  {\vec{r}} , \tau })$ are given 
by\index{random velocity}
 \begin{equation}
u_i^{\alpha} ( {\vec{r}} , \tau )
\equiv {\vec{e}}_i \cdot 
{\vec{u}}^{\alpha}({  {\vec{r}} , \tau }) 
 =  \frac{ {\vec{e}}_i \cdot  {\vec{P}}_{\vec{r}}  \Phi^{\alpha} ( {\vec{r}} , \tau ) }
 { m^{\alpha}_i}
 \label{eq:urandef}
 \; \; \; .
 \end{equation}
Here the unit vectors $\vec{e}_1 , \ldots , \vec{e}_d$ 
match the axes of the local
coordinate system attached to $\vec{k}^{\alpha}$ in which
the effective mass tensor $\tens{M}^{\alpha}$ is diagonal.
The crucial observation is now that, apart from the trivial solution
$\Phi^{\alpha} = 0$ and ${\cal{G}}^{\alpha}_1 = {\cal{G}}^{\alpha}$,
we obtain another {\it{exact}} solution of 
Eq.\ref{eq:Ansatzsu} by choosing
$\Phi^{\alpha} ( {\vec{r}}, \tau )$ and
 ${\cal{G}}^{\alpha}_1 ( {\vec{r}} , {\vec{r}} , \tau , \tau^{\prime} )$ 
such that\index{eikonal!equation}
  \begin{eqnarray}
  \left[ - {\partial}_{ \tau} -    
  {\xi}^{\alpha}_{  {\vec{P}}_{\vec{r}} } 
  \right]
  \Phi^{\alpha} ( {\vec{r}} , \tau ) & =  &
   {V}^{\alpha}  ( {\vec{r}} , \tau )
   +  [ {\vec{P}}_{\vec{r}} \Phi^{\alpha} ( {\vec{r}} , \tau )  ]
   ( 2 {\tens{M}}^{\alpha} )^{-1} 
   [ {\vec{P}}_{\vec{r}} \Phi^{\alpha} ( {\vec{r}} , \tau )  ]
  \; ,
  \nonumber
  \\
  & &
  \label{eq:difPhi}
  \end{eqnarray}
  \begin{eqnarray}
  \left[ - {\partial}_{ \tau} -    
  {\epsilon}_{ {\vec{k}}^{\alpha} + {\vec{P}}_{\vec{r}} }  + \mu
  - {\vec{u}}^{\alpha} ({  {\vec{r}} , \tau } )\cdot {\vec{P}}_{\vec{r}} 
  \right]
 {\cal{G}}^{\alpha}_1 ( {\vec{r}} , {\vec{r}}^{\prime} , \tau , \tau^{\prime} ) 
 & = & \delta ( {\vec{r}} - {\vec{r}}^{\prime} ) \delta^{\ast} ( \tau - \tau^{\prime} )
  \; .
  \nonumber
  \\
  & &
  \label{eq:difG1}
  \end{eqnarray}
Thus,
 ${\cal{G}}^{\alpha}_1 ( {\vec{r}} , {\vec{r}}^{\prime} , \tau , \tau^{\prime} )$ is
again a fermionic Green's function. 
Note that the differential equation \ref{eq:difPhi} is non-linear, but
contains only first-order derivatives.
In contrast, the original Eq.\ref{eq:Galphadifrt2} is linear but
involves second-order derivatives.
Differential equations of the type \ref{eq:difPhi} are called
{\it{eikonal equations}}\index{eikonal!equation}, and appear in many fields
of physics,
such as classical 
mechanics\footnote{Recall the Hamilton-Jacobi\index{Hamilton-Jacobi equation} 
equation \cite{Goldstein83}
$- \partial S / \partial t  = V ( {\vec{r}} , t ) + 
\frac{( \nabla S )^2 }{2m}$
for the action $S ( {\vec{r}} , t )$ of a particle
with mass $m$ that moves under the influence of an external 
potential $V ( {\vec{r}} , t)$.},
geometrical optics \cite{Landau84}, 
quantum mechanical scattering theory \cite{Sakurai94}, 
and relativistic quantum field 
theories \cite{Popov83,Fradkin66}.
The functional $\Phi^{\alpha} ( {\vec{r}} , \tau )$
is called the {\it{eikonal}}. 
In the limit $1/{m^{\alpha}_i} \rightarrow 0$ 
the eikonal equation \ref{eq:difPhi} reduces to
the corresponding equation \ref{eq:phidif} of the linearized
theory, which can be solved exactly via Fourier transformation, see
Eq.\ref{eq:Phires}.
Furthermore, in this case
the velocity ${\vec{u}}^{\alpha}({  {\vec{r}} , \tau })$  
vanishes, so that 
 ${\cal{G}}^{\alpha}_1 ( {\vec{r}} , {\vec{r}}^{\prime} , \tau , \tau^{\prime} ) 
 = 
 {{G}}^{\alpha}_{0} ( {\vec{r}} - {\vec{r}}^{\prime} , \tau - \tau^{\prime} )$.
However, if one of the masses $m^{\alpha}_i$ is finite, 
the eikonal equation \ref{eq:difPhi} is non-linear and cannot be solved exactly.
We shall discuss a method to obtain an approximate solution shortly.

The differential equation \ref{eq:difG1} 
describes the motion of a fermion under the influence of a space- and 
time-dependent {\it{random velocity}}
 ${\vec{u}}^{\alpha}({  {\vec{r}} , \tau })$. 
At the first sight it seems that this problem is
just as difficult to solve as the original 
Eq.\ref{eq:Galphadifrt2}. The crucial point is, however, that 
perturbation theory in terms of the 
{\it{derivative potential}}\index{derivative potential}
  $ {\vec{u}}^{\alpha} ({  {\vec{r}} , \tau } )\cdot {\vec{P}}_{\vec{r}} $
in  Eq.\ref{eq:difG1} is {\it{less infrared singular}}
than perturbation theory in terms of the original
random potential $V^{\alpha} ( {\vec{r}} , \tau  )$
in Eq.\ref{eq:Galphadifrt2}.
Moreover, for large effective masses $m^{\alpha}_i$
the random velocity ${\vec{u}}^{\alpha} ( {\vec{r}} , \tau )$ is small,
so that the perturbation theory
in powers of the derivative potential is justified.
Such a small parameter is absent in 
Eq.\ref{eq:Galphadifrt2}.

\begin{center}
{\bf{The eikonal equation}}
\end{center}

\noindent
Although it is impossible to solve 
the non-linear partial differential equation \ref{eq:difPhi} exactly, 
we can follow the pioneering work of 
E. S. Fradkin \cite{Fradkin66} to
obtain the solution as series in powers of $V^{\alpha}$. 
We would like to emphasize, however, that our imaginary time eikonal equation
is {\it{not identical}} with the real time eikonal equation
discussed by Fradkin \cite{Fradkin66}.  The latter
has recently been applied
by Khveshchenko and Stamp \cite{Khveshchenko93} to the problem 
of fermions coupled to gauge fields, 
and involves an additional 
time-like auxiliary variable. For a $d$-dimensional quantum system one thus has
to deal with a $d+2$-dimensional partial differential equation, which
leads to rather complicated expressions for
the higher-order terms in the eikonal expansion \cite{Fradkin66}.
In contrast, our imaginary time eikonal equation \ref{eq:difPhi}
is $d+1$-dimensional and does not depend
on additional auxiliary variables.
This facilitates the calculation of corrections
to the leading term. 
See the work \cite{Kopietzeik} for 
a detailed discussion of the real time eikonal method and a
comparison with our functional bosonization approach.

Following Fradkin \cite{Fradkin66}, we obtain the solution of Eq.\ref{eq:difPhi}
by making the ansatz
 \begin{equation}
 \Phi^{\alpha} ( {\vec{r}} , \tau ) = \sum_{n=1}^{\infty}
 \Phi^{\alpha}_n ( {\vec{r}} , \tau ) 
 \label{eq:Phiansatz}
 \; \; \; ,
 \end{equation}
where
 $\Phi^{\alpha}_n ( {\vec{r}} , \tau ) $ involves by assumption 
$n$ powers of $V^{\alpha}$. Substituting Eq.\ref{eq:Phiansatz} into
Eq.\ref{eq:difPhi} and comparing powers of $V^{\alpha}$, it is easy to see that
the $n^{\rm th}$ order term $\Phi^{\alpha}_n ( {\vec{r}} , \tau ) $  
is determined by the inhomogeneous linear differential equation
  \begin{equation}
  \left[ - {\partial}_{ \tau} -    
  {\xi}^{\alpha}_{  {\vec{P}}_{\vec{r}} } 
  \right]
  \Phi^{\alpha}_n ( {\vec{r}} , \tau ) = 
   {V}^{\alpha}_n  ( {\vec{r}} , \tau )
   \; \; \; ,
   \; \; \; n = 1,2, \ldots \; \; \; ,
   \label{eq:difPhin}
   \end{equation}
where the first order potential is simply
 \begin{equation}
 V_1^{\alpha} ( {\vec{r}} , \tau ) = 
 V^{\alpha} ( {\vec{r}} , \tau )
 \; \; \; ,
 \end{equation}
and the higher orders are
 \begin{equation}
 \hspace{-5mm}
   {V}^{\alpha}_n  ( {\vec{r}} , \tau )
   = \sum_{ n^{\prime} = 1}^{n-1}
   [ {\vec{P}}_{\vec{r}} \Phi^{\alpha}_{n^{\prime}} ( {\vec{r}} , \tau ) ]
   ( 2 {\tens{M}}^{\alpha} )^{-1} 
    [ {\vec{P}}_{\vec{r}} 
   \Phi^{\alpha}_{n- n^{\prime}} ( {\vec{r}} , \tau ) ]
   \; ,
   \; n = 2,3, \ldots \; .
  \end{equation}
Note that the inhomogeneity 
  $ {V}^{\alpha}_n  ( {\vec{r}} , \tau )$ in the differential equation \ref{eq:difPhin} 
for $\Phi^{\alpha}_n ( {\vec{r}} , \tau )$ depends only on solutions
$\Phi^{\alpha}_{n^{\prime}} ( {\vec{r}} , \tau )$ with $n^{\prime} < n$, so that
we can calculate the functionals
$\Phi^{\alpha}_n ( {\vec{r}} , \tau )$ iteratively.
Because Eq.\ref{eq:difPhin} is linear, its solution is easily obtained by means of the
Green's function of the differential operator on the left-hand side,
 \begin{equation}
  \Phi^{\alpha}_n ( {\vec{r}} , \tau ) = 
  \int \D {\vec{r}}^{\prime} \int_0^{\beta} \D \tau^{\prime}
  {G}^{\alpha}_{\rm b} ( {\vec{r}} - {\vec{r}}^{\prime} , \tau - \tau^{\prime} )
  V_n^{\alpha} ( {\vec{r}}^{\prime} , \tau^{\prime} )
  \label{eq:Phinsol}
  \; \; \; ,
  \end{equation}
where
 \begin{equation}
  {G}^{\alpha}_{\rm b} ( {\vec{r}} , \tau  )
  = \frac{1}{\beta V} \sum_q \E^{ \I ( {\vec{q}} 
  \cdot {\vec{r}} - \omega_m \tau )} 
  G^{\alpha}_{\rm b} ( q )
  \; \; \; , \; \; \; 
  G^{\alpha}_{\rm b} ( q ) =
 \frac{1}{ \I \omega_m
  - \xi^{\alpha}_{\vec{q}} }
  \label{eq:tildeGdefbos}
  \; \; \; .
  \end{equation}
This Green's function
should not be confused with the corresponding free fermionic Green's function
 \begin{equation}
 \hspace{-5mm}
  {G}^{\alpha}_0 ( {\vec{r}} , \tau  )
  = \frac{1}{\beta V} \sum_{\tilde{q}} 
   \E^{ \I ( {\vec{q}} \cdot {\vec{r}} - 
  \tilde{\omega}_n \tau ) } G^{\alpha}_0 ( \tilde{q} )
  \; \;  ,  \; \;
  G^{\alpha}_0 ( \tilde{q} ) = \frac{1}{
   \I \tilde{\omega}_n
  - \epsilon_{ {\vec{k}}^{\alpha}+{\vec{q}} } + \mu }
  \label{eq:Gdeffer}
  \;   ,
  \end{equation}
which for linearized energy dispersion and
$\epsilon_{\vec{k}^{\alpha}} = \mu$ reduces to  Eq.\ref{eq:G0res}.
Note that the Fourier transform $G^{\alpha}_{\rm b} ( q )$ 
of ${G}^{\alpha}_{\rm b} ( {\vec{r}} , \tau  )$
involves {\it{bosonic}} Matsubara frequencies
and depends on the {\it{excitation energy}}
$\xi^{\alpha}_{\vec{q}} = 
\epsilon^{\alpha}_{ \vec{k}^{\alpha} + \vec{q}}  -
\epsilon^{\alpha}_{ \vec{k}^{\alpha} }$. 
The bosonic frequencies insure that
the functional $\Phi^{\alpha} ( {\vec{r}} ,\tau )$ 
is periodic in $\tau$, so that our ansatz \ref{eq:Ansatz2}
satisfies the KMS boundary conditions (see also the discussion
in Sect.~\secref{subsec:invdiag}).
In contrast, the Fourier transform 
$G_0^{\alpha} ( \tilde{q} )$
of ${G}^{\alpha}_0 ( {\vec{r}} , \tau  )$ 
depends on fermionic frequencies
and involves the usual combination
$\epsilon^{\alpha}_{ \vec{k}^{\alpha} + \vec{q}} - \mu$. 
Recall that in general we may choose 
$\epsilon_{{\vec{k}}^{\alpha}} \neq \mu$.

\begin{figure}
\sidecaption
\psfig{figure=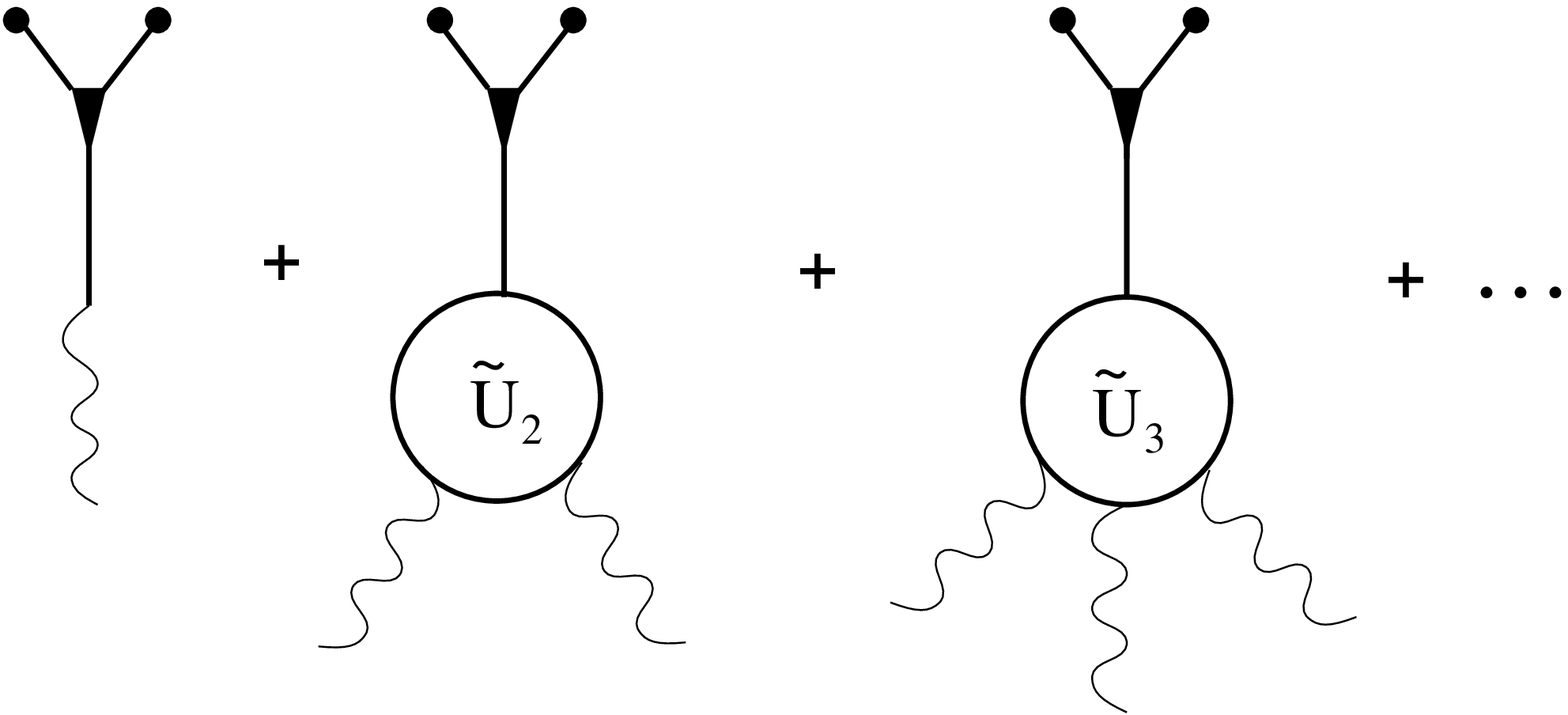,width=8cm}
\caption[
Graphical representation of the eikonal.]
{\begin{sloppypar}
Graphical representation of the 
functional
 $\Phi^{\alpha} ( {\vec{r}} , \tau ) 
 - \Phi^{\alpha} ( {\vec{r}}^{\prime} , \tau^{\prime} )$
defined in Eq.\ref{eq:eikonalpara}.
As in Fig.~\secref{fig:closedloop}, the $\phi^{\alpha}$-fields are
represented by wavy lines. The Y-shaped symbol represents
the function ${\cal{J}}^{\alpha}_{-q} 
 ( {\vec{r}} , {\vec{r}}^{\prime} , \tau , \tau^{\prime} )$ defined 
 in Eq.\ref{eq:tildeJdef}. The solid dots represent the external points
 ${\vec{r}} , \tau $ and ${\vec{r}}^{\prime} , \tau^{\prime}$.
\end{sloppypar}
}
\label{fig:eik}
\end{figure}

To carry out the above iterative procedure in practice, we find it more
convenient to work in Fourier space.
Defining the Fourier transforms $V^{\alpha}_q$ and $\Phi^{\alpha}_q$ as in
Eq.\ref{eq:Vrt}, it is easy to show that Eq.\ref{eq:difPhi} 
implies for the Fourier components
 \begin{equation}
 \left[ \I \omega_m - \xi^{\alpha}_{\vec{q}} \right] \Phi^{\alpha}_q = V^{\alpha}_q
 +  \sum_{ q^{\prime}}  ( {\vec{q}} - {\vec{q}}^{\prime} ) 
 ( 2 {\tens{M}}^{\alpha})^{-1}  {\vec{q}}^{\prime}
 \Phi^{\alpha}_{q-q^{\prime}} \Phi^{\alpha}_{q^{\prime}}
 \label{eq:Phiintegraleq}
 \; \; \; .
 \end{equation}
Keeping in mind that
$V^{\alpha}_q = \frac{\I}{\beta} \phi^{\alpha}_q$ (see Eq.\ref{eq:hatVphi}),
it is convenient to define
 \begin{equation}
 \Psi^{\alpha}_q = \frac{\beta}{\I} [ \I \omega_m - \xi^{\alpha}_{\vec{q}} ]
 \Phi^{\alpha}_q
 \label{eq:Psifunctionaldef}
 \; \; \; ,
 \end{equation}
so that Eq.\ref{eq:Phiintegraleq} can also be written as\index{eikonal!integral equation}
 \begin{equation}
 \Psi^{\alpha}_q = \phi^{\alpha}_q + \sum_{ q^{\prime} q^{\prime \prime}}
 \delta_{q , q^{\prime} + q^{\prime \prime} }
 \gamma^{\alpha}_{ q^{\prime} , q^{\prime \prime} }
 \Psi^{\alpha}_{q^{\prime}}
 \Psi^{\alpha}_{q^{\prime \prime}}
 \label{eq:Psiint}
 \; \; \; ,
 \end{equation}
where the dimensionless kernel is 
 \begin{equation}
 \gamma^{\alpha}_{ q^{\prime} , q^{\prime \prime} }
 = \frac{\I}{\beta} {\vec{q}}^{\prime}   ( 2 {\tens{M}}^{\alpha} )^{-1} {\vec{q}}^{\prime \prime} 
 {G}_{\rm b}^{\alpha} ( q^{\prime} )
 {G}_{\rm b}^{\alpha} ( q^{\prime \prime} )
 \label{eq:kernelsymm}
 \; \; \; .
 \end{equation}
Note that this kernel is symmetric under the exchange $q^{\prime} 
\leftrightarrow q^{\prime \prime}$.
The $q \equiv [ {\vec{q}} , \I \omega_m ] = 0$-term requires a special treatment.
Setting $q=0$ on both sides of Eq.\ref{eq:Phiintegraleq}, we obtain
 \begin{equation}
 0 = V^{\alpha}_0
 -  \sum_{ q^{\prime}}  {\vec{q}}^{\prime}  
 ( 2 {\tens{M}}^{\alpha})^{-1}  {\vec{q}}^{\prime}
 \Phi^{\alpha}_{-q^{\prime}} \Phi^{\alpha}_{q^{\prime}}
 \label{eq:Phiintegraleq0}
 \; \; \; .
 \end{equation}
Subtracting this from Eq.\ref{eq:Phiintegraleq}, we see that
$\Psi^{\alpha}_0 = 0$.
With the above definitions,
the eikonal can be written as 
 \begin{equation}
 \Phi^{\alpha} ( {\vec{r}} , \tau ) - \Phi^{\alpha} ( {\vec{r}}^{\prime} , \tau^{\prime} ) 
 = \sum_q {\cal{J}}^{\alpha}_{-q} 
 ( {\vec{r}} , {\vec{r}}^{\prime} , \tau , \tau^{\prime} )
 \Psi^{\alpha}_q
 \label{eq:eikonalpara}
 \; \; \; ,
 \end{equation}
where
 \begin{equation}
 {\cal{J}}^{\alpha}_{-q} 
 ( {\vec{r}} , {\vec{r}}^{\prime} , \tau , \tau^{\prime} )
 = \frac{\I}{\beta} {G}_{\rm b}^{\alpha} ( q ) 
 \left[ 
 \E^{ \I ( {\vec{q}} \cdot {\vec{r}} - \omega_m \tau ) } -
 \E^{ \I ( {\vec{q}} \cdot {\vec{r}}^{\prime} - \omega_{m} \tau^{\prime} ) } 
 \right]
 \label{eq:tildeJdef}
 \; \; \; .
 \end{equation}
Note that for linearized energy dispersion 
 ${\cal{J}}^{\alpha}_{q} ( {\vec{r}} , 0 , \tau , 0 )$ 
reduces precisely to the function
${\cal{J}}^{\alpha}_{q} ( {\vec{r}} , \tau ) $ defined in Eq.\ref{eq:Jdef}.
By iteration of the non-linear integral equation \ref{eq:Psiint} 
it is easy to obtain 
an expansion of the functional $\Psi^{\alpha}_q$ in powers of 
the Hubbard-Stratonovich field $\phi^{\alpha}_q = \frac{ \beta}{\I} V^{\alpha}_q$,
 \begin{equation}
 \Psi^{\alpha}_q = \sum_{n=1}^{\infty} \Psi^{\alpha}_{n,q} 
 \label{eq:Psinexpansion}
 \; \; \; ,
 \end{equation}
where for $q \neq 0$ the  functional $\Psi^{\alpha}_{n ,q }$ is of the form
 \begin{equation}
 \Psi^{\alpha}_{n,q} = 
 \sum_{ q_1 \ldots q_n} \delta_{q , q_1 + \ldots + q_n}
 \tilde{U}^{\alpha}_n ( q_1 \ldots  q_n )
 \phi^{\alpha}_{q_1} \cdots \phi^{\alpha}_{q_n}
 \label{eq:Psinexp}
 \; \; \; .
 \end{equation}
For practical calculations beyond the Gaussian approximation it is useful
to have a diagrammatic representation of Eq.\ref{eq:eikonalpara},
which is defined in Fig.~\secref{fig:eik}.
The dimensionless vertices $\tilde{U}^{\alpha}_n$ 
are proportional to $ (1/{m}^{\alpha})^{n-1}$.
The first three vertices are
 \begin{equation}
 \tilde{U}^{\alpha}_1 ( q_1 )  =   1
 \; \; \; , \; \; \; 
 \tilde{U}^{\alpha}_2 ( q_1 q_2 )  =   \gamma^{\alpha}_{  q_1 , q_2 } 
 \label{eq:C2vertex}
 \; \; \; ,
 \end{equation}
 \begin{equation}
 \hspace{-3mm}
 \tilde{U}^{\alpha}_3 ( q_1 q_2 q_3 )  =   
 \frac{2}{3} \left[
 \gamma^{\alpha}_{  q_1 , q_2 } 
 \gamma^{\alpha}_{  q_1 + q_2 , q_3 } 
 +
 \gamma^{\alpha}_{  q_2 , q_3 } 
 \gamma^{\alpha}_{  q_2 + q_3 , q_1 } 
 +
 \gamma^{\alpha}_{  q_3 , q_1 } 
 \gamma^{\alpha}_{  q_3 + q_1 , q_2 } 
 \right]
 \label{eq:C3vertex}
 \; .
 \end{equation}
We have used the invariance of Eq.\ref{eq:Psinexp}
under relabeling of the fields
to symmetrize the vertices\index{symmetrization} with respect to the interchange
of any two labels.
Substituting Eqs.\ref{eq:Psinexpansion} and \ref{eq:Psinexp}
into Eq.\ref{eq:eikonalpara}, we obtain the desired expansion of the
eikonal in powers of the Hubbard-Stratonovich field $\phi^{\alpha}$.
Note that each iteration involves an additional power of
$\phi^{\alpha} / m^{\alpha}$.
Because the Gaussian propagator of $\phi^{\alpha}$-field
is proportional to the RPA interaction 
(see Eq.\ref{eq:phiphiprop}), the small parameter
controlling this expansion is proportional to
$f^{{\rm RPA} , \alpha}_q / {m^{\alpha}}$.
This will become more evident 
in Sect.~\secref{sec:NGav}, where
we explicitly calculate the leading
corrections to the Gaussian approximation for the average eikonal.

\begin{center}
{\bf{The Dyson equation  for the prefactor Green's function}}
\end{center}

\begin{figure}
\sidecaption
\psfig{figure=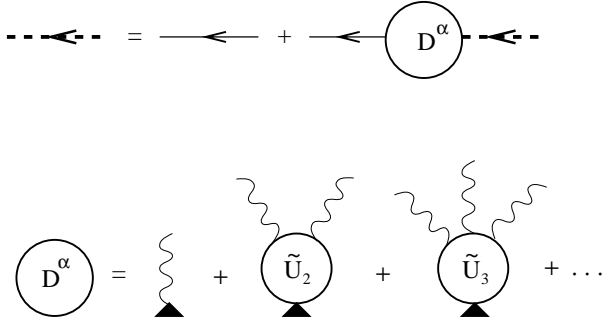,width=8cm}
\caption[
Diagrammatic representation of the Dyson equation for 
$\hat{G}^{\alpha}_1$.]
{ 
Diagrammatic representation of the Dyson equation for 
$\hat{G}^{\alpha}_1$, which is represented by a thick dashed line.
The solid triangle denotes the vertex 
 $\lambda^{\alpha}_{ \tilde{q} , \tilde{q}^{\prime} } $ defined in
Eq.\ref{eq:lambdavertex}.} 
\label{fig:Dysonderiv}
\end{figure}

\noindent
Having solved Eq.\ref{eq:Psiint} to a certain order in
$\phi^{\alpha}$, we know also the random velocity
${\vec{u}}^{\alpha}({  {\vec{r}} , \tau })$
in Eq.\ref{eq:urandef} (and hence the derivative potential
${\vec{u}}^{\alpha}({  {\vec{r}} , \tau }) \cdot {\vec{P}}_{\vec{r}}$
in Eq.\ref{eq:difG1}) to the same order in $\phi^{\alpha}$.
For practical calculations we find it again more convenient to
work in Fourier space. Defining the Fourier transform
$[ \hat{G}^{\alpha}_1 ]_{\tilde{q} \tilde{q}^{\prime} }$ 
of ${\cal{G}}^{\alpha}_1 ( \vec{r} , \vec{r}^{\prime} , \tau , \tau^{\prime} )$
as in Eq.\ref{eq:calGdef}, it is easy to see that 
in Fourier space Eq.\ref{eq:difG1} is equivalent
with the Dyson equation\index{Dyson equation!derivative potential}
 \begin{equation}
[ \hat{G}^{\alpha}_1 ]_{\tilde{q} \tilde{q}^{\prime} }
  =
  \delta_{ \tilde{q}  \tilde{q}^{\prime} }  
  G^{\alpha}_0 ( \tilde{q} )
  +
  G^{\alpha}_0 ( \tilde{q} )
  \sum_{\tilde{q}^{\prime \prime} }
 [ \hat{D}^{\alpha}]_{  \tilde{q}  \tilde{q}^{\prime \prime} } 
[ \hat{G}^{\alpha}_1 ]_{\tilde{q}^{\prime \prime} \tilde{q}^{\prime} }
 \; \; \; ,
 \label{eq:Dsoltotal}
 \end{equation}
where the matrix elements of the derivative potential are
 \begin{equation}
 [ \hat{D}^{\alpha}]_{ \tilde{q}  \tilde{q}^{\prime}  }
 =  
 \Psi^{\alpha}_{ \tilde{q} - \tilde{q}^{\prime} }
 \lambda^{\alpha}_{ \tilde{q} , \tilde{q}^{\prime} }
 \label{eq:Dmatdef}
 \; \; \; .
 \end{equation}
Here $\Psi^{\alpha}_{q}$ is defined as functional of the
$\phi^{\alpha}$-field via the non-linear integral 
equation \ref{eq:Psiint}, 
and the vertex
$ \lambda^{\alpha}_{ \tilde{q} , \tilde{q}^{\prime} } $ is given by
 \begin{equation}
 \lambda^{\alpha}_{ \tilde{q} , \tilde{q}^{\prime} } 
 = \frac{\I}{\beta}
 ( {\vec{q}} - {\vec{q}}^{\prime} )  
 ( {\tens{M}}^{\alpha} )^{-1} 
 {\vec{q}}^{\prime} 
 G_{\rm b}^{\alpha} ( \tilde{q} - \tilde{q}^{\prime} ) 
  \label{eq:lambdavertex}
  \; \; \; .
  \end{equation}
Iteration of Eq.\ref{eq:Dsoltotal} generates an expansion of
${\hat{G}}^{\alpha}_1$ in powers of the  derivative potential.
A graphical representation of Eq.\ref{eq:Dsoltotal} is shown in 
Fig.~\secref{fig:Dysonderiv}.

\subsection{Non-Gaussian averaging} 
\label{sec:NGav}

{\it{
We now average the Green's function
${\cal{G}}^{\alpha} ( {\vec{r}} , {\vec{r}}^{\prime} , \tau , \tau^{\prime} )$
with respect to the probability distribution of the 
background field.  }}

\vspace{7mm}
\noindent
To obtain the Green's function
of the many-body system, we need to
average the Green's function
 ${\cal{G}}^{\alpha} ( {\vec{r}} , {\vec{r}}^{\prime} , \tau , \tau^{\prime} )$
given in Eq.\ref{eq:Ansatz2}
with respect to the 
probability distribution ${\cal{P}} \{ \phi^{\alpha} \}$ defined in
Eq.\ref{eq:probabphidef}. Because for finite masses $m^{\alpha}_i$ the
effective  action $S_{\rm eff} \{ \phi^{\alpha} \}$
in Eqs.\ref{eq:Seffphidef} and \ref{eq:Skinphidef}
is not Gaussian, we have to use perturbation theory
to perform the averaging procedure.
Recall that the leading non-Gaussian corrections to
$S_{\rm eff} \{ \phi^{\alpha} \}$ have been explicitly calculated in 
Chap.~\secref{subsec:Explicit}, see Eq.\ref{eq:Skinprimeapprox}.
Because averaging restores translational invariance, we may set
${\vec{r}}^{\prime} = \tau^{\prime} = 0$ and calculate
 \begin{equation}
 {{G}}^{\alpha} ( {\vec{r}} , \tau ) =
 \left< {\cal{G}}^{\alpha} ( {\vec{r}} , 0 , \tau , 0 )
 \right>_{S_{\rm eff}}
 \; \; \; .
 \label{eq:Gtotalaverage}
 \end{equation}
Let us parameterize the average Green's function as
 \begin{equation}
 {{G}}^{\alpha} ( {\vec{r}} , \tau ) 
 = 
 [ {G}^{\alpha}_1 ( {\vec{r}} , \tau )  +
 {G}^{\alpha}_2 ( {\vec{r}} , \tau ) ] \E^{Q^{\alpha} ( {\vec{r}} , \tau )  } 
 \label{eq:Gtotalavparametrize}
 \; \; \; ,
 \end{equation}
where 
 \begin{equation}
 Q^{\alpha} ( {\vec{r}} , \tau )   
 = \ln 
 \left< 
 \E^{ \Phi^{\alpha} ( {\vec{r}} , \tau ) - \Phi^{\alpha} (0,0) } 
 \right>_{S_{\rm eff}}
 \label{eq:Qeikdef}
 \; \; \; ,
 \end{equation}
 \begin{equation}
  {G}^{\alpha}_1 ( {\vec{r}} , \tau )  
 = \left< 
  {\cal{G}}^{\alpha}_1 ( {\vec{r}} , 0 , \tau , 0) \right>_{S_{\rm eff}}
  \label{eq:G1avdef}
  \; \; \; ,
  \end{equation}
and the function
 ${G}^{\alpha}_2 ( {\vec{r}} , \tau )  $ 
contains all correlations between the two factors in Eq.\ref{eq:Ansatz2},
 \begin{equation}
 {G}^{\alpha}_2 ( {\vec{r}} , \tau )  
 = \frac{ \left<
  \delta  {\cal{G}}^{\alpha}_1 ( {\vec{r}} , 0 , \tau , 0) 
 \delta \E^{ \Phi^{\alpha} ( {\vec{r}} , \tau ) - \Phi^{\alpha} (0,0) } 
  \right>_{S_{\rm eff}} }
 {\left< 
 \E^{ \Phi^{\alpha} ( {\vec{r}} , \tau ) - \Phi^{\alpha} (0,0) } 
 \right>_{S_{\rm eff}}
 } 
 \; \; \; .
 \label{eq:G2def}
 \end{equation}
Here $\delta  X  = X - <X>_{S_{\rm eff}}$.
We emphasize that Eqs.\ref{eq:Gtotalavparametrize}--\ref{eq:G2def}
are an exact decomposition of the different contributions
to Eq.\ref{eq:Gtotalaverage}, the usefulness of which will become
evident shortly.

\vspace{7mm}

\begin{figure}
\sidecaption
\psfig{figure=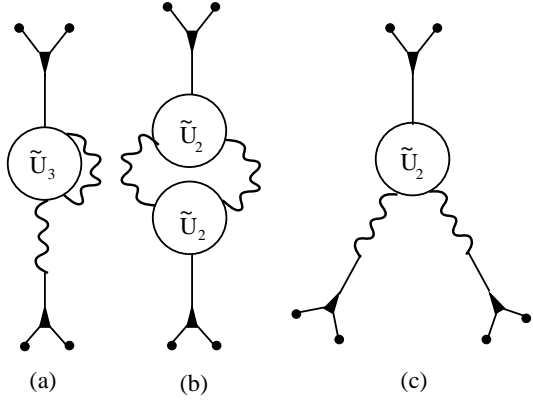,width=7cm}
\caption[
The leading non-Gaussian corrections to the average eikonal.]
{The leading non-Gaussian corrections to the average eikonal. 
The thick wavy line is the Gaussian propagator of the
$\phi^{\alpha}$-field, i.e. the RPA screened interaction.
The other symbols are defined in Fig.~\secref{fig:eik}.
The diagrams (a) and (b) 
contain two factors of ${\cal{J}}^{\alpha}$ and
hence contribute to $W_{1,2}^{\alpha}$. 
Diagrams (a) and (b) represent the first and second term in Eq.\ref{eq:W12explicit}.
Diagram (c) contains three factors of ${\cal {J}}^{\alpha}$ and
is the only contribution to 
$W_{2,2}^{\alpha}$, see Eq.\ref{eq:W22explicit}.}
\label{fig:rpaquad}
\end{figure}

Let us now consider in some detail the calculation of the function
$Q^{\alpha} ( {\vec{r}} , \tau )$ defined in Eq.\ref{eq:Qeikdef}.
By definition we have\footnote{Note
that the label $\alpha$ on the right-hand side of Eq.\ref{eq:Qeiklong} is an
external label which is not summed over.
The summation labels are denoted by $\alpha^{\prime}$.}
 \begin{eqnarray}
 Q^{\alpha} ( {\vec{r}} , \tau )
 & = &
 \nonumber \\
 & & \hspace{-20mm}
 \ln \left( 
 \frac{ \int {\cal{D}} \{ \phi^{\alpha^{\prime}} \}
 \exp \left[ - S_{\rm eff} \{ \phi^{\alpha^{\prime}} \} +
 \sum_q {\cal{J}}^{\alpha}_{-q} ( {\vec{r}} , \tau ) \phi^{\alpha}_q
 +
 {\cal{F}}^{\alpha} \{ {\cal{J}}^{\alpha} , \phi^{\alpha} \} \right] }
 {
 \int {\cal{D}} \{ \phi^{\alpha^{\prime}} \}
 \exp \left[ - S_{\rm eff} \{ \phi^{\alpha^{\prime}} \} 
 \right] }
 \right)
 \; \; ,
 \nonumber
 \\
 & &
 \label{eq:Qeiklong}
 \end{eqnarray}
where
 ${\cal{J}}^{\alpha}_{-q} ( {\vec{r}} , \tau ) 
 \equiv
 {\cal{J}}^{\alpha}_{-q} ( {\vec{r}} ,0 , \tau , 0) $, and 
the functional
 ${\cal{F}}^{\alpha} \{ {\cal{J}}^{\alpha} , \phi^{\alpha} \} $ is defined
as the sum of all terms on the right-hand side of Eq.\ref{eq:eikonalpara} 
involving more than one power of the $\phi^{\alpha}$-field.
Explicitly, the first two terms are
 \begin{eqnarray}
 {\cal{F}}^{\alpha}\{  {\cal{J}}^{\alpha} , \phi^{\alpha} \}
 & = & 
 \sum_{q,q_1, q_2} {\cal{J}}^{\alpha}_{-q} ( {\vec{r}} , \tau )
 \delta_{q , q_1 + q_2} \tilde{U}^{\alpha}_2 ( q_1 q_2 ) \phi^{\alpha}_{q_1}
 \phi^{\alpha}_{q_2}
 \nonumber
 \\
&  & \hspace{-20mm} +
 \sum_{q,q_1, q_2, q_3} {\cal{J}}^{\alpha}_{-q} ( {\vec{r}} , \tau )
 \delta_{q , q_1 + q_2 + q_3 } \tilde{U}^{\alpha}_3 ( q_1 q_2 q_3) \phi^{\alpha}_{q_1}
 \phi^{\alpha}_{q_2}
 \phi^{\alpha}_{q_3} 
  +  \ldots
 \; .
 \label{eq:Ertphi}
 \end{eqnarray}
These terms correspond precisely to the diagrams with two and three
wavy lines in Fig.~\secref{fig:eik}.
Following the procedure
outlined in Chap.~\secref{sec:beyond}, we write
 \begin{equation}
 S_{\rm eff} \{ \phi^{\alpha^{\prime}} \} = \I \sum_{\alpha^{\prime}} 
 \phi^{\alpha^{\prime}}_0 N^{\alpha^{\prime}}_0 +
 S_{{\rm eff},2} \{ \phi^{\alpha^{\prime}} \}  
 + S_{\rm kin}^{\prime} \{ \phi^{\alpha^{\prime}} \}  
 \label{eq:Seffseparate}
 \; \; \; ,
 \end{equation}
where the Gaussian part $ S_{{\rm eff},2} \{ \phi^{\alpha^{\prime}} \} $ 
of the effective action is given in Eq.\ref{eq:Seff2phigaussres}, and
the non-Gaussian part 
 $S_{\rm kin}^{\prime} \{ \phi^{\alpha^{\prime}} \}  $ is defined in Eq.\ref{eq:Skin3}.
The leading  contributions to 
 $S_{\rm kin}^{\prime} \{ \phi^{\alpha^{\prime}} \}$
are explicitly given in
Eq.\ref{eq:Skinprimeapprox}.
After eliminating the second term in the numerator of Eq.\ref{eq:Qeiklong}
by means of the shift-transformation 
 \begin{equation}
 \phi^{\alpha^{\prime}}_q \rightarrow 
 \phi^{\alpha^{\prime}}_q  + [ \underline{\tilde{f}}^{\rm RPA}_q ]^{\alpha^{\prime} \alpha}
 {\cal{J}}^{\alpha}_{q} ( {\vec{r}} , \tau )
 \label{eq:shifteik}
 \; \; \; ,
 \end{equation}
where $\underline{\tilde{f}}_q^{\rm RPA} \equiv \frac{\beta}{V}
\underline{f}_q^{\rm RPA}$ is the rescaled RPA interaction 
matrix (see Eq.\ref{eq:frpapatchdef}), we obtain 
 \begin{eqnarray}
 Q^{\alpha} ( {\vec{r}} , \tau )
 & = & \frac{1}{2} \sum_q \tilde{f}^{{\rm RPA} , \alpha}_q 
 {\cal{J}}^{\alpha}_{-q} ( {\vec{r}} , \tau )
 {\cal{J}}^{\alpha}_{q} ( {\vec{r}} , \tau )
 \nonumber
 \\
 &  & \hspace{-23mm} + \ln \left< \exp \left[
 - S_{\rm kin}^{\prime} \{ 
 \phi^{\alpha^{\prime}}_q + 
 [\underline{\tilde{f}}^{\rm RPA}_q]^{\alpha^{\prime} \alpha} 
 {\cal{J}}^{\alpha}_q  \} +
 {\cal{F}}^{\alpha} 
 \{ {\cal{J}}^{\alpha}_q , \phi^{\alpha}_q + 
 \tilde{f}^{{\rm RPA} ,\alpha}_q 
 {\cal{J}}^{\alpha}_q  
 \}
 \right] \right>_{S_{{\rm eff},2}}
 \nonumber
 \\
 &  & \hspace{-23mm} - \ln \left< \exp \left[
 - S_{\rm kin}^{\prime} \{ 
 \phi^{\alpha^{\prime}}_q  \}
 \right] \right>_{S_{{\rm eff},2}}
 \; \; \; .
 \label{eq:eikshift}
 \end{eqnarray}
We have used the notation
 $\tilde{f}^{{\rm RPA} ,\alpha}_q = 
 [\underline{\tilde{f}}^{\rm RPA}_q]^{\alpha \alpha} 
 = \frac{\beta}{V}
 [\underline{{f}}^{\rm RPA}_q]^{\alpha \alpha} $, 
see also Eq.\ref{eq:frpadiagonaldef}.
Finally, we use the linked cluster theorem \cite{Mahan81}\index{linked cluster theorem} 
and obtain, in complete analogy with
Eq.\ref{eq:generalpert1},
 \begin{eqnarray}
 Q^{\alpha} ( {\vec{r}} , \tau )
 & = & \frac{1}{2} \sum_q \tilde{f}^{{\rm RPA} , \alpha}_q 
 {\cal{J}}^{\alpha}_{-q} ( {\vec{r}} , \tau )
 {\cal{J}}^{\alpha}_{q} ( {\vec{r}} , \tau )
 + \sum_{n=1}^{\infty} \frac{ (-1)^{n}}{n} 
 \nonumber
 \\
 &  & \hspace{-21mm}  
 \times \left\{ \left< \left[
  S_{\rm kin}^{\prime} \{ 
 \phi^{\alpha^{\prime}}_q + 
 [\underline{\tilde{f}}^{\rm RPA}_q]^{\alpha^{\prime} \alpha} 
 {\cal{J}}^{\alpha}_q  \} -
 {\cal{F}}^{\alpha} 
 \{ {\cal{J}}^{\alpha}_q , \phi^{\alpha}_q + 
 \tilde{f}^{{\rm RPA} ,\alpha}_q 
 {\cal{J}}^{\alpha}_q  
 \}
 \right]^{n} \right>_{S_{{\rm eff},2}}^{\rm con}
 \right.
 \nonumber
 \\
 &  & \hspace{-16mm}  
 \left. -
 \left< \left[
  S_{\rm kin}^{\prime} \{ 
 \phi^{\alpha^{\prime}}_q  \}
 \right]^{n} \right>_{S_{{\rm eff},2}}^{\rm con}
 \right\}
 \; \; \; .
 \label{eq:eiklinked}
 \end{eqnarray}
From this expression it is obvious that in general 
the function 
$Q^{\alpha} ( {\vec{r}} , \tau )  $ can be written as  
 \begin{equation}
 Q^{\alpha} ( {\vec{r}} , \tau )   =
\sum_{n=1}^{\infty} 
Q^{\alpha}_n ( {\vec{r}} , \tau )
\label{eq:Qalphandef}
\; \; \; ,
\end{equation}
where 
$Q^{\alpha}_n ( {\vec{r}} , \tau )$ involves $n+1$ powers of the
function ${\cal{J}}^{\alpha}_q ( {\vec{r}}, \tau )$,
 \begin{eqnarray}
 Q^{\alpha}_n ( {\vec{r}} , \tau  )  & = &
 \sum_{q q_1 \ldots q_n} \delta_{q , q_1 + \ldots + q_n}
 W_n^{\alpha} (  q_1 \ldots q_n ) 
 \nonumber
 \\
 & \times &
 {\cal{J}}^{\alpha}_{-q}  ( {\vec{r}} ,  \tau  )
 {\cal{J}}^{\alpha}_{q_1}  ( {\vec{r}} ,  \tau  )
 \cdots
 {\cal{J}}^{\alpha}_{q_n}  ( {\vec{r}} , \tau  )
 \; \; \; .
 \label{eq:Qngeneral}
 \end{eqnarray}
The vertices 
 $W_n^{\alpha} (  q_1 \ldots q_n ) $
can be calculated perturbatively in powers of 
the RPA interaction.
Evidently the first term in Eq.\ref{eq:eiklinked}
corresponds to the following contribution to
$W_1^{\alpha}$, 
 \begin{equation}
 W_{1,1}^{\alpha} ( q_1 ) = \frac{1}{2}  \tilde{f}^{{\rm RPA} , \alpha}_{q_1}
 \label{eq:W11def}
 \; \; \; .
 \end{equation}
The crucial point is now that all other terms contain at least
two powers of ${f}^{{\rm RPA}, \alpha}$, so that, 
to first order in the RPA interaction, 
higher order terms can be neglected. 
In general, each vertex can be expanded as
 \begin{equation}
 W_n^{\alpha} (  q_1 \ldots q_n ) 
 =
 \sum_{m=n}^{\infty}
 W_{n,m}^{\alpha} (  q_1 \ldots q_n ) 
 \label{eq:WnmRPAexp}
 \; \; \; ,
 \end{equation}
where the second subscript gives the power of 
the RPA interaction. Because the $n^{\rm th}$-order
vertex $W_n^{\alpha}$ involves at least $n$ powers of the RPA interaction,
the $m$-sum in Eq.\ref{eq:WnmRPAexp} starts at $m=n$.
This is due to the fact that each of the higher-order diagrams
in Fig.~\secref{fig:eik} contains a single function
${\cal{J}}^{\alpha}$, but additional powers of the $\phi^{\alpha}$-field.
Actually, from 
Chap.~\secref{subsec:hidden} we expect that the true  small
parameter which controls the corrections to the Gaussian approximation
should also involve the local curvature of the
Fermi surface (i.e. the inverse effective masses $m^{\alpha}_i)$ and the
range of the interaction in momentum space.
To investigate this point,
it is convenient to visualize the
structure of the perturbation expansion for the higher-order 
terms with the help of the
graphical elements introduced in Fig.~\secref{fig:eik}.
To order $( {f}^{{\rm RPA} , \alpha} )^2$, we should retain 
 \begin{eqnarray}
 W_{1}^{\alpha} ( q_1 ) & \approx & W_{1,1}^{\alpha} ( q_1 ) + 
 W_{1,2}^{\alpha} ( q_1 )  
 \label{eq:W12retain}
 \; \; \; ,
 \\
 W_{2}^{\alpha} ( q_1 , q_2 ) & \approx & W_{2,2}^{\alpha} ( q_1 , q_2 )  
 \label{eq:W22retain}
 \; \; \; ,
 \end{eqnarray}
and neglect all $W^{\alpha}_n$ with $n \geq 3$.  
The diagrams contributing to 
$W_{1,2}^{\alpha} ( q_1 ) $ and
$W_{2,2}^{\alpha} ( q_1 , q_2 ) $ 
are shown in Fig.~\secref{fig:rpaquad}.
The explicit expressions are
 \begin{eqnarray}
 W_{1,2}^{\alpha} ( q_1 ) & = &
 3 \tilde{f}^{{\rm RPA} , \alpha}_{q_1}
 \sum_{q_2} \tilde{U}_3^{\alpha} ( q_1 , q_2 , -q_2 )
 \tilde{f}^{{\rm RPA} , \alpha}_{q_2}
 \nonumber
 \\
 &  & \hspace{-7mm} +
 \sum_{q_2} 
 \tilde{U}_2^{\alpha} ( q_2 , q_1 - q_2   )
  \tilde{U}_2^{\alpha} ( -q_2 , q_2 - q_1  )
 \tilde{f}^{{\rm RPA} , \alpha}_{q_1 - q_2}
 \tilde{f}^{{\rm RPA} , \alpha}_{q_2}
 \label{eq:W12explicit}
 \; \; \; ,
 \\
 W_{2,2}^{\alpha} ( q_1 , q_2 ) & = &
 \tilde{U}_2^{\alpha} ( q_1 , q_2 )
 \tilde{f}^{{\rm RPA} , \alpha}_{q_1}
 \tilde{f}^{{\rm RPA} , \alpha}_{q_2}
 \label{eq:W22explicit}
 \; \; \; .
 \end{eqnarray}
Because by construction $\tilde{U}^{\alpha}_n$ is proportional to
$ ( 1 / m^{\alpha} )^{n-1}$, it is clear that
$W_{1,2} \propto ( f^{{\rm RPA} , \alpha} / m^{\alpha} )^2$, while
$W_{2,2} \propto ( f^{{\rm RPA} , \alpha} )^2/ m^{\alpha} $. 
Thus, the corrections to the first order term in the average eikonal
are not only controlled by higher powers of the RPA interaction, 
but also by higher powers of the inverse effective mass $1/m^{\alpha}$.
Note that $1 / m^{\alpha}$ is a measure\footnote{
The relevant {\it{dimensionless}} parameter $C^{\alpha}$ which measures the local
curvature of the Fermi surface
has been identified in Chap.~\secref{subsec:hidden} (see Eqs.\ref{eq:g1def} 
and \ref{eq:Ccurveresult}).}
for the local curvature of the Fermi surface close 
to $\vec{k}^{\alpha}$.
Moreover, for interactions that are sufficiently well behaved for 
${\vec{q}} \rightarrow 0$ and
have a natural cutoff $q_{\rm c} \ll k_{\rm F}$ in momentum
space, each additional loop integration in Eq.\ref{eq:Qngeneral}
gives rise to a factor of $( q_{\rm c} / k_{\rm F} )^{d}$.
We therefore conclude that the leading correction to the
Gaussian approximation for the average eikonal is controlled
by the same dimensionless small parameter that appears in the
calculation of the non-Gaussian correction
to the density-density correlation function, see Eq.\ref{eq:gausscorrect}.

\begin{figure}
\sidecaption
\psfig{figure=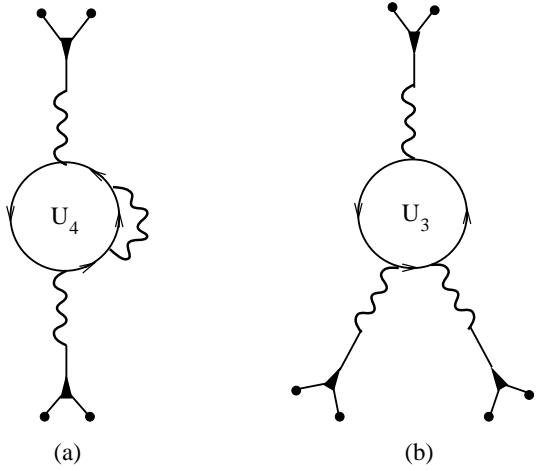,width=7cm}
\caption[
Lowest order corrections to the average eikonal due to the
non-Gaussian terms of the probability distribution.]
{ 
\begin{sloppypar}
Lowest order corrections to the average eikonal due to the
non-Gaussian terms of the probability distribution.
Both diagrams represent corrections of order
$(f^{{\rm RPA} , \alpha })^3$.
The diagram (a) involves the vertex $U_4$ 
(see Eq.\ref{eq:Skinprimeapprox}) and
renormalizes the RPA interaction
in Eq.\ref{eq:W11def}. 
The diagram (b) involves $U_3$ and leads to a renormalization of the
vertex $\tilde{U}^{\alpha}_2$ in
Eq.\ref{eq:W22explicit}.
\end{sloppypar}
}
\label{fig:nongauss}
\end{figure}

The analysis of next-to-leading terms is rather complicated.
Clearly, all contributions to the
functions $W^{\alpha}_{n,n}$ that involve only the vertices
$\tilde{U}^{\alpha}_n$ are at tree-level\index{tree-approximation} proportional
to $ (f^{{\rm RPA}, \alpha})^{n} / (m^{\alpha})^{n-1}$. 
This is due to the fact that by construction the
vertex $\tilde{U}^{\alpha}_n$ is proportional
to $1 / ( m^{\alpha} )^{n-1}$.
However, at order $(f^{{\rm RPA} , \alpha})^3$ corrections due to the
non-Gaussian part $S^{\prime}_{\rm kin} \{ \phi^{\alpha} \}$ 
of the effective action for the $\phi^{\alpha}$-field
must also be taken into account.  
These involve the vertices $U_n$  defined in Eq.\ref{eq:Uvertex}.
The leading  contributions
of this type are shown in Fig.~\secref{fig:nongauss}.
Certainly, a subset of these diagrams leads to the replacement
of the Gaussian propagator 
$ < \phi^{\alpha}_q \phi^{\alpha}_{-q} >_{S_{{\rm eff},2}} = 
\tilde{f}^{{\rm RPA} ,\alpha}_q$ 
by the exact effective propagator
$ < \phi^{\alpha}_q \phi^{\alpha}_{-q} >_{S_{\rm eff}}$,
which depends on the exact dielectric function of the many-body system.
However, the non-Gaussian part
$S^{\prime}_{\rm kin} \{ \phi^{\alpha} \}$ 
of our effective action
will also give rise to more complicated vertex corrections.

Although the vertices $U_n$ do not explicitly contain the 
curvature parameter $1 / m^{\alpha}$, 
the closed loop theorem discussed in Chap.~\secref{sec:closedloop}
implies that in the infrared limit the closed fermion loops in
Fig.~\secref{fig:nongauss} lead to large-scale cancellations, so that
we expect that also these higher order terms are proportional to
powers of the inverse effective masses. 
Fortunately, the diagrams in Fig.~\secref{fig:nongauss} are of order
$(f^{{\rm RPA }, \alpha})^3$ and therefore do not contribute to the
leading correction to the Gaussian approximation.

Finally we would like to point out that 
for models with spin the non-Gaussian corrections to the
average eikonal lead to a mixing  of the density fields with spin fields.
This implies that in the one-dimensional Tomonaga-Luttinger model
the non-linear terms in the energy dispersion
destroy the spin-charge separation \cite{Matveenko94}.
\index{spin-charge separation}

\section[The Gaussian approximation with non-linear energy dispersion]
{The Gaussian approximation \mbox{\hspace{40mm}}
with non-linear energy dispersion}
\label{sec:Gaussquad}

{\it{We now perform all averaging operations
to first order in the RPA interaction. 
We emphasize again that we do not expand in powers
of $1 / m^{\alpha}$, so that curvature effects
are taken into account non-perturbatively.}}

\subsection{The average eikonal}

\noindent
From now on we shall restrict ourselves to the first order in
the RPA interaction. Then it is sufficient to retain only
the term $Q^{\alpha}_1 ( {\vec{r}} , \tau )$ 
in Eq.\ref{eq:Qalphandef}, and approximate the vertex
$W_1^{\alpha} ( q_1 )$ by Eq.\ref{eq:W11def}.
Using the definition of ${\cal{J}}^{\alpha}_{-q} ( {\vec{r}}, \tau )$  
in Eq.\ref{eq:tildeJdef},
we have
 \begin{eqnarray}
 {\cal{J}}^{\alpha}_{-q} ( {\vec{r}}, \tau )
 {\cal{J}}^{\alpha}_{q} ( {\vec{r}}, \tau )
 & = & - \frac{2}{\beta^2} G_{\rm b}^{\alpha} ( -q ) 
 G_{\rm b}^{\alpha} ( q )
 \left[ 1 - \cos \left(  {\vec{q}} \cdot 
 {\vec{r}} - \omega_m \tau \right) \right]
 \nonumber
 \\
 & = & 
 \frac{2}{\beta^2}
 \frac{ 1 - \cos \left(  {\vec{q}} \cdot {\vec{r}} - \omega_m \tau \right) }
 { [ \I \omega_m - \xi^{\alpha}_{  {\vec{q}} } ]
 [ \I \omega_m + \xi^{\alpha}_{  - {\vec{q}}} ] }
 \label{eq:JJnonlin}
 \; \; \; ,
 \end{eqnarray}
so that
we obtain, in complete analogy with Eqs.\ref{eq:Qlondef}--\ref{eq:Slondef},
\index{Debye-Waller factor!non-linear energy dispersion}
 \begin{equation}
 Q_1^{\alpha} ( {\vec{r}}  , \tau ) = R^{\alpha}_1 -
 S_1^{\alpha} ( {\vec{r}}  , \tau ) 
 \; \; \;  ,
 \label{eq:Qlondef2}
 \end{equation}
with
 \begin{eqnarray}
 R_1^{\alpha} 
 & = & \frac{1}{\beta {{V}}} \sum_q
 \frac{ f^{{\rm RPA},\alpha}_q
   }
{ [ \I \omega_m - \xi^{\alpha}_{  {\vec{q}} } ][ \I \omega_m + 
\xi^{\alpha}_{  - {\vec{q}}} ] }
=
 S_1^{\alpha} ( 0  , 0 )
 \label{eq:Rlondef2}
 \; \; \; ,
 \\
 S_1^{\alpha} ( {\vec{r}}  , \tau )
 & = & \frac{1}{\beta {{V}}} \sum_q
 \frac{ f^{{\rm RPA},\alpha}_q
  \cos ( {\vec{q}} \cdot {\vec{r}} - \omega_m \tau )  }
{ [ \I \omega_m - \xi^{\alpha}_{  {\vec{q}} } ][ \I \omega_m + 
\xi^{\alpha}_{ - {\vec{q}}} ] }
 \label{eq:Slondef2}
 \; \; \; .
 \end{eqnarray}
Note that these expressions
contain the non-linear terms in the energy dispersion via 
$\xi^{\alpha}_{ {\vec{q}} }$ {\it{and}} 
$f^{{\rm RPA},\alpha}_q$ in a non-perturbative way.
Moreover, all problems due to the double pole\index{double pole}
in the corresponding expressions for linearized energy dispersion
(see the discussion in the introduction to Sect.~\secref{sec:eik})
have disappeared 
in Eqs.\ref{eq:Qlondef2}--\ref{eq:Slondef2}
in an almost trivial way, because
 \begin{equation}
 \xi^{\alpha}_{  - {\vec{q}} } =
 - \xi^{\alpha}_{  {\vec{q}} } +  {\vec{q}}   ( {\tens{M}}^{\alpha})^{-1} {\vec{q}} 
 \label{eq:ximinus}
 \; \; \; ,
 \end{equation}
so that
 \begin{equation}
 \hspace{-3mm}
 \frac{1}{
 [ \I \omega_m - \xi^{\alpha}_{  {\vec{q}} } ][ \I \omega_m + 
 \xi^{\alpha}_{ - {\vec{q}}} ] }
 = \frac{1}{
  {\vec{q}}  ( {\tens{M}}^{\alpha})^{-1} {\vec{q}} }
  \left[ 
  \frac{1}{
  \I \omega_m - \xi^{\alpha}_{  {\vec{q}} } } -
  \frac{1}{
  \I \omega_m + \xi^{\alpha}_{ - {\vec{q}} } } \right]
  \label{eq:partialfracmass}
  \; \; \; .
  \end{equation}
Hence, 
as long as at least one of the inverse effective masses $1/m^{\alpha}_i$ is
finite, the denominator in
Eqs.\ref{eq:Rlondef2} and \ref{eq:Slondef2}
gives only rise to {\it{simple poles}} in the complex frequency plane.
In fact, as will be discussed in more detail
in Chaps.~\secref{subsec:Thetermslin} and \secref{sec:opendis},
the double pole that appears in the Debye-Waller factor
for linearized energy dispersion
gives rise to some rather
peculiar and probably unphysical features in the
analytic structure of the Green's function
in Fourier space.

\subsection{The prefactor Green's functions\index{Green's function!prefactor}}
\label{subsec:disorder}

{\it{We use the impurity diagram technique
to calculate the leading non-trivial contributions to
the Green's functions $G^{\alpha}_1$ and $G^{\alpha}_2$
defined in Eqs.\ref{eq:G1avdef} and \ref{eq:G2def}.}}

\newpage 
\begin{center}
{\bf{Calculation of $G^{\alpha}_1$}}
\end{center}

\begin{figure}
\sidecaption
\psfig{figure=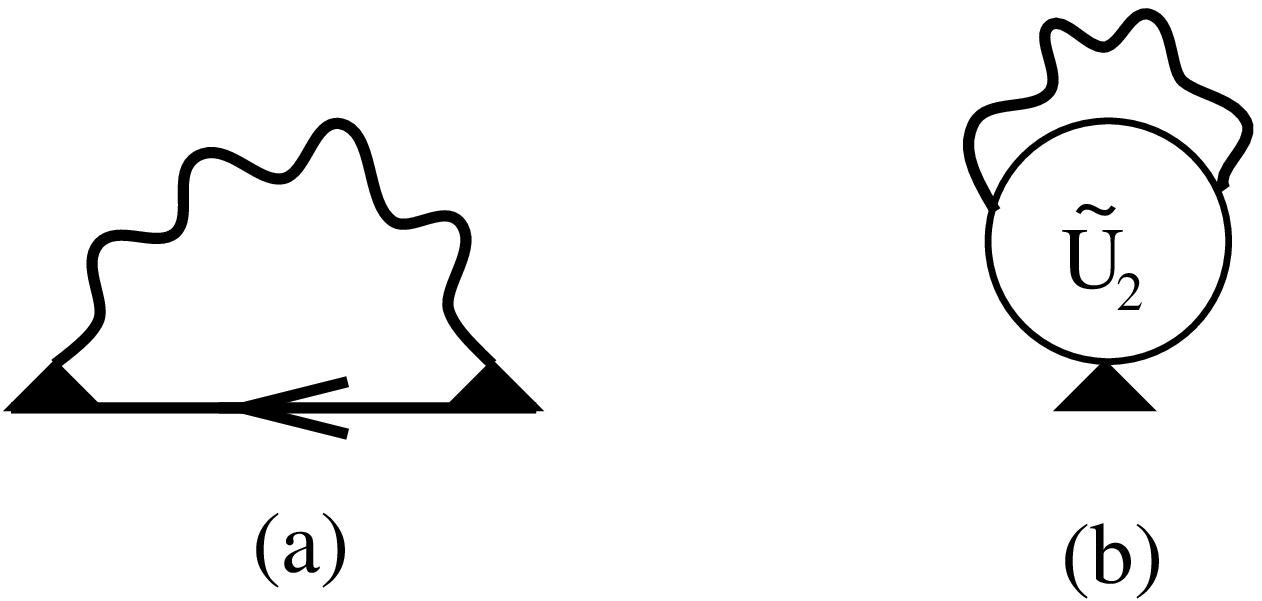,width=7cm}
\caption[
Self-consistent Born approximation for the
self-energy $\Sigma_1^{\alpha} ( \tilde{q} )$.]
{ 
(a) Self-consistent Born approximation for the
self-energy $\Sigma_1^{\alpha} ( \tilde{q} )$.
The thick solid arrow denotes 
the self-consistent Green's function
$G_{1}^{\alpha} ( \tilde{q} )$.
(b) This contribution to
$\Sigma_1^{\alpha} ( \tilde{q} )$
vanishes.
}
\label{fig:selfconBorn}
\end{figure}

\noindent
Let us first consider 
  ${G}^{\alpha}_1 ( {\vec{r}} , \tau )  
 = \left< 
  {\cal{G}}^{\alpha}_1 ( {\vec{r}} , 0 , \tau , 0) \right>_{S_{\rm eff}}$.
Naively one might try a direct expansion 
of ${G}^{\alpha}_1 ( {\vec{r}} , \tau )  $ in powers of
$f^{{\rm RPA} , \alpha}$.
The terms in this expansion are easily generated by iterating the
Dyson equation \ref{eq:Dsoltotal} and then averaging.
Because for $q \neq 0$ the Gaussian average 
$ < \phi^{\alpha}_q >_{S_{{\rm eff},2}}$ vanishes,
the leading term (of order
$f^{{\rm RPA} , \alpha}$) arises from the second 
iteration of Eq.\ref{eq:Dsoltotal}.
However, as already mentioned in Chap.~\secref{sec:thelimitations},
the direct expansion of a single-particle Green's function 
in powers of the interaction
is usually ill-defined, because a truncation at any finite order
generates unphysical multiple poles in Fourier space.
Within a perturbative approach, this
problem is avoided by calculating the irreducible self-energy
to some finite order in the interaction, and extrapolating the
perturbation series by solving the Dyson equation.
Thus, introducing the Fourier transform of
  ${G}^{\alpha}_1 ( {\vec{r}} , \tau )  $ as usual
(see Eq.\ref{eq:Gdeffer}),
 \begin{equation}
  {G}^{\alpha}_1 ( {\vec{r}} , \tau  )
  = \frac{1}{\beta V} \sum_{\tilde{q}} 
   \E^{ \I ( {\vec{q}} \cdot {\vec{r}} - 
  \tilde{\omega}_n \tau ) }
  G^{\alpha}_1 ( \tilde{q} )
  \label{eq:G1FTdef}
  \; \; \; ,
  \end{equation}
we define the irreducible self-energy $\Sigma^{\alpha}_1 ( \tilde{q} )$ via the
Dyson equation for $G^{\alpha}_1 ( \tilde{q} )$,\index{Dyson equation}
 \begin{equation}
 [{G}^{\alpha}_1 ( \tilde{q} ) ]^{-1} =  
 [{G}^{\alpha}_0 ( \tilde{q} ) ]^{-1} -
 {\Sigma}^{\alpha}_1 ( \tilde{q} ) 
 \label{eq:DysonG1}
 \; \; \; .
 \end{equation}
We now use the self-consistent Born approximation\index{Born approximation}
to calculate the self-energy
 ${\Sigma}^{\alpha}_1 ( \tilde{q} ) $.
This is a standard approximation in the theory of 
disordered systems \cite{Abrikosov63},
which is expected to be accurate if interference terms are negligible.
The corresponding Feynman diagram is shown in Fig.~\secref{fig:selfconBorn} (a), 
and yields
 \begin{equation}
 {\Sigma}^{\alpha}_1 ( \tilde{q} )
 =  - \frac{1}{\beta {{V}}} \sum_{ \tilde{q}^{\prime}}
 \lambda^{\alpha}_{   \tilde{q} ,  \tilde{q}^{\prime} }
 \lambda^{\alpha}_{   \tilde{q}^{\prime} , \tilde{q} }
 f^{{\rm RPA},\alpha}_{\tilde{q} - \tilde{q}^{\prime} }
 G_{1}^{\alpha} ( \tilde{q}^{\prime}   )
 \label{eq:sigma1res}
 \; \; \; ,
 \end{equation}
where the dimensionless vertex  
$\lambda^{\alpha}_{\tilde{q} , \tilde{q}^{\prime}}$
is defined in Eq.\ref{eq:lambdavertex}.
At the first sight it seems that the averaging 
procedure gives also rise to another contribution of order
$f^{{\rm RPA} , \alpha}$ to 
$\Sigma^{\alpha}_1$, which is shown in Fig.~\secref{fig:selfconBorn}(b).
This contribution is generated
by averaging the $\tilde{U}_2^{\alpha}$-term in Fig.~\secref{fig:Dysonderiv},
and physically describes a 
renormalization of the chemical potential. 
However, according to
Eq.\ref{eq:lambdavertex}
the vertex $\lambda^{\alpha}_{ {\tilde{q}} , \tilde{q}^{\prime} }$ 
vanishes for ${\vec{q}} = {\vec{q}}^{\prime}$, implying that
the contribution from diagram (b) in Fig.~\secref{fig:selfconBorn}
vanishes as well.
In the language of many-body theory, the diagram (a) in Fig.~\secref{fig:selfconBorn} is the
self-consistent GW diagram\index{GW approximation} 
for the self-energy associated with $G^{\alpha}_1 ( \tilde{q} )$.
Comparing Eq.\ref{eq:sigma1res} with the
expression \ref{eq:GW} for the
usual GW self-energy 
associated with  the full Green's function $G^{\alpha} ( \tilde{q} )$, we see that
the GW approximation for
$\Sigma_1^{\alpha} ( \tilde{q} )$ involves two additional powers of the
vertex $\lambda^{\alpha}_{   \tilde{q} ,  \tilde{q}^{\prime} }$
defined in Eq.\ref{eq:lambdavertex}.
The crucial point is now that 
{\it{this additional vertex makes the GW self-energy 
associated with  $G^{\alpha}_1$ less infrared singular than the
corresponding GW self-energy of the full Green's function $G^{\alpha}$.}}
To see this more clearly, we substitute
Eq.\ref{eq:lambdavertex} into Eq.\ref{eq:sigma1res} and shift
the summation variable according to $ \tilde{q} - \tilde{q}^{\prime} = - q^{\prime}$.
Then we obtain
 \begin{eqnarray}
 {\Sigma}^{\alpha}_1 ( \tilde{q} )
 & = &
 \frac{1}{\beta {{V}}} \sum_{q^{\prime}}
 \left[ {\vec{q}}^{\prime}  ({\tens{M}}^{\alpha})^{-1} ( {\vec{q}} + {\vec{q}}^{\prime} )
 \right]
 \left[ {\vec{q}}^{\prime}  ({\tens{M}}^{\alpha})^{-1}  {\vec{q}}
 \right]
 \nonumber
 \\
 & \times &
 G_{\rm b}^{\alpha} ( q^{\prime} ) G_{\rm b}^{\alpha} ( - q^{\prime} ) 
 f^{{\rm RPA} , \alpha}_{q^{\prime}} G^{\alpha}_1 ( {\tilde{q}} + q^{\prime} )
 \; \; \; .
 \label{eq:sigma1vertex}
 \end{eqnarray}
Using the symmetries of the integrand under renaming
$q^{\prime} \rightarrow - q^{\prime}$, we find that
Eq.\ref{eq:sigma1vertex} can also be written as
\index{self-energy!prefactor}
 \begin{eqnarray}
 {\Sigma}^{\alpha}_1 ( \tilde{q} )
 & = & 
 - \frac{1}{\beta {{V}}} \sum_{q^{\prime}}
 \frac{ f^{{\rm RPA},\alpha}_{q^{\prime}}}
 { [ \I \omega_{m^{\prime}} - \xi^{\alpha}_{  {\vec{q}}^{\prime} } ]
 [ \I \omega_{m^{\prime}} + 
 \xi^{\alpha}_{ - {\vec{q}}^{\prime}} ] }
 \nonumber
 \\
 & & \hspace{-13mm} \times \frac{1}{2}
 \left\{
  \left[ {\vec{q}} ({\tens{M}}^{\alpha} )^{-1} {\vec{q}}^{\prime}  \right]^2
 \left[
 G^{\alpha}_1 ( \tilde{q} + q^{\prime} )
 + 
 G^{\alpha}_1 ( \tilde{q} - q^{\prime} ) \right]
 \right.
 \nonumber
 \\
 & & \hspace{-7mm} +
 \left.
 \left[ {\vec{q}}  ({\tens{M}}^{\alpha} )^{-1} {\vec{q}}^{\prime}  \right]
 \left[ {\vec{q}}^{\prime}  ({\tens{M}}^{\alpha} )^{-1} {\vec{q}}^{\prime} \right]
 \left[ 
 G_1^{\alpha} ( \tilde{q} + q^{\prime} )
 - G_1^{\alpha} ( \tilde{q} - q^{\prime} ) \right]
 \right\}
 \label{eq:sigma1res2}
 \; \; \; .
 \end{eqnarray}
To see that the infrared behavior of
$\Sigma^{\alpha}_1 ( \tilde{q} )$  
is less singular than that of the self-energy 
$\Sigma^{\alpha} ( \tilde{q} )$  
associated the full Green's function, note that the first line
in Eq.\ref{eq:sigma1res} is (up to a sign)
identical with the factor $R^{\alpha}_1$ 
given in Eq.\ref{eq:Rlondef2}.
But we know from Sect.~\secref{sec:Identification}
that the finiteness of this factor implies
a finite quasi-particle residue.  
Conversely, non-Fermi liquid behavior should  
manifest itself via infrared divergencies in $R_1^{\alpha}$. 
The crucial point is now that the second and third lines in
Eq.\ref{eq:sigma1res2} contain  additional powers
of ${\vec{q}}^{\prime}$, so that, at least for not too singular
interactions,  
 ${\Sigma}^{\alpha}_1 ( \tilde{q} )$ 
is finite, even though the integral defining $R^{\alpha}_1$ 
does not exist. In particular, for one-dimensional
systems with regular interactions, where $R^{\alpha}_1$ is only logarithmically
divergent, ${\Sigma}^{\alpha}_1 ( \tilde{q} )$ 
does not exhibit any divergencies.

\vspace{7mm}

\begin{center}
{\bf{Calculation of $G^{\alpha}_2$}}
\end{center}

\begin{figure}
\sidecaption
\psfig{figure=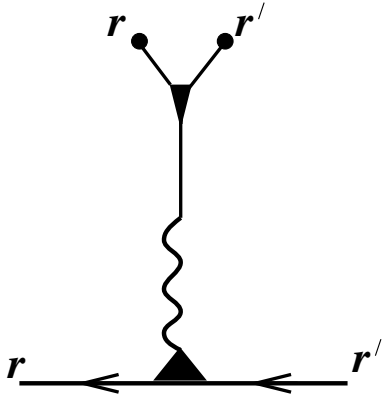,width=5cm}
\caption[
Leading contribution to $G_2^{\alpha}$.]
{ 
Leading contribution to $G_2^{\alpha}$.
Because the function ${\cal{J}}^{\alpha}_q ( {\vec{r}}, \tau )$
depends on ${\vec{r}}$ and $\tau$, the 
diagram has to be understood  as a real space, 
imaginary time diagram.
The spatial labels are 
written on the corresponding end-points.
The thick solid arrows
represent the self-consistent
average Green's function
$G_1^{\alpha}$, as defined
in Eqs.\ref{eq:G1FTdef}--\ref{eq:sigma1res}.
To lowest order in $f^{{\rm RPA} , \alpha}$ the 
thick arrows should be replaced by thin arrows
(representing the non-interacting Green's function $G_0^{\alpha}$).
However, in the spirit of the self-consistent Born approximation, we have included
disorder corrections to the Green's functions
attached to the eikonal contribution.}
\label{fig:interfere}
\vspace{1cm}
\end{figure}

\noindent
Next, let us calculate the interference contribution
$G_2^{\alpha}$ defined in Eq.\ref{eq:G2def}.
Diagrammatically this function is the sum of all
Feynman diagrams 
which combine the graphical elements
defined in Figs.~\secref{fig:eik} and \secref{fig:Dysonderiv}.
To first order in the RPA interaction
only the diagram shown in
Fig.~\secref{fig:interfere} contributes.
Evaluation of this diagram yields
 \begin{equation}
  {G}^{\alpha}_2 ( {\vec{r}} , \tau  )
  = \frac{1}{\beta V} \sum_{\tilde{q}} 
   \E^{ \I ( {\vec{q}} \cdot {\vec{r}} - 
  \tilde{\omega}_n \tau ) }
  G^{\alpha}_2 ( \tilde{q} )
  \; \; \; ,
  \; \; \;
 G^{\alpha}_2 ( {\tilde{q}} )= 
 G^{\alpha}_1 ( {\tilde{q}} )  
 Y^{\alpha} ( \tilde{q} ) 
  \; \; \; ,
  \label{eq:G2FTdef}
  \end{equation}
where, after symmetrization, the dimensionless function
$Y^{\alpha} ( \tilde{q} ) $ can be written as
 \begin{eqnarray}
 Y^{\alpha} ( \tilde{q} ) & = &  
 \frac{1}{\beta {{V}}} \sum_{q^{\prime}}
 \frac{ f^{{\rm RPA},\alpha}_{q^{\prime}}}
 { [ \I \omega_{m^{\prime}} - \xi^{\alpha}_{  {\vec{q}}^{\prime} } ]
 [ \I \omega_{m^{\prime}} + \xi^{\alpha}_{  - {\vec{q}}^{\prime}} ] }
 \nonumber
 \\
 & \times &
 \left\{
 \frac{{{\vec{q}}^{\prime}} ( {\tens{M}}^{\alpha} )^{-1} {\vec{q}}^{\prime}}{2}
 \left[
 G^{\alpha}_1 ( \tilde{q} + q^{\prime} )
 +
 G^{\alpha}_1 ( \tilde{q} - q^{\prime} ) \right]
 \right.
 \nonumber
 \\
 & & \left. \hspace{3mm}
 +
  {\vec{q}}  ({\tens{M}}^{\alpha})^{-1} {\vec{q}}^{\prime} 
 \left[ 
 G_1^{\alpha} ( \tilde{q} + q^{\prime} )
 - G_1^{\alpha} ( \tilde{q} - q^{\prime} ) \right]
 \right\}
 \label{eq:Yres}
 \; \; \; .
 \end{eqnarray}
Again we see that the integrand of $Y^{\alpha} ( \tilde{q} )$ is less
infrared singular than that of $R^{\alpha}_1$ in Eq.\ref{eq:Rlondef2}.

\vspace{7mm}

In summary, the total prefactor Green's function in
Eq.\ref{eq:Gtotalavparametrize}
can be written as
 \begin{eqnarray}
  {G}^{\alpha}_1 ( {\vec{r}} , \tau  )
  +
  {G}^{\alpha}_2 ( {\vec{r}} , \tau  )
  & = & \frac{1}{\beta V} \sum_{\tilde{q}} 
   \E^{ \I ( {\vec{q}} \cdot {\vec{r}} - 
  \tilde{\omega}_n \tau ) }
  \frac{ 1 + Y^{\alpha} ( \tilde{q} ) }
  { \I \tilde{\omega}_n - 
  \epsilon_{ {\vec{k}}^{\alpha} + {\vec{q}} } + \mu
  - \Sigma_1^{\alpha} ( \tilde{q} ) }
  \; \; \;  ,
  \nonumber
  \\
  & & 
  \label{eq:Gtotalpre}
  \end{eqnarray}
where $ \Sigma_1^{\alpha} ( \tilde{q} ) $ and
  $ Y^{\alpha} ( \tilde{q} ) $ can be calculated perturbatively in powers
of the RPA interaction. The leading contributions are given in
Eqs.\ref{eq:sigma1res2} and \ref{eq:Yres}.
In the limit of  infinite effective masses 
(corresponding to linearized energy dispersion)
the functions $\Sigma_1^{\alpha}$ and $Y^{\alpha}$ are identically zero. 
Then the right-hand side of Eq.\ref{eq:Gtotalpre}
reduces to the non-interacting Green's  function, and we recover the
result for linearized energy dispersion.
Furthermore, in the absence of interactions
$Q_1^{\alpha}$, $\Sigma_1^{\alpha}$ and $Y^{\alpha}$ vanish
identically, so that we recover the exact non-interacting
Green's function, which contains
of course infinite orders in $1/ m^{\alpha}$.
This shows that we have not performed
a naive expansion in powers of $1/m^{\alpha}$, 
as originally suggested in \cite{Haldane81}.
We would also like to emphasize that corrections
to the above expressions involve an additional power of the RPA-interaction,
so that {\it{in the weak-coupling regime}} 
we may truncate our expansion at the leading order, {\it{even
if the interaction is not dominated by small momentum transfers}}.
In other words, as long as the RPA interaction 
is finite and small, the above expressions
remain valid to first order in the interaction
even in the presence of Umklapp and  back-scattering processes!
\index{back-scattering}

Our result \ref{eq:Qlondef2} for the
leading term $Q_1^{\alpha} ( {\vec{r}} , \tau )$
in the expansion of the average eikonal 
(which can be viewed as the natural 
generalization of the Debye-Waller \ref{eq:Qlondef}--\ref{eq:Slondef}
to the case of non-linear energy dispersion)
and the corrections \ref{eq:sigma1res2} and \ref{eq:Yres} to the prefactor 
Green's function
cure all pathologies that are generated by the linearization
of the energy dispersion and the
concomitant replacement of a curved
Fermi surface by a collection of flat hyper-planes.
First of all, the singular function $\delta^{(d-1)} ( {\vec{r}}^{\alpha}_{\bot} )$
in Eq.\ref{eq:Grealtotalfull} has disappeared, because now
$G^{\alpha}_0 ( {\vec{r}} , \tau )$ is replaced by 
$G^{\alpha}_1 ( {\vec{r}} , \tau )
+ G^{\alpha}_2 ( {\vec{r}} , \tau )$. 
Due to the finite curvature term, this prefactor is a non-singular
function of all of its arguments. Of course, 
now the Fourier transformation involves a full $d+1$-dimensional integration, so
that from a numerical point of view the problem in $d> 1$ is
more difficult
than in the case of linearized energy dispersion.
Furthermore, possible problems associated with the
double pole in the expression for the
Debye-Waller factor of the linearized theory
are solved trivially, because 
the non-linear terms in the energy dispersion 
split the double pole into two isolated poles
that are separated by a distance $\vec{q} (\tens{M}^{\alpha})^{-1} \vec{q}$
on the real frequency axis (see Eq.\ref{eq:ximinus}).
\index{double pole}

\subsection{Connection with lowest order perturbation theory}
\label{sec:connectionPT}

{\it{We show that the expansion of our result for $G^{\alpha} ( {\vec{r}}, \tau )$
to first order in the RPA interaction exactly reproduces 
perturbation theory.}}

\vspace{7mm}
\noindent
By construction all corrections to
our result for the average eikonal 
in Eqs.\ref{eq:Qlondef2}--\ref{eq:Slondef2}
and the expressions
\ref{eq:sigma1res2} and \ref{eq:Yres}
for the functions $\Sigma^{\alpha}_1 ( \tilde{q} )$ and 
$Y^{\alpha} ( \tilde{q} )$ 
involve at least two powers of
$f^{{\rm RPA} , \alpha}$. 
Therefore a direct expansion of these expressions
to first order in $ f^{{\rm RPA} , \alpha}$
should {\it{exactly}} reproduce the usual perturbative result, i.e.
the GW self-energy with full non-linear energy dispersion.
In this section we show by explicit calculation that this is indeed the case.
For simplicity we shall assume here that the matrix
${\tens{M}}^{\alpha}$ is proportional to the unit matrix, so that
${\vec{q}}  ( {\tens{M}}^{\alpha} )^{-1} {\vec{q}}^{\prime} =
{\vec{q}} \cdot {\vec{q}}^{\prime} /  m^{\alpha} $.

Expanding Eq.\ref{eq:Gtotalavparametrize} to first order in the
RPA interaction, we have
 \begin{eqnarray}
 {{G}}^{\alpha} ( {\vec{r}} , \tau ) 
 & \equiv  &
 [ {G}^{\alpha}_1 ( {\vec{r}} , \tau )  +
 {G}^{\alpha}_2 ( {\vec{r}} , \tau ) ] {\E}^{Q^{\alpha} ( {\vec{r}} , \tau )  } 
 \nonumber
 \\
 &  \approx  &
 {{G}}^{\alpha}_0 ( {\vec{r}} , \tau ) 
  + 
 {{G}}^{\alpha}_0 ( {\vec{r}} , \tau ) 
 Q^{\alpha}_1 ( {\vec{r}} , \tau )   
 \nonumber
 \\
 & + &
   \frac{1}{\beta V} \sum_{\tilde{q}} 
   \E^{ \I ( {\vec{q}} \cdot {\vec{r}} - 
  \tilde{\omega}_n \tau ) }
  G_0^{\alpha} ( \tilde{q} ) \Sigma^{\alpha}_1 ( \tilde{q} ) G_0^{\alpha} ( \tilde{q} ) 
  \nonumber
  \\
   & + & \frac{1}{\beta V} \sum_{\tilde{q}} 
   \E^{ \I ( {\vec{q}} \cdot {\vec{r}} - 
  \tilde{\omega}_n \tau ) }
  G_0^{\alpha} ( \tilde{q} ) Y^{\alpha} ( \tilde{q} )
  + \ldots
  \; \; \; .
  \label{eq:Gbosexpand}
  \end{eqnarray}
Note that to first order in $f^{{\rm RPA} , \alpha}$ we may replace
$G_1^{\alpha} \rightarrow G_0^{\alpha}$ on the right-hand sides
of Eqs.\ref{eq:sigma1res2} and \ref{eq:Yres}.
On the other hand, to leading  order in the interaction we have
for the Fourier transform of the full Green's function
 \begin{eqnarray}
 G^{\alpha} ( \tilde{q} ) 
 & \equiv & \int \D {\vec{r}} \int_{0}^{\beta} \D \tau
 \E^{ - \I ( {\vec{q}} \cdot {\vec{r}} - 
 \tilde{\omega}_n \tau ) }
 {{G}}^{\alpha} ( {\vec{r}} , \tau ) 
 \nonumber
 \\
 & = & 
 G^{\alpha}_0 ( \tilde{q} ) 
 + 
 G^{\alpha}_0 ( \tilde{q} ) 
 \Sigma^{\alpha} ( \tilde{q} )
 G^{\alpha}_0 ( \tilde{q} ) 
 + \ldots
 \; \; \; .
 \label{eq:Gqiterate1}
 \end{eqnarray}
Substituting our first-order result \ref{eq:Qlondef2}
for the Debye-Waller factor
into Eq.\ref{eq:Gbosexpand},
Fourier transforming, and comparing with Eq.\ref{eq:Gqiterate1},
it is not difficult
to show that within our bosonization approach 
the first-order self-energy is approximated by
 \begin{equation}
 \Sigma^{\alpha} ( \tilde{q} )
 = \Sigma_{\rm Q}^{\alpha}  ( \tilde{q} )
 + \Sigma_1^{\alpha} ( \tilde{q} ) +
  \Sigma_{\rm Y}^{\alpha} ( \tilde{q} ) 
  \label{eq:firstordersigma}
  \; \; \; ,
  \end{equation}
where the self-energy $\Sigma_1^{\alpha} ( \tilde{q} )$
is given in Eq.\ref{eq:sigma1res}, and
the contribution 
 $ \Sigma_{\rm Q}^{\alpha}  ( \tilde{q} )$ due to the 
Debye-Waller factor on the right-hand side of Eq.\ref{eq:Gbosexpand}
is 
 \begin{eqnarray}
 \Sigma_{\rm Q}^{\alpha}  ( \tilde{q} )
 & = & \left[ \I \tilde{\omega}_n -
 \epsilon_{ {\vec{k}}^{\alpha} + {\vec{q}} } + \mu \right]^2
 \frac{1}{\beta {{V}}} \sum_{q^{\prime}}
 \frac{ f^{{\rm RPA},\alpha}_{q^{\prime}}}{
 [ \I \omega_{m^{\prime}} - \xi^{\alpha}_{{\vec{q}}^{\prime} } ]
 [ \I \omega_{m^{\prime}} + \xi^{\alpha}_{- {\vec{q}}^{\prime}} ] }
 \nonumber
 \\
 & & \hspace{-13mm} \times  \left\{
 \frac{1}{ \I \tilde{\omega}_n -
 \epsilon_{ {\vec{k}}^{\alpha} + {\vec{q}} } + \mu} 
 \right.
 \nonumber
 \\
 & & \hspace{-9mm}
 \left.
 - \frac{1}{2}
 \left[
 \frac{1}{ \I \tilde{\omega}_{n+m^{\prime}} -
 \epsilon_{ {\vec{k}}^{\alpha} + {\vec{q}} + {\vec{q}}^{\prime} } + \mu} 
 + 
 \frac{1}{ \I \tilde{\omega}_{n-m^{\prime}} -
 \epsilon_{ {\vec{k}}^{\alpha} + {\vec{q}} - {\vec{q}}^{\prime} } + \mu} 
 \right]
 \right\}
 \;  .
 \label{eq:sigmaQdef}
 \end{eqnarray}
Similarly, we obtain from \ref{eq:Yres}
for the last term in Eq.\ref{eq:firstordersigma}
 \begin{eqnarray}
 \Sigma_{\rm Y}^{\alpha}  ( \tilde{q} )
 & = & \left[ \I \tilde{\omega}_n -
 \epsilon_{ {\vec{k}}^{\alpha} + {\vec{q}} } + \mu \right]
 \frac{1}{\beta {{V}}} \sum_{q^{\prime}}
 \frac{ f^{{\rm RPA},\alpha}_{q^{\prime}}}{
 [ \I \omega_{m^{\prime}} - \xi^{\alpha}_{{\vec{q}}^{\prime} } ]
 [ \I \omega_{m^{\prime}} + \xi^{\alpha}_{- {\vec{q}}^{\prime}} ] }
 \nonumber
 \\
 & & \hspace{-16mm} \times  \left\{
 \frac{ {{\vec{q}}^{\prime}}^2 }{ 2 m^{\alpha} }
 \left[
 G^{\alpha}_0 ( \tilde{q} + q^{\prime} )
 +
 G^{\alpha}_0 ( \tilde{q} - q^{\prime} ) \right]
 +
  \frac{ {\vec{q}} \cdot {\vec{q}}^{\prime} }{ m^{\alpha} }
 \left[ 
 G_0^{\alpha} ( \tilde{q} + q^{\prime} )
 - G_0^{\alpha} ( \tilde{q} - q^{\prime} ) \right]
 \right\}
 \;  .
 \nonumber
 \\
 & &
 \label{eq:sigmaYdef}
 \end{eqnarray}
We now show that the self-energy
$\Sigma^{\alpha} ( \tilde{q} )$ given in Eq.\ref{eq:firstordersigma}
is identical with the usual GW self-energy.
At the first sight this is not at all obvious, because the three terms
on the right-hand side of Eq.\ref{eq:firstordersigma}
have no resemblance to the usual perturbative result 
for the GW self-energy, which can be written as
(see Eq.\ref{eq:GW}) 
 \begin{eqnarray}
 \Sigma_{\rm GW}^{\alpha} ( \tilde{q} )
 & = &
 - \frac{1}{\beta {{V}}} \sum_{q^{\prime}}
 f^{{\rm RPA},\alpha}_{q^{\prime}}
  \frac{1}{2}
 \left[
 \frac{1}{ \I \tilde{\omega}_{n+m^{\prime}} -
 \epsilon_{ {\vec{k}}^{\alpha} + {\vec{q}} + {\vec{q}}^{\prime} } + \mu} 
 \nonumber
 \right.
 \\
 & & 
 \left.
  \hspace{28mm} + 
 \frac{1}{ \I \tilde{\omega}_{n-m^{\prime}} -
 \epsilon_{ {\vec{k}}^{\alpha} + {\vec{q}} - {\vec{q}}^{\prime} } + \mu} 
 \right]
 \label{eq:GW2}
 \; \; \; .
 \end{eqnarray}
We have used the invariance of $f^{{\rm RPA} , \alpha}_{ q^{\prime}}$
with respect to relabeling $q^{\prime} \rightarrow - q^{\prime}$ 
to symmetrize the rest of the integrand.

Let us begin by manipulating $\Sigma_{\rm Q}^{\alpha} ( \tilde{q} )$
in precisely the same way as one would proceed in the case of
linearized energy dispersion. 
Then one would partial fraction the differences of 
two non-interacting Green's functions in the 
second line of Eq.\ref{eq:sigmaQdef}.
For linearized energy dispersion the result
can be expressed again in terms of non-interacting
Green's function\footnote{Recall the partial fraction decomposition
\ref{eq:Gpartialfrac}, which
was crucial in the proof of the closed loop theorem.}. 
For energy dispersions with a quadratic term, the
generalization of Eq.\ref{eq:Gpartialfrac} is
 \begin{equation}
 \frac{1}{ 
 \I \tilde{\omega}_n - \epsilon^{\alpha}_{  {\vec{q}} } } 
 - 
 \frac{1}{ \I \tilde{\omega}_{n+m^{\prime}} -
 \epsilon^{\alpha}_{  {\vec{q}} + {\vec{q}}^{\prime} } } 
 = 
 \frac{ \I \omega_{m^{\prime}} - \xi^{\alpha}_{ {\vec{q}}^{\prime} } - 
 \frac{ {\vec{q}} \cdot {\vec{q}}^{\prime} }{m^{\alpha} } }{
 [ \I \tilde{\omega}_n - \epsilon^{\alpha}_{ {\vec{q}} } ]
 [ \I \tilde{\omega}_{n+m^{\prime}} -
 \epsilon^{\alpha}_{ {\vec{q}} + {\vec{q}}^{\prime} } ] }
 \label{eq:partialfrac1}
 \; \; \; ,
 \end{equation}
and similarly
 \begin{equation}
 \frac{1}{ 
 \I \tilde{\omega}_n - \epsilon^{\alpha}_{  {\vec{q}} } } 
 - 
 \frac{1}{ \I \tilde{\omega}_{n-m^{\prime}} -
 \epsilon^{\alpha}_{  {\vec{q}} - {\vec{q}}^{\prime} } } 
 = 
 \frac{ - \I \omega_{m^{\prime}} - \xi^{\alpha}_{ - {\vec{q}}^{\prime} } + 
 \frac{ {\vec{q}} \cdot {\vec{q}}^{\prime} }{m^{\alpha} } }{
 [ \I \tilde{\omega}_n - \epsilon^{\alpha}_{ {\vec{q}} } ]
 [ \I \tilde{\omega}_{n-m^{\prime}} -
 \epsilon^{\alpha}_{ {\vec{q}} - {\vec{q}}^{\prime} } ] }
 \label{eq:partialfrac2}
 \; \; \; ,
 \end{equation}
where for simplicity we have introduced the notation
$\epsilon^{\alpha}_{\vec{q}} = \epsilon_{ {\vec{k}}^{\alpha} + {\vec{q}} } - \mu $.
With the help of these identities it is easy to show that Eq.\ref{eq:sigmaQdef}
can also be written as
 \begin{eqnarray}
 \Sigma_{\rm Q}^{\alpha}  ( \tilde{q} )
 & = & 
 \frac{1}{\beta {{V}}} \sum_{q^{\prime}}
 f^{{\rm RPA},\alpha}_{q^{\prime}}
 \frac{1}{2}
 \left[
 \frac{
 \I \tilde{\omega}_n - \epsilon^{\alpha}_{ {\vec{q}} }  }{
 [ \I \omega_{m^{\prime}} + \xi^{\alpha}_{- {\vec{q}}^{\prime} } ]
 [ \I \tilde{\omega}_{n+m^{\prime}} -
 \epsilon^{\alpha}_{ {\vec{q}} + {\vec{q}}^{\prime} } ] }
 \nonumber
 \right.
 \\
 & & \left.
 \hspace{28mm}
 -
 \frac{
 \I \tilde{\omega}_n - \epsilon^{\alpha}_{ {\vec{q}} }  }{
 [ \I \omega_{m^{\prime}} - \xi^{\alpha}_{ {\vec{q}}^{\prime} } ]
 [ \I \tilde{\omega}_{n-m^{\prime}} -
 \epsilon^{\alpha}_{ {\vec{q}} - {\vec{q}}^{\prime} } ] }
 \right]
 \nonumber
 \\
 &  & \hspace{-10mm} -
 \left[ \I \tilde{\omega}_n - \epsilon^{\alpha}_{ {\vec{q}} }  \right]
 \frac{1}{\beta {{V}}} \sum_{q^{\prime}}
 \frac{ f^{{\rm RPA},\alpha}_{q^{\prime}}}{
 [ \I \omega_{m^{\prime}} - \xi^{\alpha}_{{\vec{q}}^{\prime} } ]
 [ \I \omega_{m^{\prime}} + \xi^{\alpha}_{- {\vec{q}}^{\prime}} ] }
 \nonumber
 \\
 & & \hspace{10mm} \times \frac{ {\vec{q}} \cdot {\vec{q}}^{\prime} }{ 2 m^{\alpha}}
 \left[
 \frac{1}{
  \I \tilde{\omega}_{n+m^{\prime}} -
 \epsilon^{\alpha}_{ {\vec{q}} + {\vec{q}}^{\prime} }  }
 -
 \frac{1}{
  \I \tilde{\omega}_{n-m^{\prime}} -
 \epsilon^{\alpha}_{ {\vec{q}} - {\vec{q}}^{\prime} }  }
 \right]
 \label{eq:sigmaQpartial1}
 \; \; \; .
 \end{eqnarray}
Next, we use the following two exact identities,
 \begin{eqnarray}
 \frac{
 \I\tilde{\omega}_n - \epsilon^{\alpha}_{ {\vec{q}} }  }{
 [ \I \omega_{m^{\prime}} + \xi^{\alpha}_{- {\vec{q}}^{\prime} } ]
 [ \I \tilde{\omega}_{n+m^{\prime}} -
 \epsilon^{\alpha}_{ {\vec{q}} + {\vec{q}}^{\prime} } ] }
 & = &
 \frac{1}{\I \omega_{m^{\prime}} + \xi^{\alpha}_{- {\vec{q}}^{\prime} }} 
 - 
 \frac{1}{
 \I \tilde{\omega}_{n+m^{\prime}} -
 \epsilon^{\alpha}_{ {\vec{q}} + {\vec{q}}^{\prime} } }
 \nonumber
 \\
 & & \hspace{-30mm} + 
\frac{ {\vec{q}}^{\prime} \cdot ( {\vec{q}} + {\vec{q}}^{\prime} ) }{m^{\alpha} }
\frac{1}{
 [ \I \omega_{m^{\prime}} + \xi^{\alpha}_{- {\vec{q}}^{\prime} } ]
 [ \I \tilde{\omega}_{n+m^{\prime}} -
 \epsilon^{\alpha}_{ {\vec{q}} + {\vec{q}}^{\prime} } ] }
 \label{eq:partialfrac3}
 \; \; \; ,
 \\
 \frac{
 \I \tilde{\omega}_n - \epsilon^{\alpha}_{ {\vec{q}} }  }{
 [ \I \omega_{m^{\prime}} - \xi^{\alpha}_{ {\vec{q}}^{\prime} } ]
 [ \I \tilde{\omega}_{n-m^{\prime}} -
 \epsilon^{\alpha}_{ {\vec{q}} - {\vec{q}}^{\prime} } ] }
 & = &
 \frac{1}{\I \omega_{m^{\prime}} - \xi^{\alpha}_{ {\vec{q}}^{\prime} }} 
 - 
 \frac{1}{
 \I \tilde{\omega}_{n-m^{\prime}} -
 \epsilon^{\alpha}_{ {\vec{q}} - {\vec{q}}^{\prime} } }
 \nonumber
 \\
 & & \hspace{-30mm} - 
\frac{ {\vec{q}}^{\prime} \cdot ( {\vec{q}} - {\vec{q}}^{\prime} ) }{m^{\alpha} }
\frac{1}{
 [ \I \omega_{m^{\prime}} - \xi^{\alpha}_{ {\vec{q}}^{\prime} } ]
 [ \I \tilde{\omega}_{n-m^{\prime}} -
 \epsilon^{\alpha}_{ {\vec{q}} - {\vec{q}}^{\prime} } ] }
 \label{eq:partialfrac4}
 \; \; \; ,
 \end{eqnarray}
and obtain from Eq.\ref{eq:sigmaQpartial1}
 \begin{eqnarray}
 \Sigma_{\rm Q}^{\alpha}  ( \tilde{q} )
 & = &  \Sigma_{\rm GW}^{\alpha} ( \tilde{q} ) 
 - \frac{1}{\beta {{V}}} \sum_{q^{\prime}}
 f^{{\rm RPA},\alpha}_{q^{\prime}}
  \frac{1}{2}
 \left[
 \frac{1}{ \I {\omega}_{m^{\prime}} -
 \xi^{\alpha}_{ {\vec{q}}^{\prime} } } 
 - 
 \frac{1}{  \I {\omega}_{m^{\prime}} +
 \xi^{\alpha}_{  - {\vec{q}}^{\prime} }} 
 \right]
  \nonumber
  \\
  & +  & 
 \frac{1}{\beta {{V}}} \sum_{q^{\prime}}
 \frac{ f^{{\rm RPA},\alpha}_{q^{\prime}}}{
 [ \I \omega_{m^{\prime}} - \xi^{\alpha}_{{\vec{q}}^{\prime} } ]
 [ \I \omega_{m^{\prime}} + \xi^{\alpha}_{- {\vec{q}}^{\prime}} ] }
 \left[
\frac{ {\vec{q}}^{\prime} \cdot ( {\vec{q}} + {\vec{q}}^{\prime} ) }{m^{\alpha} }
 \frac{ \I \omega_{m^{\prime}} - \xi^{\alpha}_{ {\vec{q}}^{\prime} }}{ 
 \I \tilde{\omega}_{n+m^{\prime}} -
 \epsilon^{\alpha}_{ {\vec{q}} + {\vec{q}}^{\prime} } }
 \right.
 \nonumber
 \\
 & & \left. \hspace{35mm}
 +
\frac{ {\vec{q}}^{\prime} \cdot ( {\vec{q}} - {\vec{q}}^{\prime} ) }{m^{\alpha} }
 \frac{ \I \omega_{m^{\prime}} + \xi^{\alpha}_{ - {\vec{q}}^{\prime} }}{ 
 \I \tilde{\omega}_{n-m^{\prime}} -
 \epsilon^{\alpha}_{ {\vec{q}} - {\vec{q}}^{\prime} } }
 \right]
 \nonumber
 \\
 & + &
 \left[ \I \tilde{\omega}_n - \epsilon^{\alpha}_{ {\vec{q}} }  \right]
 \frac{1}{\beta {{V}}} \sum_{q^{\prime}}
 \frac{ f^{{\rm RPA},\alpha}_{q^{\prime}}}{
 [ \I \omega_{m^{\prime}} - \xi^{\alpha}_{{\vec{q}}^{\prime} } ]
 [ \I \omega_{m^{\prime}} + \xi^{\alpha}_{- {\vec{q}}^{\prime}} ] }
 \nonumber
 \\
 & & \hspace{10mm} \times \frac{ {\vec{q}} \cdot {\vec{q}}^{\prime} }{ 2 m^{\alpha}}
 \left[
 \frac{1}{
  \I \tilde{\omega}_{n+m^{\prime}} -
 \epsilon^{\alpha}_{ {\vec{q}} + {\vec{q}}^{\prime} }  }
 -
 \frac{1}{
  \I \tilde{\omega}_{n-m^{\prime}} -
 \epsilon^{\alpha}_{ {\vec{q}} - {\vec{q}}^{\prime} }  }
 \right]
 \label{eq:sigmaQpartial2}
 \; \; \; .
 \end{eqnarray}
Here the function $\Sigma_{\rm GW}^{\alpha} ( \tilde{q} )$
is given in Eq.\ref{eq:GW2}.
Finally we write in the third term on the right-hand side of Eq.\ref{eq:sigmaQpartial2}
 \begin{eqnarray}
  \I \omega_{m^{\prime}} - \xi^{\alpha}_{{\vec{q}}^{\prime} } 
  & = & -
 \left[ \I \tilde{\omega}_n - \epsilon^{\alpha}_{ {\vec{q}} }  \right]
 +
  \left[ \I \tilde{\omega}_{n+m^{\prime}} -
 \epsilon^{\alpha}_{ {\vec{q}} + {\vec{q}}^{\prime} }  \right]
 + \frac{ {\vec{q}} \cdot {\vec{q}}^{\prime}}{m^{\alpha}}
 \label{eq:decompose1}
 \; \; \; ,
 \\
  \I \omega_{m^{\prime}} + \xi^{\alpha}_{-{\vec{q}}^{\prime} } 
  & = & 
 \left[ \I \tilde{\omega}_n - \epsilon^{\alpha}_{ {\vec{q}} }  \right]
 -
  \left[ \I \tilde{\omega}_{n-m^{\prime}} -
 \epsilon^{\alpha}_{ {\vec{q}} - {\vec{q}}^{\prime} }  \right]
 + \frac{ {\vec{q}} \cdot {\vec{q}}^{\prime}}{m^{\alpha}}
 \label{eq:decompose2}
 \; \; \; ,
 \end{eqnarray}
and arrive at
 \begin{equation}
 \Sigma_{\rm Q}^{\alpha}  ( \tilde{q} )
  =   \Sigma_{\rm GW}^{\alpha} ( \tilde{q} ) -
  \Sigma_{\rm GW}^{\alpha} ( \tilde{q} = 0 )  
  - \Sigma_1^{\alpha} ( \tilde{q} )
  - \Sigma_{\rm Y}^{\alpha} ( \tilde{q} )
  \label{eq:sigmaQfinal}
  \; \; \; ,
  \end{equation}
where we have used the fact that
 \begin{eqnarray}
  \Sigma_{\rm GW}^{\alpha} ( \tilde{q} = 0 )  
  & = &
 - \frac{1}{\beta {{V}}} \sum_{q^{\prime}}
 f^{{\rm RPA},\alpha}_{q^{\prime}}
  \frac{1}{2}
 \left[
 \frac{1}{ \I {\omega}_{m^{\prime}} -
 \xi^{\alpha}_{ {\vec{q}}^{\prime} } } 
 - 
 \frac{1}{  \I {\omega}_{m^{\prime}} +
 \xi^{\alpha}_{  - {\vec{q}}^{\prime} }} 
 \right]
 \nonumber
 \\
 & = &
 - \frac{1}{\beta {{V}}} \sum_{q^{\prime}}
 f^{{\rm RPA},\alpha}_{q^{\prime}}
 \frac{ \frac{ {\vec{q}}^2 }{ 2 m^{\alpha} } }{
 [  \I {\omega}_{m^{\prime}} -
 \xi^{\alpha}_{ {\vec{q}}^{\prime} } ]
 [  \I {\omega}_{m^{\prime}} +
 \xi^{\alpha}_{ - {\vec{q}}^{\prime} } ]}
 \label{eq:SigmaGW0}
 \; \; \; .
 \end{eqnarray}
The last two terms in Eq.\ref{eq:sigmaQfinal}, which arise
due to the above partial fraction decompositions of 
$\Sigma^{\alpha}_{\rm Q}$, are exactly cancelled by the last two terms
in Eq.\ref{eq:firstordersigma}, so that the final result for
the first order self-energy calculated within our bosonization approach is
 \begin{equation}
 \Sigma^{\alpha} ( \tilde{q} ) =
 \Sigma_{\rm GW}^{\alpha} ( \tilde{q}  )  -
 \Sigma_{\rm GW}^{\alpha} ( \tilde{q} = 0 )  
 \label{eq:firstordersimafinal}
 \; \; \; .
 \end{equation}
The term
 $\Sigma_{\rm GW}^{\alpha} ( \tilde{q} = 0 )  $ subtracts 
the renormalization of the chemical potential contained in the 
first term. \index{chemical potential!renormalization}
This is in agreement with the fact that by definition
we start from the exact chemical potential of the many-body system, so that 
$\mu$ should not be renormalized.
Note, however, that in general the {\it{shape}} of the Fermi surface
will be renormalized by the interaction. This effect is lost if one linearizes
the energy dispersion.
The crucial role of 
the  terms $\Sigma_1^{\alpha} $ and $ Y^{\alpha} $ is now evident.
If we had ignored these corrections,
we would have obtained a discrepancy with
lowest order perturbation theory,
because for finite $m^{\alpha}$ the exponentiation 
$\E^{Q^{\alpha}}$  of the perturbation series 
is not quite correct. 
In a sense, we have exponentiated ``too much'', so that
it is necessary to introduce correction terms
in the prefactor.  \index{Fermi surface!renormalization of shape}

\section{Summary and outlook}
\label{sec:sumgreen}

In this chapter we have developed a new method
for calculating the single-particle Green's function of an interacting
Fermi system. Our result 
within the Gaussian approximation can be considered as the
natural generalization of the non-perturbative bosonization
solution of the Tomonaga-Luttinger model \cite{Tomonaga50,Luttinger63,Mattis65}
to arbitrary dimensions. Because in $d > 1$ the
curvature of the Fermi surface leads to qualitatively 
new effects which do not exist in $d = 1$,
we have developed a systematic method for including the non-linear terms
in the energy dispersion into the bosonization procedure in arbitrary dimensions.

Let us summarize  our main result for the Green's function
for the special case of a spherical Fermi surface 
of radius $k_{\rm F} = v_{\rm F} / m$
and a patch-independent
bare interaction $f_{q}$.  As discussed in Chap.~\secref{sec:sectors}, 
in this case it is not necessary to subdivide the Fermi surface
into several sectors -- instead, if we are interested
in the Matsubara Green's function $G ( \vec{k} , \I \tilde{\omega}_n )$
for a given ${\vec{k}}$,
we choose a special coordinate system centered at
${\vec{k}}^{\alpha}$ on the Fermi surface shown in Fig.~\secref{fig:coordgood}.
As discussed at the end of Chap.~\secref{sec:sectors},
due to the spherical symmetry, \index{spherical symmetry}
$G ( {\vec{k}} , \I \tilde{\omega}_n )$ 
depends on $\vec{k}$ exclusively via the combination
$  | {\vec{k}} | - k_{\rm F} $. Then we may write
 \begin{equation}
 G ( {\vec{k}} , \I \tilde{\omega}_n ) = G^{\alpha} ( 
  | {\vec{k}}  |
 \hat{\vec{k}}^{\alpha}  - {\vec{k}}^{\alpha} ,  \I \tilde{\omega}_n )
 \; \; \; ,
 \label{eq:Galphashiftsym}
 \end{equation}
with
 \begin{equation}
 G^{\alpha} ( {\vec{q}} , \I \tilde{\omega}_n ) =
 \int \D {\vec{r}} \int_0^{\beta} \D \tau \E^{ - \I ( 
 \vec{q} \cdot \vec{r} - \tilde{\omega}_n \tau )}
 \tilde{G}^{\alpha} ( \vec{r} , \tau ) \E^{ Q_1^{\alpha} ( \vec{r} , \tau ) }
 \label{eq:Gresultsum}
 \; \; \; ,
 \end{equation}
where the Debye-Waller factor is
 \begin{equation}
 Q_1^{\alpha} ( {\vec{r}}  , \tau )
  =  \frac{1}{\beta {{V}}} \sum_q
  f^{{\rm RPA}}_q \frac{
  1 - \cos ( {\vec{q}} \cdot {\vec{r}} - \omega_m \tau )  }
{ [ \I \omega_m - \xi^{\alpha}_{  {\vec{q}} } ][ \I \omega_m + 
\xi^{\alpha}_{ - {\vec{q}}} ] }
 \label{eq:Qlondef3}
 \; \; \; ,
 \end{equation}
and the prefactor Green's function
\index{Green's function!prefactor}
$\tilde{G}^{\alpha} ( \vec{r} , \tau )$ has the
Fourier expansion
 \begin{equation}
 \tilde{G}^{\alpha} ( {\vec{r}} , \tau  )
 = \frac{1}{\beta V} \sum_{\tilde{q}} 
 \E^{ \I ( {\vec{q}} \cdot {\vec{r}} - 
 \tilde{\omega}_n \tau ) }
 \tilde{G}^{\alpha} ( \tilde{q} ) 
 \; \; \; ,
 \label{eq:prffourier}
 \end{equation}
 \begin{equation}
 \tilde{G}^{\alpha} ( \tilde{q} ) 
 =
  \frac{ 1 + Y^{\alpha} ( \tilde{q} ) }
  { \I \tilde{\omega}_n - 
  \epsilon_{ {\vec{k}}^{\alpha} + {\vec{q}} } + \mu
  - \Sigma_1^{\alpha} ( \tilde{q} ) }
  \label{eq:Gtotalpre2}
  \; \; \; ,
  \end{equation}
with the prefactor self-energy
\index{self-energy!prefactor}
 \begin{eqnarray}
 {\Sigma}^{\alpha}_1 ( \tilde{q} )
 & = & 
 - \frac{1}{\beta {{V}}} \sum_{q^{\prime}}
 f^{{\rm RPA}}_{q^{\prime}} G_1^{\alpha} ( \tilde{q} + q^{\prime} )
 \nonumber
 \\
 & \times &
 \frac{ ( \vec{q} \cdot \vec{q}^{\prime} ) \vec{q}^{\prime 2} + 
 ( \vec{q} \cdot \vec{q}^{\prime} )^2 }{
 m^2 [ \I \omega_{m^{\prime}} - \xi^{\alpha}_{  {\vec{q}}^{\prime} } ]
 [ \I \omega_{m^{\prime}} + 
 \xi^{\alpha}_{ - {\vec{q}}^{\prime}} ] }
 \label{eq:sigma1res3}
 \; \; \; ,
 \end{eqnarray}
and the vertex function
 \begin{eqnarray}
 {Y}^{\alpha} ( \tilde{q} )
 & = & 
  \frac{1}{\beta {{V}}} \sum_{q^{\prime}}
 f^{{\rm RPA}}_{q^{\prime}} G_1^{\alpha} ( \tilde{q} + q^{\prime} )
 \nonumber
 \\
 & \times &
 \frac{ \vec{q}^{\prime 2} + 2  \vec{q} \cdot \vec{q}^{\prime}  }{
 m [ \I \omega_{m^{\prime}} - \xi^{\alpha}_{  {\vec{q}}^{\prime} } ]
 [ \I \omega_{m^{\prime}} + 
 \xi^{\alpha}_{ - {\vec{q}}^{\prime}} ] }
 \label{eq:Yres4}
 \; \; \; .
 \end{eqnarray}
Note that $Q^{\alpha}_1$, $\Sigma_1^{\alpha}$, $Y^{\alpha}$ 
are of the first order in the RPA interaction and involve
a single fermionic loop summation 
(apart from the infinite series of bubble diagrams contained in
$f^{\rm RPA}$). 
The above expressions can be considered 
as a new extrapolation of the perturbation series,
which involves a partial exponentiation 
in real space, a partial geometric resummation in Fourier space,
and an intricate mixed Fourier representation.
Our extrapolation scheme is quite different 
from the usual geometric extrapolation 
of the perturbation series for the Green's function
in momentum space, which is implicitly
performed if one first calculates the
irreducible self-energy $\Sigma ( \vec{k} , \I \tilde{\omega}_n )$ 
to some finite order
in $f^{\rm RPA}$ and then solves the Dyson equation. 
As shown in Sect.~\secref{sec:connectionPT}, 
our resummation scheme  has the important property that
the expansion of our result for the Green's function to
first order in $f^{\rm RPA}$ {\it{exactly}} 
reproduces the leading term in naive a perturbative expansion. 
Moreover, in Chap.~\secref{sec:Green1} we shall show that
in one dimension and for linearized energy dispersion
Eqs.\ref{eq:Galphashiftsym}--\ref{eq:Yres4} correctly
reproduce the exact solution of the Tomonaga-Luttinger 
model \cite{Tomonaga50,Luttinger63,Mattis65}.

In the second part of this book we shall partially evaluate 
the above expressions in some simple limiting 
cases where we can make progress without 
resorting to numerical methods.  However,
our analysis will not be complete, 
because in general 
the integrations in Eqs.\ref{eq:Galphashiftsym}--\ref{eq:Yres4}
are very difficult to perform.
It particular, the calculation of 
the full momentum- and frequency-dependent
{\bf{spectral function}}
 \begin{equation}
   A ( \vec{k} , \omega )
 = - \frac{1}{\pi} {\rm Im} G ( \vec{k} , \omega + \I 0^{+} )
 \label{eq:specfuncdef}
 \end{equation}
from Eqs.\ref{eq:Galphashiftsym}--\ref{eq:Yres4} in a non-trivial
interacting Fermi system in $d > 1$ is an interesting open problem,
which seems to require extensive numerical work.
We would like to emphasize that such a calculation would yield a 
highly  non-perturbative result for the spectral function. 
In particular, Eqs.\ref{eq:Galphashiftsym}--\ref{eq:Yres4} 
can be used to determine by direct calculation 
whether an interacting Fermi system is a Fermi liquid or not. 
In both cases these equations
are well-defined 
(at least for not too singular interactions, 
see Chap.~\secref{subsec:thelim}), and
provide an explicit expression for the
single-particle Green's function
which can serve as a basis for quantitative calculations.

Before embarking on applications of our formalism to
problems of physical interest, let us briefly mention
two  more open problems, which will not be further discussed
in this book\footnote{ 
I would like to encourage
all readers to contribute to the solution of these problems.}. 
First of all, the problem of {\bf{back-scattering}}\index{back-scattering}:
Because in
Eqs.\ref{eq:Galphashiftsym}--\ref{eq:Yres4} 
we have not made use of the patching construction and have
identified the entire momentum space with a single sector,
the restriction that the maximum momentum transfer $q_{\rm c}$
of the interaction must be smaller than the size of the sectors
(see Fig.\secref{fig:qc}) does not exclude processes with large
momentum transfer any more.  Therefore 
Eqs.\ref{eq:Galphashiftsym}--\ref{eq:Yres4}  are also valid
for short-range interactions, {\it{provided the 
dimensionless parameter $A^{\alpha}_0 $ given in
Eq.\ref{eq:gausscorrect} is small.}}
Of course, in this case we loose the small factor
$ ( q_{\rm c} / k_{\rm F} )^d$ in Eq.\ref{eq:gausscorrect}, so that
our non-perturbative expression for the Green's function
can only be accurate for sufficiently small interactions.
However, in the weak-coupling regime
Eqs.\ref{eq:Galphashiftsym}--\ref{eq:Yres4} 
can be considered as the leading term in a
non-perturbative expansion in powers of the RPA interaction,
which includes
the effect of scattering processes involving large momentum transfers,
such as back-scattering or Umklapp-scattering.

The second interesting direction for further research
is the generalization of our formalism to include
{\bf{broken symmetries}}.
Note that throughout this work we are assuming
that the electrons remain normal, i.e.  that  they do not 
undergo a phase transition to a state with spontaneously broken symmetry.
In particular, we have ignored the tendencies towards superconductivity  
and antiferromagnetism, which are known to exist
in many strongly correlated Fermi systems at sufficiently low temperatures.
It seems, however, that it is not difficult to include
these effects into our formalism, at least at the
level of the Gaussian approximation. In fact, functional
integration and Hubbard-Stratonovich
transformation are the ideal formal starting point to 
study spontaneous symmetry breaking in Fermi 
systems \cite{Negele88,Schakel89}. Therefore we expect that
it is straightforward to generalize the non-perturbative
methods developed in this book to incorporate superconductivity and
various types of itinerant magnetism.
In particular, our methods might provide 
a non-perturbative microscopic approach to nearly antiferromagnetic 
Fermi liquids \cite{Millis90}.

In this context we would also like to point out that for
systems with special spin symmetries or other
internal symmetries it might be necessary to decouple
the relevant operators by means of matrix-field Hubbard-Stratonovich
transformations which  preserve the symmetries.
This could lead to higher-dimensional generalizations {\bf{of non-abelian
bosonization}} \cite{Witten84,Affleck90}.
An attempt to develop such an approach has recently
been made by Schmeltzer \cite{Schmeltzer96}.

%
%

\part{Applications to physical systems}
%
%
%

\chapter{Singular interactions ($f_{\vec{q}} \sim | \vec{q} |^{-\eta}$)} 
\label{chap:a7sing}
\setcounter{equation}{0}

{\it{
We analyze singular density-density interactions 
that diverge in $d$ dimensions as $ | {\vec{q}} |^{- \eta }$
for $ {\vec{q}} \rightarrow 0$. 
For linearized energy dispersion we explicitly calculate
the asymptotic long-distance behavior of
$Q^{\alpha} ( {\vec{r}} , 0 )$.
For regular interactions $( \eta = 0)$ in one dimension
it is possible to calculate the full Debye-Waller factor 
$Q^{\alpha} ( {\vec{r}} , \tau )$ 
if a certain cutoff procedure is adopted.
Then we reproduce the well-known bosonization result
for the Tomonaga-Luttinger model.}}

\vspace{7mm}

\noindent
In this chapter we shall study in some detail
the Debye-Waller factor 
$Q^{\alpha} ( {\vec{r}} , \tau )$
derived in Chap.~\secref{chap:agreen}
for singular density-density interactions of the form
 \begin{equation}
 f_{\vec{q}} = \frac{g_{\rm c}^2}{ | {\vec{q}} |^{\eta} } 
 \E^{ - |{\vec{q}} | / q_{\rm c} }
 \; \; \; , \; \; \;  \eta > 0 
 \; \; \; , \; \; \;  q_{\rm c } \ll k_{\rm F} \; \; \; , 
 \label{eq:fgeneric}
 \end{equation}
where $g_{\rm c}$ is some coupling constant with the correct units.
The long-range part of the physical Coulomb interaction in $d$ dimensions corresponds
to $g_{\rm c} = -e$ (the charge of the electron), $\eta = d-1$, 
and $q_{\rm c} = \infty$, see Appendix \secref{subsubsec:Cb}.
As recently noticed by Bares and Wen \cite{Bares93},
in the more singular case $\eta = 2 ( d-1 )$ one obtains
an instability of the Fermi liquid state.
Although for general $\eta$ interactions of the above type
are unphysical, it is instructive 
study them as model systems which exhibit non-Fermi liquid behavior.

From Eq.\ref{eq:frpatot} we know that for patch-independent bare interaction
the screened interaction
$f^{{\rm RPA }, \alpha }_{ q} $ 
in Eqs.\ref{eq:Qlondef}--\ref{eq:Slondef} and 
\ref{eq:Qlondef2}--\ref{eq:Slondef2} 
can be identified with the usual 
RPA interaction $f^{\rm RPA}_q = f_{\vec{q}} [ 1 + f_{\vec{q}} \Pi_0 ( q )  ]^{-1}$.
For practical calculations it is convenient to express
$f^{\rm RPA}_{q}$  in terms of the dynamic structure 
factor\index{dynamic structure factor!within RPA}
$S_{\rm RPA} ( {\vec{q}} , \omega )$, which is 
the spectral function of the RPA polarization\footnote{
We would like to point out that the relation
\ref{eq:Srealaxis} between the imaginary part
of the polarization and the dynamic structure factor is only valid if the shape of the
Fermi surface is invariant with respect to 
inversion ${\vec{k}} \rightarrow - {\vec{k}}$. 
If we approximate the Fermi surface by a finite number
of flat patches, then
Eqs.\ref{eq:PiRPASRPA} and \ref{eq:FrpaSRPA}
are only valid if
for each patch $P^{\alpha}_{\Lambda}$ with Fermi velocity
${\vec{v}}^{\alpha}$ there exists an opposite patch
$P^{\bar{\alpha}}_{\Lambda}$ with
${\vec{v}}^{\bar{\alpha}} = - {\vec{v}}^{\alpha}$,
see Appendix~\secref{sec:finitepatch}.}
 \begin{equation}
 \Pi_{\rm RPA} ( q ) =
  \frac{ \Pi_{0} ( q ) }{ 1 + f_{\vec{q}} \Pi_{0} ( q ) }
  =  \int_{0}^{\infty} \D \omega S_{\rm RPA} ( {\vec{q}} , \omega ) \frac{  2 \omega}
  { \omega^{ 2} + \omega_{m}^2 }
  \label{eq:PiRPASRPA}
  \; \; \; ,
  \end{equation}
see Eqs.\ref{eq:Srealaxis}--\ref{eq:dynstruc} with $\beta \rightarrow \infty$. 
Hence
 \begin{eqnarray}
  f^{\rm RPA}_{q} & = & \frac{ f_{\vec{q}} }{ 1 + f_{\vec{q}} \Pi_{0} ( q ) }
  = f_{\vec{q}} - f_{\vec{q}}^2 
  \frac{ \Pi_{0} ( q ) }{ 1 + f_{\vec{q}} \Pi_{0} ( q ) }
  \nonumber
  \\
  & = & f_{\vec{q}} - f_{\vec{q}}^2 
  \int_{0}^{\infty} \D \omega S_{\rm RPA} ( {\vec{q}} , \omega ) 
  \frac{  2 \omega}
  { \omega^{2} + \omega_{m}^2 }
  \label{eq:FrpaSRPA}
  \; \; \; .
  \end{eqnarray}
The advantage of introducing the dynamic structure factor is that 
it is by construction a real non-negative function,
see Eq.\ref{eq:Sstrucspec}.
Furthermore, the qualitative behavior of the dynamic structure factor
can be understood from simple intuitive arguments \cite{Pines89}, which
is very helpful for the evaluation of complicated integrals.

\section{Manipulations with the help of the dynamic structure factor}
\label{sec:exactman}
 
{\it{By introducing the spectral function of the RPA polarization
(i.e. the dynamic structure factor), we can
perform the Matsubara sum at the very beginning of the calculation, and then make some
general statements which are valid irrespective of the precise form of the interaction.}}

\subsection{Non-linear energy dispersion\index{Debye-Waller factor!spectral representation}}

\noindent
Although in the rest of this chapter we shall for
simplicity work with linearized energy dispersion,
it is convenient to consider first the
Debye-Waller factor $Q^{\alpha} ( {\vec{r}} , \tau )$ for
quadratic energy dispersion. 
Substituting the spectral representation
\ref{eq:FrpaSRPA} into
Eqs.\ref{eq:Qlondef2}--\ref{eq:Slondef2},
the Matsubara sum over $\omega_m$
can be performed trivially, 
and we obtain\footnote{For simplicity we assume hat the effective mass
tensor ${\tens{M}}^{\alpha}$ is proportional to the unit matrix.
For general anisotropic effective mass tensor 
one should simply make the
replacement $\frac{ {\vec{q}}^2}{m^{\alpha} }
\rightarrow \xi^{\alpha}_{\vec{q}} + \xi^{\alpha}_{- {\vec{q}} }
= {\vec{q}} ({\tens{M}}^{\alpha})^{-1} {\vec{q}}$
in Eqs.\ref{eq:R2new}--\ref{eq:ImS2new}.} 
for $\beta \rightarrow \infty$
 \begin{eqnarray}
 R^{\alpha}_1 & = &   
 \frac{1}{V} \sum_{\vec{q}} f_{\vec{q}}
 \frac{  {\rm sgn} ( \xi^{\alpha}_{\vec{q}}  ) }{ 
 (- \frac{{\vec{q}}^2}{m^{\alpha}}) }
 \nonumber
 \\
 &  - & 
 \frac{1}{V} \sum_{\vec{q}} f_{\vec{q}}^2
 \int_{0}^{\infty} \D \omega
 S_{\rm RPA} ( {\vec{q}} , \omega )
 \frac{ 2 {\rm sgn} ( \xi^{\alpha}_{\vec{q}}  ) }
 { ( - \frac{ {\vec{q}}^2}{m^{\alpha}} ) ( \omega + | \xi^{\alpha}_{\vec{q}} |) }
 \label{eq:R2new}
 \; \; \; .
 \end{eqnarray}
 \begin{eqnarray}
 {\rm Re} S_1^{\alpha} ( {\vec{r}} , \tau ) & = &   
 \frac{1}{V} \sum_{\vec{q}} 
 \cos ( {\vec{q}} \cdot {\vec{r}} ) 
 f_{\vec{q}}
 \frac{ {\rm sgn} ( \xi^{\alpha}_{\vec{q}}  ) }{ (- \frac{ {\vec{q}}^2}{ m^{\alpha} }) }
 \E^{ - | {\xi}^{\alpha}_{\vec{q}} | | \tau | }
 \nonumber
 \\
 & - &
 \frac{1}{V} \sum_{\vec{q}} 
 \cos ( {\vec{q}} \cdot {\vec{r}} ) 
 f_{\vec{q}}^2
 \int_{0}^{\infty} \D \omega
 S_{\rm RPA} ( {\vec{q}} , \omega )
 \frac{ 2 {\rm sgn} ( \xi^{\alpha}_{\vec{q}}  ) }
 { ( - \frac{ {\vec{q}}^2}{ m^{\alpha} } ) ( \omega + | \xi^{\alpha}_{\vec{q}} |) }
 \nonumber
 \\
 & & \hspace{25mm} \times
 \left[
 \frac{ \omega \E^{ - | {\xi}^{\alpha}_{\vec{q}} | | \tau | }
 - | {\xi}^{\alpha}_{\vec{q}} | \E^{- \omega | \tau | } }{ \omega
 - | {\xi}^{\alpha}_{\vec{q}} | }
 \right]
 \; \; \; ,
 \label{eq:ReS2new}
 \end{eqnarray}
 \begin{eqnarray}
 {\rm Im} S_1^{\alpha} ( {\vec{r}} , \tau ) & = &   
 \frac{{\rm sgn} ( \tau ) }{V} \sum_{\vec{q}} 
 \sin ( {\vec{q}} \cdot {\vec{r}} )
 f_{\vec{q}}
 \frac{ 
 \E^{ - | {\xi}^{\alpha}_{\vec{q}} | | \tau | }}{
 (- \frac{ {\vec{q}}^2}{ m^{\alpha} }) }
 \nonumber
 \\
 & - & 
 \frac{{\rm sgn} ( \tau )}{V} \sum_{\vec{q}} 
  \sin ( {\vec{q}} \cdot {\vec{r}} ) 
 f_{\vec{q}}^2
 \int_{0}^{\infty} \D \omega
 S_{\rm RPA} ( {\vec{q}} , \omega )
 \frac{ 2 \omega  }
 { ( - \frac{ {\vec{q}}^2}{ m^{\alpha} } ) ( \omega + | \xi^{\alpha}_{\vec{q}} |) }
 \nonumber
 \\
 & & 
 \hspace{32mm} \times
 \left[
 \frac{  \E^{ - | {\xi}^{\alpha}_{\vec{q}} | | \tau | }
 - \E^{- \omega | \tau | } }{ \omega
 - | {\xi}^{\alpha}_{\vec{q}} | }
 \right]
 \; \; \; .
 \label{eq:ImS2new}
 \end{eqnarray}

\subsection{The limit of linear energy dispersion}

{\it{We now carefully take the limit $1/m^{\alpha} \rightarrow 0$
in Eqs.\ref{eq:R2new}--\ref{eq:ImS2new}.
In this way we obtain the spectral representation of the
Debye-Waller factor for linearized energy dispersion.
}}

\vspace{7mm}

\noindent
At the first sight it seems that 
Eqs.\ref{eq:R2new}--\ref{eq:ImS2new} 
diverge for
$1/m^{\alpha} \rightarrow 0$,  
because the integrand is proportional to $m^{\alpha}$.
However, this factor is cancelled when we perform the integration, 
because the contribution from the
regimes $\vec{v}^{\alpha} \cdot \vec{q} \geq 0$ and
$\vec{v}^{\alpha} \cdot \vec{q} \leq 0 $
to Eqs.\ref{eq:R2new}--\ref{eq:ImS2new}
almost perfectly cancel in such a way that
the integral is finite.
To obtain the constant part $R^{\alpha}$
of the Debye-Waller factor for linearized energy dispersion
($\xi^{\alpha}_{\vec{q}} \approx \vec{v}^{\alpha} \cdot \vec{q}$),
we expand the second term in Eq.\ref{eq:R2new}
to first order in $1 / m^{\alpha}$,
 \begin{eqnarray}
 \frac{1}{\omega + | \xi^{\alpha}_{\vec{q}} | } =
 \frac{1}{ \omega + | \vec{v}^{\alpha} \cdot \vec{q} +
 \frac{{\vec{q}}^{2}}{2 m^{\alpha} }| }
 \nonumber
 \\
 &   &  \hspace{-50mm} = 
 \frac{1}{  \omega + | \vec{v}^{\alpha} \cdot \vec{q} | }
 - 
 \frac{\vec{q}^2}{2 m^{\alpha} } 
 \frac{
 {\rm sgn} ( \vec{v}^{\alpha} \cdot \vec{q} ) }{ 
 ( \omega + | \vec{v}^{\alpha} \cdot \vec{q} |)^2 }
 + O \left( 1/ ( m^{\alpha} )^{2} \right)
 \label{eq:R1expansion}
 \; \; \; .
 \end{eqnarray}
By symmetry the first term yields a vanishing contribution
to Eq.\ref{eq:R2new}, but the contribution
from the second term in Eq.\ref{eq:R1expansion} is finite
and independent of $m^{\alpha}$.
The expansion of the term
$ {\rm sgn} (  \xi^{\alpha}_{\vec{q} }) $
in Eq.\ref{eq:R2new} to first order in $1/ m^{\alpha}$ does {\it{not}} contribute 
to the Debye-Waller factor
in the limit $|m^{\alpha}|
\rightarrow \infty$. This is perhaps not so obvious, because
the expansion of
$ {\rm sgn} (  \xi^{\alpha}_{\vec{q} }) $
in powers of $ 1/ m^{\alpha}$ produces also a term
of order $1 / m^{\alpha}$,
 \begin{eqnarray}
 {\rm sgn} \left( \vec{v}^{\alpha} \cdot \vec{q}  + \frac{{\vec{q}}^2 }{  2 m^{\alpha} } \right) 
 & = & \Theta
 \left( \vec{v}^{\alpha} \cdot \vec{q}  + \frac{{\vec{q}}^2 }{  2 m^{\alpha} } \right) 
 - \Theta
 \left( - \vec{v}^{\alpha} \cdot \vec{q}  - \frac{{\vec{q}}^2 }{  2 m^{\alpha} } \right) 
 \nonumber
 \\
 & \approx & \Theta
 ( \vec{v}^{\alpha} \cdot \vec{q}   ) 
 - \Theta
 ( - \vec{v}^{\alpha} \cdot \vec{q}  ) 
 + 
 \delta ( \vec{v}^{\alpha} \cdot \vec{q})   \frac{{\vec{q}}^2 }{   m^{\alpha} }  
 \label{eq:signexpand}
 \; \; \; .
 \end{eqnarray}
In the limit $m^{\alpha} \rightarrow \infty$
the last term in Eq.\ref{eq:signexpand} gives rise to the following contribution
to $R^{\alpha}$,
 \begin{equation}
 \delta R^{\alpha} = - \frac{1}{V} \sum_{\vec{q}} 
 \delta ( \vec{v}^{\alpha} \cdot \vec{q} )
 f_{\vec{q}} \left[ 1 -
 f_{\vec{q}} \int_{0}^{\infty} \D \omega 
 \frac{ 2 S_{\rm RPA} ( \vec{q} ,\omega )}{\omega}
 \right]
 \label{eq:deltaRexp}
 \; \; \; .
 \end{equation} 
The two terms in the square braces are due to the
first and second term in 
Eq.\ref{eq:R2new}. 
From Eq.\ref{eq:PiRPASRPA} we have
 \begin{equation}
  \int_{0}^{\infty} \D \omega \frac{ 2 S_{\rm RPA} ( {\vec{q}} , \omega )}{\omega}
  =
 \Pi_{\rm RPA} ( \vec{q} , 0 ) 
  \label{eq:PiRPASRPAstat}
  \; \; \; ,
  \end{equation}
so that $\delta R^{\alpha}$ can also be written as
 \begin{equation}
 \delta R^{\alpha} = - \frac{1}{V} \sum_{\vec{q}} 
 \delta ( \vec{v}^{\alpha} \cdot \vec{q} )  f^{\rm RPA}_{\vec{q} , 0}
 \label{eq:deltaRfinal}
 \; \; \; ,
 \end{equation}
where $ f^{\rm RPA}_{\vec{q} , 0} =
f^{\rm RPA}_{\vec{q} , \I \omega_m = 0}$ is the static RPA interaction. 
Although the contribution \ref{eq:deltaRfinal} is non-zero, 
it is exactly cancelled by a corresponding contribution 
$\delta S^{\alpha}$ that is generated by expanding
$\rm{sgn} ( \xi^{\alpha}_{\vec{q}})$ in Eq.\ref{eq:ReS2new},
 \begin{equation}
 \delta S^{\alpha} = - \frac{1}{V} \sum_{\vec{q}} 
 \delta ( \vec{v}^{\alpha} \cdot \vec{q} )  f^{\rm RPA}_{\vec{q} , 0}
 \cos ( {\vec{q}} \cdot \vec{r} )
 \label{eq:deltaSexp}
 \; \; \; .
 \end{equation} 
Noting that for linearized energy dispersion 
we may replace $\vec{r} \rightarrow r^{\alpha}_{\|} \hat{\vec{v}}^{\alpha}$
in the Debye-Waller factor
(see Eqs.\ref{eq:Gpatchreal1} and \ref{eq:Galphartreplace}), and 
using
 \begin{equation}
 \delta ( \vec{v}^{\alpha} \cdot \vec{q} )  
 \cos ( \hat{\vec{v}}^{\alpha} \cdot {\vec{q}} r^{\alpha}_{\|} )
 =
 \delta ( \vec{v}^{\alpha} \cdot \vec{q} )   
 \; \; \; ,
 \label{eq:deltareplace}
 \end{equation}
it is obvious that $\delta R^{\alpha} - \delta S^{\alpha} = 0$.
We conclude that in the limit
$1/ m^{\alpha} \rightarrow 0$ 
the constant part of the Debye-Waller factor is given by
\index{quasi-particle residue!bosonization result}
 \begin{equation}
 R^{\alpha} = - \frac{1}{V} \sum_{\vec{q}} f_{\vec{q}}^2
 \int_{0}^{\infty} \D \omega \frac{S_{\rm RPA} 
 ( {\vec{q}} , \omega )}{ ( \omega + | {\vec{v}}^{\alpha}
 \cdot {\vec{q}} | )^2}
 \label{eq:R2}
 \; \; \; .
 \end{equation}
Recall that the dynamic structure factor is real and positive by construction
(see Eq.\ref{eq:Sstrucspec}), so that it is clear that
$R^{\alpha}$ is a real negative number. 
Because for linearized energy dispersion the
quasi-particle residue is
given by $ Z^{\alpha} = \E^{ R^{\alpha} }$
(see Eq.\ref{eq:ZRrelation}),
the bosonization result for the Green's function
is for arbitrary interactions in accordance with the requirement
 \begin{equation}
 0 \leq Z^{\alpha} \leq 1
 \; \; \; .
 \label{eq:Zunity}
 \end{equation}
Note also that 
in a weak-coupling expansion the leading term
in Eq.\ref{eq:R2} is of the second order in the bare
interaction, so that
the leading interaction contribution to the quasi-particle residue
$Z^{\alpha} \approx 1 + R^{\alpha}$ is of
order $f_{\vec{q}}^2$. This is in agreement with
perturbation theory.
For non-linear energy dispersion the term $R^{\alpha}_1$
has a non-vanishing contribution that is first order in $f_{\vec{q}}$.
This is not in contradiction with perturbation theory,
because the quantity $\E^{R^{\alpha}_1}$ 
cannot be identified with the quasi-particle residue $Z^{\alpha}$ any more;
the function $Y^{\alpha} ( \tilde{q} )$ gives rise to an 
additional contribution\footnote{
From Eq.\ref{eq:sigma1res2} we see that, at least
for not too singular interactions,
$\Sigma_1^{\alpha} ( {\vec{q}} = 0 , \I \tilde{\omega}_n ) = 0$, so that
$G^{\alpha}_1$ does not renormalize the quasi-particle
residue.}
to $Z^{\alpha}$. 
From Chap.~\secref{sec:connectionPT}
we know that by construction our method produces the
correct perturbative result, so that the leading
corrections to $Z^{\alpha}$ are of the second
order in the bare interaction.

Similarly we obtain 
from Eqs.\ref{eq:ReS2new} and \ref{eq:ImS2new}
after a tedious but straightforward calculation
in the limit $1/m^{\alpha} \rightarrow  0$ 
 \begin{eqnarray}
 {\rm Re} S^{\alpha} ( r^{\alpha}_{\|} \hat{\vec{v}}^{\alpha} , \tau ) & = &
    \frac{1}{ V} \sum_{ {\vec{q} }} 
  \cos ( \hat{\vec{v}}^{\alpha} \cdot {\vec{q}} r^{\alpha}_{\|} )
 \nonumber
 \\
 & & \hspace{-25mm}
  \times \left\{ L^{\alpha}_{\vec{q}} ( \tau ) 
   - f_{\vec{q}}^2 
 \int_{0}^{\infty} \D \omega \frac{S_{\rm RPA} 
 ( {\vec{q}} , \omega )}{ ( \omega + | {\vec{v}}^{\alpha}
 \cdot {\vec{q}} | )^2}
  \right.
  \nonumber
  \\
  & &  \hspace{-10mm}
  \left.
  \times
   \frac{
 [ ( {\vec{v}}^{\alpha} \cdot {\vec{q}} )^2  + \omega^2 ] \E^{- \omega | \tau | }
 -  2 | {\vec{v}}^{\alpha} \cdot {\vec{q}} |  
 \omega \E^{ - | {\vec{v}}^{\alpha} \cdot {\vec{q}} | | \tau | } }
 { ( \omega -  | {\vec{v}}^{\alpha} \cdot {\vec{q}} | )^2 }
 \right\}
 \; \; \; ,
 \label{eq:ReS2b}
 \end{eqnarray}
 \begin{eqnarray}
 {\rm Im} S^{\alpha} ( r^{\alpha}_{\|} \hat{\vec{v}}^{\alpha} , \tau ) & = &
  \frac{{\rm sgn} ( \tau )  }{ V} \sum_{ {\vec{q} }} 
  \sin ( | \hat{\vec{v}}^{\alpha} \cdot {\vec{q}} | r^{\alpha}_{\|} )
  \nonumber
  \\
  & & \hspace{-25mm} \times
  \left\{ L^{\alpha}_{\vec{q}} ( \tau ) 
   - f_{\vec{q}}^2 
 \int_{0}^{\infty} \D \omega \frac{S_{\rm RPA} ( {\vec{q}} , 
 \omega )}{ ( \omega + | {\vec{v}}^{\alpha}
 \cdot {\vec{q}} | )^2}
 \right.
 \nonumber
 \\
 & & 
 \left.
 \hspace{-10mm} \times
 2 | {\vec{v}}^{\alpha} \cdot {\vec{q}} | \omega 
  \frac{
  \E^{ - \omega | \tau | }
 -  \E^{ - | {\vec{v}}^{\alpha} \cdot {\vec{q}} | | \tau | }  }
 { ( \omega -  | {\vec{v}}^{\alpha} \cdot {\vec{q}} | )^2 }
 \right\}
 \;  ,
 \label{eq:ImS2b}
 \end{eqnarray}
where we have defined
 \begin{equation}
 \hspace{-4mm}
 L^{\alpha}_{\vec{q}} (  \tau ) = 
  \frac{ | \tau |}{2} f_{\vec{q}} \E^{ - | {\vec{v}}^{\alpha} \cdot {\vec{q}} | | \tau | }
  \left[
  1 - f_{\vec{q}} 
 \int_{0}^{\infty} \D \omega S_{\rm RPA} ( {\vec{q}} , \omega ) 
 \frac{ 2 \omega }{   \omega^2 - ( {\vec{v}}^{\alpha}
 \cdot {\vec{q}}  )^2   }
 \right]
 \label{eq:Llintau}
 \; \; \; .
 \end{equation}
We emphasize again that
after the linearization we may replace
$\vec{r} \rightarrow r^{\alpha}_{\|} \hat{\vec{v}}^{\alpha}$
in the argument of the Debye-Waller factor, because
in this case the prefactor Green's function
$G^{\alpha}_0 ( \vec{r} , \tau )$ is
proportional to $\delta^{(d-1)} ( \vec{r}^{\alpha}_{\bot} )$
(see Eqs.\ref{eq:Gpatchreal1} and \ref{eq:Galphartreplace}).
In contrast, Eqs.\ref{eq:R2new}--\ref{eq:ImS2new} should be
considered for all $\vec{r}$.

\subsection{Finite versus infinite patch number}

\label{subsec:Thetermslin}
 
{\it{Now comes a really subtle point related to the fact that 
for linearized energy dispersion 
we cover the
Fermi surface with a finite number of patches.}}

\vspace{7mm}

\noindent
The term $L^{\alpha}_{\vec{q}} ( \tau )$ in Eqs.\ref{eq:ReS2b} and \ref{eq:ImS2b} 
is mathematically closely related to the existence of a {\it{double pole}}
in the integrand defining the Debye-Waller factor
for linear energy dispersion. \index{double pole}
When the $\omega_m$-integral in
Eqs.\ref{eq:Rlondef} and \ref{eq:Slondef}
is done by means of contour integration,
the double pole at $\I \omega_m = {\vec{v}}^{\alpha} \cdot \vec{q}$
gives rise to a contribution proportional to the
derivative of the rest of the integrand with respect the to frequency;
the resulting term is therefore proportional
to $\tau$, and can be
identified with $L^{\alpha}_{\vec{q}} ( \tau )$.
However, as long as
{\it{the Fermi surface is covered by a finite number $M$ of patches}}
we have exactly
 \begin{equation}
 L^{\alpha}_{\vec{q}} (  \tau ) = 0
 \; \; \; .
 \label{eq:Llintau0}
 \end{equation}
To prove this, we use
Eq.\ref{eq:dynstruc} to rewrite Eq.\ref{eq:Llintau} as
 \begin{eqnarray}
 L^{\alpha}_{\vec{q}} (  \tau ) & = &
  \frac{ | \tau |}{2} f_{\vec{q}} 
  \left[ 1 - f_{\vec{q}} \Pi_{\rm RPA} ( {\vec{q}} , {\vec{v}}^{\alpha} \cdot {\vec{q}}  ) \right]
  \E^{ - | {\vec{v}}^{\alpha} \cdot {\vec{q}} | | \tau | }
  \nonumber
  \\
  & = &
  \frac{ | \tau |}{2} \frac{f_{\vec{q}}}{ \epsilon_{\rm RPA} 
  ( {\vec{q}} , {\vec{v}}^{\alpha} \cdot {\vec{q}} )}
  \E^{ - | {\vec{v}}^{\alpha} \cdot {\vec{q}} | | \tau | }
  \label{eq:Llintau2}
  \; \; \; ,
  \end{eqnarray}
where the  RPA dielectric function\index{dielectric function!for finite patch number}
at frequency $\omega = {\vec{v}}^{\alpha} \cdot {\vec{q}}$
is
(see Eqs.\ref{eq:dielectricdef}, \ref{eq:Pitotdecompose} and \ref{eq:Pilong})
 \begin{equation}
 \epsilon_{\rm RPA} ( {\vec{q}} ,  {\vec{v}}^{\alpha} \cdot {\vec{q}} )  = 
 1+  f_{\vec{q}} \Pi_{0} ( {\vec{q}} , {\vec{v}}^{\alpha} \cdot {\vec{q}} )  
 \; \; \; , 
 \end{equation}
with
 \begin{equation}
  \Pi_{0} ( {\vec{q}} , {\vec{v}}^{\alpha} \cdot {\vec{q}}  )  = 
 \sum_{\alpha^{\prime} = 1 }^{M} \nu^{\alpha^{\prime}} 
 \frac{ {\vec{v}}^{\alpha^{\prime}} \cdot {\vec{q}} }
 { ( {\vec{v}}^{\alpha^{\prime}} -  {\vec{v}}^{\alpha} ) \cdot {\vec{q}} } 
 \label{eq:RPApol2}
 \; \; \; .
 \end{equation}
Evidently the term $\alpha^{\prime} = \alpha$ in Eq.\ref{eq:RPApol2} is divergent, so that 
 $\Pi_{0} ( {\vec{q}} , {\vec{v}}^{\alpha} \cdot {\vec{q}} )$ 
and hence also the dielectric function at
frequency $\omega = {\vec{v}}^{\alpha} \cdot {\vec{q}}$ are infinite.
It follows that
 \begin{equation}
  \frac{f_{\vec{q}}}{ \epsilon^{RPA} ( {\vec{q}} , {\vec{v}}^{\alpha} \cdot {\vec{q}} )}
  = 0
  \label{eq:frpaonshell}
  \; \; \; ,
  \end{equation}
so that from Eq.\ref{eq:Llintau2} we can conclude that 
$L^{\alpha}_{\vec{q}} ( \tau ) = 0$.

This proof does not go through any more if we take the limit
of an infinite number of patches,\index{patch!limit of infinite number} 
because then 
the $\alpha^{\prime}$-summation in Eq.\ref{eq:RPApol2} is for $d > 1$
replaced by an angular integration, and the singularity in the integrand
must be regularized via the usual pole prescription
${\vec{v}}^{\alpha} \cdot {\vec{q}} \rightarrow
{\vec{v}}^{\alpha} \cdot {\vec{q}} + \I 0^{+}$.
Then in $d > 1$ the function
$\Pi_{0} ( {\vec{q}} , {\vec{v}}^{\alpha} \cdot {\vec{q}} + \I 0^{+}  )$ is finite.
For example, for a spherical Fermi surface 
$\Pi_{0} ( {\vec{q}} , {\vec{v}}^{\alpha} \cdot   {\vec{q}} + \I 0^{+} ) = 
\nu g_{d} ( \hat{\vec{v}}^{\alpha} \cdot \hat{\vec{q}} + \I 0^{+}) $,
where the function $g_d ( x + \I 0^{+} )$ is
given in Eq.\ref{eq:gddef}. 
In other words, in the limit $M \rightarrow \infty$ the singularity 
in $\Pi_{0} ( {\vec{q}} , {\vec{v}}^{\alpha} \cdot {\vec{q}} )$ 
is regularized by the finite imaginary part
of the function $g_{d} ( x + i 0^{+})$ for $x<1$, see Eq.\ref{eq:Imgddef}.

The above difference between the cases $M < \infty$ and $M = \infty$
is due to qualitatively different
behavior of the dynamic structure factor in both cases.
As discussed in detail in Appendix \secref{sec:finitepatch},
for $M < \infty$  
the dynamic structure factor
$S_{\rm RPA} ( \vec{q} , \omega )$ exhibits
$M$ delta-function peaks.
For $M \rightarrow \infty$ only two of these peaks
survive and
can be identified with the undamped plasmon mode
at frequencies $\pm \omega_{\vec{q}}$, while the
other peaks merge into the particle-hole continuum. 
From the formal point of view the
procedure of substituting the
infinite-patch limit for the dynamic structure
factor into 
the Debye-Waller factor for linearized energy dispersion
(see Eqs.\ref{eq:ReS2b} and \ref{eq:ImS2b})
is certainly not satisfactory, because the approximations
used to derive these equations are only valid
as long as the sector cutoffs $\Lambda$ and $\lambda$ are
kept finite and large compared with the range $q_{\rm c}$ of the
interaction in momentum space, see Fig.\secref{fig:qc}.
But $M \rightarrow \infty$ implies that we are taking the limit
$\Lambda \rightarrow 0$, so that for  fixed $q_{\rm c}$ 
the condition $q_{\rm c} \ll \Lambda$ (see Eq.\ref{eq:qccond})
cannot be satisfied.

Obviously the problem associated with the limit of infinite
patch number does not arise in 
our more general results \ref{eq:R2new}--\ref{eq:ImS2new}
for non-linear energy dispersion,
because in this case
the dynamic structure factor exhibits the particle-hole
continuum even if we work with a finite number of patches, and
a term similar to $L^{\alpha}_{\vec{q}} ( \tau )$ that is linear
in $\tau$ simply does not appear,
because there is no double pole in the Debye-Waller factor.
The disadvantage of Eqs.\ref{eq:R2new}--\ref{eq:ImS2new} is that
these expressions are more difficult to evaluate
than the corresponding expressions 
for linearized energy dispersion.
Fortunately, at $\tau = 0$ we have 
$L^{\alpha}_{\vec{q}} ( 0 ) = 0$, so that
possible ambiguities related to the limit of infinite patch
number in the linearized theory
do not appear in all quantities involving the
{\it{static}} Debye-Waller factor $Q^{\alpha} ( r^{\alpha}_{\|} \hat{\vec{v}}^{\alpha} , 0 )$.
In this case the use of the $M = \infty$ limit for the dynamic
structure factor in the Debye-Waller factor for linearized
energy dispersion seems to be 
justified\footnote{By taking the limit $M \rightarrow \infty$
in the Debye-Waller factor, we also eliminate 
artificial nesting singularities, which are generated
if the covering of the Fermi surface contains at least
two parallel  patches, 
see Chap.~\secref{sec:nesting}. 
In this sense the limit
$M \rightarrow \infty$ is really the physical limit of interest, although
for linearized energy dispersion it is not possible to give
a formally convincing justification for this
limiting procedure. Of course, in case of ambiguities we can always
go back to our more general results \ref{eq:R2new}--\ref{eq:ImS2new}.}, 
at least as long
the patch cutoffs are small compared with $k_{\rm F}$.
In the rest of this chapter we shall therefore
focus on the
static Debye-Waller factor $Q^{\alpha} ( r_{\|}^{\alpha} 
\hat{\vec{v}}^{\alpha} , 0 )$ for linearized energy dispersion,
and use the 
$M \rightarrow \infty$ limit for the dynamic structure factor.

\section[The static Debye-Waller factor for linearized energy dispersion]
{The static Debye-Waller factor \mbox{\hspace{40mm}}
for linearized energy dispersion}
\label{sec:singular}

{\it{We now explicitly evaluate 
$Q^{\alpha} ( r^{\alpha}_{\|} \hat{\vec{v}}^{\alpha}, 0 )$
for singular interactions of the
form \ref{eq:fgeneric} 
for a spherically symmetric $d$-dimensional system.
\index{singular interactions!density-density}
We show that the Fermi liquid state is only stable
for $ \eta < 2 ( d-1 )$, but that in the interval
$2 ( d-2) < \eta < 2 ( d-1)$ the sub-leading corrections
are anomalously large. We then consider the
regime $ \eta  \geq 2 (d-1)$, and show that
for $ \eta \geq 2 ( d + 1)$
the bosonization result for the equal-time 
Debye-Waller factor $Q^{\alpha} ( r^{\alpha}_{\|} \hat{\vec{v}}^{\alpha} , 0 )$
is mathematically not well-defined.}}

\subsection{Consequences of spherical symmetry
\index{quasi-particle residue!bosonization result}}
\index{spherical symmetry}
\label{subsec:rot}

\noindent
For a spherically symmetric Fermi surface
we have $| {\vec{v}}^{\alpha} | = v_{\rm F}$ for all $\alpha$,
so that the non-interacting sector Green's function 
given in Eq.\ref{eq:Gpatchreal1} can be written as
 \begin{equation}
 G^{\alpha}_0 ( {\vec{r}}  , \tau  )
 = 
 \delta^{(d-1)} ( {\vec{r}}^{\alpha}_{\bot}  )
 {G}_{0} ( r^{\alpha}_{\|} , \tau )
 \; \; \; , \; \; \; r^{\alpha}_{\|} = \hat{\vec{v}}^{\alpha} \cdot {\vec{r}}
 \; \; \; ,
 \label{eq:Gpatchreal2}
 \end{equation}
where
 \begin{equation}
 {G}_{0} ( x , \tau )
 =
 \left( \frac{ - \I}{2 \pi} \right)
 \frac{1}
 { 
 x  
 + \I v_{\rm F}  \tau }
 \label{eq:G0patchreal2}
 \end{equation}
is the usual one-dimensional non-interacting Green's function.
Note that for a spherical Fermi surface  
the polarization $\Pi_{0} ( q )$  depends at long wavelengths only on the combination
${ \I \omega_{m}}/ ({ v_{\rm F} | {\vec{q}} | })$,
see Eq.\ref{eq:Piglobalspherical}.
It follows that the Debye-Waller factor \ref{eq:Qlondef}
is actually of the form
 $Q^{\alpha} ( r^{\alpha}_{\|} \hat{\vec{v}}^{\alpha} , \tau )
 = Q ( r^{\alpha}_{\|} , \tau )$,
where $Q ( x  , \tau )$ is the following function of two variables $x$ and $\tau$,
 \begin{equation}
 \hspace{-5mm}
 Q ( x , \tau ) = R - S ( x , \tau ) =
 \frac{1}{\beta {{V}}} \sum_{ q }  
  \frac{ f_{  \vec{q}  } 
  \left[ 1 -
  \cos ( \hat{\vec{v}}^{\alpha} \cdot {\vec{q}} x 
  - {\omega}_{m}  \tau  )  \right] }{
  \left[ 1 + f_{  \vec{q}  } \Pi_{0} ( q ) \right]
 ( \I \omega_{m} - {\vec{v}}^{\alpha} \cdot {\vec{q}} )^{2 }}
 \label{eq:DW2}
 \; \; \;  .
 \end{equation}
Due to rotational invariance, the value of the integral
is independent of the direction of the unit vector $\hat{\vec{v}}^{\alpha}$, as can
be easily seen by introducing $d$-dimensional spherical coordinates, 
see Eq.\ref{eq:angavsimplify}.
From Eqs.\ref{eq:Galphartdef} and \ref{eq:Galphartreplace}
we conclude that for rotationally invariant systems the interacting
patch Green's function\index{Green's function!real space} can be written as
 \begin{equation}
 G^{\alpha} ( {\vec{r}}  , \tau  )
 = 
 \delta^{(d-1)} ( {\vec{r}}^{\alpha}_{\bot}  )
 {G}_{0} ( r^{\alpha}_{\|} , \tau ) \E^{ Q ( r^{\alpha}_{\|}  , \tau ) }
 \; \; \; , \; \; \; r^{\alpha}_{\|} = \hat{\vec{v}}^{\alpha} \cdot {\vec{r}}
 \; \; \; .
 \label{eq:Galphart3}
 \end{equation}

\vspace{7mm}

Let us study the constant part $R$ of the Debye-Waller factor
in more detail.
The form of the RPA dynamic structure 
factor\index{dynamic structure factor!within RPA} for spherical Fermi surfaces
is discussed in detail in 
Appendix \secref{sec:dynstruc}. Using
Eqs.\ref{eq:Srpadecomp}, \ref{eq:Ssp} and \ref{eq:Scoldelta},
and taking the limit $V \rightarrow \infty$ in Eq.\ref{eq:R2}, we obtain 
 \begin{eqnarray}
 R & = & -  \int \frac{ \D {\vec{q}} }{ (2 \pi )^d}
  f_{  \vec{q}  }^2 
 \left[ \frac{ Z_{\vec{q}} }{ \left( \omega_{\vec{q}}  + 
 | {\vec{v}}^{\alpha} \cdot {\vec{q}} |  \right)^2 }
 \right.
 \nonumber
 \\
 &  & \left.
   + \frac{\nu}{\pi}
 \int_{0}^{ v_{\rm F} | {\vec{q}}|  }   \D \omega  
 \frac{1}{ \left( \omega + 
 | {\vec{v}}^{\alpha} \cdot {\vec{q}} |  \right)^2 }
 {\rm Im} \left\{ 
 \frac{g_{d} ( \frac{ \omega } {   v_{\rm F} | {\vec{q}} | } + \I 0^{+} ) } 
 { 1 + F_{ \vec{q} } 
 g_{d} ( \frac{ \omega}{   v_{\rm F} | {\vec{q}}| } + \I 0^{+})  }
 \right\}
 \right]
 \; \; \; ,
 \label{eq:Rex1}
 \end{eqnarray}
where the energy $\omega_{\vec{q}}$ and the residue $Z_{\vec{q}}$ of the 
collective  plasmon mode \index{plasmon}
are given in Eqs.\ref{eq:zerosoundsol} and \ref{eq:resdef}.
Using Eq.\ref{eq:nurelation} and the fact that according to
Eqs.\ref{eq:resman} and \ref{eq:Zddef} 
the residue of the plasmon mode is of the form
$Z_{\vec{q}} = \nu v_{\rm F} | {\vec{q}} | Z_{d} ( F_{\vec{q}} )$, 
we obtain 
 \begin{eqnarray}
 R & = & -  \frac{1}{k_{\rm F}^{d-1} \Omega_{d} }
 \int  \frac{\D {\vec{q}}}{ | {\vec{q}} | } F_{\vec{q}}^2
 \left[ \frac{ Z_{d} ( F_{\vec{q}} ) }{ \left( \frac{\omega_{\vec{q}}}{ v_{\rm F} | {\vec{q}} | }  + 
 | \hat{\vec{v}}^{\alpha} \cdot \hat{\vec{q}} |  \right)^2 }
 \right.
 \nonumber
 \\
 & & 
 \left.
 \hspace{10mm}
 +
 \int_{0}^{1} \frac{ \D x}{\pi} 
 \frac{1}{ \left(  x   +
 | \hat{\vec{v}}^{\alpha} \cdot \hat{\vec{q}} |  \right)^2 }
 {\rm Im} \left\{
 \frac{ g_{d} ( x + \I 0^{+} ) }
 { 1 + F_{\vec{q}} g_{d} ( x + \I 0^{+} ) }
 \right\} \right]
 \; \; \; ,
 \label{eq:Rex2}
 \end{eqnarray}
where  we have introduced the
usual dimensionless interaction $F_{\vec{q}} = \nu f_{\vec{q}}$.

Because by assumption $F_{\vec{q}} $  
depends only on $| {\vec{q}} |$,
the angular integration
can be expressed in terms of the function
 \begin{equation}
 h_{d} ( x ) = \left<
 \frac{ 1 }{ \left(  x + | \hat{\vec{v}}^{\alpha} \cdot \hat{\vec{q}} | \right)^2} \right>_{\hat{\vec{q}} }
 \label{eq:hddef}
 \; \; \; ,
 \end{equation}
where the angular average is defined as in Eqs.\ref{eq:angav}, \ref{eq:angavsimplify} and
\ref{eq:angavsimp1}. In $d=1$ we obtain
 \begin{equation}
 h_{1} (x ) = \frac{1}{( x + 1 )^2}
 \label{eq:h1res}
 \; \; \; ,
 \end{equation}
and in $d > 1$ 
 \begin{equation}
 h_{d} ( x ) = \gamma_{d}  \int_{0}^{\pi} \D \vartheta
 \frac{ ( \sin \vartheta )^{d-2} }{ ( x + | \cos \vartheta | )^2 }
 \label{eq:hdang}
 \; \; \; ,
 \end{equation}
with $\gamma_{d}$ given in Eq.\ref{eq:gammad2}.
In particular, in $d=2$ we have
 \begin{equation}
 h_{2} ( x ) =
 \frac{2}{\pi} \times
 \left\{ 
 \begin{array}{ll}
 \frac{1}{1 - x^2 } \left[  \frac{1}{x} - \frac{ x}{\sqrt{ 1 - x^2  } } 
 \ln \left( \frac{ 1 + \sqrt{ 1 - x^2}}{x} \right) \right] & \mbox{for $ x < 1$}
 \\
 \frac{2}{3 } & \mbox{for $ x = 1$ }
 \\
 \frac{1}{x^2 - 1 } \left[ - \frac{1}{x} + \frac{ x}{\sqrt{ x^2 - 1 } } 
 \arccos ( \frac{1}{x} ) \right] & \mbox{for $ x > 1$}
 \end{array}
 \right.
 \; \; \; ,
 \end{equation}
while in $d=3$ the result is simply
 \begin{equation} 
 h_{3} (x )  =  \frac{1}{x ( x+ 1 )}
 \; \; \; .
 \label{eq:h3angav}
 \end{equation}
For large and small $x$ we have 
 \begin{eqnarray}
 h_{d} ( x) & \sim & \frac{1}{x^2} \; \; \; , \; \; \; x \rightarrow \infty
 \; \; \; ,
 \label{eq:hdlarge}
 \\
 h_{d} ( x) & \sim & \frac{2 \gamma_{d}}{x} \; \; \; , \; \; \; x \rightarrow 0 
 \; \; \; , \; \; \; d > 1
 \label{eq:hdsmall}
 \; \; \; .
 \end{eqnarray}

We are now ready to rewrite Eq.\ref{eq:Rex2} in terms of rescaled variables.
Using Eq.\ref{eq:Zddef}
and the fact that ${\omega_{\vec{q}} }/ ({ v_{\rm F} | {\vec{q}} |})$
is according to Eq.\ref{eq:zerosoundsol} a function of $F_{  {\vec{q}}  }$,
we obtain 
 \begin{equation}
 R = -   \frac{1}{k_{\rm F}^{d-1}} \int_{0}^{\infty} \D q  q^{d-2}
 \left[ C_{d} ( F_{ q} ) + L_{d} ( F_{ q } ) \right]
 \; \; \; ,
 \label{eq:Rex3}
 \end{equation}
where the dimensionless functions 
$C_{d} (F)$ and $L_{d} ( F )$ are given by
 \begin{eqnarray}
 C_{d} ( F ) & = & 
  F^2 Z_{d} ( F )
 h_{d} \left( g_{d}^{-1} ( - \frac{1}{F} ) \right) 
 =
 \frac{ h_{d} \left( g_{d}^{-1} ( - \frac{1}{F} ) \right) }
 { g_{d}^{\prime} \left( g_{d}^{-1} ( - \frac{1}{F} ) \right) }
 \label{eq:Idcol}
 \; \; \; ,
 \\
 L_{d} ( F ) & = &
 F^2 \int_{0}^{1} \frac{ \D x}{\pi} h_{d} ( x ) 
 {\rm Im} \left\{
 \frac{ g_{d} ( x + \I 0^{+} ) }
 { 1 + F g_{d} ( x + \I 0^{+} ) }
 \right\} 
 \; ,
 \label{eq:IIdef}
 \end{eqnarray}
with $g_d ( z )$ defined in Eq.\ref{eq:gddef}.
Note that $C_d ( F )$ represents the {\it{collective mode}} contribution
to the RPA dynamic structure factor (see Eq.\ref{eq:Scoldelta}), while
$L_d ( F )$ represents the single-pair contribution due
to {\it{Landau damping}} (see Eq.\ref{eq:Ssp}). \index{Landau damping}
The asymptotic behavior of the functions $C_{d}( F )$ and $L_{d}( F )$
determines the parameter regime where the system is a Fermi liquid.
For $F \rightarrow \infty$ we have to leading order
(see Eqs.\ref{eq:gdinvlarge}, \ref{eq:gdprimeinvF} and \ref{eq:hdlarge})
 \begin{equation}
 C_{d} ( F )  \sim \frac{ \sqrt{d}}{2} \sqrt{F}
 \; \; \; , \; \; \; F \rightarrow \infty
 \; \; \; ,
 \label{eq:Gcollarge}
 \end{equation}
while the Landau damping contribution reduces to a finite constant,
 \begin{equation}
 L_{d} ( F )  \sim L_d^{\infty}  
 \equiv - \int_{0}^{1} \frac{ \D x}{\pi} h_{d} ( x )
 {\rm Im} \left\{ \frac{1}{g_{d} ( x + \I 0^{+}) } \right\}
 \; \; \; , \; \; \; F \rightarrow \infty
 \; \; \; .
 \label{eq:Gsplarge}
 \end{equation}
To see more clearly that $L_{d}^{\infty} $ is for all $d$ a finite 
{\it{positive}} constant, note that from Eqs.\ref{eq:angavsimplify} and \ref{eq:Imgddef}
 \begin{eqnarray}
 {\rm Im} \left\{ \frac{1}{g_{d} ( x + \I 0^{+}) } \right\}
 & = & - \frac{ \pi x 
 \left< \delta ( 
 \hat{\vec{q}} \cdot \hat{\vec{k}} - x ) \right>_{  \hat{\vec{k}}  }}
 { | g_{d} ( x + \I 0^{+} ) |^2 }
 \nonumber
 \\
 & = & - \frac{ \pi x 
 \gamma_{d} \int_{0}^{\pi} \D \vartheta ( \sin \vartheta )^{d-2} 
 \delta ( \cos \vartheta - x ) }
 { | g_{d} ( x + \I 0^{+} ) |^2 }
 \label{eq:ImInv}
 \; \; \; ,
 \end{eqnarray}
so that from Eq.\ref{eq:Gsplarge}
 \begin{equation}
 L^{\infty}_{d} = \gamma_{d} \int_{0}^{ \pi / 2 } \D \vartheta
 ( \sin \vartheta )^{d-2} \frac{ \cos \vartheta h_{d} ( \cos \vartheta ) }
 { | g_{d} ( \cos \vartheta + \I 0^{+} ) |^2 }
 \; \; \; .
 \label{eq:Gdinfreg}
 \end{equation}
The integrand in Eq.\ref{eq:Gdinfreg} is non-singular and positive for all
$\vartheta$, so that $0 < L^{\infty}_{d}  < \infty$.
The weak coupling behavior of $C_{d} ( F )$ is easily obtained from
Eq.\ref{eq:gdprimeinvweak},
 \begin{equation}
 C_{d} ( F ) \sim 
 \left\{
  \begin{array}{ll}
  0 & \mbox{  for  } d >3 \mbox{  and $ F < | g_{d} ( 1 ) |^{-1} $ }
  \\
  \E^{ - 2 / F } & \mbox{  for  }  d = 3
  \\
  \frac{ 2 h_{d} (1 ) }{(3-d) c_{d}}  ( c_{d} F)^{ \frac{ 5-d}{3-d}} & \mbox{  for  } d < 3
  \end{array}
  \right.
  \; ,
  \label{eq:Gdcolweak}
  \end{equation}
where the numerical constant $c_{d}$ is defined via Eq.\ref{eq:gdcloseone}.
The Landau damping part is at weak coupling proportional to $F^2$,
 \begin{equation}
 L_{d} ( F )  \sim  L_{d}^{\prime} F^2 
 \; \; \; , \; \; \; F \rightarrow 0
 \label{eq:Gspweak}
 \; \; \; ,
 \end{equation}
where the numerical constant $L_{d}^{\prime}$ is given by
 \begin{equation}
 L_{d}^{\prime} =
 \int_{0}^{\infty} \frac{\D x}{\pi} h_{d} ( x )
 {\rm Im} g_{d} ( x + \I 0^{+}) 
 \; \; \; .
 \label{eq:tildeGspdef}
 \end{equation}
Note that at strong coupling
 \begin{equation}
 \frac{ C_{d} ( F )}{L_{d} ( F ) }
 \sim \frac{ \sqrt{d}}{2 L_{d}^{\infty}  } \sqrt{F}
 \; \; \; , \; \; \; F \rightarrow \infty
 \; \; \; ,
 \label{eq:Gcoldominant}
 \end{equation}
so that the relative weight of the collective mode is always larger than
that of the Landau damping part. In the other hand, at weak coupling
it is easy to see from Eqs.\ref{eq:Gdcolweak} and \ref{eq:Gspweak} that
the Landau damping part is dominant. In particular, for $ 1 < d < 3$ we have
 \begin{equation}
 \frac{ C_{d} ( F )}{L_{d} ( F ) }
  \sim \frac{ 2 h_{d} (1 ) c_{d}^{ \frac{2}{3-d} }}{(3-d) L_{d}^{\prime}   }
    F^{ \frac{ d- 1}{3-d}} 
 \; \; \; , \; \; \; F \rightarrow 0
 \; \; \; .
 \label{eq:Gspdominant}
 \end{equation}
The important point is that for $ 1 < d < 3$ the exponent of $F$ is always positive,
so that for small $F$ the right-hand side of Eq.\ref{eq:Gspdominant}
is indeed small. Hence, the collective mode contribution is negligible at weak coupling.

\subsection{The existence of the quasi-particle residue}
\label{subsec:qpressing}

For singular interactions of the 
form \ref{eq:fgeneric}\index{singular interactions!density-density}
\index{quasi-particle residue!singular interactions}
we have
 $F_{ {\vec{q}} } = (  {\kappa } /  | {\vec{q}} | )^{\eta} 
\E^{ - | {\vec{q}} | /  q_{\rm c}  } $,
see Eq.\ref{eq:Fgeneric}.
Having determined the weak and strong coupling behavior of the
functions $C_{d} ( F )$ and $L_{d} ( F )$ in
Eq.\ref{eq:Rex3}, it is now easy to calculate the quasi-particle residue
for this type of interaction.
Introducing in Eq.\ref{eq:Rex3} the dimensionless 
integration variable $p = q / \kappa $ and setting $p_{\rm c} = q_{\rm c} / \kappa$
we obtain
 \begin{equation}
 R = - \left( \frac{ \kappa }{k_{\rm F}} \right)^{d-1} 
 \tilde{R} ( d, \eta ,  p_{\rm c}  )
 \label{eq:Rex4}
 \; \; \; ,
 \end{equation}
where the dimensionless function $\tilde{R} ( d , \eta ,  p_{\rm c}  )$ is given by
 \begin{equation}
 \tilde{R} ( d , \eta , p_{c} ) = \int_{0}^{\infty} \D p p^{d-2}
 \left[ C_{d} ( p^{- \eta} \E^{ -  p / p_{\rm c} }) + L_{d} ( p^{- \eta} \E^{ - p / p_{\rm c} } )
 \right]
 \label{eq:rdefetad}
 \; \; \; .
 \end{equation}
Because 
the functions $C_{d}( F )$ and $L_{d} ( F)$ 
do not have any singularities at finite values of $F$,
the integral in
Eq.\ref{eq:rdefetad} can only diverge due to possible infrared singularities at small
$p$, or ultraviolet singularities at large $p$.
Let us first consider the infrared limit. Because the exponent $\eta$ is positive, 
this limit is determined by the strong coupling behavior of
$C_{d} ( F )$ and $L_{d} ( F)$. 
From Eq.\ref{eq:Gcoldominant} we know that in this limit the
collective mode is dominant, so that the 
most singular contribution
arises from the first term in Eq.\ref{eq:rdefetad}. 
Using Eq.\ref{eq:Gcollarge}, it is easy to see that 
this term yields 
 \begin{eqnarray}
 \tilde{R} ( d , \eta , p_{\rm c} ) 
 & \sim & \frac{\sqrt{d}}{2} \int_{0}^{p_{\rm c}} \D p p^{d-2 - \frac{\eta }{ 2}}
 \nonumber
 \\
 & =  &
 \frac{ \sqrt{d}}{2}
 \frac{ p_{\rm c}^{d-1 - \frac{\eta }{ 2}}}{ d -1 - \frac{\eta }{ 2}}
 \; \; \; , \; \; \; \mbox{for $\eta < 2 (d - 1 )$}
 \; \; \; .
 \label{eq:BaresWen}
 \end{eqnarray}
Evidently $\tilde{R} ( d ,\eta , p_{c} ) = \infty $ for $ \eta \geq  2 ( d-1)$,
so that in this case $R = - \infty$. We conclude that
 \begin{equation}
 Z = 0 \; \; \; , \; \; \; \mbox{for $ \eta \geq \eta_{\rm ir} \equiv 2 (d-1)$ }
 \label{eq:Zbareswen}
 \; \; \; .
 \end{equation}
Therefore, the Fermi liquid is only stable for $\eta < 2 ( d-1 )$.
This result has first been derived by Bares and Wen \cite{Bares93}.
\index{singular interactions!density-density} 

Another special value for the exponent $\eta$ 
is determined by the requirement that
the integral in Eq.\ref{eq:rdefetad} is convergent
even without ultraviolet
cutoff $p_{\rm c}$. 
Assuming that we have eliminated
the high-energy degrees of freedom outside a thin shell of thickness
$\lambda$ around the Fermi surface,
we should choose $q_{\rm c} \approx \lambda$ and hence
$p_{\rm c} = \lambda / \kappa$. 
Because in practice we cannot explicitly perform the
integration over the high-energy
degrees of freedom, it is important that at the end 
of the calculation physical quantities do
not depend on $\lambda$.
This requirement is automatically satisfied if
it is possible to take the limit
$\lambda / \kappa \rightarrow \infty$, so that
the final expression for the Green's function looses its dependence on
the unphysical cutoff $\lambda$.
We now determine the range of $\eta$
where the integrand in Eq.\ref{eq:rdefetad} vanishes at large $p$ sufficiently fast
to insure convergence of the integral even without the cutoff $p_{\rm c}$. 
Because for large $p$ the arguments of the functions
$C_{d} (F ) $ and $L_{d} (F)$ in Eq.\ref{eq:rdefetad} are small,
we need to know the behavior of these functions
at weak coupling. From Eq.\ref{eq:Gspdominant} it is clear that in this regime
the Landau damping contribution $L_{d} (F )$ is dominant. 
Using Eq.\ref{eq:Gspweak}, we find that the ultraviolet behavior of
Eq.\ref{eq:rdefetad} is determined by
 \begin{equation}
 \tilde{R} ( d , \eta , p_{\rm c} )
 \sim L_{d}^{\prime} \int_{1}^{\infty}  \D p p^{d -2 - 2 \eta } \E^{ - 2 p / p_{\rm c}}
 \label{eq:ruv}
 \; \; \; .
 \end{equation}
Setting $p_{\rm c} = \infty$, we see that the integral exists only for
 \begin{equation}
 \eta > \eta_{\rm uv} \equiv \frac{ d-1}{2}
 \label{eq:etaastuv}
 \; \; \; .
 \end{equation}
If this condition is satisfied, the integrand falls off sufficiently fast to insure
convergence of the integral. Note that $\eta_{\rm uv} < \eta_{\rm ir}$, so that
there exists a finite interval for $\eta$ where 
the quasi-particle residue is finite and the ultraviolet cutoff 
can be removed.
Because we have rescaled $p = | {\vec{q}}| / \kappa$, the convergence of the integral
implies that the numerical value of the quasi-particle residue is 
determined by the regime $|{\vec{q}}  | \leqapprox \kappa$. 
In this case $\kappa$ (and not $q_{\rm c}$) acts as the relevant screening wave-vector in the
problem. In this sense an interaction
of the form \ref{eq:Fgeneric} with $\eta > \frac{ d-1}{2}$ 
and $\kappa \ll q_{\rm c}$ effectively replaces 
any unphysical ultraviolet cutoff $q_{\rm c}$ 
(which might have been generated by integrating 
out high energy modes)
by the physical cutoff $\kappa$ 
in the bosonization result for the quasi-particle residue. 
In summary, for singular density-density interactions of the form
\ref{eq:fgeneric}
\index{singular interactions!density-density}
the function $\tilde{R} ( d , \eta , \infty )$ exists  for
 \begin{equation}
 \frac{ d-1}{2} < \eta < 2 ( d-1)  
 \label{eq:etagood}
 \; \; \; .
 \end{equation}
In this interval the interaction 
falls off sufficiently fast at large $| {\vec{q}} |$
to insure convergence at short wavelengths, but diverges weak enough to
lead to a stable Fermi liquid.

\subsection{Why the Coulomb interaction is so nice}
\label{subsec:Cbnice}

As discussed in Appendix~\secref{subsubsec:Cb},
the Coulomb interaction\index{Coulomb interaction!quasi-particle residue}
\index{quasi-particle residue!Coulomb interaction}  in $1 < d \leq 3$
corresponds to $\eta =  d - 1$ and $q_{\rm c} = \infty$.
Furthermore, 
$\kappa$ can now be identified with the usual
Thomas-Fermi screening wave-vector given in Eq.\ref{eq:kappaTFdef}.
Note that $\eta = d-1$ satisfies for all $d$ the condition
\ref{eq:etagood}. 
Setting $\eta = d-1$ and $q_{\rm c} = \infty$ in Eq.\ref{eq:Rex4},
and changing variables to $F = p^{-(d-1)}$ in Eq.\ref{eq:rdefetad}, we obtain
 \begin{equation}
 R = - \left( \frac{ \kappa}{k_{\rm F}} \right)^{d-1} \frac{ \tilde{r}_{d}}{ d - 1 } 
 \label{eq:Rcbres}
 \; \; \; ,
 \end{equation}
with
 \begin{equation}
 \tilde{r}_{d} \equiv ( d - 1) \tilde{R} ( d , d-1 , \infty )
 = \int_{0}^{\infty} \frac{ \D F}{F^2}  \left[ C_{d} ( F ) + L_{d} ( F ) \right]
 \label{eq:tilderddef}
 \; \; \; .
 \end{equation}
From the previous section we know that the integral 
in Eq.\ref{eq:tilderddef} exists for all $ d > 1$.
Note also that according to Eq.\ref{eq:rsfinal} the prefactor 
$( \kappa / k_{\rm F} )^{d-1}$ is proportional
to the Wigner-Seitz radius 
$r_{\rm s}$,\index{Wigner-Seitz radius $r_{\rm s}$}\index{singular interactions!density-density}
which is the relevant small parameter in the usual high-density
expansion for the homogeneous electron gas \cite{GellMann57}.
We conclude that 
higher-dimensional bosonization predicts
for the Coulomb interaction
in dimensions $1 < d \leq 3$ a finite
result for the quasi-particle residue,
which in the limit of high densities (i.e. for $ \kappa \ll k_{\rm F})$
is close to unity and independent of the unphysical sector cutoffs.

By isolating a factor of $\frac{1}{d-1}$ in Eq.\ref{eq:Rcbres} we have anticipated that
$\tilde{r}_{d}$ has a finite limit for $ d \rightarrow 1$.
If we are only interested in the leading behavior of $R$ for
$d \rightarrow 1$, it is sufficient to calculate
$\tilde{r}_{1}$.
In this case $L_{1} ( F ) = 0$, and
the functional form of
$C_{1} ( F )$ is simply  
obtained by replacing $F_{0} \rightarrow F$ in the expression for the anomalous dimension
of the Tomonaga-Luttinger model \cite{Solyom79,Emery79,Haldane81} (see Eq.\ref{eq:anomaldimd1} below),
 \begin{equation}
 C_{1} ( F ) = 
 \frac{ F^2}{ 2 
 \sqrt{ 1 + F } \left[ \sqrt{ 1 + F } + 1 \right]^2 }
 \label{eq:g1colres}
 \; \; \; .
 \end{equation}
Substituting this into Eq.\ref{eq:tilderddef},
we obtain
 \begin{equation}
 \tilde{r}_{1} = \frac{1}{2} \int_{0}^{\infty} \D F
 \frac{1}{
 \sqrt{ 1 + F } \left[ \sqrt{ 1 + F } + 1 \right]^2 }
 = \frac{1}{2}
 \label{eq:tilder1res}
 \; \; \; .
 \end{equation}
We conclude that for $d \rightarrow 1$
 \begin{eqnarray}
 R  & = &  - \left( \frac{ \kappa}{k_{\rm F}} \right)^{d-1} \frac{ 1 }{ 2 (d - 1) } 
 + O ( 1 )
 \nonumber
 \\
 & = & - \frac{1}{2 ( d-1 )} + \frac{1}{2} \ln \left( \frac{k_{\rm F}}{\kappa} \right) + O (1)
 \; \; \; .
 \label{eq:Rd1lim}
 \end{eqnarray}
Exponentiating Eq.\ref{eq:Rd1lim}, we see that
quasi-particle residue vanishes as
 \begin{equation}
 Z \propto \left( \frac{k_{\rm F}}{\kappa} \right)^{\frac{1}{2}}
 \E^{ - \frac{1}{2 ( d-1) } }
 \; \; \; , \; \; \; d \rightarrow 1
 \; \; \; .
 \label{eq:Zres1asym}
 \end{equation}
A similar result has also been obtained by
Castellani, Di Castro and Metzner \cite{Castellani94}.

\subsection{The sub-leading corrections for $ 0 < \eta < 2 (d-1 )$}
\label{subsubsec:Anomaldamp}
\index{singular interactions!density-density}

So far we have shown that for singular interactions
of the type \ref{eq:fgeneric}
the integral defining
$R$ does not exist if $\eta \geq 2 (d-1)$. The divergence is
due to the infrared regime of the collective mode
contribution to the dynamic structure factor.
On the other hand, for $\eta  < 2 ( d-1)$ the quasi-particle residue is finite.
In this case we know from Chap.~\secref{sec:Identification}
that $S ( x , 0 )$ vanishes at large distances, 
so that in general we expect (ignoring possible logarithmic corrections)
 \begin{equation}
 S ( x , 0 ) \sim -
 \left( \frac{ \kappa}{k_{\rm F}} \right)^{d-1}
 \frac{ \tilde{S} ( d , \eta , \frac{q_{\rm c} }{ \kappa} )}{ | \kappa x |^{\zeta} }
 \; \; \; , \; \; \; x \rightarrow \infty \; \; \; , \; \; \; \zeta > 0
 \label{eq:Sxasym}
 \; \; \; ,
 \end{equation}
with some dimensionless function $\tilde{S} ( d , \eta  , p_{\rm c} )$.
In a Landau Fermi liquid we expect $\zeta = 1$, because 
otherwise the self-energy
$\Sigma ( \vec{k}^{\alpha} + \vec{q} , \omega )$ cannot have 
a power series expansion for small $\vec{q}$, see Eq.\ref{eq:sigmaexp}.
However, if $\eta$  is smaller than (but sufficiently close to)
$2 ( d-1)$, we expect an exponent $\zeta$ smaller than unity.
It turns out that there exists a critical value $\eta_{\rm c}$ such that 
$0 < \zeta < 1$ 
for $\eta_{\rm c} < \eta < 2 ( d-1 )$.
In this regime the system is
a Fermi liquid\index{Fermi liquid!anomalous sub-leading corrections} 
with anomalously large sub-leading corrections.
We now determine the critical $\eta_{\rm c}$ for singular interactions in $d > 1$.
Proceeding precisely as above,  we obtain
(see Eqs.\ref{eq:Rex2} and \ref{eq:Rex3})
 \begin{equation}
 S ( x , 0 )
 =  - \frac{1}{k_{\rm F}^{d-1} \Omega_{d}} 
 \int \frac{ \D {\vec{q}}}{ | {\vec{q}} | }
 \cos (  \hat{\vec{v}}^{\alpha} \cdot {\vec{q}} x )
 \left[ C_{d} ( F_{ {\vec{q}}} ) + L_{d} ( F_{ {\vec{q}} } ) \right]
 \; \; \; .
 \label{eq:Sequal1}
 \end{equation}
From Sect.~\secref{subsec:qpressing} we know that for singular interactions 
the integral in Eq.\ref{eq:Sequal1} is dominated by the strong-coupling limit
of the function $C_{d} ( F )$, which is given in Eq.\ref{eq:Gcollarge}.
Introducing $d$-dimensional spherical coordinates
(see Eqs.\ref{eq:angavsimplify} and \ref{eq:gammaddef}),
we obtain for the dominant part of Eq.\ref{eq:Sequal1}
after a simple rescaling \index{singular interactions!density-density}
 \begin{eqnarray}
 S ( x , 0 )
 & \sim &  - \left( \frac{\kappa}{k_{\rm F}} \right)^{d-1} 
 \frac{ \sqrt{d}}{2}  
 \frac{ {\gamma}_{d} }{ | \kappa x |^{   d -1 - \frac{\eta}{2}  } }
 \int_{0}^{q_{\rm c} | x| }  \D p p^{d-2 - \frac{ \eta}{2} } 
 \nonumber
 \\
 & \times &
 \int_{0}^{\pi} 
 \D \vartheta ( \sin \vartheta )^{d-2}
 \cos ( p  \cos \vartheta )
 \; \; \; .
 \label{eq:Sequal2}
 \end{eqnarray}
For $d -2 - \frac{\eta }{2} < 0$ the integrand vanishes for large $p$ sufficiently fast, so that
the integral is convergent even if the cutoff $q_{\rm c} $ is removed. 
In this case we obtain for $\kappa x \rightarrow \infty$ and $ q_{\rm c} \gg \kappa$
 \begin{equation}
 S ( x , 0 ) \sim -
 \left( \frac{ \kappa}{k_{\rm F}} \right)^{d-1}
 \frac{ \tilde{S} ( d , \eta , \infty) }{ | \kappa x |^{ d - 1 - \frac{\eta}{2} } }
 \; \;  , \; \; x \rightarrow \infty  \; \; ,
 \;  \; 0 < d -1 - \frac{\eta}{2} < 1  \; \; ,
 \label{eq:Sxasym2}
 \end{equation}
with
 \begin{equation} 
 \tilde{S} ( d , \eta , \infty ) = \frac{\sqrt{d}}{2} \gamma_{d}
 \int_{0}^{\infty}  \D p p^{ d - 2 - \frac{\eta}{2}} 
 \int_{0}^{\pi} \D \vartheta ( \sin \vartheta )^{d-2} \cos ( p \cos \vartheta)
 \label{eq:sdetadef}
 \; \; \; .
 \end{equation}
This is precisely the asymptotic behavior given in
Eq.\ref{eq:Sxasym}, with exponent
 $\zeta = d - 1 -  \frac{\eta}{2} <  1 $.
The integral in Eq.\ref{eq:sdetadef} can be done 
analytically \cite{Gradshteyn80,Groebner61},
and we obtain after some rearrangements
 \begin{equation}
 \tilde{S} ( d , \eta , \infty  )  =  
 - \frac{ \sqrt{ \pi d}}{4}
 \frac{ \Gamma ( \frac{d}{2} )}{
 \Gamma ( \frac{ 1 + \frac{\eta}{2}}{2} ) 
 \cos [ \frac{\pi}{2} ( d - \frac{ \eta}{2} ) ]}
 \; \; \; .
 \label{eq:sdetares2}
 \end{equation}

\index{singular interactions!density-density}
On the other hand, if the exponent $d - 2 - \frac{\eta }{ 2}$ in Eq.\ref{eq:Sequal2} 
is positive, then the integral in Eq.\ref{eq:Sequal2} depends on the cutoff $q_{\rm c}$. 
In this case we obtain for large $x$
the asymptotic behavior predicted in Eq.\ref{eq:Sxasym} with
$\zeta = 1$ and
 \begin{equation}
 \tilde{S} ( d , \eta , \frac{q_{\rm c} }{ \kappa}  ) \propto
  \frac{ 1 }{d -2 - \frac{\eta}{2}} 
  \left( \frac{q_{\rm c}}{\kappa} \right)^{d-2 - \frac{\eta }{ 2} } 
 \; \; \; , \; \; \; 
 d - 2 -  \frac{\eta}{2}  > 0
 \; \; \; .
 \end{equation}
We conclude that in the regime
 \begin{equation}
  \eta < 2 ( d - 2)  \equiv \eta_{\rm c}
 \label{eq:etacdef}
 \end{equation}
the  correction to the leading constant term of the static Debye-Waller 
factor  vanishes as $x^{-1}$ at large distances, 
so that in real space we have analyticity around $ x = \infty$.
In Fourier space this implies
analyticity around the origin,
as postulated for the self-energy in a
Landau Fermi liquid (see Eq.\ref{eq:sigmaexp}).
On the other hand, if $\eta$ lies in the regime
 \begin{equation}
 2 ( d-2) < \eta < 2 ( d-1 ) 
 \; \; \; ,
 \label{eq:etainterval}
 \end{equation}
the system is not a conventional
Landau Fermi liquid, because the corrections to the
leading constant term $R$ are not analytic.
If $\eta$ approaches the value $\eta_{\rm ir} = 2 ( d-1 )$ from below,
the constant term $R$ diverges logarithmically,
but the divergence is cancelled by
$S( x , \tau )$, so that the
total Debye-Waller factor
$Q ( x , \tau ) = R 
-S ( x , \tau ) $ remains finite. 
Similarly, we expect 
logarithmic corrections 
to the leading $x^{-1}$ decay of $S( x , 0 )$ 
at the lower limit $\eta = \eta_{\rm c} = 2(d-2) $ of the
interval in Eq.\ref{eq:etainterval}.
Interestingly, the  Coulomb 
interaction\index{Coulomb interaction!anomalous sub-leading corrections}, 
which in $d$ dimensions corresponds
to $\eta = d-1$, satisfies the condition \ref{eq:etainterval} for
$d<3$. In particular, in $d=2$ the Coulomb interaction leads to a Fermi liquid
with anomalously large sub-leading corrections.

\subsection{The regime $ \eta \geq 2 ( d-1 )$}
\label{subsec:thelim}

Finally, let us consider the regime $ \eta \geq 2 ( d-1 )$, where the
integral \ref{eq:Rex4} defining $R$ is divergent.
Clearly, if the exponent $\eta$ is chosen sufficiently large,
the divergence will be so strong that it cannot be regularized
by means of the subtraction
$Q ( x , \tau ) = R - S ( x , \tau )$ in the Debye-Waller factor.
Hence, there exists a critical value of $\eta$ where the bosonization result 
in $d$ dimensions is divergent\index{limitations of bosonization}. 
To investigate this point,
we now calculate the long-distance behavior of $Q ( x , 0 )$
for $ \eta \geq 2 ( d-1 )$.
Repeating the manipulations leading to Eq.\ref{eq:Sequal2}, we obtain
for $\frac{\eta}{2} - d + 1 \geq 0$
 \begin{eqnarray}
 Q ( x , 0 )
 & \sim &  - \left( \frac{\kappa}{k_{\rm F}} \right)^{d-1} 
 \frac{ \sqrt{d} \gamma_{d} }{2} 
 | \kappa x |^{   \frac{\eta}{2} -d + 1 } 
 \int_{0}^{q_{\rm c} | x| } \D p  p^{ - ( \frac{\eta}{2} -d + 2 )}
 \nonumber
 \\
 & \times &
 \int_{0}^{\pi} \D \vartheta
  ( \sin \vartheta )^{d-2}
 [ 1 - \cos ( p  \cos \vartheta ) ] 
 \; \; \; ,
 \label{eq:Qsingy}
 \end{eqnarray}
From this expression it is easy to show that 
precisely at $\eta = 2 ( d-1 )$
the Debye-Waller factor increases logarithmically at large
distances,
 \begin{equation}
 Q ( x , 0 ) \sim - \gamma_{\rm LL} \ln ( q_{\rm c} | x | )
 \; \; \; , \; \; \; \eta = 2 ( d-1 )
 \; \; \; ,
 \label{eq:QxLL}
 \end{equation}
with the anomalous dimension given by
 \begin{equation}
 \gamma_{\rm LL} = \frac{ \sqrt{d}}{2}  
 \left( \frac{\kappa}{k_{\rm F}} \right)^{d-1}
 \; \; \; .
 \end{equation}
The logarithmic divergence of the static Debye-Waller factor is
familiar from the one-dimensional
Tomonaga-Luttinger model (see Sect.~\secref{sec:Green1}).
As a consequence, the momentum distribution
$n_{\vec{k }}$ exhibits
an algebraic singularity at the Fermi 
surface. \index{momentum distribution!algebraic singularity}
The location of this
singularity can be used to define the Fermi surface
of the interacting system in a mathematically precise way.

\begin{figure}
\psfig{figure=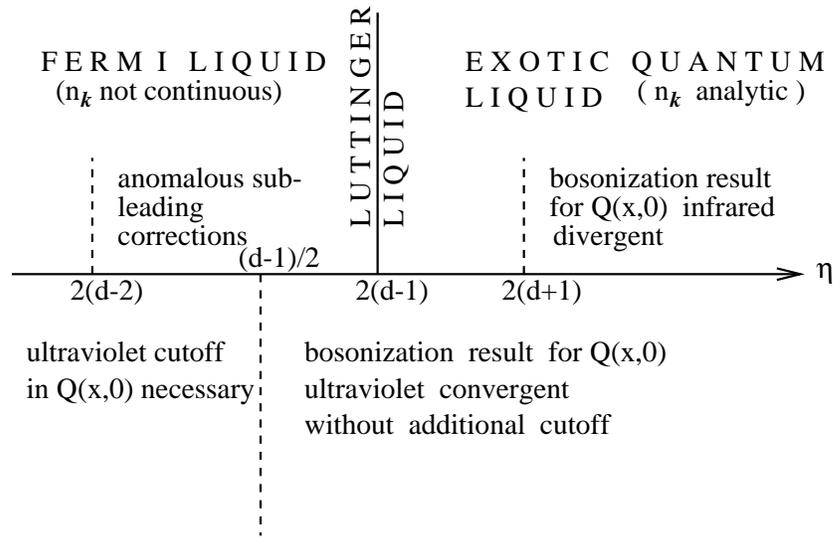,width=11cm}
\caption[Special values of $\eta$ for singular density-density 
interactions of the
type $|{\vec{q}} |^{- \eta }$.]
{
Summary of special values of the exponent $\eta$ for singular 
density-density interactions
of the type $| {\vec{q}} |^{- \eta }$ in $d$ dimensions. 
The Luttinger liquid\index{Luttinger liquid} for $\eta = 2 ( d-1 )$ corresponds to the marginal case, where
the momentum distribution $n_{\vec{k}}$ is continuous but not analytic.
Note that
$2 ( d-2 ) < \frac{ d-1}{2}$ for $ d < \frac{7}{3}$.
}
\label{fig:etasum}
\end{figure}

For $\eta > 2 ( d-1 )$ we find a {\it{stretched exponential}} divergence
\index{stretched exponential}
of the static Debye-Waller factor,
 \begin{equation}
 Q ( x , 0 ) \sim - 
 \left( \frac{\kappa}{k_{\rm F}} \right)^{d-1}
 \tilde{Q} ( d , \eta ) | \kappa x |^{ \frac{\eta}{2} - d + 1 }
 \; \; \; , \; \; \; 
 \frac{\eta}{2} - d + 1  > 0
 \; \; \; ,
 \label{eq:Qmoresing}
 \end{equation}
with
 \begin{eqnarray}
 \tilde{Q} ( d , \eta ) & = &
 \frac{ \sqrt{d} \Gamma ( \frac{d}{2} ) }{ \sqrt{\pi} [ \eta - 2 ( d-1) ] \Gamma \left( 
 \frac{ 1 + \frac{\eta}{2}}{2} \right) }
 \cos \left( \frac{\pi}{2} ( \frac{\eta}{2} - d + 1 ) \right)
 \nonumber
 \\
 & \times  & 
 \Gamma \left( 1 + ( \frac{ \eta}{2} -d + 1 ) \right)
 \Gamma \left( 1 - ( \frac{ \eta}{2} -d + 1 ) \right)
 \label{eq:Qtildeetad}
 \; \; \; .
 \end{eqnarray}
\index{singular interactions!density-density}
The important point is now that for $ \frac{\eta}{2} - d + 1 =  2$
the function $\tilde{Q} ( d , \eta )$ diverges, 
because the argument of the second $\Gamma$-function 
in Eq.\ref{eq:Qtildeetad} becomes $-1$.
Hence, for
 \begin{equation}
 \eta \geq 2 ( d + 1 )
 \label{eq:etaastdiv}
 \end{equation}
bosonization cannot cure 
the divergence due to the singular interactions. 
The physical behavior of the system in this
parameter regime cannot be discussed within the framework
of our bosonization approach.
Note that for $ \eta = 2 ( d + 1 )$ the equal-time  Debye-Waller
factor in Eq.\ref{eq:Qmoresing} would be quadratically divergent, so that
the equal-time Green's function would vanish 
like a {\it{Gaussian}} at large distances, i.e.
$Q ( x , 0 ) \propto - x^2$.
On the other hand, in the regime $ 2 ( d-1 ) < \eta < 2 ( d + 1 )$
the equal-time Green's function can be calculated via bosonization, and
vanishes like a {\it{stretched exponential}} at large distances.
We shall refer to normal Fermi systems with this property as
{\it{exotic quantum liquids}}\index{exotic quantum liquid}.
It is easy to show \cite{Kopietz95c} that
the stretched exponential decay of the static Debye-Waller factor implies that
the momentum distribution $n_{\vec{k}}$ is analytic at 
the\index{momentum distribution!analytic}
(non-interacting) Fermi surface, so that a sharp Fermi surface
of the interacting system simply cannot be defined any more.
The disappearance of a sharp Fermi surface\index{Fermi surface!destruction
due to singular interactions} in strongly correlated
Fermi systems is certainly not a special feature of the singular interactions
considered  here. For example, models with correlated hopping \cite{Hirsch89,Bariev93}
show similar behavior.
The various critical values for $\eta$ derived in this section are
summarized in Fig.~\secref{fig:etasum}.
\index{singular interactions!density-density}
The fact that exotic quantum liquids do not have a sharp Fermi surface
does {\it{not}} mean that in these systems the bosonization approach 
(which is based on the expansion of the energy dispersion for
momenta in the vicinity of the
{\it{non-interacting}} Fermi surface) is inconsistent. 
As already pointed out long time ago by Tomonaga \cite{Tomonaga50}, 
the existence of a singularity
in the momentum distribution is really not necessary for the consistency of the 
bosonization procedure as long as 
(i) the thickness of the shell where the
momentum distribution drops from unity to zero is small compared with 
the characteristic size of $k_{\rm F}$,
and (ii) the effective interaction is not too singular, so that
the Debye-Waller factor $Q ( x , 0 )$ 
is mathematically well defined\footnote{ 
The fact that the integral defining $Q ( x , 0 )$
exists does not imply the existence of $Q ( x , \tau )$ for
$\tau \neq 0$. In fact, in $d=1$ Eq.\ref{eq:etaastdiv}
tells us that the static Debye-Waller factor
is mathematically ill-defined for  $\eta \geq 4$, 
while for $\tau \neq 0$ it is easy to show from
Eq.\ref{eq:ReQ3} that the integral defining $Q ( x , \tau )$  
does not exist as soon as $\eta \geq 1$.}.
The condition (i) means that the smearing of the Fermi surface is small, so that
{\it{it does not matter 
which point ${\vec{k}}^{\alpha}$ 
within the smeared shell is
chosen as a reference point for the expansion of the
non-interacting energy dispersion}}.
For singular interactions
of the type discussed above 
the condition (ii) is satisfied as long as
$\eta < 2 ( d + 1)$.

\section{Luttinger liquid behavior in $d=1$}
\label{sec:Green1}
\index{Tomonaga-Luttinger model}

{\it{
This section does not contain
any new results, but shows that in $d=1$
our formalism correctly reproduces the well-known
bosonization result for the Green's function of the
Tomonaga-Luttinger model.}}

\hspace{7mm}

\noindent
In $d=1$ we have only two Fermi points, which may be labelled by $\alpha = + , -$. 
The associated normal vectors are
$\hat{\vec{v}}^{\alpha} = \alpha {\vec{e}}_{x}$.
Obviously the  matrix $\underline{ f}_{\vec{q}}$ is then a 
$2 \times 2$-matrix\footnote{
To distinguish the wave-vector label from the
collective label
$q = [ {\vec{q}} , \I \omega_{m} ]$, 
we shall continue to write
${\vec{q}}$ for the wave-vector, it being understood that
${\vec{q}} = q_{x} {\vec{e}}_{x}$, where ${\vec{e}}_{x}$ 
is a unit vector in the $x$-direction.}. 
The usual notation  in the literature \cite{Solyom79} 
is $[ \underline{f}_{ \vec{q}} ]^{ ++} =  
[ \underline{f}_{ \vec{q}} ]^{ --} =
g_{4} ( {\vec{q}} )$, and 
$[ \underline{f}_{ \vec{q}} ]^{ +-} =  
[ \underline{f}_{ \vec{q}} ]^{ -+} =
g_{2} ( {\vec{q}} )$. 
Because  Eqs.\ref{eq:R2}, \ref{eq:ReS2b} and \ref{eq:ImS2b} have been derived for the
special case that all matrix elements of
$\underline{ f}_{\vec{q}}$ are identical, 
these expressions should reduce
to the exact solution of the Tomonaga-Luttinger model
with interaction parameters  $ g_{4}  = g_{2}  = f_{0}$, where
$ \lim_{ {\vec{q}} \rightarrow 0 } f_{\vec{q}} = f_{0} = {\rm const}$. 
Note that in the Tomonaga-Luttinger model the energy dispersion
is linear by definition.
Writing ${\vec{r}} = r_{x} {\vec{e}}_{x}$,
it is clear from the general considerations of
Sect.~\secref{subsec:rot} that the Debye-Waller factor
depends on the sector label only via
$r^{\alpha}_{\|} = \hat{\vec{v}}^{\alpha} \cdot {\vec{r}} = \alpha r_{x}$, so that
 $Q^{\alpha} ( r^{\alpha}_{\|} \hat{\vec{v}}^{\alpha} , \tau )
 =
  Q ( x , \tau ) $, with $x = \alpha r_{x}$.

To evaluate the Debye-Waller factor from Eqs.\ref{eq:R2}, \ref{eq:ReS2b}
and \ref{eq:ImS2b},
we need the RPA dynamic structure factor in $d=1$.
From Eqs.\ref{eq:zero1},\ref{eq:Scoldelta},\ref{eq:resman} and
\ref{eq:Z1} we obtain 
 \begin{equation}
 S_{\rm RPA} ( {\vec{q}} , \omega ) = Z_{\vec{q}} \delta ( \omega - \omega_{ {\vec{q}} } )
 \; \; \; , 
 \label{eq:SRPA1}
 \end{equation}
where the residue and the collective mode are given by
 \begin{equation}
 Z_{\vec{q}} = \frac{ | {\vec{q}}| }{2 \pi \sqrt{ 1 + F_{\vec{q}} } }
 \; \; \; , \; \; \;
 \omega_{{\vec{q}}} = \sqrt{ 1 + F_{\vec{q}} } v_{\rm F} | {\vec{q}} |
 \label{eq:Zom1}
 \; \; \; ,
 \end{equation}
and $F_{\vec{q}} = \nu f_{\vec{q}} =  f_{\vec{q}} / ( \pi v_{\rm F} )$ is the usual
dimensionless interaction. Note that in $d=1$ there is no single pair contribution
to the RPA dynamic structure factor.
Furthermore, because the Fermi surface in $d=1$
is covered by $M = 2$ patches, 
we know from
the considerations of Sect.~\secref{subsec:Thetermslin}
that $L^{\alpha}_{\vec{q}} ( \tau ) = 0$ in
Eqs.\ref{eq:ReS2b} and \ref{eq:ImS2b}.
Substituting Eq.\ref{eq:SRPA1} into Eq.\ref{eq:R2},
we obtain
 \begin{eqnarray}
 R & = & - \frac{1}{V} \sum_{\vec{q}} f_{\vec{q}}^2 \frac{Z_{\vec{q}}}{ ( \omega_{\vec{q}} + 
 | {\vec{v}}^{\alpha} \cdot {\vec{q}} |)^2}
 \nonumber
 \\
 & = & - \frac{1}{V} \sum_{\vec{q}} \frac{\pi}{| q_{x}| }
 \frac{ F_{\vec{q}}^2}{ 2 
 \sqrt{ 1 + F_{\vec{q}} } \left[ \sqrt{ 1 + F_{\vec{q}} } + 1 \right]^2 }
 \; \; \; .
 \label{eq:Rpre3}
 \end{eqnarray}
In the limit $V \rightarrow \infty$ we may replace
$ \frac{1}{V} \sum_{\vec{q}} f ( | q_{x} | ) \rightarrow \int_{0}^{\infty} 
\frac{ \D q_{x}}{ \pi } f (  q_{x}  )$.
Using the identity
 \begin{equation}
 \frac{ F_{\vec{q}}^2}{ 2 
 \sqrt{ 1 + F_{\vec{q}} } \left[ \sqrt{ 1 + F_{\vec{q}} } + 1 \right]^2 }
 =
 \frac{ 1  + \frac{F_{\vec{q}}}{2} }{  \sqrt{ 1 + F_{\vec{q}} } }
 - 1 
 \label{eq:identF}
 \; \; \; ,
 \end{equation}
we finally obtain
 \begin{equation}
 R =  
 - \int_{0}^{\infty}
 \frac{ \D q_{x} }{q_{x}}
 \left[ \frac{ 1  + \frac{F_{\vec{q}}}{2} }{  \sqrt{ 1 + F_{\vec{q}} } }
 - 1 \right]
 \label{eq:R3}
 \; \; \; .
 \end{equation}
Similarly, we obtain from Eqs.\ref{eq:ReS2b} and \ref{eq:ImS2b}
 \begin{eqnarray}
 \hspace{-5mm}
 {\rm Re} S ( x , \tau )
 & = &  
 - \int_{0}^{\infty}
 \frac{ \D q_{x} }{q_{x}} 
 \cos ( q_{x} x )
 \left[ 
 \frac{ 1  + \frac{F_{\vec{q}}}{2} }{  \sqrt{ 1 + F_{\vec{q}} } }
 \E^{ - \sqrt{ 1 + F_{\vec{q}} } v_{\rm F} q_{x} | \tau | } - 
 \E^{ - v_{\rm F} q_{x} | \tau | } \right]
  \; ,
 \nonumber
 \\
 & &
 \label{eq:ReS3}
 \\
 \hspace{-5mm}
 {\rm Im} S ( x , \tau )
 & = &
 - {\rm sgn} ( \tau )
 \int_{0}^{\infty}
 \frac{ \D q_{x} }{q_{x}} \sin ( q_{x}  x   )
 \left[ 
 \E^{ - \sqrt{ 1 + F_{\vec{q}} } v_{\rm F} q_{x} | \tau | } - 
 \E^{ - v_{\rm F} q_{x} | \tau | } \right]
 \; .
 \nonumber
 \\
 & &
 \label{eq:ImS3}
 \; \; \; .
 \end{eqnarray}
Combining Eqs.\ref{eq:R3} and \ref{eq:ReS3}, we can also write
 \begin{eqnarray}
 {\rm Re} Q ( x  , \tau )
 & = & 
 R - 
 {\rm Re} S ( x , \tau ) =
 - \int_{0}^{\infty} \frac{\D q_{x}}{q_{x}}  
 \nonumber
 \\
 &  & \hspace{-25mm}  
 \times \left\{
 \frac{ 1  + \frac{F_{\vec{q}}}{2} }{  \sqrt{ 1 + F_{\vec{q}} } }
 \left[  1 - \cos ( q_{x} x )
 \E^{ - \sqrt{ 1 + F_{\vec{q}} } v_{\rm F} q_{x} | \tau | }  \right]
 - 
  \left[ 1 - \cos ( q_{x} x )
 \E^{ - v_{\rm F} q_{x} | \tau | }  \right]
 \right\}
 \; .
 \nonumber
 \\
 & &
 \label{eq:ReQ3}
 \end{eqnarray}
Eqs.\ref{eq:ReS3}--\ref{eq:ReQ3} are identical with the 
well-known bosonization result for the Green's function of an interacting Fermi system
with linearized energy dispersion \cite{Haldane81}. 

Let\index{Luttinger liquid!in one dimension} us evaluate 
Eqs.\ref{eq:ReS3}--\ref{eq:ReQ3} for interactions of the form
 $F_{\vec{q}} = F_{0} \E^{ - |{\vec{q}}| / {q_{\rm c}} }$, where
 $q_{\rm c} \ll k_{\rm F}$.
From Sect.~\secref{sec:singular} we know that
in one dimension an
interaction that approaches a constant for
$\vec{q} \rightarrow 0$   
leads to an unbounded Debye-Waller factor which grows
logarithmically at large distances.
The logarithmic singularity is
evident in Eq.\ref{eq:R3}.
Hence, according to Eq.\ref{eq:ZRrelation}
the quasi-particle residue vanishes in this case, so that the system is not a
Fermi liquid.  However, in the combination
 $R - S ( x , \tau )$ the logarithmic singularity
is removed, and we obtain a finite result for the Green's function.
Unfortunately, for interactions of the above form
the integrals in Eqs.\ref{eq:ReS3}--\ref{eq:ReQ3} cannot be
evaluated analytically. However, at length scales $x$ large
compared with the characteristic range $q_{\rm c}^{-1}$ of the interaction
the Green's function should be independent of the precise way in which the
ultraviolet cutoff is introduced. Therefore we
may regularize the $q_{x}$-integrals in any convenient way.
A standard regularization which leads to elementary integrals
is to multiply the entire integrand by a convergence factor 
$\E^{ - | \vec{q} | / q_{\rm c}}$ 
and to replace $F_{\vec{q}} \rightarrow F_{0}$ everywhere 
in the integrand \cite{Dzyaloshinskii74,Luther74}.
Although the cutoff $q_{\rm c}$ defined in this way is not identical with
the cutoff in $F_0 \E^{ - | \vec{q} | / q_{\rm c} }$, 
it can still be identified physically with
the range of the interaction in momentum space.
The relevant integrals can be found in standard 
tables \cite{Gradshteyn80,Groebner61},  and we obtain
 \begin{eqnarray}
 {\rm Re} Q ( x , \tau )
  & = &  -  \frac{ 1 + \frac{F_{0}}{2} }{ 2 \sqrt{ 1 + F_{0}}}
 \ln \left[ \frac{ x^2 + ( \tilde{v}_{\rm F} | \tau | + q_{\rm c}^{-1} )^2 }
 { q_{\rm c}^{-2} } \right]
 \nonumber
 \\
 & &
 \hspace{13mm}
 + \frac{1}{2} \ln \left[ \frac{ x^2  + ( v_{\rm F} | \tau | + q_{\rm c}^{-1} )^2 }
 {q_{\rm c}^{-2} } \right] 
 \;  ,
 \label{ReQres}
 \\
 & & \hspace{-24mm}
 {\rm Im }Q ( x , \tau )
  = 
  \frac{ {\rm sgn} ( \tau ) }{2 \I} 
 \left\{
 - \ln \left[
 \frac{ 
   x  + \I  \tilde{v}_{\rm F} | \tau | + \I q_{\rm c}^{-1} }
 {   x  - \I  \tilde{v}_{\rm F} | \tau | - \I q_{\rm c}^{-1} }
 \right]
 +
 \ln \left[
 \frac{  x + \I  v_{\rm F} | \tau | + \I q_{\rm c}^{-1}   }
 { x - \I  v_{\rm F} | \tau | - \I q_{\rm c}^{-1}   }
 \right]
 \right\}
 \; ,
 \nonumber
 \\
 & &
 \label{ImQres}
 \end{eqnarray}
where 
 $\tilde{v}_{\rm F} = \sqrt{ 1 + F_{0} } v_{\rm F}$
is the renormalized Fermi velocity. Combining the terms differently, the total
Debye-Waller factor can also be written as\index{Debye-Waller factor!in one dimension}
 \begin{eqnarray}
 Q( x , \tau ) & = & \frac{ \gamma (F_{0}) }{2} 
 \ln \left[ 
 \frac{ q_{\rm c}^{-2} }
 { x^2 + ( \tilde{v}_{\rm F}  | \tau |  + q_{\rm c}^{-1} )^2 } \right]
 \nonumber
 \\
 & + & \ln \left[ \frac{ x + \I  v_{\rm F} \tau + \I  {\rm sgn} ( \tau ) q_{\rm c}^{-1} }
 { x + \I  \tilde{v}_{\rm F} \tau + \I  {\rm sgn} ( \tau ) q_{\rm c}^{-1} }
 \right]
 \; \; \;  ,
 \label{eq:Qtot1res}
 \end{eqnarray}
where the so-called {\it{anomalous dimension}}
\index{anomalous dimension!Tomonaga-Luttinger model} is given by
 \begin{eqnarray}
 \gamma (F_{0})   & \equiv &  \frac{ 1 + \frac{F_{0}}{2} }{ \sqrt{ 1 + F_{0} }} - 1
 \nonumber
 \\
 & = &  \frac{  [ \sqrt{ 1 + F_{0}} - 1 ]^2 }
 {2 \sqrt{ 1 + F_{0} } }
 =
 \frac{ F_{0}^2}{ 2 
 \sqrt{ 1 + F_{0} } \left[ \sqrt{ 1 + F_{0} } + 1 \right]^2 }
 \label{eq:anomaldimd1}
 \; \; \; .
 \end{eqnarray}
At $\tau = 0$ we obtain from Eq.\ref{eq:Qtot1res}
 \begin{equation}
 Q ( x , 0 ) \sim - { \gamma  }  ( F_{0} )
 \ln ( q_{\rm c} | x | )
 \; \; \; , \; \; \; 
  q_{\rm c} | x | \rightarrow \infty
  \label{eq:QlargeequalLL}
  \; \; \; .
  \end{equation}
Exponentiating Eq.\ref{eq:Qtot1res} and using the expression for the
non-interacting real space Green's function given in Eq.\ref{eq:G0patchreal2}, we finally obtain
from Eq.\ref{eq:Gpatchreal2} for the interacting Green's function
(recall that $x = \alpha r_{x}$, with 
$\alpha = \pm $)\index{Green's function!Tomonaga-Luttinger model} 
 \begin{eqnarray}
 G^{\alpha} ( r_{x} , \tau )
 & = &
  \left( \frac{- \I}{2 \pi} \right) \frac{ 
 \E^{Q ( \alpha r_{x} , \tau ) }
 }{
  \alpha r_{x} + \I v_{\rm F} \tau }
  =
 \left( \frac{- \I}{2 \pi}  \right) \frac{ 1}{
  \alpha r_{x} + \I v_{\rm F} \tau }
  \nonumber
  \\
  &  & \hspace{-10mm} \times
  \left[ 
  \frac{ \alpha r_{x} + \I  {v}_{\rm F} \tau + \I {\rm sgn} ( \tau )q_{\rm c}^{-1} }
  { \alpha r_{x} + \I   \tilde{v}_{\rm F} \tau + 
  \I  {\rm sgn} ( \tau ) q_{\rm c}^{-1} }
  \right]
  \left[
  \frac{ q_{\rm c}^{-2} }
  { r_{x}^2 + ( \tilde{v}_{\rm F}  | \tau |  + q_{\rm c}^{-1} )^2 } \right]^{  \gamma / 2}
   .
  \label{eq:G1final}
  \end{eqnarray}
We now observe that for times $| \tau | \gg ( 
\tilde{v}_{\rm F} q_{\rm c})^{-1}$ or length scales
$| r_{x} | \gg q_{c}^{-1}$ 
we may neglect the cutoff $q_{c}^{-1}$ when it appears in combination with
$\tilde{v}_{\rm F} \tau$ or $x$.
{\it{In this regime}}
Eq.\ref{eq:G1final} reduces to
 \begin{equation}
 G^{\alpha} ( r_{x} , \tau )
 = \left( \frac{- \I }{2 \pi} \right) \frac{ 1}{
  \alpha r_{x} + \I \tilde{v}_{\rm F} \tau }
  \left[
  \frac{ q_{\rm c}^{-2} }
  { r_x^2 + ( \tilde{v}_{\rm F}  \tau   )^2 } \right]^{  \gamma / 2}
  \; \; \; .
  \label{eq:G1final2}
  \end{equation}
Note that this expression depends exclusively on the renormalized Fermi velocity 
$\tilde{v}_{\rm F}$.
If we rescale both space and time by a factor of $s^{-1}$, then it is obvious that
the interacting Green's function \ref{eq:G1final2} satisfies
 \begin{equation}
 G^{\alpha} ( r_{x} / s  ,  \tau / s )
 = s^{ 1 + \gamma }
 G^{\alpha} ( {r_{x}}  ,  \tau )
 \label{eq:anomalscale}
 \; \; \; .
 \end{equation}
Note that in $d$ dimensions the non-interacting sector Green's 
function \ref{eq:Gpatchreal1}
satisfies\index{anomalous scaling}
 \begin{equation}
 G^{\alpha}_{0} ( r_{x}  / s  ,  \tau / s )
 = s^{d}
 G^{\alpha}_{0} ( {r_{x}}  ,  \tau )
 \label{eq:normalscale}
 \; \; \; .
 \end{equation}
It is easy to see that in the asymptotic
long-distance and large-time limit this scaling behavior is not
changed by the interactions as long as the system is a Fermi liquid.
In this case the scaling behavior of the Green's function can be
determined by dimensional analysis.
In the renormalization group literature \cite{Ma76}, the exponent $d$ in
Eq.\ref{eq:normalscale} is called the {\it{scaling dimension}} of the
Green's function.
Because the real space Green's function has units of inverse volume, the
scaling dimension agrees with the dimensionality $d$ of the system.
The reason why the exponent $\gamma $ 
in Eq.\ref{eq:anomalscale}
is called anomalous dimension is now
clear: In $d=1$ the effect of the interactions is so drastic that the scaling behavior
of the Green's function cannot be completely determined by dimensional analysis.
There exists an {\it{anomalous}} contribution to the scaling dimension, which depends
on the strength of the interaction in a non-trivial way, as given
in Eq.\ref{eq:anomaldimd1}.

\section{Summary and outlook}

In this chapter we have studied in some detail
singular density-density interactions in $d$ dimensions that diverge
in the infrared limit  as $ | \vec{q} |^{- \eta }$.
These are perhaps the simplest model systems
for non-Fermi liquid behavior in higher dimensions.
We have confirmed the result of Bares and Wen \cite{Bares93} that
the Fermi liquid state is only stable for $ \eta < 2 ( d -1 )$.
We have also identified various other special values for
$\eta$, which are summarized in Fig.~\secref{fig:etasum}.
Unfortunately, non-Fermi liquid behavior in $d > 1$ can only be obtained
in the regime $ \eta \geq 2 ( d -1 )$, which corresponds to unphysical
super-long range interactions in real space.
For simplicity, we have restricted ourselves to the analysis
of the static Debye-Waller factor
$Q ( x , 0 )$, and have worked with linearized energy dispersion.
As discussed at the end of Sect.~\secref{subsec:Thetermslin},
we expect that 
for the density-density interactions considered
here the quadratic term in the energy dispersion will not 
qualitatively modify the long-distance behavior of
$Q ( x , 0)$.  In
Chap.~\secref{chap:arad} we shall show that this is {\it{not}} 
the case for the effective current-current
interaction mediated by transverse gauge-fields.

The asymptotic long-distance behavior of the static Debye-Waller factor 
determines the momentum distribution $n_{\vec{k}}$ for
wave-vectors close to the Fermi surface. 
In the regime $\eta > 2 ( d-1 )$ we have found that
$n_{\vec{k}}$ is analytic, so that the interactions completely wash out
the sharpness of the Fermi surface.
Our result disagrees with the works by Khodel and collaborators \cite{Khodel90},
who treated long-range interactions within
a Hartree--Fock approach and found that the
momentum distribution is completely flat in a certain finite shell
around the non-interacting Fermi surface.
This result has been criticized by Nozi\`{e}res \cite{Nozieres92}, who argued that
it is most likely an unphysical artefact of the Hartree--Fock 
approximation.
Our non-perturbative calculation supports his arguments.

An interesting unsolved problem is the 
explicit evaluation of our non-perturbative result for the  
full {\bf{momentum- and frequency-dependence of the Green's function}}
$G ( \vec{k} , \omega )$  
in the non-Fermi liquid regime
$\eta > 2  ( d-1)$.
As discussed by Volovik \cite{Volovik91}, the Green's function
of non-Fermi liquids  might exhibit some interesting topological structure
in Fourier space, which can be used for a
rather general topological definition of the Fermi surface
of an interacting Fermi system.
Recall that in Chap.~\secref{subsec:defFSint} we have defined the
Fermi surface via the singularity
in the momentum distribution. According to this definition
fermions with singular density-density interactions of the type
\ref{eq:fgeneric} with $ \eta > 2 ( d -1 )$ do not have a
Fermi surface. See \cite{Volovik91} for an alternative
topological definition of the Fermi surface, which seems to be 
general enough 
to associate a mathematically well-defined Fermi surface with a system
that has an analytic momentum 
distribution\footnote{
I would like to thank
G. E. Volovik for pointing this out to me, and for sending me
copies of the relevant references.
}. This definition requires the knowledge of the
$\vec{k}$- and $\omega$-dependence of the
Green's function
$G ( \vec{k} , \omega )$, and not only of the momentum
distribution $n_{\vec{k}}$.
As discussed in Sect.~\secref{subsec:Thetermslin},
for the calculation of
$G ( \vec{k} , \omega )$ via higher-dimensional bosonization it is
most likely necessary to retain the quadratic term in the
energy dispersion. On the other hand, the momentum distribution
$n_{\vec{k}}$
is determined by the {\it{static}} Debye-Waller factor\footnote{
This is also the reason why we expect that the
special values of the exponent $\eta $ shown in Fig.~\secref{fig:etasum}
will not be modified by the non-linear terms in the energy dispersion.},
the leading long-distance
behavior of which can be calculated correctly
with linearized energy dispersion


%
%

%
%
%

\chapter{Quasi-one-dimensional metals}
\label{chap:apatch}
\setcounter{equation}{0}

{\it{
Here comes the first experimentally relevant application of our method: 
The calculation of the
single-particle Green's function for
highly anisotropic chain-like metals.
Most of the results presented in this chapter have been
obtained in collaboration with V. Meden and 
K. Sch\"{o}nhammer \cite{Kopietz94b,Kopietz96chain}.
}}

\vspace{7mm}

\noindent
One of the main motivations for the development of the 
higher-dimensional bosonization approach is the fact 
that non-Fermi liquid behavior has been observed in the laboratory, and is
therefore an experimental reality that requires theoretical explanation. 
The most prominent example are perhaps the normal-state properties of
the high-temperature superconductors \cite{Anderson90b}, but also
experiments on quasi-one-dimensional 
conductors \cite{Dardel93,Nakamura94,Jerome91,Gweon96} 
suggest non-Fermi liquid behavior.
Note that in these highly anisotropic systems the electrons interact with
Coulomb forces, which for isotropic systems in $d > 1$ certainly do not destabilize the
Fermi liquid state. This indicates that the experimentally seen non-Fermi liquid behavior 
could be due to the spatial anisotropy of these systems.

In this chapter we shall study a simple model for a quasi-one-dimensional metal, which
consists of electrons moving in a periodic array of weakly coupled 
metallic chains embedded in three-dimensional space. 
The electrons are assumed to interact with realistic three-dimensional Coulomb forces, 
so that,
even in the absence of interchain hopping\index{interchain hopping}, 
this is not a purely one-dimensional problem.
Note that in $d=1$ the logarithmic one-dimensional 
Fourier transform of the Coulomb potential (see Eq.\ref{eq:Qpcb1})
gives rise to singularities that are stronger than in a conventional
Luttinger liquid, so that the anomalous dimension diverges \cite{Schulz83}.
However, as will be shown in Sect.~\secref{sec:chain3d}, the 
Coulomb forces between the chains remove this divergence, so that
the long-range part of the three-dimensional Coulomb interaction in an 
array of chains without interchain hopping
indeed leads to a Luttinger liquid.
It should be kept in mind, however, that in this work we shall
retain only processes with small momentum transfers, so that
possible instabilities due to back-scattering\index{back-scattering} or Umklapp-scattering  
are ignored. We are implicitly assuming that there exists a parameter
regime where these processes are irrelevant.

The problem of coupled chains without interchain hopping can also be solved by means
of the usual one-dimensional bosonization techniques \cite{Schulz83,Penc93}.
The true power of the higher-dimensional bosonization approach 
becomes apparent if we switch
on a finite interchain hopping $t_{\bot}$. In this case
conventional one-dimensional bosonization cannot be used, but
within our higher-dimensional  bosonization approach 
this problem can be handled quite easily. 
The problem of weak interchain hopping has been
discussed by Gorkov and Dzyaloshinskii \cite{Gorkov74} more
than 20 years ago.
More recently, many other authors  
have used various more or less systematic methods 
to shed more light onto this rather difficult problem 
\cite{Wen90,Bourbonnais91,Schulz91,Fabrizio92,Kusmartsev92,Castellani92,Yakovenko92,%
Finkelstein93}.

In Sects.~\secref{sec:chain3d} and \secref{sec:Finiteinter}
we shall evaluate our bosonization results
\ref{eq:R2}, \ref{eq:ReS2b} and \ref{eq:ImS2b}
for the Debye-Waller factor with
linearized energy dispersion
in the case of 
an infinite array of metallic chains. 
However, even for finite interchain hopping $t_{\bot}$
the Debye-Waller factor exhibits
an unphysical logarithmic nesting singularity, which is due to the fact
that for linearized energy dispersion 
the Fermi surface is replaced by a finite 
number of completely flat patches.
To remove this singularity, 
the artificial nesting 
symmetry of the Fermi surface has to be broken.
The simplest way to do this is to 
work with non-linear energy dispersion.
In Sect.~\secref{subsec:thefate} we shall use
our general results \ref{eq:R2new}--\ref{eq:ImS2new}
for the Debye-Waller factor with quadratic energy dispersion
to show that the logarithmic nesting singularity is indeed removed
by the curvature of the Fermi surface.
Our main result is that an arbitrarily small $t_{\bot}$ destroys the Luttinger liquid state 
and leads to a finite quasi-particle residue $Z^{\alpha}$.
We explicitly calculate $Z^{\alpha}$ for small $t_{\bot}$ and 
show that 
there exists a large intermediate regime where the
signature of characteristic Luttinger liquid 
properties is visible in physical observables,
although the system is a Fermi liquid.

\section[The Coulomb interaction in chains without interchain hopping]
{The Coulomb interaction in chains \mbox{\hspace{25mm}} 
without interchain hopping}
\label{sec:chain3d}

{\it{Before addressing the more interesting case of $t_{\bot} \neq 0$, 
it is useful to consider the three-dimensional Coulomb interaction
in metallic chains in the absence of interchain hopping.
}}

\vspace{7mm}

\begin{figure}
\sidecaption
\psfig{figure=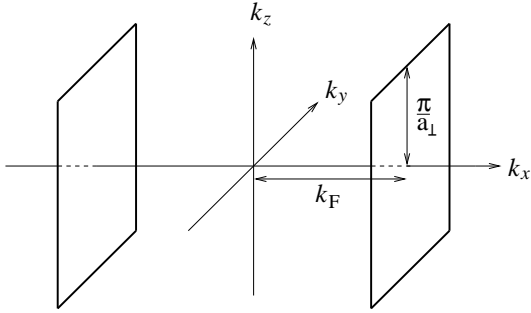,width=7cm}
\caption[Fermi surface of one-dimensional chains without interchain hopping.]
{
\begin{sloppypar}
Fermi surface\index{Fermi surface!array of chains} of a periodic array of one-dimensional chains
embedded in three-dimensional space without interchain hopping.
\end{sloppypar}
}
\label{fig:2patch}
\end{figure}

\noindent
The Fermi surface of
a periodic array of one-dimensional chains
embedded in three-dimensional space without interchain hopping
consists of two parallel completely flat
planes, as shown in Fig.~\secref{fig:2patch}.
Because the Fermi surface does not have any curvature,
it will be sufficient to work with
linearized energy dispersion. 
In this case all interaction effects 
are contained in the Debye-Waller factor
$Q^{\alpha} ( {\vec{r}} , \tau )$ given in 
Eqs.\ref{eq:R2}, \ref{eq:ReS2b} and \ref{eq:ImS2b}.
Note this these expressions
have been derived for an arbitrary geometry of the Fermi surface,
so that we simply have to substitute the parameters relevant
for the case of interest here.
Obviously the local Fermi velocities on the two sheets of the
Fermi surface in Fig.~\secref{fig:2patch} are exactly constant. 
Hence, the entire Fermi surface can be
covered with $M =2$ patches, which can be identified with 
the two sheets of the Fermi surface.
In this case the patch cutoff $\Lambda$ is given by $ 2 \pi / {a_{\bot}}$,
where $a_{\bot}$ is the distance between the chains.
Because the number of patches is finite,
we know from the general analysis given in Chap.~\secref{subsec:Thetermslin}
that $L^{\alpha}_{\vec{q}} ( \tau ) = 0 $
in Eqs.\ref{eq:ReS2b} and \ref{eq:ImS2b}. 
Let us label the right patch
in Fig.~\secref{fig:2patch} by $\alpha = +$, and the left one by $\alpha = -$. 
The associated local Fermi velocities
are ${\vec{v}}^{\alpha} = \alpha v_{\rm F} {\vec{e}}_{x}$.
The non-interacting linearized energy dispersion close to the Fermi surface
is then $\xi^{\alpha}_{\vec{q}} = 
{\vec{v}}^{\alpha} \cdot {\vec{q}} = \alpha v_{\rm F} q_{x}$,
and
the local density of states\index{density of states!array of chains} is
 \begin{equation}
 \nu^{\alpha} =  \int_{- \lambda}^{\lambda} \frac{\D q_{x}}{ 2 \pi } 
 \int_{ - \frac{ \pi}{a_{\bot}} }^{ \frac{ \pi}{a_{\bot}} }  \frac{ \D q_{y}  }{  2 \pi }
 \int_{ - \frac{ \pi}{a_{\bot}} }^{ \frac{ \pi}{a_{\bot}} }  \frac{   \D q_{z} }{  2 \pi }
 \delta ( \alpha v_{\rm F} q_{x} ) = 
 \frac{1}{2 \pi v_{\rm F} a_{\bot}^2 }
 \; \; \; ,
 \label{eq:localdens}
 \end{equation}
where $\lambda$ is some radial cutoff that should be chosen large compared
with the Thomas-Fermi wave-vector $\kappa$ given in
Eq.\ref{eq:Cbcont} below.
Hence, the non-interacting polarization at long wave-lengths is
 \begin{equation}
 \Pi_{0} ( q ) = \sum_{\alpha = \pm } \nu^{\alpha} \frac{ {\vec{v}}^{\alpha} \cdot {\vec{q}} }
 {  {\vec{v}}^{\alpha} \cdot {\vec{q}} - \I \omega_{m} }
 = \nu g_{1} ( \frac{ \I \omega_{m}}{ v_{\rm F} | q_{x} | } )
 \label{eq:Pi0patch}
 \; \; \; ,
 \end{equation}
where the total density of states is given by
 \begin{equation}
 \nu = \sum_{\alpha = \pm } \nu^{\alpha} = 
 \frac{1}{ \pi v_{\rm F} a_{\bot}^2 }
 \label{eq:nuglobalpatch}
 \; \; \; ,
 \end{equation}
and the function $g_{1} (  \I y )$ is
defined in Eq.\ref{eq:g1y}.
The one-dimensional form of the polarization implies that the RPA 
dynamic structure factor\index{dynamic structure factor!array of chains ($t_{\bot}=0$)}
is formally identical with  the one-dimensional expression \ref{eq:SRPA1}, with
the  collective mode and the residue given by
\begin{eqnarray}
\omega_{{\vec{q}}} & = & \sqrt{ 1 + F_{\vec{q}} } v_{\rm F} | q_{x} |
\; \; \; ,
\label{eq:col1patch}
\\
Z_{\vec{q}} & = & 
= \frac{ \nu v_{\rm F} | q_{x} |}{2 \sqrt{ 1 + F_{\vec{q}} } }
=
\frac{ | q_{x} | }{2 \pi a_{\bot}^2 \sqrt{ 1 + F_{\vec{q}} } }
\label{eq:Zom1patch}
\; \; \; .
\end{eqnarray}
Here $F_{\vec{q}} = \nu f_{\vec{q}}  $
is the usual dimensionless interaction.
Compared with the one-dimensional
result given in Eq.\ref{eq:Zom1}, the residue \ref{eq:Zom1patch} has an extra
factor of $a_{\bot}^2$ in the denominator, because we are now dealing with
a three-dimensional system.
At length scales large compared with the lattice spacing $a_{\bot}$, we may replace the
Fourier transform of the Coulomb potential by its continuum approximation, so that in this
case\index{Thomas-Fermi wave-vector}
  \begin{equation}
  F_{\vec{q}} = \frac{ \kappa^2}{ {\vec{q}}^2 } \; \; \; , \; \; \; 
  \kappa^2 = \frac{4 e^2}{v_{\rm F} a_{\bot}^2}
  \; \; \; , \; \; \; \mbox{for $|  {\vec{q}} | a_{\bot}  \ll 1$}
  \; \; \; .
  \label{eq:Cbcont}
  \end{equation}
Given the fact that $q_{y}$ and $q_{z}$ 
appear in the Debye-Waller factor 
only via $F_{\vec{q}}$, and that the
dynamic structure factor has the one-dimensional form, it is now easy
to see that the frequency integration 
in Eqs.\ref{eq:R2}, \ref{eq:ReS2b} and \ref{eq:ImS2b}
is exactly the same as in the one-dimensional case, so that we can simply
copy the results of Chap.~\secref{sec:Green1}.
From Eqs.\ref{eq:R3}--\ref{eq:ImS3} we obtain
 \begin{equation}
 R =  
 - \int_{0}^{\infty}
 \frac{ \D q_{x} }{q_{x}}
  \left< \frac{F_{\vec{q}}^2}
 {2 \sqrt{ 1 + F_{\vec{q}} } [ \sqrt{1+F_{\vec{q}} } + 1 ]^2 }
 \right>_{\rm BZ}
 \label{eq:R3patch}
 \; \; \; ,
 \end{equation}
 \begin{eqnarray}
 {\rm Re} S ( x , \tau )
 & = &  
 - \int_{0}^{\infty}
 \frac{ \D q_{x} }{q_{x}} 
 \cos ( q_{x} x )
 \nonumber
 \\
 & & \times
 \left[ 
 \left<
 \frac{ 1  + \frac{F_{\vec{q}}}{2} }{  \sqrt{ 1 + F_{\vec{q}} } }
 \E^{ - \sqrt{ 1 + F_{\vec{q}} } v_{\rm F} q_{x} | \tau | } 
 \right>_{\rm BZ}
 - \E^{ - v_{\rm F} q_{x} | \tau | } \right]
  \; ,
 \label{eq:ReS3patch}
 \\
 {\rm Im} S ( x , \tau )
 & = &
 - {\rm sgn} ( \tau )
 \int_{0}^{\infty}
 \frac{ \D q_{x} }{q_{x}} \sin ( q_{x}  x   )
 \nonumber
 \\
 & & \times
 \left[ 
 \left< \E^{ - \sqrt{ 1 + F_{\vec{q}} } v_{\rm F} q_{x} | \tau | } \right>_{\rm BZ}
 - \E^{ - v_{\rm F} q_{x} | \tau | } \right]
 \; \; ,
 \label{eq:ImS3patch}
 \end{eqnarray}
where for any function $f ( {\vec{q}} )$
the symbol $ < f ( {\vec{q}}  ) >_{\rm BZ} $ denotes averaging over the
transverse Brillouin zone\index{Brillouin zone average}, 
 \begin{equation}
 \left< f ( {\vec{q}} ) \right>_{\rm BZ} = 
  \frac{ a_{\bot}^2 }{  ( 2 \pi)^2 }  
 \int_{ - \frac{ \pi}{a_{\bot}} }^{ \frac{ \pi}{a_{\bot}} }   \D q_{y} 
 \int_{ - \frac{ \pi}{a_{\bot}} }^{ \frac{ \pi}{a_{\bot}} }     \D q_{z} 
 f ( {\vec{q}} )
 \label{eq:BZavdef}
 \; \; \; .
 \end{equation}
The above expression for the Green's function can also be derived by 
means of standard one-dimensional bosonization techniques \cite{Schulz83,Penc93}. 
However, as will be shown in
Sect.~\secref{sec:Finiteinter},
our derivation via higher-dimensional bosonization
has the advantage that it can be generalized to the case of finite interchain hopping.

For $\tau = 0$ we can make
progress analytically in the regime where the Thomas-Fermi screening length $\kappa^{-1}$ is
large compared with the transverse lattice spacing $a_{\bot}$,
i.e. for $ \kappa a_{\bot} \ll 1$.
Because  in this case all wave-vector integrals are dominated by the regime
$| {\vec{q}} | \leqapprox \kappa$, it is allowed to use  in Eq.\ref{eq:Cbcont}
the continuum approximation for the Fourier transform of the Coulomb potential. 
Note that 
 \begin{equation}
 \kappa a_{\bot} = \sqrt{ \frac{ 4 e^2}{v_{\rm F}} }
 \; \; \; ,
 \label{eq:kappaabot}
 \end{equation}
so that the condition $\kappa a_{\bot} \ll 1$ means that the
dimensionless coupling constant $e^{2} / v_{F}$ should be small compared with
unity. Unfortunately at experimentally relevant densities 
this parameter is of the order of unity,
so that in this case the continuum approximation for the Fourier transform of the
Coulomb potential cannot be used. To reach the experimentally
relevant parameter regime, one should therefore take in $F_{\vec{q}}$
the discreteness of the lattice 
in the transverse direction into account. 
In \cite{Kopietz94b} this was done by means of an Ewald summation technique \cite{Ziman64}.
Here we would like to restrict ourselves to the regime $\kappa a_{\bot} \ll 1$.

For $ \tau = 0$ we need to calculate the following Brillouin zone average
 \begin{equation}
 \hspace{-2mm}
 \gamma_{\rm cb} ( {q_{x}} ) =
 \left< \frac{F_{\vec{q}}^2}
 {2 \sqrt{ 1 + F_{\vec{q}} } [ \sqrt{1+F_{\vec{q}} } + 1 ]^2 }
 \right>_{\rm BZ}
 =
  \left< \frac{ 1  + \frac{F_{\vec{q}}}{2} }{  \sqrt{ 1 + F_{\vec{q}} } } \right>_{\rm BZ}
 - 1 
 \label{eq:tildeFqparallel}
 \; .
 \end{equation}
Substituting  Eq.\ref{eq:Cbcont} into Eq.\ref{eq:tildeFqparallel} and using
Eq.\ref{eq:kappaabot}, the integration is elementary, and we obtain 
for $\kappa a_{\bot} \ll 1$
 \begin{equation}
 \gamma_{\rm cb} ( {q_{x}} ) =
 \frac{e^{2}}{2 \pi v_{\rm F}} \frac{ 1}{ \left[ \frac{| q_{x}| }{ \kappa } + 
 \sqrt{ 1 + ( \frac{q_{x} }{ \kappa} )^2 } \right]^2 }
 \; \; \; .
 \label{eq:Ftildeqpares}
 \end{equation}
The asymptotic behavior for large and small $| q_{x} |$ is
 \begin{equation}
 \gamma_{\rm cb} ( q_{x} ) \sim \frac{e^2}{ 2 \pi v_{\rm F}} \times
 \left\{
 \begin{array}{lr}
 1 & \mbox{for $ |q_{x} | \ll \kappa $ }
 \\
 ( \frac{ \kappa}{ 2 q_{x} })^2
 & \mbox{ for $ | q_{x}| \gg \kappa$ }
 \end{array}
 \right.
 \; \; \; .
 \label{eq:Ftildeasym}
 \end{equation}
Because $\gamma_{\rm cb} ( q_{x} )$ has a finite limit 
for $q_{x} \rightarrow 0$, the integral \ref{eq:R3patch}
defining $R$ is logarithmically divergent, so that the
system is a Luttinger liquid\index{Luttinger liquid!array of chains $(t_{\bot}=0)$}. 
Moreover, for $q_{x} \gg \kappa$ the function
$\gamma_{\rm cb}( {q_{x}})$ vanishes sufficiently fast to insure 
ultraviolet convergence of the integral defining
$Q ( x , 0)$.
Recall that in the one-dimensional Tomonaga-Luttinger model
(see Chap.~\secref{sec:Green1})
it was necessary to introduce an ultraviolet cutoff $q_{\rm c}$ to make the
integrals convergent. The precise physical origin
of this cutoff has remained somewhat obscure.
In the present problem, however, the effective
ultraviolet cutoff can be identified with the Thomas-Fermi screening wave-vector.
To calculate the anomalous dimension, we consider the
long-distance behavior of the static Debye-Waller factor.
Using Eqs.\ref{eq:R3patch},\ref{eq:ReS3patch} and \ref{eq:Ftildeqpares}, 
and introducing the dimensionless integration variable $p = q_{x} / \kappa$,
we obtain
 \begin{equation}
 Q ( x , 0 )
 = - \frac{e^2}{ 2 \pi v_{\rm F}} 
 \int_{0}^{\infty} \frac{\D p}{p}
  \frac{ 1 - \cos ( p \kappa {x} ) }
 { \left[ p + 
 \sqrt{ 1 + p^2 } \right]^2 }
 \; \; \; .
 \label{eq:Qstatpatch}
 \end{equation}
To calculate the asymptotic behavior of the integral
for large $\kappa x$, we write
 \begin{equation}
  \frac{ 1 - \cos ( p \kappa x  ) }{p}
 = \frac{\D}{ \D p} \left[ \ln p - {\rm Ci} ( p \kappa x  ) \right]
 \label{eq:IBPq}
 \; \; \; ,
 \end{equation}
where \cite{Abramowitz70}
 \begin{equation}
 {\rm Ci} ( x) = - \int_{x}^{\infty} \D t \frac{ \cos t}{t}
 \; \; \; .
 \end{equation}
An integration by parts yields
 \begin{eqnarray}
 & & \hspace{-10mm}
 Q ( x , 0 )
  =  - \frac{e^2}{ 2 \pi v_{\rm F}} 
 \left[
 \lim_{p \rightarrow 0} [ {\rm Ci} ( p \kappa  |x|  ) - \ln p ]
 + 2 \int_{0}^{\infty} \D p \frac{ \ln p }{ \sqrt{1 + p^2 } [ p + \sqrt{ 1 + p^2} ]^2 }  \right]
 \nonumber
 \\
 & + & 
 \frac{e^2}{  \pi v_{\rm F}} 
 \int_{0}^{\infty} \D p \frac{ {\rm Ci} ( p \kappa |x| ) }{ \sqrt{1 + p^2 } [ p + \sqrt{ 1 + p^2} ]^2 } 
 \label{eq:IBPq2}
 \; \; \; .
 \end{eqnarray}
Using the fact that \cite{Abramowitz70}
 \begin{equation}
 \lim_{p \rightarrow 0} [ {\rm Ci} ( p \kappa | x | ) - \ln p ]
 = \ln ( \kappa | x | ) + \gamma_{\rm E} \; \; \; ,
 \end{equation}
where $\gamma_{\rm E} \approx 0.577 $ is the Euler constant, and noting that the last term in
Eq.\ref{eq:IBPq2} vanishes as $\frac{\ln ( \kappa |x| )}{ \kappa | x |}$ as $x \rightarrow \infty$, we 
finally obtain
 \begin{equation}
 Q ( x , 0 )
 \sim - \frac{e^2}{ 2 \pi v_{\rm F}} 
 \left[ \ln ( \kappa |x | ) + b + 
 O \left( \frac{ \ln ( \kappa | x | ) }{ \kappa | x | } \right)
 \right]
 \; \; \; ,
 \label{eq:Qpatchres}
 \end{equation}
where the numerical constant $b$ is  given by
 \begin{equation}
 b = \gamma_{\rm E} +
  2 \int_{0}^{\infty} \D p \frac{ \ln p }{ \sqrt{1 + p^2 } [ p + \sqrt{ 1 + p^2} ]^2 } 
  \label{eq:k1constdef}
  \; \; \; .
  \end{equation}
We conclude that the interacting equal-time Green's function vanishes at large
distances as\index{Green's function!coupled chains ($t_{\bot} \neq 0$)}
 \begin{eqnarray}
 G^{\alpha} ( {\vec{r}} , 0 ) & = &
 G^{\alpha}_{0} ( {\vec{r}} , 0 ) 
 \left( \frac{ \E^{- b } }{ \kappa | r_{x}  | } \right)^{\gamma_{\rm cb}}
 \nonumber
 \\
 & = & \delta ( r_{y} ) \delta ( r_{z} )
 \left( \frac{- \I }{2 \pi } \right) \left(
 \frac{ \E^{ - b }}{ \kappa} \right)^{\gamma_{\rm cb} }
 \frac{1}{ | r_{x} |^{1 + \gamma_{\rm cb} } }
 \; \; \; ,
 \label{eq:Gpatchres}
 \end{eqnarray}
where the anomalous 
dimension\index{anomalous dimension!array of chains ($t_{\bot} = 0$)} 
$\gamma_{\rm cb}$ is given by
 \begin{equation}
 \hspace{-4mm}
 \gamma_{\rm cb}  \equiv  
  \lim_{q_{x} \rightarrow 0}
 \gamma_{\rm cb} ( q_{x} ) 
 =
  \lim_{q_{x} \rightarrow 0}
 \left< \frac{F_{\vec{q}}^2}
 {2 \sqrt{ 1 + F_{\vec{q}} } [ \sqrt{1+F_{\vec{q}} } + 1 ]^2 }
 \right>_{\rm BZ}
=
  \frac{e^2}{ 2 \pi v_{\rm F} }  
 \; .
 \label{eq:gammacbres}
 \end{equation}
We would like to emphasize again that this expression is only 
valid for $e^{2} / v_{\rm F} \ll 1$, so that it would be incorrect
to extrapolate Eq.\ref{eq:gammacbres} to the
experimentally relevant regime $e^2 / v_{\rm F} = O (1)$. In this regime 
the simple continuum approximation for the Fourier transform
of the Coulomb potential is not sufficient, and one has to
use numerical methods to calculate the anomalous dimension.
This numerical calculation has been performed in
\cite{Kopietz94b}, with the result that
in the experimentally relevant regime
the anomalous dimension is indeed of the order of unity.
Recent photoemission studies \cite{Dardel93,Nakamura94,Gweon96} 
of quasi-one-dimensional conductors
suggest values for the anomalous dimension in the range
$1.0 \pm 0.2$, which is in agreement
with our result.
However, the comparison of the experimental result with
Eq.\ref{eq:gammacbres} is at least problematic, because
our calculation was based on several idealizations which
are perhaps not satisfied in the realistic experimental system.
First of all, the experiments are certainly not performed on
perfectly clean systems. Because any finite disorder
changes the algebraic decay 
in Eq.\ref{eq:Gpatchres} into an exponential one
(see Chap.~\secref{chap:adis}), 
the Luttinger liquid behavior
is completely destroyed by impurities.
Therefore one cannot exclude the possibility that
the experiments do not measure the Luttinger liquid nature of the system, but 
are essentially determined by impurities.
Another possibly unjustified idealization in our calculation is the
neglect of processes with large momentum transfers, which might
favour charge-density wave instabilities or other broken symmetries. The associated
pseudo-gaps in the excitation spectrum will certainly lead to a further suppression
of the momentum integrated spectral function 
in the vicinity of the Fermi energy, which competes with
the suppression inherent in the Luttinger liquid state.
Nevertheless, in spite of all these caveats, 
we believe that the large 
value of $\gamma_{\rm cb}$ 
due to long-range Coulomb forces 
can give rise to an important contribution
to the suppression of the spectral weight seen in the experiments.

At finite $\tau$ we have not been able to calculate the integral defining
$Q ( x , \tau )$ analytically.
In \cite{Kopietz94b} the numerical
method developed by Meden and Sch\"{o}nhammer \cite{Meden92} was 
used to calculate the full momentum- and frequency-dependent spectral function.
More detailed numerical calculations
can be found in the thesis by Meden \cite{Meden96}.
In contrast to our present discussion,
in the works \cite{Kopietz94b,Kopietz96chain,Meden96}
the spin degree of freedom was also taken into account,
and the phenomenon of spin-charge separation 
was studied.\index{spin-charge separation}
The fact that the spin and the charge excitations 
manifest themselves 
with different velocities in the single-particle
Green's function is one of the fundamental
characteristics of a Luttinger liquid \cite{Haldane81}.

\section{Finite interchain hopping}
\label{sec:Finiteinter}

\begin{figure}
\sidecaption
\psfig{figure=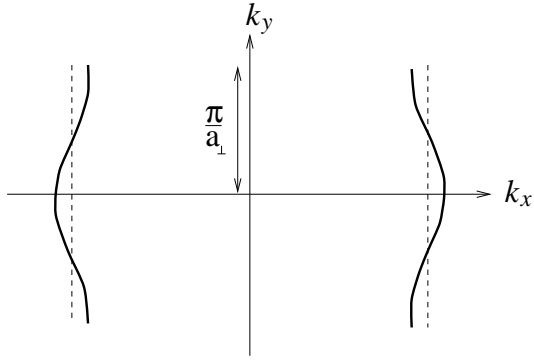,width=7cm}
\caption[Fermi surface of a quasi-one-dimensional conductor.]
{ 
Fermi surface of a periodic array of 
chains\index{Fermi surface!coupled chains ($t_{\bot} \neq 0$)}
with interchain hopping. Only the intersection with the plane $k_{z} = 0$ is shown. }
\label{fig:quasi1d}
\end{figure}

\noindent
Experimentally the interchain  hopping 
$t_{\bot}$\index{interchain hopping} can never be completely
turned off. Realistic Fermi surfaces of quasi-one-dimensional conductors have therefore
the form shown in Fig.~\secref{fig:quasi1d}. 
The amplitude of the modulation of the Fermi surface sheets
is proportional to the interchain hopping $t_{\bot}$. Because
the intrachain hopping $t_{\|}$ is of the order of $E_{\rm F}$, 
the relevant small dimensionless parameter which measures the
quasi-one-dimensionality of the system is
 \begin{equation}
 \theta = \frac{| {t}_{\bot} |}{E_{\rm F}}
 \; \; \; .
 \label{eq:Thetasmallpardef}
 \end{equation}
From the previous section we know that for $\theta = 0$ the system
is a Luttinger liquid. We now calculate the
Green's function of the system for small but finite $\theta$,
assuming transverse hopping only in the 
$y$-direction. This approximation is justified
for materials where the interchain hopping
$t_{\bot} = t_{y}$ in the $y$-direction is large compared
with the interchain hopping in the $z$-direction. 
As discussed in \cite{Jerome91}, 
this condition is satisfied for some experimentally studied materials.

\subsection{The $4$-patch model\index{four-patch model}}
\label{subsec:4patchmodel}

{\it{We now break the symmetry of the Fermi
surface by deforming the flat sheets into wedges, so
that we obtain a model with four patches.}} 

\vspace{7mm}

\noindent
For simplicity let us first assume 
that the Fermi surface has the shape shown in Fig.~\secref{fig:4patch}:
\begin{figure}
\psfig{figure=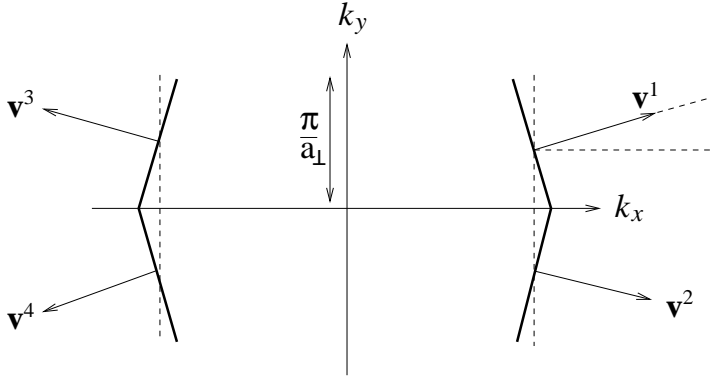,width=11cm}
\caption[Fermi surface of the $4$-patch model.]
{ 
Fermi surface\index{four-patch model!Fermi surface} of the $4$-patch model. Only the intersection
with a plane of constant $k_{z}$ is shown. }
\label{fig:4patch}
\end{figure}
it consists of four perfectly flat patches,
which are obtained by replacing the sine modulation in Fig.~\secref{fig:quasi1d}
by a triangle.
Because the coefficients in the Fourier decomposition of a triangular wave 
vanish rather slowly,
the microscopic origin for
such a Fermi surface is a particular superposition of long-range hoppings. 
We shall refer to our hopping model as the  $4$-{\it{patch model}}.
In a sense, this is the simplest example for a 
non-trivial model in $d > 1$ which can be
discussed within the framework of higher-dimensional bosonization.
The patches are now labelled by $\alpha = 1,2,3,4$.
Because the curvature of the patches vanishes by construction,
the local Fermi velocities are constant on a given patch.
From Fig.~\secref{fig:4patch} we see that
 \begin{equation}
 \begin{array}{ccl}
 {\vec{v}}^{1} & = &  ( {\vec{e}}_{x} \cos \theta + {\vec{e}}_{y} \sin \theta ) v_{\rm F} \\
 {\vec{v}}^{2} & = & ( {\vec{e}}_{x} \cos \theta - {\vec{e}}_{y} \sin \theta ) v_{\rm F}
 \\
 {\vec{v}}^{3} & = & ( - {\vec{e}}_{x} \cos \theta + {\vec{e}}_{y} \sin \theta ) v_{\rm F}
 \\
 {\vec{v}}^{4} & = & ( - {\vec{e}}_{x} \cos \theta - {\vec{e}}_{y} \sin \theta ) v_{\rm F}
 \end{array}
 \; \; \; .
 \label{eq:velocities4patch}
 \end{equation}
To calculate the Green's function, we simply repeat the steps of the previous section.
The non-interacting polarization is now
 \begin{equation}
 \Pi_{0} ( q ) = \frac{ \nu}{4}\sum_{\alpha = 1 }^{4}  
 \frac{ {\vec{v}}^{\alpha} \cdot {\vec{q}} }
 { {\vec{v}}^{\alpha} \cdot {\vec{q}}  - \I \omega_{m} }
 = \frac{ \nu}{2} \sum_{\alpha = 1}^{2} 
 \frac{ ( {\vec{v}}^{\alpha} \cdot {\vec{q}})^2 }
 { ( {\vec{v}}^{\alpha} \cdot {\vec{q}})^2 + \omega_{m}^2 }
 \label{eq:Pi04patch}
 \; \; \; ,
 \end{equation}
where the global density of states 
 $\nu = \sum_{\alpha =1 }^{4} \nu^{\alpha} $ is for small $\theta$ identical
with Eq.\ref{eq:nuglobalpatch}.
As usual, the collective modes are obtained by solving Eq.\ref{eq:plasmondef2},
which for $M = 4$ patches leads to the bi-quadratic equation
 \begin{equation}
  z^4 - 
 \left( 1 + \frac{F_{\vec{q}} }{2} \right) 
 \left( \xi_{\vec{q}}^2 + \tilde{\xi}_{\vec{q}}^2 \right)
  z^2
 + 
  ( 1 + F_{\vec{q}} )  \xi_{\vec{q}}^2 \tilde{\xi}_{\vec{q}}^2
  = 0
 \; \; \; ,
 \label{eq:biquad4patch}
 \end{equation}
where we have introduced the notation
 \begin{eqnarray}
 \xi_{\vec{q}} & = & {\vec{v}}^{1} \cdot {\vec{q}} = v_{\rm F} 
 ( q_{x} \cos \theta + q_{y} \sin \theta )
 \; \; \; ,
 \label{eq:xi4patchdef}
 \\
 \tilde{\xi}_{\vec{q}} & = & {\vec{v}}^{2} \cdot {\vec{q}} = v_{\rm F} 
 ( q_{x} \cos \theta - q_{y} \sin \theta )
 \label{eq:barxi4patchdef}
 \; \; \; .
 \end{eqnarray}
The bi-quadratic equation \ref{eq:biquad4patch} is easily solved.\index{four-patch model!collective modes}
The two solutions are
 \begin{eqnarray}
   \omega_{\vec{q}}^2  & = &
  \left( 1 + \frac{F_{\vec{q}}}{2} \right) 
 \frac{ \xi_{\vec{q}}^2 + \tilde{\xi}_{\vec{q}}^2 }{2}
 \nonumber
 \\
 & &
 + \frac{1}{2} \left[  F_{\vec{q}}^2
 \left( \frac{ \xi_{\vec{q}}^2 + \tilde{\xi}_{\vec{q}}^2 }{2} \right)^2 +
 ( 1 + F_{\vec{q}} ) ( \xi_{\vec{q}}^2 - \tilde{\xi}_{\vec{q}}^2 )^2 \right]^{1/2}
 \; ,
 \label{eq:omega4patch}
 \\
   \tilde{\omega}_{\vec{q}}^2  & = &
  \left( 1 + \frac{F_{\vec{q}}}{2} \right) 
 \frac{ \xi_{\vec{q}}^2 + \tilde{\xi}_{\vec{q}}^2 }{2}
 \nonumber
 \\
 & &
 - \frac{1}{2}
 \left[  F_{\vec{q}}^2
 \left( \frac{ \xi_{\vec{q}}^2 + \tilde{\xi}_{\vec{q}}^2 }{2} \right)^2 +
 ( 1 + F_{\vec{q}} ) ( \xi_{\vec{q}}^2 - \tilde{\xi}_{\vec{q}}^2 )^2 \right]^{1/2}
 \;  .
 \label{eq:baromega4patch}
 \end{eqnarray}
Note that for small $\theta$
 \begin{eqnarray}
 \frac{ \xi_{\vec{q}}^2 + \tilde{\xi}_{\vec{q}}^2 }{2}  
 & \approx & v_{\rm F}^2 ( q_{x}^2 + \theta^2 q_{y}^2 )
 \; \; \; ,
 \label{eq:xisumsmall}
 \\
  \xi_{\vec{q}}^2 - \tilde{\xi}_{\vec{q}}^2  & \approx & 4 \theta v_{\rm F}^2   q_{x} q_{y}
  \; \; \; ,
  \label{eq:xidifsmall}
  \\
  \xi_{\vec{q}} \tilde{\xi}_{\vec{q}}  & \approx &  v_{\rm F}^2   ( q_{x}^2 - \theta^2 q_{y}^2 )
  \; \; \; ,
  \end{eqnarray}
and that
 \begin{equation}
 \omega_{\vec{q}}^2 - \tilde{\omega}_{\vec{q}}^2 = 
 \left[  F_{\vec{q}}^2
 \left( \frac{ \xi_{\vec{q}}^2 + \tilde{\xi}_{\vec{q}}^2 }{2} \right)^2 +
 ( 1 + F_{\vec{q}} ) ( \xi_{\vec{q}}^2 - \tilde{\xi}_{\vec{q}}^2 )^2 \right]^{1/2}
 \; \; \; .
 \label{eq:omegadif}
 \end{equation}
The right-hand side of Eq.\ref{eq:baromega4patch} is non-negative because
 \begin{eqnarray}
  \left[ \left( 1 + \frac{F_{\vec{q}}}{2} \right) 
 \frac{ \xi_{\vec{q}}^2 + \tilde{\xi}_{\vec{q}}^2 }{2} \right]^2
 - \frac{1}{4}
 \left[  F_{\vec{q}}^2
 \left( \frac{ \xi_{\vec{q}}^2 + \tilde{\xi}_{\vec{q}}^2 }{2} \right)^2 +
 ( 1 + F_{\vec{q}} ) ( \xi_{\vec{q}}^2 - \tilde{\xi}_{\vec{q}}^2 )^2 \right]
 & & 
 \nonumber
 \\
 & & \hspace{-100mm} =  ( 1 + F_{\vec{q}} ) \xi_{\vec{q}}^2 \tilde{\xi}_{\vec{q}}^2 \geq 0
 \label{eq:positivecheck}
 \; \; \; .
 \end{eqnarray}
Therefore both modes $\omega_{\vec{q}}$ and $\tilde{\omega}_{\vec{q}}$ are not damped and give
rise to $\delta$-function peaks in the dynamic structure factor.
The dielectric function is then given by\index{four-patch model!dielectric function}
 \begin{equation} 
 \epsilon_{\rm RPA} ( {\vec{q}} , \omega ) \equiv
 1 + F_{\vec{q}} \Pi_{0} ( {\vec{q}} , \omega ) = 
 \frac{ ( \omega^2 -  \omega_{\vec{q}}^{2} )
 ( \omega^2 - \tilde{\omega}_{\vec{q}}^2 )}
 {
 ( \omega^2 -  \xi_{\vec{q}}^{2} )
 ( \omega^2 - \tilde{\xi}_{\vec{q}}^2 )}
 \label{eq:col4patchfac}
 \; \; \; ,
 \end{equation}
so that the RPA polarization is simply 
 \begin{equation}
 \Pi_{\rm RPA} ( {\vec{q}} , \omega ) = \nu 
 \frac{ \xi_{\vec{q}}^2 \tilde{\xi}_{\vec{q}}^2 - \omega^2 
 \frac{ \xi_{\vec{q}}^2 + \tilde{\xi}_{\vec{q}}^2}{2} }
 {
 ( \omega^2 -  \omega_{\vec{q}}^{2} )
 ( \omega^2 - \tilde{\omega}_{\vec{q}}^2 )}
 \; \; \; .
 \label{eq:Pirpa4patch}
 \end{equation}
Note that $\epsilon_{\rm RPA} ( {\vec{q}} , \xi_{\vec{q}} ) =
\epsilon_{\rm RPA} ( {\vec{q}} , \tilde{\xi}_{\vec{q}} ) = \infty$, in agreement
with Eq.\ref{eq:frpaonshell}.
The RPA dynamic structure factor\index{four-patch model!dynamic structure factor}
is then easily calculated from
Eq.\ref{eq:SPi}. For $\omega > 0$ we obtain
 \begin{equation}
 S_{\rm RPA} ( {\vec{q}} , \omega ) = Z_{\vec{q}} \delta ( \omega - \omega_{\vec{q}} )
 + \tilde{Z}_{\vec{q}} \delta ( \omega - \tilde{\omega}_{\vec{q}} )
 \label{eq:SRPA4patch}
 \; \; \; ,
 \end{equation}
with the residues given by
 \begin{eqnarray}
 Z_{\vec{q}} & = & \frac{\nu}{2 \omega_{\vec{q}} }
 \frac{ 
 \omega_{\vec{q}}^2 \frac{ \xi_{\vec{q}}^2 + \tilde{\xi}_{\vec{q}}^2 }{2}  - \xi_{\vec{q}}^2
 \tilde{\xi}_{\vec{q}}^2 }{ \omega_{\vec{q}}^2 - \tilde{\omega}_{\vec{q}}^2 }
 \; \; \; ,
 \label{eq:Z1patch}
 \\
 \tilde{Z}_{\vec{q}} & = &  \frac{\nu}{2 \tilde{\omega}_{\vec{q}} }
 \frac{ 
 \xi_{\vec{q}}^2
 \tilde{\xi}_{\vec{q}}^2 
 - \tilde{\omega}_{\vec{q}}^2 \frac{ \xi_{\vec{q}}^2 + \tilde{\xi}_{\vec{q}}^2 }{2}   }
 { \omega_{\vec{q}}^2 - \tilde{\omega}_{\vec{q}}^2 }
 \label{eq:Z2patch}
 \; \; \; .
 \end{eqnarray}
In the limit $\theta \rightarrow 0$ we have $\tilde{\xi}_{\vec{q}} \rightarrow \xi_{\vec{q}}
\rightarrow v_{\rm F} q_{x}$, so that
$\omega_{\vec{q}} \rightarrow \sqrt{ 1 + F_{\vec{q}} } v_{\rm F} | q_{x} |$ and
$\tilde{\omega}_{\vec{q}} \rightarrow v_{\rm F} | q_{x} |$.
It is also easy to see that the residue $Z_{\vec{q}}$ reduces in this limit
to the result \ref{eq:Zom1patch} without interchain hopping, while the residue
$\tilde{Z}_{\vec{q}}$ vanishes. 

To calculate the Green's function, we substitute
Eq.\ref{eq:SRPA4patch} into Eqs.\ref{eq:R2}, \ref{eq:ReS2b} and \ref{eq:ImS2b}. Because
the dynamic structure factor consists of a sum of two $\delta$-functions, the frequency
integration is trivial. 
As before $L^{\alpha}_{\vec{q}} ( \tau ) = 0$, because
we have covered the Fermi surface with a finite number of patches.
To see whether the interchain hopping destroys the Luttinger liquid state, 
it is sufficient to calculate the static Debye-Waller factor.
Substituting Eq.\ref{eq:SRPA4patch} into Eqs.\ref{eq:R2} and \ref{eq:ReS2b},
we obtain
 \begin{eqnarray}
 Q^{\alpha} ( r_{\|}^{\alpha}  \hat{\vec{v}}^{\alpha} , 0 )
 & = & 
 R^{\alpha} -
 S^{\alpha} ( r_{\|}^{\alpha}  \hat{\vec{v}}^{\alpha} , 0 )
 \nonumber
 \\
  &  &  \hspace{-23mm} =
      - \frac{1}{V} \sum_{\vec{q}} 
   \left[ 1 - \cos ( \hat{\vec{v}}^{\alpha} \cdot \vec{q} 
   r^{\alpha}_{\|} ) \right]
   f_{\vec{q}}^2 
 \left[ \frac{Z_{\vec{q}}}{ ( \omega_{\vec{q}} + 
 | {\vec{v}}^{\alpha} \cdot {\vec{q}} |)^2}
 +  \frac{\tilde{Z}_{\vec{q}}}{ ( \tilde{\omega}_{\vec{q}} + 
 | {\vec{v}}^{\alpha} \cdot {\vec{q}} |)^2}
 \right]
 \;  .
 \nonumber
 \\
 & &
 \label{eq:R4patch1}
 \end{eqnarray}
To evaluate Eq.\ref{eq:R4patch1}, we need to simplify the above expressions for the
collective modes and the associated residues.
Depending on the
relative order of magnitude
of $F_{\vec{q}}$ and $\theta$,
three regimes have to be distinguished,
 \begin{equation}
 \begin{array}{lc}
(a) : &  \theta \ll 1 \ll F_{\vec{q}}   \\
(b) : & \theta \ll F_{\vec{q}} \ll 1 \\
(c) :  & F_{\vec{q}} \ll \theta \ll 1 
\end{array}
\label{eq:regimeschain1d}
\; \; \; .
\end{equation}
Note that in the weak coupling regime $(b)$ 
the energy scale set by the interaction 
is smaller than the intrachain hopping energy $t_{\|}$,
but still large compared with the 
interchain hopping $t_{\bot}$.
In the
second weak coupling regime $(c)$ the interaction 
is even smaller than the 
kinetic energy associated with transverse hopping.
Because for $| {\vec{q}} | \leqapprox \kappa$ the dimensionless Coulomb interaction
$F_{\vec{q}}$
is large compared with unity, in the present problem
only the strong coupling regime $(a)$ is of interest.

We begin with the evaluation of the first term in Eq.\ref{eq:R4patch1}
involving the mode $\omega_{\vec{q}}$.
We then show that
the contribution of the second term, which involves
the other mode $\tilde{\omega}_{\vec{q}}$,
grows actually logarithmically for $r_{\|}^{\alpha} \rightarrow \infty$,
signaling  Luttinger liquid behavior.
However, this is due to an unphysical
nesting symmetry inherent in our $4$-patch model
with flat patches; in Sect.~\secref{subsec:thefate}
we shall slightly deform our patches so that
they have a finite curvature, and show
that in this case the contribution from the second mode 
remains bounded for all $r_{\|}^{\alpha}$ and
is negligible compared with the contribution from the first mode.

\begin{center}
{\bf{The plasmon mode}}
\end{center}

\noindent
From Eqs.\ref{eq:omega4patch} and \ref{eq:baromega4patch} it is easy to see that,
up to higher orders in $\theta / F_{\vec{q}}$, the collective density mode
$\omega_{\vec{q}}$ in the strong coupling regime 
can be approximated by
 \begin{eqnarray}
 \omega_{\vec{q}} & \approx & v_{\rm F} 
 \sqrt{ 1 + F_{\vec{q}} }  \sqrt{ q_{x}^2 + \theta^2 q_{y}^2 }
 \label{eq:om4SC}
 \; \; \; .
 \end{eqnarray}
Note that for $\theta \rightarrow 0$ this mode
reduces to the plasmon mode \ref{eq:col1patch}
in the absence of interchain hopping. 
Therefore we shall refer to the collective mode $\omega_{\vec{q}}$
as the plasmon mode.
Substituting Eq.\ref{eq:om4SC} into Eq.\ref{eq:Z1patch}, we obtain
for the associated residue
 \begin{eqnarray}
 Z_{\vec{q}} & \approx & \frac{\nu v_{\rm F} \sqrt{ q_{x}^2 + \theta^2 q_{y}^2}}{2 \sqrt{1+F_{\vec{q}}} }
 \label{eq:Z1patchSC}
 \; \; \; ,
 \end{eqnarray}
which should be compared with Eq.\ref{eq:Zom1patch}.
Note that the 
only effect of the interchain hopping is the replacement
$| q_{x} | \rightarrow \sqrt{ q_{x}^2 + \theta^2 q_{y}^2 }$.
The contribution $R^{\alpha}_{\rm pl}$ of the plasmon mode
to the constant part $R^{\alpha}$
of the Debye-Waller factor 
\ref{eq:R4patch1} is then for small $\theta$ given by
 \begin{equation}
 R^{\alpha}_{\rm pl} =  - \int_{0}^{ \infty } \D q_{x}
 \left< 
 \frac{1}{ \sqrt{ q_{x}^2 + \theta^2 q_{y}^2 }} 
  \frac{F_{\vec{q}}^2 }
 { 2 \sqrt{ 1 + F_{\vec{q}} } 
 \left[ \sqrt{ 1 + F_{\vec{q}} } + 1
 \right]^2 }
 \right>_{\rm BZ}
 \label{eq:R4patch2}
 \; \; \; .
 \end{equation}
Although $R^{\alpha}_{\rm pl}$ is to this order in $\theta$ independent
of $\alpha$, we shall keep the patch index here.
If we set $\theta = 0$ in this expression, we recover the 
previous result \ref{eq:R3patch}
in the absence of interchain hopping, which is
logarithmically divergent. This divergence is due to the fact
that for $\theta = 0$ the first factor in Eq.\ref{eq:R4patch2}
can be pulled out of the averaging bracket. However, for any
finite $\theta$ the $q_{x}$- and $q_{y}$-integrations are correlated, 
so that it is not possible to factorize the integrations. Hence,
any non-zero value of $\theta$ couples the phase space 
of the ${\vec{q}}$-integration.
Because for $\theta \rightarrow 0$ the integral in Eq.\ref{eq:R4patch2} is logarithmically
divergent, the coefficient of the leading logarithmic 
term can be extracted by ignoring the $q_{x}$-dependence of
the second factor in the averaging bracket.
Then we obtain to leading logarithmic order
 \begin{eqnarray}
 R^{\alpha}_{\rm pl} & \sim &  - \int_{0}^{ \kappa } \D q_{x}
 \left< 
 \frac{1}{ \sqrt{ q_{x}^2 + \theta^2 q_{y}^2 }} 
 \lim_{q_{x} \rightarrow 0}
 \left[ \frac{F_{\vec{q}}^2 }
 { 2 \sqrt{ 1 + F_{\vec{q}} } 
 \left[ \sqrt{ 1 + F_{\vec{q}} } + 1
  \right]^2 }
 \right]
 \right>_{\rm BZ}
 \nonumber
 \\
 & = & - \left< 
 \ln \left( \frac{\kappa}{\theta | q_{y} | } \right)
 \lim_{q_{x} \rightarrow 0}
 \left[ \frac{F_{\vec{q}}^2 }
 { 2 \sqrt{ 1 + F_{\vec{q}} } 
 \left[ \sqrt{ 1 + F_{\vec{q}} } + 1
 \right]^2 }
 \right] \right>_{\rm BZ}
 \nonumber
 \\
 & = &
 - \gamma_{\rm cb} \left[ \ln \left( \frac{1}{\theta} \right) + b_{1} \right]
 \label{eq:Rres4patch}
 \; \; \; ,
 \end{eqnarray}
where $\gamma_{\rm cb}$ is given in Eq.\ref{eq:gammacbres}, and
$b_{1}$ is a numerical constant of the order of unity.

The contribution 
$S^{\alpha}_{\rm pl} ( r^{\alpha}_{\|} \hat{\vec{v}}^{\alpha} , 0 )$
of the plasmon  mode to the spatially varying part 
of the Debye-Waller factor at equal times 
can be calculated analogously.
Note that
$ r^{\alpha}_{\|} = \hat{\vec{v}}^{\alpha} \cdot {\vec{r}}
= \pm r_{x} \cos \theta  \pm r_{y} \sin \theta$.
Repeating the steps leading to Eq.\ref{eq:R4patch2}, we obtain
 \begin{eqnarray}
 S^{\alpha}_{\rm pl}  ( r^{\alpha}_{\|} \hat{\vec{v}}^{\alpha} , 0 ) 
 & = & - \int_{0}^{ \infty } \D q_{x}
 \cos ( q_{x} r^{\alpha}_{\|} )
 \nonumber
 \\
 & & \times
 \left< 
 \frac{ \cos ( \theta q_{y} r^{\alpha}_{\|} )}{ \sqrt{ q_{x}^2 + \theta^2 q_{y}^2 }} 
  \frac{F_{\vec{q}}^2 }
 { 2 \sqrt{ 1 + F_{\vec{q}} } 
 \left[ \sqrt{ 1 + F_{\vec{q}} } + 1
  \right]^2 }
 \right>_{\rm BZ}
 \; \; \; .
 \nonumber
 \\
 & &
 \label{eq:S4patch2}
 \end{eqnarray}
Because the Thomas-Fermi wave-vector $\kappa$ acts as an effective
ultraviolet cutoff, 
the value of the integral \ref{eq:S4patch2} is determined
by the regime $|{\vec{q}} | \leqapprox \kappa$. 
For $\theta \kappa | r^{\alpha}_{\|} | \ll 1$ we may approximate
in this regime
 $ \cos (\theta q_{y} r^{\alpha}_{\|} ) \approx 1$ under the integral sign. 
Furthermore, for $\kappa | r^{\alpha}_{\|}| \gg 1$ the oscillating 
factor $\cos ( q_{x} r^{\alpha}_{\|} )$ effectively replaces $\kappa$ by
$|r^{\alpha}_{\|}|^{-1}$ as relevant ultraviolet cutoff. 
We conclude that  in the  parametrically large  intermediate regime 
 \begin{equation}
 \kappa^{-1} \ll | r^{\alpha}_{\|} | \ll  ( \theta \kappa )^{-1}
 \label{eq:regimeinterm}
 \end{equation}
we have 
to leading logarithmic order 
 \begin{equation}
 S^{\alpha}_{\rm pl} ( r^{\alpha}_{\|} \hat{\vec{v}}^{\alpha} , 0 ) \sim
 - \gamma_{\rm cb} \left[ \ln \left( \frac{1}{\theta 
 \kappa | r^{\alpha}_{\|} | } \right) + b_{2} \right]
 \; \; \; ,
 \label{eq:Sres4patch}
 \end{equation}
where $b_{2}$ is another numerical constant.

\begin{center}
{\bf{The nesting mode\index{nesting}}}
\end{center}

\begin{figure}
\sidecaption
\psfig{figure=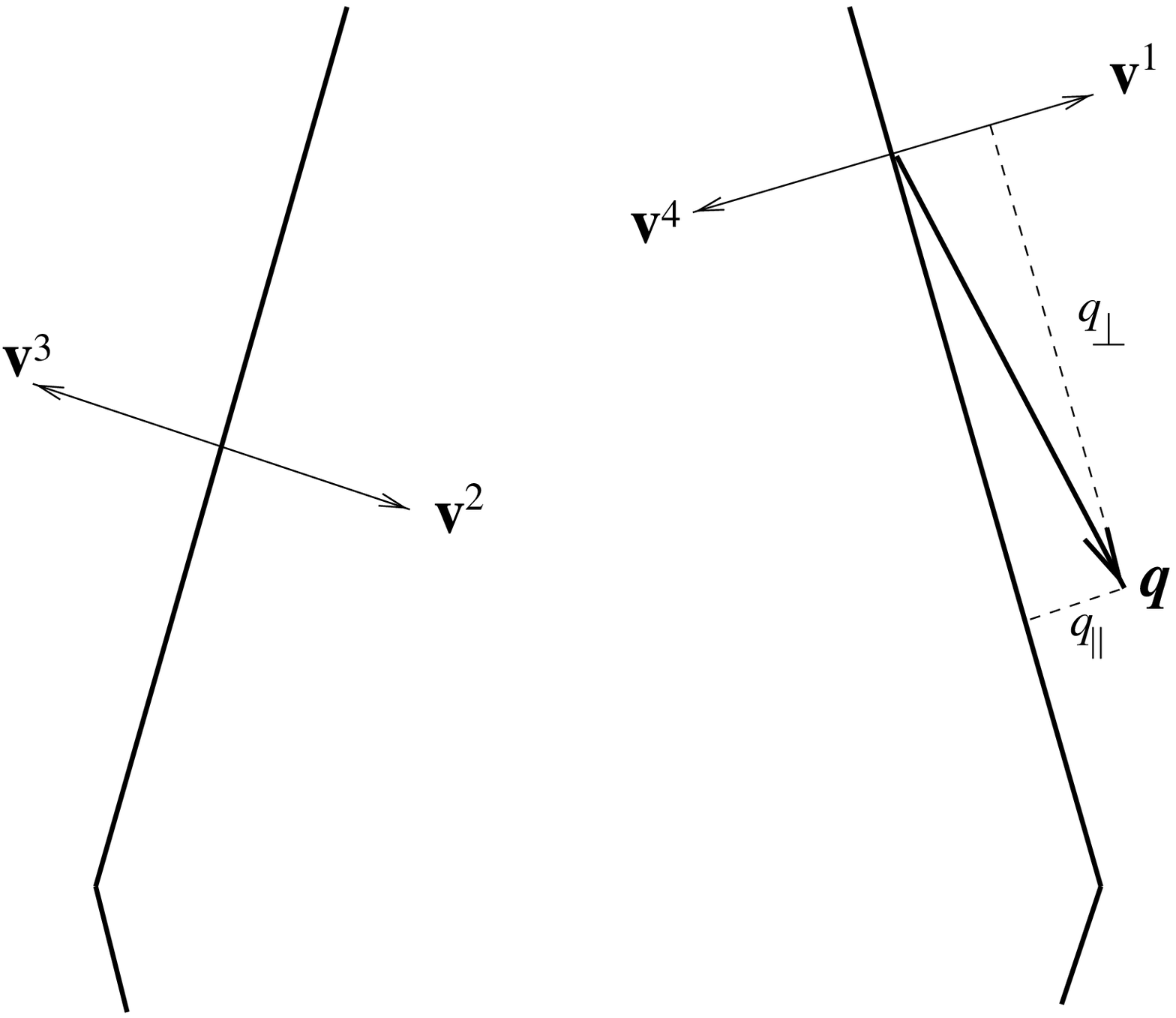,width=6cm}
\caption[Wave-vector that contributes to the nesting instability.]
{
Wave-vector ${\vec{q}}$ that
contributes to the nesting instability
in the $4$-patch model.
The direction of ${\vec{q}}$ is almost perpendicular to
${\vec{v}}^1$ and ${\vec{v}}^4$, so that
$|q_{\|}| \leqapprox \theta |q_{\bot}|$ and hence
$|\xi_{\vec{q}} | \leqapprox | \tilde{\xi}_{\vec{q}} |$.
}
\label{fig:patchproject}
\end{figure}

\noindent
Let us now focus on the contribution from the second
term in Eq.\ref{eq:R4patch1}, which involves the collective
mode 
$\tilde{\omega}_{\vec{q}}$. 
With the help of Eq.\ref{eq:positivecheck} the 
dispersion of this mode can also be written as
 \begin{equation}
   \tilde{\omega}_{\vec{q}}^2   = 
  \left( 1 + \frac{F_{\vec{q}}}{2} \right) 
 \frac{ \xi_{\vec{q}}^2 + \tilde{\xi}_{\vec{q}}^2 }{2}
 \left[
 1 - \left(
 1 -  G_{\vec{q}}
  \right)^{1/2} \right]
 \;  ,
 \label{eq:baromega4patch2}
 \end{equation}
 \begin{equation}
 G_{\vec{q}} \equiv
 \frac{ ( 1 + F_{\vec{q}} ) }{
 (1 + \frac{  F_{\vec{q}}}{2})^2}
\frac{ 4  \xi_{\vec{q}}^2 \tilde{\xi}_{\vec{q}}^2
 }{ 
 \left(  \xi_{\vec{q}}^2 + \tilde{\xi}_{\vec{q}}^2  \right)^2 }
 \; \; \; .
 \label{eq:Gqdef}
 \end{equation}
For $| G_{\vec{q}} | \ll 1$
this implies to leading order
 \begin{equation}
   \tilde{\omega}_{\vec{q}}  \approx
 \sqrt{\frac{  1 + F_{\vec{q}}  }{ 1 +  \frac{F_{\vec{q}}}{2} }}
 \frac{
 | \xi_{\vec{q}}| | \tilde{\xi}_{\vec{q}} |
 }{  \sqrt{ \xi_{\vec{q}}^2 + \tilde{\xi}_{\vec{q}}^2}}  
 \label{eq:baromegaregimecrit}
 \; \; \; .
 \end{equation}
From this expression it is obvious that $\tilde{\omega}_{\vec{q}}$ vanishes 
on the planes defined by
$  \xi_{\vec{q}} = 0 $ or 
$  \tilde{\xi}_{\vec{q}} = 0$.
These equations define
precisely the set of points on the Fermi surface.
The vanishing of the collective mode
$\tilde{\omega}_{\vec{q}}$ on the Fermi surface
is due to the fact that by construction 
the curvature of the patches is exactly zero,
so that the Fermi surface has a nesting symmetry\index{nesting}:
patches $1$ and $4$ (or $2$ and $3$) in 
Fig.~\secref{fig:4patch} can be connected by 
vectors in the directions of ${\vec{v}}^{1}$
(or ${\vec{v}}^{2}$) that can be attached to an arbitrary point on the patches.
For realistic Fermi surfaces of the type shown in Fig.~\secref{fig:quasi1d} 
this nesting symmetry is absent, so that the associated zero modes do not exist. 
The vanishing of the mode $\tilde{\omega}_{\vec{q}}$ gives
rise to an unphysical singularity in Eq.\ref{eq:R4patch1}. 
To see this more clearly, it is necessary to
calculate the residue $\tilde{Z}_{\vec{q}}$
in the regime $| G_{\vec{q}} | \ll 1$.
Expanding the square root in Eq.\ref{eq:baromega4patch2} to second order in $G_{\vec{q}}$,
we obtain
 \begin{equation}
   \tilde{\omega}_{\vec{q}}^2  =
 \frac{  1 + F_{\vec{q}}  }{ 1 +  \frac{F_{\vec{q}}}{2} }
 \frac{
  \xi_{\vec{q}}^2  {\tilde{\xi}_{\vec{q}}}^2
 }{  \xi_{\vec{q}}^2 + \tilde{\xi}_{\vec{q}}^2}
 \left[ 1 + \frac{G_{\vec{q}} }{4} + O ( G_{\vec{q}}^2 ) \right]
 \; \; \; .
 \label{eq:Cqexpansion}
 \end{equation}
The numerator in the expression for
the associated residue $\tilde{Z}_{\vec{q}}$ (see Eq.\ref{eq:Z2patch}) 
can then be written as
\begin{equation}
 \xi_{\vec{q}}^2
 \tilde{\xi}_{\vec{q}}^2 
 - \tilde{\omega}_{\vec{q}}^2 \frac{ \xi_{\vec{q}}^2 + \tilde{\xi}_{\vec{q}}^2 }{2}  
 \approx
 \frac{
  \xi_{\vec{q}}^2  {\tilde{\xi}_{\vec{q}}}^2
 }{  2 + F_{\vec{q}} }
 \left[ 1 - ( 1 + F_{\vec{q}} ) \frac{G_{\vec{q}}}{4} \right]
 \label{eq:numeratorZ2}
 \; \; \; .
 \end{equation} 
For simplicity let us first consider the
regime $F_{\vec{q}} \gg 1$.
From the definition \ref{eq:Gqdef}
it is clear that in this case
the condition $| G_{\vec{q}} | \ll 1$
is valid for all values of $\xi_{\vec{q}}$ and $\tilde{\xi}_{\vec{q}}$.
Because the terms of order
$G_{\vec{q}}^2$ that have been ignored in Eq.\ref{eq:Cqexpansion} 
are proportional to $F_{\vec{q}}^{-2}$,
it is consistent to expand the right-hand side of
Eq.\ref{eq:numeratorZ2} to first order in $F_{\vec{q}}^{-1}$, in
which case we obtain
\begin{equation}
 \xi_{\vec{q}}^2
 \tilde{\xi}_{\vec{q}}^2 
 - \tilde{\omega}_{\vec{q}}^2 \frac{ \xi_{\vec{q}}^2 + \tilde{\xi}_{\vec{q}}^2 }{2}  
 \approx
  \frac{\xi_{\vec{q}}^2  {\tilde{\xi}_{\vec{q}}}^2}{
  F_{\vec{q}}} 
  \left[
 \frac{ ( \xi_{\vec{q}}^2 - \tilde{\xi}_{\vec{q}}^2 )^2}{
 ( \xi_{\vec{q}}^2 + \tilde{\xi}_{\vec{q}}^2 )^2}
 + O ( F_{\vec{q}}^{-1} ) \right]
 \label{eq:numeratorZ22}
 \; \; \; .
 \end{equation}
Substituting this expression into Eq.\ref{eq:Z2patch}
and using Eqs.\ref{eq:baromegaregimecrit} and \ref{eq:om4SC},
we obtain for $F_{\vec{q}} \gg 1$
 \begin{equation}
 \tilde{Z}_{\vec{q}}
 \approx \frac{\nu}{\sqrt{2} F_{\vec{q}}^2}
 \frac{ | \xi_{\vec{q}} | |\tilde{ \xi}_{\vec{q}} |}{ \sqrt{
 \xi_{\vec{q}}^2 + \tilde{ \xi}_{\vec{q}}^2 } }
 \frac{ ( \xi_{\vec{q}}^2 - \tilde{\xi}_{\vec{q}}^2 )^2}{
 ( \xi_{\vec{q}}^2 + \tilde{\xi}_{\vec{q}}^2 )^2}
 \label{eq:Zbarapproxfinal}
 \; \; \; .
 \end{equation}
Note that this expression correctly vanishes if we set $\theta = 0$.
We conclude that for $F_{\vec{q}} \gg 1$
the second term in Eq.\ref{eq:R4patch1} involves the integrand
 \begin{equation}
 \hspace{-6mm}
  \frac{\tilde{Z}_{\vec{q}}}{ ( \tilde{\omega}_{\vec{q}} + 
 | {\vec{v}}^{\alpha} \cdot {\vec{q}} |)^2}
 \approx
 \frac{\nu
 | \tilde{\xi}_{\vec{q}} | 
 }{\sqrt{2} F_{\vec{q}}^2 | \xi_{\vec{q}} | 
  \sqrt{\xi_{\vec{q}}^2 + \tilde{ \xi}_{\vec{q}}^2 } }
 \frac{ ( \xi_{\vec{q}}^2 - \tilde{\xi}_{\vec{q}}^2 )^2}{
 ( \xi_{\vec{q}}^2 + \tilde{\xi}_{\vec{q}}^2 )^2}
 \left[ 
 \frac{ \sqrt{2} | {\tilde{\xi}}_{\vec{q}} | }{ 
 \sqrt{\xi_{\vec{q}}^2 + \tilde{ \xi}_{\vec{q}}^2 }}
 + 1 \right]^{-2} 
 \; ,
 \label{eq:Sbarcontrib}
 \end{equation}
where,
without loss of generality, we have set 
${\vec{v}}^{\alpha} \cdot {\vec{q}} = \xi_{\vec{q}}$.
To discuss the singularities of this integrand, it is
convenient to
choose the integration variables 
$q_{\|} = {\hat{\vec{v}}}^{1} \cdot {\vec{q}} = q_x \cos \theta + q_y \sin \theta $ and
$q_{\bot} = - q_x \sin \theta + q_y \cos \theta$. 
Then $\xi_{\vec{q}} = v_{\rm F} q_{\|}$ and 
$\tilde{\xi}_{\vec{q}} = v_{\rm F} ( q_{\|} - 2 \theta q_{\bot} )$ to leading order
in $\theta$.
Hence,
 \begin{equation}
 \frac{ ( \xi_{\vec{q}}^2 - \tilde{\xi}_{\vec{q}}^2 )^2}{
 ( \xi_{\vec{q}}^2 + \tilde{\xi}_{\vec{q}}^2 )^2}
 \sim
 \left\{
 \begin{array}{cc}
 \theta^2 q_{\bot}^2 / q_{\|}^2
 & \mbox{for $ | q_{\|} | \geqapprox \theta | q_{\bot} | $} \\
 1
 & \mbox{for $ | q_{\|} | \leqapprox \theta | q_{\bot} | $} 
 \end{array}
 \right.
 \; \; \; .
 \label{eq:casesqq}
 \end{equation}
Note that the condition
$\mbox{$ | q_{\|} | \leqapprox \theta | q_{\bot} | $}$
is equivalent with $| \xi_{\vec{q}} | \leqapprox | \tilde{\xi}_{\vec{q}} |$.
Geometrically this means that the wave-vector ${\vec{q}}$ is almost
parallel to the surface of the first and fourth patch, so that
its projection $ q_{\|} $
onto the local normals $\hat{\vec{v}}^1$ and $\hat{\vec{v}}^4$
is much smaller than the projection 
onto the normals $\hat{\vec{v}}^{2}$ 
and $\hat{\vec{v}}^{3}$ 
of the other two patches, see Fig.~\secref{fig:patchproject}.
The contribution from the
regime $| q_{\|}| \geqapprox \theta | q_{\bot} |$ 
to Eq.\ref{eq:R4patch1} is finite and 
small in the strong coupling limit of interest here. 
On the other hand, in the
regime $| q_{\|}| \leqapprox \theta | q_{\bot} |$ 
we have
 \begin{equation}
  \frac{\tilde{Z}_{\vec{q}}}{ ( \tilde{\omega}_{\vec{q}} + 
 | {\vec{v}}^{\alpha} \cdot {\vec{q}} |)^2}
 \approx
 \frac{ \nu }{ 
 v_{\rm F} | q_{\|} | } \frac{1}{
 \sqrt{2}( \sqrt{2} + 1 )^2 F_{\vec{q}}^2  }
  \; \; \; ,
  \; \; \; \mbox{for $F_{\vec{q}} \gg 1 $}
  \; \; \; .
  \label{eq:nestsing}
  \end{equation}
Substituting this expression into Eq.\ref{eq:R4patch1}, we see that
the contribution of the nesting mode to the
constant part $R^{\alpha}$ 
of our Debye-Waller factor
leads to the logarithmically divergent integral
$   \int_{0}^{ \theta |q_{\bot}| } \frac{\D q_{\|}}{  q_{\|}  }$.
Of course, in the
combination $R^{\alpha} - S^{\alpha} ( r^{\alpha}_{\|}
\hat{\vec{v}}^{\alpha} , \tau )$
this divergence is removed, 
and we obtain a Debye-Waller factor that grows 
logarithmically at large distances.  This is precisely the
Luttinger liquid behavior
discussed in Chap.~\secref{sec:Green1},
so that our $4$-patch model is a higher-dimensional example
for a Luttinger liquid.
However, the logarithmic growth of the static Debye-Waller factor at large
distances is {\it{not}} due to the collective mode
$\omega_{\vec{q}}$ which in the limit $\theta \rightarrow 0$
can be identified with the plasmon without interchain hopping;
instead, in our $4$-patch model the Luttinger liquid behavior
is generated by the new nesting mode $\tilde{\omega}_{\vec{q}}$,
which disappears at $\theta = 0$.

Clearly, the non-Fermi liquid behavior
of our $4$-patch  model is due to the
artificial nesting symmetry of the Fermi surface,
which manifests itself for
$| q_{\|} | \leqapprox \theta | q_{\bot} |$.
In this regime 
the dimensionless parameter
$G_{\vec{q}} $ in Eq.\ref{eq:baromega4patch2}
is small compared with unity 
for all $F_{\vec{q}}$,  so that it is easy to 
repeat the above calculations for arbitrary $F_{\vec{q}}$. 
We obtain from Eq.\ref{eq:baromegaregimecrit} to leading order
 \begin{equation}
   \tilde{\omega}_{\vec{q}}  \approx
 \sqrt{\frac{  1 + F_{\vec{q}}  }{ 1 +  \frac{F_{\vec{q}}}{2} }}
 v_{\rm F} | q_{\|} |
 \; \; \; , \; \; \; | q_{\|} | \leqapprox \theta | q_{\bot} |
 \; \; \; ,
 \label{eq:omegabarcrit}
 \end{equation}
and from Eqs.\ref{eq:Z2patch} and \ref{eq:numeratorZ2}
 \begin{equation}
 \tilde{Z}_{\vec{q}} \approx
 \frac{ \nu v_{\rm F} | q_{\|} |}{ 4   
 [ 1 + \frac{F_{\vec{q}}}{2} ]^{\frac{3}{2} }
 [ 1 + F_{\vec{q}} ]^{\frac{1}{2}} }
 \; \; \; , \; \; \; | q_{\|} | \leqapprox \theta | q_{\bot} |
 \label{eq:Z2patchcrit}
 \; \; \; .
 \end{equation}
In the limit $F_{\vec{q}} \gg 1$
this expression agrees with 
Eq.\ref{eq:Zbarapproxfinal} 
if we restrict ourselves to the regime $| \xi_{\vec{q}} | \leqapprox | \tilde{\xi}_{\vec{q}} |$.
We conclude that
for  $| q_{\|} | \leqapprox \theta | q_{\bot} |$
 \begin{equation}
  \frac{\tilde{Z}_{\vec{q}}}{ ( \tilde{\omega}_{\vec{q}} + 
 | {\vec{v}}^{\alpha} \cdot {\vec{q}} |)^2}
 \approx
 \frac{ \nu }{
 4 v_{\rm F} | q_{\|} | 
 [ 1 + \frac{F_{\vec q}}{2} ]^{\frac{3}{2} }
 [ 1 + F_{\vec{q}} ]^{\frac{1}{2}} 
 \left[ \sqrt{ \frac{ 1 + F_{\vec{q}} }{ 1 + \frac{ F_{\vec{q}}}{2} } }  + 1 \right]^2}
 \; \; \; . 
  \label{eq:nestsing2}
  \end{equation}
It is now obvious that the nesting singularity  exists for arbitrary coupling strength.
However, this singularity is an unphysical feature
of our $4$-patch model, and does not exist for
realistic Fermi surfaces shown in Fig.~\secref{fig:quasi1d}.
We shall now refine our model
by giving the patches a finite curvature.
We then use our bosonization results 
for non-linear energy dispersion 
derived in Chap.~\secref{sec:eik}
to show that the contribution from the nesting mode becomes negligible
compared with the contribution from the plasmon mode
$\omega_{\vec{q}}$.

\subsection{How curvature kills the nesting singularity\index{nesting}}
\label{subsec:thefate}

{\it{We consider a generalized $4$-patch model with curved patches,
and first give a simple intuitive argument
why the quadratic term in the energy dispersion 
removes the nesting singularity.
We then use our result \ref{eq:R2new} for the
Debye-Waller factor with non-linear energy dispersion
to confirm this argument by explicit calculation.}}

\begin{figure}
\sidecaption
\psfig{figure=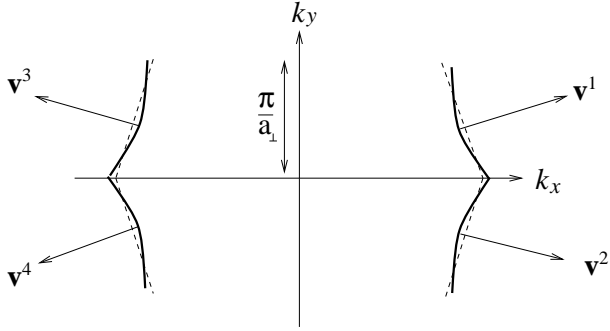,width=8cm}
\caption[Fermi surface of the $4$-patch model with curved patches]
{
Fermi surface\index{four-patch model!Fermi surface} of the $4$-patch model
with curved patches. 
If the component of ${\vec{q}}$ perpendicular to ${\vec{v}}^{\alpha}$
is denoted by $q_{\bot}$,
the patches can be described by energy dispersions
of the form \ref{eq:curveng}
with negative effective mass $m_{\bot}$.
}
\label{fig:4patchcurved}
\end{figure}

\vspace{7mm}
\noindent
It is physically clear that any finite curvature of the patches
will destroy the nesting symmetry and hence remove the logarithmic divergence
in the Debye-Waller factor.
Let us therefore replace the
completely flat patches of Fig.~\secref{fig:4patch} by
the slightly curved patches
shown in Fig.~\secref{fig:4patchcurved}.
The corresponding energy dispersions are
 \begin{equation}
 \xi^{\alpha}_{\vec{q}} = {\vec{v}}^{\alpha} \cdot {\vec{q}} 
 + \frac{  q_{\bot}^2 }{2  m_{\bot}  }
 \; \; \; , \; \; \;
 q_{\bot} =
  {\hat{\vec{v}}}^{\alpha}_{\bot} \cdot {\vec{q}}
 \; \; \; ,
 \label{eq:curveng}
 \end{equation}
where 
${\hat{\vec{v}}}^{\alpha}_{\bot}$ is a unit vector perpendicular to
${\vec{v}}^{\alpha}$, and the effective mass $m_{\bot}$ is negative. 
Note that terms of the form
${q_{\|}^2}/({2 m_{\|} })$ do not describe the
curvature of the patches and can be ignored for our
purpose (recall the discussion 
after Eq.\ref{eq:xisphere} in Chap.~\secref{sec:eik}).
Let us first estimate the effect of curvature in a simple
qualitative way, which leads to exactly the same result
as the explicit evaluation of the bosonization
expression for curved Fermi surfaces.
Obviously, the curvature term in the energy dispersion becomes important
for $v_{\rm F} | q_{\|}| \leqapprox  {q_{\bot}^2}/(2 | m_{\bot}   |)$.
Therefore we expect that for curved patches 
the lower limit for the $q_{\|}$-integral will
effectively be replaced by
${q_{\bot}^2}/({2 | m_{\bot} | v_{\rm F}  })$.
We conclude that the effect of curvature 
can be qualitatively taken into account by substituting
 \begin{equation}
\int_{0}^{ \theta |q_{\bot}| } \frac{\D q_{\|}}{  q_{\|}  }
\rightarrow
\int_{
\frac{q_{\bot}^2}{2 | m_{\bot} | v_{\rm F}  }
}^{ \theta |q_{\bot}| } \frac{d q_{\|}}{  q_{\|}  }
= \ln \left( \frac{ 2 | m_{\bot} | v_{\rm F} \theta}{ | q_{\bot} | } \right)
\label{eq:curvaturesub}
\; \; \; .
\end{equation}
In physically relevant cases we expect 
$| m_{\bot} | \approx m_{\|} / \theta = k_{\rm F} / ( v_{\rm F} \theta )$, 
so that the right-hand side of Eq.\ref{eq:curvaturesub} reduces to the
integrable logarithmic factor $\ln ( 2 k_{\rm F} / | q_{\bot}| )$.
Note that the above argument is only consistent if
${q_{\bot}^2}/({2 | m_{\bot}| v_{\rm F}  }) \ll \theta | {{q}}_{\bot} |$,
even for the largest relevant $q_{\bot}$.
Keeping in mind that the effective ultraviolet-cutoff
for the $q_{\bot}$-integral is the
Thomas-Fermi wave-vector $\kappa$ 
(see Eq.\ref{eq:kappaabot}), this condition reduces to $\kappa \ll k_{\rm F}$.
Combining Eqs.\ref{eq:nestsing} and \ref{eq:curvaturesub}, it is easy to see
that the (regularized) nesting mode
simply renormalizes the numerical constant $b_1$ in Eq.\ref{eq:Rres4patch}. 
We therefore conclude that the leading small-$\theta$ behavior of $R^{\alpha}$
is entirely due to the plasmon mode $\omega_{\vec{q}}$.

We now confirm the above argument with the help of the 
bosonization result for the Green's function
for non-linear energy dispersion
derived in Chap.~\secref{sec:eik}.
With finite curvature we should replace the expression \ref{eq:R4patch1} for 
the constant part of the Debye-Waller factor of the $4$-patch model
by $R^{\alpha}_1$ given in Eq.\ref{eq:R2new}. 
Of course, we should now use
the dynamic structure factor
$S_{\rm RPA} ( {\vec{q}} , \omega )$ 
corresponding to the Fermi surface
shown in Fig.~\secref{fig:4patchcurved}. 
Due to the curvature of patches, 
$S_{\rm RPA} ( {\vec{q}} , \omega )$ 
is now more complicated than in Eq.\ref{eq:SRPA4patch}.  
Apart from a $\delta$-function peak representing
the physical plasmon mode, we expect that, due to Landau damping, the 
peak associated with the nesting mode $\tilde{\omega}_{\vec{q}}$
is now spread out into a continuum in a finite frequency interval. 
However, in order to estimate the fate of the nesting mode 
in the presence of curvature, it is sufficient to
substitute the dynamic structure factor 
\ref{eq:SRPA4patch} for flat patches into Eq.\ref{eq:R2new}. 
Certainly, if in this approximation the 
nesting singularity is removed, 
then more accurate
approximations for $S_{\rm RPA} ( {\vec{q}} , \omega )$ 
will lead to the same result, because
the curvature terms smooth out the sharpness of the nesting mode.
Combining then Eqs.\ref{eq:SRPA4patch} and \ref{eq:R2new},
we see that the contribution of the nesting mode to
$R_1^{\alpha}$ is given by
 \begin{equation}
 R^{\alpha}_{1, {\rm nest}} \approx 
   -  
 \frac{1}{V} \sum_{\vec{q}} 
 f_{\vec{q}}^2
 \tilde{Z}_{\vec{q}}
 \frac{ 2 {\rm sgn} ( \xi^{\alpha}_{\vec{q}}  ) }
 {  \frac{ q_{\bot}^2}{ |m_{\bot}|  }  
 ( \tilde{\omega}_{\vec{q}} + | \xi^{\alpha}_{\vec{q}} |) }
 \label{eq:R2nest}
 \; \; \; ,
 \end{equation}
with $\tilde{Z}_{\vec{q}}$ and $\tilde{\omega}_{\vec{q}}$ given in
Eqs.\ref{eq:Z2patch} and \ref{eq:baromega4patch2}.
From Sect.~\secref{subsec:4patchmodel} we know that
possible nesting singularities 
are due to the regime $| q_{\|} | \leqapprox \theta | q_{\bot} |$.
Thus, restricting the integral in Eq.\ref{eq:R2nest} to this
regime, we have  
from Eqs.\ref{eq:Zbarapproxfinal} and
\ref{eq:baromegaregimecrit}
in the strong coupling limit
$ f_{\vec{q}}^2
 \tilde{Z}_{\vec{q}}  \approx  v_{\rm F} | q_{\|} | / ( {\sqrt{2} \nu})$
and
 $\tilde{\omega}_{\vec{q}}  \approx  \sqrt{2} v_{\rm F} | q_{\|} |$.
Recall that for the three-dimensional Coulomb interaction
the strong-coupling condition $\nu f_{\vec{q}} \gg 1$
is equivalent with $| {\vec{q}} | \ll \kappa$, 
where the Thomas-Fermi wave-vector $\kappa$ is given in 
Eq.\ref{eq:kappaabot}.
It is useful to introduce again the integration variables
$q_{\|} = \hat{ {\vec{v}}}^{\alpha} \cdot {\vec{q}}$ and
$q_{\bot} = \hat{ {\vec{v}}}^{\alpha}_{\bot} \cdot {\vec{q}}$.
Putting everything together, we 
find that the contribution from the
critical regime $| q_{\|} | \leqapprox \theta | q_{\bot} |$ to
Eq.\ref{eq:R2new} can be written as
 \begin{equation}
 \hspace{-4mm}
 R^{\alpha}_{1, {\rm nest}} \approx 
 - \frac{\sqrt{2} \kappa }{\pi^3 \nu} \int_{0}^{\kappa}
 \D q_{\bot} 
 \frac{ | m_{\bot} | }{ q_{\bot}^2 }
 \int_{- \theta | q_{\bot} |}^{\theta | q_{\bot} | } \D q_{\|}
 \frac{ 
  | q_{\|} | {\rm sgn} ( q_{\|} - \frac{ q_{\bot}^2}{2 
 | m_{\bot}| v_{\rm F} } ) }{  \sqrt{2} | q_{\|} | + \left| q_{\|} -
 \frac{ q_{\bot}^2}{ 2 | m_{\bot} | v_{\rm F}  } \right|  }
 \label{eq:R2nestcrit}
 \; .
 \end{equation}
The $q_{\|}$-integration can now be performed analytically.
The integral is proportional to
${ q_{\bot}^2}/ { | m_{\bot}| }$, which cancels 
the singular factor of $ | m_{\bot} | / { q_{\bot}^2 }$ 
in Eq.\ref{eq:R2nestcrit}.
We obtain
\begin{equation}
 R^{\alpha}_{1, {\rm nest}} \approx 
 - \frac{\sqrt{2} \kappa }{ 2 ( \sqrt{2} + 1 )^2 \pi^3 \nu v_{\rm F}} 
 \int_{0}^{\kappa} \D q_{\bot} \left[ \ln \left( \frac{ 2 |m_{\bot}| v_{\rm F }
 \theta}{ q_{\bot} } \right) + b_3 \right]
 \; \; \; ,
 \label{eq:R2nestcrit2}
 \end{equation}
where $b_3$ is a numerical constant of the order of unity.
This is the same type of 
integral as in Eq.\ref{eq:curvaturesub}, so that 
our simple intuitive arguments given above
are now put on a more solid basis.
As already mentioned, in physically relevant cases we expect 
$| m_{\bot} | v_{\rm F} \theta \approx k_{\rm F}$. Using then 
Eqs.\ref{eq:nuglobalpatch} and \ref{eq:kappaabot}, 
we finally obtain
\begin{equation}
 R^{\alpha}_{1, {\rm nest}} \approx 
 - \gamma_{\rm cb} b_4
 \label{eq:R2nestfinal}
 \; \; \; ,
 \end{equation}
where $\gamma_{\rm cb} = e^2 / (2 \pi v_{\rm F})$ (see Eq.\ref{eq:gammacbres}), and
$b_4$ is another numerical constant of the order of unity.
Thus, for patches with finite curvature the contribution
of the nesting mode is finite. 
It is also easy to see that the curvature terms do
{\it{not}} modify the logarithmic small-$\theta$ behavior
of $R^{\alpha}$ given in Eq.\ref{eq:Rres4patch}. 
This is so because
the leading $\ln ( 1/ {\theta} )$-term in Eq.\ref{eq:Rres4patch}
is generated by the energy scale $v_{\rm F} \theta | q_{\bot} |$, which is
by assumption larger than the curvature energy ${ q_{\bot}^2}/({ 2 | m_{\bot} | })$.

\subsection{Anomalous scaling in a Fermi liquid}
\label{subsec:anomalous}

{\it{Now comes the important conclusion about 
the physical system of interest (to be
distinguished from the $4$-patch model discussed in 
Sect.~\secref{subsec:4patchmodel}).}}

\vspace{7mm}

\noindent
Comparing Eq.\ref{eq:R2nestfinal}
with the corresponding contribution 
\ref{eq:Rres4patch} 
to the Debye-Waller factor that is due to
the plasmon mode $\omega_{\vec{q}}$, we conclude that for small $\theta$
the Debye-Waller factor is dominated by the
plasmon mode. In particular, for realistic Fermi surfaces
of the form shown in Fig.~\secref{fig:quasi1d} the
constant part $R^{\alpha}$ of the Debye-Waller factor is finite.
To leading logarithmic order for small $\theta$  
we may therefore
approximate $R^{\alpha} \approx R^{\alpha}_{\rm pl}$, 
where $R^{\alpha}_{\rm pl}$ is given in Eq.\ref{eq:Rres4patch}.
We conclude that
for any non-zero $\theta$
the system is a Fermi liquid, 
with quasi-particle residue\index{four-patch model!quasi-particle residue}
\index{quasi-particle residue!four-patch model}
 \begin{equation}
 Z^{\alpha} = \E^{R^{\alpha}} \propto \theta^{\gamma_{\rm cb}}
 \label{eq:QPres4patch}
 \; \; \; ,
 \end{equation}
where $\gamma_{\rm cb}$ is given in
Eq.\ref{eq:gammacbres}.
Thus, for $\theta \rightarrow 0$ the quasi-particle residue
vanishes with a non-universal power of $\theta$, which 
{\it{can be identified with the anomalous dimension of the corresponding
Luttinger liquid that would exist for $\theta = 0$ at the same value of the
dimensionless coupling constant $  e^2 /  v_{\rm F}$.}}
Combining Eqs.\ref{eq:Rres4patch} and
\ref{eq:Sres4patch}, we obtain for the
total static Debye-Waller factor to leading logarithmic order in $\theta$
 \begin{eqnarray}
 Q^{\alpha} ( r^{\alpha}_{\|} \hat{\vec{v}}^{\alpha} , 0 ) & = &
 R^{\alpha} -
 S^{\alpha} ( r^{\alpha}_{\|} \hat{\vec{v}}^{\alpha} , 0 )
 \nonumber
 \\
 &  & \hspace{-20mm}  = - \gamma_{\rm cb} 
 \left[ \ln ( \kappa | r^{\alpha}_{\|} | )  + O (1) \right]
 \;  \; ,
 \; \; \mbox{ $ \kappa^{-1}  \ll | r^{\alpha}_{\|} | \ll  ( \theta \kappa )^{-1}$ }
 \; \; .
 \label{eq:Qres4patch}
 \end{eqnarray}
Exponentiating this expression, we see that the
interacting Green's function satisfies the anomalous scaling 
relation,\index{anomalous scaling!four-patch model}
 \begin{equation}
 \hspace{-4mm}
 G^{\alpha} ( {\vec{r}} / s  ,  0 ) = s^{3 + \gamma_{\rm cb}}
 G^{\alpha} ( {\vec{r}}  ,  0 )
 \; \; \; ,
 \; \; \; \mbox{ $\kappa^{-1}  \ll | r^{\alpha}_{\|} | \; , \; 
 |r^{\alpha}_{\|} | / s   \ll  (\theta \kappa )^{-1}$ }
 \; .
 \end{equation}
Thus, in spite of the fact that the system is a Fermi liquid, there exists for small
$\theta$ a parametrically large intermediate regime where the interacting Green's function
satisfies the  anomalous scaling law typical for Luttinger liquids,
as discussed in Chap.~\secref{sec:Green1}. Moreover, 
the effective anomalous exponent 
{\it{is precisely given by the
anomalous dimension of the Luttinger liquid that would exist for $\theta = 0$}}.
This is a very important result, 
because in realistic experimental systems the interchain hopping $t_{\bot}$ 
is never exactly zero.
We thus arrive at the important conclusion that for small $\theta$ the
anomalous dimension of the Luttinger liquid is in principle measurable,
although strictly speaking the system is a Fermi liquid.
The relevance of $t_{\bot}$ in an infinite array of  
weakly coupled chains has also been discussed in
\cite{Bourbonnais91,Castellani92} by means of a perturbative expansion
to lowest order in $t_{\bot}$. 
In contrast, our approach
is non-perturbative in $t_{\bot}$.

\subsection{The nesting singularity for general Fermi surfaces\index{nesting}}
\label{sec:nesting}

{\it{We show that quite generally 
the nesting symmetries
introduced  via the patching construction
give rise to logarithmic singularities and hence
to unphysical Luttinger liquid behavior.
}}

\vspace{7mm}

\noindent
The nesting singularity discussed in 
Sect.~\secref{subsec:4patchmodel} is not a special feature
of our $4$-patch model. 
Singularities of this type will appear in any model
where the Fermi surface is covered by a finite number $M$ of flat patches,
such that at least some of the patches have a nesting symmetry.
The simplest analytically tractable case 
is perhaps a square Fermi surface $(M=4)$ in two dimensions,
which has first been discussed by Mattis \cite{Mattis87}, and 
more recently by
Hlubina \cite{Hlubina94} and by Luther \cite{Luther94}.
However, unless there exists a real physical
nesting symmetry in the problem, 
these nesting singularities are 
artificially generated by approximating  
a curved Fermi surface by a collection of completely flat patches.

There are several ways to cure this problem.
The simplest one is perhaps to choose the patches such that
nesting symmetries do not exist. For example, 
in the case of a circular Fermi surface in $d=2$
we avoid artificial nesting symmetries
by choosing an odd number of identical patches
(see Fig.~\secref{fig:patch3} for $M=5$).
The disadvantage of this construction is that
it explicitly breaks the inversion symmetry of the
Fermi surface, so that the negative frequency part of the
dynamic structure factor has to be treated 
separately\footnote{
As already mentioned in the first footnote of Chap.~\secref{chap:a7sing}, 
in this case the relation \ref{eq:Srealaxis} between
the imaginary part of the polarization and the dynamic structure factor 
(and all equations derived from Eq.\ref{eq:Srealaxis}) are not correct.
In particular, the expressions derived in Chap.~\secref{sec:exactman}
cannot be used.}.

The second possibility is to take the limit
$M \rightarrow \infty$ at some intermediate point in the
calculation, for example 
in the Debye-Waller factor given in Eqs.\ref{eq:Qlondef}--\ref{eq:Slondef}.
Because for finite $M$ the residue of the
nesting mode in the dynamic structure factor 
is proportional to $M^{-1}$, its contribution vanishes in the limit
$M \rightarrow \infty$. 
To see this, suppose that we approximate
a spherical Fermi surface with an even number $M$
of identical patches (see Fig.~\secref{fig:patch2dexample}
for $M=12$ in two dimensions).
The corresponding non-interacting
polarization $\Pi_{0} ( {\vec{q}} , z )$ is
given in Eq.\ref{eq:Pi0Meven}. 
From the discussion of the nesting mode 
in the $4$-patch model in Sect.~\secref{subsec:4patchmodel}
we expect that for some directions of ${\vec{q}}$
there will exist one particular patch ${P}^{\mu}_{\Lambda}$
such that
the energy $ | {\vec{v}}^{{\mu}} \cdot {\vec{q}} |$ is much
smaller than all the other energies
$ | {\vec{v}}^{\alpha} \cdot {\vec{q}} |$,
$\alpha \neq {\mu}$. 
Furthermore, we expect that for sufficiently small
$q_{\|}^{\mu} \equiv \hat{\vec{v}}^{{\mu}} \cdot {\vec{q}}$ 
the nesting mode $\omega_{\vec{q}}^{{\mu}}$ 
gives rise to a
$\delta$-function peak in the dynamic structure factor
with
$\omega_{\vec{q}}^{{\mu}} \propto |{\vec{v}}^{ {\mu}} 
\cdot {\vec{q}}|$. 
Because for sufficiently small $q_{\|}^{\mu}$ this energy is
much smaller than
$ | {\vec{v}}^{\alpha} \cdot {\vec{q}} |$
with $\alpha \neq {\mu}$, 
the energy dispersion of the nesting  mode can be
approximately calculated by setting $z^2=0$ in all terms 
with $\alpha \neq {\mu}$ in the expression \ref{eq:Pi0Meven}
for the non-interacting polarization $\Pi_0 ( \vec{q} , z )$
for finite patch number.
This yields for the polarization in the regime of wave-vectors ${\vec{q}}$ satisfying 
$ | {\vec{v}}^{{\mu}} \cdot {\vec{q}} | \ll
 | {\vec{v}}^{\alpha} \cdot {\vec{q}} |$ for  $\alpha \neq {\mu}$
 \begin{equation}
 \Pi_{0} ( {\vec{q}} , z ) \approx
 \frac{\nu}{M}
 \left[ M-2 + \frac{ 2 ( {\vec{v}}^{{\mu}} \cdot {\vec{q}} )^2}{
  ( {\vec{v}}^{ {\mu} } \cdot {\vec{q}} )^2 - z^2}
  \right]
  \label{eq:nestpolarization}
  \; \; \; .
  \end{equation}
The collective mode equation \ref{eq:plasmondef2} is
then easily solved, with the result that the dispersion of the
nesting mode is given by
 \begin{equation}
 \omega_{\vec{q}}^{ {\mu}} = 
 \sqrt{ \frac{1 + F_{\vec{q}} }{1 +
 \frac{M-2}{M} F_{\vec{q}} }} | {\vec{v}}^{{\mu}} \cdot {\vec{q}} |
 \label{eq:nestM}
 \; \; \; .
 \end{equation}
For the associated residue
we obtain with the help of Eq.\ref{eq:residuemode}
 \begin{eqnarray}
 Z_{\vec{q}}^{{\mu}} & \equiv & \frac{\nu^2}{F_{\vec{q}}^2
 \left. \frac{ \partial}{\partial z} \Pi_{0} ( {\vec{q}} , z )
 \right|_{ z = \omega_{\vec{q}}^{{\mu}} }}
  =  \frac{ \nu | {\vec{v}}^{{\mu}} \cdot {\vec{q}} |}{
 M \left[ 1 + \frac{M-2}{M} F_{\vec{q}} \right]^{\frac{3}{2}}
 \left[ 1 + F_{\vec{q}} \right]^{\frac{1}{2}} }
 \label{eq:nestres}
 \;  .
 \end{eqnarray}
If we set $M=4$ we recover the corresponding expressions 
\ref{eq:omegabarcrit} and \ref{eq:Z2patchcrit}
for the nesting mode in the $4$-patch model. 
From Eq.\ref{eq:R2} it is now obvious that in the thermodynamic
limit the constant part $R^{{\mu}}$ of the Debye-Waller factor
is proportional to $ \frac{1}{M} \int_{0} \frac{ \D q_{\|}}{q_{\|}}$. 
Clearly, the logarithmic divergence for finite $M$ is removed
if the take the limit $M \rightarrow \infty$.
It should be kept in mind, however, that
taking the limit $M \rightarrow \infty$ 
at intermediate stages of the calculation\index{patch!limit of infinite number} 
is not quite satisfactory,
because $M \rightarrow \infty$ implies 
that the patch cutoff $\Lambda$ vanishes.
As discussed in Chap.~\secref{subsec:Thetermslin},
in this case it is difficult to formally justify our derivation 
of the Green's function with linearized energy dispersion 
given in Chap.~\secref{sec:Derivation},
because the condition $q_{\rm c} \ll \Lambda$
is violated in this limit (see
Fig.~\secref{fig:qc}).
In Sect.~\secref{subsec:thefate} we have 
solved this problem  with the help of our
background field method developed in
Chap.~\secref{sec:eik}, which leads to a simple way for
including the effect of the curvature of the Fermi surface
into the bosonization procedure.

\section{Summary and outlook}

In this chapter we have used our non-perturbative
higher-dimensional bosonization approach 
to calculate the single-particle Green's function of
weakly coupled metallic chains.
This problem is not only of current experimental interest,
but its solution via higher-dimensional bosonization also nicely 
illustrates the 
approximations inherent in this approach.
In particular, we have shown that the replacement of 
a curved Fermi surface by a finite number of flat patches 
leads to unphysical logarithmic singularities and to Luttinger behavior in the
Green's function when at least two opposite patches are parallel.
In this respect we agree with 
the works by Mattis \cite{Mattis87} and by Hlubina\cite{Hlubina94},
who studied this problem in the special case of a square Fermi surface.
However, we have also shown that for more realistic curved Fermi surfaces
these logarithmic singularities disappear. Any finite
value of the interchain hopping $t_{\bot}$ leads then
to a bounded Debye-Waller factor, signalling
Fermi liquid behavior.

Very recently
the singularities generated by flat regions on opposite sides of a 
two-dimensional Fermi surface 
have been analyzed by Zheleznyak {\it{et al.}} \cite{Zheleznyak96}
with the help of the parquet approximation \cite{Diatlov57,Gorkov74}.
These authors obtained results which are, at least at the first sight, 
at variance with our finding (as well as with \cite{Mattis87,Hlubina94}).
Note, however, that
in our approach we have ignored the spin degree of freedom as well as
scattering processes involving large momentum transfers. 
In particular, we have not taken into account the instabilities towards
charge- or spin-density wave order, which
according to the authors of \cite{Zheleznyak96} 
become essential at sufficiently low 
temperatures. It is therefore not surprising that
we obtain a different result than Zheleznyak {\it{et al.}} \cite{Zheleznyak96}.
Our calculation
is restricted to a parameter regime where
the low energy physics is dominated by forward scattering.
The existence of such a regime is by no means obvious \cite{Gorkov74},
and we have assumed that for some
range of temperature, interchain hopping, and interaction strength
the instabilities mentioned above can indeed be ignored.

Our finding that
any finite value of the interchain hopping 
leads to a Fermi liquid is supported
by lowest order perturbation theory \cite{Castellani92}. 
However, there have
been recent claims in the literature \cite{Clarke96,Tsvelik96} that
coupled chains with finite $t_{\bot}$ can remain Luttinger liquids 
if the interaction is sufficiently strong, 
so that the anomalous dimension characterizing the
Luttinger liquid at $t_{\bot} = 0$ 
exceeds a certain critical value.
It is important to realize that
this result can only be obtained within an approach that
{\it{allows for a change in the shape of the Fermi surface as  the
interaction is turned 
on\footnote{I would like to thank Steven Strong for his detailed explanations
of this point.}}}.
Unfortunately, 
higher-dimensional bosonization with linearized energy dispersion
cannot describe the renormalization of the shape of the Fermi surface
due to the interactions, 
because after the linearization the relative position of the flat patches 
on the Fermi surface remains completely rigid\footnote{Recall
in this context our discussion 
at the end of Chap.~\secref{sec:Identification}.
concerning the absence of effective mass renormalizations
for linearized energy dispersion.}.
On the other hand, our more general bosonization result
for the Green's function with non-linear energy dispersion
derived in Chap.~\secref{sec:eik} certainly incorporates also the
renormalization of the shape of the Fermi surface due to the interactions.
Thus, an extremely interesting open problem is the 
{\bf{full analysis of 
the higher-dimensional 
bosonization result for the Green's function of coupled chains with non-linear
energy dispersion.}}
Note that within the Gaussian approximation one should not only
calculate the Debye-Waller factor 
$Q^{\alpha}_1 ( \vec{r} , \tau )$ in Eqs.\ref{eq:Qlondef2}--\ref{eq:Slondef2}, but also
the prefactor self-energy $\Sigma_1^{\alpha} ( \tilde{q} )$ 
and the vertex function $Y^{\alpha} ( \tilde{q} )$ given in
Eqs.\ref{eq:sigma1vertex} and \ref{eq:Yres}.
Furthermore, although for $t_{\bot} = 0$ and for linearized energy
dispersion the Gaussian approximation is exact and correctly reproduces the
solution of the Tomonaga-Luttinger model (see Chap.~\secref{sec:Green1}),
it is not clear whether for finite $t_{\bot}$ 
the Gaussian approximation is still sufficient, 
so that it might be necessary to include at least certain sub-classes of
the non-Gaussian corrections discussed in Chap.~\secref{sec:eik}.
Obviously, the problem of coupled chains is far from being solved.
We hope that the methods developed in this book  
will help to shed more light onto this very interesting problem.

Finally, we would like to point out that the problem of 
calculating the Green's function of an {\it{infinite
array}} of coupled chains is
very different from the problem of
{\it{two}} coupled 
chains \cite{Fabrizio92,Kusmartsev92,Yakovenko92,Finkelstein93}.
The two-chain problem is it not so easy
to solve by means of higher-dimensional bosonization, because
in this case
the Fermi surface consists of four isolated points, which
evidently cannot be treated as a simple higher-dimensional surface.
It turns out that even for long-range Coulomb interactions 
it is impossible to map the
two-chain system onto a pure forward scattering problem.
In fact, the calculation of the
Green's function in the two-chain system can be mapped onto an
effective back-scattering problem in one dimension \cite{Bartosch96,Shannon96}, 
which in general
cannot be solved exactly. However, if one assumes 
certain special values of the
interchain and intrachain interaction,
the Green's function of an arbitrary number
of coupled chains can be calculated exactly \cite{Bartosch96}. 
Although these special interactions are perhaps unphysical, 
it is interesting to note that in these models
Luttinger liquid behavior coexists with coherent interchain 
hopping \cite{Shannon96}.
This seems to disagree with the result
of Clarke, Strong, and Anderson \cite{Clarke94,Clarke96},
who claim that
Luttinger liquid behavior necessarily destroys
coherent interchain hopping.

%
%

%
%
%

\chapter{Electron-phonon interactions}
\label{chap:aph}
\setcounter{equation}{0}

{\it{We couple electrons to phonons via Coulomb forces, and show that
for isotropic three-dimensional systems the long-range part of the
Coulomb interaction cannot destabilize the Fermi liquid state.
However, Luttinger liquid behavior in three dimensions can be
due to quasi-one-dimensional anisotropy in the electronic band structure
or in the phonon frequencies. A brief account of the
results presented in this chapter has been published in \cite{Kopietz95pho}.}}

\vspace{7mm}

\noindent
The interplay between
the vibrations of the ionic lattice in a solid and
the interactions
between the conduction electrons
still lacks a complete understanding \cite{Kim89,Kulic94,Grilli94}.
Conventionally this problem is approached
perturbatively, which is possible as long as
the mass $M_{\rm i}$ of the
ions is much larger than the effective mass $m$ of the electrons.
In this case a theorem due to Migdal \cite{Migdal58}\index{Migdal theorem} 
guarantees that, to leading order
in $\sqrt{ {m}/{M_{\rm{i}}}}$,
the electron-phonon vertex is not renormalized by phonon corrections.
However, in heavy fermion systems  the parameter 
$\sqrt{ {m}/{M_{\rm{i}}}} $ is not necessarily small,
so that Migdal's theorem may not be valid. Then
the self-consistent renormalization of the phonon energies due to the coupling to the
electrons cannot be neglected \cite{Pines89,Bardeen55}.
In diagrammatic approaches 
it is often tacitly assumed that the phonons remain
well defined collective modes \cite{Fetter71,Engelsberg63}.
Moreover, 
an implicit assumption in the proof of Migdal's theorem is that
the electronic system is a Fermi liquid.
In view of the experimental evidence of  non-Fermi liquid behavior 
in the normal state of some of 
the high-temperature superconductors \cite{Anderson90b},
it is desirable to study the coupled electron-phonon system by means of a method
which does not assume {\it{a priori}} a Fermi liquid. 
Our functional bosonization approach fulfills this requirement,
so that it offers a new non-perturbative way 
to study coupled electron-phonon systems in $d > 1$. 
In one dimension the problem of electron-phonon interactions has recently been 
analyzed via bosonization in the works \cite{Meden94,Loss94}. 

We would like to emphasize, however, that we shall
retain only processes involving small
momentum transfers and neglect superconducting instabilities.
Recall that in BCS superconductors the phonons 
mediate an effective attractive 
interaction between the electrons, which
at low enough temperatures overcomes the repulsive Coulomb interaction
and leads to superconductivity \cite{Schrieffer71}.
Thus,
the analysis presented below is restricted to the parameter regime where the electronic
system is in the normal metallic state. 
However, we do {\it{not}} assume that the electronic system is a Fermi liquid. 

Throughout this chapter we shall work with linearized energy dispersion,
because we shall focus on the calculation 
of the {\it{static}} Debye-Waller factor $Q^{\alpha} ( {\vec{r}}, 0 )$. 
As discussed in Chap.~\secref{subsec:Thetermslin},
the long-distance behavior of this quantity should only be weakly
affected by the non-linear terms in the energy dispersion.
Note that this approximation is most likely not sufficient  for
the calculation of
$Q^{\alpha} ( \vec{r} , \tau )$ for $\tau \neq 0$,
because in this case the double pole 
that appears in the bosonization result for the
Debye-Waller factor with linearized energy dispersion leads to some
unphysical features (see the discussion at the beginning
of Chap.~\secref{sec:eik} and in Chap.~\secref{subsec:Thetermslin}).
In this case one should retain the non-linear terms in the
energy dispersion.

This chapter is subdivided into four main sections. 
In Sect.~\secref{sec:theeffective} we define the 
coupled electron-phonon system in the language of functional integrals.
By integrating over the phonon degrees of freedom, we then derive the
effective action for the electrons, and 
determine the precise form of the effective retarded
density-density interaction between the electrons mediated
by the phonons.
Because this interaction is of the density-density type discussed 
in Chap.~\secref{chap:agreen}, 
we obtain in Sect.~\secref{sec:bospho}
a non-perturbative expression for the electronic Green's function
by simply substituting the
proper effective interaction $f_{q}^{{\rm RPA} , \alpha}$ into
Eqs.\ref{eq:Rlondef} and \ref{eq:Slondef}.  
In Sect.~\secref{subsec:phodamp} we show that
our approach takes also the renormalization of the
phonon spectrum due to the coupling to the electrons
into account.
Finally,
in Sect.~\secref{sec:resphonon}
we shall explicitly calculate the quasi-particle residue and 
examine the conditions under which
the residue can become small or even vanishes.
In particular, we 
discuss one-dimensional phonons with dispersion
$\Omega_{\vec{q}} = c_{\rm s} | q_{x} |$
that are coupled to three-dimensional electrons with
a spherical Fermi surface.  We show that in this case
the quasi-particle residue vanishes at the
points ${\vec{k}}^{\alpha} = \pm k_{\rm F} {\vec{e}}_{x}$ on the Fermi sphere,
and that close to these special points the single-particle Green's function
exhibits Luttinger liquid behavior.

\section{The effective interaction}
\label{sec:theeffective}

{\it{
We introduce a simple model for electrons that are coupled to 
longitudinal acoustic phonons and derive the 
associated retarded electron-electron interaction by means of
functional integration.}}

\subsection{The Debye model}

Following the classic textbook by Fetter and Walecka \cite{Fetter71},
we use the Debye model\index{Debye model}
to describe the interaction between electrons and 
longitudinal acoustic (LA) phonons\index{LA phonons}. 
In this model the ionic background charge
is approximated by a homogeneous elastic medium. 
Although the ions in real solids form a lattice, 
the discrete lattice structure is unimportant for
LA phonons with wave-vectors $| {\vec{q}}| \ll k_{\rm F}$. 
For a detailed description of this model and its physical justification
see chapter 12 of the book by Fetter and Walecka \cite{Fetter71}.
However, some subtleties concerning screening and phonon energy
renormalization have been ignored in \cite{Fetter71}.
To clarify these points, we first give a careful derivation of the effective
electron-electron interaction in this model via functional integration.

In our Euclidean functional integral approach, the dynamics of the isolated phonon 
system is described via the action 
 \begin{equation}
 S_{\rm ph} \{ b \} = \beta \sum_{q} [ - \I \omega_{m} + \Omega_{\vec{q}} ]  b^{\dagger}_{q} b_{q} 
 \label{eq:Sphdef}
 \; \; \; ,
 \end{equation}
where $b_{q}$ is a complex field representing the
phonons in the coherent state functional integral. 
For simplicity let us first 
assume isotropic acoustic phonons, with dispersion
relation $\Omega_{\vec{q}} = c_{\rm s} | {\vec{q}} |$,
where $c_{\rm s}$ is the {\it{bare}} velocity of sound,
which is determined by the {\it{short-range part}} of the Coulomb potential and 
all other non-universal forces between the ions.
The {\it{long-range}} part of the Coulomb potential 
will be treated explicitly\footnote{
From Appendix~\secref{subsubsec:Cb} it is clear 
that the boundary between the
long- and short-wavelength regimes is defined  by the
Thomas-Fermi wave-vector\index{Thomas-Fermi wave-vector}
$\kappa = ( 4 \pi e^2 \nu )^{{1}/{2}}$.}. 
In Eq.\ref{eq:Sphdef} and
all subsequent expressions involving phonon variables 
it is understood that wave-vector summations 
are cut off when the phonon frequency reaches the Debye frequency \cite{Ashcroft76}.
As before, the electronic degrees of freedom are represented by a Grassmann field
$\psi$, so that the total action of the interacting electron-phonon system is
\index{effective action!electron-phonon system}
 \begin{equation}
 S \{ \psi , b \} = S_{0} \{ \psi \} + S_{\rm ph} \{ b \} + S_{\rm int} \{ \psi , b \}
 \label{eq:Stotph}
 \; \; \; .
 \end{equation}
Here $S_{0} \{ \psi \}$ describes the dynamics of the non-interacting
electron system (see Eq.\ref{eq:S0psidef}), and
$S_{\rm int} \{ \psi , b \}$ represents the
Coulomb energy associated with all charge fluctuations in the system,
 \begin{equation}
S_{\rm int} \{ \psi , b \} = \frac{e^2}{2} \int_{0}^{\beta} \D \tau
\int \D {\vec{r}} \int \D {\vec{r}}^{\prime} 
\frac{ \rho^{\rm tot} ( {\vec{r}} , \tau ) \rho^{\rm tot} ( {\vec{r}}^{\prime} , \tau) }
{ | {\vec{r}} - {\vec{r}}^{\prime} | }
\label{eq:Stotpsib}
\; \; \; ,
\end{equation}
where 
 \begin{equation}
 \rho^{\rm tot} ( {\vec{r}} , \tau ) = \psi^{\dagger} ( {\vec{r}} , \tau )
 \psi ( {\vec{r}} , \tau )
 - \rho^{\rm ion} ( {\vec{r}} , \tau )
 \label{eq:rhototdecomp}
 \end{equation}
represents the total density of charged particles
at point ${\vec{r}}$ and imaginary time $\tau$.
The ionic density $\rho^{\rm ion} ( {\vec{r}} , \tau )$
is of the form
 \begin{equation}
 \rho^{\rm ion} ( {\vec{r}} , \tau ) =  z \frac{N}{V}  + \delta
 \rho^{\rm ion} ( {\vec{r}} , \tau ) 
 \label{eq:rhoion}
 \; \; \; ,
 \end{equation}
where the first term represents the charge density  of the uniform
background charge, which in the absence of phonons 
exactly compensates the total charge of the conduction electrons.
Here $z \geq 1 $ is the valence of the ions and $z N$ is the total number of conduction electrons.
The fluctuating component of the ionic charge density
is related to the bosonic field $b_{q}$ via
 \begin{equation} 
 \delta \rho^{\rm ion} ( {\vec{r}} , \tau ) 
 = - z \frac{N}{V}
  \nabla \cdot {\vec{d}} ( {\vec{r}} , \tau )
 \label{eq:rhoionfluc}
 \; \; \; ,
 \end{equation}
where the displacement field ${\vec{d}} ( {\vec{r}} , \tau )$ is
given by \cite{Fetter71}
 \begin{equation}
 {\vec{d}} ( {\vec{r}} , \tau )  = 
 \frac{- \I }{\sqrt{N}} \sum_{q}
 \frac{  \hat{\vec{q}}  }{\sqrt{ 2 M_{\rm{i}} \Omega_{\vec{q}} } }
 \left[ b_q \E^{ \I ( {\vec{q}} \cdot {\vec{r}} - \omega_{m} \tau ) } 
 -
 b_q^{\dagger} \E^{ - \I ( {\vec{q}} \cdot {\vec{r}} - \omega_{m} \tau ) } \right]
 \; \; \; ,
 \label{eq:displacefield}
 \end{equation}
so that
 \begin{equation}
 \nabla \cdot {\vec{d}} ( {\vec{r}} , \tau )  = 
 \frac{1}{\sqrt{N}} \sum_{q}
 \frac{ | {\vec{q}} | }{\sqrt{ 2 M_{\rm{i}} \Omega_{\vec{q}} } }
 \left[ b_q \E^{ \I ( {\vec{q}} \cdot {\vec{r}} - \omega_{m} \tau ) } 
 +
 b_q^{\dagger} \E^{ - \I ( {\vec{q}} \cdot {\vec{r}} - \omega_{m} \tau ) } \right]
 \; \; \; .
 \label{eq:divdisplacefield}
 \end{equation}
Substituting Eq.\ref{eq:rhototdecomp} into Eq.\ref{eq:Stotpsib},
we obtain  three contributions, which after Fourier transformation
can be written as
 \begin{equation}
 S_{\rm int} \{ \psi , b \} = 
 S_{\rm int}^{\rm el} \{ \psi \} +
 S_{\rm int}^{\rm el-ph} \{ \psi , b  \} +
 S_{\rm int}^{\rm ph} \{ b \} 
 \; \; \; ,
 \end{equation}
with 
 \begin{eqnarray}
 S_{\rm int}^{\rm el} \{ \psi \} & = &
 \frac{\beta }{2 V} \sum_{q} f_{\vec{q}}^{\rm cb} \rho_{-q} \rho_{q}
 \label{eq:Sintel}
 \; \; \; ,
 \\
 S_{\rm int}^{\rm el-ph} \{ \psi , b  \} & = &
 - \frac{\beta }{2 V} \sum_{q} f_{\vec{q}}^{\rm cb} 
 \left[ \rho_{-q} \rho_{q}^{\rm ion} + \rho_{- q}^{\rm ion}
 \rho_{q}   \right]
 \label{eq:Sintelion}
 \; \; \; ,
 \\
 S_{\rm int}^{\rm ph} \{ b  \} & = &
 \frac{\beta }{2 V} \sum_{q} f_{\vec{q}}^{\rm cb} \rho^{\rm ion}_{-q} \rho^{\rm ion}_{q}
 \label{eq:Sintion}
 \; \; \; ,
 \end{eqnarray}
where we have defined
 \begin{equation}
 f_{\vec{q}}^{\rm cb} =
 \left\{
 \begin{array}{cc}
 \frac{ 4 \pi e^2}{ {\vec{q}}^2 } & \mbox{ for ${\vec{q}} \neq 0 $ } \\
 0 & \mbox{ for $ {\vec{q}} = 0 $}
 \end{array}
 \right.
 \; \; \; .
 \end{equation}
The Fourier coefficients of the densities can be expressed in terms of the
Fourier coefficients $\psi_{k}$ and $b_{q}$ of the electron and phonon fields,
 \begin{eqnarray}
 \rho_{q} & = & \sum_{k} \psi_{k}^{\dagger} \psi_{k+q}
 \label{eq:rhoqFourierph}
 \; \; \; ,
 \\
 \rho_{q}^{\rm ion} & = & - z \sqrt{N} \frac{ | {\vec{q}} | }{\sqrt{2 M_{\rm{i}} \Omega_{\vec{q}} }}
 \left[ b_{q} + b^{\dagger}_{-q} \right]
 \label{eq:rhoionqFourierph}
 \; \; \; .
 \end{eqnarray}
The part of the action involving the phonon degrees of freedom can then be written as
 \begin{eqnarray}
  S_{\rm ph} \{ b \} + S_{\rm int}^{\rm ph} \{ b \} + 
 S_{\rm int}^{\rm el-ph} \{ b , \psi \}
  & =  &   \beta \sum_{q} \left[ ( - \I \omega_{m} + \Omega_{\vec{q}} ) b^{\dagger}_{q} b_{q}
  \right.
  \nonumber
  \\
  & & \hspace{-40mm}
  \left.
 + \frac{{W}_{\vec{q}}}{4} ( b_{q} + b^{\dagger}_{-q} ) ( b_{-q} + b^{\dagger}_{q} )
 + g_{{\vec{q}}} \rho_{-q} ( b_{q} + b^{\dagger}_{-q} ) \right]
 \label{eq:Sphtot}
 \; \; \; ,
 \end{eqnarray}
with
 \begin{eqnarray}
 W_{\vec{q}} & = &  
 \left[ \frac{z^2 N}{V}  \frac{ {\vec{q}}^2 }{M_{\rm{i}} }  \right]
 \frac{ f_{\vec{q}}^{\rm cb}}{  \Omega_{\vec{q}} }
 \label{eq:Wqdef}
 \; \; \; ,
 \\
 g_{\vec{q}} & = & 
 \left[ \frac{ z^2 N}{V}  \frac{ {\vec{q}}^2 }{ M_{\rm{i}} }  \right]^{1/2}
 \frac{ f_{\vec{q}}^{\rm cb}}{  \sqrt{ 2 V  \Omega_{\vec{q}} } }
 \label{eq:gqdef}
 \; \; \; .
 \end{eqnarray}
At this point Fetter and Walecka \cite{Fetter71} make the following two approximations:
(a) the bare Coulomb interaction $f_{\vec{q}}^{\rm cb}$ in
$S_{\rm int}^{\rm el-ph} \{ \psi , b \}$ 
is replaced by the static screened interaction, 
$ { 4 \pi e^2}/{ \vec{q}^2 } \rightarrow { 4 \pi e^2}/{ \kappa^2}$,
and (b) the contribution
$S_{\rm int}^{\rm ph} \{ b \}$ is simply dropped. 
We shall see shortly that the approximation (b) amounts to
ignoring the self-consistent renormalization of the phonon frequencies \cite{Bardeen55,Pines89}.
Although Fetter and Walecka \cite{Fetter71} argue that these approximations correctly
describe the physics of screening, it is not quite satisfactory 
that one has to rely here on words and not on calculations. 
Because in our bosonization method screening can be derived 
from first principles, we do not follow the
``screening by hand''\index{screening!by hand} procedure of \cite{Fetter71},
and retain at this point all terms in Eqs.\ref{eq:Sintel}--\ref{eq:Sintion}
with the bare Coulomb interaction.

\subsection{Integration over the phonons}

{\it{In this way we obtain the effective electron-electron interaction mediated by the phonons.}}

\vspace{7mm}

\noindent
We are interested in the exact electronic Green's function of the interacting many-body
system. The Matsubara Green's function can be written as a functional
integral average
 \begin{equation}
 G ( k )
 = - \beta \frac{ 
 \int {\cal{D}} \left\{ \psi \right\} 
 {\cal{D}} \left\{ b \right\}
 \E^{- {S} \{ \psi , b \} } 
 \psi_{k} \psi^{\dagger}_{k}
 }
 { \int {\cal{D}} \left\{ \psi \right\} 
 {\cal{D}} \left\{ b \right\}
 \E^{- S \{ \psi , b \} } }
 \; \; \; .
 \label{eq:Gphdef}
 \end{equation}
Evidently the $b$-integration in Eq.\ref{eq:Gphdef} is Gaussian, and can therefore
be carried out {\it{exactly}}. After a straightforward integration we  obtain
the following exact expression for the interacting Green's function
 \begin{equation}
 G ( k )
 = - \beta \frac{ 
 \int {\cal{D}} \left\{ \psi \right\} 
 \E^{- {S}_{\rm eff} \{ \psi \} } 
 \psi_{k} \psi^{\dagger}_{k}
 }
 { \int {\cal{D}} \left\{ \psi \right\} 
 \E^{- S_{\rm eff} \{ \psi  \} } }
 \; \; \; ,
 \label{eq:Gph2}
 \end{equation}
with 
 \begin{equation}
 S_{\rm eff} \{ \psi \} = S_{0} \{ \psi \} + S_{\rm int}^{\rm el} \{ \psi \} 
 - \beta \sum_{q} 
 \left[
 \frac{ g_{\vec{q}}^2 \Omega_{\vec{q}} }{ \omega_{m}^2 +  \Omega_{\vec{q}}^2 + 
  \Omega_{\vec{q}} W_{\vec{q}} } \right] 
 \rho_{-q} \rho_{q}
 \; .
 \label{eq:Seffdefph}
 \end{equation}
The last term is the effective interaction between the electrons mediated by the
phonons. Combining the last two terms in Eq.\ref{eq:Seffdefph} and using the
above definitions of $W_{\vec{q}}$ and $g_{\vec{q}}$, we finally arrive at
 \begin{equation}
 S_{\rm eff} \{ \psi \}
 = S_{0} \{ \psi \} + \frac{\beta}{2 V}
 \sum_{q} f_{q}
 \rho_{-q} \rho_{q}
 \; \; \; ,
 \label{eq:Seffph2}
 \end{equation}
where the total effective interaction is given by
 \begin{equation}
  f_{q} = f_{\vec{q}}^{\rm cb} \left[ 1 -    
 \frac{  f_{\vec{q}}^{\rm cb} \frac{z^2 N {\vec{q}}^2 }{VM_{\rm{i}}} }
 {
 \omega_{m}^2 +  \Omega_{\vec{q}}^2  + 
  f_{\vec{q}}^{\rm cb} 
 \frac{ z^2 N {\vec{q}}^2}{V M_{\rm{i}}}    }
 \right]
 =
 f_{\vec{q}}^{\rm cb} \frac{ \omega_{m}^2 +  \Omega_{\vec{q}}^2 }
 {
 \omega_{m}^2 + \Omega_{\vec{q}}^2  + 
  f_{\vec{q}}^{\rm cb} 
 \frac{ z^2 N {\vec{q}}^2}{V M_{\rm{i}}}    }
 \;  .
 \label{eq:fefftotdef}
 \end{equation}
Defining the electron-phonon coupling constant\index{electron-phonon coupling} 
$\gamma$ via
 \begin{equation}
\frac{ z^2 N {\vec{q}}^2}{VM_{\rm{i}}} 
\equiv \nu^2 \gamma^2  \Omega_{\vec{q}}^2
 \label{eq:gammaphcoupdef}
 \; \; \; ,
 \end{equation}
where $\nu$ is the density of states, we see that
Eq.\ref{eq:fefftotdef} can also be written as
 \begin{equation}
  f_{q} = \frac{f_{\vec{q}}^{\rm cb} }{ 1 + \nu^2 \gamma^2 f_{\vec{q}}^{\rm cb}    
  \frac{  \Omega_{\vec{q}}^2}{ \omega_{m}^2 +  \Omega_{\vec{q}}^2 } }
 \label{eq:fefftot2}
 \; \; \; .
 \end{equation}
It is instructive to compare Eq.\ref{eq:fefftot2} with the
expression that would result from the
``screening by hand''\index{screening!by hand} procedure described above.
The approximation (a) amounts to the replacement
 \begin{equation}
 g_{\vec{q}}^2  \Omega_{\vec{q}} \rightarrow  \frac{ z^2 N 
 {\vec{q}}^2}{VM_{\rm{i}}}
 \left( \frac{4 \pi e^2}{\kappa^2} \right)^2 \frac{1}{2 V}
 \label{eq:replacebyhand}
 \end{equation}
in
Eq.\ref{eq:Seffdefph}, while (b) is equivalent with
$W_{\vec{q}} \rightarrow 0$.
Using the fact that $\kappa^2 = 4 \pi e^2 \nu$, 
it is easy to see that in this approximation
the effective interaction $f_{q}$ in Eq.\ref{eq:fefftotdef}
is replaced by
 \begin{equation}
 f_{q} \rightarrow f_{\vec{q}}^{\rm cb} - \gamma^2 \frac{    \Omega_{\vec{q}}^2 }
 { \omega_{m}^2 +   \Omega_{\vec{q}}^2 }
 \; \; \; .
 \label{eq:FetterWal}
 \end{equation}
For consistency, we should
also replace $ { 4 \pi e^2}/{ {\vec{q}}^2}
\rightarrow { 4 \pi e^2}/{\kappa^2}$ in the direct Coulomb interaction, which amounts to
setting $f_{\vec{q}}^{\rm cb} \rightarrow {1}/{\nu}$ in the first term of
Eq.\ref{eq:FetterWal}. Evidently the phonon contribution in Eq.\ref{eq:FetterWal} 
can be obtained from an expansion of the exact result 
\ref{eq:fefftot2} {\it{to first order}} in $\gamma^2$ and the subsequent
replacement $f_{\vec{q}}^{cb} \rightarrow \frac{1}{\nu}$ in the phonon part.
By performing these replacements one implicitly
neglects the self-consistent renormalization of the phonon frequencies \cite{Pines89}. 
Therefore one should also replace in Eq.\ref{eq:FetterWal}
$\Omega_{\vec{q}} \rightarrow \tilde{\Omega}_{\vec{q}}$, where
the {\it{renormalized}} phonon frequencies $\tilde{\Omega}_{\vec{q}}$ include by definition 
the effect of the electronic degrees of freedom on the
phonon dynamics.
In this way  one arrives at the usual form of the
electron-phonon interaction that is frequently used in the literature.
Evidently in the conventional
``screening by hand'' approach 
the renormalization of the phonon dispersion due to
interactions with the electrons remains unknown, and it
is implicitly assumed that a 
{\it{self-consistent}} calculation 
would lead to an effective interaction of
the form \ref{eq:FetterWal}, with well-defined phonon modes.  
Note that the coupling to the electronic system will certainly lead to
a finite damping of the phonon mode\index{damping!phonon}\index{phonon!damping}, which  
is not properly described by Eq.\ref{eq:FetterWal}.
In contrast,
the effective interaction in Eq.\ref{eq:fefftot2} is an {\it{exact}}
consequence of the microscopic model
defined in Eqs.\ref{eq:Sphdef}--\ref{eq:Stotpsib}.
In fact, as will be shown in Sect.~\secref{subsec:phodamp},
the phonon energy shift\index{phonon!energy shift} 
and damping {\it{can be derived}} from this expression!

\section{The Debye-Waller factor} 
\label{sec:bospho}

{\it{Given the effective frequency-dependent density-density
interaction \ref{eq:fefftot2},
it is now easy to obtain a non-perturbative
expression for the single-particle Green's function, 
which is valid even if the system is not a Fermi liquid.}}

\vspace{7mm}
\index{Debye-Waller factor!phonons}

\noindent
Because the phonons simply modify the effective density-density interaction,
we can obtain a non-perturbative expression  for the
interacting Green's function by substituting the
interaction \ref{eq:fefftot2} into our general bosonization formula 
for linearized energy dispersion
given in Eqs.\ref{eq:Qlondef}--\ref{eq:Slondef} 
and \ref{eq:Galphartdef}--\ref{eq:Galphaqtildedef}.
Because the interaction $f_{q}$ in Eq.\ref{eq:fefftot2} does not depend on the patch indices,
the effective interaction $f_{q}^{{\rm RPA} , \alpha}$ in Eqs.\ref{eq:Rlondef} and \ref{eq:Slondef} is 
the usual RPA interaction,
so that the Debye-Waller factor associated with patch $\alpha$ is 
given by
 \begin{equation}
 Q^{\alpha} ( {\vec{r}} , \tau )  =
 \frac{1}{\beta {{V}}} \sum_{ q }  f^{\rm RPA}_{q}
  \frac{ 1 -
  \cos ( {\vec{q}} \cdot  {\vec{r}} 
  - {\omega}_{m}  \tau  ) 
 }
 {
 ( \I \omega_{m} - {\vec{v}}^{\alpha} \cdot {\vec{q}} )^{2 }}
 \label{eq:DWph}
 \; \; \; ,
 \end{equation}
with
 \begin{equation}
 f^{\rm RPA}_{q} =
  \frac{ f_{q} }{ 1 + f_{q} \Pi_{0} ( q ) }
  = 
  \frac{ f^{\rm cb}_{\vec{q}} }{ 1 + f^{\rm cb}_{\vec{q}}  {\Pi}_{\rm ph} ( q ) }
  \label{eq:frpaph}
  \; \; \; ,
  \end{equation}
where
 \begin{equation}
 {\Pi}_{\rm ph} ( q ) = \Pi_{0} ( q ) +   \nu  \tilde{\gamma}^2
 \frac{ 1 }{ 1 +  \omega_{m}^2 / \Omega_{\vec{q}}^2  }
 \label{eq:Piphdef}
 \end{equation}
is the {\it{dressed inverse phonon propagator\index{phonon!propagator}}} \cite{Engelsberg63}.
Here $\tilde{\gamma}^2$ is the
dimensionless measure for the strength of the electron-phonon 
coupling\index{electron-phonon coupling!dimensionless}, 
 \begin{equation}
\tilde{\gamma}^2 = \nu \gamma^{2} = \frac{ z^2 N}{V M_{\rm{i}} \nu c_{\rm s}^2}
\; \; \; .
\label{eq:tildegammadef}
\end{equation}
Using Eq.\ref{eq:nurelation},
this
reduces for a spherical three-dimensional Fermi surface to\footnote{ 
Note that the total number of conduction electrons is
now $z N$, so that we should replace in Eq.\ref{eq:nurelation} 
$N \rightarrow zN$.}
 \begin{equation}
 \tilde{\gamma}^2 =
 \frac{z}{3} \frac{m}{M_{\rm{i}}} \left( \frac{v_{\rm F}}{ c_{\rm s} } \right)^2
 \label{eq:tildegammasphere}
 \; \; \; .
 \end{equation}
We conclude that the phonons simply give rise to an additive contribution 
to the non-interacting polarization\index{polarization!phonons}.
Assuming that the Fermi surface is spherically symmetric, we can also write
 \begin{equation}
 {\Pi}_{\rm ph} ( q ) = \nu {g}_{\rm ph} ( {\vec{q}} , \I \omega_{m} )
 \; \; \; ,
 \label{eq:Piphalso}
 \end{equation}
where the dimensionless function
${g}_{\rm ph} ( {\vec{q}} , \I \omega_{m} )$ is given by
 \begin{equation}
 {g}_{\rm ph} ( {\vec{q}} , \I \omega_{m} ) =
 g_{3}
 ( \frac{ \I \omega_{m}}{ v_{\rm F} | {\vec{q}} | } ) +   
 \tilde{\gamma}^2  g_{1} ( \frac{ \I \omega_{m}}{ \Omega_{\vec{q}} } )    
 \label{eq:tildeg3defph}
 \; \; \; ,
 \end{equation}
and the functions $g_{1} ( \I y )$ and  $g_{3} ( \I y  )$ 
are defined in Eqs.\ref{eq:g1y} and \ref{eq:g3y}.
Note that the phonon part of Eq.\ref{eq:tildeg3defph} involves the dimensionless
function $g_{1} ( \I y )$ that 
appears in the polarization of the
one-dimensional electron gas, see Eq.\ref{eq:Pi0patch}.
Of course, here the origin for this function is the
coupling of the electron system to {\it{another well defined  collective mode}},
whereas in the chain-model it was essentially due to the {\it{shape of the Fermi surface}}.
However, the appearance of the one-dimensional polarization function in Eq.\ref{eq:tildeg3defph}
suggests the possibility that a quasi-one-dimensional phonon dispersion
$\Omega_{\vec{q}}$ might lead to Luttinger behavior even 
if the electron dispersion is three-dimensional.
We shall confirm this expectation in Sect.~\secref{subsec:Aniso}.

Because all effects due to the phonons are contained in the function
${g}_{\rm ph} ( {\vec{q}} , \I \omega_m )$,
the general expressions for the various
contributions to the Debye-Waller factor derived in Chap.~\secref{sec:exactman} remain valid.
We simply have to use the corresponding RPA dynamic structure 
factor\index{dynamic structure factor!phonons},
 \begin{equation}
 S_{\rm RPA} ( {\vec{q}} , \omega ) = \frac{\nu }{\pi} {\rm Im} \left\{ 
 \frac{ {g}_{\rm ph} ( {\vec{q}} , 
 \omega + \I 0^{+} ) }{ 1 +  \left( \frac{\kappa}{\vec{q}}  \right)^2
 {g}_{\rm ph} ( {\vec{q}} , \omega + \I 0^{+} )   } \right\}
 \label{eq:Piphdyn}
 \; \; \; .
 \end{equation}
In the following section we shall discuss the form of
 $S_{\rm RPA} ( {\vec{q}} , \omega )$ in some detail.

\section{Phonon energy shift\index{phonon!energy shift} 
and phonon damping\index{phonon!damping}}
\label{subsec:phodamp}

{\it{
We show that the dynamic structure factor \ref{eq:Piphdyn}
contains the
the self-consistent renormalization of the phonon dynamics
due to the coupling to the electronic system.}}

\vspace{7mm}

\begin{figure}
\sidecaption
\psfig{figure=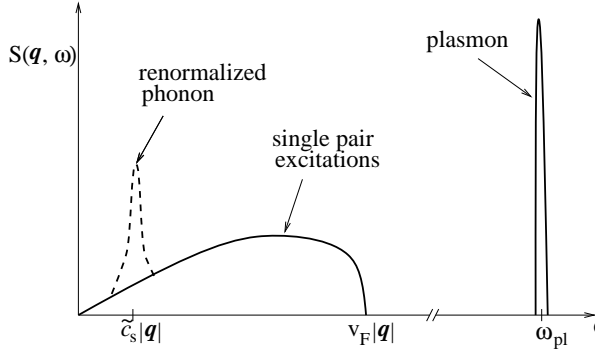,width=8cm}
\caption[The dynamic structure factor for an electron-phonon system.]
{A rough sketch of the various contributions to the RPA 
dynamic structure factor\index{dynamic structure factor!phonons}
\ref{eq:Piphdyn} in the regime where the phonon mode is well defined.
Here $\tilde{c}_{\rm s}$ is the renormalized phonon velocity, see Eq.\ref{eq:csren}.}
\label{fig:phofig}
\end{figure}

\noindent
The renormalization of the phonon spectrum due to the coupling to the
electrons can be obtained from the phonon peak of $S_{\rm RPA} ( {\vec{q}} , \omega )$.
The qualitative behavior of the dynamic structure factor 
can be determined from simple physical considerations \cite{Pines89},
and is shown in Fig.~\secref{fig:phofig}.
%
%
In the absence of phonons, 
 $S_{\rm RPA} ( {\vec{q}} , \omega )$
consists of a sum of two terms, which are discussed in detail in 
Appendix~\secref{sec:dynstruc}.
The first term $S_{\rm RPA}^{\rm col} ( {\vec{q}} , \omega )$ 
is a $\delta$-function peak due to
the collective plasmon mode. From Eq.\ref{eq:colmodcbres} we see that
for the Coulomb interaction in $d=3$ the plasmon
approaches  at long wavelengths
a finite value, the plasma frequency\index{plasma frequency} 
$\omega_{\rm pl} = {v_{\rm F} \kappa }/{ \sqrt{3}}$.
Within the RPA this mode is not damped, so that its contribution 
to the dynamic structure factor is
 \begin{equation}
 S_{\rm RPA}^{\rm col} ( {\vec{q}} , \omega ) = Z_{ {\vec{q}}}
 \delta ( \omega - \omega_{\rm pl} )
 \; \; \; ,
 \label{eq:Splasmon3}
 \end{equation}
with the residue $Z_{\vec{q}}$ given in Eq.\ref{eq:Zcbres3}.
For small $  z {m}/{M_{\rm{i}}} $ this contribution is only weakly affected by phonons.
This follows from the fact that at the plasma frequency 
the ratio of the phonon to the electron contribution 
in Eq.\ref{eq:tildeg3defph} is for $ |{\vec{q}} | \ll \kappa$ given by
 \begin{equation}
 \frac{ \tilde{\gamma}^{2} g_{1} ( \frac{ \omega_{\rm pl}}{  \Omega_{\vec{q}}  } ) }
 {  g_{3} ( \frac{ \omega_{\rm pl} }{ v_{\rm F} | {\vec{q}} |  } ) }
 \approx 3  \tilde{\gamma}^2 \left( \frac{  \Omega_{\vec{q}}}{ v_{\rm F} {\vec{q}} }  \right)^2
 = 3 \tilde{\gamma}^2 \left( \frac{c_{\rm s} }{v_{\rm F}} \right)^2
 = z \frac{ m}{M_{\rm{i}}}
 \label{eq:colsmall1}
 \; \; \; .
 \end{equation}
Evidently we may ignore the effect of the phonon on the
plasmon mode provided
 \begin{equation}
  \tilde{\gamma}  \frac{c_{\rm s} }{v_{\rm F}} =  
  \sqrt{ \frac{z m}{3 M_{\rm{i}}}} \ll 1
  \label{eq:plasmasmall}
  \; \; \; .
  \end{equation}
Note that the validity of the Migdal theorem in a Fermi liquid 
is based on precisely this condition.
In addition to the plasmon mode, the dynamic structure factor 
is non-zero in the regime
$\omega \leq v_{\rm F} | {\vec{q}} |$.
In the absence of phonons, $S_{\rm RPA} ( {\vec{q}} , \omega )$ 
is here a rather featureless function, 
representing the decay of collective density
fluctuations into particle hole pairs, 
i.e. Landau damping\index{Landau damping}
(see Eq.\ref{eq:Ssp}).
Mathematically, the Landau damping arises from the finite imaginary part
of the function $g_{3} ( \frac{ \omega}{v_{\rm F} | {\vec{q}} | } + \I 0^{+} )$ for 
$ \omega < v_{\rm F} | {\vec{q}} |$.
As long as the {\it{renormalized}} phonon 
velocity\index{phonon!renormalized} is small
compared with $v_{\rm F}$ and phonon damping is negligible, we expect
that phonons give rise to an additional narrow peak 
that sticks out of the smooth background due to
Landau damping.  This is the renormalized phonon mode.

We now confirm this picture
by explicitly  calculating the approximate form of the dynamic structure factor in the vicinity
of the phonon peak. 
To determine the {\it{renormalized}} phonon frequency, we look for
solutions of the collective mode equation
 \begin{equation}
 1 + \left( \frac{ \kappa}{  {\vec{q}} } \right)^2 g_{\rm ph} ( {\vec{q}} , z ) = 0
 \label{eq:colmodeph}
 \; \; \; .
 \end{equation}
Anticipating that this equation has a solution 
with $ | z | \ll v_{\rm F} | {\vec{q}} |$, we may approximate
the function $g_3 ( z )$ in Eq.\ref{eq:tildeg3defph}
by the expansion of
$g_{3} ( x + \I 0^{+} )$ 
for small $x$, which is according to 
Eqs.\ref{eq:gammapart} and \ref{eq:Imgddefsmall} given by
 \begin{equation}
 g_{3} ( x + \I 0^{+} ) \approx 1 + \I \frac{ \pi}{2} x
 \label{eq:g3approxim}
 \; \; \; .
 \end{equation}
Substituting this approximation for $g_{3}$ into Eq.\ref{eq:tildeg3defph} and
using Eq.\ref{eq:g1x}, 
we find the following cubic equation for the 
dressed phonon frequency,
 \begin{equation}
 z^2 -  \Omega_{\vec{q}}^2 \left[ 1 + 
 \frac{ \tilde{\gamma}^2}{ 1 + ( \frac{ {\vec{q}}}{\kappa} )^{2}  } \right]
 + \I \frac{\pi}{2} \frac{1}
 { [ 1 + ( \frac{\vec{q}}{\kappa} )^2 ] }
 \frac{ z }{v_{\rm F} | {\vec{q}}  | } 
  \left[ z^2 -  \Omega_{\vec{q}}^2 \right] = 0
 \; \; \; .
 \label{eq:cubicphonon}
 \end{equation}
If we ignore the damping term, this equation has
a solution at
 $z = \tilde{{\Omega}}_{\vec{q}}$, 
where the renormalized phonon frequency is 
 \begin{equation}
 \tilde{{\Omega}}_{\vec{q}} = \Omega_{\vec{q}} 
 \sqrt{ 1 + 
 \frac{ \tilde{\gamma}^2}{ 1 + \left( \frac{ {\vec{q}} }{\kappa} \right)^2 } }
 \label{eq:Omegaren}
 \; \; \; .
 \end{equation}
For  $  \tilde{\Omega}_{\vec{q}}  \ll {v_{\rm F} | {\vec{q}} | }$  
the cubic term in Eq.\ref{eq:cubicphonon} can be treated perturbatively.
This term shifts the solution to $z = \tilde{{\Omega}}_{\vec{q}}
- \I \tilde{\Gamma}_{\vec{q}}$, with the damping given by
 \begin{equation}
 \tilde{{\Gamma}}_{\vec{q}} = \frac{\pi}{4} \frac{  \Omega_{\vec{q}}^2}{v_{\rm F} | {\vec{q}} | }
 \frac{\tilde{\gamma}^2}{  [  1 + ( \frac{ {\vec{q}} }{\kappa})^2 ]^2 }
 \label{eq:Gammaphonon}
 \; \; \; .
 \end{equation}
Note that 
 \begin{eqnarray}
 \frac{\tilde{\Gamma}_{\vec{q}} }{ \tilde{\Omega}_{\vec{q}}}
 & = & \frac{\pi}{4} \frac{ c_{\rm s} }{v_{\rm F}}
 \frac{ \tilde{\gamma}^2 }{ [ 1 + ( \frac{ {\vec{q}} }{\kappa} )^2 ]^{\frac{3}{2}} 
 [  1 +   \tilde{\gamma}^2 + ( \frac{ {\vec{q}} }{\kappa} )^2  ]^{ \frac{1}{2} } }
 \nonumber
 \\
 & \approx &
 \frac{\pi}{4} \frac{ c_{\rm s} }{v_{\rm F}} \frac{ \tilde{\gamma}^2}{ \sqrt{ 1 + \tilde{\gamma}^2 }}
 \; \; \; , \; \; \; \mbox{ for $| {\vec{q}} | \ll \kappa$ }
 \; \; \; ,
 \end{eqnarray}
so that the collective phonon mode is always well defined 
as long as the condition
\ref{eq:plasmasmall} is satisfied.
Thus, in the regime $ \tilde{\gamma} \ll {v_{\rm F}}/{c_{\rm s}}$
there is a well defined narrow peak with frequency
$\tilde{{\Omega}}_{\vec{q}}$ and width 
$\tilde{\Gamma}_{\vec{q}}$ in the dynamic structure factor, which
sticks out of the smooth background due to the particle 
hole continuum (see Fig.~\secref{fig:phofig}).
This corresponds to the renormalized phonon mode.
\index{phonon!renormalized}
Using Eq.\ref{eq:Omegaren} we may define a wave-vector-dependent phonon velocity 
 \begin{equation}
 \tilde{\Omega}_{\vec{q}} = \tilde{c}_{\rm s} ( {\vec{q}} )  | {\vec{q}} |
 \; \; \; , \; \; \; \tilde{c}_{\rm s} ( {\vec{q}} ) = c_{\rm s} 
 \sqrt{ 1 + 
 \frac{ \tilde{\gamma}^2}{ 1 + \left( \frac{ {\vec{q}} }{\kappa}\right)^2 } }
 \label{eq:csren}
 \; \; \; .
 \end{equation}
The renormalization of the phonon velocity is obviously a screening effect.
At short length scales there is no screening charge around the phonon, so that
it propagates with the bare velocity.
At long wavelengths, however, the phonon has to drag along the screening
cloud, so that its velocity is modified.
For large $\tilde{\gamma}$ the renormalized phonon velocity
reduces at long wavelengths to
 $\tilde{c}_{\rm s} ( 0 ) \approx c_s \tilde{\gamma}$.
For a spherical three-dimensional Fermi surface
we may use Eq.\ref{eq:tildegammasphere} to  rewrite
this as\index{Bohm-Staver relation}
 \begin{equation}
 \tilde{c}_{\rm s} ( 0 ) \approx \sqrt{ \frac{z}{3} \frac{m}{M_{\rm{i}}}}
 v_{\rm F}
 \; \; \; .
 \label{eq:BohmStarv}
 \end{equation}
This well-known result is called the
Bohm-Staver relation \cite{Ashcroft76,Bohm50}.
Note that the renormalized phonon velocity \ref{eq:BohmStarv}
is independent of the bare velocity $c_{\rm s}$.

To calculate the dynamic structure factor in the vicinity
of the phonon peak, we also need the height of the peak.
Expanding the denominator in
Eq.\ref{eq:Piphdyn} around $\omega = \tilde{\Omega}_{\vec{q}}$, we obtain
for the residue associated with the phonon peak
 \begin{eqnarray}
 Z_{ {\vec{q}}}^{\rm ph} & = & \frac{\nu}{ ( \frac{ \kappa}{ | {\vec{q}}| } )^4 
 \left. \frac{ \partial}{\partial \omega} 
 g_{\rm ph} ( {\vec{q}} , \omega + \I 0^{+} )  
 \right|_{\omega = \tilde{\Omega}_{\vec{q}} } }
 \nonumber
 \\
 & = &
 \frac{ \nu}{2} \tilde{\Omega}_{\vec{q}} \left( \frac{ | \vec{q} | }{\kappa} \right)^4 
 \frac{ \tilde{\gamma}^2}{ [ 1 + ( \frac{ {\vec{q}} }{\kappa} )^2 ] 
 [  1 +  \tilde{\gamma}^2 + ( \frac{ {\vec{q}} }{\kappa} )^2  ] }
 \label{eq:Zqph}
 \; \; \; .
 \end{eqnarray}
Compared with the residue of the plasmon peak in Eq.\ref{eq:Zcbres3},
the phonon residue
is at long wavelengths smaller by a factor of 
 \begin{equation}
 \left( \frac{\vec{q}}{\kappa} \right)^2 
 \frac{\Omega_{\vec{q}} }{ \omega_{\rm pl} }
 \frac{\tilde{\gamma}^2 }{ {\sqrt{ 1 + \tilde{\gamma}^2 }}} 
 \; \; \; .
 \label{eq:factorsmallphonon}
 \end{equation}
Note that this is a small parameter even at ${\vec{q}}^2 \approx \kappa^2$
provided Eq.\ref{eq:plasmasmall} is satisfied.
In summary,  
for $\tilde{\gamma} \ll {v_{\rm F}} / c_{\rm s}$ 
the total dynamic structure factor  
can be approximated by
 \begin{equation}
 S_{\rm RPA} ( {\vec{q}} , \omega )  = S_{\rm RPA}^{\rm col} ( {\vec{q}} , \omega ) + 
 S_{\rm RPA}^{\rm sp} ( {\vec{q}} , \omega )
 +  \frac{ Z_{{ \vec{q}}}^{\rm ph} }{\pi}
 \frac{ \tilde{\Gamma}_{\vec{q}} }{ ( \omega - \tilde{\Omega}_{\vec{q}} )^2 
 + \tilde{\Gamma}_{\vec{q}}^2 }
 \label{eq:Srpacloseph}
 \; \; ,
 \end{equation}
with $\tilde{\Omega}_{\vec{q}}$, $\tilde{\Gamma}_{\vec{q}}$ and $Z_{{\vec{q}}}^{\rm ph}$ given in 
Eqs.\ref{eq:Omegaren}, \ref{eq:Gammaphonon} and \ref{eq:Zqph}.
The plasmon contribution $S_{\rm RPA}^{\rm col} ( {\vec{q}} , \omega )$
is given in Eq.\ref{eq:Splasmon3}, while the single pair contribution
$S_{\rm RPA}^{\rm sp} ( {\vec{q}} , \omega )$
is given in Eq.\ref{eq:Ssp}.

\section{The quasi-particle residue\index{quasi-particle residue!phonons}}
\label{sec:resphonon}

{\it{
We now calculate the quasi-particle residue $Z^{\alpha}$ 
and determine the conditions under which $Z^{\alpha}$ becomes small or even
vanishes\index{phonon!quasi-particle residue}. 
}}

\vspace{7mm}

\noindent
According to Eqs.\ref{eq:ZRrelation} and \ref{eq:R2}, the quasi-particle
residue associated with patch $P^{\alpha}_{\Lambda}$ on the Fermi surface is
$Z^{\alpha} = \E^{R^{\alpha}}$, where the constant part $R^{\alpha}$ 
of the Debye-Waller  can be written as
 \begin{equation}
 R^{\alpha} = - \int \frac{ \D {\vec{q}}}{ ( 2 \pi )^3} ( f^{\rm cb}_{\vec{q}})^2
 \int_{0}^{\infty} \D \omega \frac{S_{\rm RPA} 
 ( {\vec{q}} , \omega )}{ ( \omega + | {\vec{v}}^{\alpha}
 \cdot {\vec{q}} | )^2}
 \label{eq:R2ph}
 \; \; \; ,
 \end{equation}
with $S_{\rm RPA} ( {\vec{q}} , \omega )$ given in Eq.\ref{eq:Piphdyn}.
We would like to emphasize that this expression is valid 
for arbitrary strength of the electron-phonon interaction.
In particular, it is valid for $\tilde{\gamma} \geqapprox {v_{\rm F}}/{c_{\rm s}}$, where
the phonon mixes with the plasmon and
the decomposition \ref{eq:Srpacloseph} of the dynamic structure factor
is not valid. 
In this case we should use Eq.\ref{eq:Piphdyn}.
It is not difficult to see that the integral exists
for arbitrary values of $\tilde{\gamma}$  provided  neither the electron
dispersion nor the phonon dispersion is one-dimensional.
Therefore phonons that couple to electrons via long-range Coulomb forces cannot
destabilize the Fermi liquid state.

In order to make progress analytically, we shall
restrict ourselves from now on to the regime
$\tilde{\gamma} \ll {v_{\rm F}}/{c_{\rm s}}$. Then the phonons can be
considered as well defined
collective modes, so that the dynamic structure factor can be approximated
by Eq.\ref{eq:Srpacloseph}.
As shown in Chap.~\secref{subsec:Cbnice}  (see Eq.\ref{eq:Rcbres}),
the contribution of the first two terms in Eq.\ref{eq:Srpacloseph}
to $R^{\alpha}$
can be written as $- ( \frac{ \kappa }{ k_{\rm F}})^2 \frac{\tilde{r}_{3}}{2}$, where
the numerical constant $\tilde{r}_{3}$ is given in Eq.\ref{eq:tilderddef}.
Because  by assumption $\tilde{\Gamma}_{\vec{q}}  \ll \tilde{\Omega}_{\vec{q}}$,
the last term in Eq.\ref{eq:Srpacloseph} acts
under the integral in Eq.\ref{eq:R2ph} like a
$\delta$-function, so that
 \begin{equation}
 R^{\alpha} = - \left( \frac{ \kappa}{k_{\rm F} } \right)^2 \frac{ \tilde{r}_{3}}{2}
 + R^{\alpha}_{\rm ph}
 \; \; \; ,
 \label{eq:Ralphaelphotot}
 \end{equation}
with
 \begin{equation}
  R^{\alpha}_{\rm ph}
   =  - \frac{ \tilde{\gamma}^2 }{2 \nu } 
   \int \frac{ \D {\vec{q}} }{ ( 2 \pi )^3}
  \frac{ \tilde{\Omega}_{\vec{q}} }
  { 
  \left[ \tilde{\Omega}_{\vec{q}} + | {\vec{v}}^{\alpha} \cdot {\vec{q}} | \right]^2 
 \left[ 1 + ( \frac{ {\vec{q}} }{\kappa} )^2 \right] 
 \left[  1 + \tilde{\gamma}^2 + ( \frac{ {\vec{q}} }{\kappa} )^2  \right]}
 \label{eq:Ralpharespho}
 \; .
 \end{equation}

\subsection{Isotropic phonon dispersion}
\label{subsec:Iso}

Let us first evaluate Eq.\ref{eq:Ralpharespho} for
the isotropic phonon dispersion
$\Omega_{\vec{q}} = c_{\rm s} | {\vec{q}} |$.
Using Eq.\ref{eq:nurelation} we obtain 
 \begin{eqnarray}
 & & \hspace{-7mm}
  R^{\alpha}_{\rm ph}
    =   - \frac{ \tilde{\gamma}^2 }{2 k_{\rm F}^2 } 
   \frac{c_{\rm s}}{v_{\rm F}}  
   \int \frac{ \D {\vec{q}} }{  4 \pi }
  \frac{   | {\vec{q}} |  }{  
  \left[ \frac{\tilde{c}_{\rm s} ( {\vec{q}} ) }{v_{\rm F}}  | {\vec{q}} |
  + | \hat{\vec{v}}^{\alpha} \cdot {\vec{q}} | \right]^2 }
  \frac{1}{
 \left[ 1 + ( \frac{ {\vec{q}} }{\kappa} )^2 \right]^{ \frac{3}{2} } 
 \left[  1 + \tilde{\gamma}^2 + ( \frac{ {\vec{q}} }{\kappa} )^2  \right]^{ \frac{1}{2} } }
 \;  .
 \nonumber
 \\
 & &
 \label{eq:Ralpharespho3}
 \end{eqnarray}
Because according to Eq.\ref{eq:csren} the renormalized phonon velocity
$\tilde{c}_{\rm s} ( {\vec{q}} )$ depends only on $| {\vec{q}} |$,
the angular integration can now be done exactly. The relevant integral
is just the function $h_{3} ( x)$ given in Eq.\ref{eq:h3angav}.
After a simple rescaling we obtain
 \begin{eqnarray}
 & & \hspace{-8mm} R^{\alpha}_{\rm ph} =
 - \frac{\tilde{\gamma}^2 }{4} \left( \frac{ \kappa}{k_{\rm F}} \right)^2
 \int_{0}^{\infty} \D x
 \frac{ 1}{ \left[ 1 + \tilde{\gamma}^2 + x \right] 
 \left[ 1 + x  + \frac{c_{\rm s}}{v_{\rm F}}  ( 1 + x )^{\frac{1}{2} }
 ( 1 + \tilde{\gamma}^2 + x)^{\frac{1}{2}}  \right]}
 \; .
 \nonumber
 \\
 & &
 \label{eq:Rphonon4}
 \end{eqnarray}
Clearly, in the regime \ref{eq:plasmasmall}
we may ignore the term proportional to ${c_{\rm s} }/{ v_{\rm F}}$
in the denominator of Eq.\ref{eq:Rphonon4}. The integral is then elementary,
 \begin{equation}
 \int_{0}^{\infty} \D x
 \frac{ 1}{ \left[ 1 + \tilde{\gamma}^2 + x \right] \left[ 1 + x  \right] }
 = \frac{1}{\tilde{\gamma}^2} \ln ( 1 + \tilde{\gamma}^2 )
 \; \; \; ,
 \end{equation}
so that we finally obtain
 \begin{equation}
 R^{\alpha}_{\rm ph} =
 - \frac{1}{4} \left( \frac{ \kappa}{k_{\rm F}} \right)^2  \ln ( 1 +  \tilde{\gamma}^2 )
 \; \; \; .
 \label{eq:Rphonon5}
 \end{equation}
Note that the small parameter ${c_{\rm s} }/{ v_{\rm F}}$ 
has disappeared in the prefactor,
so that the final result depends only on the dimensionless strength
of the electron-phonon coupling $\tilde{\gamma}^2$.
Combining Eqs.\ref{eq:Rphonon5} and \ref{eq:Ralphaelphotot},
and using the fact that $ ( {\kappa }/{ k_{\rm F}} )^2 = {2 e^2}/ ({\pi v_{\rm F}})$
(see Appendix~\secref{subsubsec:Cb}), we obtain
 \begin{equation}
 R^{\alpha} = - \frac{ e^2}{ \pi v_{\rm F} } 
 \left[  \tilde{r}_{3}
 + \frac{1}{2}  \ln ( 1 + \tilde{\gamma}^2 )  \right]
 \; \; \; .
 \label{eq:Ralphacomb}
 \end{equation}
In the regime $\kappa \ll k_{\rm F}$ where our  bosonization approach
is most accurate, the prefactor ${e^2} / ({ \pi v_{\rm F}})$ in
Eq.\ref{eq:Ralphacomb} is a small number,
see Eq.\ref{eq:kappaTFrel}.
For weak electron-phonon coupling $\tilde{\gamma}^2$ we may expand
$\ln ( 1 + \tilde{\gamma}^2 ) \approx \tilde{\gamma}^2$.
Because $\tilde{r}_{3}$ is a number of the order of unity, 
the phonon contribution to the
quasi-particle residue is then negligible.
On the other hand, for large $\tilde{\gamma}^2$ the
phonon contribution is dominant.
Exponentiating Eq.\ref{eq:Ralphacomb}
we obtain for the quasi-particle residue\index{phonon!quasi-particle residue}
 \begin{equation}
 Z^{\alpha} = \left[ \frac{ \E^{ - \tilde{r}_{3} } }{ \sqrt{ 1 + \tilde{\gamma}^2 } } 
 \right]^{  \frac{e^2}{\pi v_{\rm F}}  } 
 \; \; \; .
 \label{eq:Zresphoiso}
 \end{equation}
If we take the high-density limit $v_{\rm F} \rightarrow \infty$ at fixed
$\tilde{\gamma}$, the quasi-particle residue approaches unity.
On the other hand, if we keep the density fixed but increase
the electron-phonon coupling $\tilde{\gamma}$,
we obtain
 \begin{equation}
 Z^{\alpha} = \left[ \frac{ \E^{ - \tilde{r}_{3} } }{ \tilde{\gamma} } 
 \right]^{ \frac{e^2}{ \pi v_{\rm F}} } 
 \; \; \; , \; \; \; 1 \ll \tilde{\gamma}^2 \ll \left( \frac{ v_{\rm F}}{c_{\rm s}} \right)^2
 \; \; \; .
 \label{eq:Zresphoisolarge}
 \end{equation}

\subsection{Quasi-one-dimensional electrons or phonons}
\label{subsec:Aniso}

It is straightforward to 
generalize our results for anisotropic systems.
For example, for strictly one-dimensional
electron dispersion the polarization in Eq.\ref{eq:Piphdef} is given by
 \begin{equation}
 \Pi_{0} (q ) = \nu \frac{ ( v_{\rm F} q_{x} )^2}{\omega_{m}^2 + ( v_{\rm F} q_{x} )^2}
 = \nu g_{1} ( \frac{ \I \omega_{m}}{ v_{\rm F} | q_{x} | } )
 \label{eq:Pianiso1}
 \; \; \; .
 \end{equation}
In this case it is not difficult to show that Eq.\ref{eq:Ralpharespho}
gives rise to Luttinger liquid behavior {\it{even if the
phonon dispersion is three-dimensional}}.

Alternatively, we may couple
one-dimensional phonons\index{phonon!anisotropic} to three-dimensional electrons.
Then  we should set 
 $\Omega_{\vec{q}} =  c_{\rm s} | q_{x} |$
in Eqs.\ref{eq:Omegaren} and \ref{eq:Ralpharespho}, 
while choosing for $\Pi_{0} ( q )$ the usual three-dimensional
polarization. 
Let us examine this possibility more closely.
From Eq.\ref{eq:Ralpharespho} we obtain in this case 
 \begin{eqnarray}
 & & \hspace{-8mm}
  R^{\alpha}_{\rm ph}
   =  - \frac{\tilde{\gamma}^2 }{2 k_{\rm F}^2 } 
   \frac{c_{\rm s}}{v_{\rm F}}  
   \int \frac{ d {\vec{q}} }{  4 \pi }
  \frac{    | q_{x} |  }{  
  \left[ \frac{c_{\rm s} ( {\vec{q}} ) }{v_{\rm F}}  | q_{x} |
  + | \hat{\vec{v}}^{\alpha} \cdot {\vec{q}} | \right]^2 
 \left[ 1 + ( \frac{ {\vec{q}} }{\kappa} )^2 \right]^{ \frac{3}{2} } 
 \left[  1 + \tilde{\gamma}^2 + ( \frac{ {\vec{q}} }{\kappa} )^2  \right]^{ \frac{1}{2} } }
 \; .
 \nonumber
 \\
 & &
 \label{eq:Ralpharespho1d}
 \end{eqnarray}
The crucial observation is now that for $\hat{\vec{v}}^{\alpha} = \pm {\vec{e}}_{x}$ 
we have $ | \hat{\vec{v}}^{\alpha} \cdot {\vec{q}} | = | q_{x} |$, so that the 
{\it{phase space for the $q_{x}$-integration is decoupled from the 
remaining phase space and the integral is logarithmically divergent}}\footnote{
We have encountered precisely the same 
situation before in our analysis of metallic chains without
interchain hopping, see Chap.~\secref{sec:chain3d}.}.
For  all other directions $\hat{\vec{v}}^{\alpha} \neq \pm {\vec{e}}_{x}$, the phase space
for the ${\vec{q}}$-integration is coupled, so that the logarithmic divergence
is cut off and the quasi-particle residue
is finite. Although for
$\hat{\vec{v}}^{\alpha} = \pm {\vec{e}}_{x}$
the integral in Eq.\ref{eq:Ralpharespho1d} is logarithmically divergent,
the total Debye-Waller factor $Q^{x} ( r_{x} {\vec{e}}_{x} , \tau )$ is
finite\footnote{We use the label $\alpha = x$ for the
patch with ${\vec{k}}^{\alpha} = k_{\rm F} {\vec{e}}_{x}$.}. 
Because the divergence in $R^{x}_{\rm ph}$ is logarithmic, we expect 
Luttinger\index{Luttinger liquid!anisotropic phonons}
liquid behavior. To obtain the anomalous dimension, it is sufficient to
calculate the leading logarithmic term in the large-distance expansion of
$Q^{x} ( r_{x} {\vec{e}}_{x} , 0 )$. 
Introducing the dimensionless integration variable
${\vec{p}} = {\vec{q}}  / \kappa $, we obtain from Eq.\ref{eq:Ralpharespho1d}
 \begin{eqnarray}
  Q^{x} ( r_{x} {\vec{e}}_{x} , 0 )
  & = &  - \frac{e^2}{2 \pi^2 v_{\rm F} } \frac{c_{\rm s}}{v_{\rm F}}
  \tilde{\gamma}^2  
  \int_{0}^{\infty} \D p_{x}
  \frac{ 1 - \cos ( p_{x} \kappa r_{x} )}{ p_{x} }
   \int_{- \infty}^{\infty}  \D p_{y} 
   \int_{- \infty}^{\infty}  \D p_{z} 
  \nonumber
  \\
  &   \times &
   \frac{1}{ \left[ 1 + \frac{c_{\rm s}}{v_{\rm F}}
   \sqrt{ 1 + \frac{\tilde{\gamma}^2 }{ 1 + {\vec{p}}^2 }} \right]^2
   \left[ 1 + {\vec{p}}^2 \right]^{\frac{3}{2}} 
   \left[ 1 + \tilde{\gamma}^2 + {\vec{p}}^2 \right]^{\frac{1}{2} }}
 \label{eq:RalphaLLph}
 \; \; \; .
 \end{eqnarray}
In the regime \ref{eq:plasmasmall}
where the phonon mode is well defined we may
again ignore the term  proportional to
${c_{\rm s}}/{v_{\rm F}}$ in the first factor of the
second line in Eq.\ref{eq:RalphaLLph}.
Furthermore, to extract the leading logarithmic term, we may  set $p_{x} = 0$
in the second line of Eq.\ref{eq:RalphaLLph}. 
The $p_{y}$- and $p_{z}$-integrations can then easily 
be performed in circular coordinates, so that we finally obtain
 \begin{equation}
  Q^{x} ( r_{x} {\vec{e}}_{x} , 0 ) \sim - \gamma_{\rm ph} \ln ( \kappa | r_{x} | )
  \; \; \; , \; \; \; \kappa |r_{x} | \rightarrow \infty
  \; \; \; ,
  \end{equation}
where the anomalous dimension\index{anomalous dimension!one-dimensional phonons} is
 \begin{equation}
 \gamma_{\rm ph} = \frac{e^2}{\pi v_{\rm F}} \frac{ c_{\rm s}}{v_{\rm F}}
 \left[ \sqrt{ 1 + \tilde{\gamma}^2 } - 1 \right]
 \label{eq:anomalphdef}
 \; \; \; .
 \end{equation}
Note that for weak electron-phonon coupling the anomalous dimension $\gamma_{\rm ph}$ is proportional
to $ \tilde{\gamma}^2$, while in the strong coupling limit it is
of order $\tilde{\gamma}$. However, one should keep in mind that Eq.\ref{eq:anomalphdef} 
has been derived for $\tilde{\gamma} \ll
{v_{\rm F}}/ c_{\rm s} $
(see Eq.\ref{eq:plasmasmall}), so that in the regime 
of validity of Eq.\ref{eq:anomalphdef} the anomalous dimension 
is always small compared with unity.

It is also interesting to calculate the
quasi-particle residue in the vicinity of the 
Luttinger liquid  points 
${\vec{k}}^{\alpha} = \pm k_{\rm F} {\vec{e}}_{x}$
on the Fermi surface.
A quantitative measure for the vicinity  to these points is the
parameter $\delta  = 1 - 
| \hat{\vec{v}}^{\alpha} \cdot {\vec{e}}_{x} |$.
Obviously $\delta = 0$ corresponds to the
Luttinger liquid points,
so that for small enough $\delta$ 
we should obtain a Fermi liquid with small quasi-particle residue.
A simple calculation shows that for
$0< \delta \ll { c_{\rm s}}/{ v_{\rm F} }$ the constant part
$R^{\alpha}_{\rm ph}$ of the Debye-Waller factor
is finite, and behaves as
 \begin{equation}
 R^{\alpha}_{\rm ph} \sim - \gamma_{\rm ph} 
 \left[ \ln \left( \frac{ c_{\rm s}}{v_{\rm F} \delta } \right)
 + c 
 + O ( \frac{ v_{\rm F} \delta }{c_{\rm s}} ) \right]
 \; \; \; , \; \; \; \delta \ll \frac{c_{\rm s}}{v_{\rm F}}
 \; \; \; ,
 \label{eq:Ralphaphsmall}
 \end{equation}
where $c = O (1)$ is a numerical constant.
Hence, for $\delta \rightarrow 0$ the quasi-particle residue vanishes as
 \begin{equation}
 Z^{\alpha}_{\rm ph} \propto \left[ \frac{ v_{\rm F} \delta }{c_{\rm s} } 
 \right]^{\gamma_{\rm ph}}
 \; \; \; , \; \; \; \delta \ll \frac{c_{\rm s}}{v_{\rm F}}
 \; \; \; .
 \label{eq:Zvanishpho}
 \end{equation}
Note that the exponent is given by the anomalous
dimension of the Luttinger liquid that would exist for $\delta = 0$.
Recall that the quasi-particle residue 
of weakly coupled chains 
discussed in Chap.~\secref{sec:Finiteinter}
shows a very similar behavior.
Obviously the parameter 
$\theta$ in
Eq.\ref{eq:QPres4patch}, which measures 
the closeness of the coupled chain system
to one-dimensionality,
corresponds 
to $v_{\rm F} \delta / c_{\rm s}$ in Eq.\ref{eq:Zvanishpho}.
Both parameters are a dimensionless measure
for the distance to the Luttinger liquid points in a suitably 
defined parameter space.
From Eq.\ref{eq:Ralphaphsmall} it is also clear that
in the present problem the 
vicinity to the Luttinger liquid points 
$\vec{k}^{\alpha} = \pm k_{\rm F} \vec{e}_x$
becomes
only apparent in the regime $\delta \ll {c_{\rm s} }/{v_{\rm F}}$.
For $ \delta \geqapprox {c_{\rm s}}/{v_{\rm F}}$
the correction term of order
${ v_{\rm F}} \delta /{c_{\rm s}}$ in Eq.\ref{eq:Ralphaphsmall} cannot be ignored. 
In the extreme case $\delta = 1$ the integration in Eq.\ref{eq:Ralpharespho1d}
gives rise to a factor
 \begin{equation}
 \int_{0}^{ \kappa} \D q_{y}
  \frac{    | q_{x} |  }{  
  \left[ \frac{c_{\rm s}   }{v_{\rm F}}  | q_{x} |
  +  q_{y}  \right]^2 } \propto \frac{v_{\rm F}}{ c_{\rm s} }
  \label{eq:qyint}
  \; \; \; ,
  \end{equation}
so that outside a small neighborhood of 
the points $\vec{k}^{\alpha} = \pm k_{\rm F} \vec{e}_x$
the prefactor of $R^{\alpha}_{\rm ph}$ has the same order of magnitude as
in the isotropic case, see Eq.\ref{eq:Rphonon5}.

\section{Summary and outlook}

In this chapter we have studied the Debye-model for
electron-phonon interactions  with the help of
our non-perturbative bosonization approach.
The Debye-model 
has been discussed and physically motivated in the classic
textbook by Fetter and Walecka \cite{Fetter71}. However,
these authors did not treat
the screening problem in a formally convincing manner
(although the physical content of their
``screening-by-hand''approach is correct).
In Sect.~\secref{sec:theeffective} we have
shown by means of functional integration that the screening of the
Coulomb interaction in the Debye-model can be derived 
in a very simply way from first principles.

Higher-dimensional bosonization predicts that
long-wavelength
isotropic LA phonons that couple to the electrons via long-range
Coulomb forces
can never destabilize the Fermi liquid state
in $d > 1$.
On the other hand, anisotropy in the phonon dispersion 
can lead to small quasi-particle residues 
at corresponding  patches of the Fermi surface, while the
shape of the Fermi surface remains spherical.
Of course, in realistic materials the phonon dispersion 
cannot be strictly one-dimensional on general 
grounds\footnote{Even at $T=0$ a one-dimensional harmonic crystal is not stable.
For example, the mean square displacement of a given site diverges logarithmically
with the size of the system. At $T > 0$ the divergence is even linear.
I would like to thank Roland Zeyher for pointing this out to me.},
but we know from
Chap.~\secref{sec:Finiteinter} that
the vicinity to the Luttinger 
liquid point in a suitably defined parameter 
space\index{Luttinger liquid!anisotropic phonons} is sufficient to lead to
characteristic Luttinger liquid features in the spectral
function of a Fermi liquid.
More generally,
our calculation suggests that
the coupling between electrons and {\it{any well defined 
quasi-one-dimensional collective mode can 
lead to Luttinger liquid behavior in three-dimensional
Fermi systems.}} 

Finally, let us again point out some open research problems.
So far we have explicitly evaluated the static
Debye-Waller factor $Q^{\alpha} ( \vec{r} , 0 )$ in the regime
$\tilde{\gamma} \ll v_{\rm F} / c_{\rm s}$
(see Eq.\ref{eq:plasmasmall}) where 
phonons and plasmons involve different energy scales.
Although we have convinced ourselves that 
the Fermi liquid remains stable in the
strong coupling regime $\tilde{\gamma} \geqapprox
v_{\rm F} / c_{\rm s}$ (where Migdal's theorem does not apply),
the calculation of the
{\bf{Debye-Waller factor for strong electron-phonon coupling}}
still remains to be done. 
Let us emphasize that our non-perturbative result
for the Green's function is also valid in this case,
but its explicit evaluation most likely
requires considerable numerical work.
An even more interesting (but also more difficult) problem is the
evaluation of our non-perturbative result 
for the Green's function
of electrons with non-linear energy dispersion
given in Eqs.\ref{eq:Galphashiftsym}--\ref{eq:Yres4}
for our coupled electron-phonon system.

Another direction for further research is
based on the expectation that, at
sufficiently low temperatures, the 
retarded interaction mediated by the phonons will
drive the Fermi system into a
{\bf{superconducting state}}.
As already mentioned in Chap.~\secref{sec:sumgreen},
with the help of a Hubbard-Stratonovich field
that couples to the relevant order parameter \cite{Schakel89}
it should not be too difficult to incorporate 
superconductivity into our functional bosonization formalism.
In this way our approach might offer  a non-perturbative way
to study superconducting symmetry breaking in correlated Fermi systems.

%
%

%
%
%

\chapter{Fermions in a stochastic medium}
\label{chap:adis}
\setcounter{equation}{0}

{\it{We use our background field method to calculate the
disorder\index{disorder} averaged single-particle Green's function
of fermions subject to a time-dependent random potential
with long-range spatial correlations. 
We show that
bosonization  provides a 
microscopic basis for the description of the quantum dynamics of an 
interacting many-body system via an effective stochastic model
with Gaussian probability distribution.
In the limit of static disorder
our method is equivalent with conventional perturbation
theory based on the lowest order Born approximation.
We also critically discuss the linearization  of the energy dispersion,
and give a simple example where this approximation leads to an unphysical result.
Some of the calculations described in this chapter have been published in
\cite{Kopietz95dis}.
}}

\vspace{7mm}

\noindent
The complicated quantum dynamics of a many-body system of interacting electrons
can sometimes by modeled by an effective non-interacting system that
is coupled to a dynamic random potential with a suitably 
defined probability distribution \cite{Girvin79}.
Although the precise form of the probability distribution
is in principle completely determined by the 
nature of the degrees of freedom that couple to the electrons 
(for example photons, phonons, or magnons),
one usually has to rely on perturbation theory to characterize
the random potential of the effective stochastic model. 
In this chapter we shall show that for random potentials with sufficiently long-range
spatial correlations our bosonization approach allows us to
relate the probability distribution of the effective stochastic model
in a very direct and essentially non-perturbative way to the underlying many-body system.

The dynamic random potential could also be due to some non-equilibrium
external forces. In this case the identification with an underlying many-body
system is meaningless. 
The motion of a single isolated electron in an externally given time-dependent
random potential has recently been discussed by many 
authors \cite{Ovchinnikov74,Madhukar77,Inaba81,Jaynnavar82,Henrichs84,Kardar87,Lebedev95}.
Here we would like to focus on the
problem of calculating the average Green's function of electrons
\index{Green's function!disorder average}
in the presence of a filled Fermi sea.
We shall show that our functional integral formulation of higher-dimensional
bosonization offers a new non-perturbative approach to this problem in arbitrary
dimensions.

Although within the conventional operator approach this connection between bosonization 
and random systems seems rather surprising,
it is obvious within our functional bosonization approach: 
In Chap.~\secref{chap:agreen} the calculation of the Green's function of the
interacting system has been mapped via a Hubbard-Stratonovich transformation
onto the problem of calculating the  average Green's function of an effective
non-interacting system in a dynamic random potential
$V^{\alpha} ( {\vec{r}} , \tau )$, 
see Eqs.\ref{eq:avphi3}, \ref{eq:Galphadifrt}, and \ref{eq:Galphadifrt2}.
As shown in Chap.~\secref{sec:Derivation}, 
for linearized energy dispersion and for sufficiently
long-range potentials $V^{\alpha} ( \vec{r} , \tau )$
it is possible to calculate the Green's function
${\cal{G}}^{\alpha} ( {\vec{r}} , {\vec{r}}^{\prime} , \tau , \tau^{\prime} )$ 
for a given realization of the random potential without
resorting to perturbation theory.
The translationally invariant  Green's function of the many-body system is
then obtained by averaging 
${\cal{G}}^{\alpha} ( {\vec{r}} , {\vec{r}}^{\prime} , \tau , \tau^{\prime} )$ 
over all realizations of the random potential $V^{\alpha} ( {\vec{r}} , \tau )$.
Of course, in the interacting many-body system the probability distribution
for this averaging is determined by the nature of the interaction and the kinetic energy
(see Eqs.\ref{eq:avphi}--\ref{eq:Skinphidef}), while 
in the stochastic model the probability distribution of the random potential
has to be specified externally.
However,  in our calculation of the Green's function
${\cal{G}}^{\alpha} ( {\vec{r}} , {\vec{r}}^{\prime} , \tau , \tau^{\prime} )$ 
for frozen random potential the
nature of the probability distribution is irrelevant, so that 
the method described in Chap.~\secref{chap:agreen} can be directly applied to
disordered systems.

\section{The average Green's function}
\label{sec:dis1}

{\it{
We introduce a  model of non-interacting fermions
subject to a general dynamic random potential
and derive a non-perturbative 
expression for the average Green's function by 
translating the results of Chap.~\secref{chap:agreen} into the
language of disordered systems.}}

\subsection{Non-interacting disordered fermions}

\noindent
The Green's function 
${\cal{G}} ( {\vec{r}} , {\vec{r}}^{\prime} , \tau , \tau^{\prime} )$ 
of non-interacting fermions moving under the influence of an 
imaginary time 
random potential\index{random potential}
$U ( {\vec{r}} , \tau )$ is defined via the usual equation
\index{Green's function!definition in dynamic random potential}
 \begin{equation}
 \hspace{-4mm}
 \left[ - \partial_{\tau} -  \frac{ ( - \I \nabla_{\vec{r}}  )^2}{2m} + \mu  - U ( {\vec{r}} , \tau )
 \right] 
 {\cal{G}} ( {\vec{r}} , {\vec{r}}^{\prime} , \tau , \tau^{\prime} ) 
 =
 \delta ( {\vec{r}} - {\vec{r}}^{\prime} ) \delta^{\ast} ( \tau - \tau^{\prime} )
 \label{eq:Grrttdef}
 \; .
 \end{equation}
We assume that the random potential  has a Gaussian probability 
distribution\index{probability distribution!random potential} with
zero average and general covariance function\index{covariance function}
 $C ( {\vec{r}} - {\vec{r}}^{\prime} , \tau - \tau^{\prime} )$, i.e.
 \begin{eqnarray}
 \overline{ U ( {\vec{r}} , \tau ) } & = & 0
 \; \; \; ,
 \label{eq:Uranav}
 \\
 \overline{ U ( {\vec{r}} , \tau )
 U ( {\vec{r}}^{\prime} , \tau^{\prime} ) }
 & = & C  ( {\vec{r}} - {\vec{r}}^{\prime} , \tau - \tau^{\prime} )
 \label{eq:gcordef}
 \; \; \; ,
 \end{eqnarray}
where the over-bar denotes averaging over the probability distribution
${\cal{P}} \{ U \}$
of the random potential $U$.
Explicitly, the probability distribution is given by
 \begin{equation}
 {\cal{P}} \left\{ {U} \right\}
 = \frac{ 
 \E^{ - \frac{1}{2 \beta V } \sum_{q} 
   C^{-1}_{q} 
 {U}_{-q} {U}_{q} }
  }
 { \int {\cal{D}} \left\{ U \right\}
 \E^{ - \frac{1}{2 \beta V } \sum_{q} 
 {C}^{-1}_{q} 
 {U}_{-q} {U}_{q}
 }}
 \label{eq:propdef}
 \; \; \; ,
 \end{equation}
where 
the Fourier components of the random potential
and the covariance function are
 \begin{equation}
 {U}_{q}  =   
 \int_{0}^{\beta} \D \tau  \int \D {\vec{r}}
 \E^{ - \I ( {\vec{q}} \cdot {\vec{r}} - \omega_{m} \tau )} U ( {\vec{r}} , \tau ) 
 \; \; \; ,
 \label{eq:Uqran}
 \end{equation}
 \begin{equation}
 C_{q}  =   
 \int_{0}^{\beta} \D \tau  \int \D {\vec{r}}
 \E^{ - \I ( {\vec{q}} \cdot {\vec{r}} - \omega_{m} \tau )} C ( {\vec{r}} , \tau ) 
 \; \; \; .
 \end{equation}
Hence,
 \begin{equation}
 \overline{ {U}_{q} {U}_{-q} }
 \equiv \int {\cal{D}} \left\{ U \right\} 
 { \cal{P}} \left\{  U \right\}
  {U}_{q} {U}_{-q}  = 
  \beta V {C}_{q} 
  \label{eq:Cqcov}
  \; \; \; .
  \end{equation}
All statistical properties of our model are contained in
the covariance function $C_{q} = C_{ {\vec{q}} , \I \omega_{m} }$.
If we would like to describe an underlying many-body system in thermal
equilibrium \cite{Girvin79}, then 
it is (at least in principle) possible to continue
the covariance function to real frequencies,
so that the average real time dynamics corresponding to
Eq.\ref{eq:Grrttdef} can be obtained by analytic continuation.
On the other hand, for an externally specified non-equilibrium 
potential $U ( {\vec{r}} , \tau )$ there is in general no simple
relation between real and imaginary time dynamics\footnote{
However, for some special cases the analytic continuation is
certainly possible. For example, in 
Sect.~\secref{subsec:Gaussianwhite} we shall discuss 
the Gaussian white noise\index{Gaussian white noise} limit, 
where $C_{q}$ is a frequency-independent constant,
so that the analytic continuation is trivial.}.

We are interested in the average Green's function
 \begin{equation}
 G ( {\vec{r}} - {\vec{r}}^{\prime} , \tau - \tau^{\prime} )
 = \overline{ 
 {\cal{G}} ( {\vec{r}} , {\vec{r}}^{\prime} , \tau , \tau^{\prime} ) 
 }
 \; \; \; .
 \label{eq:Ginterested}
 \end{equation}
For an exact calculation of the average Green's function 
one should first solve the differential equation \ref{eq:Grrttdef} 
for an arbitrary realization of the random potential, and then average the result 
with the probability distribution \ref{eq:propdef}.
Usually this an impossible task,  
so that one has to use some approximate method.
A widely used perturbative  approach, which works very well
for {\it{time-independent}} random potentials, is based on the impurity 
diagram technique \cite{Abrikosov63}.
In the metallic regime  it is often sufficient to calculate the self-energy in lowest order 
Born approximation.  For static disorder 
the average Green's function
is then found to vanish at 
distances large compared with the correlation range of the
covariance function\index{Green's function!disorder average} 
as \cite{Abrikosov63,Lee85,Altshuler85,Fukuyama85}
 \begin{equation}
 G ( {\vec{r}} - {\vec{r}}^{\prime} , \tau - \tau^{\prime} )
 = G_{0} ( {\vec{r}} - {\vec{r}}^{\prime} , \tau - \tau^{\prime} )
 \E^{ - \frac{ | {\vec{r}} - {\vec{r}}^{\prime} |}{2 \ell } }
 \; \; \; ,
 \label{eq:Gavstatdelta}
 \end{equation}
where $G_{0}$ is the Green's function of the clean system, and
the length $\ell$ is called the elastic mean 
free path\index{elastic mean free path}.
In Fourier space Eq.\ref{eq:Gavstatdelta} becomes
 \begin{equation}
 G( k ) = \frac{1}{ \I \tilde{\omega}_{n} - ( \frac{ {\vec{k}}^2 }{2m} - \mu ) 
 +  {\rm sgn} ( \tilde{\omega}_{n} ) \frac{\I }{2 \tau } }
 \; \; \; ,
 \label{GMatsubaradis}
 \end{equation}
where $ \tau = \ell / v_{\rm F}$ is the elastic lifetime\index{elastic lifetime}.
The extra factor of
$ \E^{ - \frac{ | {\vec{r}} - {\vec{r}}^{\prime} |}{2 \ell } }$ in
Eq.\ref{eq:Gavstatdelta} is nothing but  
the usual Debye-Waller factor that arises in the Gaussian averaging procedure.
Below we shall show that this factor can also be obtained
as a special case of the
Debye-Waller that is generated via bosonization.

\vspace{7mm}

Within our bosonization approach the average Green's function is
calculated in the most direct way: First we obtain the exact
Green's function 
for a given realization of the random potential, and then this expression is averaged.
As explained in detail in Chap.~\secref{sec:Derivation}, 
our approach is most accurate if there exists a cutoff 
$q_{\rm c} \ll k_{\rm F}$ such that 
for $|{\vec{q}}| \geqapprox q_{\rm c}$
the Fourier components $U_{q}$ of the random potential (and hence also
the Fourier components $C_q$ of
the covariance function) become negligibly small. 
In other words, 
we should restrict ourselves to random potentials
with sufficiently long-range spatial 
correlations\index{random potential!long-range correlated}.
Evidently the most popular model of static $\delta$-function correlated disorder 
does not fall into this category.
This would correspond to
$C_{q} =  \gamma_{0} \beta \delta_{\omega_{m} ,0 }$,
where the parameter $\gamma_{0}$ is related to the elastic lifetime $\tau$ via
$\gamma_{0} = ( 2 \pi \tau \nu )^{-1}$.
However, in view of the fact that 
a random potential with a finite correlation range $q_{\rm c}^{-1}$ is expected
to lead for distances $ | {\vec{r}} | \gg q_{\rm c}^{-1}$ to qualitatively
identical results for single-particle properties as a 
$\delta$-function correlated random potential,
the restriction to long-range correlations seems not to be very serious.

To model the disorder, we simply add the term
 \begin{equation}
 S_{\rm dis} \{ \psi , U \}
 = \beta \sum_{q} \sum_{\alpha} {U}_{-q} \rho^{\alpha}_{q}
 \end{equation}
to the action \ref{eq:Sdecoupdef}
in our Grassmannian functional integral \ref{eq:Gpatchdef}.
Here $\rho^{\alpha}_{q}$ is the sector density defined in Eq.\ref{eq:rhoalpha}.
The average Green's function can now be calculated 
by repeating the steps described in Chap.~\secref{sec:Derivation}.
For simplicity, in this chapter we shall work with
linearized energy dispersion. In Sect.~\secref{sec:opendis} we shall further comment
on the accuracy of this approximation in the present context.
Thus,
after subdividing the Fermi surface into patches as described
in Chap.~\secref{subsec:patch}, we linearize the
energy dispersion locally and thus replace Eq.\ref{eq:Grrttdef} by  
a {\it{linear}} partial differential equation for
the sector Green's function
 ${\cal{G}}^{\alpha} ( {\vec{r}} , {\vec{r}}^{\prime} , \tau , \tau^{\prime} ) $ 
 (see Eq.\ref{eq:Galphadifrt})
 \begin{equation}
 \left[ - \partial_{\tau} + \I {\vec{v}}^{\alpha} 
 \cdot \nabla_{\vec{r}}    - U ( {\vec{r}} , \tau )
 \right] 
 {\cal{G}}^{\alpha} ( {\vec{r}} , {\vec{r}}^{\prime} , \tau , \tau^{\prime} ) 
 =
 \delta ( {\vec{r}} - {\vec{r}}^{\prime} ) \delta^{\ast} ( \tau - \tau^{\prime} )
 \label{eq:Grrttpatch}
 \; \; \; .
 \end{equation}
As shown in Chap.~\secref{subsec:invdiag},
the exact solution of this linear differential equation 
is given by Schwinger's 
ansatz \cite{Schwinger62}\index{Schwinger ansatz}, 
and can be written as
(see Eqs.\ref{eq:Ansatz}, \ref{eq:G0res} and \ref{eq:Phires})
 \begin{eqnarray}
 {\cal{G}}^{\alpha} ( {\vec{r}} , {\vec{r}}^{\prime} , \tau , \tau^{\prime} ) 
 & = &
 G_{0}^{\alpha} ( {\vec{r}} - {\vec{r}}^{\prime} , \tau - \tau^{\prime} )
 \nonumber
 \\
 & \times &
 \exp \left[ \frac{1}{\beta V}  \sum_{q} U_{q} \frac{ \E^{\I 
 ( {\vec{q}} \cdot {\vec{r}} - \omega_{m} \tau )}
 -  \E^{ \I ( {\vec{q}} \cdot {\vec{r}}^{\prime} - \omega_{m} \tau^{\prime} )}}
 { \I \omega_{m} - {\vec{v}}^{\alpha} \cdot {\vec{q}} } \right]
 \label{eq:Grealizationdis}
 \; \; \; .
 \end{eqnarray}
The Gaussian average of this expression is now trivial and yields the
usual Debye-Waller factor,\index{Debye-Waller factor!disorder}
 \begin{eqnarray}
 \overline{ {\cal{G}}^{\alpha} ( {\vec{r}} , {\vec{r}}^{\prime} , \tau , \tau^{\prime} ) }
& \equiv  &
 G^{\alpha} ( {\vec{r}} - {\vec{r}}^{\prime} , \tau - \tau^{\prime} )
 \nonumber
 \\
 & = &
 G_{0}^{\alpha} ( {\vec{r}} - {\vec{r}}^{\prime} , \tau - \tau^{\prime})
  e^{  Q_{\rm dis}^{\alpha} ( {\vec{r}} - {\vec{r}}^{\prime} , \tau - \tau^{\prime} ) }
  \; \; \; ,
  \label{eq:Gavdis}
  \end{eqnarray}
with
 \begin{eqnarray}
   Q_{\rm dis}^{\alpha} ( {\vec{r}} - {\vec{r}}^{\prime} , \tau - \tau^{\prime} ) & = &
  - \frac{1}{2 ( \beta V )^2 } \sum_{q} 
 \overline{ U_{q} U_{-q}} 
 \frac{ \left| \E^{\I ( {\vec{q}} \cdot {\vec{r}} - \omega_{m} \tau )}
 -  \E^{ \I ( {\vec{q}} \cdot {\vec{r}}^{\prime} - \omega_{m} \tau^{\prime} )} \right|^2 }
 { ( \I \omega_{m} - {\vec{v}}^{\alpha} \cdot {\vec{q}})^2  }
 \nonumber
 \\
 &  &  \hspace{-15mm} =
    -   \frac{1}{\beta V } \sum_{q} {C}_{q}
   \frac{ 1 - \cos [ {\vec{q}} \cdot ( {\vec{r}} - {\vec{r}}^{\prime} ) - 
   \omega_{m} ( \tau - \tau^{\prime} ) ] }
   { ( \I \omega_{m} - {\vec{v}}^{\alpha} \cdot {\vec{q}} )^2 }
  \;  .
  \label{eq:QdisDW}
  \end{eqnarray}
The average Matsubara Green's function 
can then be written as (see Eqs.\ref{eq:Galphartdef}--\ref{eq:Galphaqtildedef})
\index{Green's function!bosonization result for disorder}
 \begin{equation}
 \hspace{-4mm}
 G (k)  =  \sum_{\alpha} \Theta^{\alpha} ( {\vec{k}} )
 \int \D {\vec{r}} \int_{0}^{\beta} \D \tau 
 \E^{ - \I [ ( {\vec{k}} - {\vec{k}}^{\alpha} ) \cdot  {\vec{r}}
 - \tilde{\omega}_{n}  \tau  ] }
 {{G}}^{\alpha}_{0} ( {\vec{r}}  , \tau  )
 \E^{ Q^{\alpha}_{\rm dis} ( {\vec{r}} , \tau ) }
 \; .
 \label{eq:Gkresdis}
 \end{equation}
This completes the solution of the non-interacting problem.

\subsection{Interacting disordered fermions}
\label{subsec:Intdis}

\begin{figure}
\sidecaption
\psfig{figure=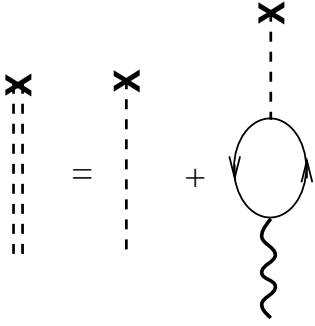,width=4cm}
\caption[Screening of the impurity potential]
{
Screening of the impurity potential.
The bare impurity potential is denoted by a
dashed line with a cross, and the thick wavy line represents
the RPA interaction (see Fig.~\secref{fig:bubblecor} (d)).
The effective screened disorder potential $U_q^{\rm RPA}$ is denoted by a double dashed line with cross.
In \cite{Kopietzper} we have discussed these diagrams
in a different context.
}
\label{fig:screenran}
\end{figure}

\noindent
Disorder and interactions are treated on equal footing
in our bosonization approach,
so that it is easy to include electron-electron interactions
into the above calculation.
Eq.\ref{eq:Grrttpatch}
should then be replaced by
 \begin{eqnarray}
 \left[ - \partial_{\tau} + \I {\vec{v}}^{\alpha} \cdot \nabla_{\vec{r}}    
 - U ( {\vec{r}} , \tau )
 - V^{\alpha} ( {\vec{r}} , \tau )
 \right] 
 {\cal{G}}^{\alpha} ( {\vec{r}} , {\vec{r}}^{\prime} , \tau , \tau^{\prime} ) 
  & = &
  \nonumber
  \\
  & & \hspace{-20mm}
 \delta ( {\vec{r}} - {\vec{r}}^{\prime} ) \delta^{\ast} ( \tau - \tau^{\prime} )
 \; \; \; ,
 \label{eq:Grrttpatch2}
 \end{eqnarray}
where $V^{\alpha} ( {\vec{r}} , \tau )$ is the
same Hubbard-Stratonovich field as in Eq.\ref{eq:Galphadifrt}.
The solution of this equation is again of the
form \ref{eq:Grealizationdis}, with $U_q$ replaced by
$U_q + (\beta V)  V^{\alpha}_q$, where
$V^{\alpha}_q$ are the Fourier components\footnote{
The additional factor of $\beta V$ is due to the different
normalizations of the Fourier transformations,
compare Eqs.\ref{eq:Vrt} and \ref{eq:Uqran}.}
of $V^{\alpha} ( {\vec{r}} , \tau )$.
Given the exact solution
of Eq.\ref{eq:Grrttpatch2}, we obtain the translationally invariant
average Green's function 
of the interacting many-body system by averaging over the disorder and
over the Hubbard-Stratonovich field. Explicitly,
 \begin{eqnarray}
 G^{\alpha} ( {\vec{r}} - {\vec{r}}^{\prime} , \tau - \tau^{\prime} )
 & =  &
 \nonumber
 \\
 & &
 \hspace{-20mm}
 \int {\cal{D}} \left\{ U \right\} 
 { \cal{P}} \left\{  U \right\}
 \int {\cal{D}} \left\{ \phi^{\alpha} \right\} 
 { \cal{P}} \left\{   \phi^{\alpha} , U  \right\}
 {\cal{G}}^{\alpha} ( {\vec{r}} , {\vec{r}}^{\prime} , \tau , \tau^{\prime} ) 
 \label{eq:avUphi}
 \; \; \; ,
 \end{eqnarray}
with ${ \cal{P}} \{  U \}$ given in Eq.\ref{eq:propdef}. 
 The probability distribution
$ { \cal{P}} \{   \phi^{\alpha} , U \}$ has exactly the same form
as in Eqs.\ref{eq:probabphidef}--\ref{eq:Skinphidef}, 
the only modification being that
the elements of the infinite matrix $\hat{V}$ 
in Eq.\ref{eq:Skinphidef}
are now given by
 \begin{equation}
 [ \hat{V} ]_{ k k^{\prime} }   =
 \frac{\I}{\beta}
 \sum_{\alpha} 
   \Theta^{\alpha} ( {\vec{k}} ) \left[ \phi^{\alpha}_{k- k^{\prime}}
   - \frac{\I}{V} U_{k - k^{\prime}} \right]
   \label{eq:hatVphiran}
 \; \; \; . 
 \end{equation}
Recall that according to Eq.\ref{eq:hatVphi}
the Fourier components $V^{\alpha}_q$ of the
potential $V^{\alpha} ( \vec{r} , \tau )$
in Eq.\ref{eq:Grrttpatch2} are related to 
the Fourier components $\phi^{\alpha}_q$ of our
Hubbard-Stratonovich field via
  $V^{\alpha}_{q} =
      \frac{\I}{\beta} 
  {\phi}^{\alpha}_{q}$.
For long-range random-potentials the closed loop theorem
guarantees that the Gaussian approximation is very accurate, so that
we may approximate
 \begin{eqnarray}
 {\cal{P}} \{ \phi^{\alpha} , U \} 
 & \approx &
 \nonumber
 \\
 & & \hspace{-10mm}
 \frac{  
 \exp \left[ - {S}_{{\rm eff},2 } \{ \phi^{\alpha} \} 
 - \frac{\I}{V} \sum_{q \alpha \alpha^{\prime}}
 [ \underline{\tilde{\Pi}}_0 (q) ]^{\alpha \alpha^{\prime} }
 \phi^{\alpha}_{-q} U_q
 \right]  
 }
 {
 \int {\cal{D}} \left\{ \phi^{\alpha} \right\} 
 \exp \left[  - {S}_{{\rm eff},2 } \{ \phi^{\alpha} \} 
 - \frac{\I }{V} \sum_{q \alpha \alpha^{\prime}}
 [ \underline{\tilde{\Pi}}_0 (q) ]^{\alpha \alpha^{\prime} }
 \phi^{\alpha}_{-q} U_q
 \right]  }
 \; \; \; ,
 \label{eq:probabphiUdef}
 \end{eqnarray}
where the Gaussian action
 ${S}_{{\rm eff},2 } \{ \phi^{\alpha} \} $ is given in Eq.\ref{eq:Seff2phigaussres}, and
the matrix elements 
$[ \underline{\tilde{\Pi}}_0 (q) ]^{\alpha \alpha^{\prime} }$
of the rescaled sector polarization are defined in Eq.\ref{eq:U2res}.
Note that by construction 
 \begin{equation}
 \int {\cal{D}} \left\{ \phi^{\alpha} \right\} 
 {\cal{P}} \{ \phi^{\alpha} , U \} 
 = 1
 \label{eq:probabphiUnorm}
 \; \; \; ,
 \end{equation}
i.e.  for any given realization
of the random potential $U$ the distribution
$ {\cal{P}} \{ \phi^{\alpha} , U \} $ is properly normalized.
Because the random potential 
$U$ in Eq.\ref{eq:probabphiUdef} appears also in the denominator,
it seems at the first sight that the averaging over 
${ \cal{P}} \{  U \}$
in Eq.\ref{eq:avUphi} cannot be directly performed, so that 
one has to use the replica approach \cite{Kleinert96}\index{replica approach}. 
Fortunately this is not the case, 
because we have the freedom of integrating
first over the $\phi^{\alpha}$-field before
averaging over the disorder.
Then it is easy to see that
the $U$-dependence of the denominator in Eq.\ref{eq:probabphiUdef}
is exactly cancelled 
by a corresponding factor in the numerator, so that
the averaging can be carried out exactly,
{\it{without resorting to the replica approach}}.
Thus, after performing the trivial Gaussian integrations we obtain
for the average sector Green's function of the interacting many-body system
 \begin{equation}
 G^{\alpha} ( {\vec{r}}  , \tau )
 = 
 G^{\alpha}_0 ( {\vec{r}}  , \tau )
 \exp \left[ 
  Q^{\alpha}( {\vec{r}} , \tau ) 
 +
 \tilde{Q}^{\alpha}_{\rm dis} ( {\vec{r}} , \tau ) \right]
 \label{eq:Gavranintfinal}
 \; \; \; ,
 \end{equation}
where 
\index{Debye-Waller factor!disorder and interactions}
the Debye-Waller factor
$Q^{\alpha} ( {\vec{r}} , \tau ) $ due
to the interactions is given in Eqs.\ref{eq:Qlondef}--\ref{eq:Slondef},
and the modified Debye-Waller factor
 $\tilde{Q}^{\alpha}_{\rm dis} ( {\vec{r}} , \tau )$ due to disorder is
obtained from 
${Q}^{\alpha}_{\rm dis} ( {\vec{r}} , \tau )$ 
in Eq.\ref{eq:QdisDW} by replacing
the bare covariance function $C_q$ by the
{\it{screened}} covariance function
 \begin{equation}
 {C}_q^{\rm RPA} = \frac{C_q}{ \left[ 1 + \Pi_0 ( q ) f_q \right]^2}
 \label{eq:covscrdef}
 \; \; \; .
 \end{equation}
Diagrammatically this expression
describes the screening\index{screening!of impurity potential} 
of the impurity potential
by the electron-electron interaction. The corresponding
Feynman diagrams are shown in Fig.~\secref{fig:screenran}.
Note that in Fourier space 
the screening correction in 
Fig.~\secref{fig:screenran} 
is $- U_q \Pi_0 (q) f^{\rm RPA}_q $, 
which should be added to the bare disorder potential $U_q$.  Hence, 
the total screened disorder potential has the Fourier components
 \begin{equation}
 U_q^{\rm RPA} = U_q - U_q 
 \frac{\Pi_0 (q) f_q }{ 1  + \Pi_0 ( q ) f_q } = \frac{U_q}{ 1 +
 \Pi_0 (q) f_q }
 \; \; \; .
 \label{eq:UimpRPAscreen}
 \end{equation}
In $d=1$ a result similar to Eq.\ref{eq:Gavranintfinal}
has also been obtained by Kleinert \cite{Kleinert96}, and by Hu and Das Sarma \cite{Hu93}. 
However, Kleinert has obtained his result by combining functional bosonization \cite{Lee88}
with the replica approach to treat the disorder averaging. 
As shown in this section,
there is no need for introducing replicas
if one integrates over the Hubbard-Stratonovich field {\it{before}}
averaging over the disorder.
In the expression derived by Hu and Das Sarma \cite{Hu93} 
the screening of the random potential is not explicitly taken into account.

\section{Static disorder\index{disorder!static}}
\label{sec:dis2}

{\it{We show that for static random potentials
with sufficiently long-range correlations Eq.\ref{eq:Gkresdis}
agrees precisely with the usual perturbative result.}}

\vspace{7mm}

\noindent
According to Eq.\ref{eq:Uqran} 
the Fourier coefficients $U_q$ of
a time-independent random potential\index{random potential!static} 
$U ( {\vec{r}} )$  are
 \begin{equation}
 U_{q} =  \beta \delta_{\omega_{m} , 0 }
 U_{\vec{q}} 
 \; \; \; , \; \; \; 
 U_{\vec{q}}  =
 \int \D {\vec{r}} \E^{- \I {\vec{q}} \cdot {\vec{r}} } U ( {\vec{r}} )
 \; \; \; .
 \end{equation}
For simplicity let us assume that the Fourier transform
of the static correlator has a simple separable 
form\footnote{
Any other cutoff function
(for example $\E^{ - {\vec{q}}^2 / q_{\rm c}^2}$)
yields qualitatively identical results.
Our choice leads to particularly simple integrals.},
 \begin{equation}
 C_{q} = \frac{\beta}{V} 
 \overline{ U_{\vec{q}} U_{- {\vec{q}} } }
 =
 \beta \delta_{\omega_{m} , 0 }  \gamma_{\vec{q}} 
 \; \; \; , \; \; \; \gamma_{\vec{q}} = \gamma_{0}
 \E^{- | \vec{q} |_{1}  / q_{\rm c} } 
 \; \; \; ,
 \label{eq:Cqformstat}
 \end{equation}
where $| {\vec{q}} |_{1} = \sum_{i=1}^{d} | q_{i} |$.
As discussed in Chap.~\secref{subsec:Greal},
for linearized energy dispersion
we may set
$ {\vec{r}} = r^{\alpha}_{\|} \hat{\vec{v}}^{\alpha} $ 
in the argument of the Debye-Waller factor, because the
function $G_{0}^{\alpha} ( {\vec{r}} , \tau )$
is proportional to $\delta^{(d-1)} ( {\vec{r}}_{\bot}^{\alpha} )$, see Eq.\ref{eq:Gpatchreal1}.
Then we obtain from Eq.\ref{eq:QdisDW} for $ V \rightarrow \infty$
 \begin{equation}
 Q_{\rm dis}^{\alpha} ( r^{\alpha}_{\|} \hat{\vec{v}}^{\alpha} , \tau )
 = -  \frac{\gamma_{0}}{  | {\vec{v}}^{\alpha} |^2} 
 \int \frac{\D {\vec{q}} }{ ( 2 \pi )^d}
 \E^{ - |{\vec{q}}|_1 / {q_{\rm c}} }
   \frac{ 1 - \cos ( {\hat{\vec{v}}^{\alpha}} \cdot {\vec{q}}  r^{\alpha}_{\|}  ) }
   { ( \hat{\vec{v}}^{\alpha} \cdot {\vec{q}} )^2 }
  \; \; \; .
  \label{eq:Qdisstat1}
  \end{equation}
Note that for a spherical Fermi surface $| {\vec{v}}^{\alpha} | = v_{\rm F}$ is independent
of the patch index, but in general
${\vec{v}}^{\alpha}$ depends on $\alpha$.
The Debye-Waller factor is independent of $\tau$ because we have
assumed a static random potential. For $|r^{\alpha}_{\|} q_{\rm c} | \gg 1$
the integral in Eq.\ref{eq:Qdisstat1} is easily done and yields
\index{Debye-Waller factor!static disorder}
 \begin{equation}
 Q_{\rm dis}^{\alpha} ( r^{\alpha}_{\|} \hat{\vec{v}}^{\alpha} , \tau )
 \sim - \frac{ | r^{\alpha}_{\|} | }{ 2 \ell^{\alpha} }
 \; \; \; , \; \; \; |r^{\alpha}_{\|} q_{\rm c} | \gg 1
 \; \; \; ,
 \label{eq:Qdisstatres}
 \end{equation}
where the inverse elastic mean free path\index{elastic mean free path} 
$\ell^{\alpha}$ is given by
 \begin{equation}
 \frac{1}{\ell^{\alpha}} = 
 \left( \frac{ q_{\rm c}}{  {\pi}} \right)^{d-1}
 \frac{\gamma_{0}}{ | {\vec{v}}^{\alpha} |^2 }
 \; \; \; .
 \label{eq:ellalphadef}
 \end{equation}
We conclude that at large distances
 \begin{equation}
 G^{\alpha} ( {\vec{r}} , \tau ) = 
 G_{0}^{\alpha} ( {\vec{r}} , \tau )  \exp \left[ { - \frac{ |  
 \hat{\vec{v}}^{\alpha}  \cdot {\vec{r}} | }{2 \ell^{\alpha} }} \right]
 \; \; \; .
 \label{eq:Gpatchavresstat}
 \end{equation}
The complete averaged real space Green's function is then according
to Eq.\ref{eq:Grealtotal} given by
 \begin{equation}
 G ( {\vec{r}} , \tau ) =  \sum_{\alpha}   
 \E^{  \I  {\vec{k}}^{\alpha}  \cdot {\vec{r}} }
 G_{0}^{\alpha} ( {\vec{r}} , \tau )  \exp \left[ { - \frac{ |  
 \hat{\vec{v}}^{\alpha} \cdot {\vec{r}} | }{2 \ell^{\alpha} }} \right]
 \; \; \; .
 \label{eq:Grealtotal2}
 \end{equation}
From Eq.\ref{eq:Qdisstatres} it is evident that in $d=1$ any finite
static disorder 
destroys the Luttinger liquid features in the 
momentum distribution \cite{Hu93}.
Recall that regular interactions in one-dimensional Fermi systems
give rise to a contribution to the Debye-Waller factor 
that grows only logarithmically as 
$r^{\alpha}_{\|} \rightarrow \infty$, see Eq.\ref{eq:QlargeequalLL}.
At sufficiently large distances this logarithmic divergence is completely
negligible compared with the linear divergence due to disorder in
Eq.\ref{eq:Qdisstatres}.
Note also that the linear growth of the 
Debye-Waller factor in Eq.\ref{eq:Qdisstatres}
is independent of the dimensionality 
of the system, and implies  that the
momentum distribution\index{momentum distribution!with disorder}
$n_{ {\vec{k}}^{\alpha} + {\vec{q}}}$
is for small ${\vec{q}}$ an analytic function of ${\vec{q}}$.
Thus, any finite disorder washes out the singularities 
in the momentum distribution.

\begin{figure}
\sidecaption
\psfig{figure=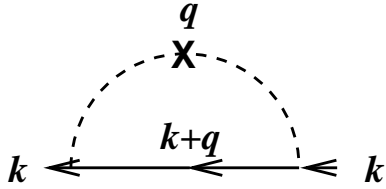,width=5cm}
\caption[Lowest order Born approximation for the elastic lifetime]
{
Lowest order Born approximation for the average self-energy
of non-interacting fermions in a static random potential.
The dashed line with the cross denotes the average $\overline{U_{\vec{q}}  U_{ - {\vec{q}}} }$.
}
\label{fig:Born}
\end{figure}

For a comparison with the usual perturbative result, 
let us also
calculate the Fourier transform of Eq.\ref{eq:Gpatchavresstat}.
Shifting the coordinate origin 
to point ${\vec{k}}^{\alpha}$ on the Fermi surface
by setting ${\vec{k}} = {\vec{k}}^{\alpha} + {\vec{q}}$, 
and choosing $| {\vec{q}}  | \ll q_{\rm c}$,
it is easy to show that Eq.\ref{eq:Gpatchavresstat} implies for
the averaged Matsubara Green's function
 \begin{equation}
 G ( {\vec{k}}^{\alpha} + {\vec{q}} , \I \tilde{\omega}_{n} )
 = 
 G^{\alpha} ( {\vec{q}} , \I \tilde{\omega}_{n} ) 
 = \frac{1}{ \I \tilde{\omega}_{n} -  {\vec{v}}^{\alpha} \cdot {\vec{q}}
 +  {\rm sgn} ( \tilde{\omega}_{n} ) \frac{\I}{2 \tau^{\alpha} } }
 \; \; \; ,
 \label{eq:GMatsubaradisbos}
 \end{equation}
where the inverse elastic lifetime\index{elastic lifetime} 
associated with sector $\alpha$ is given by
 \begin{equation}
 \frac{1}{\tau^{\alpha}} = 
 \frac{ | {\vec{v}}^{\alpha} | }{ \ell^{\alpha}} =
 \left( \frac{ q_{\rm c}}{  {\pi}} \right)^{d-1}
 \frac{\gamma_{0}}{ | {\vec{v}}^{\alpha} |}
 \; \; \; .
 \label{eq:elmeanres}
 \end{equation}
Eqs.\ref{eq:GMatsubaradisbos} and \ref{eq:elmeanres}
agree with the usual perturbative result of the lowest order 
Born approximation\index{Born approximation}
for the average self-energy.
The relevant diagram is shown in Fig.~\secref{fig:Born}, and yields
for the imaginary part of the self-energy
 \begin{eqnarray}
 {\rm Im} \Sigma (  {\vec{k}} )  & = & \frac{1}{V^2}  \sum_{\vec{q}} \overline{ U_{\vec{q}} U_{ - \vec{q}} }
 {\rm Im} G ( {\vec{k}} + {\vec{q}} , - \I 0^{+} )
 \nonumber
 \\
 & = &   \frac{\gamma_{0}}{V} \sum_{\vec{q}} 
 \E^{ -  | {\vec{q}}|_{1} / {q_{\rm c}} }
 {\rm Im} G ( {\vec{k}} + {\vec{q}} , - \I 0^{+} )
 \; \; \; .
 \label{eq:Imsigmadis}
 \end{eqnarray}
Because the random potential is static, the self-energy does not depend
on the frequency.
Note that for $q_{\rm c} = \infty$, corresponding to a random potential
with $\delta$-function correlation in real space,  
we may shift ${\vec{q}} + {\vec{k}} \rightarrow {\vec{q}}$ in
Eq.\ref{eq:Imsigmadis}, so that the self-energy is independent of ${\vec{k}}$. 
Then Eq.\ref{eq:Imsigmadis} reduces to the usual
result $\frac{1}{2 \tau } = {\rm Im}  \Sigma 
= \pi \gamma_{0} \nu $.
As already mentioned, the approximations leading to
Eq.\ref{eq:Grealtotal2} are not accurate in this case,
because the correlator involves also large momentum transfers.
On the other hand, for $q_{\rm c} \ll k_{\rm F}$ only wave-vectors  
$| {\vec{q}}| \ll k_{\rm F}$ contribute
in Eq.\ref{eq:Imsigmadis}, so that we may linearize 
\index{linearization of energy dispersion}
the energy dispersion.
Then we obtain
 \begin{eqnarray}
 \frac{1}{ \tau^{\alpha}} & \equiv &
 2 {\rm Im } \Sigma ( {\vec{k}}^{\alpha} ) 
 = 2 \pi \gamma_{0} \int \frac{ \D {\vec{q}} }{ ( 2 \pi )^d}
 \E^{ -  |{\vec{q}}|_1 / {q_{\rm c}} }
 \delta ( {\vec{v}}^{\alpha} \cdot {\vec{q}} )
 \nonumber
 \\
 & = & \frac{\gamma_{0}}{ ( 2 \pi )^{d-1}} ( 2 q_{\rm c} )^{d-1} \frac{1}{ | {\vec{v}}^{\alpha} | }
 \label{eq:dampdisres}
 \; \; \; ,
 \end{eqnarray}
which agrees precisely with Eq.\ref{eq:elmeanres}.
We conclude that for static disorder with long-range correlations
our bosonization approach reproduces the lowest order Born approximation
for the elastic lifetime.

\section{Dynamic disorder\index{random potential!dynamic}\index{disorder!dynamic}}
\label{sec:dis3}

{\it{We first derive a
strikingly simple relation between interacting
Fermi systems and effective stochastic models with
time-dependent disorder. We then explicitly
evaluate the average Green's function in some simple cases.}}

\vspace{7mm}

\noindent
The case of a static random potential is  not very exciting, 
because we have simply reproduced the perturbative result. 
New interesting physics emerges if we consider a general dynamic random 
potential. To calculate the average Green's function, we should
specify the dynamic covariance function $C_{q}$ in
Eq.\ref{eq:Cqcov} and then evaluate the Debye-Waller factor \ref{eq:QdisDW}. 
If we would like to describe with our stochastic model an underlying interacting
many-body system in thermal equilibrium, then the form of $C_{q}$ is
determined by the nature of the interaction. 
In the case of the coupled electron-phonon system at high
temperatures an explicit microscopic calculation
of $C_{q}$ has been given by Girvin and Mahan \cite{Girvin79},
who found that
the disorder can be modeled by a 
white noise dynamic random potential, corresponding to a 
frequency-independent $C_{q}$.\index{stochastic model for interactions}
The identification of $C_{q}$ with
the parameters of the underlying many-body system given in
\cite{Girvin79} is based on a perturbative calculation
of the self-energy at high temperatures.

In contrast,
our functional bosonization approach allows us 
to relate the covariance function $C_{q}$ of the 
random system at low temperatures in a direct and essentially
non-perturbative way to the underlying many-body system.
Evidently, the requirement that the average Green's function
of the random system  should be identical with the
Green's function of the interacting many-body system
without disorder is equivalent with
the postulate that the corresponding Debye-Waller factors should be identical.
Comparing then $Q^{\alpha}_{\rm dis} ( {\vec{r}} , \tau )$
in Eq.\ref{eq:QdisDW} with the Debye-Waller factor
$Q^{\alpha} ( {\vec{r}} , \tau )$ due to a general density-density interaction given
in Eqs.\ref{eq:Qlondef}--\ref{eq:Slondef}, we conclude that
we should identify
 \begin{equation}
 C_{q} = -  f_{q}^{\rm RPA}
 =
 -  f_{{q}} + f_{{q}}^2 
  \int_{0}^{\infty} \D \omega S_{\rm RPA} ( {\vec{q}} , \omega ) 
  \frac{  2 \omega}
  { \omega^{2} + \omega_{m}^2 }
  \label{eq:CrpaSRPA}
  \; \; \; ,
  \end{equation}
where $f_{q}$ is the bare
interaction of the underlying many-body system, and
we have used Eq.\ref{eq:FrpaSRPA} to express 
$f^{\rm RPA}_q$ in terms of the dynamic structure factor.
Eq.\ref{eq:CrpaSRPA} is the link between the phenomenological
stochastic model and the microscopic many-body system. 
In spite of its apparent
simplicity, Eq.\ref{eq:CrpaSRPA} is a highly non-trivial result, because it 
is based on a non-perturbative resummation
of the entire perturbation series of the many-body problem.

\subsection{Gaussian white noise\index{Gaussian white noise}}
\label{subsec:Gaussianwhite}

Even if the  random potential
is determined by 
some non-equilibrium external forces, 
it is useful to decompose the 
covariance function $C_{q}$ as in Eq.\ref{eq:CrpaSRPA}, because then
we can simply use the results of Chap.~\secref{sec:exactman} to
evaluate the Debye-Waller factor.
Let us first consider the case of  Gaussian
white noise random potential
with covariance given by\footnote{
Note that the constant $C_{0}$ has units of volume $\times$ energy,
just like the usual Landau interaction parameters.}
 \begin{equation}
 C_{q} = 
  C_{0} \E^{ - | {\vec{q}} |_{1}/ {q_{\rm c}} }
 \; \; \; .
 \label{eq:Cqwhite}
 \end{equation}
Because a white noise random potential involves fluctuations on all energy
scales with equal weight, the covariance function
$C_{q}$ is independent of the frequency.
Comparing Eq.\ref{eq:Cqwhite} with Eq.\ref{eq:CrpaSRPA}, 
it is clear that the corresponding Debye-Waller factor can be simply
obtained from Eqs.\ref{eq:R2}, \ref{eq:ReS2b}, and \ref{eq:ImS2b}
by setting $f_{\vec{q}} = 
  - C_{0} \E^{ - | {\vec{q}} |_{1} / {q_{\rm c}} }$
and $S_{\rm RPA} ( {\vec{q}} , \omega ) = 0$. 
From Eq.\ref{eq:R2} it is then obvious that in this case
the constant part 
of the Debye-Waller factor,
\begin{equation}
 R^{\alpha}_{\rm dis} = - \frac{1}{\beta V} \sum_{q} 
 \frac{C_{q}}{ ( \I \omega_{m} - {\vec{v}}^{\alpha} \cdot {\vec{q}} )^2}
 \; \; \; ,
 \end{equation}
vanishes for $\beta \rightarrow \infty$. 
This is in sharp contrast to the static random potential, where
$R^{\alpha}_{\rm dis}$ is divergent, see Eq.\ref{eq:Qdisstat1}.
For the space- and time-dependent contribution we obtain from
Eqs.\ref{eq:ReS2b} and \ref{eq:ImS2b}
 \begin{eqnarray}
 {\rm Re} S^{\alpha}_{\rm dis} ( r^{\alpha}_{\|} \hat{\vec{v}}^{\alpha} , \tau ) & = &
   - C_0 \frac{ | \tau | }{ 2 V} \sum_{ {\vec{q} }} 
  \cos ( \hat{\vec{v}}^{\alpha} \cdot {\vec{q}} r^{\alpha}_{\|} )
   \E^{ - { | {\vec{q}} |_{1}}/{q_{\rm c}} }
   \E^{ - | {\vec{v}}^{\alpha} \cdot {\vec{q}} | | \tau | }
  \; \; \; ,
  \label{eq:ReSdis}
  \\
 {\rm Im }S^{\alpha}_{\rm dis} ( r^{\alpha}_{\|} \hat{\vec{v}}^{\alpha} , \tau ) & = &
  -  C_0 \frac{\tau }{ 2 V} \sum_{ {\vec{q} }} 
  \sin ( | \hat{\vec{v}}^{\alpha} \cdot {\vec{q}} | r^{\alpha}_{\|} )
   \E^{ - { | {\vec{q}} |_{1}}/{q_{c}} }
     \E^{ - | {\vec{v}}^{\alpha} \cdot {\vec{q}} | | \tau | }
  \; \; \; .
  \label{eq:ImSdis}
 \end{eqnarray}
Note that in this limit only the term $L^{\alpha}_{\vec{q}} ( \tau )$ in 
Eqs.\ref{eq:ReS2b} and \ref{eq:ImS2b} survives.
With the above simple form of $C_{q}$ the 
${\vec{q}}$-integration is trivial. We obtain for the total Debye-Waller 
factor\index{Debye-Waller factor!dynamic disorder, Gaussian white noise}
 \begin{equation}
 Q_{\rm dis}^{\alpha} ( r^{\alpha}_{\|} \hat{\vec{v}}^{\alpha} , \tau )
 = 
 - S_{\rm dis}^{\alpha} ( r^{\alpha}_{\|} \hat{\vec{v}}^{\alpha} , \tau )
 = 
 \frac{ \I  W \tau }{ r^{\alpha}_{\|} + \I | {\vec{v}}^{\alpha} | \tau 
 + \I \; {\rm sgn}( \tau ) q_{\rm c}^{-1} }
 \label{eq:Qdisres}
 \; \; \; ,
 \end{equation}
where we have defined
 \begin{equation}
 W = 
 \frac{C_{0}}{2 \pi  } \left( \frac{ q_{\rm c}}{\pi} \right)^{d-1}
 \label{eq:tildec0def}
 \; \; \; .
 \end{equation}
Note that $W$ has units of velocity.
We conclude that the average sector Green's function is given by
\index{Gaussian white noise}
 \begin{equation}
 G^{\alpha} ( {\vec{r}} , \tau ) = 
 G_{0}^{\alpha} ( {\vec{r}} , \tau )  
 \exp \left[ 
 \frac{ \I  W \tau }
 { \hat{\vec{v}}^{\alpha} \cdot {\vec{r}} + \I | {\vec{v}}^{\alpha} | \tau 
 + \I \; {\rm sgn}( \tau ) q_{\rm c}^{-1} } \right]
 \; \; \; .
 \label{eq:Gpatchavresdyn}
 \end{equation}
Because the Debye-Waller factor vanishes at $\tau = 0$, we have
 $G^{\alpha} ( {\vec{r}} , 0 ) = 
 G_{0}^{\alpha} ( {\vec{r}} , 0 )  $, so that
the white noise dynamic random potential does not affect the momentum distribution. 
Hence, the average momentum distribution exhibits the same
jump discontinuity as in the absence of randomness.
In contrast, a static random potential completely washes out any singularities in the 
average momentum distribution. 

In Fourier space Eq.\ref{eq:Gpatchavresdyn}
looks rather peculiar. 
Let us first calculate the imaginary frequency Fourier transform,
 \begin{equation}
 G^{\alpha} ( {\vec{q}} , \I \tilde{\omega}_{n} ) =
 \int_{0}^{\beta} \D \tau \E^{ \I \tilde{\omega}_{n} \tau } G^{\alpha} ( {\vec{q}} , \tau )
 \; \; \; ,
 \label{eq:Gdisqomega}
 \end{equation}
where
 \begin{eqnarray}
 G^{\alpha} ( {\vec{q}} , \tau ) & = &
 \int \D {\vec{r}} \E^{ - \I {\vec{q}} \cdot {\vec{r}} } G^{\alpha} ( {\vec{r}} , \tau )
 \nonumber
 \\
 &  & \hspace{-15mm} =
 \frac{- \I}{ 2 \pi } \int_{- \infty}^{\infty} \D x 
 \frac{ \E^{ - \I q^{\alpha}_{\|} x }}{ x + \I | {\vec{v}}^{\alpha} | \tau}
 \exp \left[  
 \frac{ \I  W \tau }{ x + \I | {\vec{v}}^{\alpha} | \tau 
 + \I \; {\rm sgn}( \tau ) q_{\rm c}^{-1} } \right]
 \; \; \; ,
 \label{eq:Gdisqtau}
 \end{eqnarray}
with $q^{\alpha}_{\|} = \hat{\vec{v}}^{\alpha} \cdot {\vec{q}}$.
Because the argument of the exponential in the last factor of
Eq.\ref{eq:Gdisqtau} is always finite, we may 
expand the exponential in an infinite series and exchange the order of integration and
summation. For $\beta \rightarrow \infty$ the resulting integrals 
can then be done by means of contour integration\index{contour integration}.
Assuming for simplicity $ q^{\alpha}_{\|} \geq 0$ and $\tau > 0$, 
the relevant residue is
 \begin{eqnarray}
 \lefteqn{
 {\rm Res} \left[
 \frac{ \E^{ - \I q^{\alpha}_{\|} z }}{ [ z + \I 
 | {\vec{v}}^{\alpha}| \tau ] [ z + \I | {\vec{v}}^{\alpha} | \tau + 
  \I q_{\rm c}^{-1} ]^{n} }
 \right]_{z = - \I | {\vec{v}}^{\alpha}|  \tau - \I q_{\rm c}^{-1} }
 }
 \nonumber
 \\
 & = & - \E^{ - | {\vec{v}}^{\alpha} | q^{\alpha}_{\|} \tau } 
 \E^{ - { q^{\alpha}_{\|}}/{q_{\rm c}} }
 \sum_{k=0}^{\infty} \sum_{m = 0}^{\infty}
 \delta_{n , k+ m + 1 } \frac{ ( - \I q^{\alpha}_{\|} )^k ( - \I q_{\rm c} )^{m+1} }{ k ! }
 \; \; \; .
 \end{eqnarray}
After some straightforward algebra we obtain
 \begin{equation}
 G^{\alpha} ( {\vec{q}} , \tau )  = 
 - \E^{ -  | {\vec{v}}^{\alpha} | q^{\alpha}_{\|}  \tau } 
 \E^{ - {q^{\alpha}_{\|} }/{ q_{\rm c}} }
 \sum_{n = 0}^{\infty} \frac{ ( \frac{q^{\alpha}_{\|} }{ q_{\rm c}} )^{n}}{ n ! }
 \sum_{m = 0}^{n} \frac{ ( W q_{\rm c} \tau )^{m}}{m!}
 \label{eq:Gdisseries}
 \; \; \; .
 \end{equation}
Substituting this expression into Eq.\ref{eq:Gdisqomega}, 
the $\tau$-integration is trivial, so that we obtain
in the limit $\beta \rightarrow \infty$
 \begin{equation}
 G^{\alpha} ( {\vec{q}} , \I \omega )  = 
 - \frac{\E^{ - {q^{\alpha}_{\|} }/{ q_{\rm c}} }}{   {\vec{v}}^{\alpha}  
 \cdot {\vec{q}} - \I \omega }
 \sum_{n = 0}^{\infty} \frac{ ( \frac{q^{\alpha}_{\|} }{ q_{\rm c}} )^{n}}{ n ! }
 \sum_{m = 0}^{n} \left[ \frac{  W q_{\rm c} }
 { {\vec{v}}^{\alpha} \cdot {\vec{q}} - \I \omega }
 \right]^{m}
 \label{eq:Gdisseries2}
 \; \; \; .
 \end{equation}
The summations are now elementary, and we finally obtain
\index{Gaussian white noise}
 \begin{eqnarray}
 G^{\alpha} ( {\vec{q}} , i \omega )  & = &
 \frac{1}{ W q_{\rm c} + 
 \I \omega - {\vec{v}}^{\alpha} \cdot {\vec{q}}  }
 \nonumber
 \\
 &  \times & 
 \left\{  1 +
 \frac{ W q_{\rm c} \E^{ - { \hat{\vec{v}}^{\alpha} \cdot {\vec{q}} }/{q_{\rm c}} }}
 { \I \omega - {\vec{v}}^{\alpha} \cdot {\vec{q}}   }
 \exp \left[   -    W
  \frac{ \hat{\vec{v}}^{\alpha} \cdot {\vec{q}} }
 { \I \omega - {\vec{v}}^{\alpha} \cdot {\vec{q}}   } \right]
 \right\}
 \label{eq:Gdisqomegares}
 \; .
 \end{eqnarray}
Recall that we have assumed $\hat{\vec{v}}^{\alpha} \cdot {\vec{q}} \geq 0$.
For $ | \omega | \ll W q_{\rm c}$  and
$  q^{\alpha}_{\|}  \ll {\rm min} \{ q_{\rm c} , 
{W q_{\rm c}} / { | {\vec{v}}^{\alpha} |}   \}$ 
this reduces to
 \begin{equation}
 G^{\alpha} ( {\vec{q}} , \I \omega )   \sim
 \frac{1}{ \I \omega - {\vec{v}}^{\alpha} \cdot {\vec{q}} }
 \exp \left[ - W \frac{ \hat{\vec{v}}^{\alpha} \cdot {\vec{q}} }
 { \I \omega - {\vec{v}}^{\alpha} \cdot {\vec{q}} } \right]
 \label{eq:Gdisqomegareslim}
 \; \; \; .
 \end{equation}
If we now analytically continue\index{analytic continuation} 
this expression to real frequencies
by replacing $ \I \omega \rightarrow \omega + \I 0^{+}$, we encounter
an essential singularity at $\omega = {\vec{v}}^{\alpha} \cdot {\vec{q}}$.
As will be explained in
Sect.~\secref{sec:opendis},
we believe that this singularity is an artefact of the linearization
of the energy dispersion.

\subsection{Finite correlation time}
 
A  dynamic random potential 
with a finite correlation time can be modeled by
the covariance function
 \begin{equation}
 C_{q} = Z_{\vec{q}} \frac{2 \Omega_{\vec{q}}}{ \omega_{m}^2 + 
 \Omega_{\vec{q}}^2 }
 \; \; \; ,
 \label{eq:Cqtaufinite}
 \end{equation}
with some residue $Z_{\vec{q}}$ and frequency $\Omega_{\vec{q}}$. In the time
domain this implies for $\beta \rightarrow \infty$
 \begin{equation}
 C ( {\vec{q}} , \tau ) \equiv  \frac{1}{\beta } \sum_{m} C_{q} 
 \E^{ - \I \omega_{m} \tau }
 =  Z_{\vec{q}} \E^{ - \Omega_{\vec{q}} | \tau | }
 \; \; \; .
 \label{eq:Cqtau}
 \end{equation}
Note that we can rewrite Eq.\ref{eq:Cqtaufinite} as
 \begin{equation}
 C_{q} = \int_{0}^{\infty} \D \omega S_{\rm col} ( {\vec{q}} , \omega )
 \frac{ 2 \omega}{ \omega_{m}^2 + \omega^2 }
 \; \; \; ,
 \label{eq:Cqspec}
 \end{equation}
with 
 \begin{equation}
 S_{\rm col} ( {\vec{q}} , \omega ) = Z_{\vec{q}} \delta ( \omega - \Omega_{\vec{q}} )
 \; \; \; .
 \label{eq:Scoldef}
 \end{equation}
Comparison with Eq.\ref{eq:CrpaSRPA} shows that the 
exponentially decaying imaginary time correlator in Eq.\ref{eq:Cqtau}
corresponds to an undamped collective mode of 
an underlying many-body system. 
To calculate the Green's function, we simply
compare Eqs.\ref{eq:CrpaSRPA} and \ref{eq:Cqspec}, and note that
both expressions agree if we set
$f_{q} \rightarrow 0$ and
$f_{q}^2 S_{\rm RPA} ( {\vec{q}} , \omega )
\rightarrow S_{\rm col} ( {\vec{q}} , \omega )$.
Hence we can obtain the spectral representation of the
Debye-Waller factor by making these replacements in Eqs.\ref{eq:R2},
\ref{eq:ReS2b} and \ref{eq:ImS2b}.
The constant part is given by
 \begin{equation}
 R^{\alpha}_{\rm dis} = - \int \frac{ \D {\vec{q}}}{ ( 2 \pi )^d}
 \frac{ Z_{\vec{q}}}{ ( \Omega_{\vec{q}} + | {\vec{v}}^{\alpha} \cdot {\vec{q}} | )^2 }
 \; \; \; .
 \end{equation}
In contrast to the Gaussian white noise random potential, the finite correlation time
leads to a renormalization of the quasi-particle residue. 
Similarly, $S^{\alpha}_{\rm dis} ( r^{\alpha}_{\|} 
\hat{\vec{v}}^{\alpha} , \tau )$ can be obtained
by substituting Eq.\ref{eq:Scoldef} into Eqs.\ref{eq:ReS2b} and \ref{eq:ImS2b}.
For simplicity we shall assume here that the frequency
$\Omega_{\vec{q}} = \Omega_{0} $ is dispersionless and larger than
all other energy scales in the problem, 
and choose
$Z_{\vec{q}} = Z_{0} \E^{ - { | {\vec{q}} |_{1}}/{q_{\rm c}} }$.
Keeping the next-to-leading order in $\Omega_{0}^{-1}$ 
we obtain\index{Debye-Waller factor!dynamic disorder, finite correlation time}
 \begin{eqnarray}
 Q_{\rm dis}^{\alpha} ( r^{\alpha}_{\|} \hat{\vec{v}}^{\alpha} , \tau )
 & \approx &
 \frac{ Z_{0}}{ \pi | {\vec{v}}^{\alpha} | \Omega_{0} } 
 \left( \frac{q_{\rm c}}{\pi} \right)^{d-1}
 \left[ 
 \frac{ \I  | {\vec{v}}^{\alpha} | \tau }{ r^{\alpha}_{\|} + \I | {\vec{v}}^{\alpha} | \tau 
 + \I \; {\rm sgn}( \tau ) q_{\rm c}^{-1}  }
 \right.
 \nonumber
 \\
 & & \left. + \frac{ | {\vec{v}}^{\alpha} | q_{\rm c}}{\Omega_{0}} \left(
  1 -
 \frac{ \E^{ - \Omega_{0} | \tau | }}{ 1 + (r^{\alpha}_{\|} q_{\rm c} )^2 } 
 \right)
 \right]
 \label{eq:Qdistaures}
 \; \; \; .
 \end{eqnarray}
If we take the limit $\Omega_{0} \rightarrow \infty$ while keeping
${Z_{0}}/{ \Omega_{0}} $ constant, we recover
Eq.\ref{eq:Qdisres},
with $W = ({Z_{0}}/ ({\pi  \Omega_{0}}) )
( {q_{\rm c}}/{\pi} )^{d-1}$.
Because the leading term in Eq.\ref{eq:Qdistaures} has the same structure as 
Eq.\ref{eq:Qdisres}, the spectral function exhibits again an essential singularity
at $\omega = {\vec{v}}^{\alpha} \cdot {\vec{q}}$.
Therefore the essential singularity in Eq.\ref{eq:Gdisqomegareslim}
is {\it{not}} a special feature of the Gaussian white noise limit.

\section{Summary and outlook}
\label{sec:opendis}
 
In this chapter we have used our background field method
developed in Chap.~\secref{chap:agreen} to calculate
the average Green's function of electrons 
subject to a long-range random potential.
For simplicity, 
we have worked with linearized energy dispersion.
Although for static disorder we have correctly reproduced
the usual perturbative result of the Born approximation, 
for time-dependent disorder we have
obtained the rather peculiar expression \ref{eq:Gdisqomegareslim}
for the Fourier transform of the Green's function, which
involves an essential singularity on resonance (i.e. for
$\omega = {\vec{v}}^{\alpha} \cdot \vec{q}$).
We believe that this singularity is an artefact
of the linearization of the energy dispersion.
This is based on the observation that
in the white noise limit considered
in Sect.~\secref{subsec:Gaussianwhite}
the long-distance behavior of the 
Debye-Waller factor is completely determined by the
term $L^{\alpha}_{ \vec{q} } ( \tau )$ 
of Eqs.\ref{eq:ReS2b} and \ref{eq:ImS2b}. 
As discussed in detail in
Chap.~\secref{subsec:Thetermslin},
this term is generated by the
{\it{double pole}} in the Debye-Waller 
factor for
linearized energy dispersion.
On the other hand,
for non-linear energy dispersion
this double pole is split into two separate poles, so that a term
similar to $L^{\alpha}_{ \vec{q} } ( \tau )$  does not appear.
Thus, an interesting open problem is
the {\bf{evaluation of the Debye-Waller factor due to dynamic disorder 
for non-linear energy dispersion}}.
In this context we would also like mention
that a numerical analysis \cite{Metzner95num}
of the higher-dimensional bosonization result
for the Green's function
{\it{with linearized energy dispersion}}
(see Eqs.\ref{eq:Qlondef}--\ref{eq:Slondef} and
\ref{eq:Galphartdef}--\ref{eq:Galphaqtildedef})
indicates that also for generic density-density interactions
there exists some kind of unphysical singularity
in the spectral function close to the resonance 
$ \omega = {\vec{v}}^{\alpha} \cdot \vec{q}$.
We believe that this singularity has precisely the same
origin as the singularity in Eq.\ref{eq:Gdisqomegareslim}, namely
the double pole in the Debye-Waller factor for linearized energy
dispersion.

Another interesting unsolved problem is the  correct 
description of the {\bf{diffusive motion}} of the
electrons within the framework of higher-dimensional bosonization\footnote{
The case of one dimension \cite{Giamarchi88} is special,
because, 
at least in the absence of interactions, 
the localization length of
one-dimensional disordered fermions 
has the same order of magnitude as the elastic 
mean free path \cite{Altshuler85,Fukuyama85}. 
Therefore the diffusive regime does not exist in $d=1$.}.
The signature of diffusion is
known to manifest itself also in the
low-energy behavior of the single-particle Green's function
of an {\it{interacting}} disordered Fermi system.
Evidently our result \ref{eq:Gavranintfinal} 
for the average Green's function 
in the presence of electron-electron interactions
does not contain interference terms describing 
the interplay between disorder and  electron-electron interactions.
Note that 
the perturbative calculation of the
average Green's function for disordered electrons
in the presence of electron-electron interactions
leads to singular terms due to multiple
impurity scattering. These appear even at the first 
order in the effective electron-electron interaction and
involve the so-called
{\it{Diffuson}} \index{Diffuson propagator} and {\it{Cooperon}} 
propagators \cite{Lee85,Altshuler85,Fukuyama85}.\index{Cooperon
propagator}
While the Cooperon involves 
momentum transfers of the order of $2 k_{\rm F}$,
the Diffuson is most singular for small momentum transfers. 
Because our approach attempts 
to treat the complete forward scattering 
problem non-perturbatively,
the Diffuson
should not be neglected.
In fact, it is well known that the Diffuson qualitatively  modifies the 
effective screened interaction at long wavelengths \cite{Altshuler85}. 
Furthermore, the so-called $g_1$-contribution to the 
self-energy \cite{Fukuyama85},
which to lowest order in the electron-electron interaction
involves two Diffuson propagators,
can be viewed as an
effective long-range interaction between the electrons.
This interaction is
generated by many successive impurity scatterings
and is a consequence of the
diffusive  motion of the electrons
in a disordered metal.  Obviously,
such a motion cannot be correctly described
within the approximations 
inherent in higher-dimensional bosonization at the level of the
Gaussian approximation.
However, 
in Chap.~\secref{sec:eik} we have developed a general
method for calculating the Green's function
beyond the Gaussian approximation, 
which  might lead to a new non-perturbative
approach to the problem of 
electron-electron interactions in disordered Fermi systems.

%
%

%
%

\chapter{Transverse gauge fields}
\label{chap:arad}
\setcounter{equation}{0}

{\it{
We generalize our functional bosonization approach
to the case of fermions that are
coupled to transverse abelian gauge fields.
This is perhaps the physically 
most important application of higher-dimensional bosonization,
because transverse gauge fields 
appear in effective low-energy theories for
strongly correlated electrons
and quantum Hall systems.
In this chapter we shall restrict ourselves to the 
formal development of the methods.
An important physical application to the 
quantum Hall effect is given in the Letter \cite{Kopietz96che}.
For linearized energy dispersion 
we have discussed the gauge field problem in
the work \cite{Kopietz95b}. 
It turns out, however, that in physically relevant cases 
quantitatively correct results
for the single-particle Green's function
can only be obtained if one retains the
quadratic terms in the expansion of the
energy dispersion close to the Fermi surface.
}}

\vspace{7mm}

\noindent
As shown in the classic textbook by Feynman and Hibbs \cite{Feynman63},
the static Coulomb interaction
${ 4 \pi e^2}/ { {\vec{q}}^2 }$ between
electrons can be obtained by coupling the electronic density to the
scalar potential $\phi$ of the Maxwell field \index{Maxwell field}and
integrating in the functional integral over all complexions of $\phi$.
The transverse radiation field\index{transverse radiation field} 
${\vec{A}}$ is usually neglected in condensed matter,
because the coupling between the current density and the transverse
radiation field involves an extra factor of
${v_{\rm F}}/{c}$.
At metallic densities the Fermi velocity $v_{\rm F}$ is
two orders of magnitude smaller than the velocity of light $c$, 
so that  for all practical applications it is justified to ignore the radiation field.
The leading correction to the static Coulomb interaction
is an effective retarded interaction between paramagnetic current densities,
\index{current-current interaction}
mediated by the transverse radiation field.
Within the RPA the propagator of the transverse radiation field
is in Coulomb gauge  and for frequencies
$| \omega_m | \ll v_{\rm F} |{\vec{q}} | $
given by (see Eq.\ref{eq:Hrparad2} below)
 \begin{equation}
 h^{{\rm RPA} , \alpha}_{q}
 = -  \frac{1}{\nu} \left( \frac{v_{\rm F}}{c} \right)^2
 \frac{ 1 - ({\hat{\vec{k}}}^{\alpha} \cdot {\hat{\vec{q}}} )^2}
 { \left( \frac{{\vec{q}} }{ \kappa } \right)^2 + 
 \frac{\pi}{4} 
 ( \frac{v_{\rm F}}{c} )^2
 \frac{| \omega_{m}| }{ v_{\rm F} | {\vec{q}} | } 
  }
 \label{eq:Hrparad2b}
 \; \; \; .
 \end{equation}
Here $\nu$ is the density of states at the Fermi surface,
and $\kappa$ is the usual Thomas-Fermi wave-vector in three dimensions.
In 1973 Holstein, Norton and Pincus \cite{Holstein73}
showed that the associated effective current-current interaction 
gives rise to logarithmic singularities 
in the perturbative expansion of the
electronic self-energy, and
concluded that the low-energy behavior of the
single-particle Green's function is not of the
Fermi liquid type.
However, they also showed that due to the smallness of the parameter
${v_{\rm F}}/{c}$ the deviations from conventional
Fermi liquid behavior are 
beyond experimental resolution, so that they have
little practical consequences. 
Later the behavior of the electrodynamic field in metals was studied
in more detail by Reizer \cite{Reizer89}. A nice pedagogical
discussion of this problem can be found in the textbook by
Tsvelik \cite{Tsvelik95}.

The recent excitement about the unusual normal-state properties of the
high-temperature 
superconductors \cite{Anderson90b,Baskaran87,Lee89,Ioffe89,Blok93} 
as well as half-filled quantum 
Hall systems \cite{Lopez91,Kalmeyer92,Halperin93} has revived the interest in the problem
of electrons coupled to gauge fields from a more  general point of view.
In theoretical models for these systems the transverse gauge field is not necessarily the
Maxwell field,\index{Maxwell field}
so that in principle the magnitude of the velocity associated with the
gauge field can be comparable with the Fermi velocity,
and the effective coupling
constant can be of the order of unity.
Moreover, the effective dimensionality is not necessarily $d=3$.
Thus, we are led to the general problem of fermions
in $d$ dimensions that
are coupled to transverse gauge fields with 
RPA propagator given by
 \begin{equation}
 h^{{\rm RPA} , \alpha}_{q}
 = -  \frac{1}{\nu  } 
 \frac{ 1 - ({\hat{\vec{k}}}^{\alpha} \cdot {\hat{\vec{q}}} )^2}
 { \left( \frac{ | {\vec{q}}| }{ q_{\rm c} } \right)^{\eta} + 
  \lambda_d \frac{| \omega_{m}| }{ v_{\rm F} | {\vec{q}} | } 
  }
 \label{eq:Hrparad3}
 \; \; \; .
 \end{equation}
Here $\eta >0 $ is some exponent, 
$\lambda_d$ is a numerical constant, and $q_{\rm c}$ is some 
characteristic momentum scale.
We recover  Eq.\ref{eq:Hrparad2b}
by setting $\eta =2$ , $\lambda_3 = {\pi}/{4}$, and
$q_{\rm c} =  {v_{\rm F}} \kappa/ c$.
On the other hand, the gauge field propagator
in the two-dimensional Maxwell-Chern-Simons action\index{Maxwell-Chern-Simons theory} 
(which is believed to describe the low-energy
physics of composite Fermions in the half-filled Landau level \cite{Halperin93})
corresponds to the choice $\eta = 1$, $\lambda_2 = 1$, and
$q_{\rm c} = ( 2 k_{\rm F} )^2 / \kappa$, where $\kappa$ is in this case
the Thomas-Fermi wave-vector in $d=2$, see Eq.\ref{eq:kappaTFdef}.

The low-energy behavior
of the Green's function of fermions 
that are coupled to transverse gauge fields with propagator \ref{eq:Hrparad3}
has recently been studied 
with the help of a variety of non-perturbative techniques, such as
renormalization group and scaling methods \cite{Gan93,Nayak94,Onoda95,Chakravarty95},
a ${1}/{N}$-expansion \cite{Ioffe94}, higher-dimensional 
bosonization \cite{Kwon94,Houghton94,Kopietz96che,Kopietz95b}, 
a quantum Boltzmann equation \cite{Kim95},
and other non-perturbative resummation 
schemes \cite{Lee96,Plochinski94,Khveshchenko93,Castellani94b}. 
According to Ioffe {\it{et al.}} \cite{Ioffe94} 
as well as Castellani and Di Castro \cite{Castellani94b}, 
in the case of transverse gauge fields it is not allowed to
locally linearize the energy dispersion (thus approximating
the Fermi surface by a collection of flat patches)
because the effective interaction mediated by the gauge field is dominated by
momentum transfers parallel to the Fermi surface.
In fact, the method used by Ioffe {\it{et al.}} \cite{Ioffe94}
produces results that are in disagreement with the
predictions of higher-dimensional bosonization with
linearized energy dispersion \cite{Kwon94,Kopietz95b}.
Because the linearization of the energy dispersion is one
of the main (and a priori uncontrolled) approximations
inherent in earlier formulations of higher-dimensional
bosonization \cite{Houghton93,Castro94,Houghton94,Castro95,Kopietz94,Kopietz95,Kopietz95d},
one might suspect that the linearization is at least partially responsible
for this disagreement.

\begin{figure}
\sidecaption
\psfig{figure=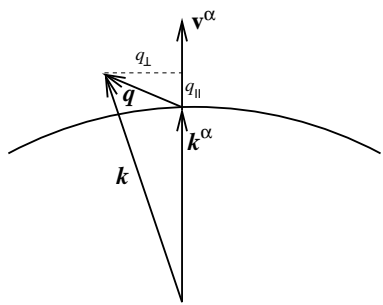,width=5cm}
\caption[Local coordinate system on the Fermi surface.]
{Local coordinate system on the Fermi surface and definition
of the components $q_{\|}$ and ${q}_{\bot}$ of
$\vec{q} = \vec{k} - \vec{k}^{\alpha}$.
}
\label{fig:localgauge}
\end{figure}

Let us give a simple argument why 
for the effective interaction mediated by transverse gauge fields the curvature of the
Fermi surface might indeed be more important 
than in the case of the conventional density-density interactions
discussed in Chap.~\secref{chap:a7sing}.
Consider a fermion with momentum $\vec{k} = \vec{k}^{\alpha}
+ \vec{q}$ such that $| \vec{q} | \ll k_{\rm F}$.
From Eq. \ref{eq:Hrparad3} we see that the typical momentum 
$q_{\omega}$ transfered by the gauge field in a low-energy process with energy
$\omega$ is determined by
$( q_{\omega} / q_{\rm c} )^{\eta} = \lambda_d \omega / ( v_{\rm F} q_{\omega } )$,
so that
 \begin{equation}
 q_{\omega} = \left( \frac{ \lambda_d q_{\rm c}^d }{v_{\rm F}} \right)^{\frac{1}{1 + \eta}}
 \omega^{\frac{1}{1 + \eta}}
 \label{eq:qomegatyp}
 \; \; \; .
 \end{equation}
Because the factor 
 $1 - ({\hat{\vec{k}}}^{\alpha} \cdot {\hat{\vec{q}}} )^2$ in
Eq.\ref{eq:Hrparad3} is maximal for wave-vectors $\vec{q}$ that are
perpendicular to $\vec{k}^{\alpha}$ (see Fig.~\secref{fig:localgauge}),
we conclude that the
typical momentum transfer $q_{\bot}$ parallel to the Fermi surface
is of the order of $q_{\omega}$. 
On the other hand, for an energy dispersion of the form\footnote{
To study curvature effects, we may omit the term 
quadratic in $q_{\|} = \vec{q} \cdot \hat{\vec{v}}^{\alpha}$,
i.e. $q_{\|}^2 / ( 2 m_{\|} )$. 
As discussed in Chaps.~\secref{sec:eik} and \secref{subsec:thefate}, 
this term does not describe the curvature of the Fermi surface
and is irrelevant. For convenience we have also
omitted the patch index on $q_{\|}^{\alpha}$ and $q_{\bot}^{\alpha}$.}
$\xi^{\alpha}_{\vec{q}} = v_{\rm F} q_{\|} + {q_{\bot}^2}/({ 2 m_{\bot}})$, 
the curvature term is negligible provided
 \begin{equation}
 \frac{q_{\bot}^2}{ 2 m_{\bot}  
 v_{\rm F} q_{\|} } \ll 1 
 \label{eq:curvatureneglect}
 \; \; \; .
 \end{equation} 
Setting $q_{\bot} \approx q_{\omega}$ and using the fact that
$ v_{\rm F} q_{\|}   \approx \omega$ close to the poles of the Green's function, we
see that Eq.\ref{eq:curvatureneglect} reduces to
 \begin{equation}
 \frac{1}{ 2 m_{\bot}} 
 \left( \frac{ \lambda_d q_{\rm c}^d }{v_{\rm F}} \right)^{\frac{2}{1 + \eta}}
 \omega^{\frac{1 - \eta }{1 + \eta}}
 \ll 1
 \label{eq:curvcrit}
 \; \; \; .
 \end{equation}
For $\eta < 1$ this condition is always satisfied 
at sufficiently low energies, so that in this case 
curvature should be irrelevant.
On the other hand, for $\eta > 1$ 
the left-hand side becomes arbitrarily large for small
$\omega$, so that we expect that in the low-energy regime 
the curvature of the Fermi surface will become important.
Of course, the above arguments are rather hand-waving, so that
more rigorous methods are necessary to examine the role of curvature
in the bosonization approach to the gauge field problem.
Having developed a non-perturbative method to include curvature
effects into higher-dimensional 
bosonization (see Chap.~\secref{sec:eik}), we 
shall in this chapter examine the role of curvature by 
explicitly calculating the effect of the 
quadratic term in the energy dispersion on the
gauge field contribution to the Debye-Waller factor.

\section{Effective actions}
\label{sec:Nonrel}

{\it{We define a general field theory for 
non-relativistic electrons that are coupled to transverse abelian
gauge fields. 
This theory contains the usual Maxwell action as a special case.
We discuss in some detail the effective matter action that is obtained
by integrating first over the gauge field, and
the effective gauge field action that results from the integration 
over the matter degrees of freedom.}}

\subsection{The coupled matter gauge field action}
\label{subsec:Definitionofmodel}

{\it{Transverse gauge fields can be viewed 
as Hubbard-Stratonovich fields that couple
to the fermionic current density.
We show how to obtain the propagators via functional integration
and how to impose the Coulomb gauge constraint with
the help of the Fadeev-Popov method.}}

\vspace{7mm}

\noindent
Measuring wave-vectors with respect to local
coordinate systems centered at the Fermi surface,
the Euclidean Maxwell action \cite{Kapusta89} can be written as
\index{effective action!coupled matter gauge-field}
\index{Maxwell field}
 \begin{equation}
 {S}  \{ \psi , \phi^{\alpha} , {\vec{A}}^{\alpha} \}
 = {S}_{0} \{ \psi \} + {S}_{1} \{ \psi , \phi^{\alpha} , {\vec{A}}^{\alpha} \}
 + {S}_{2} \{ \phi^{\alpha} , {\vec{A}}^{\alpha} \}
 \label{eq:Sgf}
 \; \; \; ,
 \end{equation}
where the matter action $S_{0} \{ \psi \}$ is defined in Eq.\ref{eq:S0psidef}, and
 \begin{eqnarray}
 {S}_{1} \{ \psi , \phi^{\alpha} , {\vec{A}}^{\alpha} \}
 & = &   \sum_{ q } \sum_{\alpha} \left[
 \I \rho^{\alpha}_{q}  \phi_{ -q}^{\alpha}
 - {\vec{j}}_{ q}^{\alpha}  \cdot {\vec{A}}_{ -q}^{\alpha}
 \right]
 \; \; \; ,
 \label{eq:S1gf}
 \\
 & &
 \hspace{-25mm}
 {S}_{2} \{ \phi^{\alpha} , {\vec{A}}^{\alpha}  \}  = 
 \frac{1}{2} \sum_{q} \sum_{\alpha \alpha^{\prime} }
 \left[ [ \underline{\tilde{f}}_{ {{q}} }^{-1} ]^{ \alpha \alpha^{\prime} }
 \phi_{-q}^{\alpha} \phi_{q}^{\alpha^{\prime}}
 + [ \underline{\tilde{h}}_{q}^{-1} ]^{ \alpha \alpha^{\prime}}
 {\vec{A}}_{-q}^{\alpha} \cdot {\vec{A}}_{q}^{\alpha^{\prime}} \right]
 \label{eq:S2gf}
 \;  .
 \end{eqnarray}
For convenience we have used the {\it{Coulomb gauge}}\index{Coulomb gauge},
 \begin{equation}
 {\vec{q}} \cdot  {\vec{A}}^{\alpha}_{q} = 0
 \label{eq:Cbgaugelocal}
 \; \; \; ,
 \end{equation}
because then the longitudinal and transverse components of the gauge field
$A_{\mu}^{\alpha} = [ \phi^{\alpha} , {\vec{A}}^{\alpha} ]$ in
Eq.\ref{eq:S2gf} are decoupled. 
The sector density $\rho^{\alpha}_{q}$
is defined in Eq.\ref{eq:rhoalpha}, and the gauge invariant sector
current density ${\vec{j}}^{\alpha}_{q}$ has a para- and a diamagnetic contribution,
\index{current density}
 \begin{equation}
 {\vec{j}}_{ q}^{\alpha}
  =   {\vec{j}}^{{\rm para} , \alpha}_{ q} +
 {\vec{j}}^{{\rm dia} , \alpha}_{ q} 
 \label{eq:jalphapatch}
 \; \; \; ,
 \end{equation}
with
 \begin{eqnarray}
 {\vec{j}}_{ q}^{{\rm para} , \alpha} & = & \sum_{k} 
 \Theta^{\alpha} ( {\vec{k}} ) 
 \frac{  ( {\vec{k}} + {\vec{q}}/ 2 ) }{m c }
 \psi^{\dagger}_{k} \psi_{k+q}
 \; \; \; ,
 \label{eq:jparaalpha}
 \\
 {\vec{j}}_{  q}^{{\rm dia }, \alpha} & = &  
 - \frac{ 1}{ 2 m c^2 \beta} \sum_{ q^{\prime}}
 {\vec{A}}^{\alpha}_{ q-q^{\prime} } \rho^{\alpha}_{q^{\prime}}
 \label{eq:jdiaalpha}
 \; \; \; .
 \end{eqnarray}
For arbitrary matrices
$\underline{\tilde{f}}_{{q}}$ and $\underline{\tilde{h}}_{q}$ 
in Eq.\ref{eq:S2gf} the above action is more
general than the usual Maxwell action.
The latter can be obtained 
by choosing the matrix elements of
$\underline{\tilde{f}}_{q}$  and $\underline{\tilde{h}}_{q}$  
to be independently of the sector indices given 
by\footnote{
Of course, matrices with all equal elements are not invertible, so that we should
regularize $\underline{\tilde{f}}^{-1}_q$ and 
$\underline{\tilde{h}}^{-1}_q$ 
in some convenient way.
As already mentioned in Chap.~\secref{subsec:HS1sub2},
our final results for physical quantities
can be entirely expressed in terms of the
original matrices
$\underline{\tilde{f}}_{q}$ and $\underline{\tilde{h}}_q$,
so that for our purpose it is sufficient to
assume at intermediate stages that
$\underline{\tilde{f}}^{-1}_q$ and 
$\underline{\tilde{h}}^{-1}_q$ have been properly regularized.}
 \begin{equation}
 [ \underline{ \tilde{f}}_{q} ]^{\alpha \alpha^{\prime} }
  =  \tilde{f}_{\vec{q}} 
 = \frac{ \beta}{{V}} \frac{ 4 \pi e^2}{ {\vec{q}}^2 }
 \; \; \; , \; \; \; 
 \; \; \; 
 [ \underline{ \tilde{h}}_{q} ]^{\alpha \alpha^{\prime} }  = 
 \tilde{h}_{{q}}  
 = \frac{ \beta}{{V}} \frac{ 4 \pi e^2 }{ 
 {\vec{q}}^2  + ( \frac{ \omega_{m} }{  c } )^2 } 
 \; \; \; .
 \label{eq:tildehdefgf}
 \end{equation}
If we set $\vec{A}^{\alpha} = 0$ in
Eq.\ref{eq:Sgf}, then
$ {S}  \{ \psi , \phi^{\alpha} , 0 \}$ agrees
precisely with the action given 
$S \{ \psi , \phi^{\alpha} \}$ defined in Eq.\ref{eq:Sdecoupdef},
which has been obtained from the original density-density interaction by means
of the Hubbard-Stratonovich transformation
discussed in Chap.~\secref{sec:HS1}.
Evidently the gauge field
$A^{\alpha}_{\mu}$ can be viewed as a generalized Hubbard-Stratonovich field
which couples to the (gauge invariant) current density.
The Maxwell-Chern-Simons action\index{Chern-Simons theory}, which plays an
important role in the theory of the half-filled Landau level \cite{Halperin93}, 
contains an additional term involving the
coupling between the $\phi^{\alpha}$- and the ${\vec{A}}^{\alpha}$-field \cite{Kwon94}. 
This coupling is due to the fact that in these theories 
density fluctuations are effectively mapped onto fluctuations of the gauge field.
By a proper choice of the propagator of the $\phi^{\alpha}$-field, one can
therefore control the value of the exponent $\eta$ 
that characterizes the dispersion of the gauge field propagator
in Eq.\ref{eq:Hrparad3}.
Because in this section we would like
to discuss the basic concepts,
we shall ignore at this point the
Chern-Simons coupling between $\phi^{\alpha}$- and ${\vec{A}}^{\alpha}$-field.
From the general structure of the final result for the single-particle
Green's function the modifications arising from 
the  Chern-Simons coupling will become obvious.

In complete analogy with Eq.\ref{eq:Gpatchdef}, the exact single-particle Green's function
is now given by
 \begin{equation}
 G ( k )
 = - \beta \frac{ 
 \int {\cal{D}} \{ \psi \} 
 {\cal{D}} \{ \phi^{\alpha} \}
 {\cal{D}} \{ {\vec{A}}^{\alpha} \}
 \E^{- {S} \{ \psi , \phi^{\alpha} , {\vec{A}}^{\alpha} \} } 
 \psi_{k} \psi^{\dagger}_{k}
 }
 { \int {\cal{D}} \{ \psi \} 
 {\cal{D}} \{ \phi^{\alpha} \}
 {\cal{D}} \{ {\vec{A}}^{\alpha} \}
 \E^{- {S} \{ \psi , \phi^{\alpha} , {\vec{A}}^{\alpha} \} }}
 \label{eq:Gpatchgfdef}
 \; \; \; ,
 \end{equation}
where the functional integration over the ${\vec{A}}^{\alpha}$-field is subject to the
Coulomb gauge condition \ref{eq:Cbgaugelocal}.
Although in a gauge theory the single-particle propagator is in general
gauge dependent \cite{Tsvelik95}, we expect
physical quantities derived from it 
to be gauge invariant\footnote{In particular, 
the imaginary part of the retarded Green's function
can be directly related to the photoemission spectrum
as long as certain standard approximations
(which are discussed in detail
in \cite{Almbladh83}) 
are assumed to be correct.  Thus,
we expect that ${\rm Im} G ( \vec{k} , \omega + \I 0^{+} )$
is to a large extent gauge invariant.
I would like to thank C. K\"{u}bert and A. Muramatsu for pointing this 
out to me.}.
Moreover, as recently shown by Sylju{\aa}sen \cite{Syljuasen96},
for a particular class of gauge choices\index{gauge invariance}, the most singular
part of the fermionic self-energy in non-relativistic
quantum electrodynamics is independent of the gauge.

The matrices $\underline{\tilde{f}}_{q}$ and $\underline{ \tilde{h}}_{q}$
determine the free propagator of the gauge field,\index{gauge field propagator!free}
 \begin{equation}
  D^{ \alpha \alpha^{\prime} }_{0, \mu \nu } ( q )  =  
  \frac{  
  \int
 {\cal{D}} \{ \phi^{\alpha} \}
 {\cal{D}} \{ {\vec{A}}^{\alpha} \}
 \E^{- {S}_{2} \{ \phi^{\alpha} , {\vec{A}}^{\alpha} \} } 
 A_{q , \mu}^{\alpha} A_{-q , \nu}^{\alpha^{\prime}}
 }
 { 
  \int
 {\cal{D}} \{ \phi^{\alpha} \}
 {\cal{D}} \{ {\vec{A}}^{\alpha} \}
 \E^{- {S}_{2} \{ \phi^{\alpha} , {\vec{A}}^{\alpha} \} } 
 }
 \label{eq:D0gfdef}
 \; \; \; ,
 \end{equation}
where we use the convention that $A^{\alpha}_{q,0} = \phi^{\alpha}_{q}$.
{\it{In Coulomb gauge}}\index{Coulomb gauge} the longitudinal and transverse components do not mix, so that
 \begin{eqnarray}
 D^{\alpha \alpha^{\prime} }_{0,00} (q) & = & [ \underline{\tilde{f}}_{{q}} ]^{\alpha \alpha^{\prime}}  
 \; \; \; ,
 \label{eq:D00alpha}
 \\
 D^{\alpha \alpha^{\prime}}_{0,0 i} (q) 
 & = & D^{\alpha \alpha^{\prime} }_{0,i0} ( q ) = 0 \; \; \; , \; \; \;  i = 1 , \ldots , d
 \; \; \; ,
 \label{eq:D0ialpha} 
 \\
 D^{\alpha \alpha^{\prime} }_{0, i j } ( q ) & =  &
 [ \underline{\tilde{h}}_{q} ]^{\alpha \alpha^{\prime}} 
 \left[ \delta_{ij} -  ( {\vec{e}}_{i} \cdot \hat{\vec{q}}  )
 ( {\vec{e}}_{j} \cdot  \hat{\vec{q}}  ) \right]
  \; \; \; , \; \; \; i,j = 1, \ldots , d
 \label{eq:D0gfij}
 \; \; \; ,
 \end{eqnarray}
where $\hat{\vec{q}} = {\vec{q}} / | {\vec{q}} |$.
A simple way to derive Eq.\ref{eq:D0gfij} 
is to impose the Coulomb gauge condition 
\ref{eq:Cbgaugelocal} in the functional integral by means of the 
Fadeev-Popov method \cite{Popov83}\index{Fadeev-Popov method}.
A very nice pedagogical discussion of this method 
can be found in the recent textbook by Sterman \cite[\mbox{page $190$}]{Sterman93}.
In the problem at hand, the Fadeev-Popov method amounts
to inserting the following
integral representation of the functional $\delta$-function
into the integrand of the denominator and the numerator of
Eq.\ref{eq:D0gfdef},
 \begin{equation}
 \prod_{\alpha } \delta \{ \nabla \cdot {\vec{A}}^{\alpha} ( {\vec{r}} , \tau ) \}  =  
 \int {\cal{D}} \left\{ \lambda^{\alpha} \right\}
 \E^{ - \sum_{q \alpha} {\vec{A}}_{- q}^{\alpha} \cdot {\vec{q}} \lambda_{q}^{\alpha} }
 \label{eq:Coulombconstraint2}
 \; \; \; ,
 \end{equation}
and then treating the ${\vec{A}}^{\alpha}$-integrations as unrestricted.
The integration over the auxiliary fields $\lambda^{\alpha}_q$, 
$\alpha = 1 , \ldots , M$ 
enforces the Coulomb gauge condition \ref{eq:Cbgaugelocal}
for each sector.
Shifting\index{shift transformation}
 \begin{equation}
 {\vec{A}}^{\alpha}_{q} \rightarrow {\vec{A}}^{ \alpha}_{q}
 - \sum_{\alpha^{\prime }}  [ \underline{\tilde{h}}_{q} ]^{ \alpha \alpha^{\prime} }
 {\vec{q}} \lambda^{\alpha^{\prime}}_{q}
 \label{eq:shiftFP}
 \; \; \; ,
 \end{equation}
and using
$ [ \underline{\tilde{h}}_{q} ]^{\alpha \alpha^{\prime} }
= [ \underline{\tilde{h}}_{-q} ]^{\alpha^{\prime} \alpha}$
(see also Eq.\ref{eq:ftildeherm}),
 we replace
 \begin{eqnarray}
 \frac{1}{2} \sum_{q} \sum_{\alpha \alpha^{\prime} }
  [ \underline{\tilde{h}}_{q}^{-1} ]^{ \alpha \alpha^{\prime}}
 {\vec{A}}_{-q}^{\alpha} \cdot {\vec{A}}_{q}^{\alpha^{\prime}} 
 + \sum_{q \alpha} {\vec{A}}^{\alpha}_{-q  } \cdot {\vec{q}} \lambda^{\alpha}_{q }
 &    \rightarrow &
 \nonumber
 \\
 & & \hspace{-73mm} 
 \frac{1}{2} \sum_{q} \sum_{\alpha \alpha^{\prime} }
  [ \underline{\tilde{h}}_{q}^{-1} ]^{ \alpha \alpha^{\prime}}
 {\vec{A}}_{-q}^{  \alpha} \cdot {\vec{A}}_{q}^{ \alpha^{\prime}} 
  + \frac{1}{2} \sum_{q} \sum_{\alpha \alpha^{\prime} }
  \lambda^{\alpha}_{-q} {\vec{q}}^2  [ \underline{\tilde{h}}_{q}  ]^{\alpha \alpha^{\prime}}
  \lambda^{\alpha^{\prime}}_{q}
  \label{eq:Sgfshift}
  \; \; \; .
  \end{eqnarray}
Using the fact that the integration measure is invariant with respect to 
shift transformations \cite{Popov87}, we obtain
 \begin{eqnarray}
 D^{\alpha \alpha^{\prime}}_{0, i j } ( q )   & =  &
  \left\{
 \int {\cal{D}} \left\{ \lambda^{\alpha} \right\}
 {\cal{D}} \left\{ {\vec{A}}^{\alpha} \right\}
 \E^{- \tilde{S}_{2} \{ {\vec{A}}^{\alpha} , \lambda^{\alpha} \} }
 \right\}^{-1}
 \nonumber
 \\
 & \times & 
 \int {\cal{D}} \left\{ \lambda^{\alpha} \right\}
 {\cal{D}} \left\{ {\vec{A}}^{\alpha} \right\}
 \E^{- \tilde{S}_{2} \{ {\vec{A}}^{\alpha} , \lambda^{\alpha} \} }
 \nonumber
 \\
 &     \times  &
 \left[ A^{\alpha}_{q,i} 
  A^{\alpha^{\prime}}_{-q,j} - q_{i} q_{j} 
  \sum_{\alpha_{1} \alpha_{2}} 
  [ \underline{\tilde{h}}_{q} ]^{\alpha \alpha_{1}}
  \lambda_{ q}^{\alpha_{1}} \lambda_{-q}^{\alpha_{2}} 
   [ \underline{\tilde{h}}_{q} ]^{\alpha_{2} \alpha^{\prime} }
  \right]
 \; \; \; ,
 \label{eq:D1}
 \end{eqnarray}
where
 \begin{equation}
 \hspace{-4mm}
 \tilde{S}_{2} \{ {\vec{A}}^{\alpha} , \lambda^{\alpha} \} 
 =
 \frac{1}{2} \sum_{q } \sum_{\alpha \alpha^{\prime}} \left[
 [ \underline{\tilde{h}}_{q}^{-1}]^{\alpha \alpha^{\prime}}
 {\vec{A}}^{\alpha}_{-q} 
 \cdot {\vec{A}}^{\alpha^{\prime}}_{q}  
 + {\vec{q}}^2 [ \underline{\tilde{h}}_{q}]^{\alpha \alpha^{\prime}}  
 \lambda^{\alpha}_{- q} \lambda^{\alpha^{\prime}}_{q}
 \right] 
 \; .
 \label{eq:Stilde2A}
 \end{equation}
The unrestricted Gaussian integrations are now easily done, and we finally
arrive at Eq.\ref{eq:D0gfij}.

\subsection{The effective matter 
action\index{effective action!matter action}}

{\it{ \ldots can be obtained by integrating first over the
gauge field.}} 

\vspace{7mm}

\noindent
To see the connection with the
conventional many-body approach more clearly,
it is instructive to calculate the effective interaction 
between the matter degrees of freedom mediated by the gauge field.
Performing in Eq.\ref{eq:Gpatchgfdef} the integration over the gauge field first, 
we obtain an expression
of the same form as Eq.\ref{eq:GmatFourier}, with matter action
$ S_{\rm mat} \{ \psi \} = 
S_{0} \{ \psi \} + S_{\rm int} \{ \psi \} $, where now
 \begin{equation}
 \hspace{-3mm}
 S_{\rm int} \{ \psi \}  = 
 - \ln \left( 
 \int {\cal{D}} \{ \phi^{\alpha}  \} 
 \int {\cal{D}} \{ {\vec{A}}^{\alpha}  \} 
 \E^{ -
 S_{1} \{ \psi , \phi^{\alpha} , {\vec{A}}^{\alpha} \} - 
 S_{2} \{ \phi^{\alpha} , {\vec{ A}}^{\alpha} \} }
 \right)
 \label{eq:Sintdef}
 \; \; \; .
 \end{equation}
{\it{In Coulomb gauge}} the integration over
the longitudinal component $\phi^{\alpha}$ is trivial, and corresponds
just to undoing the Hubbard-Stratonovich transformation  
of Chap.~\secref{sec:HS1}.
Hence,
 \begin{equation}
 S_{\rm int} \{ \psi \}  =  \frac{ 1 }{2} \sum_{{q}}  \sum_{ \alpha \alpha^{\prime} }
 {\tilde{f}}_{{q}}^{\alpha \alpha^{\prime}}  
 {\rho}_{ - {q}}^{\alpha} {\rho}_{q}^{\alpha^{\prime}} 
 + S_{\rm int}^{\rm rad} \{ \psi \}   
 \; \; \; ,
 \label{eq:Sintdecomp}
 \end{equation}
with
 \begin{equation}
 S_{\rm int}^{\rm rad} \{ \psi \}  = 
 - \ln \left( \int {\cal{D}} \{ {\vec{A}}^{\alpha} \} 
 \E^{ - S_{3} \{ \psi , {\vec{A}}^{\alpha} \}  }
 \right)
 \label{eq:Sintraddef}
 \; \; \; ,
 \end{equation}
where the integration is subject to the constraint ${\vec{q}} \cdot {\vec{A}}_{q}^{\alpha} = 0$, and
 \begin{eqnarray}
  S_{3} \{ \psi , {\vec{A}}^{\alpha} \}  & = &
  \frac{1}{2} \sum_{q } \sum_{\alpha \alpha^{\prime}}
   [ \underline{\tilde{h}}_{q}^{-1}]^{\alpha \alpha^{\prime}}  
  {\vec{A}}^{\alpha}_{-q} \cdot {\vec{A}}^{\alpha^{\prime}}_{q} 
  -  \sum_{q} \sum_{\alpha} {\vec{j}}^{{\rm para}, \alpha}_{q} \cdot {\vec{A}}^{\alpha}_{-q} 
  \nonumber
  \\
   & + & \frac{1}{2mc^2 \beta} \sum_{ q  q^{\prime}} \sum_{\alpha} 
   \rho^{\alpha}_{q - q^{\prime}} 
  {\vec{A}}^{\alpha}_{-q} \cdot {\vec{A}}^{\alpha}_{q^{\prime}} 
  \label{eq:Srad}
  \; \; \; .
  \end{eqnarray}
Because the diamagnetic part of the current density gives rise to a term
in Eq.\ref{eq:Srad}
which is not diagonal in momentum space, the functional integration 
in Eq.\ref{eq:Sintraddef}
cannot be carried out exactly.
The higher order diamagnetic contributions\index{diamagnetism}
generate also {\it{retarded density-density interactions}}, which
should be combined with the static Coulomb interaction due to the
longitudinal component of the gauge field.
For the Maxwell field
these corrections are of higher order in
${e^2}/{c} \approx \frac{1}{137}$,
so that it is allowed to ignore them.
Calculating the action
$S_{\rm int}^{\rm rad} \{ \psi \} $ perturbatively,
we find to leading order that the transverse gauge field 
generates the following
effective action for the matter degrees of freedom,
 \begin{eqnarray}
 S_{\rm int}^{\rm rad} \{ \psi \}   & \approx &
 - \frac{1}{2} \sum_{q} \sum_{\alpha \alpha^{\prime}} \sum_{ij }  
 {{j}}^{{\rm para}, \alpha}_{-q,i} 
 D^{\alpha \alpha^{\prime}}_{0, ij} (q)
  {{j}}^{{\rm para} , \alpha^{\prime}}_{q,j}  
  \nonumber
  \\
 &  & \hspace{-13mm} =
 -  \frac{1}{2} \sum_{q}  \sum_{\alpha \alpha^{\prime} }
 \sum_{ij}
  j^{{\rm para} ,\alpha}_{-q,i} 
  [ \underline{\tilde{h}}_{ {{q}} }  ]^{\alpha \alpha^{\prime} } 
 \left[ \delta_{ij} -  (  \hat{\vec{e}}_{i} \cdot \hat{\vec{q}} )
 (  \hat{\vec{e}}_{j}  \cdot \hat{\vec{q}} ) \right]
  j^{{\rm para} ,\alpha^{\prime}}_{q,j}
 \label{eq:Sradpert}
 \; .
 \end{eqnarray}
Hence, the coupling between radiation field and matter gives rise to an 
effective interaction between the transverse parts of the paramagnetic
current densities \cite{Pethick89,Baym90}.
\index{singular interactions!current-current}
For $d=3$ we may use the vector product to rewrite the second line in Eq.\ref{eq:Sradpert} as
 \begin{equation}
 S_{\rm int}^{\rm rad} \{ \psi \}    \approx 
 - \frac{1}{2} \sum_{q} \sum_{\alpha \alpha^{\prime}} 
  [ \underline{\tilde{h}}_{ {{q}} }  ]^{\alpha \alpha^{\prime} } 
  ( \hat{\vec{q}} \times {\vec{j}}^{{\rm para} , \alpha }_{-q} ) \cdot
 ( \hat{\vec{q}} \times {\vec{j}}^{{\rm para} , \alpha^{\prime}}_{ q} )
 \; \; \; .
 \end{equation}
In a conventional many-body approach, one would now treat the effective
two-body
interactions in $S_{\rm int} \{ \psi  \}$ perturbatively \cite{Reizer89}.
However, a priori such an expansion cannot be justified,
because the interaction
becomes arbitrary large for small wave-vectors and frequencies.
In the case of the longitudinal component of the gauge field the physics 
of screening comes as a rescue. 
By performing an infinite resummation of a formally divergent series
(which is of course nothing but the
RPA for the effective density-density 
interaction \cite{Fetter71,Mattuck67}),
it is possible to formulate the perturbative expansion such that
at high densities the effective expansion parameter is small.
Unfortunately, this strategy fails in the case of the
effective current-current  interaction mediated by the transverse
radiation field, because in the static limit\footnote{ 
Note that at {\it{finite frequencies}} 
transverse gauge fields are dynamically screened,
see Eq.\ref{eq:epsilontrans3} below. 
For the Maxwell field this is called the
{\it{skin effect}}\index{skin effect}.}
transverse gauge fields
are not screened as long as the gauge invariance is
not spontaneously broken\index{screening!and gauge invariance}. 
Therefore the conventional many-body approach
fails as far as the perturbative calculation
of the effect of $S_{\rm int}^{\rm rad} \{ \psi \}$ 
on the single-particle Green's function is concerned.
This has first been noticed by Holstein, Norton and Pincus \cite{Holstein73},
and has been discussed later in more detail by Reizer \cite{Reizer89}.

\subsection{The effective gauge field 
action\index{effective action!gauge field}} 
 
{\it{\ldots can be obtained by integrating first over the matter field.
This is what we need for our functional bosonization approach.}}

\vspace{7mm}

\noindent
If we integrate in Eq.\ref{eq:Gpatchgfdef} first over the Grassmann fields, we obtain, 
in complete analogy with Eqs.\ref{eq:avphi}--\ref{eq:Skinphidef},
 \begin{eqnarray}
 G( k )  & = &
 \int {\cal{D}} \{ \phi^{\alpha}  \} 
  {\cal{D}} \{  {\vec{A}}^{\alpha} \} 
 {\cal{P}} \{ \phi^{\alpha} , {\vec{A}}^{\alpha} \} 
 [ \hat{G} ]_{kk}
 \equiv 
 \left< [ \hat{G}]_{  k  k }   \right>_{ S_{\rm eff} }
 \label{eq:avphigf}
 \;  ,
 \end{eqnarray}
where the probability distribution is now
 \begin{eqnarray}
 {\cal{P}} \{ \phi^{\alpha} , {\vec{A}}^{\alpha}  \} 
 & =  &
 \frac{  
 \E^{ - {S}_{\rm eff} \{ \phi^{\alpha} , {\vec{A}}^{\alpha} \} }  }
 {
 \int {\cal{D}} \{ \phi^{\alpha} \} 
  {\cal{D}} \{  {\vec{A}}^{\alpha} \} 
 \E^{ - {S}_{\rm eff} \{ \phi^{\alpha} , {\vec{A}}^{\alpha} \} }  }
 \; \; \; ,
 \label{eq:probabphidefgf}
 \end{eqnarray}
\index{probability distribution!gauge field}
with
 \begin{eqnarray}
 {S}_{\rm eff} \{ \phi^{\alpha} , {\vec{A}}^{\alpha} \} 
 & = & {S}_{2} \{ \phi^{\alpha} , {\vec{A}}^{\alpha} \} 
 +  {S}_{\rm kin} \{ \phi^{\alpha} , {\vec{A}}^{\alpha} \} 
 \label{eq:Seffphidefgf}
 \; \; \; .
 \end{eqnarray}
The potential energy part ${S}_{2} \{ \phi^{\alpha} , {\vec{A}}^{\alpha} \} $ 
of the effective action is given in
Eq.\ref{eq:S2gf}, and the kinetic energy contribution is 
  \begin{equation}
  {S}_{\rm kin} \{ \phi^{\alpha} , {\vec{A}}^{\alpha} \} 
   =   - {\rm Tr} \ln [ 1 - \hat{G}_{0} \hat{V} ]  
  \label{eq:SkingfTrLog}
  \; \; \; .
  \end{equation}
The matrix elements of $\hat{V}$ are 
 \begin{equation}
 [ \hat{V} ]_{ k k^{\prime} }    = 
 \sum_{\alpha} 
   \Theta^{\alpha} ( {\vec{k}} ) V^{\alpha}_{k- k^{\prime}}
   \; \; \; ,
 \end{equation}
 \begin{equation}
 V^{\alpha}_{q }  =  \frac{1}{\beta} \left[ \I \phi^{\alpha}_{q } 
 - {\vec{u}}^{\alpha} \cdot 
 {\vec{A}}^{\alpha}_{q   } +
 \frac{ 1}{2 mc^2 \beta} \sum_{q^{\prime \prime}} 
 {\vec{A}}^{\alpha}_{-q^{\prime \prime}} \cdot {\vec{A}}^{\alpha}_{q + q^{\prime \prime}}
 \right]
 \label{eq:Valphadef2}
 \; \; \; .
 \end{equation}
Here ${\vec{u}}^{\alpha}
=  {\vec{k}}^{\alpha} / ({mc})$
is a dimensionless vector with magnitude of the order of $v_{\rm F} / c$.
The infinite matrix $\hat{G}$ is defined as in
Eq.\ref{eq:GhatDysonshift}, with $V^{\alpha}_{q}$ now given
in Eq.\ref{eq:Valphadef2}.
For the calculation of the kinetic energy contribution to the
effective gauge field action we shall use the
Gaussian approximation. Note that
the generalized closed loop theorem
discussed in Chap.~\secref{sec:closedloop} implies that,
at least in certain parameter regimes (at high densities 
and at long wavelengths), the corrections to the Gaussian approximation
are small.  In this case the
expansion of the logarithm in Eq.\ref{eq:SkingfTrLog} 
can be truncated at the second order, so that
we may approximate (see Eq.\ref{eq:Trloggauss})
 \begin{eqnarray}
  {S}_{\rm kin} \{ \phi^{\alpha} , {\vec{A}}^{\alpha} \} 
  & \approx &
   {\rm Tr} \left[ \hat{G}_{0} \hat{V} \right]
  + \frac{1}{2}
  {\rm Tr} \left[ \hat{G}_{0} \hat{V} \right]^2
  \nonumber
  \\
  & \equiv &
  {S}_{{\rm kin},1} \{ \phi^{\alpha} , {\vec{A}}^{\alpha} \} 
  +
  {S}_{{\rm kin},2} \{ \phi^{\alpha} , {\vec{A}}^{\alpha} \} 
 \label{eq:Trloggaussgf}
 \; \; \; .
 \end{eqnarray}
The first term yields
 \begin{eqnarray}
 {S}_{{\rm kin} , 1} \{ \phi^{\alpha} , {\vec{A}}^{\alpha} \}
   & = & \sum_{\alpha} N^{\alpha}_{0} V^{\alpha}_{0}
   \nonumber
   \\
  &  & \hspace{-13mm} = \sum_{\alpha}
  N^{\alpha}_{0}
  \left[ \I  \phi^{\alpha}_{0} -  {\vec{u}}^{\alpha} \cdot {\vec{A}}^{\alpha}_{0}
  + \frac{ 1}{2 m c^2 \beta} \sum_{q} 
  {\vec{A}}^{\alpha}_{-q} \cdot {\vec{A}}^{\alpha}_{q}
  \right]
  \label{eq:Skin1gf}
  \; ,
  \end{eqnarray}
where $N^{\alpha}_{0}$ is the number of occupied states in sector
$K^{\alpha}_{\Lambda , \lambda}$, see Eq.\ref{eq:N0def}. 
If we neglect the terms with the transverse gauge field, 
Eq.\ref{eq:Skin1gf} reduces to $S_{{\rm kin},1} \{ \phi^{\alpha} \}$, 
see Eq.\ref{eq:Skin1}.
As already mentioned in the first footnote of Chap.~\secref{chap:a4bos},
the terms involving $\phi_0^{\alpha}$ and ${\vec{A}}^{\alpha}_{0}$ do not
contribute to fermionic correlation functions at zero temperature, and 
can be ignored for our purpose.
Note, however, that the last term in Eq.\ref{eq:Skin1gf} is quadratic and has to be retained
within the Gaussian approximation. This {\it{diamagnetic}}
contribution to the effective gauge field action can be written as
\index{effective action!diamagnetic contribution to gauge field}
 \begin{equation}
 {S}_{{\rm kin},1}^{\rm dia} \{ {\vec{A}}^{\alpha} \} 
 = \frac{1}{2} \sum_{q} \sum_{\alpha} \tilde{\Delta}^{\alpha}
  {\vec{A}}^{\alpha}_{-q} \cdot {\vec{A}}^{\alpha}_{q}
 \label{eq:SkindiaA}
 \; \; \; , \; \; \; 
 \tilde{\Delta}^{\alpha} = \frac{ N^{\alpha}_{0}}{\beta mc^2} 
 \; \; \; .
 \end{equation}
The second order term in Eq.\ref{eq:Trloggaussgf} is 
 \begin{equation}
 {S}_{{\rm kin} , 2} \{ \phi^{\alpha} , {\vec{A}}^{\alpha} \}
 = - \frac{ \beta^2  }{2}
    \sum_{q} \sum_{\alpha} \tilde{\Pi}^{\alpha}_{0} (q) V^{\alpha}_{-q} V^{\alpha}_{q}
  \label{eq:tr2gf}
  \; \; \; ,
  \end{equation}
where
 $\tilde{\Pi}^{\alpha}_{0} (q) = \frac{V}{\beta}
 {\Pi}^{\alpha}_{0} (q)$ 
is the dimensionless sector
polarization\footnote{See Eqs.\ref{eq:U2res} and \ref{eq:Pilong};
for simplicity we have assumed sufficiently small 
$| \vec{q} | / k_{\rm F}$ and large
sectors $K^{\alpha}_{\Lambda , \lambda}$, so that
only the diagonal element of Eq.\ref{eq:U2res} has to be retained.}.
From Eq.\ref{eq:Valphadef2} it is clear that Eq.\ref{eq:tr2gf}
contains also terms that are cubic and quartic in the fields.
The origin for these non-Gaussian terms are the diamagnetic fluctuations
described by the last term in Eq.\ref{eq:Valphadef2}. 
Within the Gaussian approximation we shall simply ignore these terms.
It is important to stress, however, that the closed loop
theorem does not imply the cancellation of these terms,
because it applies to the total field
$V^{\alpha}_{q}$.
Thus, within the Gaussian approximation we have
\index{Gaussian approximation!for gauge field action}
\index{effective action!paramagnetic contribution to gauge field}
 \begin{equation}
 {S}_{{\rm kin} , 2} \{ \phi^{\alpha} , {\vec{A}}^{\alpha} \}
 \approx 
 {S}_{{\rm kin},2 } \{ \phi^{\alpha} \}
 + {S}_{{\rm kin},2}^{\rm para} \{ {\vec{A}}^{\alpha} \}
 + {S}_{{\rm kin},2}^{\rm mix} \{ \phi^{\alpha} , {\vec{A}}^{\alpha} \}
 \label{eq:Skin2gauss}
 \; \; \; ,
 \end{equation}
where
 ${S}_{{\rm kin},2} \{ \phi^{\alpha} \}$ is given
in Eq.\ref{eq:Skin2phires} (with
$\tilde{\Pi}^{\alpha \alpha^{\prime}}_{0} (q ) \approx
\delta^{\alpha \alpha^{\prime}} \tilde{\Pi}^{\alpha}_{0} ( q )$)
and
 \begin{eqnarray}
 {S}_{{\rm kin},2}^{\rm para} \{ {\vec{A}}^{\alpha} \}
 & = &
 - \frac{1}{2} 
 \sum_{q} \sum_{\alpha } 
 \tilde{\Pi}_{0}^{\alpha } (q)
 ( {\vec{u}}^{\alpha} \cdot {\vec{A}}^{\alpha}_{-q}   )
 ( {\vec{u}}^{\alpha} \cdot  {\vec{A}}^{\alpha}_{q}  )
 \label{eq:SkinparaA}
 \; \; \; ,
 \\
 {S}_{{\rm kin},2}^{\rm mix} \{ \phi^{\alpha} , {\vec{A}}^{\alpha} \} & = &
 \I 
 \sum_{q} \sum_{\alpha } 
 \tilde{\Pi}_{0}^{\alpha } (q)
 ( {\vec{u}}^{\alpha} \cdot {\vec{A}}^{\alpha}_{-q}   )
 \phi^{\alpha}_{q}
 \; \; \; .
 \label{eq:Skinmix}
 \end{eqnarray}
Collecting all quadratic terms, we obtain for the
effective gauge field action defined in Eq.\ref{eq:Seffphidefgf}
within the Gaussian approximation
 \begin{equation}
 {S}_{{\rm eff},2} \{ \phi^{\alpha} , {\vec{A}}^{\alpha} \} =
 {S}_{{\rm eff},2} \{ \phi^{\alpha} \} +
 {S}_{{\rm eff},2} \{ {\vec{A}}^{\alpha} \} 
 +
 {S}_{{\rm kin},2}^{\rm mix} \{ \phi^{\alpha} , {\vec{A}}^{\alpha} \} 
 \; \; \; ,
 \label{eq:Sgau}
 \end{equation}
with
 \begin{eqnarray}
 {S}_{{\rm eff},2} \{ \phi^{\alpha} \} 
 & = &
 \frac{1}{2} \sum_{q} \sum_{\alpha \alpha^{\prime}}
 \phi^{\alpha}_{-q} [ 
 ( {\underline{\tilde{f}}_{q}^{\rm RPA} } )^{-1} ]^{\alpha \alpha^{\prime} } 
 \phi^{\alpha^{\prime}}_{q}
 \; \; \; ,
 \label{eq:Seffphi}
 \\
 {S}_{{\rm eff},2} \{ {\vec{A}}^{\alpha} \} 
 & = &
 \frac{1}{2} \sum_{q} \sum_{\alpha \alpha^{\prime}} \sum_{ij }
 A^{\alpha}_{-q,i} [ ( { \underline{\tilde{h}}_{q}^{\rm RPA} })^{-1} 
 ]^{\alpha \alpha^{\prime} }_{ij} 
 A^{\alpha^{\prime}}_{q,j}
 \label{eq:SeffA}
 \; \; \; .
 \end{eqnarray}
Here the matrix $\underline{\tilde{f}}^{\rm RPA}_{q} = \frac{\beta}{V}
\underline{{f}}^{\rm RPA}_{q}$
is the rescaled RPA interaction matrix
(see also Eqs.\ref{eq:gaupropphi} and \ref{eq:frpapatchdef}),
 \begin{equation}
 [ ({ \underline{\tilde{f}}^{\rm RPA}_{q} })^{-1} ]^{ \alpha \alpha^{\prime} }
 = [ \underline{\tilde{f}}_{{q}}^{-1} ]^{\alpha \alpha^{\prime} }
 + \delta^{ \alpha \alpha^{\prime} } \tilde{\Pi}_{0}^{\alpha} (q) 
 \; \; \; ,
 \label{eq:frpatildedef}
 \end{equation}
and $ ({\underline{\tilde{h}}^{\rm RPA}_{q} })^{-1}$ is the
following  matrix in the sector and coordinate labels, 
\index{gauge field propagator!RPA}
 \begin{equation}
 [ ({  \underline{\tilde{h}}^{\rm RPA}_{q} })^{-1} ]^{ \alpha \alpha^{\prime} }_{ij}
 = \delta_{ij} [ \underline{\tilde{h}}_{{q}}^{-1} ]^{\alpha \alpha^{\prime} }
 + \delta^{ \alpha \alpha^{\prime} } \left[
 \delta_{ij} \tilde{\Delta}^{\alpha} - u^{\alpha}_{i} u^{\alpha}_{j} \tilde{\Pi}_{0}^{\alpha} (q) 
 \right]
 \; \; \; ,
 \label{eq:hrpadef}
 \end{equation}
where
 \begin{equation}
 u^{\alpha}_{i} = { \vec{e}}_{i} \cdot {\vec{u}}^{\alpha} 
 \; \; \; , \; \; \; 
{\vec{u}}^{\alpha} = \frac{ \vec{k}^{\alpha} }{  m c }
\; \; \; .
\label{eq:ualphaidefdef}
\end{equation}
The diamagnetic term $\tilde{\Delta}^{\alpha}$ in Eq.\ref{eq:hrpadef} 
represents the increase in energy due to diamagnetic fluctuations of the 
transverse gauge field\index{diamagnetism}, while the last term represents the lowering of the
energy due to paramagnetism\index{paramagnetism}. In Sect.~\secref{sec:TransDW}
we shall show that in the static limit there exists an exact
cancellation between these two terms, so that the transverse gauge field is not screened.
The action
 ${S}_{{\rm kin},2}^{\rm mix} \{ \phi^{\alpha} , {\vec{A}}^{\alpha} \} $ 
 describes the mixing between
longitudinal and transverse components of the gauge field, which arises due to the presence
of the matter degrees of freedom. Note that {\it{in Coulomb gauge}}
the isolated gauge field action
${S}_{2} \{ \phi^{\alpha} , {\vec{A}}^{\alpha} \}$ does not contain such a mixing term.
In Sect.~\secref{sec:TransDW} we shall show that in the special case
when the elements of the interaction matrices 
$\underline{ \tilde{f}}_{q}$ and
$\underline{ \tilde{h}}_{q}$ are constants independent of the
patch indices, this mixing term does not contribute 
to the final expression for the Green's function. 

\section{The Green's function in Gaussian approximation}
\label{sec:Derivationrad}

{\it{Using our background field method 
described in Chap.~\secref{chap:agreen},
we derive a non-perturbative expression for the 
single-particle Green's function in 
Coulomb gauge. Gauge fixing is again
imposed with the help of the Fadeev-Popov method.
We use the Gaussian approximation, but work with non-linear energy dispersion.
}}

\subsection{The Green's function for fixed gauge field} 

{\it{For simplicity let us first consider the case of linearized energy dispersion,
and then discuss the modifications due to the quadratic term
in the energy dispersion.}}

\vspace{7mm}

\noindent
For linearized energy dispersion
we may copy the results of Chap.~\secref{sec:Derivation}.
To obtain the Green's function from
Eq.\ref{eq:avphigf}, we first need to calculate the 
diagonal matrix elements $[ \hat{G} ]_{kk}$ for a fixed
configuration of the gauge fields.
Obviously this can be done in precisely the same way as
described in Chap.~\secref{subsec:invdiag};
we simply should substitute
the modified form \ref{eq:Valphadef2} of the
potential $V^{\alpha}_{q}$ into the expression for 
$\hat{G}^{-1}$ given in Eq.\ref{eq:GhatDysonshift}.
Using Eqs.\ref{eq:Galphaexpinv}, \ref{eq:Ansatz} and \ref{eq:Gkres1},
we obtain for the
interacting Matsubara Green's function 
within the Gaussian approximation
 \begin{eqnarray}
 G (k) & = & \sum_{\alpha} \Theta^{\alpha} ( {\vec{k}} )
 \int \D {\vec{r}} \int_{0}^{\beta} \D \tau 
 \E^{ - \I [ ( {\vec{k}} - {\vec{k}}^{\alpha} ) \cdot  {\vec{r}}
 - \tilde{\omega}_{n}  \tau  ] }
 \nonumber
 \\
 & \times &
 {{G}}^{\alpha}_{0} ( {\vec{r}}  , \tau  )
 \left< \E^{ \Phi^{\alpha} ( {\vec{r}} , \tau ) - \Phi^{\alpha} ( 0 , 0 ) } 
 \right>_{S_{{\rm eff},2}}
 \; \; \; ,
 \label{eq:Gkres3}
 \end{eqnarray}
where now, in complete analogy with Eq.\ref{eq:Phidifer},
 \begin{eqnarray}
 \Phi^{\alpha} ( {\vec{r}} , \tau ) - \Phi^{\alpha} ( 0 , 0 ) &  = &
  \nonumber
  \\
  & & \hspace{-30mm}  
  \sum_{q} 
  {\cal{J}}^{\alpha}_{-q} ( {\vec{r}} , \tau )
  \left[ \phi^{\alpha}_{q} + \I {\vec{u}}^{\alpha} \cdot {\vec{A}}^{\alpha}_{q}
  -   \frac{\I }{2 mc^2 \beta} \sum_{q^{\prime \prime }}
  {\vec{A}}^{\alpha}_{ - q^{\prime \prime } } \cdot {\vec{A}}^{\alpha}_{q+q^{\prime \prime}}
  \right]
  \label{eq:Phidifer2}
  \; ,
  \end{eqnarray}
with ${\cal{J}}^{\alpha}_{q} ( {\vec{r}} , \tau )$ given in Eq.\ref{eq:Jdef}.
The last term in Eq.\ref{eq:Phidifer2} it is not diagonal in momentum space,
and represents higher order diamagnetic
fluctuations beyond the RPA.
Because in our derivation of the effective 
gauge field action we have already ignored these higher order 
fluctuations, it is consistent to drop this term here as well.

From Chap.~\secref{sec:backfixed} we know that 
for fermions with {\it{quadratic energy dispersion}}
Eq.\ref{eq:Gkres3} should be replaced by
 \begin{eqnarray}
 G (k) & = & \sum_{\alpha} \Theta^{\alpha} ( {\vec{k}} )
 \int \D {\vec{r}} \int_{0}^{\beta} \D \tau 
 \E^{ - \I [ ( {\vec{k}} - {\vec{k}}^{\alpha} ) \cdot  {\vec{r}}
 - \tilde{\omega}_{n}  \tau  ] }
 \nonumber
 \\
 & \times &
 \left< {\cal{G}}^{\alpha}_{1} ( {\vec{r}} ,0 , \tau ,0 )
 \E^{ \Phi^{\alpha} ( {\vec{r}} , \tau ) - \Phi^{\alpha} ( 0 , 0 ) } 
 \right>_{S_{{\rm eff},2}}
 \; \; \; ,
 \label{eq:Gkres5}
 \end{eqnarray}
where the functional $ \Phi^{\alpha} ( {\vec{r}} , \tau )$ 
satisfies the eikonal equation \ref{eq:difPhi}, and the Green's function
 ${\cal{G}}^{\alpha}_{1} ( {\vec{r}} , {\vec{r}}^{\prime} , \tau , \tau^{\prime} )$
is the solution of the differential equation \ref{eq:difG1}.
The potential $V^{\alpha} ( {\vec{r}} , \tau )$ in these expressions should now be
identified with the Fourier transform of Eq.\ref{eq:Valphadef2}.
Because in this chapter we would like to restrict ourselves to the
Gaussian approximation, it is consistent to truncate the 
eikonal expansion \ref{eq:Phiansatz} at the first order. 
In this case
$\Phi^{\alpha} ( {\vec{r}} , \tau ) - \Phi^{\alpha} ( 0 , 0 ) $ 
is formally identical with Eq.\ref{eq:Phidifer2}, except that
${\cal{J}}^{\alpha}_{q} ( {\vec{r}} , \tau )$ is now defined in Eq.\ref{eq:tildeJdef}.

\subsection{Gaussian averaging} 

{\it{
We would like to emphasize again that we do not linearize the energy dispersion,
because later we shall show that
in the case of transverse gauge fields the curvature of the Fermi surface
qualitatively changes the
long-distance behavior of the Debye-Waller factor.}}

\vspace{7mm}

\noindent
Let us begin with the calculation of the average eikonal
$Q^{\alpha} ( \vec{r} , \tau )$.
Within the Gaussian
approximation $Q^{\alpha} ( \vec{r} , \tau )$ is given by
(see Eq.\ref{eq:Qeikdef})
 \begin{equation}
 \E^{Q^{\alpha} ( \vec{r} ,\tau )} =
 \left< \E^{ \Phi^{\alpha} ( {\vec{r}} , \tau ) - \Phi^{\alpha} ( 0 , 0 ) } 
 \right>_{S_{{\rm eff},2}} 
 \; \; \; .
 \label{eq:Qeikgaugegauss}
 \end{equation}
It is convenient to integrate first over the longitudinal
field $\phi^{\alpha}$ before averaging over the transverse
components $\vec{A}^{\alpha}$ of the gauge field.
Because of the coupling between the
longitudinal and transverse fields in
${S}^{\rm mix}_{{\rm kin},2} \{ \phi^{\alpha} , {\vec{A}}^{\alpha} \}$, 
the integration over the 
$\phi^{\alpha}$-field generates also a contribution to the effective
action for the transverse gauge fields.
From Eq.\ref{eq:Skinmix} we have
 \begin{eqnarray}
 \lefteqn{
  \sum_{q} 
 {\cal{J}}^{\alpha}_{-q} ( {\vec{r}} , \tau ) \phi^{\alpha}_{q}
 - {S}_{{\rm kin},2}^{\rm mix} \{ \phi^{\alpha} , {\vec{A}}^{\alpha} \} 
 }
 \nonumber
 \\
 & = &
 \sum_{q} \sum_{\alpha^{\prime}} 
 \left[
 \delta^{\alpha^{\prime} \alpha } 
 {\cal{J}}^{\alpha}_{-q}  ( {\vec{r}} , \tau )
 - \I \tilde{\Pi}_{0}^{\alpha^{\prime}} ( q ) 
 {\vec{u}}^{\alpha^{\prime}} \cdot {\vec{A}}^{\alpha^{\prime}}_{-q}
 \right] 
 \phi^{\alpha^{\prime}}_{q}
 \; \; \; ,
 \label{eq:mixJ}
 \end{eqnarray}
so that the $\phi^{\alpha}$-integration in Eq.\ref{eq:Qeikgaugegauss} yields
 \begin{eqnarray}
 & & \hspace{-7mm}
  \int {\cal{D}} \{ \phi^{\alpha} \}
 \exp \left[
  - {S}_{{\rm eff},2} \{ \phi^{\alpha} \} - 
  {S}_{{\rm kin},2}^{\rm mix} \{ \phi^{\alpha} , {\vec{A}}^{\alpha} \}
 + \sum_{q} 
 {\cal{J}}^{\alpha}_{-q} ( {\vec{r}} , \tau ) \phi^{\alpha}_{q} \right]
  =  {\rm const } \times
 \nonumber
 \\
 & & \hspace{-2mm}
  \exp \left\{ 
 \frac{1}{2} \sum_{q} \sum_{\alpha^{\prime} \alpha^{\prime \prime}}
 \left< \phi^{\alpha^{\prime}}_{q} 
 \phi^{\alpha^{\prime \prime}}_{-q} \right>_{S_{{\rm eff},2}}
 \left[
 \delta^{ \alpha^{\prime} \alpha } 
 {\cal{J}}^{\alpha}_{-q}  ( {\vec{r}} , \tau )
 - \I \tilde{\Pi}_{0}^{\alpha^{\prime}} ( q ) 
 {\vec{u}}^{\alpha^{\prime}} \cdot {\vec{A}}^{\alpha^{\prime}}_{-q}
 \right] 
 \right.
 \nonumber
 \\
 & & \hspace{24mm}
 \left.
 \times
 \left[
 \delta^{ \alpha^{\prime \prime } \alpha} 
 {\cal{J}}^{\alpha}_{q}  ( {\vec{r}} , \tau )
 - \I \tilde{\Pi}_{0}^{\alpha^{\prime \prime }} ( q ) {\vec{u}}^{\alpha^{\prime \prime }} 
 \cdot {\vec{A}}^{\alpha^{\prime \prime }}_{q}
 \right] 
 \right\}
 \label{eq:DebeyeWaller2}
 \;  .
 \end{eqnarray}
From Eq.\ref{eq:phiphiprop} we know that
the Gaussian propagator of the $\phi^{\alpha}$-field is simply given by the
rescaled RPA interaction (see also Eq.\ref{eq:frpatildedef}),
so that
 \begin{eqnarray}
 \left< \E^{ \Phi^{\alpha} ( {\vec{r}} , \tau ) - \Phi^{\alpha} ( 0 , 0 ) } 
 \right>_{S_{{\rm eff},2}}
  & = &  \E^{ Q^{\alpha}_1 ( {\vec{r}} , \tau ) } 
  \nonumber
  \\
  & & \hspace{-40mm} \times
 \frac{ \int {\cal{D}} \left\{ {\vec{A}}^{\alpha} \right\} 
 \exp \left[ { - {S}_{{\rm eff},2}^{\prime} \{ {\vec{A}}^{\alpha} \}  
 + \I \sum_{q \alpha^{\prime} i}   
 {\cal{K}}^{\alpha \alpha^{\prime}}_{-q,i} (  {\vec{r}} , \tau ) 
  {{A}}^{\alpha^{\prime}}_{q, i}  } \right]
 } 
 {
 \int {\cal{D}} \{ {\vec{A}}^{\alpha} \} 
 \exp \left[ { - {S}_{{\rm eff},2}^{\prime} \{ {\vec{A}}^{\alpha} \} } \right]  }
 \; \; \; ,
 \label{eq:gauss1}
 \end{eqnarray}
where the Debye-Waller factor $Q^{\alpha}_1 ( {\vec{r}} , \tau )$
due to the longitudinal component of the gauge field
is given in Eqs.\ref{eq:Qlondef2}--\ref{eq:Slondef2}, and
 \begin{equation}
 {\cal{K}}^{\alpha \alpha^{\prime}}_{q,i} (  {\vec{r}} , \tau )  = 
 {\cal{J}}^{\alpha}_{q} (  {\vec{r}} , \tau ) 
 U_{q , i}^{\alpha \alpha^{\prime}} 
 \label{eq:Kappadef}
 \; \; \; ,
 \end{equation}
with
 \begin{equation}
 U_{q,i}^{\alpha \alpha^{\prime}} 
  = 
 u^{\alpha}_{i} \delta^{\alpha \alpha^{\prime}}
  -
  u^{\alpha^{\prime}}_{i}
  \tilde{\Pi}^{\alpha^{\prime} }_{0} (q) 
  [ \underline{\tilde{f}}_{q}^{\rm RPA} ]^{\alpha \alpha^{\prime} }
  \; \; \; .
  \label{eq:Ualphadef}
 \end{equation}
Note that the label $\alpha$  of
 ${\cal{K}}^{\alpha \alpha^{\prime}}_{-q,i} (  {\vec{r}} , \tau ) $ 
in Eq.\ref{eq:gauss1} is an external label, and {\it{not}} a summation label.
The renormalized Gaussian action 
 ${S}_{{\rm eff},2}^{\prime} \{ {\vec{A}}^{\alpha} \} $
differs from the action
 ${S}_{{\rm eff},2} \{ {\vec{A}}^{\alpha} \} $
given in Eq.\ref{eq:SeffA} by an additional term that is generated
because of the coupling between the ${\phi}^{\alpha}$- and  ${\vec{A}}^{\alpha}$-fields
in $S_{{\rm kin},2}^{\rm mix} \{ \phi^{\alpha} , {\vec{A}}^{\alpha} \}$,
\begin{eqnarray}
 {S}_{{\rm eff},2}^{\prime} \{ {\vec{A}}^{\alpha} \} 
  & =  &
 {S}_{{\rm eff},2} \{ {\vec{A}}^{\alpha} \} 
  \nonumber
  \\
 & + & \frac{1}{2} \sum_{q} \sum_{\alpha \alpha^{\prime}}
  [ \underline{\tilde{f}}_{q}^{\rm RPA} ]^{\alpha \alpha^{\prime} }
  \tilde{\Pi}^{\alpha }_{0} (q) 
  \tilde{\Pi}^{\alpha^{\prime} }_{0} (q) 
  ( {\vec{u}}^{\alpha} \cdot {\vec{A}}^{\alpha}_{-q} )
  ( {\vec{u}}^{\alpha^{\prime}} \cdot {\vec{A}}^{\alpha^{\prime}}_{ q} )
  \nonumber
  \\
  & \equiv &
 \frac{1}{2} \sum_{q} \sum_{\alpha \alpha^{\prime}} \sum_{ij }
 [  \underline{\tens{H}}_{q}^{-1}   ]^{\alpha \alpha^{\prime} }_{ij} 
 A^{\alpha}_{-q,i} 
 A^{\alpha^{\prime}}_{q,j}
 \label{eq:SeffAprime}
 \; \; \; ,
 \end{eqnarray}
\index{effective action!gauge field}
where we have defined
 \begin{equation}
 [  \underline{\tens{H}}_{q} ^{-1} ]^{\alpha \alpha^{\prime} }_{ij}
  = 
 [ { \underline{\tilde{h}}_{q}^{\rm RPA} }^{-1} ]^{\alpha \alpha^{\prime} }_{ij}
 + u^{\alpha}_{i} u^{\alpha^{\prime}}_{j} \tilde{\Pi}_{0}^{\alpha} (q)
 \tilde{\Pi}_{0}^{\alpha^{\prime} } ({q}) 
 [ \underline{\tilde{f}}^{\rm RPA}_{q} ]^{\alpha \alpha^{\prime} }
 \label{eq:hRdef}
 \; \; \; .
 \end{equation}

Next, let us integrate over the transverse
gauge field in Eq.\ref{eq:gauss1}.
The Gaussian integration generates 
another Debye-Waller factor, so that \index{Debye-Waller factor!transverse gauge fields}
 \begin{equation}
 \left< \E^{ \Phi^{\alpha} ( {\vec{r}} , \tau ) - \Phi^{\alpha} ( 0 , 0 ) } 
 \right>_{S_{{\rm eff},2}}
  =  \E^{ Q^{\alpha}_1 ( {\vec{r}} , \tau ) }
   \E^{ Q^{\alpha}_{\rm tr} ( {\vec{r}} , \tau ) }
   \label{eq:avfinal2}
 \; \; \; ,
 \end{equation}
where
 \begin{equation}
 \hspace{-5mm}
 Q^{\alpha}_{\rm tr} ( {\vec{r}} , \tau ) 
 = - \frac{1}{2} \sum_{q} \sum_{\alpha^{\prime} \alpha^{\prime \prime}}
 \sum_{ij} 
\left< A^{\alpha^{\prime}}_{q,i} A^{\alpha^{\prime \prime}}_{-q , j } 
\right>_{ S_{{\rm eff},2}^{\prime} } 
 {\cal{K}}^{ \alpha \alpha^{\prime}}_{-q , i} (  {\vec{r}} , \tau )
 {\cal{K}}^{ \alpha \alpha^{\prime \prime}}_{q , j} (  {\vec{r}} , \tau )
 \; ,
 \label{eq:Qtrans}
 \end{equation}
with
 \begin{eqnarray}
\left< A^{\alpha}_{q,i} A^{\alpha^{ \prime}}_{-q , j } \right>_{ S_{{\rm eff},2}^{\prime} } 
& \equiv &
 [ D^{\rm RPA} (q) ]^{\alpha \alpha^{\prime} }_{ij }     
\nonumber
\\
  & =  &
  \frac{ 
 \int {\cal{D}} \left\{ {\vec{A}}^{\alpha} \right\} 
 \E^{- {S}_{{\rm eff},2}^{\prime}    
 \{ {\vec{A}}^{\alpha} \} }
 A_{q , i}^{\alpha} A_{- q , j}^{\alpha^{\prime}}
 } 
 { \int {\cal{D}} \left\{ {\vec{A}}^{\alpha}   \right\} 
 \E^{- {S}_{{\rm eff},2}^{ \prime}   
 \{ {\vec{A}}^{\alpha} \} }
 }  
 \label{eq:Drpaalphadef}
 \; \; \; .
 \end{eqnarray}
To calculate this propagator, we impose again the Coulomb gauge condition
by inserting functional $\delta$-functions in the
form \ref{eq:Coulombconstraint2} and then shifting
\index{Fadeev-Popov method}
\index{shift transformation}
 \begin{equation}
 A^{\alpha}_{q,i} \rightarrow  A^{ \alpha}_{q,i}
 - \sum_{\alpha^{\prime }} \sum_{j} [ \underline{\tens{H}}_{q} ]^{ \alpha \alpha^{\prime} }_{ij}
 q_{j} \lambda^{\alpha^{\prime}}_{q}
 \label{eq:shift2}
 \; \; \; .
 \end{equation}
This leads to the replacement
 \begin{eqnarray}
 S_{{\rm eff},2}^{{\prime} }  \{ {\vec{A}}^{\alpha} \} 
 + \sum_{q} \sum_{ \alpha} {\vec{A}}^{\alpha}_{-q  } \cdot {\vec{q}} \lambda^{\alpha}_{q }
  \rightarrow 
  \nonumber
  \\
  & & \hspace{-40mm}
  {S}_{{\rm eff},2}^{{\prime} }  \{ {\vec{A}}^{ \alpha} \} 
  + \frac{1}{2} \sum_{q} \sum_{\alpha \alpha^{\prime} }
  \lambda^{\alpha}_{-q} [ ( {\vec{q}} 
  \underline{\tens{H}}_{q}  {\vec{q}}) ]^{\alpha \alpha^{\prime}}
  \lambda^{\alpha^{\prime}}_{q}
  \label{eq:Seffshiftgf}
  \; \; \; ,
  \end{eqnarray}
where $ ( {\vec{q}}  \underline{\tens{H}}_{q}  {\vec{q}}) $ is a matrix in the patch labels, with
elements given by
 \begin{equation}
  [ ( {\vec{q}}  \underline{\tens{H}}_{q}  {\vec{q}} ) ]^{\alpha \alpha^{\prime}}
  = \sum_{i j} q_{i} [ \underline{\tens{H}}_{q} ]^{\alpha \alpha^{\prime} }_{ij} q_{j}
  \label{eq:notation}
  \; \; \; .
  \end{equation}
Performing the independent Gaussian integrations we finally obtain
\index{gauge field propagator!RPA}
 \begin{equation}
 \hspace{-5mm}
 [ D^{\rm RPA} ( q ) ]_{ij}^{\alpha \alpha^{\prime} } = [ \underline{\tens{H}}_{q} 
 ]^{\alpha \alpha^{\prime}}_{ij}
 -  
 \left[ ( {\vec{e}}_{i}  \underline{\tens{H}}_{-q}  {\vec{q}} )
 (  {\vec{q}}  \underline{\tens{H}}_{q}  {\vec{q}} )^{-1} 
 ( {\vec{q}}  \underline{\tens{H}}_{-q} {\vec{e}}_{j} ) \right]^{\alpha \alpha^{\prime}}
 \; ,
 \label{eq:gaugeprop}
 \end{equation}
where the product in the last term 
should be understood as a product of matrices in the patch indices.
Using Eq.\ref{eq:tildeJdef}, the transverse part of the average
eikonal can also be written as
 \begin{equation}
 Q_{\rm tr}^{\alpha} ( {\vec{r}}  , \tau ) = R^{\alpha}_{\rm tr} -
 S_{\rm tr}^{\alpha} ( {\vec{r}}  , \tau ) 
 \; \; \;  ,
 \label{eq:Qraddef}
 \end{equation}
with
 \begin{eqnarray}
 R_{\rm tr}^{\alpha} 
 & = & \frac{1}{\beta {{V}}} \sum_q
 \frac{ h^{{\rm RPA},\alpha}_q
   }
{ [ \I \omega_m - \xi^{\alpha}_{  {\vec{q}} } ][ \I \omega_m + 
\xi^{\alpha}_{  - {\vec{q}}} ] }
=
 S_{\rm tr}^{\alpha} ( 0  , 0 )
 \label{eq:Rraddef}
 \; \; \; ,
 \\
 S_{\rm tr}^{\alpha} ( {\vec{r}}  , \tau )
 & = & \frac{1}{\beta {{V}}} \sum_q
 \frac{ h^{{\rm RPA},\alpha}_q
  \cos ( {\vec{q}} \cdot {\vec{r}} - \omega_m \tau )  }
{ [ \I \omega_m - \xi^{\alpha}_{  {\vec{q}} } ][ \I \omega_m + 
\xi^{\alpha}_{ - {\vec{q}}} ] }
 \label{eq:Sraddef}
 \; \; \; .
 \end{eqnarray}
The effective interaction is
 \begin{equation}
  h^{{\rm RPA}, \alpha}_{q} =  -  \frac{V}{\beta} 
  \sum_{\alpha^{\prime} \alpha^{\prime \prime} } \sum_{ij} 
  U^{ \alpha \alpha^{\prime}}_{-q,i} 
  [ D^{\rm RPA} ( q ) ]^{\alpha^{\prime} \alpha^{\prime \prime}}_{ij} 
  {U}^{ \alpha \alpha^{\prime \prime }  }_{q,j} 
  \; \; \; .
  \label{eq:hrpaalpha}
  \end{equation}
Note that these equations are valid for arbitrary patch geometry and arbitrary
patch-dependent bare interactions $[\underline{\tilde{f}}_{{q}} ]^{\alpha \alpha^{\prime} }$
and $[\underline{\tilde{h}}_{q}]^{\alpha \alpha^{\prime} }$. 
In deriving Eq.\ref{eq:hrpaalpha} 
we have assumed that the effective mass tensor is 
proportional to the unit matrix.
To discuss quasi-one-dimensional anisotropic systems, it is necessary to allow
for different effective masses $m^{\alpha}_{i}$, $i = 1 , \ldots , d$. 
In this case Eq.\ref{eq:hrpaalpha} is still correct, provided
we take the different effective masses in the definition
of ${\vec{v}}^{\alpha}$ into account. 
Then we have
${\vec{v}}^{\alpha}  = ( {\tens{M}}^{\alpha} )^{-1} {\vec{k}}^{\alpha}$,
where the effective mass tensor $\tens{M}^{\alpha}$ is
defined in Eq.\ref{eq:Meffdef}.
Hence we should replace in Eq.\ref{eq:hrpadef}
 \begin{equation}
 \delta_{ij} \tilde{\Delta}^{\alpha}
  \rightarrow  \delta_{ij} \tilde{\Delta}^{\alpha}_{i} 
 \; \;  ,  \; \; \tilde{\Delta}^{\alpha}_{i} = 
 \frac{N_{0}^{\alpha}}{\beta m_{i}^{\alpha} c^2 }
 \; \; \; , \; \; \; 
 {\vec{u}}^{\alpha}
 \rightarrow ({\tens{M}}^{\alpha})^{-1} \frac{{\vec{k}}^{\alpha}}{  c}
 \label{eq:ualphairedef}
 \;  .
 \end{equation}
For the special case of patch-independent bare interactions
and {\it{linear energy dispersion}} (i.e. 
$\xi^{\alpha}_{\vec{q}} \rightarrow {\vec{v}}^{\alpha} \cdot {\vec{q}}$)
Eqs.\ref{eq:Qraddef}--\ref{eq:hrpaalpha} 
are equivalent with the 
expression given by
Kwon, Houghton and Marston \cite{Kwon94}.
However, as will be shown in
Sect.~\secref{sec:TransDWMax},
for physically relevant forms of the gauge field propagator
the linearization of the energy dispersion 
is not allowed. 
In fact, in \cite{Kopietz96che}
we have shown that 
in the case of the two-dimensional 
Chern-Simons action the low-energy behavior of the 
spectral function is completely dominated by the
prefactor self-energy $\Sigma_1^{\alpha} ( \tilde{q} )$
and vertex function $Y^{\alpha} ( \tilde{q} )$ discussed
in  Chap.~\secref{subsec:disorder},
which are ignored for linearized energy 
dispersion.

Comparing Eqs.\ref{eq:Qraddef}--\ref{eq:Sraddef} with 
\ref{eq:Qlondef2}--\ref{eq:Slondef2}, it is obvious that at the level
of the Gaussian approximation the contributions from the
longitudinal and transverse components to the
total Debye-Waller factor 
are additive and formally
identical; we simply have to use the corresponding RPA screened propagators.
Of course, this is only true in Gaussian approximation, which
produces the first order term in an expansion in powers
of the RPA interaction. 
Clearly, the leading contributions to the prefactor Green's function 
\index{Green's function!prefactor}
are also additive
so that we may simply copy the relevant equations
from Chap.~\secref{subsec:disorder}.
Thus, in complete
analogy with Eqs.\ref{eq:Gtotalavparametrize}
and \ref{eq:Gtotalpre} we obtain
\index{Green's function!bosonization result for gauge fields}
 \begin{eqnarray}
 \left< {\cal{G}}^{\alpha}_{1} ( {\vec{r}} ,0 , \tau ,0 )
 \E^{ \Phi^{\alpha} ( {\vec{r}} , \tau ) - \Phi^{\alpha} ( 0 , 0 ) } 
 \right>_{S_{{\rm eff},2}}
 & = &
  \tilde{G}^{\alpha} ( {\vec{r}} , \tau )  
 \E^{Q^{\alpha}_1 ( {\vec{r}} , \tau )  
 + Q^{\alpha}_{\rm tr} ( {\vec{r}} , \tau )  
 } 
 \label{eq:Gtotalavparametrize2}
 \; \; ,
 \end{eqnarray}
with
 \begin{eqnarray}
  \tilde{G}^{\alpha} ( {\vec{r}} , \tau )  
 & \equiv  &
  {G}^{\alpha}_1 ( {\vec{r}} , \tau  )
  +
  {G}^{\alpha}_2 ( {\vec{r}} , \tau  )
  \nonumber
  \\
  &  & \hspace{-15mm} =
  \frac{1}{\beta V} \sum_{\tilde{q}} 
   \E^{ \I ( {\vec{q}} \cdot {\vec{r}} - 
  \tilde{\omega}_n \tau ) }
  \frac{ 1 +  Y^{\alpha} ( \tilde{q} ) + Y^{\alpha}_{\rm tr} ( \tilde{q} )  }
  { \I \tilde{\omega}_n - 
  \epsilon_{ {\vec{k}}^{\alpha} + {\vec{q}} } + \mu
  - \Sigma_1^{\alpha} ( \tilde{q} ) - \Sigma_{1,{\rm tr}}^{\alpha} ( \tilde{q} ) }
  \label{eq:Gtotalpretr}
  \; \; \; .
  \end{eqnarray}
Here $\Sigma_1^{\alpha} ( \tilde{q} )$  
and $Y^{\alpha} ( \tilde{q} )$ are 
given in Eqs.\ref{eq:sigma1res2} and \ref{eq:Yres}, while
$\Sigma_{1,{\rm tr}}^{\alpha} ( \tilde{q} )$  
and $Y^{\alpha}_{\rm tr} ( \tilde{q} )$  
can be obtained by replacing in these equations $f^{{\rm RPA} , \alpha}_q \rightarrow
h^{{\rm RPA} , \alpha}_q$. 
In the special case of a spherical Fermi surface with radius
$k_{\rm F} = m v_{\rm F}$ 
the prefactor self-energy due to the transverse gauge field
is explicitly given by \index{self-energy!prefactor}
(see Eq.\ref{eq:sigma1res3})
 \begin{eqnarray}
 {\Sigma}^{\alpha}_{1, {\rm tr}} ( \tilde{q} )
 & = & 
 - \frac{1}{\beta {{V}}} \sum_{q^{\prime}}
 h^{{\rm RPA}, \alpha }_{q^{\prime}} G_1^{\alpha} ( \tilde{q} + q^{\prime} )
 \nonumber
 \\
 & \times &
 \frac{ ( \vec{q} \cdot \vec{q}^{\prime} ) \vec{q}^{\prime 2} + 
 ( \vec{q} \cdot \vec{q}^{\prime} )^2 }{
 m^2 [ \I \omega_{m^{\prime}} - \xi^{\alpha}_{  {\vec{q}}^{\prime} } ]
 [ \I \omega_{m^{\prime}} + 
 \xi^{\alpha}_{ - {\vec{q}}^{\prime}} ] }
 \label{eq:sigma1res3tr}
 \; \; \; ,
 \end{eqnarray}
with $h^{{\rm RPA} , \alpha}_q$ given in Eq.\ref{eq:hrpaalpha}.
The corresponding vertex function is
(see Eq.\ref{eq:Yres4})
 \begin{eqnarray}
 {Y}^{\alpha}_{\rm tr} ( \tilde{q} )
 & = & 
  \frac{1}{\beta {{V}}} \sum_{q^{\prime}}
 h^{{\rm RPA} , \alpha }_{q^{\prime}} G_1^{\alpha} ( \tilde{q} + q^{\prime} )
 \nonumber
 \\
 & \times &
 \frac{ \vec{q}^{\prime 2} + 2  \vec{q} \cdot \vec{q}^{\prime}  }{
 m [ \I \omega_{m^{\prime}} - \xi^{\alpha}_{  {\vec{q}}^{\prime} } ]
 [ \I \omega_{m^{\prime}} + 
 \xi^{\alpha}_{ - {\vec{q}}^{\prime}} ] }
 \label{eq:Yres4tr}
 \; \; \; .
 \end{eqnarray}
As discussed in Chaps.~\secref{sec:sectors} and \secref{sec:sumgreen}, 
for spherical Fermi surfaces  
there is no need to introduce several patches. 
Then the index $\alpha$ simply indicates that
all wave-vectors are measured with respect to a
point $\vec{k}^{\alpha}$ on the Fermi surface, as shown in
Fig.~\secref{fig:coordgood}.
{\it{In this case there are no uncontrolled
corrections due to around-the-corner processes\index{around-the-corner processes} 
to the above expressions.}}
In the following section we shall simplify 
$h^{{\rm RPA} , \alpha}_q$ such that we see more clearly
that it contains the physics of transverse screening.

\section{Transverse screening\index{screening!transverse}}
\label{sec:Transscreen}
\label{sec:TransDW}

{\it{
Assuming patch-independent bare interactions,
we derive from Eq.\ref{eq:hrpaalpha} the transverse dielectric tensor and 
show that in the static limit the transverse gauge field is not screened.
We then discuss in some detail the form of $h^{{\rm RPA} ,\alpha}_{q}$
for spherical $d$-dimensional Fermi surfaces, where the effective mass
tensor is isotropic.}}

\subsection{The transverse dielectric tensor\index{polarization!transverse}}

\begin{figure}
\sidecaption
\psfig{figure=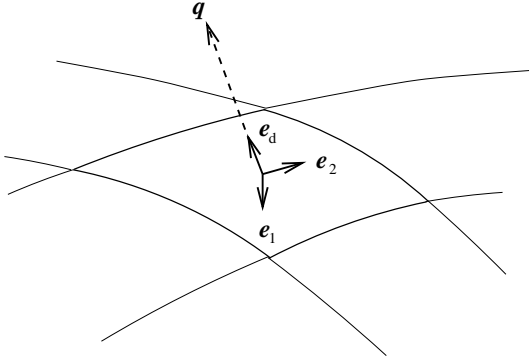,width=7cm}
\caption[Local coordinate system in which the transverse dielectric tensor is diagonal.]
{Local coordinate system associated with a patch
on the Fermi surface
in which the transverse dielectric tensor ${\tens{E}}_{q}$ is
diagonal.}
\label{fig:coordorient}
\end{figure}

\noindent
Here and in the following section we assume that
the effective masses $m_i$ are independent of the patch index, i.e.
$[ \tens{M}^{\alpha}]_{ij} = m_i \delta_{ij}$.
The expression for
$h^{{\rm RPA}, \alpha}_{q}$ in Eq.\ref{eq:hrpaalpha} can be simplified
if we assume that all elements of the bare matrix $\underline{\tilde{h}}_{q}$ 
are identical, 
 $[ \underline{\tilde{h}}_{q}  ]^{\alpha \alpha^{\prime} }=
 \tilde{h}_{q} \equiv \frac{ \beta}{V} h_{q}$.
Using the same method as in Eq.\ref{eq:Neumann2}, 
we find that in this case also the matrix
$\underline{ \tens{H}}_{q}$ is independent of the patch indices,
 \begin{equation}
[ \underline{\tens{H}}_{q} ]_{ij}^{\alpha \alpha^{\prime}} =  \tilde{h}_{q} 
 [  {\tens{E}}_{q}^{-1} ]_{ij}
 \; \;  \; ,
 \label{eq:epstransverse1}
 \end{equation}
where ${\tens{E}}_{q}  $ 
is a matrix in the spatial indices, with matrix elements
given by
 \begin{equation}
 [ {\tens{E}}_{q} ]_{ij}  =   \delta_{ij} + h_{q} 
 \left[
 \delta_{ij} \Delta_{i} - P_{ij} ( q ) + 
  ( {\vec{e}}_{i} \cdot {\vec{\Pi}}_{q}   ) 
  ( {\vec{e}}_{j} \cdot {\vec{\Pi}}_{q}   )
  f^{\rm RPA}_{q}
  \right]
 \label{eq:epsilonij}
 \; \; \; ,
 \label{eq:Piijdef}
 \end{equation}
and
 \begin{eqnarray}
 {\Delta}_{i} & = & \frac{\beta}{V} \sum_{\alpha} \tilde{\Delta}^{\alpha}_{i}
 = 
 \frac{N}{V m_{i} c^2 }  
 \; \; \; ,
 \label{eq:diamagdef1}
 \\
 P_{ij} ( q) & = & \sum_{\alpha} 
 ( {\vec{e}}_{i} \cdot {\vec{u}}^{\alpha}  ) 
 ( {\vec{e}}_{j} \cdot {\vec{u}}^{\alpha}  ) 
 \Pi^{\alpha}_{0} ( q )
 \label{eq:Pijdef}
 \; \; \; ,
 \\
 {\vec{\Pi}}_{q}  & = & \sum_{\alpha} 
  {\vec{u}}^{\alpha}  
 \Pi^{\alpha}_{0} ( q)
 \; \; \; .
 \label{eq:Pi0vecdef}
 \end{eqnarray}
The first term in Eq.\ref{eq:Piijdef}
is the diamagnetic\index{diamagnetism}
transverse polarization tensor, the second term is the 
paramagnetic\index{paramagnetism} one, and
the last term describes the coupling between the longitudinal
and the transverse fluctuations. 
Substituting Eq.\ref{eq:epstransverse1} into the general
expression for the gauge field propagator in Eq.\ref{eq:gaugeprop},
we obtain\index{gauge field propagator!RPA}
 \begin{equation}
 [ D^{\rm RPA} ( q ) ]^{\alpha \alpha^{\prime}}_{ij}  = \tilde{h}_{q}
 \left[ [{\tens{E}}^{-1}_{q} ]_{ij} - 
 \frac{ \sum_{kl=1}^{d} [ {\tens{E}}_{q}^{-1} ]_{i k} q_{k}
 q_{l} [ {\tens{E}}^{-1}_{q} ]_{l j} }
 {  {\vec{q}}  {\tens{E}}^{-1}_{q} {\vec{q}} }
 \right]
 \label{eq:gaugeprop2}
 \; \; \; ,
 \end{equation}
where we have used the same notation as in Eq.\ref{eq:Ansatzsu}.
From Eqs.\ref{eq:hrpaalpha} and Eq.\ref{eq:gaugeprop2}
we finally obtain 
 \begin{equation}
 h^{{\rm RPA}, \alpha}_{q} = - h_{q} \left[
 {\vec{u}}^{\alpha}_{q}  {\tens{E}}_{q}^{-1}  {\vec{u}}^{\alpha}_{q} -
 \frac{ ( {\vec{u}}^{\alpha}_{q}  {\tens{E}}_{q}^{-1}  {\vec{q}} )
 ( {\vec{q}}   {\tens{E}}_{q}^{-1}  {\vec{u}}^{\alpha}_{q} ) }
 {  {\vec{q}}  {\tens{E}}^{-1}_{q}  {\vec{q}} }
 \right]
 \; \; \; ,
 \label{eq:grpares2}
 \end{equation}
where ${\vec{u}}^{\alpha}_{q}  = 
 {\vec{u}}^{\alpha} - {\vec{\Pi}}_{q} f^{\rm RPA}_{q}$. 
Eq.\ref{eq:grpares2} can be further simplified by 
choosing an appropriate coordinate system.
Because the scalar products are independent of the 
choice of the coordinate system and
$q \equiv [ \vec{q} , \I \omega_m ]$ appears as an external parameter, 
we may choose the orientation of the coordinate system
such that one of its axis
(${\vec{e}}_{d}$, for example)
matches the direction of ${\hat{ \vec{q}} }$, as shown in Fig.~\secref{fig:coordorient}. 
Because for small ${\vec{q}}$ the function
$\Pi^{\alpha}_{0} ( q )$ in Eq.\ref{eq:Pi0vecdef} depends 
on ${\vec{q}}$ only via ${\vec{v}}^{\alpha} \cdot {\vec{q}}$, it is easy to see that
in the long wavelength limit
 $\hat{ \vec{q}}_{\bot} \cdot {\vec{\Pi}}_{q} = 0$
for any direction $\hat{\vec{q}}_{\bot}$ that is orthogonal to ${\vec{q}}$.
Note that by construction
the $d-1$ directions ${\vec{e}}_{i}$, $i=1, \ldots , d-1$ are all perpendicular to
${\vec{q}}$, so that the last term in Eq.\ref{eq:Piijdef} 
does not contribute to  Eq.\ref{eq:grpares2}.
For the same reason 
 $P_{ij} ( q ) = \delta_{ij} P_{ii} ( q ) $.
It follows that in this basis
the matrix ${\tens{E}}_{q}$ is diagonal. 
The eigenvalues corresponding to the
$d-1$ directions orthogonal to $\hat{\vec{q}}$ 
are simply given by
 \begin{equation}
 \epsilon_{i} ( q ) = [ {\tens{E }}_{q} ]_{ii}
 = 1 + h_{q}  \Pi_{i} ( q )
 \; \; \; , \; \; \; i = 1, \ldots , d-1
 \; \; \; ,
 \end{equation}
where the  transverse polarization in direction ${\vec{e}}_{i}$ is
 \begin{equation}
 \Pi_{i} ( q ) =
  \Delta_{i} - 
  \sum_{\alpha} ( {\vec{e}}_{i} \cdot {\vec{u}}^{\alpha}  )^2
 \Pi^{\alpha}_{0} ( q ) 
 \; \; \; , \; \; \; i = 1, \ldots , d-1
 \; \; \; .
 \label{eq:Pitrans}
 \end{equation}
We finally obtain for the effective screened interaction
 \begin{equation}
 h^{{\rm RPA}, \alpha}_{q} = - h_{q}  \sum_{i=1}^{d-1} 
 \frac{  ( {\vec{e}}_{i} \cdot {\vec{u}}^{\alpha}  )^2 }
 {\epsilon_{i} ( q )}
 \label{eq:hqfinal}
 \; \; \; .
 \end{equation}
Note that
Eq.\ref{eq:hqfinal} involves only the
transverse eigenvalues of $\tens{E}_q$,
because $h^{ {\rm{RPA}} , \alpha}_q$ is by construction
the propagator of the {\it{transverse}} components of the gauge field.
The dimensionless functions $\epsilon_{i} ( q )$ are called the
{\it{transverse dielectric functions}}\index{dielectric function!transverse}.

\subsection{Screening\index{screening!and gauge invariance} and gauge invariance}

According to Eqs.\ref{eq:Pilong}, \ref{eq:pi0tot}, \ref{eq:diamagdef1} and \ref{eq:Pitrans}
the longitudinal and transverse polarizations 
are for small ${\vec{q}}$ and arbitrary frequencies given by
 \begin{eqnarray}
 \Pi_{0} ( q ) & = & \sum_{\alpha} 
 \nu^{\alpha}
 \frac{ {\vec{v}}^{\alpha} \cdot {\vec{q} } }
 { {\vec{v}}^{\alpha} \cdot {\vec{q}} - \I \omega_{m} }
 \label{eq:Pilon1}
 \\
 \Pi_{i} ( q ) & = & 
 \frac{1}{m_{i} c^2 }  \sum_{\alpha} \left[
 \frac{N^{\alpha}_{0}}{V} - m_{i} 
  (  {\vec{v}}^{\alpha} \cdot {\vec{e}}_{i} )^2 
 \nu^{\alpha} 
 \frac{ 
 {\vec{v}}^{\alpha} \cdot {\vec{q} } }
 { {\vec{v}}^{\alpha} \cdot {\vec{q}} - \I \omega_{m} }
 \right]
 \label{eq:Pitrans1}
 \; \; \; ,
 \end{eqnarray}
where ${\vec{e}}_{i} \cdot {\vec{q}} = 0$, and we have used the fact that
$ m_i {\vec{v}}^{\alpha} \cdot {\vec{e}}_{i} = 
  {\vec{k}}^{\alpha}  \cdot {\vec{e}}_{i}$ (see Eq.\ref{eq:ualphairedef}).
In the static limit we have
 $\Pi_{0} ( {\vec{q}} , 0 ) = \sum_{\alpha} \nu^{\alpha} = \nu$,
where the total density of states 
is given in Eq.\ref{eq:nudefinf}.
A finite value of the longitudinal polarization implies that
long-range interactions are screened. For example, for the three-dimensional
Coulomb interaction the static longitudinal dielectric function 
is within the RPA given by
 $\epsilon_{\rm RPA} ( {\vec{q}} , 0 ) = 1 +  {\kappa^2}/{  {\vec{q}}^2  }$,
see Eqs.\ref{eq:dielectricdef} and \ref{eq:Fqdef2}.
For wave-vectors $ |{\vec{q}} | \ll \kappa$ the longitudinal
dielectric function is large compared with unity, 
so that the interaction is screened at
length scales larger than the Thomas-Fermi length $\kappa^{-1}$.
On the other hand, as long as the gauge symmetry is not
spontaneously broken, the transverse gauge field is not screened 
in the static limit.
Using Eq.\ref{eq:Thetasum}, it is easy to see that the transverse polarization 
in a direction ${\vec{e}}_{i}$ orthogonal to ${\vec{q}}$
can in the static limit and for small $| {\vec{q}} | $ be written as 
 \begin{equation}
 \Pi_{i} ( {\vec{q}} , 0 ) = 
 \frac{1}{m_{i} c^2 } \int \frac{ \D {\vec{k}} }{ (2 \pi )^d }
 \left[ \Theta ( \mu - \epsilon_{\vec{k}} ) - 
 \frac{k_{i}^2}{m_{i}} \delta ( \mu - \epsilon_{\vec{k}} )
 \right]
 \label{eq:Pitrans2}
 \; \; \; .
 \end{equation}
But
 \begin{equation}
 \delta ( \mu - \epsilon_{\vec{k}} ) = - \frac{ m_{i} }{  k_{i}}
 \frac{ \partial}{\partial k_{i}} \Theta ( \mu - \epsilon_{\vec{k}} )
 \; \; \; ,
 \label{eq:trick1}
 \end{equation}
so that we obtain after an integration by parts
 \begin{equation}
 \Pi_{i} ( {\vec{q}} , 0 ) = 
 \frac{1}{m_{i} c^2 } \int \frac{ \D {\vec{k}} }{ (2 \pi )^d }
 \left[ \Theta ( \mu - \epsilon_{\vec{k}} ) + 
 k_{i} \frac{ \partial }{\partial k_{i}} \Theta ( \mu - \epsilon_{\vec{k}} )
 \right]
 = 0
 \label{eq:Pitrans3}
 \; \; \; .
 \end{equation}
The vanishing of the transverse polarization tensor in the static limit is due to
a perfect cancellation between the dia- and paramagnetic contributions.
The fundamental symmetry which is responsible for this cancellation
is gauge invariance, which insures that
the transverse gauge field remains massless in the presence of matter.
Hence, as long as the gauge symmetry is not spontaneously broken, the transverse
gauge field is not screened in the static limit. 
However, as shown by Kohn and Luttinger \cite{Kohn65},
\index{Kohn-Luttinger effect}
any interacting Fermi system shows at very low temperatures a superconducting
instability (Kohn-Luttinger effect), 
so that gauge invariance is in fact broken at very low temperatures, and the
transverse gauge field is eventually screened. 
This instability is not included in our calculation.

\subsection{The transverse dielectric function \mbox{\hspace{40mm}}
for  spherical Fermi surfaces}

For spherical Fermi surfaces the
effective mass tensor is proportional to the unit matrix.
The $d-1$ transverse eigenvalues of ${\tens{E}}_{q}$
are then degenerate, and are called 
the {\it{transverse dielectric function}} \cite{Pines89},
\index{dielectric function!transverse}
 \begin{equation}
 \epsilon_{\bot} ( q )  = 1 + h_{q}  
 \Pi_{\bot} ( q )
 \label{eq:epsilontrans2}
 \; \; \; .
 \end{equation}
From Eq.\ref{eq:Pitrans} we see that
the transverse polarization $\Pi_{\bot} ( q )$ 
is within the RPA given by
 \begin{equation}
 \Pi_{\bot} ( q ) =
 \Delta -
  \sum_{\alpha} ( \vec{e}_i \cdot {\vec{u}}^{\alpha}  )^2 
  \Pi_{0}^{\alpha} ( q )
  \label{eq:Pibotdef}
  \; \; \; ,
  \end{equation}
with $\Delta = N / ( V mc^2)$ (see Eq.\ref{eq:diamagdef1}).
Here ${\vec{e}}_{i} $ is any of the $d-1$  unit vectors perpendicular
to ${\hat{\vec{q}}} = {\vec{e}}_{d} $.
From Eq.\ref{eq:Pitrans1} it is easy to show that for
a spherical Fermi surface
\index{dielectric function!transverse}
 \begin{equation}
 \Pi_{\bot} ( q ) = 
 \left( \frac{v_{\rm F}}{c} \right)^2 \nu  
 {\Lambda }_{d} \left( \frac{ \I \omega_{m}}{v_{\rm F} | {\vec{q}} | } \right)
 \; \; \; ,
 \label{eq:Pitransbot}
 \end{equation}
where the dimensionless function $\Lambda_d ( z )$ is
 \begin{equation}
 {\Lambda }_{d} \left( z \right)
  = 
 \frac{1}{d} -
 \left< ( {{\vec{e}}}_{i} \cdot \hat{\vec{k}} )^2
 \frac{ \hat{\vec{q}} \cdot \hat{\vec{k}} }
 { \hat{\vec{q}} \cdot \hat{\vec{k}} - z }  
 \right>_{\hat{\vec{k}}}
 \label{eq:gdbardef}
 \; \; \; .
 \end{equation}
Here the angular average is defined as in Eq.\ref{eq:angav}.
By symmetry, the average is independent of the choice of ${\vec{e}}_{i}$.
Using the fact that 
 $\left< ( {{\vec{e}}}_{i} \cdot \hat{\vec{k}} )^2
 \right>_{\hat{\vec{k}} } = { 1 }/{d}$
it is easy to see that Eq.\ref{eq:gdbardef}
can also be written as
\begin{equation}
 {\Lambda}_{d} \left( z \right)
  =  - z 
 \left< \frac{ 
 ( {{\vec{e}}}_{i} \cdot \hat{\vec{k}} )^2 
 }
 { \hat{\vec{q}} \cdot \hat{\vec{k}} - z }  
 \right>_{\hat{\vec{k}}} 
 \label{eq:gdbar2}
 \; \; \; .
 \end{equation}
Because all transverse directions are equivalent,
we may replace in the average
 \begin{equation}
 ( {{\vec{e}}}_{i} \cdot \hat{\vec{k}} )^2 
 \rightarrow \frac{\sum_{i=1}^{d-1}
 ( {{\vec{e}}}_{i} \cdot \hat{\vec{k}} )^2 }{d-1}
 = \frac{ 1 - ( {\hat{\vec{q}}} \cdot {\hat{\vec{k}}} )^2 }{ d-1}
 \label{eq:replaceav}
 \; \; \; ,
 \end{equation}
so that
\begin{equation}
 {\Lambda}_{d} \left( z \right)
  =  - \frac{z }{d-1}
 \left< \frac{ 
 1 -
 ( \hat{\vec{q}} \cdot \hat{\vec{k}} )^2 
 }
 { \hat{\vec{q}} \cdot \hat{\vec{k}} - z }  
 \right>_{\hat{\vec{k}}} 
 \label{eq:gdbar3}
 \; \; \; .
 \end{equation}
From this expression we find
 \begin{eqnarray}
 {\rm Im} {\Lambda}_{d} ( x + \I 0^{+} ) & = & - \frac{ \pi x (1 - x^2)}{d-1} \left< 
 \delta ( 
 \hat{\vec{q}} \cdot \hat{\vec{k}} - x ) \right>_{ \hat{\vec{k}} }
 \nonumber
 \\
 & \sim & - \frac{ \pi \gamma_{d}}{d-1} x
 \; \; \; , \; \; \; \mbox{for $|x| \ll 1$ }
 \label{eq:Imgbardef}
 \; \; \; ,
 \end{eqnarray}
with the numerical constant $\gamma_{d}$ given in Eq.\ref{eq:gammaddef}.
Note that, in contrast to $\gamma_{d}$, the quantity
 \begin{equation}
 \tilde{\gamma}_{d} \equiv \frac{ \gamma_{d}}{d-1} = 
 \frac{ \Gamma ( \frac{d}{2} ) }{ (d-1 ) \sqrt{\pi}  \Gamma ( \frac{d-1}{2} ) }
 \label{eq:gammatildeddef}
 \end{equation}
has a finite limit as $d \rightarrow 1$.
In particular,
 \begin{equation}
 \tilde{\gamma}_{1} = \frac{1}{2} \; \; \; , \; \; \; 
 \tilde{\gamma}_{2} = \frac{1}{\pi} \; \; \; , \; \; \; 
 \tilde{\gamma}_{3} = \frac{1}{4}
 \label{eq:gammatildepart}
 \; \; \; .
 \end{equation}
On the imaginary axis Eq.\ref{eq:Imgbardef} implies
 \begin{equation}
 {\Lambda }_{d} \left( \I y \right)
 \sim \lambda_d |y| 
 \; \; \; ,
 \; \; \; 
 \mbox{for
 $ | y | \ll  1 $}
 \; \; \; ,
 \label{eq:Landaudamptilde}
 \end{equation}
where 
 \begin{equation}
 \lambda_d = \pi \tilde{\gamma}_d = \frac{\pi \gamma_d}{d-1}
 \label{eq:chidgaugedef}
 \; \; \; .
 \end{equation}

 \vspace{7mm}

For the Maxwell action discussed in 
Sect.~\secref{subsec:Definitionofmodel} 
the bare interaction $h_q$ is given in Eq.\ref{eq:tildehdefgf}.
Then we obtain
from Eqs.\ref{eq:epsilontrans2} and \ref{eq:Pitransbot}
for the transverse dielectric function
\index{dielectric function!transverse}
 \begin{equation}
 \epsilon_{\bot} (q) = 1 + \left( \frac{v_{\rm F}}{c} \right)^2
 \frac{ {\Lambda}_{d} ( \frac{ \I \omega_{m}}{ v_{\rm F} | {\vec{q}} | } ) }
 { ( \frac{ {\vec{q} } }{\kappa} )^2 + ( \frac{ \omega_{m} }{ c \kappa } )^2 }
 \label{eq:epsilontrans3}
 \; \; \; ,
 \; \; \; 
 \kappa^2 = 4 \pi e^2 \nu
 \; \; \; .
 \end{equation}
After some simple rescalings 
we obtain for the effective interaction 
\ref{eq:Sraddef}
 \begin{equation}
  h^{{\rm RPA} , \alpha}_{q}
 = -  \frac{1}{\nu} \left( \frac{v_{\rm F}}{c} \right)^2
 \frac{ 1 - ({\hat{\vec{k}}}^{\alpha} \cdot {\hat{\vec{q}}} )^2}
 { \left( \frac{{\vec{q}} }{ \kappa } \right)^2 + 
 \left( \frac{v_{\rm F}}{c} \right)^2 \left[
 ( \frac{\omega_{m} }{ v_{\rm F} \kappa  } )^2 + 
 {\Lambda}_{d} \left( \frac{\I \omega_{m} }{ v_{\rm F} | {\vec{q}} | } \right)
 \right] }
 \label{eq:Hrparad}
 \; \; \; .
 \end{equation}
In the regime $|\omega_{m} | \ll v_{\rm F} | {\vec{q}} |$
we obtain from Eq.\ref{eq:Landaudamptilde}
 \begin{equation}
 {\Lambda}_{d} \left( \frac{\I \omega_{m} }{ v_{\rm F} | {\vec{q}} | } \right)
  \sim  
  \lambda_{d} \frac{ | \omega_{m} |}{v_{\rm F} | {\vec{q}} | }
 \; \; \; . 
 \label{eq:gbarlow}
 \end{equation}
To this order in $\omega_{m}$ the term proportional to
$\omega_{m}^2$ in the denominator of Eq.\ref{eq:Hrparad} is negligible, so that
the effective interaction can be approximated by
 \begin{equation}
 \hspace{-3mm}
 h^{{\rm RPA} , \alpha}_{q}
 \approx -  \frac{1}{\nu} \left( \frac{v_{\rm F}}{c} \right)^2
 \frac{ 1 - ({\hat{\vec{k}}}^{\alpha} \cdot {\hat{\vec{q}}} )^2}
 { \left( \frac{{\vec{q}} }{ \kappa } \right)^2 + 
 \lambda_{d}  ( \frac{v_{\rm F}}{c} )^2
 \frac{| \omega_{m}| }{ v_{\rm F} | {\vec{q}} | } 
  }
 \;  \; , 
 \;  \; \mbox{for $ | \omega_{m} | \ll v_{\rm F} | {\vec{q}} | $ }
 \label{eq:Hrparad2}
 \; .
 \end{equation}
For $d=3$ we recover Eq.\ref{eq:Hrparad2b}.
The term proportional to $ | \omega_m | / ( v_{\rm F}  | \vec{q} | )$ 
in the denominator describes the dynamical screening of the
fluctuations of the gauge field due to Landau damping. \index{Landau damping}
This term is responsible for the dynamical screening
of the magnetic field in a clean metal, i.e. the
{\it{anomalous skin effect}} \cite{Tsvelik95}.
\index{skin effect!anomalous}

\section{The transverse Debye-Waller factor}
\label{sec:TransDWMax}

{\it{We now analyze the
transverse Debye-Waller factor 
$ Q_{\rm tr}^{\alpha} ( {\vec{r}} ,  \tau )$  
in Eq.\ref{eq:Qraddef} in more detail.
We determine the parameter regime where 
$ Q_{\rm tr}^{\alpha} ( {\vec{r}} ,  \tau )$ is 
bounded for large distances or times, and where the 
non-linear terms in the energy dispersion
must be retained in order to obtain qualitatively correct results.}}
\index{Debye-Waller factor!transverse gauge fields}

\vspace{7mm}

\noindent
In Sect.~\secref{sec:Derivationrad} we have derived 
a non-perturbative expression for the
single-particle Green's function $G ( k )$ in
Coulomb gauge (see Eqs.\ref{eq:Gkres5} and \ref{eq:Gtotalavparametrize2}).
The effect of  
the Gaussian fluctuations of the gauge field
is parameterized in terms of three distinct contributions
to the Green's function:
the Debye-Waller factor
$Q^{\alpha}_{\rm tr} ( \vec{r} , \tau )$
in Eqs.\ref{eq:Qraddef}--\ref{eq:Sraddef},
the prefactor self-energy
$\Sigma_{1 , {\rm tr}}^{\alpha} ( \tilde{q} )$
in Eq.\ref{eq:sigma1res3tr}, and the prefactor
vertex $Y^{\alpha}_{\rm tr} ( \tilde{q} )$ in
Eq.\ref{eq:Yres4tr}.
Thus,
the calculation of $G ( k )$ and the
resulting spectral function
has been reduced
to the purely mathematical problem of doing the relevant integrations.
Unfortunately, it is impossible to
perform these integrations analytically,
so that a complete  analysis of our non-perturbative result
for $G ( k )$
requires extensive numerical work, which is
beyond the scope of this book.
In the recent Letter \cite{Kopietz96che}  
we have made some progress in this
problem. In particular, we have
shown that in physically relevant cases the low-energy behavior
of the spectral function is essentially determined by
the functions
$\Sigma_{1 , {\rm tr}}^{\alpha} ( \tilde{q} )$
and
$Y^{\alpha}_{\rm tr} ( \tilde{q} )$, and {\it{not}}
by the Debye-Waller factor
$Q^{\alpha}_{\rm tr} ( \vec{r} , \tau )$.
Because for linearized energy dispersion
both functions
$\Sigma_{1 , {\rm tr}}^{\alpha} $ and  $Y^{\alpha}_{\rm tr} $
vanish, the problem of fermions that are coupled to gauge fields
can only be studied via
higher-dimensional bosonization if the
quadratic terms in the expansion of the energy dispersion
close to the Fermi surface are retained\footnote{Because of the formal
similarity between higher-dimensional bosonization and the
leading term in the conventional eikonal 
expansion \cite{Kopietzeik}, it seems that this is true for
any eikonal type of approach to this 
problem \cite{Khveshchenko93,Lee96}.}.
To see more clearly why  
the curvature of the Fermi surface 
is so important in the present problem, we
shall in this section study the Debye-Waller factor
$Q^{\alpha}_{\rm tr} ( \vec{r} , \tau )$ in some detail.

\subsection{Exact rescalings}

\noindent
Let us consider a spherical Fermi surface in $d$ dimensions and
a general gauge field propagator of the form
 \begin{equation}
 h^{{\rm RPA} , \alpha}_{q}
 = -  \frac{1}{\nu  } 
 \frac{ 1 - ({\hat{\vec{k}}}^{\alpha} \cdot {\hat{\vec{q}}} )^2}
 { \left( \frac{ | {\vec{q}}| }{ q_c } \right)^{\eta} + 
  {\Lambda}_d ( 
  \frac{ \I \omega_{m} }{ v_{\rm F} | {\vec{q}} | }  ) }
 \label{eq:Hrparad4}
 \; \; \; ,
 \end{equation}
where ${\Lambda}_d ( \I y  ) \sim \lambda_d | y |$ 
for small $|y|$, see Eq.\ref{eq:gdbar3}.
Substituting Eq.\ref{eq:Hrparad4}
into Eq.\ref{eq:Qraddef}, we obtain
 \begin{equation}
 \hspace{-6mm}
 Q_{\rm tr}^{\alpha} ( {\vec{r}} ,  \tau )  = 
 -   \frac{1}{\beta V \nu}
  \sum_{q}
 \frac{ 1 - ({\hat{\vec{k}}}^{\alpha} \cdot {\hat{\vec{q}}} )^2}
 { \left( \frac{ |{\vec{q}}| }{ q_{\rm c} } \right)^{\eta} + 
 {\Lambda}_{d} \left( \frac{\I \omega_{m} }{ v_{\rm F} | {\vec{q}} | } \right)
  }
 \frac{ 1 -
  \cos ( {\vec{q}} \cdot  {\vec{r}}  
  - {\omega}_{m}  \tau  ) }
 {
 ( \I \omega_{m} - {\xi}^{\alpha}_{\vec{q}} )
 ( \I \omega_{m} + {\xi}^{\alpha}_{- \vec{q}} )
 }
 \label{eq:Qbot2}
 \; .
 \end{equation}
As discussed in Chap.~\secref{subsec:Greal}, 
for {\it{linearized energy dispersion}}
we may replace ${\vec{r}} \rightarrow  r^{\alpha}_{\|} \hat{\vec{v}}^{\alpha}$ 
in Eq.\ref{eq:Qbot2},
because the sector Green's function 
$G_0^{\alpha} ( \vec{r} , \tau )$
is proportional to
$\delta^{(d-1)} ( \vec{r}^{\alpha}_{\bot} )$,
see Eq.\ref{eq:Gpatchreal1}.
Although for non-linear energy dispersion we should
consider $Q_{\rm tr}^{\alpha} ( {\vec{r}} ,  \tau ) $ for all ${\vec{r}}$,
we shall restrict ourselves here to the 
direction ${\vec{r}} = r^{\alpha}_{\|} \hat{\vec{v}}^{\alpha}$. 
This is sufficient for investigating whether the
non-linear terms in the energy dispersion
qualitatively modify the
result obtained for linearized energy dispersion.
Obviously, for
${\vec{r}} = r^{\alpha}_{\|} \hat{\vec{v}}^{\alpha}$
the $\vec{q}$-dependence of
the right-hand side of Eq.\ref{eq:Qbot2} 
involves only the absolute value of $\vec{q}$ and
the component\footnote{ 
Note that for a spherical Fermi surface $\hat{\vec{v}}^{\alpha} = 
\hat{\vec{k}}^{\alpha}$.}
$q^{\alpha}_{\|} = {\hat{\vec{v}}}^{\alpha} \cdot {\vec{q}}$.
Then the $d+1$-dimensional integration in Eq.\ref{eq:Qbot2}
can be reduced to a
three-dimensional one with the help of $d$-dimensional
spherical coordinates: for
$V \rightarrow \infty$ and $\beta \rightarrow \infty$
we have for any function
 $f ( |{\vec{q}} | , {\hat{\vec{q}}} \cdot 
 \hat{\vec{v}}^{\alpha} , \I \omega_{m} )$
\index{spherical coordinates}
 \begin{eqnarray}
 \frac{1}{\beta V \nu} \sum_{q} 
 f ( |{\vec{q}} | , {\hat{\vec{q}}} \cdot \hat{\vec{v}}^{\alpha} , \I \omega_{m} )
 & \rightarrow &
 \nonumber
 \\
 & & \hspace{-45mm}
 \frac{v_{\rm F}}{k_{\rm F}^{d-1}} \gamma_{d} 
 \int_{0}^{\infty} \D q q^{d-1} 
 \int_{0}^{\pi} \D \vartheta ( \sin \vartheta )^{d-2}
 \int_{- \infty}^{\infty} \frac{ \D \omega}{2 \pi}
 f ( q , \cos \vartheta , \I \omega )
 \label{eq:thermtemplim}
 \; \; \; ,
 \end{eqnarray}
where the numerical constant $\gamma_{d}$ is 
given in Eq.\ref{eq:gammad2}, and we have used
Eq.\ref{eq:nurelation}.
Introducing the dimensionless integration variables 
 \begin{equation}
 p = \frac{q }{ q_{\rm c}}
 \; \; \; \; , \; \; \; \;
 y = \frac{\omega }{v_{\rm F} q} = \frac{ \omega}{v_{\rm F} q_{\rm c} p }
 \; \; \; ,
 \end{equation}
and noting that $Q^{\alpha}_{\rm tr} 
( r^{\alpha}_{\|} {\hat{\vec{v}}}^{\alpha} , \tau )$  
depends on the sector index $\alpha$ only via
$r^{\alpha}_{\|} = \hat{\vec{v}}^{\alpha} \cdot {\vec{r}}$,
we may write
 \begin{equation}
 Q_{\rm tr}^{\alpha} ( r^{\alpha}_{\|} \hat{ {\vec{v}}}^{\alpha} ,  \tau ) =
 Q_{\rm tr} ( q_{\rm c} {r}^{\alpha}_{\|} , v_{\rm F} q_{\rm c} {\tau} )
 \; \; \; ,
 \end{equation}
where the function 
 $Q_{\rm tr} ( \tilde{x} , \tilde{\tau} )$ is given by
 \begin{eqnarray}
 Q_{\rm tr} ( \tilde{x} ,  \tilde{\tau} ) & = &
  -  
  \frac{\gamma_{d} g^{d-1}}{2 \pi}
 \int_{0}^{\infty} \D p p^{d-2} 
 \int_{0}^{\pi} \D \vartheta ( \sin \vartheta )^d
 \int_{- \infty}^{\infty} { \D y}
 \nonumber
 \\
 & \times &
 \frac{ 1 - \cos \left[ p  ( \tilde{x} \cos \vartheta  -  \tilde{\tau} y ) \right] }
 { \left[ p^{\eta} +
 {\Lambda}_{d} ( \I y ) \right]  
 \left[ \I y - \cos \vartheta  - \frac{g}{2}p \right]
 \left[ \I y - \cos \vartheta  + \frac{g}{2}p \right]
   }
 \label{eq:Qbot3}
 \; \; \; ,
 \end{eqnarray}
and the dimensionless coupling constant $g$ is simply
 \begin{equation}
 g = \frac{q_{\rm c}}{k_{\rm F}}
 \label{eq:ggdef}
 \; \; \; .
 \end{equation}
Note that the linearization of the energy dispersion corresponds to setting
$g=0$ in the integrand of Eq.\ref{eq:Qbot3}. 
The evaluation of Eq.\ref{eq:Qbot2}  
for ${\vec{r}} = r^{\alpha}_{\|} \hat{\vec{v}}^{\alpha}$
is now reduced to the three-dimensional integration. 
Possible non-Fermi liquid behavior due to the coupling
between fermions and the gauge field should be due to the regime
$|y |=  { | \omega_m |}/( { v_{\rm F} | {\vec{q}} | })
{ \raisebox{-0.5ex}{$\; \stackrel{<}{\sim} \;$}} 1$, because
here the gauge field propagator is most singular.
To further investigate this point, we may
approximate 
  ${\Lambda}_{d} ( \I y ) \approx {\lambda}_{d} |y|$
(see Eqs.\ref{eq:Landaudamptilde} and \ref{eq:chidgaugedef}).
Of course, when substituting this expression into Eq.\ref{eq:Qbot3},
we should restrict the $y$-integration to the regime
$|y | \leq  y_{\rm c} = O (1)$.
Moreover, physically it is clear that 
the power-law $ ( | {\vec{q}} | / q_{\rm c} )^{\eta}$
of the gauge field propagator
in Eq.\ref{eq:Hrparad4} can only be valid up to some
finite cutoff $Q_{\rm c}$,
because at short wavelengths
non-universal short-range interactions will dominate\footnote{
In the case of the three-dimensional Coulomb interaction
we should choose $Q_{\rm c }= q_{\rm c} \approx \kappa$ (the Thomas-Fermi wave-vector),
so that $p_{\rm c} \approx 1$.
In general, however, $Q_{\rm c}$ and $q_{\rm c}$ need not be equal.
For example, in  Sect.~\secref{subsec:ChernSimons} we shall show that
in the two-dimensional Chern-Simons theory for the half-filled Landau level
$Q_{\rm c}$ can be much larger than
$q_{\rm c}$.}.
Assuming that the form \ref{eq:Hrparad4} remains valid up to
$|{\vec{q}} | \leq Q_{\rm c}$, we should impose
a cutoff $p_{\rm c} = Q_{\rm c }/ q_{\rm c}$ 
on the $p$-integration in Eq.\ref{eq:Qbot3}.
With these cutoffs the integration volume is finite, 
so that possible non-Fermi liquid behavior 
must be due to infrared singularities.

To exhibit the infrared behavior of the
integrand in Eq.\ref{eq:Qbot3} more clearly, it is advantageous to
perform another rescaling of the integration variables,
substituting
 \begin{equation}
 y = | \cos \vartheta | u 
 \; \; \; , \; \;  \; 
 p = | \cos \vartheta |^{\frac{1}{ \eta}} k
 \; \; \; .
 \end{equation}
Then we obtain (taking the above ultraviolet cutoffs into account)
 \begin{eqnarray}
 Q_{\rm tr} ( \tilde{x} ,  \tilde{\tau} ) & = &
 \nonumber
 \\ & & \hspace{-23mm}
  -  
  \frac{\gamma_{d} g^{d-1}}{2 \pi}
 \int_{0}^{\pi} \D \vartheta ( \sin \vartheta )^d
 | \cos \vartheta |^{ \frac{d-1}{\eta} -2 }
 \int_{0}^{ p_{\rm c} | \cos \vartheta |^{ - \frac{1}{\eta}}}  \D k k^{d-2} 
 \int_{- y_{\rm c}  | \cos {\vartheta} |^{-1} }^{
 y_{\rm c} | \cos \vartheta |^{-1} } { \D u}
 \nonumber
 \\
 &  & \hspace{-23mm} \times
 \frac{ 1 - \cos \left[ k  | \cos \vartheta |^{ \frac{1}{\eta} + 1} 
  ( \tilde{x} s_\vartheta  -  \tilde{\tau} u ) \right] }
 { \left[ k^{\eta} +
 \lambda_{d} | u| \right]  
 \left[ \I u - s_\vartheta  - \frac{g}{2}{ k 
 | \cos \vartheta |^{ \frac{1}{\eta} -1 }}  \right]
 \left[ \I u - s_\vartheta  + \frac{g}{2}{ k 
 | \cos \vartheta |^{ \frac{1}{\eta} -1 }}  \right]
   }
 \label{eq:Qbot4}
 \;  ,
 \end{eqnarray}
where we have defined
 $ s_{\vartheta} = {\rm sgn} ( \cos \vartheta ) $.
From Eq.\ref{eq:Qbot4} it is now evident
that the regime $\vartheta \approx {\pi}/{2}$ 
can give rise to singular behavior
(in the sense that the $\vartheta$-integral diverges if  we retain only
the space- and time-independent contribution $R_{\rm tr}$),
because the integral over the factor
$| \cos \vartheta |^{ \frac{d-1}{\eta} -2 }$ 
does not exist for
$\frac{d-1}{\eta} -2  < -1$.
In fact, let us {\it{assume}} for the moment that the rest of the integrand does not
modify the small-$\vartheta$ behavior of the integral. 
Evidently, this assumption will be correct provided
it is allowed to set $g=0$ in the rest of the integral, 
corresponding to the linearization of the energy dispersion.
In this case
the angular integration is free of singularities 
as long as the integral
 \begin{equation}
 A_{d, \eta}  =    
 \int_{0}^{\pi } \D \vartheta 
 ( \sin \vartheta )^d | \cos \vartheta |^{ \frac{ d-1}{\eta} -2 }
 =
 \frac{ \Gamma ( \frac{ d+1}{2} ) 
 \Gamma ( \frac{ d-1 - \eta }{2 \eta} ) }{  \Gamma ( \frac{ (1 + \eta )d -1 }{2 \eta} ) }
 \label{eq:adeta}
 \end{equation}
is finite. This is the case for
$\frac{d-1}{\eta} -2 > -1$, or
 \begin{equation}
 \eta < d - 1
 \label{eq:Adetaexist}
 \; \; \; .
 \end{equation}
This is precisely the
criterion for the existence of the quasi-particle residue
that is obtained for linearized energy dispersion \cite{Kopietz95b}, where
one sets $g=0$ in the integrand of Eq.\ref{eq:Qbot4}.
In particular, for linearized energy dispersion 
higher-dimensional bosonization
predicts 
for the three-dimensional Maxwell action ($\eta = 2$)
and the two-dimensional Maxwell-Chern-Simons action ($\eta = 1$)
non-Fermi liquid behavior due to a logarithmic divergence of
$A_{d , \eta }$.
As a consequence, the momentum distribution exhibits an algebraic singularity 
at the Fermi surface \cite{Kopietz95b}, just like in 
the one-dimensional Tomonaga-Luttinger model.
The crucial point is, however, 
that for  $\eta > 1$ the assumption that the rest of the 
$\vartheta$-dependence of the second line in
Eq.\ref{eq:Qbot4} does not modify the
infrared behavior of the integrand is {\it{not}} correct, because for
$\eta > 1$ and 
any finite $g$ the curvature terms in the denominator of
Eq.\ref{eq:Qbot4} become arbitrarily large for 
$\cos \vartheta \rightarrow 0$. {\it{Hence, for $\eta > 1$
the non-linear terms in the expansion of the energy dispersion
close to the Fermi surface cannot be ignored!}}
On the other hand, for $\eta <  1$ these terms vanish
for $\cos \vartheta \rightarrow 0$, so that in this case
the criterion \ref{eq:Adetaexist} is valid. 
Only then the finiteness of $A_{d ,\eta}$ implies the
existence of $R^{\alpha}_{\rm tr}$, 
so that the system shows Fermi liquid behavior.
But in the physically interesting cases of the
Maxwell  action ($d=3$, $\eta =2$) and the Maxwell-Chern-Simons action
\index{Maxwell-Chern-Simons theory}
($d =2$, $\eta = 1$) the curvature term in the denominator
of Eq.\ref{eq:Qbot4} cannot be neglected. 
Interestingly, for $\eta = 1$ there terms
are independent of $\vartheta$, so that their
relevance cannot be determined by simple power counting.
We shall come back to this point in Sect.~\secref{subsec:ChernSimons},
where we show by explicit evaluation of the relevant integral that even then the curvature terms 
are essential.
Note that the above analysis confirms our intuitive
arguments based on the simple estimate \ref{eq:curvcrit}.
In the following section we shall study the effect of the curvature terms 
more carefully.

\subsection{The relevance of curvature\index{curvature!relevance}}
\label{subsec:londisrad}

{\it{
We study the 
effect of the quadratic term in the energy dispersion on
the constant part $R_{\rm tr}$ of the
Debye-Waller factor. This is sufficient to see  whether the
curvature of the Fermi surface is relevant or not.}}  

\vspace{7mm}

\noindent
According
to Eq.\ref{eq:Qbot4} the constant part $R_{\rm tr}$ of the
Debye-Waller factor can be written as
 \begin{eqnarray}
 R_{\rm tr}  & = &
  -  
  \frac{(d-1) g^{d-1}}{ \pi^2  }
 \int_{0}^{\pi / 2 } \D \vartheta ( \sin \vartheta )^d
 | \cos \vartheta |^{ \frac{d-1}{\eta} -2 }
 \nonumber
 \\
 &  \times  &
 \int_{0}^{ p_{\rm c} | \cos \vartheta |^{ - \frac{1}{\eta}}}  \D k k^{d-2} 
 F (  \lambda_d^{-1} k^{\eta}  , gk( \cos \vartheta )^{\frac{1}{\eta} -1 }  )
 \label{eq:Rbot4}
 \; \; \; ,
 \end{eqnarray}
with 
 \begin{eqnarray}
 F ( E , \gamma )  & = &  
 \int_{- \infty}^{\infty} { \D u}
 \frac{ 1 }{ \left[ E + | u |  \right] 
 \left[ \I u - 1 - \frac{\gamma}{2} \right]
 \left[ \I u - 1 + \frac{\gamma}{2} \right]}
 \nonumber
 \\
 & =  &
 \frac{2}{\gamma}
 \left\{ 
 ( 1 - \frac{\gamma}{2} )
 \int_{0}^{\infty} { \D u}
 \frac{ 1 }{ \left[ u+ E  \right] 
 \left[ u^2 + ( 1 - \frac{\gamma}{2} )^2 \right] }
 \right.
 \nonumber
 \\
 &  &
 \hspace{5mm} - \left.
 ( 1 + \frac{\gamma}{2} )
 \int_{0}^{\infty} { \D u}
 \frac{ 1 }{ \left[ u+ E  \right] 
 \left[ u^2 + ( 1 + \frac{\gamma}{2} )^2 \right] }
 \right\}
 \label{eq:I12pp}
 \; \; \; .
 \end{eqnarray}
In deriving the prefactor in Eq.\ref{eq:Rbot4} we have used 
$\gamma_d / \lambda_d = (d-1) / \pi$,
see Eq.\ref{eq:chidgaugedef}.
Because we are interested in the singularities
of the integrand for small $\cos \vartheta$, we have 
replaced the upper cutoff $\pm y_{\rm c} / | \cos \vartheta |$ for the 
$u$-integration by $\pm \infty$. 
We shall verify a posteriori that
the integral without cutoff is convergent,
so that this procedure is justified.
Note, however, that we retain the cutoff $p_{\rm c}$ for the $k$-integration
in Eq.\ref{eq:Rbot4},
because for the  two-dimensional Maxwell-Chern-Simons theory
the value of the integral will crucially  depend on this cutoff (see 
Sect.~\secref{subsec:ChernSimons} below).
Using
 \begin{equation}
 \int_{0}^{\infty} \D u \frac{1}{ [ u +  E ] [ u^2 + a^2 ]}
  =  \frac{1}{a^2 + E^2} \left[ \frac{ \pi E}{2 a} - \ln ( \frac{E}{a} ) \right]
  \; \; \; ,
 \end{equation}
the integrations in Eq.\ref{eq:I12pp} are easily done, and we obtain
 \begin{eqnarray}
 F ( E , \gamma ) & = &
 \frac{\pi E}{\gamma} \left\{ 
 \frac{ {\rm sgn} ( 1 - \frac{ \gamma}{2} )}{ E^2 + ( 1 - \frac{\gamma}{2} )^2 } -
 \frac{ 1 }{ E^2 + ( 1 + \frac{\gamma}{2} )^2 } \right\}
 \nonumber
 \\
 & + & \frac{2}{\gamma} \left\{
 \frac{ ( 1 - \frac{\gamma}{2} ) \ln \left[ \frac{ 
 |1 - \frac{\gamma}{2} |}{E} \right]}{ E^2 + ( 1 - \frac{\gamma}{2} )^2 }
 -
 \frac{ ( 1 + \frac{\gamma}{2} ) \ln \left[ \frac{ 
 |1 + \frac{\gamma}{2} |}{E} \right]}{ E^2 + ( 1 + \frac{\gamma}{2} )^2 }
 \right\}
 \label{eq:FEGdef}
 \; \; \; .
 \end{eqnarray}
It is easy to show that the function 
$F ( E , \gamma )$ has a finite limit as $\gamma \rightarrow 0$,
which is given by \cite{Kopietz95b}
 \begin{equation}
 F ( E , 0 ) = 2 \frac{ \pi E - E^2 -1 + ( E^2 -1 ) \ln E }{ ( 1 + E^2 )^2}
 \; \; \; .
 \label{eq:FEGlim}
 \end{equation}
The cancellation of the singular prefactor $1/ \gamma$ in Eq.\ref{eq:FEGdef}
can be traced back to the factor ${\rm sgn} ( \xi^{\alpha}_{\vec{q}} )$ 
in our general spectral representation given in Eq.\ref{eq:R2new}. In fact, it is
instructive to re-derive Eqs.\ref{eq:Rbot4} and \ref{eq:FEGdef}
from Eq.\ref{eq:R2new}.
Therefore we simply note that the gauge field propagator in Eq.\ref{eq:Hrparad4} 
can also be written as
 \begin{equation}
 h^{{\rm RPA} , \alpha}_{q} = - ( h^{\alpha}_{\vec{q}} )^2 \int_{0}^{\infty} \D {\omega}
 S_{\rm RPA} ( {\vec{q}} ,\omega ) \frac{ 2 \omega}{\omega^2 + \omega_m^2}
 \label{eq:hrpaspec1}
 \; \; \; ,
 \end{equation}
with
 \begin{equation}
 ( h^{\alpha}_{\vec{q}} )^2 \equiv 
 \frac{ 1 - ({\hat{\vec{k}}}^{\alpha} \cdot {\hat{\vec{q}}} )^2}{ \nu^2 }
 \left( \frac{q_{\rm c}}{ | {\vec{q}} | } \right)^{\eta}
 \label{eq:halphaaux}
 \; \; \; ,
 \end{equation}
and
 \begin{equation}
 S_{\rm RPA} ( {\vec{q}} ,\omega ) = \frac{ \nu}{\pi}
 {\rm Im} \left\{  \frac{1}{1 + ( \frac{q_{\rm c}}{ | {\vec{q}} | } )^{\eta} 
 {\Lambda}_d ( \frac{  \omega  }{ v_{\rm F} | {\vec{q}} | } + \I 0^{+}) }
 \right\}
 \label{eq:Shdef}
 \; \; \; .
 \end{equation}
For $| \omega_m | \ll v_{\rm F} | {\vec{q}} |$ 
we may approximate (see Eq.\ref{eq:Imgbardef})
\begin{equation}
 {\Lambda}_d ( \frac{ \omega  }{ v_{\rm F} | {\vec{q}} | } + \I 0^{+} ) 
 \approx - \I \lambda_d 
  \frac{ \omega }{ v_{\rm F} | {\vec{q}} | }
  \; \; \; ,
  \end{equation}
so that in this regime Eq.\ref{eq:Shdef} reduces to the 
usual dynamic structure factor due to an overdamped mode, 
\index{dynamic structure factor!transverse gauge fields}
 \begin{equation}
 S_{\rm RPA} ( {\vec{q}} ,\omega ) = \frac{ \nu}{\pi}
 \frac{ \omega \Gamma_{\vec{q}} }{ \omega^2 + \Gamma_{\vec{q}}^2 }
 \label{eq:Shdamp}
 \; \; \; , \; \; \; 
 \Gamma_{\vec{q}} = \frac{
 v_{\rm F} | {\vec{q}} |^{ 1 + \eta} }{ \lambda_d  q_{\rm c}^{\eta} }
 \label{eq:dampedmodedef}
 \; \; \; .
 \end{equation}
Substituting Eq.\ref{eq:hrpaspec1} for the gauge field propagator
into Eq.\ref{eq:Qbot2}, and taking the limit $\beta \rightarrow \infty$,
the $\omega_m$-integral is easily done.
The result is formally identical with 
Eqs.\ref{eq:R2new}--\ref{eq:ImS2new},
except that we should replace
$f_{\vec{q}}^2 \rightarrow (h_{\vec{q}}^{\alpha} )^2$ and
omit the terms proportional to
$f_{\vec{q}}$.
After rescaling the integration variables as above, we obtain
the following alternative expression for $R_{\rm tr}$,
 \begin{eqnarray}
 R_{\rm tr}   & =  &
  -  
  \frac{2 (d-1) g^{d-2}}{ \pi^2}
 \int_{0}^{\pi} \D \vartheta ( \sin \vartheta )^d
 \nonumber
 \\
 & \times &
 \int_{0}^{p_{\rm c}} \D p p^{d-3}
 \left[ \Theta ( - \cos \vartheta - \frac{gp}{2} ) -
 \Theta (  \cos \vartheta + \frac{gp}{2} ) \right]
 \nonumber
 \\
 & \times & \int_{0}^{\infty} \D x 
 \frac{x}{ [ x^2 + \lambda_d^{-2} p^{2 \eta}  ][  x + | \cos \vartheta +
 \frac{gp}{2} | ] }
 \label{eq:Rtralternative}
 \; \; \; .
 \end{eqnarray}
The $x$-integration is easily performed using
 \begin{equation}
 \int_{0}^{\infty} \D x \frac{x}{ [ x^2 + E^2 ] [ x + a ]}
  =  \frac{E}{E^2 + a^2} \left[ \frac{ \pi }{2 } + 
  \frac{a}{E} \ln ( \frac{a}{E} ) \right]
  \; \; \; , \; \; a > 0 \; ,
  \label{eq:Iab2}
 \end{equation}
and after rescaling
$p =  | \cos \vartheta |^{\frac{1}{\eta} } k$ 
we recover Eq.\ref{eq:Rbot4}.

\begin{figure}
\sidecaption
\psfig{figure=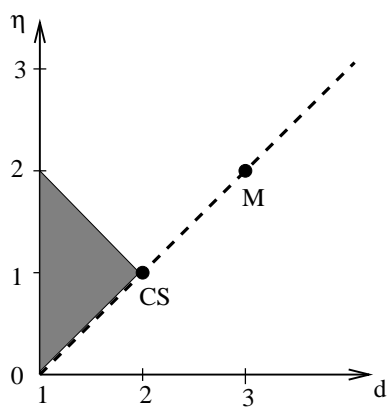,width=5.0cm}
\caption[Parameter regime where the transverse Debye-Waller factor is bounded.]
{The shaded triangle is the parameter regime in the $d-\eta$-plane
where the long-distance and large-time behavior of the Debye-Waller 
$Q^{\alpha}_{\rm tr} ( r^{\alpha}_{\|} \hat{\vec{v}}^{\alpha} , \tau )$
due to transverse gauge fields
gives rise to non-Fermi liquid behavior.
Note that for linearized energy dispersion one incorrectly 
obtains non-Fermi liquid behavior for all points above the dashed line 
$\eta = d - 1$.
The points M and CS correspond to the three-dimensional Maxwell theory
and the two-dimensional Maxwell-Chern-Simons theory, respectively.} 
\label{fig:shaded}
\end{figure}

From Eqs.\ref{eq:Rbot4} and \ref{eq:FEGdef} it is easy to 
determine whether $R_{\rm tr}$ is finite or not.
First of all, if we linearize the energy dispersion, we
effectively replace the function 
$F (  \lambda_d^{-1} k^{\eta}  , gk( \cos \vartheta )^{\frac{1}{\eta} -1 }  )$ in
Eq.\ref{eq:Rbot4} by $F (  \lambda_d^{-1} k^{\eta}  , 0 )$.
Because according to Eq.\ref{eq:FEGlim} this function is non-singular for
small $k$, the existence of $R_{\rm tr}$ is determined 
by the singularity in the remaining $\vartheta$-integration. 
In this way we recover the
criterion \ref{eq:Adetaexist}. 
On the other hand, for finite $g$ and $\eta > 1$ it is clear that
the singularity of the integrand 
of Eq.\ref{eq:Rbot4}
for small $\cos \vartheta$ is determined by the
large-$\gamma$  behavior of the function $F  ( E , \gamma)$, which is given by
 \begin{equation}
 F ( E , \gamma ) \sim \frac{ 4 \ln \gamma}{ \gamma^2} 
 \; \; \; , \;  \; \; \gamma \rightarrow \infty
 \; \; \; .
 \label{eq:GEGlarge}
 \end{equation}
Obviously the curvature term in the energy dispersion
gives rise to an additional factor of $ (\cos \vartheta )^{2 - \frac{2}{\eta} } / g^2$,
so that the most singular part of
the integral in Eq.\ref{eq:Rbot4} becomes
 \begin{equation}
 R_{\rm tr}^{\rm sing}    =  
  -  B_{d,\eta} g^{d-3}
 \int_{0}^{\pi/2} \D \vartheta ( \sin \vartheta )^d
 ( \cos \vartheta )^{ \frac{d-3}{\eta} }
 \; \; \; , \; \; \; \eta > 1
 \; \; \; ,
 \label{eq:Rpowereta}
 \end{equation}
where $B_{d,\eta}$ is some numerical constant which depends
on $d$ and $\eta$ in a complicated way, but remains finite as long as $\eta >1$.
By simple power counting, we see that this integral exists for
$\frac{d-3}{\eta} > -1$, i.e.
 \begin{equation}
 \eta > 3- d
 \label{eq:etaexist}
 \; \; \; .
 \end{equation}
Combining this result with the criterion \ref{eq:Adetaexist}
for $\eta < 1$, and using the fact that a finite value
of $R_{\rm tr}$ implies the boundedness of 
the total Debye-Waller factor $Q_{\rm tr} ( \tilde{x} , \tilde{\tau} )$ for
all $\tilde{x} $ and $ \tilde{\tau}$, we conclude that outside the
shaded region shown in Fig.~\secref{fig:shaded}
the contribution of the transverse gauge fields to the
Debye-Waller factor remains bounded.
Hence, non-Fermi liquid behavior 
due to 
$Q^{\alpha}_{\rm tr} ( r^{\alpha}_{\|} \hat{\vec{v}}^{\alpha} , \tau )$
is only possible in the shaded regime shown in Fig.~\secref{fig:shaded}.
In particular, for the three-dimensional Maxwell action
($\eta = 2 $, $d = 3$) higher-dimensional bosonization with 
linearized energy dispersion predicts  
that the static Debye-Waller factor grows logarithmically
with distance (as in the one-dimensional
Tomonaga-Luttinger model), while the
inclusion of curvature leads to a bounded Debye-Waller factor!

In the marginal case 
$\eta = 1$ our simple
power-counting analysis is not sufficient,
and we cannot avoid explicitly performing the relevant integrations.
Because in two dimensions the case $\eta = 1$ 
is of particular physical interest in connection with 
the half-filled Landau level, we shall 
analyze this case in some detail in the
following section.

\subsection{Two-dimensional Maxwell-Chern-Simons 
theory \index{Maxwell-Chern-Simons theory}}
\label{subsec:ChernSimons}

{\it{We evaluate the constant part $R_{\rm tr}$ of
the Debye-Waller factor given in Eq.\ref{eq:Rbot4}
for the special case of the 
two-dimensional Maxwell-Chern-Simons theory ($\eta = 1, d=2$).}}

\vspace{7mm}

\noindent
Two-dimensional electron systems in strong external magnetic fields
are difficult to handle within the framework of conventional many-body theory.
In fact, in this problem the most successful theories 
are based on variational wave-functions, and do not make use  
of the standard methods of second quantization \cite{Laughlin83}.
But also functional methods have been very 
fruitful \cite{Zhang89,Lopez91,Fradkin91}.
Of particular recent interest has been
the case
when the areal density of the electron gas and the strength of the 
external magnetic field are such that the lowest Landau level is
exactly half-filled\footnote{This means that
$N {\phi_0}/{\phi} = {1}/{2}$, 
where $N$ is the number of electrons in the system, $\phi_0 = {hc}/{e}$ is the
flux quantum, and $\phi$ is the total magnetic flux through the 
system.}. Then the two-dimensional electron system is mathematically
equivalent to a system of fermions interacting with a 
Chern-Simons gauge field \index{Chern-Simons theory} 
such that the average gauge field acting
on the fermions is zero \cite{Lopez91,Kalmeyer92,Halperin93}.
The Chern-Simons field effectively attaches two flux
quanta to each electron. The resulting spinless fermions are called
{\it{composite fermions}}\index{composite fermions} \cite{Jain89}.
At the mean-field level, where fluctuations of the Chern-Simons gauge
field are ignored, the magnetic field generated by the Chern-Simons field
exactly cancels the external magnetic field, so that
mean-field theory predicts that composite fermions
in the half-filled Landau level should behave like free spinless
fermions without magnetic field, with Fermi wave-vector $k_{\rm F} =
( 4 \pi n_{\rm e} )^{1/2}$. Here $n_{\rm e}$ is the areal density
of the two-dimensional electron gas.
Although the existence of a well-defined Fermi surface 
in the half-filled Landau level has been confirmed
by several 
experiments \cite{Willett93,Leadley94,Manoharan94,Coleridge95,Smet96}
and there exists general agreement that experimentally composite fermions
manifest themselves as well-defined quasi-particles,
theoretically the situation is less clear.
For a summary of 
the current status of the fermionic Chern-Simons description of 
quantum Hall systems
see the recent review by Halperin \cite{Halperin97}.

In the usual perturbative approach 
the leading correction to the Green's function of
composite fermions due to the
fluctuations of the Chern-Simons field 
is obtained from the GW self-energy
\index{GW approximation!for gauge fields}
(see Eq.\ref{eq:GW})
 \begin{equation}
 \Sigma_{\rm GW, tr}^{\alpha} ( \tilde{q} ) =
 - \frac{1}{\beta {{V}}} \sum_{q^{\prime}}
 h^{{\rm RPA}, \alpha }_{q^{\prime}} G_0^{\alpha} ( \tilde{q} + q^{\prime} )
 \label{eq:GWtr}
 \; \; \; .
 \end{equation}
In $d=2$
this expression is easily evaluated if one introduces
circular coordinates centered at $\vec{k}^{\alpha}$  
and first performs the angular integration
exactly \cite{Kopietz96che}. 
In the regime where $ v_{\rm F} | \vec{q} |$ is not much larger than
$| \tilde{\omega}_n | $, one finds
to leading order for small frequencies \cite{Halperin93,Kim94}
 \begin{equation}
 \Sigma_{\rm GW, tr}^{\alpha} ( \tilde{q} ) \propto - \I 
  \tilde{\omega}_n  \ln \left( 
 \frac{ v_{\rm F} q_{\rm c} }{ | \tilde{\omega}_n | } \right)
 \label{eq:sigmaGWgfres}
 \; \; \; .
 \end{equation}
After analytic continuation to real frequencies,
the real part of the self-energy
vanishes  as $\omega \ln ( v_{\rm F} q_{\rm c} / | \omega | )$
for $\omega \rightarrow 0$,
implying a logarithmically vanishing quasi-particle 
residue\footnote{
In this section we shall consider only the case
$\eta = 1$ (the
Maxwell-Chern-Simons theory\index{Maxwell-Chern-Simons theory}),
corresponding to the
unscreened Coulomb interaction.
In \cite{Kopietz96che} 
we have studied the general Chern-Simons
theory with $\eta > 1$. \index{Chern-Simons theory}
Then 
 $\Sigma_{\rm GW, tr}^{\alpha} ( \tilde{q} ) \propto - \I {\rm sgn}
 ( \tilde{\omega}_n ) | \tilde{\omega}_n |^{\frac{2}{ 1 + \eta } }$, so that
 the quasi-particle residue vanishes like a power law 
 for $\omega \rightarrow 0$.
Note that $\eta = 2$
describes experiments where
the long-range part of the Coulomb interaction is
screened by metal plates.  
}.
Because $Z^{\alpha} $ vanishes, the momentum distribution
does not exhibit a step-discontinuity at the Fermi surface 
(see Eq.\ref{eq:momentumZ}).
Thus, lowest order perturbation theory suggests that
composite fermions are {\it{not}} well defined quasi-particles.
This seems to disagree with the experimental 
evidence \cite{Willett93,Leadley94,Manoharan94,Coleridge95,Smet96}
that composite fermions behave like
well-defined quasi-particles.

\index{Maxwell-Chern-Simons theory}
Because the leading perturbative correction
completely changes the mean-field picture,
the single-particle Green's function
can only be calculated by means of non-perturbative methods, which
sum infinite orders in perturbation theory.
However, controlled non-perturbative methods
in $d > 1$ are rare, and it is at least controversial
whether the methods applied so far 
to the problem of composite fermions
in the half-filled Landau level
are really valid. These include
the so-called eikonal approximation \cite{Khveshchenko93},
a $1/N$-expansion \cite{Ioffe94}, and higher-dimensional
bosonization with linearized energy dispersion \cite{Kwon94,Houghton94}.
All of these methods predict some kind of non-Fermi liquid behavior, 
but there is no general agreement on the 
detailed form of the Green's function.
Although this might be related  to the 
gauge-dependence of
the single-particle Green's function
(see, however, the work \cite{Syljuasen96} and the 
footnote after Eq.\ref{eq:Gpatchgfdef}),
the discrepancies between the various approaches
could also be related to uncontrolled approximations inherent
in each of the different resummation schemes\footnote{For 
example, in the work \cite{Kopietzeik} we have
shown that in one dimension 
the usual eikonal approximation \cite{Fradkin66} does not correctly reproduce
the exact solution of the Tomonaga-Luttinger model.}.
Moreover, as we shall show in this section, even in 
the marginal case  of $\eta = 1$ 
higher-dimensional bosonization {\it{with linearized energy dispersion}}
does not correctly resum the leading singularities in the
perturbation series.
Note that in two dimensions the case  $\eta = 1$ corresponds 
to the right corner CS of the shaded triangle in Fig.~\secref{fig:shaded}.
Naively, one might expect that precisely on the boundary of the
triangle one obtains logarithmic singularities in the transverse Debye-Waller factor,
which are correctly predicted by bosonization with linearized energy dispersion \cite{Kwon94}.
We now show that at the special point $d=2$ and $\eta =1$ this is not the case.

\vspace{7mm}

Using the fact that in two dimensions $\lambda_2 = 1$
(see Eqs.\ref{eq:gammatildepart} and \ref{eq:chidgaugedef}),
we obtain from Eq.\ref{eq:Rbot4} for $\eta = 1$
 \begin{equation}
 R_{\rm tr}    =  
  -  
  \frac{ g}{ \pi^2}
 \int_{0}^{\pi/2} \D \vartheta \frac{( \sin \vartheta )^2}{
  \cos \vartheta  }
 \int_{0}^{ p_{\rm c}  ( \cos \vartheta )^{ - 1}}  \D k  
 F (  k  , gk  )
 \label{eq:RCS}
 \; \; \; .
 \end{equation}
To determine the proper values of our dimensionless constant
$g$, we note 
that in case of the Chern-Simons propagator the wave-vector $q_{\rm c}$
in Eq.\ref{eq:Hrparad3} is given by $q_{\rm c} = ( 2 k_{\rm F} )^2 / \kappa$, where
$\kappa =  2 \pi e^2 \nu = e^2 m $ is the Thomas-Fermi 
wave-vector in two dimensions \cite{Halperin93}. 
Hence we obtain from Eq.\ref{eq:ggdef} 
 \begin{equation}
 g = \frac{4 k_{\rm F} }{ \kappa} = \frac{4 v_{\rm F} }{ e^2}
 \label{eq:gCS}
 \; \; \; .
 \end{equation}
To obtain the relevant ultraviolet cutoff $p_{\rm c}$, we note that
the dimensionless Coulomb interaction becomes 
larger than unity for
$| {\vec{q}} |
{ \raisebox{-0.5ex}{$\; \stackrel{<}{\sim} \;$}} \kappa$, see
Appendix~\secref{subsubsec:Cb}. Hence,
 \begin{equation}
 p_{\rm c} = \frac{\kappa}{q_{\rm c}} = \left( \frac{\kappa}{2 k_{\rm F}} \right)^2
 = \frac{4}{  g^2}
 \label{eq:pcCS}
 \; \; \; .
 \end{equation}
Note that \index{Maxwell-Chern-Simons theory}
the charge $e$ in Eq.\ref{eq:gCS} should be understood as effective
screened charge, which takes the dielectric screening
due to the background into account.
Halperin, Lee and Read \cite{Halperin93} have estimated $v_{\rm F} / e^2 \approx 0.3$
in the experimentally relevant regime, so that
$g \approx 1.2$. 
Let us also emphasize that
the prefactor of $g = 4 k_{\rm F} / \kappa$ in Eq.\ref{eq:RCS}
is the {\it{inverse}} of the
prefactor of $\kappa /k_{\rm F}$ that appears 
in the constant part $R^{\alpha}$ of the Debye-Waller factor
for conventional density-density interactions
in $d=2$, see Eq.\ref{eq:Rex4}.
Hence the wave-vector scale $q_{\rm c} = ( 2 k_{\rm F} )^2 / \kappa$
now plays the same role as the
Thomas-Fermi wave-vector $\kappa$ in the case of density-density interactions.
Note, however, that $g = q_{\rm c} / k_{\rm F}$ is the
only small parameter which formally justifies the truncation of the
eikonal expansion \ref{eq:Qalphandef} at the first 
order\index{hidden small parameter}\footnote{
Recall that in Chap.~\secref{subsec:hidden} we have
shown by explicit calculation of corrections to the density-density correlation
function beyond the RPA that the loop integrations give rise
to additional powers of $q_{\rm c} / k_{\rm F}$, see
Eq.\ref{eq:gausscorrect}.
Because the non-Gaussian corrections
$Q_n^{\alpha} ( {\vec{r}} , \tau )$, $n \geq 2$,
to the average eikonal involve additional powers of the
interaction, these corrections are
controlled by higher orders in $q_{\rm c} / k_{\rm F}$.
Note that for a spherical Fermi surface the curvature parameter
$C^{\alpha}$ in Eq.\ref{eq:gausscorrect} is of the order of unity.
Furthermore, for the Chern-Simons action the value of the
relevant dimensionless effective interaction is not small,
which leaves us with
$q_{\rm c} / k_{\rm F}$ as the only small parameter in the problem.}.
We therefore conclude that Eq.\ref{eq:RCS}
can only be qualitatively correct for $g \ll 1$.
If this condition is satisfied,
the higher-order corrections $Q_n^{\alpha} ( {\vec{r}} , \tau )$, $n \geq 2$,
to the average eikonal (see Eq.\ref{eq:Qalphandef})
are controlled by higher
powers of $g$, which are generated by additional loop 
integrations.
Obviously, in the experimentally relevant regime \cite{Willett93}
the condition $g \ll 1$ is not satisfied, so that
for an accurate quantitative comparison with experiments
it is not sufficient to retain only the leading term
in the eikonal expansion.

In order to perform a controlled calculation, we 
shall restrict ourselves from now on to the regime $g \ll 1$, 
with the hope that  the qualitative behavior
of the Green's function does not change for larger $g$.
Naively one might be tempted to replace the upper limit
for $k$-integration in Eq.\ref{eq:RCS} by infinity, because
$p_{\rm c} = 4 / g^2 \gg 1 $ for small
$g$, and 
the $\vartheta$-integration seems to be dominated by 
the regime  $ \cos \vartheta \ll 1$. 
If we {\it{linearize}} the energy dispersion, such a procedure is indeed
correct, because in this case the $k$-integration yields a finite 
number, which is according to 
Eq.\ref{eq:FEGlim} given by
 \begin{equation}
 \int_{0}^{\infty} \D k F ( k , 0)
  = 2 \int_{0}^{\infty} \D k \frac{ \pi k - k^2 -1 + ( k^2 -1 ) \ln k }{ ( 1 + k^2 )^2}
 \; \;  \; .
 \label{eq:FEGlimint}
 \end{equation}
To perform the integration, we need \cite{Gradshteyn80,Groebner61}
 \begin{eqnarray}
 I_{ \mu } & =  &
 \int_{0}^{\infty} \D x \frac{ x^{2 \mu -1} }{ [ 1 + x^2 ]^2 }
  =  \frac{1}{2} \Gamma ( \mu ) \Gamma ( 2 - \mu )
 \; \; \; ,  \; \; \;  0 < \mu < 2
 \; \; \; ,
 \label{eq:I1def}
 \\
 \tilde{I}_{ \mu } & = & \int_{0}^{\infty} \D x \frac{ x^{2 \mu -1} \ln x }{ [ 1 + x^2]^2 }
 \nonumber
 \\
 & = &
 \left\{
 \begin{array}{ll}
 \frac{( \mu - 1) \pi }{ 4 \sin ( \pi \mu ) }
 \left[ \cot( \pi \mu ) - \frac{ 1}{\mu -1} \right]
 &
 \; \; , \; \; 0 < \mu < 2 \; , \; \mu \neq 1 \\
 0 &
 \; \; , \; \;  \mu = 1 
 \end{array}
 \right.
 \; \; \; .
 \end{eqnarray}
Hence,
 \begin{equation}
 \int_{0}^{\infty} \D k F ( k , 0)
 = 2 \left[
  \pi I_{ \frac{ 3}{4}  }
 -I_{\frac{ 3}{2}  } 
 - I_{ \frac{ 1}{2}  } +
 {\tilde{I}}_{\frac{ 3}{2} } -
  \tilde{I}_{\frac{ 1}{2} }
  \right]
 \label{eq:bdres}
 \; \; \; .
 \end{equation}
With
 $I_{1}  = 1/2$, $ I_{ \frac{1}{2} } = I_{ \frac{3}{2} } = { \pi}/{4}$,
 $\tilde{I}_{ \frac{1}{2} } = - {\pi}/{4}$, and  $\tilde{I}_{ \frac{3}{2} } = {\pi}/{4}$
we finally obtain
 \begin{equation}
 \int_{0}^{\infty} \D k F ( k , 0 ) = \pi
 \; \; \; .
 \label{eq:Fkg0}
 \end{equation}
It is now easy to see that $R_{\rm tr}$ is logarithmically divergent.
Of course, in this case we should consider the total Debye-Waller factor
$Q_{\rm tr} ( \tilde{x} , \tilde{\tau} )$
in Eq.\ref{eq:Qbot4}, which grows logarithmically for large
$\tilde{x}$ or $\tilde{\tau}$. Setting
for simplicity $\tilde{\tau} = 0$, it is easy to show 
from Eqs.\ref{eq:Qbot4} and \ref{eq:RCS} that, 
to leading logarithmic order
for large $\tilde{x}$,
one obtains with linearized energy dispersion
 \begin{equation}
 \hspace{-4mm}
 Q_{\rm tr} ( \tilde{x} , 0 )     \sim
  -  
  \frac{ g}{ \pi}
 \int_{0}^{\pi/2} \D \vartheta \frac{( \sin \vartheta )^2}{
  \cos \vartheta  }
  \left[ 1 - \cos [ ( \cos \vartheta )^2 \tilde{x} ] \right]
  \sim - \frac{g}{2 \pi} \ln \tilde{x}
 \label{eq:QCS}
 \;  .
 \end{equation}
This implies anomalous scaling characteristic for Luttinger liquids,
with anomalous dimension given by $\gamma_{\rm CS} = {g}/({2 \pi})$.

\vspace{7mm}

The crucial point is now that the above result is completely changed by 
the quadratic term in the energy dispersion, because 
even for small $g$ it is {\it{not}} allowed to set $g=0$ in the
integrand $F ( k , gk)$ of Eq.\ref{eq:RCS}. 
To see this, consider the function
 \begin{equation}
 J (  g , h ) = \int_{0}^{h} \D k F ( k , gk )
 \label{eq:Igldef}
 \; \; \; , \;  \; \; g > 0
 \; \; \; .
 \end{equation}
According to Eq.\ref{eq:RCS}
the constant part $R_{\rm tr}$ of the Debye-Waller factor
is determined by $J ( g , p_{\rm c} (\cos \vartheta )^{-1} )$, where
$p_{\rm c} \propto g^{-2}$, see Eq.\ref{eq:pcCS}.
For $g=0$ we have from Eq.\ref{eq:bdres} 
$\lim_{h \rightarrow \infty} J ( 0 , h ) = \pi$.
To evaluate $J ( g , h )$ for finite $g$, 
we use Eq.\ref{eq:FEGdef} to write
 $J (  g , h) = \sum_{n=1}^{4}
 J_n (  g , h )$, where\footnote{
Although $J_3$ and $J_4$ are logarithmically divergent, 
the divergence cancels in the sum $J_3 + J_4$, which is the only
relevant combination.}
\begin{eqnarray}
 & &  \hspace{-22mm}
 J_1 (  g , h )  = 
 \frac{\pi }{g} 
 \left[
 \int_{0}^{2/g} \D k 
 \frac{ 1}{ k^2 + ( 1 - \frac{gk}{2} )^2 } 
 -
 \int_{2/g}^{h} \D k 
 \frac{ 1}{ k^2 + ( 1 - \frac{gk}{2} )^2 } 
 \right]
 \label{eq:I1gl}
 \; \; \; ,
 \\
 J_2 (  g , h ) & = &
 - \frac{\pi }{g} 
 \int_{0}^{h} \D k 
 \frac{  ( 1 + \frac{ gk}{2} )}{ k^2 + ( 1 + \frac{gk}{2} )^2 } 
 \label{eq:I2gl}
 \; \; \; ,
 \\
 J_3 (  g , h ) & = &
 \frac{2}{g}
 \int_{0}^{h} \frac{\D k}{k}
 \frac{ ( 1 - \frac{gk}{2} ) \ln \left[ \frac{ 
 |1 - \frac{gk}{2} |}{k} \right]}{ k^2 + ( 1 - \frac{gk}{2} )^2 }
 \label{eq:I3gl}
 \; \; \; ,
 \\
 J_4 (  g , h ) & = &
 - \frac{2}{g}
 \int_{0}^{h} \frac{\D k}{k}
 \frac{ ( 1 + \frac{gk}{2} ) \ln \left[ \frac{ 
 |1 + \frac{gk}{2} |}{k} \right]}{ k^2 + ( 1 + \frac{gk}{2} )^2 }
 \label{eq:I4gl}
 \; \; \; .
 \end{eqnarray}
From Eq.\ref{eq:pcCS} we see that the upper limit for 
the $k$-integration in Eq.\ref{eq:Rbot4}
is large compared with $g/2$, so that we may assume $h > g/2$.
In the first integral on the right-hand side of
Eq.\ref{eq:I1gl} and 
in Eq.\ref{eq:I3gl}
we substitute $ x = {k}/({ 1 - \frac{gk}{2} })$
(so that $\frac{\D x } {x^2} = \frac{\D k }{ k^2}$), and in the
second integral of Eq.\ref{eq:I1gl} we set
$ x = {k}/ ( { \frac{gk}{2} -1 })$ (so 
that $\frac{\D x }{ x^2} = - \frac{\D k }{ k^2}$). Similarly, in
the above expressions for $J_2$ and $J_4$ we substitute
$ x = {k}/ ({ 1 + \frac{gk}{2} })$
(so that again $\frac{\D x }{ x^2} = \frac{\D k }{ k^2}$).
With these substitutions it is easy to show that
\index{boring integrals}
 \begin{eqnarray}
 J_1 (  g , h) 
 + J_2 (  g , h )  & =  &
 \frac{\pi}{g} \int_{ 
 \frac{2 }{g + {2}/{ h} } 
 }^{
 \frac{2 }{g - {2}/{ h} }} \D x \frac{1}{1 + x^2}
 \nonumber
 \\
 &  & \hspace{-20mm} = \frac{\pi}{g}
 \left\{ \arctan 
 \left[ \frac{2}{g ( 1 - \frac{2}{g h} )} \right]
 -
 \arctan 
 \left[ \frac{2}{g ( 1 + \frac{2}{g h}) } \right]
 \right\}
 \label{eq:J1J2res}
 \; \; \; ,
 \end{eqnarray}
and
\begin{equation}
 J_3 (  g , h ) 
 + J_4 (  g , h )   =  
 - \frac{2}{g} \int_{ 
 \frac{2 }{g + {2}/{ h} } 
 }^{
 \frac{2 }{g - {2}/{ h} }} \D x \frac{ \ln x }{x (1 + x^2)}
 \label{eq:J3J4res}
 \; \; \; .
 \end{equation}
In the limit of interest ($g \ll 1$, $g h \gg 1)$
the width of the interval of integration is small,
 \begin{equation}
 \frac{2}{g ( 1 - \frac{2}{ g h} )} 
 -
 \frac{2}{g ( 1 + \frac{2}{g h} )} 
 \approx \frac{8}{g^2 h}
 \; \; \; .
 \label{eq:integralwidth}
 \end{equation}
Hence, to leading order, the integrals can be approximated by 
the product of the value of the integrand at
$x = 2/g$ and the width of the interval
of integration.
Then we obtain to leading order
 \begin{eqnarray}
 J_1 (  g , h ) 
 + J_2 (  g , h )  & \approx &
 \frac{2 \pi}{g h}  
 \label{eq:J1J2approx}
 \; \; \; ,
 \\
 J_3 (  g , h ) 
 + J_4 (  g , h )  & \approx &
 - \frac{4 \ln g^{-1} }{ h}  
 \label{eq:J3J4approx}
 \; \; \; .
 \end{eqnarray}
Note that for small $g$ the contribution $J_1+J_2$ is dominant.
Taking the limit $h \rightarrow \infty$, we obtain\index{boring integrals}
$\lim_{h \rightarrow \infty} J ( g , h ) = 0$, 
so that we conclude that
 \begin{equation}
 \int_{0}^{\infty} \D k F ( k , gk ) = 0
 \; \; \; , \; \; \; g > 0
 \; \; \; ,
 \label{eq:Fkg1}
 \end{equation}
which should be compared with Eq.\ref{eq:Fkg0}.
Because Eq.\ref{eq:J1J2approx} depends
on the {\it{product}} of the small parameter $g$ and the
large parameter $h$, it is clear that
for any finite $g$ the limiting behavior
of the integral $J ( g , h )$ for large $h$ 
is very different from $\lim_{h \rightarrow \infty}
J ( 0 , h ) = \pi$.
This is the mathematical reason why the linearization of the energy dispersion
in the two-dimensional Chern-Simons theory 
is not allowed.
Using Eq.\ref{eq:Fkg1}, 
we see that Eq.\ref{eq:RCS} can be  rewritten as
 \begin{equation}
 R_{\rm tr}    =  
   \frac{ g}{ \pi^2}
 \int_{0}^{\pi/2} \D \vartheta \frac{( \sin \vartheta )^2}{
  \cos \vartheta  }
 \int_{ p_{\rm c}  ( \cos \vartheta )^{ - 1}}^{\infty}  \D k  
 F (  k  , gk  )
 \label{eq:RCS2}
 \; \; \; .
 \end{equation}
From this expression it is evident that the cutoff-dependence
of the $k$-integral gives rise to an additional power
of $\cos \vartheta$ in the numerator, which removes the logarithmic
divergence that has been artificially generated by linearizing
the energy dispersion.
According to Eq.\ref{eq:pcCS} we should choose $p_{\rm c} = 4 / g^2$, 
so that we obtain to leading order for small $g$
(see Eq.\ref{eq:J1J2approx})
 \begin{equation}
 \int_{ \frac{4}{ g^{2}}( \cos \vartheta )^{ - 1}}^{\infty}  \D k  
 F (  k  , gk  ) \sim - \frac{\pi}{2} g \cos \vartheta
 \label{eq:Fgasymres}
 \; \; \; .
 \end{equation}
Hence Eq.\ref{eq:RCS2}
reduces to
 \begin{equation}
 R_{\rm tr}      \sim
  - \frac{1}{8}g^2
  \; \; \; , \; \; \;
  g \ll 1
  \; \; \; .
  \label{eq:RCS3}
  \end{equation}
The  precise numerical value of the prefactor $1/8$ 
is the result of our special choice of the cutoff $p_{\rm c}$ in Eq.\ref{eq:pcCS} and 
has no physical significance.
However, Eqs.\ref{eq:QCS} and \ref{eq:RCS3}
imply that in the case of the two-dimensional Chern-Simons theory
it is {\it{not}} allowed to linearize the energy dispersion \cite{Kwon94}.
Physically  Eq.\ref{eq:RCS3} represents a contribution 
from gauge field fluctuations with wavelengths 
large compared with the Thomas-Fermi screening length
$ \kappa^{-1}$
to the reduction of the quasi-particle residue. 
While for linearized energy dispersion one finds that these fluctuations
wash out any step-discontinuity at the Fermi surface,
the quadratic term in the energy dispersion drastically changes this
scenario: in the regime $g \ll 1$ under consideration
the right-hand-side of
Eq.\ref{eq:RCS3} is very small, so that this
term can be safely ignored and certainly
does not modify the mean-field prediction of 
a step discontinuity at the Fermi surface.

\section{Summary and outlook}
\label{sec:opengauge}

In this chapter we have generalized our non-perturbative background field
method for calculating the single-particle Green's function
to the case of
fermions that are coupled to transverse gauge fields.
Let us summarize  again our main result for the special case of
a spherical Fermi surface in $d$ dimensions.
As discussed in Chaps.~\secref{sec:sectors} and \secref{sec:sumgreen},
in this case
it is {\it{not}} necessary to partition the
Fermi surface into several patches, 
so that uncontrolled corrections due to
the around-the-corner processes discussed in Chap.~\secref{subsec:proper}
simply do not arise.\index{around-the-corner processes}
The Matsubara Green's function
can then be written as
 \begin{equation}
 G ( {\vec{k}}^{\alpha} + \vec{q} , \I \tilde{\omega}_n ) = 
 \int \D {\vec{r}} \int_0^{\beta} \D \tau \E^{ - \I ( 
 \vec{q} \cdot \vec{r} - \tilde{\omega}_n \tau )}
 \tilde{G}^{\alpha} ( \vec{r} , \tau ) \E^{ 
 Q^{\alpha} ( \vec{r} , \tau )}
 \label{eq:Gresultsum2}
 \; \; \; ,
 \end{equation}
 \begin{equation}
 Q^{\alpha} ( \vec{r} , \tau ) =
 Q_1^{\alpha} ( \vec{r} , \tau ) + 
 Q_{\rm tr}^{\alpha} ( \vec{r} , \tau )
 \label{eq:DWtotresrad}
 \; \; \; ,
 \end{equation}
where the longitudinal
Debye-Waller factor $Q_1^{\alpha} ( \vec{r} , \tau )$
is given in Eqs.\ref{eq:Qlondef2}--\ref{eq:Slondef2},
and the contribution 
 $Q_{\rm tr}^{\alpha} ( \vec{r} , \tau )$ from
 the transverse gauge field to the Debye-Waller factor is
given in Eqs.\ref{eq:Qraddef}--\ref{eq:Sraddef}.
The prefactor Green's function \index{Green's function!prefactor}
 $\tilde{G}^{\alpha} ( \vec{r} , \tau ) $ 
has the following Fourier expansion,
 \begin{equation}
 \tilde{G}^{\alpha} ( {\vec{r}} , \tau  )
 = \frac{1}{\beta V} \sum_{\tilde{q}} 
 \E^{ \I ( {\vec{q}} \cdot {\vec{r}} - 
 \tilde{\omega}_n \tau ) }
 \tilde{G}^{\alpha} ( \tilde{q} ) 
 \; \; \; ,
 \label{eq:prffouriertr}
 \end{equation}
 \begin{equation}
 \tilde{G}^{\alpha} ( \tilde{q} ) 
 =
  \frac{ 1 + Y^{\alpha} ( \tilde{q} ) + Y^{\alpha}_{\rm tr} ( \tilde{q} ) }
  { \I \tilde{\omega}_n - 
  \epsilon_{ {\vec{k}}^{\alpha} + {\vec{q}} } + \mu
  - \Sigma_1^{\alpha} ( \tilde{q} ) 
  - \Sigma_{1 , {\rm tr}}^{\alpha} ( \tilde{q} ) 
  }
  \label{eq:Gtotalpre2tr}
  \; \; \; ,
  \end{equation}
where the self-energies and the vertex functions are given in
Eqs.\ref{eq:sigma1res3}, \ref{eq:Yres4}, \ref{eq:sigma1res3tr}, and
\ref{eq:Yres4tr}.
Due to the spherical symmetry,\index{spherical symmetry}
it is sufficient to 
evaluate Eq.\ref{eq:Gresultsum2} for external wave-vectors of the
form $\vec{q} = q^{\alpha}_{\|} \hat{\vec{k}}^{\alpha}$, and then
replace $q^{\alpha}_{\|} \rightarrow | {\vec{k}} | - k_{\rm F}$ in the
final result, see also
Eqs.\ref{eq:Galphashiftsym},\ref{eq:Gresultsum}
and the discussion in Chap.~\secref{sec:sectors}.

In Sect.~\secref{sec:TransDWMax} we have 
shown that for the calculation of the transverse
Debye-Waller factor $Q^{\alpha}_{\rm tr} ( \vec{r} , \tau )$ 
it is essential to retain the quadratic term
in the expansion of the energy dispersion close to the
Fermi surface. In physically relevant cases one obtains then a
{\it{bounded}} Debye-Waller factor, 
which does not 
lead to a breakdown of the Fermi liquid state. 
This is in sharp contrast with the results of
higher-dimensional bosonization with linearized 
energy dispersion
\cite{Kwon94,Houghton94,Kopietz95b}. 
We would like to emphasize
that the quadratic term in the
energy dispersion is {\it{irrelevant}} in the renormalization group sense.
However, it is relevant
in the sense that
the {\it{exponentiation of the perturbation series for the
real-space Green's function}},
which in arbitrary dimensions is the characteristic feature of bosonization with
linearized energy dispersion,
does not resum the dominant singularities.

One of the most interesting 
problems for further research is the
{\bf{evaluation of the prefactor Green's function}}
$\tilde{G}^{\alpha} ( \tilde{q} )$ in the case of the
Chern-Simons theory for the half-filled Landau level. 
Because by construction our approach
exactly reproduces the  leading term in a naive 
expansion of the Green's function in
powers of the effective interaction (see Chap.~\secref{sec:connectionPT}),
the perturbatively obtained signature \ref{eq:sigmaGWgfres} 
of non-Fermi liquid behavior is certainly 
contained in Eqs.\ref{eq:Gresultsum2}--\ref{eq:Gtotalpre2tr}.  
Very recently 
Castilla and the present author \cite{Kopietz96che}
have made considerable progress in evaluating 
the above expressions for the case $\eta > 1$.
Because we know from Sect.~\secref{sec:TransDWMax}
that the contribution from the 
Debye-Waller factor is finite and small,
the low-energy behavior of the total spectral function 
is essentially determined
by the imaginary part of the prefactor
Green's function $\tilde{G}^{\alpha} ( \vec{q} , \omega + \I 0^{+} )$.
Most importantly,  
we have shown in the Letter \cite{Kopietz96che} that the Gaussian fluctuations of the
Chern-Simons gauge field do not invalidate
the quasi-particle picture for the composite fermions
in the half-filled Landau level.
On other words,
our non-perturbative approach predicts
a narrow peak in the spectral function,
with a width that vanishes faster than
the quasi-particle energy as $\vec{q} \rightarrow 0$.
This clearly demonstrates
that lowest order perturbation theory is not reliable, 
and explains the experimental fact
that composite fermions in
half-filled quantum Hall systems
behave like well-defined
quasi-particles \cite{Willett93,Leadley94,Manoharan94,Coleridge95,Smet96}.

In our opinion,
the calculation of the Green's function
of fermions that are coupled to gauge fields 
is the physically most interesting and important application
of the non-perturbative method developed in this book. 
Gauge fields in non-relativistic condensed matter systems
arise not only in connection with the quantum Hall effect,
but also in effective low-energy theories
for strongly correlated Fermi
systems \cite{Anderson90b,Baskaran87,Lee89,Ioffe89,Blok93}.
Because the gauge field problem cannot be 
analyzed within perturbation theory,
controlled non-perturbative methods are necessary. 
In the absence of any other small parameter,
the Gaussian approximation employed in our 
background field approach is justified 
for  $g = q_{\rm c} / k_{\rm F} \ll 1 $ (see
Eq.\ref{eq:ggdef}).
In this case
the closed loop theorem
discussed in Chap.~\secref{sec:closedloop}
guarantees that the corrections to the
Gaussian approximation
involve higher powers of our small parameter $g$.

It seems that the potential of our approach 
is far from being exhausted.
Let us point out two obvious directions for 
further research.
First of all, 
the combination of the methods developed in Chap.~\secref{chap:adis} with the
results of the present chapter might  lead to a new non-perturbative
approach to the {\bf{random gauge field problem}}.\index{random gauge fields}
Random gauge fields and the related problem of random magnetic fields
have  recently been analyzed with the help of many 
different  
methods \cite{Altshuler92,Khveshchenko93b,Aronov94,Zhang94,Ludwig94}.
Of course, this problem is 
interesting 
in connection with the quantum Hall effect,
because experimental systems always have a finite 
amount of disorder.
Another interesting and only partially solved problem is the
{\bf{explicit calculation of the non-Gaussian
corrections}} to our
non-perturbative result for the single-particle Green's function. 
Recall that in Chap.~\secref{sec:beyond}
we have performed such a calculation 
for the density-density correlation function.
Although
in Chap.~\secref{sec:eik} we have derived explicit
expressions for the non-Gaussian corrections
to the average eikonal
(see Eqs.\ref{eq:Qalphandef}--\ref{eq:W22retain}), 
a detailed analysis of the leading
correction to the Gaussian approximation still remains
to be done.
At this point we cannot exclude the possibility that, although the leading term
in the expansion of the average eikonal (i.e.
the Debye-Waller factor $Q^{\alpha}_{\rm tr} ( \vec{r} , \tau )$
given in Eq.\ref{eq:Qbot2})
remains bounded for all 
$\vec{r} $ and $\tau$, the higher order terms
$Q_n^{\alpha} ( {\vec{r}} , \tau )$, $n \geq 2$, 
in Eq.\ref{eq:Qngeneral}
exhibit singularities which lead to non-Fermi liquid behavior in the
spectral function.

%
%
%

%
%
%

\appendix
\chapter*{Appendix: Screening and collective modes}
\label{chap:ascr}
\setcounter{equation}{0}
\setcounter{footnote}{0}
\renewcommand{\theequation}{A.\arabic{equation}}
\renewcommand{\thechapter}{A}
\markboth{Appendix: Screening and collective modes}{}
\addcontentsline{toc}{chapter}{Appendix: Screening and collective modes}

\setcounter{equation}{0}

{\it{
We summarize some useful expressions for the
polarization, the dynamic structure factor and the
long wavelength behavior of the collective plasmon mode within the RPA.
The results presented in this chapter are not new, but a systematic
discussion of the above quantities as function
of dimensionality seems not to exist in the literature.
}}

\section[The non-interacting polarization for spherical Fermi surfaces]
{The non-interacting polarization \mbox{\hspace{35mm}}
for spherical Fermi surfaces}
\label{subsec:lonscr}

{\it{\ldots which we need in order to calculate the dynamic 
structure factor within the RPA.
Here and in the following two sections we assume
spherical symmetry. 
}} 

\vspace{7mm}

\noindent
For a spherical Fermi surface \index{spherical symmetry}
in $d$ dimensions
it is easy to show from
Eqs.\ref{eq:Pitotdecompose} and \ref{eq:Pilong} that
the non-interacting polarization\index{polarization!spherical Fermi surface} is 
in the limit $V , \beta \rightarrow \infty$ 
and $| {\vec{q}} | \ll k_{\rm F}$ given by
 \begin{equation}
 \Pi_{0} ( q ) = \nu g_{d} \left( { \frac{\I 
 \omega_{m} }{ v_{\rm F} | {\vec{q}} | } } \right)
 \; \; \; ,
 \label{eq:Piglobalspherical}
 \end{equation}
where the density of states\index{density of states!total} 
at the Fermi energy is (see Eq.\ref{eq:nudef})
 \begin{equation}
 \nu =  \int \frac{ \D {\vec{k}}}{ ( 2 \pi )^d} \delta ( \epsilon_{\vec{k}} - \mu )
 \; \; \; ,
 \label{eq:nudefinf}
 \end{equation}
and the dimensionless function $g_{d} ( z )$ is defined by
 \begin{equation}
 g_{d} ( z ) = \left< \frac{ \hat{\vec{q}} \cdot \hat{\vec{k}} }{ \hat{\vec{q}} \cdot \hat{\vec{k}} - z } 
 \right>_{{\hat{\vec{k}}}}
 \label{eq:gddef}
 \; \; \; .
 \end{equation}
Here $\hat{\vec{k}} = {\vec{k}} / | {\vec{k}} |$, $\hat{\vec{q}} = {\vec{q}} / | {\vec{q}} |$, and
$ < \ldots >_{\hat{\vec{k}}}$ denotes angular average over the
surface of the $d$-dimensional unit sphere in ${\vec{k}}$-space, i.e.
for any function $f ( \hat{\vec{k}} )$ 
 \begin{equation}
 \left<  f ( \hat{\vec{k}} ) \right>_{\hat{\vec{k}} } =
 \frac{ \int \D \Omega_{ \hat{\vec{k}}} f ( \hat{\vec{k}} ) }{
 \int \D \Omega_{ \hat{\vec{k}}} }
 \label{eq:angav}
 \; \; \; ,
 \end{equation}
where $\D \Omega_{ \hat{\vec{k}} }$ is the differential solid angle
at point $\hat{\vec{k}}$ on the unit sphere.
Note that by construction
 $g_{d} ( 0 ) = 1$.
For a system of $N$ spinless electrons with mass $m$ in a $d$-dimensional volume $V$
the density of states can be written as
 \begin{equation}
 \nu = 
 \frac{d}{2 \mu } \frac{N}{V } = \frac{ \Omega_{d}}{ (2 \pi )^d } 
 \frac{ k_{\rm F}^{d-1}}{  v_{\rm F} }
 = \frac{ \Omega_{d}}{ ( 2 \pi )^{d} } m k_{\rm F}^{d-2}
 \label{eq:nurelation}
 \; \; \; ,
 \end{equation}
where $\Omega_{d}$ is the surface area of the unit sphere in
$d$ dimensions, 
 \begin{equation}
 \Omega_{d} = \int \D \Omega_{\hat{\vec{k}}} = 
 \frac{ 2 \pi^{\frac{d}{2}} }{ \Gamma ( \frac{d}{2} ) }
 \; \; \; .
 \label{eq:omegad}
 \end{equation}
The integrand in Eq.\ref{eq:gddef} depends only on 
$\cos \vartheta = \hat{\vec{q}} \cdot \hat{\vec{k}}$.
For this type of functions it is convenient to use $d$-dimensional 
spherical coordinates\index{spherical coordinates}, 
 \begin{equation}
 \left<  f ( \hat{\vec{q}} \cdot \hat{\vec{k}} ) \right>_{\hat{\vec{k}}} =
 \gamma_{d}
 \int_{0}^{\pi} \D \vartheta ( \sin \vartheta )^{d-2} f ( \cos \vartheta ) 
 \label{eq:angavsimplify}
 \; \; \; , \; \; \; \mbox{for $d > 1$}
 \; \; \; ,
 \end{equation}
 \begin{equation}
 \left<  f ( \hat{\vec{q}} \cdot \hat{\vec{k}} ) \right>_{\hat{\vec{k}}} =
 \frac{1}{2} \left[ f (1) + f (-1) \right]
 \; \; \; , \; \; \; \mbox{for $d = 1$}
 \label{eq:angavsimp1}
 \; \; \; .
 \end{equation}
Here the numerical constant $\gamma_{d}$ is defined by
 \begin{equation}
 \gamma_{d} = 
 \left< \delta ( \hat{\vec{q}} \cdot \hat{\vec{k}} ) \right>_{\hat{\vec{k}}}  =
 \left[  \int_{0}^{\pi} \D \vartheta ( \sin \vartheta )^{d-2} \right]^{-1}
 \; \; \; ,
 \label{eq:gammaddef}
 \end{equation}
and can be identified with the ratio
of the surfaces of the unit spheres in $d-1$ and $d$ dimensions,
 \begin{equation}
 \gamma_{d} = \frac{ \Omega_{d-1}}{\Omega_{d} } =
 \frac{ \Gamma ( \frac{d}{2} )}{\sqrt{\pi} \Gamma ( \frac{ d-1 }{2} ) }
 \label{eq:gammad2}
 \; \; \; .
 \end{equation}
In particular,
 \begin{equation}
 \gamma_{1} = 0 \; \; \; , \; \; \; 
 \gamma_{2} = \frac{1}{\pi} \; \; \; , \; \; \; 
 \gamma_{3} = \frac{1}{2}
 \label{eq:gammapart}
 \; \; \; .
 \end{equation}
For $z=  \I y $ and real $y$ the function $g_{d} ( \I y )$ is an even
and positive function of $y$, and is in $d=1,2,3$ explicitly given by
 \begin{eqnarray}
 {g}_{1} ( \I y ) & = & 1- \frac{y^2}{ 1+ y^2} 
 = \frac{ 1}{  1 + y^2}
 \; \; \; ,
 \label{eq:g1y}
 \\
 {g}_{2} ( \I y ) & = & 1- \frac{|y|}{\sqrt{1 + y^2 }} 
 \; \; \; ,
 \label{eq:g2y}
 \\
 {g}_{3} ( \I y ) & = & 1- |y| \arctan \left( \frac{1}{|y|}  \right)
 \label{eq:g3y}
 \; \; \; .
 \end{eqnarray}
On the real axis we have
 \begin{eqnarray}
 {g}_{1} ( x+ \I 0^{+} ) & = & \frac{1}{  1 - ( x+ \I 0^{+} )^2} 
 \; \; \; ,
 \label{eq:g1x}
 \\
 {g}_{2} ( x + \I 0^{+} ) & = & 1- \frac{x}{\sqrt{(x + \I 0^{+} )^2 - 1}} 
 \; \; \; ,
 \label{eq:g2x}
 \\
 {g}_{3} ( x + \I 0^{+} ) & = & 1-  \frac{x}{2} \ln 
 \left( \frac{ x + \I 0^{+} + 1}{x + \I 0^{+} - 1}  \right)
 \label{eq:g3x}
 \; \; \; .
 \end{eqnarray}
For $|x| < 1$ the function
${g}_{d} ( x + \I 0^{+} )$ has in $d > 1$ a finite imaginary part. 
In the expression for the RPA dynamic structure factor discussed below this imaginary part 
describes the decay of density fluctuations into particle-hole
excitations, i.e. Landau damping \cite{Pines89}\index{Landau damping}.
From Eq.\ref{eq:gddef} it is easy to show that 
 \begin{equation}
 {\rm Im} {g}_{d} ( x + \I 0^{+} ) =  \pi x \left< \delta ( 
 \hat{\vec{q}} \cdot \hat{\vec{k}} - x ) \right>_{  \hat{\vec{k}}  }
 \label{eq:Imgddef}
 \; \; \; ,
 \end{equation}
so that
 \begin{equation}
 {\rm Im} {g}_{d} ( x + \I 0^{+} ) =  \pi \gamma_{d} x 
 \; \; \; ,
 \; \; \; 
 \mbox{for
 $ | x | \ll  1 $}
 \label{eq:Imgddefsmall}
 \; \; \; .
 \end{equation}
Keeping in mind that 
$g_{d} ( 0 ) = 1$,
this implies on the imaginary axis
 \begin{equation}
 {g}_{d} \left(  \I y \right)
 =
 1 -  \pi \gamma_{d} |y| 
 \; \; \; ,
 \; \; \; 
 \mbox{for
 $ | y | \ll  1 $}
 \; \; \; .
 \label{eq:Landaudamp}
 \end{equation}
For large $|z|$ we have in any dimension
  \begin{equation}
  g_{d} ( z) \sim - \frac{1}{d z^2} \; \; \; , \; \; \; \mbox{for $|z| \gg 1$}
  \; \; \; .
  \label{eq:gdlimlarge}
  \end{equation}
Finally, on the real axis we have in the vicinity of unity
to leading order in $\delta  = x - 1 > 0 $
 \begin{equation}
 g_{d} ( 1 + \delta )  
 \sim  \left\{
 \begin{array}{lc}
  g_{d} ( 1) <  0 & 
    \mbox{for $ d > 3 $} \\
 - \frac{1}{2} \ln ( 1/ {\delta  } ) &
     \mbox{for $ d = 3 $} \\
 - c_{d} /  \delta^{ \frac{3-d}{2} }  &
   \mbox{for $ d< 3 $}
 \end{array}
 \right.
 \; \; \; ,
 \label{eq:gdcloseone}
 \end{equation}
where $c_{d}$ is a positive numerical constant. In particular, 
 $c_{1} = \frac{1}{2}$ and $c_{2} = \frac{1}{ \sqrt{2}}$.

\section[The dynamic structure factor for spherical Fermi surfaces]
{The dynamic structure factor \mbox{\hspace{40mm}}
for spherical Fermi surfaces
\index{dynamic structure factor!spherical Fermi surface}}
\label{sec:dynstruc}

{\it{Within the RPA the dynamic structure 
factor\index{random-phase approximation!dynamic structure 
factor}\index{dynamic structure factor!within RPA}
consists of two contributions: The first one is a featureless
function and describes the decay of density fluctuations into 
particle-hole pairs, i.e. Landau damping; \index{Landau damping}
the second one is a 
$\delta$-function peak due to the collective plasmon mode.}}

\vspace{7mm}

\noindent
For simplicity we shall assume 
in the rest of this chapter that
the bare interaction is frequency-independent, i.e.
$f_{q} = f_{\vec{q}}$.
Introducing the dimensionless interaction 
 \begin{equation}
 F_{\vec{q}} = \nu f_{\vec{q}} 
 \label{eq:Fqdef}
 \; \; \; ,
 \end{equation}
we obtain from Eqs.\ref{eq:PiRPA} and \ref{eq:Piglobalspherical} 
for the RPA density-density correlation function 
in the long-wavelength limit
 \begin{equation}
 \Pi_{\rm RPA} ( q ) = \nu 
 \frac{g_{d} ( \frac{ \I \omega_{m}}{  v_{\rm F} | {\vec{q}} |  }) } 
 { 1 + F_{\vec{q}} 
 g_{d} ( \frac{ \I \omega_{m}}{  v_{\rm F} | {\vec{q}} | })  }
 \; \; \; .
 \label{eq:RPAdensres}
 \end{equation}
According to Eq.\ref{eq:SPi} the RPA dynamic structure factor
is then given by
 \begin{equation}
 S_{\rm RPA} ( {\vec{q}} , \omega ) = \frac{ \nu}{\pi} 
 {\rm Im}  \left\{ 
 \frac{g_{d} ( \frac{ \omega } {   v_{\rm F} | {\vec{q}} | } + \I 0^{+} ) } 
 { 1 + F_{\vec{q}} 
 g_{d} ( \frac{ \omega}{   v_{\rm F} | {\vec{q}}| } + \I 0^{+})  }
 \right\}
 \label{eq:Srpadef}
 \; \; \; .
 \end{equation}
From the properties of the function $g_{d} ( z)$ discussed above it is clear that
there exist two separate contributions to the imaginary part
in Eq.\ref{eq:Srpadef},
 \begin{equation}
 S_{\rm RPA} ( {\vec{q}} , \omega ) =  
 S_{\rm RPA}^{\rm sp} ( {\vec{q}} , \omega ) +
 S_{\rm RPA}^{\rm col} ( {\vec{q}} , \omega )  
 \; \; \; .
 \label{eq:Srpadecomp}
 \end{equation}
The first term
 $S_{\rm RPA}^{\rm sp} ( {\vec{q}} , \omega )  $ 
describes the creation and annihilation of a single particle-hole pair \cite{Pines89}.
This process, which is called {\it{Landau damping}}\index{Landau damping},
is only possible in $d > 1$
and for energies $ 0 < \omega  \leq   v_{\rm F} | {\vec{q}} |$. 
Mathematically Landau damping is due to the
finite imaginary part
of $g_{d} ( x + \I 0^{+} )$ for $x < 1$. Thus
 \begin{equation}
 S_{\rm RPA}^{\rm sp} ( {\vec{q}} , \omega ) = 
 \Theta \left( 1- \frac{\omega}{  v_{\rm F} | {\vec{q}} | } \right) 
 \frac{ \nu}{\pi} {\rm Im} \left\{ 
 \frac{g_{d} ( \frac{ \omega } {   v_{\rm F} | {\vec{q}} | } + \I 0^{+} ) } 
 { 1 + F_{\vec{q}} 
 g_{d} ( \frac{ \omega}{   v_{\rm F} | {\vec{q}}| } + \I 0^{+})  }
 \right\}
 \label{eq:Ssp}
 \; \; \; .
 \end{equation}
The second term
 $S_{\rm RPA}^{\rm col} ( {\vec{q}} , \omega ) $ 
arises from the poles of Eq.\ref{eq:RPAdensres}, which
define the dispersion relation $\omega_{\vec{q}}$ of the
collective plasmon mode\index{plasmon},
 \begin{equation}
 1 + F_{\vec{q}} g_{d} \left( \frac{ \omega_{\vec{q}}  }{  v_{\rm F} | {\vec{q}} | } \right) = 0
 \; \; \; .
 \label{eq:zerosounddisp}
 \end{equation}
The formal solution of Eq.\ref{eq:zerosounddisp} is
 \begin{equation}
 \frac{ \omega_{\vec{q}}  }{  v_{\rm F} | {\vec{q}} | } 
 = g_{d}^{-1} \left( - \frac{1}{F_{\vec{q}} } \right)
 \label{eq:zerosoundsol}
 \; \; \; ,
 \end{equation}
where $g_{d}^{-1} (x )$ is the inverse of the function $g_{d} ( x )$,
i.e. $g_{d}^{-1} ( g_{d} ( x ) ) = x$.
Because of the simple form of $g_{1} ( x )$ and $g_{2} ( x )$,
the solution of Eq.\ref{eq:zerosoundsol} in $d=1$ and $d=2$ can be
calculated analytically,
 \begin{eqnarray}
 \frac{ \omega_{\vec{q}}}{  v_{\rm F} | {\vec{q}} | }
 & = & \sqrt{ 1 + F_{\vec{q}} } 
 \; \; \; , \; \; \; \mbox{ for $d=1$} \; \; \;  ,
 \label{eq:zero1}
 \\
 \frac{ \omega_{\vec{q}}}{  v_{\rm F} | {\vec{q}} | }
 & = & \sqrt{ 1 + \frac{ F_{\vec{q}}^2}{1 + 2 F_{\vec{q}} } } =
 \frac{ | 1 + F_{\vec{q}} | }{ \sqrt{ 1 + 2 F_{\vec{q}} }}
 \; \; \; , \; \; \; \mbox{for $d=2$}
 \; \; \; .
 \label{eq:zero2}
 \end{eqnarray}
Note that $\omega_{\vec{q}}$ is real, so that the
plasmon mode is not damped. 
It is easy to see that for repulsive interactions
in arbitrary dimensions the plasmon mode is not damped within the RPA \cite{Pines89},
so that it gives rise to a
$\delta$-function contribution to the RPA dynamic structure factor,
 \begin{equation}
 S_{\rm RPA}^{\rm col} ( {\vec{q}} , \omega )   = Z_{\vec{q}} \delta ( \omega - \omega_{\vec{q}} )
 \label{eq:Scoldelta}
 \; \; \; ,
 \end{equation}
with
 \begin{equation}
 Z_{\vec{q}} = 
 \frac{1}{f_{\vec{q}}^2
 \left. \frac{ \partial}{\partial z} \Pi_{0} ( {\vec{q}} , z )
 \right|_{ z = \omega_{\vec{q}} }}
 =
 \frac{ \nu}{ F_{\vec{q}}^2} \frac{  v_{\rm F} | {\vec{q}} | }
 {g_{d}^{\prime} ( \frac{\omega_{\vec{q}}}{  v_{\rm F} | {\vec{q}} | } ) }
 \; \; \; ,
 \label{eq:resdef}
 \end{equation}
where $g_{d}^{\prime} ( z)$ is the derivative of the function $g_{d} ( z )$.
Because the dispersion relation of the collective mode satisfies
Eq.\ref{eq:zerosoundsol},
 $g_{d}^{\prime} ( \frac{\omega_{\vec{q}}}{  v_{\rm F} | {\vec{q}} | } )$
can be considered as function of $F_{\vec{q}}$. 
 We conclude that $Z_{\vec{q}}$ is of the form
 \begin{equation}
 Z_{\vec{q}} =  \nu  v_{\rm F} | {\vec{q}} | Z_{d} ( F_{\vec{q}} )
 \; \; \; ,
 \label{eq:resman}
 \end{equation}
where the function $Z_{d} ( F )$ is given by
 \begin{equation}
 Z_{d} ( F ) = \frac{1}{F^2 g_{d}^{\prime} ( g_{d}^{-1} ( - \frac{1}{F} ) ) }
 \label{eq:Zddef}
 \; \; \; .
 \end{equation}
In $d=1,2,3$ we have explicitly
 \begin{eqnarray} 
 Z_{1} (F)
 & = & \frac{ 1}{2 \sqrt{ 1 + F } }
 \; \; \; ,
 \label{eq:Z1}
 \\
 Z_{2} (F)
 & = & \frac{ F }{ (  1 + 2 F )^{\frac{3}{2}} }
 \; \; \; ,
 \label{eq:Z2}
 \\
 Z_{3} (F)
 & = & 
 \frac{ g_{3}^{-1} (- \frac{1}{ F} ) }{ F^2}
 \left[
 \frac{ 1}{ [ g_{3}^{-1} ( - \frac{1}{F} )]^2 - 1} -  \frac{1}{F}  \right]^{-1}
 \label{eq:Z3}
 \; \; \; . 
 \end{eqnarray} 

The strong and weak coupling behavior can be obtained analytically 
in any dimension. The collective mode for large
$F_{\vec{q}}$ is determined by the asymptotic behavior of $g_{d} ( x )$ for
large $x$. From Eq.\ref{eq:gdlimlarge} it is easy to show that to leading order
 \begin{equation}
 g_{d}^{-1} ( - \frac{1}{F} )  \sim  \sqrt{ \frac{F}{d} } 
 \; \; \; , \; \; \;  F \gg 1 
 \; \; \; ,
 \label{eq:gdinvlarge}
 \end{equation}
and that
 \begin{equation}
 g_{d}^{\prime} ( x )  \sim  \frac{ 2}{d x^3}
 \; \; \; , \; \; \;  x \gg 1 
 \; \; \; .
 \label{eq:gdprimelarge}
 \end{equation}
Then it is easy to see that for $F \gg 1$
 \begin{equation}
 g_{d}^{\prime} ( g_{d}^{-1} ( - \frac{1}{F} ) )  \sim   \frac{ 2 \sqrt{d}}{F^{{3}/{2} }}
 \label{eq:gdprimeinvF}
 \; \; \; .
 \end{equation}
It follows that the leading behavior at strong coupling is 
 \begin{eqnarray}
 \omega_{\vec{q}} & \sim & \frac{ v_{\rm F} | {\vec{q}} | }{\sqrt{d}} \sqrt{ F_{\vec{q}} }
 \; \; \; , \; \; \; F_{\vec{q}} \gg 1
 \label{eq:omegastrong}
 \\
 Z_{d} ( F_{\vec{q}} ) =
 \frac{Z_{\vec{q}}}{\nu v_{\rm F} | {\vec{q}} | }
 & \sim   &  \frac{ 1 }{2 \sqrt{d} \sqrt{F_{\vec{q}}} }
 \; \; \; , \; \; \; F_{\vec{q}} \gg 1
 \; \; \; .
 \label{eq:Zstrong}
 \end{eqnarray}

The dispersion of the plasmon mode at weak coupling is determined by the
behavior of the function $g_{d} ( 1 + \delta )$ for small positive
$\delta$, which is given in Eq.\ref{eq:gdcloseone}.
Because $g_{d} ( 1 )$  is finite for $d > 3$, the
collective mode equation \ref{eq:zerosounddisp} does not
have any solution for $F_{\vec{q}} < 1 / | g_{d} ( 1 ) |$ in
$d > 3$.
In this case there is at weak coupling no collective mode contribution
to the dynamic structure factor.
For $d \leq 3$ we
find to leading order
for small $F$
 \begin{equation}
 g_{d}^{-1} ( - \frac{1}{F} ) \sim
 \left\{ \begin{array}{lc}
 1 + e^{-2/F} & 
    \mbox{for $ d = 3 $} \\
 1 + ( c_{d} F )^{ \frac{2}{3-d} } &
    \mbox{for $ d < 3 $}
  \end{array}
  \right.
  \label{label:eq:gdinvweak}
  \; \; \; ,
  \end{equation}
 \begin{equation}
 g_{d}^{\prime} ( g_{d}^{-1} ( - \frac{1}{F} ) ) \sim
 \left\{ \begin{array}{lc}
 \frac{1}{2} \E^{2/F} & 
    \mbox{for $ d = 3 $} \\
 \frac{3-d}{2} c_{d} ( c_{d} F )^{ - \frac{5-d}{3-d} } &
   \mbox{for $ d < 3 $}
  \end{array}
  \right.
  \label{eq:gdprimeinvweak}
  \; \; \; ,
  \end{equation}
so that at weak coupling the collective mode and the residue are
 \begin{equation}
 \frac{ \omega_{\vec{q}}}{ v_{\rm F} | {\vec{q}} | }  \sim
 \left\{ \begin{array}{lc}
 1 + \E^{-2/F_{\vec{q}} } & 
    \mbox{for $ d = 3 $} \\
 1 + ( c_{d} F_{\vec{q}} )^{ \frac{2}{3-d} } &
   \mbox{for $ d < 3 $}
  \end{array}
  \right.
  \; \; \; , \; \; \; F_{\vec{q}} \ll 1
  \label{label:eq:collweak}
  \; \; \; ,
  \end{equation}
 \begin{equation}
 Z_{d} ( F_{\vec{q}} ) = \frac{Z_{{\vec{q}}}}{ \nu v_{\rm F} | {\vec{q}} | }  \sim
 \left\{ \begin{array}{lc}
 \frac{2  }{F_{\vec{q}}^2} e^{-2/F_{\vec{q}}} & 
    \mbox{for $ d = 3 $} \\
 \frac{2  }{3-d} c_{d} ( c_{d} F_{\vec{q}} )^{  \frac{d-1}{3-d} } &
   \mbox{for $ d < 3 $}
  \end{array}
  \right.
  \; \; \; , \; \; \; F_{\vec{q}} \ll 1
  \label{label:eq:Zqweak}
  \; \; \; .
  \end{equation}

\section{Collective modes for singular interactions}

{\it{Here we explicitly calculate the dispersion relation of the
plasmon mode and the associated residue for singular interactions that diverge
in $d$ dimensions as $ | {\vec{q}} |^{- \eta }$ for $ {\vec{q}} \rightarrow 0$
(see Chap.~\secref{chap:a7sing}).
We start with the physically most important Coulomb interaction
and then discuss the general case.}}

\subsection{The Coulomb interaction in $ 1 \leq d \leq 3$ }
\label{subsubsec:Cb}

\noindent
The bare Coulomb potential\index{Coulomb interaction} between two charges separated 
by a distance
${\vec{r}}$ is $e^{2} / | {\vec{r}} |$ in any dimension.
For $ 1 < d \leq 3$
the Fourier transformation to momentum space is easily calculated using $d$-dimensional spherical
coordinates (see Eq.\ref{eq:angavsimplify}), with the result
 \begin{equation}
 f_{\vec{q}}   
 = \int \D \vec{r} \E^{\I \vec{k} \cdot \vec{r} } \frac{e^2}{ | \vec{r} | }
 =
 \frac{ \Gamma ( d-1) \Omega_{d} e^{2} }{ | {\vec{q}} |^{d-1} }
 \; \; \; , \; \; \; d > 1
 \label{eq:Qpcb}
 \; \; \; .
 \end{equation}
In $d=1$ the integral in Eq.\ref{eq:Qpcb} is logarithmically divergent, and must be regularized.
Introducing a short-distance cutoff $a$, one obtains
 \begin{equation}
 f_{\vec{q}} =  2 e^{2} \ln \left( \frac{1}{ | {\vec{q}}| a } \right)
 \; \; \; , \; \; \; d = 1
 \label{eq:Qpcb1}
 \; \; \; .
 \end{equation}
In dimensions $d > 1$ the 
{\it{Thomas-Fermi screening wave-vector}}\index{Thomas-Fermi wave-vector} $\kappa$ is defined by
 \begin{equation}
 \kappa^{d-1} =  \nu \Gamma ( d - 1 ) \Omega_{d} e^2
 \; \; \; ,
 \label{eq:kappaTFdef}
 \end{equation}
to that with the help of Eq.\ref{eq:nurelation} we obtain
 \begin{equation}
 \left( \frac{ \kappa}{k_{\rm F} } \right)^{d-1} = \frac{ \Gamma ( d-1 ) \Omega^2_{d}}{ ( 2 \pi)^d}
 \frac{ e^2}{v_{\rm F}}
 \; \; \; .
 \label{eq:kappaTFrel}
 \end{equation}
Thus, the requirement that the Thomas-Fermi screening wave-vector should be
small compared with $k_{\rm F}$ 
is equivalent with 
$ {e^{2}} / {v_{\rm F}} \ll 1$.
Note that 
 $e^{2} / {v_{\rm F}}  = 
  ( k_{\rm F} a_{\rm B} )^{-1} =
 \alpha c / {v_{\rm F}} $ 
where
$a_{\rm B} = 1/ ( m e^2)$ is the Bohr radius,
and
 $\alpha = e^2 / c \approx \frac{1 }{ 137}$ 
is the fine structure constant
\index{fine structure constant}\footnote{ 
Recall that we have set $\hbar = 1$.
In conventional Gaussian units we have
$a_{\rm B} = \hbar^2 / (m e^2)$ and
$\alpha = e^2 / (\hbar c)$.}. 
Up to a numerical factor of the order of unity,
the parameter $( \kappa / k_{\rm F} )^{d-1}$
can be identified 
with the usual dimensionless Wigner-Seitz radius 
$r_{\rm s}$\index{Wigner-Seitz radius $r_{\rm s}$}, 
which is a measure for the density
of the electron gas. In $d$ dimensions $r_{\rm s}$ is defined by
 $V/N = V_{d} (a_{\rm B} r_{\rm s} )^{d}$
where 
 \begin{equation}
 V_{d} = \frac{\Omega_{d} }{ {d}} = 
 \frac{ 2 \pi^{\frac{d}{2}}  }{  d \Gamma ( \frac{d}{2} )}
 \end{equation}
is the volume of the $d$-dimensional unit sphere.
Using the fact that 
the density of spinless fermions
in $d$ dimensions 
can be written as $N / V = V_d k_{\rm F}^d / ( 2 \pi )^d$, we 
obtain in $d$ dimensions
 \begin{equation}
 r_{\rm s} =  \left( \frac{  1}{V_d} \right)^{\frac{2}{d}}  
 \frac{ 2 \pi  e^2}{  v_{\rm F} }
 \label{eq:rsddim}
 \; \; \; .
 \end{equation}
Combining this with Eq.\ref{eq:kappaTFrel}, we conclude that
 \begin{equation}
 \left( \frac{ \kappa}{k_{\rm F} } \right)^{d-1} = \frac{  \Gamma ( d-1 ) \Omega_d^2 }{ ( 2 \pi )^{d+1}}
  V_{d}^{\frac{2}{d}} r_{\rm s}
  \label{eq:rsfinal}
  \; \; \; .
  \end{equation}
In particular, in $d=2$ we have 
 $\kappa / k_{\rm F}  =   e^2 / v_{\rm F} =  r_{\rm s} /2$,
and in three dimensions
 $( \kappa / k_{\rm F} )^2  =  2 e^2 / ( \pi  v_{\rm F} )
 \approx 0.263 r_{\rm s}$.

With the above definitions,
the dimensionless Coulomb interaction $F_{\vec{q}} = \nu f_{\vec{q}}$ 
can be written as
 \begin{equation}
 F_{\vec{q}} = \left( \frac{ \kappa }{ | {\vec{q}} | } \right)^{d-1}
 \label{eq:Fqdef2}
 \; \; \; .
 \end{equation}
Because $F_{\vec{q}} \gg 1$ for $ | {\vec{q}} | \ll  \kappa$, the
Thomas-Fermi screening wave-vector $\kappa$ defines the boundary
between the long and short wavelength regimes, 
and can therefore be identified with the cutoff $q_{\rm c}$ introduced in
Chap.~\secref{subsec:proper}.
It follows that for the Coulomb problem
the bosonization approach is most accurate at high densities,
where $r_{\rm s} \ll 1$ and hence $\kappa \ll k_{\rm F}$.
We would like to emphasize that bosonization is {\it{not an expansion 
in powers of}} $r_{\rm s}$ \cite{GellMann57};
the condition $r_{\rm s} \ll 1$ is necessary
to make the higher-dimensional
bosonization approach consistent. 

For the Coulomb potential the dimensionless RPA interaction can be written as
 \begin{equation}
 F_{q}^{\rm RPA}  \equiv \nu f_{q}^{\rm RPA} = \frac{1}{ ( \frac{| {\vec{q}}| }{\kappa} )^{d-1} + 
 g_{d}  ( \frac{\I \omega_{m}}{v_{\rm F} | {\vec{q}} | } )}
 \label{eq:Frpacb}
 \; \; \; .
 \end{equation}
Because $F_{\vec{q}}$
diverges as ${\vec{q}} \rightarrow 0$, 
the behavior of the collective mode for $| {\vec{q}} | \ll \kappa$ is
determined by the strong coupling limit $F_{\vec{q}} \gg 1$, which 
is given in Eqs.\ref{eq:omegastrong} and \ref{eq:Zstrong}.
Hence, for the Coulomb interaction in $d$ dimensions
the collective plasmon mode and its weight are 
at long wavelengths given by\index{Coulomb interaction!plasmon mode}
 \begin{eqnarray}
 \omega_{\vec{q}} & = &
 \frac{  v_{\rm F} \kappa}{\sqrt{d}} 
 \left( \frac{ | {\vec{q}} | }{\kappa} \right)^{\frac{3-d}{2}}
 \; \; \; ,
 \label{eq:colmodcbres}
 \\
 Z_{\vec{q}} & = & 
 \frac{  \nu v_{\rm F} \kappa}{2 \sqrt{d}} 
 \left( \frac{ | {\vec{q}} | }{\kappa} \right)^{\frac{d+1}{2}}
 \label{eq:Zcbres}
 \; \; \; .
 \end{eqnarray}
In three dimensions this yields
 \begin{eqnarray}
 \omega_{\vec{q}} & = &
 \frac{  v_{\rm F} \kappa}{\sqrt{3}} \equiv \omega_{\rm pl}
 \; \; \; , \; \; \; d=3
 \; \; \; ,
 \label{eq:colmodcbres3}
 \\
 Z_{\vec{q}} & = & 
 \frac{  \nu }{2} \omega_{\rm pl}  
 \left(  \frac{  {\vec{q}}  }{\kappa}  \right)^2
 \; \; \; , \; \; \; d=3
 \label{eq:Zcbres3}
 \; \; \; .
 \end{eqnarray}
Thus, in $d=3$ the plasmon mode approaches at long wavelengths a 
constant value $\omega_{\rm pl}$,
the {\it{plasma frequency}}\index{plasma frequency}.

\subsection{General singular interactions} 
\label{subsec:Generalsing}

Finally, let us consider general singular
interactions of the form \ref{eq:fgeneric}.
Defining the screening wave-vector
 \begin{equation}
  \kappa = ( g_{\rm c}^2 \nu )^{1/ \eta}
  \; \; \; ,
  \label{eq:kappascdef2}
  \end{equation}
we see that the dimensionless interaction corresponding to
Eq.\ref{eq:fgeneric} is
 \begin{equation}
 F_{\vec{q}} \equiv \nu f_{\vec{q}}=  \left( \frac{ \kappa}{|{\vec{q}} |} 
 \right)^{\eta} \E^{- | {\vec{q}} | / q_{\rm c} }
 \label{eq:Fgeneric}
 \; \; \; .
 \end{equation}
The dimensionless RPA interaction can be written as
 \begin{equation}
 F_{q}^{\rm RPA}  = \frac{1}{ ( \frac{| {\vec{q}}| }{\kappa} )^{\eta} \E^{|{\vec{q}}| / q_{c} }+ 
 g_{d}  ( \frac{\I \omega_{m}}{v_{\rm F} | {\vec{q}} | } )}
 \label{eq:Frpagen}
 \; \; \; .
 \end{equation}
Assuming that $\kappa \ll q_{\rm c}$, we see that 
$F_{\vec{q}} \gg 1$ for $|{\vec{q}} | \ll \kappa$.
In this regime the collective mode and the associated residue
are easily obtained from Eqs.\ref{eq:omegastrong} and \ref{eq:Zstrong},
 \begin{eqnarray}
 \omega_{\vec{q}} & = & \frac{ v_{\rm F} \kappa}{\sqrt{d}} \left( \frac{ | {\vec{q}} |}{\kappa} 
 \right)^{ 1 - \eta / 2 }
 \; \; \; ,
 \label{eq:omegaeta}
 \\
 Z_{\vec{q}} & = & \frac{ \nu v_{\rm F} \kappa}{2 \sqrt{d}} \left( \frac{ | {\vec{q}} |}{\kappa} 
 \right)^{ 1 + \eta / 2 }
 \label{eq:Zeta}
 \; \; \; .
 \end{eqnarray}
For $ \eta = d-1$
these expressions reduce to
Eqs.\ref{eq:colmodcbres} and \ref{eq:Zcbres}.

\section{Collective modes for finite patch number}
\label{sec:finitepatch}

{\it{We discuss the polarization and the dynamic structure factor
for Fermi surfaces that consist of
a finite number $M$ of flat patches. 
The calculations in this section are valid for arbitrary 
Fermi surface geometries,
i.e. we do not assume that for $M \rightarrow \infty$
the Fermi surface approaches a sphere.}}

\vspace{7mm}

\noindent
A crucial step in
higher-dimensional bosonization with linearized energy dispersion
is the replacement of
an arbitrarily shaped Fermi surface by a finite
number of flat patches $P^{\alpha}_{\Lambda}$.
Let us assume that the number of patches is {\it{even}}, 
and that for each patch $P^\alpha_{\Lambda}$
with local Fermi velocity
${\vec{v}}^{\alpha}$ and density of states $\nu^{\alpha}$ there exists
an opposite patch $P^{\bar{\alpha}}_{\Lambda}$ with
${\vec{v}}^{\bar{\alpha}} = - {\vec{v}}^{\alpha}$ 
and $\nu^{\bar{\alpha}} = \nu^{\alpha}$. 
This guarantees that the
inversion symmetry of the Fermi surface is not
artificially broken by the patching 
construction (see the first footnote in Chap.~\secref{chap:a7sing}).
For simplicity let us also assume that all patch densities of states
$\nu^{\alpha}$ are identical, so that 
$\nu^{\alpha} = \nu / M$, where $\nu = \sum_{\alpha = 1}^{M} \nu^{\alpha}$
is the global density of states 
(see Eqs.\ref{eq:nualphadef} and  \ref{eq:nudef}).
Then 
the non-interacting polarization\index{polarization!finite patch number} 
$\Pi_{0} ( {\vec{q}} , z )$ is at long wave-lengths
given by (see Eqs.\ref{eq:Pilong} and \ref{eq:pi0tot})
 \begin{eqnarray}
 \Pi_{0} ( {\vec{q}} , z ) & = & \frac{ 2 \nu}{M} \sum_{\alpha = 1}^{M/2}
 \frac{ ( {\vec{v}}^{\alpha} \cdot {\vec{q}} )^2 }{
 ( {\vec{v}}^{\alpha} \cdot {\vec{q}})^2 - z^2 }
 = \nu \frac{ {P}_{M-2} ( {\vec{q}} , z )}{Q_{M} ( {\vec{q}}, z ) }
 \label{eq:Pi0Meven}
 \; \; \; ,
 \\
 Q_{M} ( {\vec{q}} , z ) & = & \prod_{\alpha = 1}^{M/2}
 ( z^2 - ( {\vec{v}}^{\alpha} \cdot {\vec{q}} )^2 )
 \label{eq:QMdefeven}
 \; \; \; ,
 \\
 P_{M-2} ( {\vec{q}} , z ) & = & \frac{ 2 }{M} \sum_{\alpha = 1}^{M/2}
 ( {\vec{v}}^{\alpha} \cdot {\vec{q}} )^2
 \left[
 \prod_{ \stackrel{ \scriptstyle \alpha^{\prime} = 1}{  \alpha^{\prime} \neq \alpha}}^{M/2}
 ( ({\vec{v}}^{\alpha^{\prime}} \cdot {\vec{q}})^2 - z^2 )
 \right]
 \label{eq:PMdefeven}
 \; \; \; ,
 \end{eqnarray}
where it is understood that the sums are over all patches with
${\vec{v}}^{\alpha} \cdot {\vec{q}} \geq 0$,
and in the special case $M=2$ the product in Eq.\ref{eq:PMdefeven}
should be replaced by unity.
The RPA polarization can then be written as
 \begin{equation}
 \Pi_{\rm RPA} ( {\vec{q}} ,  z ) 
 = \nu \frac{ P_{M-2} ( {\vec{q}} , z )}
 { Q_{M} ( {\vec{q}} , z ) +
 F_{\vec{q}} P_{M-2} (  {\vec{q}} , z )}
 \; \; \; ,
 \label{eq:PiRPArat}
 \end{equation}
where as usual $F_{\vec{q}} = \nu f_{\vec{q}}$.
Thus, the RPA condition 
for the collective density modes,
 \begin{equation}
 1 + f_{\vec{q}} \Pi_0 ( {\vec{q}}, z ) = 0
 \label{eq:plasmondef2}
 \; \; \; ,
 \end{equation}
is equivalent with
 \begin{equation}
  Q_{M} ( {\vec{q}} , z ) +
 F_{\vec{q}} P_{M-2} (  {\vec{q}} , z ) = 0
 \label{eq:PolynomM}
 \; \; \; .
 \end{equation}
Because the left-hand side of this equation is a polynomial in 
$z^2$ with degree $M/2$, for a given ${\vec{q}}$ 
we obtain $M/2$ roots 
in the complex $z^2$-plane.
The locations of the roots is easily obtained graphically by 
plotting the right-hand side of Eq.\ref{eq:Pi0Meven}
as function of real $z^2$ and looking for the intersections
with $- 1 / f_{\vec{q}}$. For generic $\vec{q}$ all
$( {\vec{v}}^{\alpha} \cdot \vec{q})^2$ are different and positive,
and we can order the energies such that
 \begin{equation}
 0 < ( \vec{v}^{\alpha_1} \cdot \vec{q} )^2 <
 ( \vec{v}^{\alpha_2} \cdot \vec{q} )^2 < \ldots
 <
 ( \vec{v}^{\alpha_{{M}/{2}}} \cdot \vec{q} )^2 
 \; \; \; .
 \label{eq:rootsorder}
 \end{equation}
A repulsive interaction leads then $M / 2$ to real roots 
$( \omega_{\vec{q}}^{2} )^{ ( \alpha )}$, $\alpha = 1 , \ldots , M/2$,
  of the polynomial \ref{eq:PolynomM} 
(considered as function of $z^2$), which are
located between the unperturbed poles,
 \begin{eqnarray}
 \hspace{-4mm}
 0 & < & ( \vec{v}^{\alpha_1} \cdot \vec{q} )^2 <
 ( \omega_{\vec{q}}^{2} )^{(1)}  <
 ( \vec{v}^{\alpha_2} \cdot \vec{q} )^2  
 <
 ( \omega_{\vec{q}}^{2} )^{(2)}  <
 \nonumber
 \\
 & & 
 \hspace{24mm}
 \ldots
 <
 ( \vec{v}^{\alpha_{M/2}} \cdot \vec{q} )^2 
 <
 ( \omega_{\vec{q}}^{2} )^{(M/2)}  
 \; \; \; .
 \label{eq:rootsmode}
 \end{eqnarray}
Because the roots are on the positive real axis in the complex $z^2$-plane,
they represent undamped collective modes,
which
give rise to $\delta$-function peaks in the
RPA dynamic structure factor. \index{dynamic structure factor!finite patch number}
Hence, for $\omega > 0 $ 
the dynamic structure factor has the following form\footnote{
The above simple proof that for finite $M$ and repulsive interactions
the RPA dynamic structure factor consists
only of $\delta$-function peaks can be found in the work \cite{Kopietz96chain}, and
is due to Kurt Sch\"{o}nhammer.}
 \begin{equation}
 S_{\rm RPA}  ( {\vec{q}} , \omega ) = \sum_{\alpha = 1}^{M/2}
 Z^{\alpha}_{\vec{q}} \delta ( \omega - \omega^{\alpha}_{\vec{q}} )
 \; \; \; ,
 \end{equation}
with the residues given by (see also Eq.\ref{eq:resdef})
 \begin{equation}
 Z_{\vec{q}}^{{\alpha}} = \frac{1}{f_{\vec{q}}^2
 \left. \frac{ \partial}{\partial z} \Pi_{0} ( {\vec{q}} , z )
 \right|_{ z = \omega_{\vec{q}}^{{\alpha}} }}
 \label{eq:residuemode}
 \; \; \;  .
 \end{equation}
In the limit $M \rightarrow \infty$
and (at least) for sufficiently strong coupling\footnote{
Recall that in Sect.~\secref{sec:dynstruc}
we have shown that for spherical Fermi surfaces in $d > 3$
the collective plasmon mode exists only for
sufficiently strong interactions.}
the mode $\omega^{{M/2}}_{\vec{q}}$
with the largest energy 
survives as a $\delta$-function peak, and can be identified with the
collective plasmon mode\index{plasmon} $\omega_{\vec{q}}$,
see Eqs.\ref{eq:zerosoundsol}--\ref{eq:Scoldelta}. 
All other modes represent a quasi-continuum in the sense
that they merge for $M \rightarrow \infty$ into the
particle-hole continuum described by
$S_{\rm RPA}^{\rm sp} ( \vec{q} , \omega )$ in Eq.\ref{eq:Ssp}.
For non-generic $\vec{q}$ such that
$\vec{v}^{\alpha_i} \cdot \vec{q} = 0$ 
for some $\alpha_i$ or $( \vec{v}^{\alpha_i} \cdot \vec{q})^2  = 
( \vec{v}^{\alpha_j} \cdot \vec{q})^2  $ for some
$\alpha_i \neq \alpha_j$, the
number of distinct modes in the quasi-continuum  is reduced.

In the strong coupling limit it is easy to obtain an analytic
expression for the collective plasmon mode and the associated residue.
From Eq.\ref{eq:omegastrong} we expect that
for $F_{\vec{q}} \equiv \nu f_{\vec{q}} \gg 1$ there exists
one real solution $\omega_{\vec{q}}$ 
with $\omega_{\vec{q}}^2 = O ( F_{\vec{q}} )$.
For $z^2$ close to $\omega_{\vec{q}}^2$ we may therefore
expand $\Pi_{0} ( {\vec{q}} , z )$ in powers of $z^{-2}$.
The leading terms are
 \begin{eqnarray}
 \hspace{-6mm}
\Pi_{0} ( {\vec{q}} , z ) & = &
-  \nu \left[
\frac{ 2}{M} \sum_{\alpha = 1}^{M/2}
 ( {\vec{v}}^{\alpha} \cdot {\vec{q}} )^2 \right]
 z^{-2}
  - \nu 
\left[
\frac{2}{M} \sum_{\alpha = 1}^{M/2}
 ( {\vec{v}}^{\alpha} \cdot {\vec{q}} )^4 \right]
 z^{-4}
 + O ( z^{-6} )
 \;  .
 \nonumber
 \\
 \label{eq:Pi0zlargeexp}
 \end{eqnarray}
Substituting this approximation into Eq.\ref{eq:plasmondef2},
it is easy to show that the dispersion of the plasmon mode is
for large $F_{\vec{q}}$ given by
 \begin{equation}
 \omega_{\vec{q}}^2 =
 F_{\vec{q}} 
\frac{ 2}{M} \sum_{\alpha = 1}^{M/2}
 ( {\vec{v}}^{\alpha} \cdot {\vec{q}} )^2
 + \frac{
\frac{2}{M} \sum_{\alpha = 1}^{M/2}
 ( {\vec{v}}^{\alpha} \cdot {\vec{q}} )^4 }
 { 
\frac{ 2}{M} \sum_{\alpha = 1}^{M/2}
 ( {\vec{v}}^{\alpha} \cdot {\vec{q}} )^2 }
 + O ( F_{\vec{q}}^{-1} ) 
 \label{eq:plasmonpatchlarge}
 \; \; \; .
 \end{equation}
Using the fact that
for a spherical Fermi surface
 \begin{equation}
 \lim_{M \rightarrow \infty}
\frac{ 2}{M} \sum_{\alpha = 1}^{M/2}
 ( {\vec{v}}^{\alpha} \cdot {\vec{q}} )^2 
 = v_{\rm F}^2 {\vec{q}}^2 \langle ( \hat{\vec{v}}^{\alpha} \cdot \hat{\vec{q}} )^2 
 \rangle_{\hat{\vec{q}} } = \frac{ v_{\rm F}^2 {\vec{q}}^2}{d}
 \label{eq:Minflim}
 \; \; \; ,
 \end{equation}
the leading term in Eq.\ref{eq:plasmonpatchlarge}
reduces for $M \rightarrow \infty$
to Eq.\ref{eq:omegastrong}. 
For energies $z$ close to $\omega_{\vec{q}}$ we may write
 \begin{equation}
 \Pi_{\rm RPA} ( {\vec{q}} , z ) \approx - \frac{ Z_{\vec{q}} }{ z - \omega_{\vec{q}}  }
 \label{eq:PiRPAclose}
 \; \; \; ,
 \end{equation}
with 
 \begin{equation}
 Z_{\vec{q}} 
  \approx  \frac{\nu}{2 \sqrt{F}_{\vec{q}} }
 \left[ 
\frac{ 2}{M} \sum_{\alpha = 1}^{M/2}
 ( {\vec{v}}^{\alpha} \cdot {\vec{q}} )^2  \right]^{1/2}
 \label{eq:Zqclose}
 \; \; \; ,
 \end{equation}
where in the second line we have retained the leading term in the 
expansion for large large $F_{\vec{q}}$.
For $M \rightarrow \infty$  and spherical Fermi surfaces
we may use again Eq.\ref{eq:Minflim}  
and recover our previous result \ref{eq:Zstrong} for $Z_{\vec{q}}$.

%
%


\renewcommand{\thechapter}{\arabic{chapter}}

\chapter*{References}
\markboth{References}{References}
\addcontentsline{toc}{chapter}{References}

\renewcommand{\thechapter}{\Roman{chapter}}
\renewcommand{\thechapter}{A}


\newpage
\addcontentsline{toc}{chapter}{Index}

\printindex
\end{document}